\documentclass[a4paper,12pt,openany]{book}
\usepackage{fancyhdr}
\usepackage{ifthen}
\usepackage{amsmath}
\usepackage{amssymb}
\usepackage{bm}
\usepackage{graphicx}
\usepackage{epsfig}
\usepackage{rotating}
\usepackage{ogonek}

\fancypagestyle{plain}{%
 \fancyhf{}
 \fancyhead[LE]{\makebox[1cm][l]{\thepage} }
 \fancyhead[RE]{\nouppercase{\leftmark}}
 \fancyhead[RO]{\makebox[1cm][r]{\thepage} }
 \fancyhead[LO]{\nouppercase{\rightmark}}
}
\newcommand{\REM}[1]{\ifthenelse{1=1}{}{#1}}

\newcommand{\EQN}[1]{\begin{eqnarray}#1\end{eqnarray}}
\newcommand{\SEQN}[2]{\begin{subequations}#1\begin{eqnarray}#2\end{eqnarray}\end{subequations}}
\newcommand{\EQNa}[1]{\begin{equation}\begin{array}{rcl} #1 \end{array}\end{equation}}
\newcommand{\ENUM}[1]{\begin{enumerate}{#1}\end{enumerate}}
\newcommand{\ITEM}[1]{\begin{itemize}{#1}\end{itemize}}

\newcommand{\order}[1]{\ensuremath{\mathcal{O}\left({#1}\right)}}
\newcommand{\op}[1]{\widehat{#1}}
\newcommand{\dagop}[1]{\widehat{#1}{}^{\dagger}}
\newcommand{\ket}[1]{\left|{#1}\right\rangle}
\newcommand{\bra}[1]{\left\langle{#1}\right|}
\newcommand{\braket}[2]{\left.\left\langle{#1}\right|{#2}\right\rangle}

\newcommand{\average}[1]{\left\langle{{#1}}\right\rangle_{\rm stoch}}
\newcommand{\tr}[1]{{\rm Tr}\left[{#1}\right]}
\newcommand{\vari}[1]{{\rm var}\left[{#1}\right]}
\newcommand{\re}[1]{{\rm Re}\left\{{#1}\right\}}
\newcommand{\im}[1]{{\rm Im}\left\{{#1}\right\}}
\newcommand{\dada}[2]{\frac{\partial{#1}}{\partial{#2}}}
\newcommand{\dd}[2]{\frac{d{#1}}{d{#2}}}

\newcommand{\bo}[1]{\mathbf{#1}}
\newcommand{\wt}[1]{\widetilde{#1}}
\newcommand{\mc}[1]{\mathcal{#1}}
\renewcommand{\bar}{\overline}
\newcommand{\ul}[1]{\hspace{0.1em}\underline{#1}\hspace{0.1em}}

\newcommand{\etal}{{\it\ et al\,}}
\newcommand{\half}{\ensuremath{\text{$\scriptstyle\frac{1}{2}$}}}
\newcommand{\Half}{\ensuremath{\frac{1}{2}}}

\newcommand{\matri}[1]{\left[\begin{array}{c} #1 \end{array}\right]}
\renewcommand{\matrix}[1]{\left[\begin{array}{c@{\quad}c} #1 \end{array}\right]}
\newcommand{\matrixx}[1]{\left[\begin{array}{c@{\quad}c@{\quad}c} #1 \end{array}\right]}

\newcommand{\mm}[2]{{}^{#1}_{#2}}

							\let\footnotecopyof\footnote
\renewcommand{\footnote}[1]%
  {\footnotecopyof{#1}}

\newcommand{\hfilll}{\hspace*{\fill}}
\begin{document}
\ifx\href\undefined\else\hypersetup{linktocpage=true}\fi 
\pagestyle{empty}
\begin{titlepage}
\begin{center}{\scshape\Huge First-principles Quantum \\
Simulations of Many-mode \\
Open Interacting Bose Gases\\
Using Stochastic Gauge Methods\\}
\vfill
{\scshape A thesis submitted for\\
the degree of Doctor of Philosophy\\
at the University of Queensland in\\
June 2004}
\vfill
\fbox{\hspace{80mm}\rule{0mm}{60mm}}
\vfill
Piotr Pawe\l\ Deuar, BSc. (Hons)\\
\vfill
{School of Physical Sciences}
\vfill\mbox{}
\end{center}
\end{titlepage}

\mbox{}\clearpage

\mbox{}\vfill\begin{minipage}{12cm}
{\bf Statement of Originality}\\
Except where acknowledged in the customary manner, the material presented in this thesis is, to the best of my knowledge 
and belief, original and has not been submitted in whole or in part for a degree in any university.\\
\vspace{1.5cm}\\
\rule{5cm}{0.5pt}\\
Piotr P. Deuar
\vspace{1.5cm}\\
\rule{5cm}{0.5pt}\\
Peter D. Drummond 
\end{minipage}\vfill\mbox{}\vfill\clearpage

\mbox{}\vfill\begin{minipage}{12cm}
{\bf Statement of Contribution by Others}\\

The original concepts of the gauge P representation and of (drift) stochastic gauges were due to Peter Drummond, as was the suggestion to investigate boundary term removal and thermodynamics of uniform 1D gases. The concepts in Section~\ref{CH5Comparison} and Subsection~\ref{CH6FirstManymode}, are also due to Peter Drummond, but are included because they are important for a background understanding.

The XMDS program [http://www.xmds.org/] (at the time, in 2001, authored by Greg Collecutt and Peter Drummond) was used for the calculations of Chapter~\ref{CH6}, while the calculations in the rest of the thesis were made with programs written by me but loosely based on  that 2001 version of XMDS.

The exact Yang \& Yang solutions in Figure~\ref{FIGUREdenseper} were calculated with a program written by Karen Kheruntsyan.

\mbox{}\vspace{1.5cm}\\
\rule{5cm}{0.5pt}\\
Piotr P. Deuar
\vspace{1.5cm}\\
\rule{5cm}{0.5pt}\\
Peter D. Drummond 
\end{minipage}\vfill\mbox{}\vfill\clearpage

\chapter*{Acknowledgements}
\pagenumbering{roman}
\thispagestyle{empty}

Above all I thank Prof. Peter Drummond, my principal Ph.D. supervisor. 
You led by example, and showed me what true scientific work is all about --- along with the sheer glee of it. 
Our many physics discussions were always extremely illuminating and productive. Thank you for your patience (which I hope not to have stretched too thin) and positive attitude.

I also owe much to Drs. Bill Munro and Karen Kheruntsyan, my secondary supervisors in the earlier and later parts of my candidature, respectively. Your assistance through this whole time has been invaluable.
I would also like to thank Dr. Margaret Reid, who first introduced me to real scientific work in my Honours year. 

My thanks go also to my fellow Ph.D. students (mostly now Drs.) at the physics department, for the innumerable discussions and intellectually stimulating atmosphere. Especially Damian Pope, Timothy Vaughan, Joel Corney, and Greg Collecutt by virtue of the greater similarity of our research. It is much easier to write a thesis with a good example --- thank you Joel. I am also conscious that there are many others at the department who have assisted me in innumerable ways.  

From outside of U.Q., I thank especially Prof. Ryszard Horodecki from Gda\'{n}sk University both for your outstanding hospitality and the opportunity of scientific collaboration. I also thank Dr. Marek Trippenbach and Jan Chwede\'{n}czuk from Warsaw University for stimulating discussions during the final stages of my thesis. 

More personally, I thank my wife Maria, for her support and above all her extreme tolerance of the husband engrossed in physics, particularly while I was writing up. 

\clearpage

\chapter*{Publications by the Candidate Relevant to the Thesis but not forming part of it}
\thispagestyle{empty}
Some of the research reported in this thesis has been published in the following refereed publications:
\begin{enumerate}
\item P.~Deuar and P.~D. Drummond.
 Stochastic gauges in quantum dynamics for many-body simulations.
 {\em Computer Physics Communications} \textbf{ 142},  442--445 (Dec. 2001).
\item P.~Deuar and P.~D. Drummond.
\newblock Gauge {P}-representations for quantum-dynamical problems: Removal of
  boundary terms.
\newblock {\em Physical Review A} \textbf{ 66},  033812 (Sep. 2002).
\item P.~D. Drummond and P.~Deuar.
\newblock Quantum dynamics with stochastic gauge simulations.
\newblock {\em Journal of Optics B --- Quantum and Semiclassical Optics}
  \textbf{5},  S281--S289 (June 2003).
\item P.~D. Drummond, P.~Deuar, and Kheruntsyan~K. V.
\newblock Canonical {B}ose gas simulations with stochastic gauges.
\newblock {\em Physical Review Letters} \textbf{ 92},  040405  (Jan. 2004).
\end{enumerate}
This is noted in the text where appropriate.
\clearpage

\chapter*{Abstract}
\thispagestyle{empty}

The quantum dynamics and grand canonical thermodynamics of many-mode (one-, two-, and three-dimensional) interacting Bose gases are simulated from first principles. The model uses a lattice Hamiltonian based on a continuum second-quantized model with two-particle interactions, external potential, and interactions with an environment, with no further approximations. 
The interparticle potential can be either an (effective) delta function as in Bose-Hubbard models, or extended with a shape resolved by the lattice. 

Simulations are of a set of stochastic equations that in the limit of many realizations correspond exactly to the full quantum evolution of the many-body systems. These equations describe the evolution of samples of the gauge P distribution of the quantum state, details of which are developed. 

Conditions under which general quantum phase-space representations can be used to derive stochastic simulation methods are investigated in detail, given the criteria:
1) The simulation corresponds exactly to quantum mechanics in the limit of many trajectories.
2) The number of equations scales linearly with system size, to allow the possibility of efficient first-principles quantum mesoscopic simulations.
3) All observables can be calculated from one simulation.
4) Each stochastic realization is independent to allow straightforward use of parallel algorithms.
Special emphasis is placed on allowing for simulation of open systems.
In contrast to typical Monte Carlo techniques based on path integrals, the phase-space representation approach can also be used for dynamical calculations.
\enlargethispage{0.5cm}

Two major (and related) known technical stumbling blocks with such stochastic simulations are instabilities in the stochastic equations, and pathological trajectory distributions as the boundaries of phase space are approached. These can (and often do) lead to systematic biases in the calculated observables. The nature of these problems are investigated in detail. 

Many phase-space distributions have, however, more phase-space freedoms than the minimum required for exact correspondence to quantum mechanics, and these freedoms can in many cases be exploited to overcome the instability and boundary term problems, recovering an unbiased simulation. The stochastic gauge technique, which achieves this in a systematic way, is derived and heuristic guidelines for its use are developed. 

The gauge P representation is an extension of the positive P distribution, which uses coherent basis states, but allows a variety of useful stochastic gauges that are used to overcome the stability problems. Its properties are investigated, and the resulting equations to be simulated for the open interacting Bose gas system are derived. 

The dynamics of the following many-mode systems are simulated as examples:
1) Uniform one-dimensional and two-dimensional Bose gases after the rapid appearance of significant two-body collisions (e.g. after entering a Feshbach resonance). 
2) Trapped bosons, where the size of the trap is of the same order as the range of the interparticle potential.
3) Stimulated Bose enhancement of scattered atom modes during the collision of two Bose-Einstein condensates.
The grand canonical thermodynamics of 
uniform one-dimensional Bose gases is also calculated for a variety of temperatures and collision strengths.
Observables calculated include first to third order spatial correlation functions (including at finite interparticle separation) and momentum distributions.  The predicted phenomena are discussed.
  
\enlargethispage{0.5cm}
Improvements over the positive P distribution and other methods are discussed, and simulation times are analyzed 
for Bose-Hubbard lattice models from a general perspective. To understand the behavior of the equations, and subsequently optimize the gauges for the interacting Bose gas, single- and coupled two-mode dynamical and thermodynamical models of interacting Bose gases are investigated in detail. 
  Directions in which future progress 
can be expected are considered.

Lastly, safeguards are necessary to avoid biased averages when exponentials of Gaussian-like trajectory 
distributions  are used (as here), and these are investigated.
  

\clearpage
\tableofcontents
\listoffigures
\addcontentsline{toc}{chapter}{List of Figures}
\listoftables
\addcontentsline{toc}{chapter}{List of Tables}


\pagestyle{plain}
\chapter*{Thesis Rationale and Structure}
\addcontentsline{toc}{chapter}{Thesis Rationale and Structure}
\pagenumbering{arabic}

\section*{Rationale}
\addcontentsline{toc}{section}{Rationale}

It is a common view that first-principles quantum simulations of mesoscopic dynamics are intractable 
because of the complexity and astronomical size of the relevant Hilbert space. The following quotes illustrate the significance of the problem:
include\cite{Feynman82,Ceperley99}:
\begin{quote}
{\it ``Can a quantum system be probabilistically simulated by a classical universal computer? \dots the answer is 
  certainly, No!'' }(Richard P. Feynman, 1982).
\label{feynmanquote}  

%
{\it ``One is forced to either simulate very small systems (i.e. less than five particles) or to make serious approximations'' }
(David M. Ceperley, 1999).
\label{ceperleyquote}  
\end{quote} 
  This is certainly  true if one wishes to follow all the intricate details of a wavefunction that 
completely specifies the state of the system. Hilbert space size grows exponentially as more subsystems (e.g. particles or modes) are added, and methods that calculate state vectors or density matrix elements bog down very quickly. Path integral Monte Carlo methods also fail because of the well-known destructive interference between paths that occurs when one attempts dynamics calculations.

  Such a situation appears very unfortunate because for many complex physical systems a reliable simulation method is 
often the only way to obtain accurate quantitative predictions or perhaps even a well-grounded understanding. This is particularly so in situations where several length/time/energy scales or processes are of comparable size/strength, or when non-equilibrium phenomena are important. The need for reliable quantum dynamics simulations can be expected to become ever more urgent as more mesoscopic systems displaying quantum behavior are accessed experimentally. A pioneering system in this respect are the Bose-Einstein condensates of alkali-atom gases realized in recent years\cite{Anderson-95,Davis-95,Bradley-95,Fried-98}. 

There is, however, a very promising simulation method using phase-space representations that works around this complexity problem.  In brief, a correspondence is made between the full quantum state and a distribution over separable operator kernels, each of which can be specified by a number of variables {\it linear} in the number of subsystems (e.g. modes). If one then samples the operator kernels according to their distribution, then as the number of samples grows, observable averages of these operators 
approach the exact quantum values, i.e. ``emerge from the noise''. In principle one could reach arbitrary accuracy, but in practice computer power often severely limits the number of samples. Nevertheless, if one concentrates only on bulk properties, and is prepared to sacrifice precision beyond (typically) two to four significant digits, many first-principles quantum mesoscopic dynamics results can be obtained. The mesoscopic region can be reached because simulation time scales only log-linearly\footnote{This is because discrete Fourier transforms, usually required for kinetic energy evaluation, can be calculated on timescales proportional to $N_{\rm var}\log N_{\rm var}$. For some particularly demanding models, the scaling may be log-polynomial in $N_{\rm var}$ due to an increased number of terms in the equation for each variable if there is complicated coupling between all subsystem pairs, triplets, etc. In any case, simulation time never scales exponentially, as it would for brute force methods based on density matrix or state vector elements.}  ($N_{\rm var}\log N_{\rm var}$) with variable number $N_{\rm var}$, and so still scales log-linearly with system size. Some initial examples of calculations with this method, particularly with the positive P representation\cite{Chaturvedi-77,DrummondGardiner80} based on a separable coherent state basis, are
 many-mode quantum optics calculations in Kerr dispersive media\cite{Carter-87,Drummond-93} 
evaporative cooling of Bose gases with repulsive delta-function interactions\cite{Steel-98,DrummondCorney99}, and breathing of a trapped one-dimensional Bose gas\cite{Carusotto-01}.  
   
In summary, first-principles mesoscopic quantum dynamics is hard, but some progress can (and has) been made in recent years. 
In broad terms, {\it the aim of the work reported in this thesis is to advance the phase-space simulation methods some more .}  The investigation here is carried out in several directions:
\pagebreak
\ENUM{
\item \textbf{Stochastic gauges} Many phase-space distributions have more phase-space degrees of freedom than the minimum required for exact correspondence to quantum mechanics. Each such degree of freedom leads to possible modifications of the stochastic equations of motion by the insertion of arbitrary functions or {\it gauges} in an appropriate way. While the choice of these gauge functions does not influence the correspondence to quantum mechanics in the limit of infinitely many samples, it can have an enormous effect on the efficiency and/or statistical bias of a simulation with a finite number of samples. (Hence the use of the word ``gauge'', in analogy with the way electromagnetic gauges do not change the 
physical observables but can have an important effect on the ease with which a calculation proceeds). In particular, non-standard choices of these gauge functions can lead to enormous improvements in simulation efficiency, or can be essential to  allow any unbiased simulation at all. 

Here, a systematic way to include these freedoms is derived and ways of making an advantageous gauge choice are considered in some detail. As a corollary, some results present in the literature\cite{Plimak-01,Carusotto-01,DeuarDrummond01} are found to be examples of non-standard gauge choices.

The gauge P representation, which is  a generalization of the successful positive P representation based on a coherent state basis to allow a variety of useful gauges, is explained. Application of it to interacting Bose gases is developed.

\item \textbf{Removal of systematic biases using stochastic gauges}. 
Two major (and related) stumbling blocks for phase-space distribution methods have been  instabilities in the stochastic equations, and pathological trajectory distributions as the boundaries of phase space are approached. These occur for nonlinear systems and can (in fact, apart from special cases, {\it do}) lead to overwhelming noise or systematic biases in the calculated observables, preventing dependable simulations. To date this has been ``problem number one'' for these methods. 

In this thesis it is shown how appropriate stochastic gauges can be used to overcome the instability and boundary term problems in a wide range of models, recovering an unbiased simulation, and opening the way for reliable simulations. Heuristic ways of achieving this in general cases are considered in detail, and examples are given for  known cases in the literature. 

\item \textbf{Improvement of efficiency using stochastic gauges}  
The stochastic gauge method also has the potential to significantly improve the efficiency of simulations when appropriate gauges are chosen. In particular, the time for which useful precision is obtained can be extended in many cases by retarding the growth of noise. 
This is investigated  for the case of gauge P simulations of interacting Bose gases and some useful gauges obtained. The regimes in which improvements can be seen with the gauges developed here are characterized.
Heuristic guidelines for gauge choice in more general representations and models are also given. Example simulations are made.
 
\item \textbf{General requirements for usable distributions in open systems} 
Another aim here is to determine what are the bare necessities for a phase-space distribution approach to be successful, so that other details of the representation used (e.g. choice of basis) can be tailored to the model in question. This can be essential to get any meaningful results, as first-principles mesoscopic simulations are often near the limit of what can be tractably calculated. To this end, general features of the correspondence between quantum mechanics and stochastic equations
for the variables specifying the operator kernels are considered in some detail.
 
Particular concern is given to simulating open systems, as most experimentally realizable 
systems do not exist in isolation and will have significant thermal and also particle-exchange interactions with external environments. These must usually be  taken into account for an accurate description. (The alternative is to make wholesale 
approximations to the model whose precise effect is often difficult to ascertain).
 
\item \textbf{Application to thermodynamic calculations}
  A separate issue are static calculations of thermodynamic equilibrium ensembles. These have traditionally been the domain 
of quantum Monte Carlo methods, the most versatile of which have been those based on the path integral approach. 
While path integral methods are generally not useful for dynamical calculations because destructive interference between the paths occurs extremely rapidly, masking any dynamics, good results can be obtained for static thermodynamic ensemble calculations. If the model is kind, even \order{10^4} particles or modes can be successfully simulated, if one again concentrates on bulk properties and only several significant digits. 

Phase-space representations that include dynamically changing trajectory weights can also be used for thermodynamics calculations. The thermal density matrix is evolved with respect to inverse temperature after starting with the known high temperature state. (Such simulations are sometimes said to be ``in imaginary time'' because of a similarity between the resulting equations and the Schr\"{o}dinger equation after multiplying time by $i$). This approach appears competitive with path integral Monte Carlo methods in terms of efficiency, but offers two distinct advantages:
Firstly, all observables can in principle be calculated in a single simulation run, which is not the case in  path integral methods. These latter require separate algorithms for e.g. observables in position space, momentum space, or observables not diagonal in either. Secondly, a single simulation run gives results for a range of temperatures, while in path integral methods a new simulation is needed for each temperature value.

\item \textbf{Demonstration with non-trivial examples}
Lastly, but perhaps most importantly, one wishes to demonstrate  that the results obtained actually are useful in non-trivial cases.  Simulations of dynamics and thermodynamics in a variety of mesoscopic interacting systems   are carried out, and their physical implications considered.  The emphasis will be foremost on simulations of interacting Bose gases. It is chosen to concentrate on these because they are arguably 
the systems where quantum effects due to collective motion of atoms are most clearly seen and most commonly investigated 
by contemporary experiments. This refers, of course, to the celebrated Bose-Einstein condensate experiments on cold,
trapped, rarefied, alkali-metal gases. (For a recent overview see, for example, the collection of articles 
in Nature \textit{Insight on ultracold matter}\cite{Southwell02,Chu02,AnglinKetterle02,Burnett-02,Monroe02}.) 
As these gases are an extremely dynamic field of research the need for first-principle simulation methods appears urgent here. 
}

The ``holy grail'' of quantum simulations, 
which would be a universally-applicable black-box tractable simulation 
method\footnote{In Newtonian dynamics this is just the usual ``start with initial conditions and integrate the differential equations''.},  probably does not exist. However, simulations are of such fundamental importance to 
reliable predictions in mesoscopic physics that any significant progress has the potential to be a catalyst for 
far-ranging discoveries (and in the past often has). 

\section*{Structure}
\addcontentsline{toc}{section}{Structure}

This thesis begins with two introductory chapters, which explain in more detail 
the background issues. That is, Chapter~\ref{CH1} discusses the fundamentals of why many-body
quantum simulations are difficult, compares several approaches, and motivates the choice to work on the mode-based phase-space representation simulation methods considered in this thesis. Chapter~\ref{CH2} summarizes the interacting Bose gas model that will be considered in all the simulation examples, and discusses under what circumstances a first-principles rather than a semiclassical calculation is needed to arrive at reliable predictions.

The body of the thesis is then divided into three parts of a rather different nature. Part A investigates what general properties of a phase-space representation are needed for a successful simulation, and explains the gauge P representation, which will be used in later parts. Part B calibrates this method on some toy problems relevant to the interacting Bose gas case, while Part C applies them to non-trivial mesoscopic systems (dynamics and thermodynamics of interacting Bose gases).

Accordingly, Chapter~\ref{CH3} presents a generalized formalism for phase-space representations of mixed quantum 
states and characterizes the necessary properties for an exact correspondence between quantum mechanics and 
the stochastic equations. The stochastic gauge technique, which forms the basis of the rest of the developments in this thesis, is developed in Chapter~\ref{CH4}. In Chapter~\ref{CH5}, properties of the gauge P representation are investigated, and its application to  interacting Bose gases developed. 
In Chapter~\ref{CH6} it is explained how stochastic gauges can be used to overcome ``technical difficulty number one'':
the systematic ``boundary term'' biases that otherwise prevent or hinder many attempted phase-space distribution simulations. Some relevant technical issues regarding stochastic simulations have been relegated to Appendices.

One- and two-mode toy models are used in Part B  to check correctness of the method, and to optimize 
the gauge functions in preparation for the target aim of simulating multi-mode systems. These models are the
single-mode interacting Bose gas dynamics in Chapter~\ref{CH7}, the dynamics of two such Rabi-coupled modes in Chapter~\ref{CH8}, and grand canonical single-mode thermodynamics in Chapter~\ref{CH9}.

Finally, Chapters~\ref{CH10} and \ref{CH11} give examples of nontrivial many-mode simulations of many-mode interacting Bose gas dynamics and grand canonical thermodynamics, respectively. 

Chapter~\ref{CH12} summarizes and concludes the work along with some speculation on fruitful directions of 
future research. 


\chapter[Introduction: Quantum many-body simulations]
{Introduction:\\ Why are quantum many-body simulations difficult and what can be done about it?}
\label{CH1}

  Here it is considered how stochastic methods can allow first-principles quantum simulations despite the seemingly intractably large Hilbert space required for such a task. The main approaches: path integral Monte Carlo, quantum trajectories, and phase-space distributions are briefly compared (without going into mathematical details of the methods themselves). This thesis develops and applies the stochastic gauge technique to make improvements to the phase-space distribution methods, and the choice to concentrate on these methods is motivated by their greater versatility. 
The emphasis throughout this thesis on methods applicable to open systems (leading to mostly mode-based formulations) is also discussed.

\section{Hilbert space complexity}
\label{CH1Hilbert}
In classical mechanics, an exhaustive simulation of an interacting $N$-particle system is (in principle) computable\footnote{``Computable'' here having the usual meaning that simulation time scales linearly, or (grudgingly) polynomially with $N$. Naturally if $N$ is truly macroscopic such a system still cannot be simulated in practice, but in such large classical systems the primary hindrance to accurate predictions is usually deterministic chaos rather than simulation speed.} on a 
classical computer. For the trivial case of a non-interacting three-dimensional (3D) gas in a potential, one would start from initial conditions in $6N$ real variables\footnote{Position and momentum in each degree of freedom.}, and then generically simulate by solving $6N$ differential equations. For particles with ranged binary interactions\footnote{As in the majority of first-principles classical models.} there are $6N$ real differential equations, each with \order{N} terms, so simulation times scale as $N^2$ --- grudgingly computable.

  Let us now look at the quantum case. Consider a generic mesoscopic system, divided up (arbitrarily) into $N$ subsystems. Most typically these subsystems would be $N$ particles, $N$ spatial or momentum modes, or $N$ orbitals. If we allow each mode or orbital to be occupied by up to $P_{\rm max}$ particles ($P_{\rm max}=1$ for fermions, and more for bosons), then the full state vector of a pure system in such a model has $(P_{\rm max}+1)^N$ complex number components. In the other popular formulation with a set number of particles $N$, one would discretize $\mc{D}$-dimensional space into $M_{\rm latt}$ lattice points in each dimension, leading to $M_{\rm latt}^\mc{D}$ lattice points in all. The state vector of such a system would then have $M_{\rm latt}^{N\mc{D}}$ complex number components. Both of these cases are {\it exponential} in the system size $N$, and the amount of memory needed to hold the state (let alone the time needed to evolve the corresponding number of coupled differential equations) becomes intractable very rapidly. This is the Hilbert space complexity problem, and is generic to all mesoscopic quantum calculations. 

For example, if one had 8000 Gigabytes of memory just to hold the state vector\footnote{Assuming 4 byte floating point data.}, one could have 40 fermion modes, or 12 boson modes with each one occupied by at most 9 atoms. In the set particle number formulation, one could get away with 4 particles in a (very sparse!) $10\times10\times10$ three-dimensional lattice, or 12 particles in such a one-dimensional lattice. This indicates where the estimate of  about ``five particles'' quoted in the Thesis Rationale on page~\pageref{ceperleyquote} came from.

   For open systems, one needs even more data space to hold the state, as the 
density matrix $\op{\rho}$ will have approximately half the square of the number of components in a state vector --- exponential in system size again, but with twice the exponent.
 The general conclusion is that for mesoscopic (and even many microscopic) systems a brute force approach is doomed to failure, except for perhaps a few special cases that are separable\footnote{e.g. an ideal gas}.

\section{Bulk properties are the key}
\label{CH1Bulk}
  The exponential growth of Hilbert space mentioned above appears prohibitive, but let us first stand back from the quantum problem for a moment. In classical mechanics there are enormous amounts of variables to keep track of once a system has macroscopic numbers of particles --- Avogadro's number is very big! 
The reason that statistical mechanics is so successful at dealing with this issue is that for a large system, one is only  really interested in {\it bulk} properties (temperature, pressure, two-particle correlations, \dots).  The exact position or momentum  of particle number 9501292851748 is not of interest to anyone. Even more importantly, knowledge of such a ``raw'' observable calculated from a model has no physical predictive power, because in such a large system it is not feasible to measure the initial conditions exhaustively. Only a statistical description (based on bulk properties) can be input as initial conditions into a model and because of this, only bulk statistical properties of the model will correspond accurately 
to the behavior of the physical system under study.

   In quantum systems, we have (in a general sense) an analogous problem, but it is much more acute because it turns up when we have \order{5} subsystems, not \order{10^5}. This is simply because in quantum mechanics there is much more to keep track of. The full classical description contains 
the positions and momenta of all the particles (e.g. system variables record things like ``the position and momentum of particle 753024''), whereas a full quantum description contains the amplitude and phase {\it of each one of all the possible configurations} between the entire set of particles\footnote{Or, more generally,  ``subsystems'', particularly when particles are not conserved.}. (e.g. a quantum mechanics variable would be 
expressing something like ``the probability and phase (relative to a general reference) of 
the particular configuration of particles for which: number 1 is at $x_1$, \dots, number 753024 is at $x_{753024}$, \dots, and number $N$ is at $x_N$'').  In classical mechanics the entire description is just one of these possible correlations.

  Conceptually, what lets one vanquish the enormous numbers of variables is similar in both cases: We're only interested in {\it bulk properties}, be they themselves classical or quantum in nature\footnote{An experimentally accessible example of a macroscopic property of a ``quantum'' nature is the condensate phase at a given position in a Bose-Einstein Condensate.}.

\section{Sampling system configurations}
\label{CH1Configurations}

There are two kinds of conceptually distinct kinds of correlations to sample (rather than characterize exhaustively)
to estimate the position of the quantum system in Hilbert space. 

  Firstly, in models that include interactions with an external environment, the system is described fully by a density matrix rather than a pure state vector. Here, the mixture of pure states that combine to the reduced density matrix description of the system must be sampled. While this kind of sampling reduces the number of variables substantially (We go from some huge number to its square root), it is not enough to make a qualitative difference. The pure state vectors still reside in a space whose dimension scales exponentially with system size $N$. 

An example of an approach that carries out only this first kind of sampling is the quantum trajectories method for open systems. (Some reviews of this approach are e.g. the articles of Zoller and Gardiner\cite{ZollerGardiner97}, and Plenio and Knight\cite{PlenioKnight98}, or Wiseman and Milburn\cite{WisemanMilburn93} for continuous variable measurements.). This method can be successful in simulations of several particles, especially when microscopic details of their entanglement are important  --- for example, simulations with up to $24$ qubits have been reported\cite{Carlo-03}.

  What is really needed to attack at least mesoscopic systems, is to change the scaling with number of subsystems from exponential down to at least polynomial. This is achieved by going from the pure system states (which are in general superpositions of various separable system configurations) to a {\it sample of the separable system configurations}. (examples of such separable system configurations are: the set of occupations of all modes, or the set of positions of all the particles.) The essence of the two sampling steps can be schematically expressed as
\EQNa{
\op{\rho}\quad &\longrightarrow&\quad \sum_{j'=1}^{\mc{S}'} \ket{\psi_{j'}}\langle\wt{\psi}_{j'}|, \\
\ket{\psi_{j'}} \quad&\longrightarrow&\quad \sum_{j=1}^{\mc{S}} \otimes_{k=1}^N |\,C^{(j)}_k\rangle_k
,}
where the sums are over $\mc{S}$ and $\mc{S}'$ samples, and the tensor products over $N$ subsystems, having 
local subsystem parameters $C_k$ ($C$ stands for separable Configuration). The pure states $\ket{\psi_{j'}}$ and $|\wt{\psi}_{j'}\rangle$ are not necessarily separable product states, but the samples $\otimes_k|\,C^{(j)}_k\rangle_k$ are.  In the limit of many samples ($\mc{S}$, $\mc{S}'\longrightarrow\infty$), the correspondence becomes exact.  

This second sampling is really the crucial step to allow any kind of first-principles quantum simulation of realistic mesoscopic models\footnote{%
Obviously an enormous amount of physical prediction can be made with mean field, or other semiclassical methods, but by first-principles models I refer to those where no semiclassical approximations are made.}, as the number of variables to keep track of now finally scales proportionally to $N$.
\enlargethispage{1cm}

  The unavoidable price paid is loss of accuracy. Fortunately most of this price gets transferred onto the detailed
non-bulk properties, 
which one is not interested in anyway. This occurs almost automatically, because there are just so many more of these detailed properties. In many cases bulk properties of quite large systems can be calculated with useful precision, where no simulation at all was possible with brute force methods.
The calculation time required for a given uncertainty $\Delta$ now scales (by the central limit theorem) roughly proportionally to $N/\Delta^2$, whereas 
for a brute force non-stochastic method it would be\footnote{The time required for multiplication operations on floating point numbers with $d_{10}\approx-\log_{10}{\Delta}$ digits scales as ${d_{10}}^2$} proportional to $e^N(\log\Delta)^2$.
Clearly, one won't be getting more than two- or three-digit accuracy with the stochastic methods in most cases, but this is 
often sufficient for a good comparison between theory and experiment.

\section{Path integral Monte Carlo and static thermal calculations}
\label{CH1PIMC}

   The most widely used approach for many-mode/body calculations is path integral Monte Carlo. The basis for these methods were Feynman's imaginary time path integrals\cite{Feynman53}. 
This method, however, is only  successful for static calculations of thermal equilibrium states (reasons explained below).

In the static thermal case, the un-normalized density matrix at temperature $T$ is $\op{\rho}_u(T) = \exp\left[-\op{H}/k_B T\right]$, and path integrals make use of the simple (and exact) property that $\op{\rho}_u(T) = \op{\rho}_u(\mc{M}T)^{\mc{M}}$. If $\mc{M}$ is sufficiently large that $\mc{M}k_BT\gg\langle\op{H}\rangle$, then one can use a high-temperature approximation\footnote{%
Typically one uses the Feynman-Kacs or Trotter formula\cite{Trotter59} 
$\exp\left[-(\op{H}_V+\op{H}_K)/k_BT\right] = \lim_{\mc{M}\to\infty}\left( \exp\left[-\op{H}_V/\mc{M}k_B T \right] \exp\left[-\op{H}_K/\mc{M}k_B T \right]\right)^{\mc{M}}$
to evaluate the effect of the potential $\op{H}_V$ and kinetic $\op{H}_K$ terms of the Hamiltonian separately.}
 to evaluate $\op{\rho}_u(\mc{M}T)$. 
The density matrix $\op{\rho}_u(T)$ can be written as a convolution of the high temperature density matrix elements (in complete bases of separable system configurations $C$ labeled by $C_{(0)},\dots,C_{(\mc{M})}$):
\EQN{
\op{\rho}_u(T) &=& \int\dots\int dC_{(0)} dC_{(1)}\dots dC_{(\mc{M})}\\
&& \bra{C_{(0)}}\op{\rho}_u(\mc{M}T)\ket{C_{(1)}}\bra{C_{(1)}}
\op{\rho}_u(\mc{M}T)\ket{C_{(2)}}\cdots\bra{C_{(\mc{M}-1)}}\op{\rho}_u(\mc{M}T)\ket{C_{(\mc{M})}}\nonumber
.}
   To obtain properties diagonal in the configuration basis chosen, one takes samples while setting $C_{(0)}=C_{(\mc{M})}$, and
ends up sampling a ``ring polymer'' of configurations, the number of variables in a full sample scaling as $\mc{M}N$. 
The distribution of the samples is given by the product of the expectation values: 
\EQN{
P(C_{(0)},C_{(1)},\dots,C_{(\mc{M})}) = 
\otimes_{j=1}^{\mc{M}} \bra{C_{(j-1)}}\op{\rho}_u(\mc{M}T)\ket{C_{(k)}},
}
which is real positive, or largely so.
Off-diagonal properties require separate calculations with slightly different sample properties.
An extensive discussion of the details of the path integral approach can be found e.g. in Ceperley\cite{Ceperley95,Ceperley96LN} and references therein.

While, formally, this approach might appear to be also applicable for dynamical evolution  via $\op{\rho}(t) = \op{\rho}(0)\exp\left[-i\op{H}t/\hbar\right]$, in practice this does not work. The reason is that now 
$P(C_{(j)})$ is primarily a product of complex phase factors, rather than real positive values. One obtains a very wide range of configurations, all with similar weighting, but mutually canceling phase factors. In the limit of infinite trajectories, almost all of these configurations would destructively interfere leaving just the observed dynamics, but for finite numbers of samples in a complex system there is no automatic way to tell which configurations are the important ones.  The end result is that no sensible observable averages emerge from the noise even for very short times. 

\section{phase-space evolution methods}
\label{CH1Phasespace}
The density matrix of a system can often be written in a form 
\EQN{\label{configurationP}
\op{\rho} = \int P(C)\,\op{\Lambda}(C)\,dC
,}
 where $C$ is a separable system configuration\footnote{%
The variables of $C$ can be discrete or continuous. In the former case the integral over $C$ is replaced by a sum.}, containing some number of variables linear in $N$ (the number of subsystems). $P(C)$ is a positive function, and $\op{\Lambda}(C)$, the {\it kernel}, is a projector or similar tensor product operator parameterized by $C$. 

In a generic case where pure quantum states of each subsystem (labeled $k$) can be fully described by the set of parameters $C_k$, a viable kernel would be $\op{\Lambda}(C) = e^{i\theta}\otimes_{k}\ket{C_k}_k\langle\wt{C}_k|_k$. Here $\ket{C_k}_k$ is the pure state of subsystem $k$ given by parameters $C_k$, and $\theta$ is a phase. The full configuration then would be $C=\{\theta,C_1,\dots,C_N,\wt{C}_1,\dots,\wt{C}_N\}$.

The basic idea is that $P(C)$ can be interpreted as a probability distribution
of the configuration $C$. Then, the evolution equation for $\op{\rho}$ (with time or with temperature) produces a 
corresponding evolution equation for $P$, and finally evolution equations for the configuration samples $C$. 
Depending on the system and the kernel $\op{\Lambda}$ chosen, this procedure can often be carried out, and 
evolution equations for the configuration samples of the system derived. 
In particular, {\it dynamical evolution can be simulated} while the number of variables remains linear or polynomial in $N$.

In fact phase-space distribution methods are much more versatile than path integral approaches on a number of counts:
\ENUM{
\item Dynamics can be simulated. 
\item Results are given for a range of times (or temperatures) by one simulation
\item All observables can be calculated in one simulation
}
In both phase-space distribution and path integral approaches, the amount of computer effort scales linearly with system size $N$. More details of how this is achieved are given in Chapter~\ref{CH3}.
See also a tabular comparison of the various quantum many-body/mode simulation approaches in Tables~\ref{TABLEDynamics} and~\ref{TableThermodynamics}.

Unfortunately, a major stumbling block for phase-space distribution methods is that the equations for the configuration variables in $C$ often have stability problems, which can lead to large noise and/or statistical bias.  This is one of the main reasons that these methods have not been as widely investigated nor used as path integral Monte Carlo, despite their other advantages.

This thesis concerns itself with methods of overcoming this stability problem using the stochastic gauge technique, and some demonstration of subsequent applications. It is also intended to show that when stability issues are taken care of, these phase-space distribution methods can  be competitive with path integrals for static thermal ensemble calculations as well.

\section{Open systems and the mode formulation}
\label{CH1Open}

  Actual physical systems do not exist in isolation, and there is always some degree of coupling to the external environment. This coupling becomes ever more pronounced as one moves from microscopic to mesoscopic systems, and is considered responsible for the reduction of quantum phenomena to classical behavior as the macroscopic limit is reached.

\begin{table}[t]
\caption[Comparison of quantum dynamics simulation methods]{\label{TABLEDynamics}\footnotesize
A comparison of different methods to exactly simulate quantum dynamics from first principles. As applicable to many-body/mode simulations with $N$ subsystems (e.g. particles, modes).
\normalsize}\vspace*{3pt}
\hspace{-0.85cm}\begin{minipage}{\textwidth}\begin{tabular}{|l|c|c|c|c|}
\hline
Variety:		& \scshape Explicit  	&\scshape Quantum	&\scshape Path		&\scshape Phase-space	\\
			& \scshape Density Matrix&\scshape Trajectories	&\scshape Integral	&\scshape Distributions	\\
\hline\hline
Stochastic		& No			& Yes			& Yes			& Yes			
\\\hline
Sample size		& $\propto e^{2N}$	& $\propto e^{N}$	& $\propto\mc{M}N$	& $\propto N$		
\\\hline
Phase factor 		&			&			&			&  	\\	
catastrophe\footnote{Where samples of similar weight but different phase factors interfere, masking any physically meaningful results.}
			& No			& No			& Yes			& No\footnote{At long times some gauged schemes may develop interfering phase factors. This occurs only for some gauges in some systems.} 
\\\hline
Unstable equations	&&&&\\
or sampling bias 	& No			& No			& No			& Often\footnote{However, in a wide range of cases the instability can be removed using the stochastic gauge technique, as shown in this thesis.}		
\\\hline
Observables from &&&Limited&\\
 one calculation	& All			& All			& subset	& All	
\\\hline		
Simulation scope	& Range of $t$		& Range of $t$		& Single $t$		& Range of $t$		
\\\hline
\end{tabular}\end{minipage}
\end{table}

  Traditionally in most many-body studies the system has been taken to be composed of a set number of $N$ particles. 
A different formalism is required if one is to investigate open systems that can exchange particles with a reservoir having chemical potential $\mu$, or for dynamical calculations if particles are lost or injected into the system. A natural choice is to consider lattice models where rather than dividing the system into $N$ subsystems consisting of a particle, we divide the system up into $N$ spatial (or momentum) lattice points that can be occupied by a variable number of particles. The kernel $\op{\Lambda}$ then becomes a separable tensor product of lattice-point states, described by some local parameters. Their total number for a system configuration sample is again proportional to $N$. Because we want the methods developed in this thesis to be applicable to systems that allow particle exchange with an environment, this is the kind of model that we will mostly concentrate on here. The general methods developed are, however, by no means restricted to mode-based models.
\begin{table}[t]
\caption[Comparison of quantum thermodynamics calculation methods]{\label{TableThermodynamics}\footnotesize
A comparison of different methods to exactly calculate static quantum properties of equilibrium thermodynamic ensembles from first principles. As applicable to many-body/mode simulations with $N$ subsystems (e.g. particles, modes).
\normalsize}\vspace*{3pt}
\hspace{-1.5cm}\begin{minipage}{\textwidth}\begin{tabular}{|l|c|c|c|c|}
\hline
Variety:		& \scshape Explicit  	&\scshape Quantum	&\scshape Path		&\scshape Phase-space	\\
			& \scshape Density Matrix&\scshape Trajectories	&\scshape Integral	&\scshape Distributions	\\
\hline\hline
Stochastic		& No			& Yes			& Yes			& Yes			
\\\hline
Sample size		& $\propto e^{2N}$	& $\propto e^{N}$	& $\propto\mc{M}N$	& $\propto N$		
\\\hline
Phase factor 		&			&			& For fermions\footnote{%
For fermion calculations the ``fermion sign problem'' appears after $k_BT\lesssim \Delta E_{FB}$, where $\Delta E_{FB}$ is the free energy difference between the fermion system, and an analogous system of bosons. For a review see Schmidt and Kalos\cite{SchmidtKalos84}}
												& Can occur \\		
catastrophe\footnote{Where samples of similar weight but different phase factors interfere, masking any physically meaningful results.}
			& No			& No			& $k_BT\lesssim \Delta E_{FB}$									  			& for low $T$
\\\hline
Unstable equations	&&&&\\
or sampling bias 	& No			& No			& No			& Often\footnote{However, in a wide range of cases the instability can be removed using the stochastic gauge technique, as shown in this thesis.}		
\\\hline
Observables from &&&Limited&\\
 one calculation	& All			& All			&  subset	& All			
\\\hline
Simulation scope	& Range of $T$		& Range of $T$		& Single $T$		& Range of $T$		
\\\hline
\end{tabular}\end{minipage}
\end{table}

  A feature to be aware of in such a model is that in general the state will be in a superposition or mixture of pure states with different numbers of total particles. 
This is not expected to hinder comparison with experiment, however. A subsystem such as e.g. a trapped gas is perfectly entitled to be in a superposition of number states, and e.g. evaporative cooling of Bose atoms will produce BECs of different sizes in each individual experimental run. 

For open systems with particle transfer to the environment, the boundary where the system ends and the environment starts will be at best hazy, and particles will take some time to travel from one to the other. There will always be a gray area, and in it some particles, where the labeling of particles as belonging to the system or environment is dubious. Superpositions and/or mixtures of different particle number states are actually trying to reflect this labeling difficulty which is encountered in experiments.

Also, in many situations the effect of such superpositions will be small: A reasonably accurate description of a realistic mesoscopic system in most cases requires a fairly large number of lattice points to resolve spatial and momentum variations in the system. In such a case, for statistical reasons,  the variation in particle number of the whole system will be small in comparison with the total number of particles.

Lastly, it should also be pointed out that if desired, one can always simply turn off the external couplings, and 
obtain results for closed number-conserving Hamiltonian models, which are just special cases with external interaction strength zero. (Although a simulation method that is ``hard-wired'' to a conservative system is likely to be somewhat more efficient than an open systems method.)


\chapter{System of choice: Open interacting Bose gases}
\label{CH2}
  Simulations presented in this thesis mostly\footnote{Although not exclusively, see for example Chapter~\ref{CH6}.} concern themselves with simulations of interacting Bose gases in a lattice model with two-particle interactions\footnote{In the most commonly used cold alkali-metal gas models these interactions are local as in a Bose-Hubbard model, and are then often referred to as ``delta-function'' interactions.}. Here those basic details of the model that are repeatedly referred to in the body of the thesis are introduced. Also, this choice of mesoscopic system from the wide variety whose simulations could have been attempted is motivated.

This chapter does not attempt to present a review of the state of knowledge on interacting Bose gases, as this is currently very extensive and doing so would unnecessarily lengthen the thesis without introducing any additional novelty. Instead, I recommend to the reader the excellent reviews on both theoretical and experimental aspects of Bose Einstein condensation in cold alkali gases by Leggett\cite{Leggett01}, Dalfovo\etal\cite{Dalfovo-99}, Parkins and Walls\cite{ParkinsWalls98}. Topical also are the 1998 Nobel lectures of Chu\cite{Chu98}, Cohen-Tannoudji\cite{Cohen-Tannoudji98}, and Phillips\cite{Phillips98} on laser cooling and magneto-optical trapping, as well as the reviews of evaporative cooling by Ketterle and van Druten\cite{KetterlevanDruten96} and Walraven\cite{Walraven96}. An overview of recent developments can be found in e.g. the collection of articles 
in Nature \textit{Insight on ultracold matter}\cite{Southwell02,Chu02,AnglinKetterle02,Burnett-02,Monroe02}.  Physical properties of interacting Bose gases pertinent to the analysis or interpretation of the simulations in Parts B and C of this thesis are given in the relevant chapters. 

\section{Motivation}
\label{CH2Motivation}

\begin{figure}[t]
\center{\includegraphics[width=8.6cm]{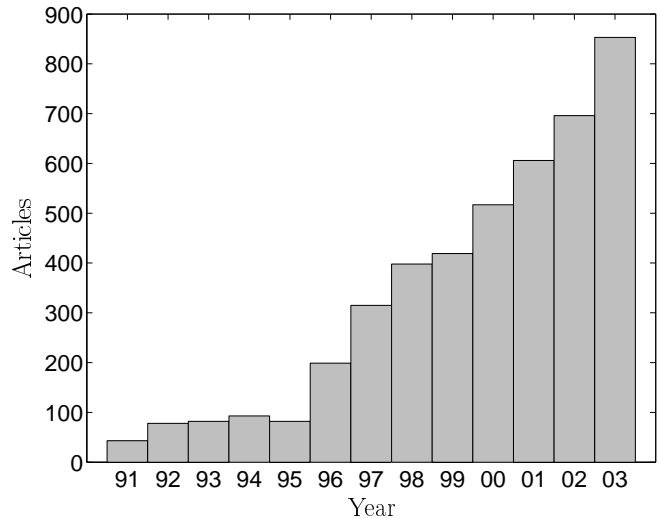}}\vspace{-0.5cm}\par
\caption[Articles on BECs per year]{\label{FIGURELiterature}\footnotesize
\textbf{Papers published in refereed journals} (by year) on the topic of Bose-Einstein condensation. Based on a search for the keywords ``Bose Einstein condens*\,'' or ``Bose condens*\,'' through titles and abstracts in the ISI ``Web of Science'' citation database ( http://wos.isiglobalnet.com/ ) on 11 Feb 2004. Note the rapid increase after the first successful BEC experiments in 1995.
\normalsize}
\end{figure}

 The choice to concentrate on cold dilute interacting Bose gas models when trying out the simulation methods developed in this thesis is motivated by the following considerations:
\ITEM{

\item The prime motivation is that mesoscopic systems that are very well described by these models 
are experimentally accessible. These are the cold dilute alkali-metal gases at temperatures below or near 
the Bose-Einstein condensation (BEC) temperature. 
In the last several years production of cold Bose gases of alkali-metal atoms has become almost routine, occurring in many tens of labs around the world. The largest interest (see Figure~\ref{FIGURELiterature}) has been in gases below the condensation temperature where quantum effects come to dominate the system even though it is mesoscopic or even approaching macroscopic in size. These were first realized in 1995 by Anderson\etal\cite{Anderson-95} (${}^{87}$Rb), Davis\etal\cite{Davis-95} (${}^{23}$Na), and Bradley\etal\cite{Bradley-95} (${}^7$Li), by laser cooling the gas in a trap, and subsequently lowering the temperature even further by evaporative cooling. Three years later the first signatures of BEC were seen by Fried\etal\cite{Fried-98} in the somewhat different case of hydrogen. Presently, many variants of these systems are being investigated experimentally. For example, 
quasi-one-dimensional gases  have been achieved\cite{Gorlitz-01,Schreck-01,Greiner-01}, and 
coherent four-wave mixing between atom clouds has been observed\cite{Vogels-02}, just to name a few of the experimental directions relevant to the models considered later in this thesis.

\item A related argument is that there are phenomena in the experimental systems that are not precisely understood and appear quite resistant to conclusive theoretical analysis by approximate methods. This invites a first-principles investigation.
 
Most of these occur in situations where there is a transition or interplay between disordered high temperature behavior well above condensation temperature $T_c$, and the pure many-body quantum states approached as $T\to0$. The first can be treated with kinetic theories, while the latter is very well described by the Gross-Pitaevskii (GP)\cite{Gross61,Gross63,Pitaevskii61} mean-field equations, or Bogoliubov theory\cite{Bogoliubov47}\footnote{At finite temperatures , but significantly below $T_c$}. As usual in physical systems, the intermediate regime is the most complicated and is the place where first-principles simulations are the most needed.

Examples of such issues where first-principles simulations can make a step forward to full understanding are, for example: 
\ENUM{
\item The behavior of the system during initial condensation, when the $T\approx T_c$ barrier is crossed. This includes issues such as ``does the always condensate form in the ground state, or is (at least metastable) condensation into excited states possible?'' There are some indications that condensates may occur e.g. in states of non-zero centre-of-mass motion\cite{DrummondCorney99}.
\item Is the exact ground state of a BEC the same as that obtained by semiclassical GP methods?
\item Dynamics of condensates during atom interferometry.
\item The decoupling of atoms from the trapped condensate in an atom laser arrangement. How is the  
emitting condensate disturbed, and in what state are the pulses emitted?
\item Scattering of atoms into empty momentum modes when condensates collide\cite{Vogels-02}.
\item Heating and other processes that eventually destroy the condensate in experiments.
}
It is worth noting that most of the above processes usually involve coupling of the condensate to an external environment. This is further indication that a simulation method capable of handling open systems is desirable.

\item Not only is the field of Bose-Einstein condensation experimentally accessible, but the sheer amount of research on the topic has been growing (and continues to grow) at a great rate (see Figure~\ref{FIGURELiterature}). This would indicate that 
some of the predictions made may be verified rapidly.

\item Lastly, but importantly, many-mode simulations of similar systems have been successful with a phase-space method (the positive P distribution\cite{Chaturvedi-77,DrummondGardiner80}). Some examples of this include simulations of evaporative cooling of a Bose gas to the beginning of condensation\cite{Steel-98,DrummondCorney99,Corney99}, as well as simulations of optical soliton propagation in nonlinear Kerr media\cite{Carter-87,Drummond-93}, which have a similar form of nonlinearity to \eqref{deltaH} --- quadratic in local particle density. The gauge distributions developed in this thesis include the positive P distribution as a special case, and improve on it. This previous success is an indication that the newer methods are likely to lead to significant results with these systems.
}
\vfill

\section{Field model with binary interactions}
\label{CH2Hamiltonian}

  Consider a $\mc{D}$-dimensional boson field $\op{\Psi}(\bo{x})$ undergoing two-particle interactions. After second quantization the Hamiltonian of the system can be written as
\EQN{\label{hamiltonian}
\op{H} &=& \int d^\mc{D}\bo{x}\left\{ \frac{\hbar^2}{2m}\dada{\dagop{\Psi}(\bo{x})}{\bo{x}} \dada{\op{\Psi}(\bo{x})}{\bo{x}} + V^{\rm ext}(\bo{x})\dagop{\Psi}(\bo{x})\op{\Psi}(\bo{x}) \right\}\nonumber\\
&&\quad+\quad\frac{1}{2}\int d^\mc{D}\bo{x}d^\mc{D}\bo{y} U(\bo{x}-\bo{y})\dagop{\Psi}(\bo{x})\dagop{\Psi}(\bo{y})\op{\Psi}(\bo{x})\op{\Psi}(\bo{y})
.}
  The bosons have mass $m$, move in $\mc{D}$-dimensional space whose coordinates are $\bo{x}$, experience an external conservative potential $V_{\rm ext}(\bo{x})$, and an interparticle potential $U(\bo{r})=U(-\bo{r})$ where $\bo{r}$ is the spacing between two particles. The field operators $\dagop{\Psi}(\bo{x})$ and $\op{\Psi}(\bo{x})$ are creation and destruction operators on bosons at $\bo{x}$, and obey the usual commutation relations
\EQN{
\left[\op{\Psi}(\bo{x}), \dagop{\Psi}(\bo{y})\right] = \delta^\mc{D}(\bo{x}-\bo{y})
.}   
The total number of bosons is 
\EQN{
\op{N} = \int d^\mc{D}\bo{x} \dagop{\Psi}(\bo{x})\op{\Psi}(\bo{x})
.}
In a typical Bose-Einstein condensation experiment with cold trapped alkali-metal gases the trapping potential $V_{\rm ext}$ is parabolic, in general non-spherical.

  Three-body\footnote{Strictly speaking, all $N>2$ body.} processes are not included in the model. While these can be neglected for most contemporary experiments with cold rarefied alkali-metal gases or BECs, they can become significant when density is high enough, or when one operates near a Feshbach resonanace, 
requiring modifications to the model. (For example, three-body recombination into molecules\cite{Kagan-85,Kagan-88} can play an important role in loss of atoms from a BEC).

\section{Reduction to a lattice model}
\label{CH2Lattice}
To allow computer simulations, \eqref{hamiltonian} must be reduced to a lattice form.  
Provided the lattice is sufficiently fine to resolve all features, no change in physical interpretation occurs. 

One issue to keep in mind, however, is that some care must be taken when going from a continuum to a lattice model to make sure that the behavior of the latter at length scales longer than the lattice spacing is the same  --- i.e. that the lattice model has been renormalized.

One way to do this is to impose:
\ENUM{
\item A spatial box size in each dimension $L_d$,
\item A momentum cutoff in Fourier space $k_d^{\rm max}$,
}
where spatial dimensions have been labeled by $d=1,\dots,\mc{D}$. In this case one proceeds by expressing all spatially-varying functions (generically called $f(\bo{x})$) in Fourier modes as
\EQN{\label{fourierexpansion}
  f(\bo{x}) = \left(\prod_d\frac{\sqrt{2\pi}}{L_d}\right)\sum_{\wt{\bo{n}}} \wt{f}_{\wt{\bo{n}}} e^{i\bo{k}_{\wt{\bo{n}}}\cdot\bo{x}},
}
where the index $\wt{\bo{n}} = \left\{\wt{n}_1,\dots,\wt{n}_\mc{D}\right\}$ has an integer component $\wt{n}_d=0,\pm 1,\pm 2,\dots$ for each dimension. The wave vector $\bo{k}_{\wt{\bo{n}}}=\left\{k_1(\wt{n}_1),\dots,k_\mc{D}(\wt{n}_\mc{D})\right\}$  also contains $\mc{D}$ components 
\EQN{\label{kd}
k_d(\wt{n}_d) = 2\pi\wt{n}_d/L_d. 
}

This has imposed the spatial box size, and now we also impose a momentum cutoff $k_d^{\rm max}$ by truncating all terms in \eqref{fourierexpansion} for which any $|k_d(\wt{n}_d)| > k_d^{\rm max}$. That is,
\EQN{\label{fouriertruncation}
  f(\bo{x})\to f_{\rm cut}(\bo{x}) =  \left(\prod_d\frac{\sqrt{2\pi}}{L_d}\right)
\sum_{|\wt{n}_1|\le L_1k_1^{\rm max}/2\pi}\cdots 
\sum_{|\wt{n}_\mc{D}|\le L_\mc{D}k_\mc{D}^{\rm max}/2\pi}
 \wt{f}_{\wt{\bo{n}}} e^{i\bo{k}_{\wt{\bo{n}}}\cdot\bo{x}}.
}
This leaves an $M_d$ point lattice in each dimension, with $M_d=1+2\,\text{int}\left[L_dk_d^{\rm max}/2\pi\right]$, when $\text{int}[\,\cdot\,]$ gives the integer value (rounded down). 

The lattice points in Fourier space $k_d$ have been defined by \eqref{kd}, the function values there are $\wt{f}_{\wt{\bo{n}}}$, while the function values in normal space $f_{\bo{n}}$ are given by the discrete inverse Fourier transform
of the $\wt{f}_{\wt{\bo{n}}}$:
\EQN{\label{latticefunctions}
  f_{\bo{n}} = f_{\rm cut}(\bo{x}_{\bo{n}}).
}
 Here the spatial lattice is $\bo{x}_{\bo{n}}=\left\{x_1(n_1),\dots,x_\mc{D}(n_\mc{D})\right\}$ given by 
\EQN{\label{xd}
x_d(n_d) = n_d\Delta x_d
,}
  with spacing $\Delta x_d=L_d/M_d$ . The index $\bo{n}$ is composed of non-negative integers $\bo{n}=\{n_1,\dots,n_\mc{D}\}\,:\,n_d=0,1,\dots,M_d-1$. 

The lattice model is physically equivalent to the continuum field model provided that
\ENUM{
\item No features occur on length scales $\approx L_d$ or greater.
\item The truncated Fourier components ($\wt{f}_{\wt{\bo{n}}}$ corresponding to any $k_d>k_d^{\rm max}$) are negligible so that for all practical purposes $f(\bo{x})=f_{\rm cut}(\bo{x})$. In other words, no features occur on length scales $\approx \Delta x_d$ or smaller.
}
Conversely, in cases when some processes occur entirely on length scales smaller than the lattice spacings $\Delta x_d$, 
the lattice model may be renormalized in non-trivial ways (so that $f_{\bo{n}}\neq f(\bo{x}_{\bo{n}})$\ ) to still be physically equivalent to the continuum model on lattice spacing length scales. An example is outlined in Section~\ref{CH2Local}.

Proceeding in this manner for the case of extended interparticle interactions \eqref{hamiltonian}, and using the lattice notation of \eqref{fourierexpansion}, \eqref{fouriertruncation}, and \eqref{latticefunctions}, let us  
 define the lattice annihilation operators 
\EQN{\label{adefpsi}
\hat{a}_{\bo{n}} = \op{\Psi}_{\bo{n}}\sqrt{\prod_d \Delta x_d}
,}
which obey the boson commutation relations
\EQN{
\left[\op{a}_{\bo{n}}, \dagop{a}_{\bo{m}}\right] = \delta_{\bo{n}\bo{m}}
.}
The particle number operator at lattice point $\bo{n}$ is
\EQN{
\op{n}_{\bo{n}} = \dagop{a}_{\bo{n}}\op{a}_{\bo{n}}
.}
With these, one obtains the expression:
\EQN{\label{latticeH}
\op{H} \to \sum_{\bo{n}\bo{m}} \left[ \hbar\omega_{\bo{n}\bo{m}}\dagop{a}_{\bo{n}}\op{a}_{\bo{m}} + \frac{1}{2}u_{\bo{n}\bo{m}}\op{n}_{\bo{n}}\left(\op{n}_{\bo{m}}-\delta_{\bo{nm}}\right) \right]
.}
  In this normally ordered expression, the frequencies $\omega_{\bo{nm}}=\omega_{\bo{mn}}^*$ 
come from the kinetic energy and external potential. They produce a local particle number dependent energy, and linear coupling to other sites, the latter arising only from kinetic processes. The frequencies $\omega_{\bo{nm}}$ are explicitly
\EQN{\label{omegadef}
\omega_{\bo{nm}} = \omega_{\bo{mn}}^* = \frac{\delta_{\bo{nm}}}{\hbar}V^{\rm ext}_{\bo{n}} + \frac{\hbar}{2m}\left(\prod_d\frac{\Delta x_d}{\sqrt{2\pi}}\right)K^{(2)}_{\bo{n}-\bo{m}}
,}
where the function $K^{(2)}$ is the $\mc{D}$-dimensional discrete inverse Fourier transform of $|\bo{k}|^2=\sum_d k_d^2$.  That is, 
\EQN{\label{k2def}
K^{(2)}_{\bo{n}} = \left(\prod_d\frac{\sqrt{2\pi}}{L_d}\right)\sum_{\wt{\bo{n}}}|\bo{k}_{\wt{\bo{n}}}|^2 e^{i\bo{k}_{\bo{\wt{n}}}\cdot\bo{x}_{\bo{n}}}
.}

The interparticle potential can also be discretized on the lattice, however there are a few subtleties caused by the finite box lengths $L_d$.  Let us define the lattice mutual interaction strength $U_{\bo{n}}=U_{\rm cut}(\bo{x}_{\bo{n}})$  in the notation of \eqref{latticefunctions} and \eqref{fouriertruncation}. At first glance, the lattice scattering terms  might be $u_{\bo{nm}}=U_{\bo{n}-\bo{m}}$ since $x_d(n_d)\propto n_d$, but there are two issues that complicate matters:
\ENUM{
\item One has only non-negative lattice labels $\bo{n}$ and coordinates $x_d(n_d)$, as seen from \eqref{xd}.
\item The Fourier decomposition on a box of side lengths $L_d$ implicitly assumes periodic boundary conditions, so e.g. particles at lattice points $\bo{n}=0$ and $\bo{m}=\{M_1-1,\dots,M_{\mc{D}}-1\}$ are effectively in very close proximity.
}
  In the continuum field model the interparticle potential is symmetric ($U(\bo{r})=U(-\bo{r})$), so to impose periodicity on length scales $L_d$, the lattice interparticle potential should obey $U_{\bo{n}} = U_{\{M_1-n_1,n_2,\dots,n_{\mc{D}}\}} = \dots = U_{\{n_1,n_2,\dots,M_{\mc{D}}-n_{\mc{D}}\}}$. This symmetry then lets us write the scattering terms as 
\EQN{\label{uuu}
u_{\bo{nm}}=u_{\bo{mn}} = U_{|\bo{n}-\bo{m}|}
,}
where the notation means $|\bo{n}-\bo{m}|=\{|n_1-m_1|\,,\dots,|n_{\mc{D}}-m_{\mc{D}}|\}$.
For the lattice model to be equivalent to an open continuum field model also requires $U_{\bo{n}}$ to be negligible when any $n_d\approx\order{M_d/2}$.

\section{Locally interacting lattice model}
\label{CH2Local}

  The majority of mesoscopic experiments with interacting Bose gases where quantum mechanical features  have been seen 
use alkali-metal atoms (Li, Na, K, Rb, Cs, and also H) in or near the BEC regime (see also Section~\ref{CH2Motivation}), where a significant simplification of the lattice Hamiltonian \eqref{latticeH} can be made.

Broadly speaking, what is required is that only binary collisions be relevant, and that these occur on length scales much smaller than all other relevant length scales of the system, {\it including the lattice spacings $\Delta x_d$}. 
The simple-minded continuum to lattice procedure of the previous Section~\ref{CH2Lattice} no longer suffices, because the interparticle interactions occur at momenta well above $k_d^{\rm max}$. Detailed consideration to this renormalization is given in  Section IV of the aforementioned review by Leggett\cite{Leggett01}, and in even more detail by Dalibard\cite{Dalibard99} and Wiener\etal\cite{Wiener-99}. The reasoning for single-species cold alkali-metal Bose gases is (in brief) as follows:
\ITEM{
\item The leading term in inter-atomic potentials at long range $\gtrsim 5$\AA\ for the alkali-metal atoms is the 
van der Waals interaction $\propto 1/|\bo{r}|^6$. 
\item This potential gives a van der Waals length $r_{\rm vdw}$ and energy $E_{\rm vdw}\sim\hbar^2/mr_{\rm vdw}^2$ characteristic of the atom species, which are the typical extent and binding energy of the largest bound state. For the alkali-metals $E_{\rm vdw}$ is of order $0.1-1$mK, except for hydrogen, which has $E_{\rm vdw}\approx 3$K.
\item The binding energy of these states is much higher than thermal energies at or near the BEC condensation temperature for these atoms (typically, $T_c\approx\order{100{\rm nK}}$). For atom pairs with relative orbital angular momentum $l_{\rm rel}\neq0$, the scattering strength is 
smaller than $s$-wave scattering ($l_{\rm rel}=0$) by a factor of $(k_BT/E_{\rm vdw})^{l_{\rm rel}}$\cite{Leggett01}. Conclusion: for $k_BT\ll E_{\rm vdw}$ only $s$-wave scattering interactions need be considered.
\item For van der Waals potentials, $s$-wave scattering at low energies can be characterized by a single parameter, the $s$-wave scattering length $a_s$ (see e.g. Landau and Lifshitz\cite{LandauLifshitz59}, Sec. 108). This is typically 
of similar size to $r_{\rm vdw}$. (e.g. for ${}^{87}$Rb, $a_s=5.77$nm \cite{Boesten-97}), and can be considered to be roughly the radius scale of an ``equivalent'' hard sphere (For indistinguishable bosons, the scattering cross-section is $8\pi a_s^2$). 
\item If the $s$-wave scattering length is {\it much smaller than all other relevant length scales of the system}, then 
the exact inter-atom interaction can be replaced by a local lattice interaction where $u_{\bo{nm}}\propto\delta_{\bo{nm}}$ in the formalism of \eqref{latticeH}. Relevant length scales include the de Broglie wavelength of the fastest atoms, the interparticle spacing,  the trap size, the transverse thickness (in the case of 2D or 1D systems), and {\it the lattice spacing} $\Delta x_d$. 
}

Following this broad line of argument, it can be shown\cite{Dalibard99,Wiener-99,Leggett01} that for two indistinguishable bosons  in 3D, the  scale-independent coupling constant $g$ is given by
\EQN{\label{gcoup}
g = \frac{4\pi\hbar^2 a_s}{m}
.}

This then leads to a lattice Hamiltonian of the form 
\EQN{\label{deltaH}
\op{H} = \sum_{\bo{n}\bo{m}} \hbar\omega_{\bo{n}\bo{m}}\dagop{a}_{\bo{n}}\op{a}_{\bo{m}} +\hbar\chi\sum_{\bo{n}}\op{n}_{\bo{n}}(\op{n}_{\bo{n}}-1)
,}
with the (scale-dependent) lattice self-interaction strength being
\EQN{\label{chidef}
\chi = \frac{g}{2\hbar\prod_d\Delta x_d} 
.}

In the case of an effectively one- or two-dimensional gas of indistinguishable bosons, the coupling constant 
is instead given by 
\EQN{\label{gdef}
g=\frac{4\pi\hbar^2 a_s}{\lambda_0 m}
} 
where $\lambda_0$ is, respectively,  the effective thickness or cross-section in the transverse (collapsed) dimensions.

\section{Open dynamic equations of motion}
\label{CH2Dynamics}

For a wide range of physical situations the interaction of an open system with its environment can be considered Markovian, and in this situation can be modeled with a master equation, where the evolution of the density matrix is dependent only on its present value, with no time lag effects. Broadly speaking, this means that any information about the system received by the environment via interactions is dissipated much faster than the system dynamics we perceive. Hence, no feedback into the system occurs. 

Under these conditions the equation of motion for the density matrix of the system can be written (in Linblad form) as:
\EQN{\label{dynamixmaster}
\dada{\op{\rho}}{t} &=& \frac{1}{i\hbar}\left[ \op{H},\op{\rho}\, \right] - \Half\sum_j\left[ \dagop{L}_j\op{L}_j \op{\rho} + \op{\rho}\dagop{L}_j\op{L}_j\right] +\sum_j \op{L}_j\op{\rho}\dagop{L}_j 
.}
  The Linblad operators $\op{L}_j$ represent various coupling processes to the environment, and their number depends on the details of the system-environment interactions. Note that $\op{\rho}$ is the reduced interaction picture density matrix of the (Bose gas + environment) system, traced over the environment.
Some examples of common processes are:

Single-particle losses to a standard\footnote{%
The Hamiltonian of the heat bath is taken to be $\op{H} = \hbar\omega_{\rm bath}\sum_j\left(\dagop{b}_j\op{b}_j + \half\right)$ 
with $\op{b}_j$ being the annihilation operator for the $j$th heat bath subsystem. 
}
 zero temperature heat bath with rates $\gamma_{\bo{n}}$ at $\bo{x}_{\bo{n}}$:
\EQN{\label{zerotemplosses}
\op{L}_{\bo{n}} = \op{a}_{\bo{n}}\sqrt{\gamma_{\bo{n}}}
.}

Exchange with a finite temperature heat bath with coupling strength  proportional to $\gamma_{\bo{n}}$ at $\bo{x}_{\bo{n}}$:
\EQNa{\label{templosses}
\op{L}_{\bo{n}} &=& \op{a}_{\bo{n}}\sqrt{\gamma_{\bo{n}}(1+\bar{n}_{\rm bath})},\\
\op{L}'_{\bo{n}} &=& \dagop{a}_{\bo{n}}\sqrt{\gamma_{\bo{n}}\bar{n}_{\rm bath}}
,}
Where the sum in \eqref{dynamixmaster} is taken over both terms in $\op{L}_{\bo{n}}$ and $\op{L}'_{\bo{n}}$.
For the standard  heat bath with energy $\hbar\omega_{\rm bath}$ per particle for all modes, $\bar{n}_{\rm bath}$, the mean number of particles per mode, is given by the Bose-Einstein distribution expression 
\EQN{\label{bedistribution}
\bar{n}_{\rm bath}=1/[\exp(\hbar\omega_{\rm bath}/k_BT)-1]
,}
 and hence for a zero temperature bath $\bar{n}_{\rm bath}=0$.

Two-particle (at a time) losses to the standard zero temperature heat bath\cite{DrummondGardiner80} with rate $\gamma^{(2)}_{\bo{n}}$ at $\bo{x}_{\bo{n}}$:
\EQN{\label{twoparticlelosses}
\op{L}_{\bo{n}} = \op{a}^2_{\bo{n}}\sqrt{\gamma_{\bo{n}}^{(2)}}
.}

Coherent gain from a driving field $\varepsilon_{\bo{n}}$ at $\bo{x}_{\bo{n}}$ can be modeled by adding a term of the form 
\EQN{\label{coherentgain}
\op{H}_c &=& i\hbar\sum_{\bo{n}}\left[\varepsilon_{\bo{n}}\dagop{a}_{\bo{n}} - \varepsilon^*_{\bo{n}}\op{a}_{\bo{n}}\right]
}
to the Hamiltonian.

Details of the derivation of the loss/gain  expressions (like those above, or more general) can be found e.g. in Gardiner\cite{Gardiner83}, Chapter 10, or Louisell\cite{Louisell70,Louisell74}.

\section{Thermodynamic equations of motion}
\label{CH2Thermodynamics}
The (un-normalized) density matrix of a grand canonical ensemble in contact with a bath at temperature $T$ and with chemical potential $\mu$ is given by
\EQN{
  \op{\rho}_u = \exp\left[-(\op{H}-\mu\op{N})/k_BT\right] = e^{-\op{K}\tau}
.}
  In the second expression the ``imaginary time''\footnote{
So called because of the similarity of the left equality of \eqref{thermomasterstub} to \eqref{dynamixmaster} with $t$ replaced by $i\hbar\tau/2$, and $\op{L}_j=0$.
} 
 $\tau=1/k_BT$ and the ``Kamiltonian'' $\op{K}$ have been introduced.

The rate of change of $\op{\rho}_u$ with $\tau$ can be written as\footnote{%
 $[\op{A},\op{B}]_+ = \op{A}\op{B}+\op{B}\op{A}$. }
\EQN{\label{thermomasterstub}
\dada{\op{\rho}_u}{\tau} = -\Half\left[ \dada{\op{K}\tau}{\tau}, \op{\rho}_u\right]_+ = -\dada{\op{K}\tau}{\tau}\op{\rho}_u
,}
provided that 
\EQN{\label{thermomastercondition}
  \left[ \dada{\op{K}}{\tau}, \op{K}\right] = 0
.}

  For the Hamiltonian  here (\,\eqref{latticeH} or \eqref{deltaH}\,), the equation of motion can be written
\EQN{\label{thermomaster}
\dada{\op{\rho}_u}{\tau} = \left[  \mu_e(\tau)\op{N} - \op{H} \right] \op{\rho}_u
,}
where the ``effective'' chemical potential is
\EQN{\label{muedef}
\mu_e(\tau) = \dada{\left[\tau\mu(\tau)\right]}{\tau}
.}
The condition \eqref{thermomastercondition} implies, in this case, that only the chemical potential $\mu(T)$ can  
be temperature dependent\footnote{%
Strictly speaking, a constant (in space) external potential $V^{\rm ext}$ may also be temperature dependent, but this is physically equivalent to a correction to $\mu$. Quantities such as $g$ and $V^{\rm ext}(\bo{x})$ must be constant with $T$ for \eqref{thermomaster} to hold.
}

The utility of this equation stems from the fact that the grand canonical ensemble at $\tau=0$ (i.e. high temperature $T\to\infty$) is known, and given by the simple expression
\EQN{\label{rhouo}
  \op{\rho}_u(0) = \exp\left[-\lambda_n\op{N}\right]
,}
  where 
\EQN{\label{lambdandef}
\lambda_n = -\lim_{\tau\to0}\left[\tau\mu(\tau)\right]
,}
 and the initial mean number of particles per lattice site is 
\EQN{\label{barnexpr}
\bar{n}_0 = \left[e^{\lambda_n}-1\right]^{-1}
.}
This leads to initial conditions for the stochastic simulation that can be efficiently sampled in most cases. Finite temperature results are then obtained by evolving the system forward in $\tau$.

\part{Phase-space distribution methods for many-mode quantum mechanics}
\chapter{Generalized quantum phase-space representations}
\label{CH3}

\section{Introduction}
\label{CH3Intro}
In this chapter a formalism encompassing very general phase-space distributions describing quantum states and their evolution is presented. The correspondence made is between a quantum density operator and a distribution of operators, or ``kernels'', as in \eqref{configurationP}.  The motivation for this is to investigate what distribution properties are essential for tractable mesoscopic first-principles simulations, and to provide a systematic framework in which comparison between methods is simplified. To this end, the following two conditions on the distribution will be kept in mind throughout:
\ITEM{
\item\textbf{Exact correspondence:} A statistical sample of the kernel operators chosen according to the phase-space distribution must approach the exact quantum density matrix in the limit of infinite samples. This property must be maintained 
during evolution via stochastic equations. 
\item\textbf{Variable number linear in system size:} The number of variables needed to specify a kernel (and hence the number of stochastic equations needed to model quantum evolution) must scale linearly with $N$, the number of subsystems. 
}

The formalism is introduced in Section~\ref{CH3Representation}, and correspondence between density matrix and distribution considered. Subsequently, it is shown how the density matrix corresponds to a statistical sample of kernel operators in 
Section~\ref{CH3Stochastic}, while in~\ref{CH3Equations} it is shown how the quantum evolution corresponds to a set of stochastic equations for these.  Along the way, examples are given using the positive P representation, commonly used in quantum optics.  The resulting requirements on the operator kernels for such a scheme are summarized in Section~\ref{CH3Requirements}.

It is also shown that the phase-space distribution methods have two generic properties that are convenient for practical simulations:
\ITEM{ 
\item\textbf{Parallel sample evolution:} The individual stochastic realizations of the kernel operator (which are later to be averaged to obtain observables) evolve independently of each other. This allows straightforward and efficient parallel computation (more detail in Section~\ref{CH3Parallel}).
\item\textbf{All observables in one simulation:} One algorithm is  capable of giving estimates for any/all  observable averages (see Section~\ref{CH3StochasticMoments}).
}

 Some previous generalizations of phase-space representations such as generalized P representations\cite{DrummondGardiner80}, and the (already quite general) discussion of phase-space representations in the article by Drummond and Deuar\cite{DrummondDeuar03}, are contained as special cases of the discussion in this chapter.

\section{Representation of a density matrix}
\label{CH3Representation}

We expand the density matrix 
\EQN{\label{basicform}
  \op{\rho} = \int P(C) \op{\Lambda}(C) dC
,} 
as in \eqref{configurationP}, where $C$ is a set of configuration variables specifying the kernel operator $\op{\Lambda}(C)$. 
  The idea is that if $P$ is real and positive, it can be considered a probability distribution, and we then can approximate the density matrix using $\mc{S}$ samples $\op{\Lambda}(C)$. Each such sample is fully defined by its set of configuration parameters $C$.  

\subsection{Properties of the distribution}
\label{CH3RepresentationProperties}

To be able to interpret the function $P(C)$ as a probability distribution, we require $P$ to be
\ITEM{
\item {\scshape Real.}
\item {\scshape Non-negative:} $P\ge 0$.
\item {\scshape Normalizable:} i.e. $\int P(C) dC$ converges to a finite value.
\item {\scshape Non-singular.} The primary reason why a singular distribution  is a problem is that it may be incapable of being sampled in an unbiased way by a finite number of samples. For example, if one initially has a non-singular distribution $P(t)$ but singularities arise through some dynamical process after a time $t_{\rm sing}$, then a set of samples that were unbiased estimators of $P(t)$ initially ($t\ll t_{\rm sing}$), will generally be incapable of sampling the singularities at $t\ge t_{\rm sing}$ correctly. 

Some singular behavior can, however, be tolerated in initial distributions (e.g. $P(v)=\sum_{j\ll\mc{S}}\delta(v-v^{(j)}_0)$ for some real variable $v$) provided the number of singularities is finite and much less than the number of samples $\mc{S}$. This can even be desirable to achieve a starting set of samples that is compact.
}

If these conditions are satisfied, then 
when each sample (labeled by  $j$) is defined by its own configuration parameters $C^{(j)}$, 
\EQN{\label{limitmanysamples}
  \lim_{\mc{S}\to\infty} \sum_{j=1}^{\mc{S}} \op{\Lambda}(C^{(j)}) = \op{\rho}
.}

 Many well-known distributions of the general form \eqref{basicform} do not satisfy these conditions for general quantum states. For example the Wigner distribution\cite{Wigner32} is commonly negative in some regions of phase space when the system exhibits nonclassical statistics, the complex P distribution\cite{DrummondGardiner80} is not real, and the Glauber-Sudarshan P distribution\cite{Glauber63,Sudarshan63} can be singular\cite{KlauderSudarshan70} (also when nonclassical statistics are present). These distributions are often very useful, but not for many-body stochastic simulations. 

\subsection{Splitting into subsystems}
\label{CH3RepresentationProduct}

   A mesoscopic system will consist of some number $N$ of subsystems. In this thesis the subsystems are usually 
spatial or momentum modes, although other common approaches are to label as subsystems individual particles (if their number is conserved), or orbitals. 
    
A crucial aim, as pointed out at the beginning of this chapter and in Section~\ref{CH1Configurations}, is to have system configurations $C$ specifying each sample contain a number of parameters (variables) that is only linear in $N$. Otherwise any calculations with macro- or mesoscopic $N$ will become intractable due to the sheer number of variables.  Hence, the kernel $\op{\Lambda}$ should be a separable tensor product of subsystem operators\footnote{%
Strictly speaking, $\op{\Lambda}$ could also be a finite sum of ($\ll N$) tensor products of subsystem operators. This may be useful in some situations.}: 
\EQN{\label{separable}
\op{\Lambda} = f_{\rm glob}(C_{\rm glob}) \otimes_{k=1}^N  \op{\Lambda}_k (C_k,C_{\rm glob})
.}
Here the {\it local kernel} for each $k$th subsystem is described by its own set of local configuration variables $C_k$. There may also be some additional ``global'' variables (of number $\ll\order{N}$) in the set $C_{\rm glob}$ affecting global factors $f_{\rm glob}(C_{\rm glob})$ or several subsystems. The full configuration variable set for the kernel is $C=\{C_{\rm glob},C_1,\dots,C_M\}$. 

A separate issue altogether is the choice of basis (i.e. the choice of $\op{\Lambda}_k$) for each subsystem. This  chapter aims to stay general, and choosing $\op{\Lambda}_k$ is deferred to Chapter~\ref{CH5} and later, apart from discussing general features of the $\op{\Lambda}$s necessary for a successful simulation.

\subsection{Dual configuration space of off-diagonal kernels}
\label{CH3RepresentationDual}

Here, arguments will be put forward that for non-trivial simulations the local kernels $\op{\Lambda}_k$ are best chosen to  include off-diagonal operators. 

A density matrix can always be diagonalized in some orthogonal basis, however there are some basic reasons why such a basis 
is not usually suitable for many-mode simulations of the type considered here. 
\ENUM{
\item Firstly, a general quantum state will contain entanglement between subsystems, which precludes writing it as a distribution in separable form \eqref{separable} with only diagonal kernels. 
\item Secondly, while for some situations the state will, after all, be separable, and could be written using only locally diagonal kernel operators, generally the basis that diagonalizes the density matrix will change with time in a nontrivial way. The kernels, on the other hand, do not change. This means that while the initial separable state could, in this case,  be sampled with diagonal local kernels, the subsequent exact quantum evolution could not be simulated using those kernels. 
\item Thirdly,  exceptions to the above arguments occur if one has a system that remains separable while it evolves. Alternatively, if its inseparability has simple time-evolution  that could be found exactly by other means, then this time-evolution could be hardwired into a time-dependent kernel $\op{\Lambda}(t)$ in such a way that \eqref{separable} continues to hold with diagonal (now-time-dependent) local kernels. In such a case, however, there is no point in carrying out time-consuming stochastic simulations of the whole system, when one could just investigate each subsystem separately.   
}

  Still, the above arguments are not a rigorous proof, primarily because non-orthogonal basis sets have not been considered. 
However, no diagonal distribution that can be used to represent completely general quantum states is presently known.
For example the Glauber-Sudarshan P distribution\cite{Glauber63,Sudarshan63}, which has $\op{\Lambda}_k$ as projectors onto local coherent states (which are non-orthogonal), is defined for all quantum states, but it has been shown that when some nonclassical states are present, the distribution $P$ becomes singular\cite{KlauderSudarshan70}, and not amenable to unbiased stochastic simulations.

In summary, it appears that to allow for off-diagonal entangling coherences between subsystems, the local kernels should be of an off-diagonal form
\EQN{\label{localkernel}
\op{\Lambda}_k(C_k,C_{\rm glob}) = \ket{C'_k,C_{\rm glob}}\langle\wt{C}'_k,C_{\rm glob}|
,}
where the $C'_k$ and $\wt{C}'_k$ are subsets of {\it different} independent parameters, and the local parameter set for the kernel is $C_k=\{C'_k, \wt{C}'_k\}$.

Finally, an exception to this off-diagonal conjecture might be the dynamics of closed systems --- which remain as pure states for the whole duration of a simulation. In such a case, one might try to expand the state vector (rather than the density matrix) $\ket{\psi}$ directly, along the lines of 
\EQN{\label{vectorexpansion}
\ket{\psi} = \int P_{\psi}(C_{\psi},\theta)\, e^{i\theta}\ket{C_{\psi}} dC_{\psi}d\theta
.}
  The global phase factor $e^{i\theta}$ is required to allow for superpositions with a real positive $P_{\psi}$.

\subsection{An example: the positive P distribution}
\label{CH3RepresentationPP}

So that the discussion does not become too opaque, let us make a  connection to how this looks in a concrete example, the positive P distribution\cite{Chaturvedi-77,DrummondGardiner80}. This distribution has widely used  with success in quantum optics, and with Bose atoms as well\cite{DrummondCorney99,Drummond-04} in mode-based calculations. It also forms the basis of the gauge P distribution, which will be explained and investigated in Chapter~\ref{CH5} and used throughout this thesis. The positive P local kernel at a lattice point (spatial mode --- of which there are $N$) is
\EQN{\label{ppkernel}
\op{\Lambda}_k = \ket{\alpha_k}_k\bra{\beta^*_k}_k\exp\left(-\alpha_k\beta_k+\Half|\alpha_k|^2+\Half|\beta_k|^2\right)
,}
 where the states $\ket{\alpha_k}_k$ are normalized  coherent states at the $k$th lattice point with amplitude $\alpha_k$, their form given by
\EQN{
 \ket{\alpha_k}_k = \exp\left[ \alpha_k \op{a}_k -\Half|\alpha_k|^2\right] \ket{0}_k
.}
Note: the subscript $k$ on state vectors indicates that they are local to the $k$th subsystem, and $\ket{0}_k$ is the vacuum.
The global function is just $f_{\rm glob}(C_{\rm glob})=1$, and the entire kernel can be written in terms of vectors of parameters $\bm{\alpha}=\{\alpha_1,\dots,\alpha_N\}$ using 
\EQN{
\ket{\bm{\alpha}}=\otimes_{k=1}^N \ket{\alpha_k}_k
} as
\EQN{\label{pplambda}
  \op{\Lambda} = \ket{\bm{\alpha}}\bra{\bm{\beta}^*}\exp\left(
-\bm{\alpha}\cdot\bm{\beta}+\Half|\bm{\alpha}|^2+\Half|\bm{\beta}|^2\right).
}
The sets of parameters are $C_{\rm glob}=\{\,\}, C_k=\{\alpha_k,\beta_k\}$. The kernel is in ``dual'' off-diagonal operator space, and it has been shown\cite{DrummondGardiner80} that any density matrix can be written with real, positive $P(C)$
using this kernel. It has also been shown constructively\cite{DrummondGardiner80} that a non-singular positive P distribution exists, and is given by
\EQN{\label{pprho}
  P_+(\bm{\alpha},\bm{\beta}) = \frac{1}{(4\pi^2)^N}\exp\left(-\frac{1}{4}|\bm{\alpha}-\bm{\beta}^*|^2\right)\bra{\frac{\bm{\alpha}+\bm{\beta}^*}{2}}\op{\rho}\,\ket{\frac{\bm{\alpha}+\bm{\beta}^*}{2}}
.}
There are four real variables per lattice point.

\section{Stochastic interpretation of the distribution}
\label{CH3Stochastic}

To efficiently sample the quantum state we make two correspondences. Firstly, as discussed in Section~\ref{CH3Representation}, the distribution $P(C)$ corresponds to the density matrix $\op{\rho}$. 
The second correspondence, as per \eqref{limitmanysamples}, is between the distribution $P(C)$ that exists in some high-dimensional space, and $\mc{S}$  samples $C^{(j)}$ distributed according to $P(C)$.

\subsection{Calculating observables}
\label{CH3StochasticMoments}

Quantum mechanics concerns itself with calculating expectation values of observables, so the equivalence between it and the stochastic equations for the variables in $C$ rests solely on obtaining the same evolution of observable averages.

Suppose one wishes to calculate the expectation value of observable $\op{O}$ given a set of $\mc{S}$ operator samples $\{\op{\Lambda}(C^{(j)})\}$, with $j=1,\dots,\mc{S}$. (Operationally, one actually has the set of $\mc{S}$ operator parameter sets $\{ C^{(j)} \}$.) For a normalized density matrix, the expectation value is 
$\langle\op{O}\rangle = \tr{\op{O}\op{\rho}}$, however e.g. in the thermodynamic evolution of Section~\ref{CH2Thermodynamics} one has un-normalized density matrices, in which case 
\EQN{
\langle\op{O}\rangle = \frac{\tr{\op{O}\op{\rho}_u}}{\tr{\op{\rho}_u}}
.} 
Using the representation \eqref{basicform} this leads to
\EQN{
\langle\op{O}\rangle = \frac{\int P(C) \tr{\op{O}\op{\Lambda}(C)} dC }{\int P(C) \tr{\op{\Lambda}(C)} dC}
.} 
We can make use of the Hermitian nature of both density matrices and observables to get extra use out of non-Hermitian (off-diagonal) kernels, giving
\EQN{\label{obsintegral}
\langle\op{O}\rangle = \frac{\int P(C) \left( \tr{\op{O}\op{\Lambda}(C)}  + \tr{\op{O}\dagop{\Lambda}(C^*)}\right)dC }
{\int P(C)\tr{\op{\Lambda}(C)+\dagop{\Lambda}(C^*)} dC}
.} 
Actually, one could impose hermiticity on the kernel at the representation level via $\op{\Lambda}\to\half(\op{\Lambda}+\dagop{\Lambda})$, but this can complicate the resulting stochastic equations for $C$ --- so, let us impose this only at the level of the observable moments. 

When samples are taken $P(C)$ is interpreted as a probability distribution, and, lastly, noting that the expectation values of observables are real (since $\op{O}$ is Hermitian), one obtains that the expectation value of an arbitrary observable $\op{O}$ can be estimated from the samples by the quantity
\EQN{\label{observables}
   \bar{O} = \frac{ \average{\re{\tr{\op{O}\op{\Lambda}}}} + \average{\re{\tr{\op{O}\dagop{\Lambda}}}}}
{2\average{\re{\tr{\op{\Lambda}}}}}
,}
where stochastic averages over samples are indicated by $\average{\cdot}$. The correspondence\footnote{Provided there are no boundary term errors (see Chapter~\ref{CH6}).} is 
\EQN{
\lim_{\mc{S}\to\infty} \bar{O} = \langle\op{O}\rangle
.}

\subsection{Assessing estimate accuracy}
\label{CH3StochasticAccuracy}

One expects that the accuracy of the averages in the \eqref{observables} expression using $\mc{S}$ samples will improve as $\sqrt{\mc{S}}$ via the Central Limit Theorem (CLT), since they are just normalized sums over $\mc{S}$ terms. The uncertainty in a mean value $\bar{v}=\average{v}$ calculated with $\mc{S}$ samples can then be estimated by
\EQN{\label{uncertainty}
\Delta \bar{v} = \sqrt{\frac{\average{v^2}-\average{v}^2}{\mc{S}}}
}
at the one $\sigma$ confidence level.

There is, however, a subtlety when several averages over the same variables $C$ are combined as in \eqref{observables}, because the quantities averaged may be correlated. Then the accuracy estimate \eqref{uncertainty} may be either too large or too small. A way to overcome this is \textbf{subensemble averaging}. While the best estimate $\bar{O}$ is still calculated from the full ensemble $\mc{S}$ as in \eqref{observables}, the $\mc{S}$ samples are also 
binned into $\mc{S}_E$ subensembles with $s$ samples in each ($\mc{S}=s\mc{S}_E$). The $s$ elements of each subensemble 
are used as in \eqref{observables} to obtain independent estimates of the observable average $\{ \bar{O}^{(1)},\bar{O}^{(2)},\dots,\bar{O}^{(\mc{S}_E)}\}$, distributed around $\bar{O}$.  One has (approximately)
\EQN{\label{subensembleaverage}
\bar{O} \approx \frac{1}{\mc{S}_E}\sum_j \bar{O}^{(j)}
.}
Given this, the CLT can be applied to the $\bar{O}^{(j)}$ to estimate the uncertainty in the observable estimate at the one $\sigma$ level:
\EQN{\label{subensembleuncertainty}
\Delta\bar{O} = \Delta\langle\op{O}\rangle = \sqrt{\frac{\frac{1}{\mc{S}_E}\sum_j\left(\bar{O}^{(j)}\right)^2 - \frac{1}{\mc{S}_E^2}\left(\sum_j\bar{O}^{(j)}\right)^2}{\mc{S}_E}}
.}

A practical issue to keep in mind is that both the number of samples in a subensemble $s$ and the number of subensembles themselves should be large enough so that both: 1) The subensemble estimates $\bar{O}^{(j)}$ are reasonably close to the accurate value $\bar{O}\approx\langle\op{O}\rangle$ so that \eqref{subensembleaverage} is true, and also 2) that there is enough of them that the right hand side of \eqref{subensembleaverage} has an approximately Gaussian distribution, and the CLT can be applied to obtain \eqref{subensembleuncertainty}.

An issue that arises when $\bar{O}$ involves a quotient of random variable averages as in \eqref{observables}, is that the subensemble size $s$ should be large enough that the denominator is far from zero for all subensembles. Otherwise subensembles for which this denominator is close to zero have an inordinate importance in the final estimate of the mean. More details in Appendix~\ref{APPC}.

\subsection{Calculating non-static observables}
\label{CH3StochasticNonstatic}

The observable estimate \eqref{observables} of Section~\ref{CH3StochasticMoments} implicitly assumes that the explicit form of $\op{O}$ is known {\it a priori}. When comparing with experiment, these are usually all the observable averages one needs. 

Nevertheless, in many theoretical works some other observables that do not fit this mould are considered. Perhaps the most common example of these is the mutual fidelity between two states $\op{\rho}_1$ and $\op{\rho}_2$
\EQN{
F(\op{\rho}_1,\op{\rho}_2) = \tr{\op{\rho}_1\op{\rho}_2} = \frac{\tr{\op{\rho}_{1u}\op{\rho}_{2u}}}{\tr{\op{\rho}_{1u}}\tr{\op{\rho}_{2u}}}
,}
which lies between zero (for two orthogonal pure states) and unity for pure $\op{\rho}_1=\op{\rho}_2$. A commonly considered special case is the purity $F_0(\op{\rho})=F(\op{\rho},\op{\rho})$. 

A difficulty with evaluating such quantities as $F_0(\op{\rho})$ or $F(\op{\rho}_1,\op{\rho}_2)$ on states calculated with stochastic methods is that 
the density matrices are not known exactly, only their estimates in the form of $\mc{S}$ kernel samples. However, one can still proceed by expanding using \eqref{basicform} as $\op{\rho}_j=\int P_j(C_j)\op{\Lambda}_j(C_j)dC_j$ ($j=1,2$) to give
\EQN{
F(\op{\rho}_1,\op{\rho}_2) &=& \frac{\int P_1(C_1) P_2(C_2) \tr{\op{\Lambda}_1(C_1)\op{\Lambda}_2(C_2)}dC_1 dC_2}{\int P_1(C_1) \tr{\op{\Lambda}_1(C_1)}dC_1 \int P_2(C_2) \tr{\op{\Lambda}_2(C_2)}dC_2}
.}
(Note that the kernels ($\op{\Lambda}_1$ and $\op{\Lambda}_2$, respectively) used for representing the two density matrices do not have to be of the same form. The utility of doing so, however, depends entirely on whether the trace of the product of the two kinds of kernels can be evaluated in closed form for arbitrary sets of parameters $C_1$ and $C_2$.)

As before, off-diagonal kernels can be used twice due to the hermiticity of density matrices, giving
\EQN{
F(\op{\rho}_1,\op{\rho}_2) &=&\frac{\int P_1(C_1) P_2(C_2) \tr{(\op{\Lambda}_1(C_1)+\dagop{\Lambda}_1(C_1^*))(\op{\Lambda}_2(C_2)+\dagop{\Lambda}_2(C_2^*))}dC_1 dC_2}
{\int P_1(C_1) \tr{\op{\Lambda}_1(C_1)+\dagop{\Lambda}_1(C_1^*)}dC_1 \int P_2(C_2) \tr{\op{\Lambda}_2(C_2)+\dagop{\Lambda}_2(C_2^*)}dC_2}
\nonumber\\&&.} 
The $P_1(C_1)$ and $P_2(C_2)$ can be interpreted as probability distributions of {\it independently realized} configuration samples. 
Denoting the first set of $\mc{S}_1$ samples as $\{C_1^{(j_1)}\}$, and the second set of $\mc{S}_2$ as $\{C_2^{(j_2)}\}$, one obtains an estimate of the mutual fidelity between the two states represented by those two sets of samples as 
\EQN{\label{fidelityestimate}
\bar{F} &=&\frac{\sum_{j_1,j_2} \re{\tr{\op{\Lambda}_1(C_1^{(j_1)})\left\{\op{\Lambda}_2(C_2^{(j_2)})+\dagop{\Lambda}_2([C_2^{(j_2)}]^*)\right\}}}}
{2\sum_{j_1}\re{\tr{\op{\Lambda}_1(C_1^{(j_1)})}}\sum_{j_2}\re{\tr{\op{\Lambda}_2(C_2^{(j_2)})}}}
}
(remembering that observable averages are real), with the large sample limit
\EQN{
  \lim_{\mc{S}_1\to\infty, \mc{S}_2\to\infty} \bar{F} = F(\op{\rho}_1,\op{\rho}_2)
.}
Note that the sums in \eqref{fidelityestimate} are carried out over {\it all pairs} of samples $C_1^{(j_1)}$ and $C_2^{(j_2)}$. 

If the two states whose mutual fidelity is to be calculated arise in the same simulation, then one can ensure independence of the two sets of samples labeled by $j_1$ and $j_2$ by setting aside half of all simulated samples for the set $\{C_1^{(j_1)}\}$, the other for $\{C_2^{(j_2)}\}$. This separation of samples is always necessary if one wants to estimate purity $F_0$.

Uncertainty estimates can be obtained in a similar subensemble averaging manner as outlined in Section~\ref{CH3StochasticAccuracy}, by working out \eqref{fidelityestimate} for each subensemble, then using the CLT to estimate uncertainty in the mean of the subensemble estimates, by the same procedure as in \eqref{subensembleuncertainty}.

Finally, a major disadvantage of trying to work out fidelity estimates in the above manner from such stochastic simulations is that one must keep all the trajectories in computer memory, so that after all samples have been produced they can be combined in all possible pairs to evaluate the quantities $\tr{\op{\Lambda}_1(C_1^{(j_1)})\op{\Lambda}_2(C_2^{(j_2)})}$. Compared to a simulation that only considers static observables as per \eqref{observables}, this increases the space required by a factor of $\mathcal{S}$.

\subsection{Overcomplete vs. orthogonal bases}
\label{CH3StochasticOvercomplete}

Consider observable calculations when using local orthogonal bases for the local kernel operators $\op{\Lambda}_k$. In a mode formulation, this would imply that the local parameters $C_k$ consist of discrete quantum numbers for the $k$th mode (e.g. occupation), while for a particle formulation $C_k$ could, for example, consist of (continuous) positions of the $k$th particle, since all position eigenstates are orthogonal. Denoting this basis as $\ket{C'_k}_k$, and remembering that off-diagonal kernels should be used (see Section~\ref{CH3RepresentationDual}), a typical local kernel (omitting global parameters $C_{\rm glob}$) will have the form $\op{\Lambda}_k= \ket{C'_k}_k\langle\wt{C}'_k|_k$ as in \eqref{localkernel}, and the local parameter set is $C_k=\{C'_k,\wt{C}'_k\}$.  

Suppose one wants to calculate the expectation value of a local observable $\op{O}=\op{O}_k\otimes_{k'\neq k}\op{I}_{k'}$
that is diagonal\footnote{Or that contains some components diagonal in the basis $\ket{C'_k}_k$.} in this $\ket{C'_k}_k$ basis. The observable estimate expression \eqref{observables} is proportional to averages of quantities like 
\EQN{\label{traceol}
\tr{\op{O}\op{\Lambda}}=\tr{\op{O}_k\op{\Lambda}_k}f_{\rm glob}(C_{\rm glob}) \prod_{k'\neq k}\tr{\op{\Lambda}_{k'}}
.}

If {\it any one} of the local kernel samples $\op{\Lambda}_{k'}$ or $\op{\Lambda}_k$ are non-diagonal, then the whole $N$-subsystem sample contributes nothing to the observable estimate, even if the off-diagonal kernel samples are for a different subsystem than the one in which the local observable is being considered ($k$). In a large system, {\it practically all} samples are likely to have at least one subsystem in which the local kernel is non-diagonal\footnote{Apart from some special cases, of course.}. For example, if a proportion $p_{\rm od}\le 1$ of samples of each subsystem are off-diagonal then only a proportion $(1-p_{\rm od})^N$ will contribute to observable estimates.  The result --- no reasonable estimate of observables despite lots of calculation.  
Similar effects occur for off-diagonal local observables (in this case for a sample to contribute all $k'\neq k$ subsystems must have diagonal kernel samples, while the $k$th subsystem sample must have diagonal $\op{O}_k\op{\Lambda}_k$.), 
and for observables $\op{O}=\otimes_j\op{O}_j$ spanning several subsystems (in this case $\op{O}_j\op{\Lambda}_j$ must be diagonal for all $j$).

A related issue arises in path integral methods (based on particle positions, for example), and it is the reason that the $\mc{M}$ configurations in one ``sample'' must form a ring polymer structure ($C_{(0)}=C_{(\mc{M})}$) if one wants to calculate diagonal observables.  Off-diagonal observables (e.g. momentum distributions in position eigenstate bases) require a separate simulation, which contains some open polymer structures. 

This inefficient sampling is the main reason why simulations based on particle positions, or occupation numbers at lattice sites, are not usually successful in phase-space distribution based methods. In both cases the position eigenstates or Fock occupation number states used are orthogonal. 

To deal with this sampling problem, one can use overcomplete basis sets (in which the basis vectors are not orthogonal) for the description of the subsystems. Then the local off-diagonal kernels $\op{\Lambda}_k$ can usually be explicitly normalized so that $\tr{\op{\Lambda}_k}=1$. The traces averaged in \eqref{observables} become simply 
\EQN{\label{traceoln}
\tr{\op{O}\op{\Lambda}}=\tr{\op{O}_k\op{\Lambda}_k} f_{\rm glob}(C_{\rm glob})
,}
and only local kernel parameters $C_k$ are relevant for the calculation of local observables. (Possibly apart from some global factors, which do not lead to sample contributions decaying exponentially with $N$). 

The majority of this thesis considers methods based on coherent state expansions, which are of this kind.
A non-orthogonal approach for particle-conserving systems that does use a position wavefunction is to write the density matrix as a distribution over off-diagonal projectors onto orbitals (with arbitrary wavefunctions, not necessarily orthogonal) occupied by all $N$ particles, as recently developed by Carusotto\etal\cite{Carusotto-01}. 

Overcomplete expansions go a long way towards working around the problem, but not all non-orthogonal kernels that one might want to use can be normalized. A case in point are kernels that have zero trace for a set of parameters of measure zero. In such a case one should definitely {\it not} normalize (at least not fully) as boundary term errors tend to result (see Chapter~\ref{CH6}). A partial normalization that avoids any singularities in the resulting kernel (potentially caused by normalizing a kernel whose trace tends to zero) may be successful in such cases. However, the inefficient sampling can then  recur. The reason is that there is then an effective ``overlap'' range in configuration space, such that if the ``ket'' and ``bra'' parameters $C'_k$ and $\wt{C}'_k$ differ by more than this range, the resulting kernel will  have minimal weight in observable averages in comparison with more diagonal kernels. If the phase-space distributions of $C'_k$ and $\wt{C}'_k$ have spreads significantly wider than the overlap range, a proportion $p_{\rm od}\approx\order{1}$ of samples again have largely zero weight  in moment calculations, leading again to only $\approx(1-p_{\rm od})^N$ significant samples overall. This situation can be additionally insidious because the weight of the samples is never exactly zero. This leads superficially to  finite averages in the observable expression \eqref{observables}, however if $\mc{S}(1-p_{\rm od})^N\lesssim 1$ there will be rare ``very diagonal'' configurations that are not sampled at all, but contribute a great deal to the averages in the limit $\mc{S}\to\infty$.  Conclusion: Care must be taken with un-normalized kernels to make sure no sampling bias is introduced.

Summarizing, barring possible developments that may be able to work around the sampling problem discussed in this section, it appears to be necessary to use overcomplete basis sets for the local kernels $\op{\Lambda}_k$ to avoid wasting a lot of computing resources on samples that do not end up contributing to observable estimates. It also appears desirable to make the trace of the local kernels normalized to unity if possible.

\subsection{Positive P distribution example continued}
\label{CH3StochasticPP}
Returning to the positive P distribution example of Section~\ref{CH3RepresentationPP}, all operators 
of supported states can be written as linear combinations of the moments of the local annihilation and creation operators 
$\op{a}_k,\,\dagop{a}_k$. Thus, to evaluate any observable it suffices to know how to evaluate an expectation value of a general product of the form 
$\otimes_j \dagop{a}_{L_j}\otimes_{k} \op{a}_{L'_{k}}$ (The $L_j$ and $L'_{k}$ are subsystem labels (not necessarily unique), while the $j$ and $k$ are ``subsystem label counters'').

The coherent states 
forming the overcomplete kernels  are eigenstates of the annihilation operators $\op{a}_k$:
\EQN{\label{ppaket}
\op{a}_k\ket{\alpha_k}_k = \alpha_k\ket{\alpha_k}_k
,} 
and kernels are normalized:
\EQN{
\tr{\op{\Lambda}_k}=1
.}
Now the same procedure as in Section~\ref{CH3StochasticMoments} for \eqref{observables} if followed, omitting only taking the real part of the numerator and denominator because $\otimes\dagop{a}\otimes\op{a}$ is (for convenience) not necessarily a strictly Hermitian observable. This leads to the expectation value estimate
\EQNa{
\langle \otimes_j \dagop{a}_{L_j}\otimes_{k} \op{a}_{L'_{k}} \rangle &=& \lim_{\mc{S}\to\infty} \bar{\{L_j,L'_{k}\}}\\
\bar{\{L_j,L'_{k}\}} &=& \label{ppobservables}\Half\average{ \prod_j\beta_{L_j} \prod_{k}\alpha_{L'_{k}} + \prod_j \alpha_{L_j}^* \prod_{k} \beta_{L'_{k}}^*}
.}
Hermitian observables are constructed by combining the operator products and stochastic averages of \eqref{ppobservables} with their adjoints and complex conjugates, respectively.

For the fidelity calculations of Section~\ref{CH3StochasticNonstatic}, one finds
\SEQN{\label{trlamlam}}{
\tr{\op{\Lambda}(\bm{\alpha}_1,\bm{\beta}_1)\op{\Lambda}(\bm{\alpha}_2,\bm{\beta}_2)} &=& \prod_k
\tr{\op{\Lambda}_{k}(\alpha_{1k},\beta_{1k})\op{\Lambda}_{k}(\alpha_{2k},\beta_{2k})}\\
\tr{\op{\Lambda}_{k}(\alpha_{1k},\beta_{1k})\op{\Lambda}_{k}(\alpha_{2k},\beta_{2k})}
&=& \exp\left[ -(\alpha_{1k}-\alpha_{2k})(\beta_{1k}-\beta_{2k})\right]\\
\tr{\op{\Lambda}_{k}(\alpha_{1k},\beta_{1k})\dagop{\Lambda}_{k}(\alpha^*_{2k},\beta^*_{2k})}
&=& \exp\left[ -(\alpha_{1k}-\beta^*_{2k})(\beta_{1k}-\alpha^*_{2k})\right]
}
for two kernel samples described by variables $\alpha_{1k},\,\beta_{1k}$ and $\alpha_{2k},\,\beta_{2k}$ respecively.

The coherent state basis is non-orthogonal: 
\EQN{\label{cohorthog}
\braket{\beta^*}{\alpha}=e^{\alpha\beta-\half|\alpha|^2-\half|\beta|^2}
,} and overcomplete: 
\EQN{\label{cohovercomp}
\int \ket{\alpha}\!\bra{\alpha}d^2\alpha = \op{I}\pi
.}
 (Note: the notation $d^2v=d\re{v}d\im{v}$ is used.)

\subsection{Non-uniqueness of distributions}
\label{CH3StochasticNonunique}

  When kernels with distinct parameters are non-orthogonal, as is the case when using local overcomplete basis sets, a given quantum state $\op{\rho}$ can be described by a whole family of different distributions $P(C)$. 

  A simple example is the vacuum state with the aforementioned positive P distribution. 
Any density matrix that satisfies 
\EQN{\label{vacuumidentity}
  \bra{0}\op{\rho}\ket{0} = \tr{\op{\rho}}
}
must be the vacuum state $\ket{0}\bra{0}$, since \eqref{vacuumidentity} specifies that the population of any non-vacuum states must be zero. Expanding as in \eqref{basicform} gives
\EQN{
  \int P(C) \bra{0}\op{\Lambda}(C)\ket{0} dC = \int P(C) \tr{\op{\Lambda}(C)} dC
,}
which for the positive P distribution  is (using $\int P(C) dC = 1$) 
\EQN{\label{vacuumpp}
  \int P_+(\bm{\alpha},\bm{\beta})\,e^{-\bm{\alpha}\cdot\bm{\beta}}\, d^{2N}\!\bm{\alpha}\,d^{2N}\!\bm{\beta} = 1
,}
A class of P distributions that satisfy this are Gaussian ensembles of $\alpha_k=\beta_k$ for each subsystem $k$:
\EQNa{\label{vacpp}
  P_+(\bm{\alpha},\bm{\beta}) &=& \prod_k P_k(\alpha_k,\beta_k),\nonumber\\
  P_{+k}(\alpha_k,\beta_k) &=& \delta^2(\alpha_k-\beta_k) \left(\frac{1+2\sigma_k^2}{2\pi\sigma_k^2}\right)e^{-|\alpha_k|^2/2\sigma^2_k}
}
for arbitrary positive real standard deviations $\bm{\sigma}=\{\sigma_1,\dots,\sigma_N\}$. This can be checked by substitution into \eqref{vacuumpp}. The most compact such distribution is of course in the limit of $\bm{\sigma}\to0$, when
\EQN{
  P_+(\bm{\alpha},\bm{\beta}) = \delta^{2N}(\bm{\alpha})\delta^{2N}(\bm{\beta})
.}

One consequence of this is that the variables $\alpha_k$ and $\beta_k$ do not necessarily correspond to any physical observable. For example the ``coherent state amplitudes'' $\alpha_k$ may be highly nonzero, despite representing the vacuum.  
Only when they appear in the combinations allowed by \eqref{observables} is there a physical meaning. As an example, the estimator of occupation number $\langle\op{n}_k\rangle$ is given, from \eqref{observables}, by
\EQN{
\bar{n}_k &=& \average{\,\re{\alpha_k\beta_k}}\nonumber\\
&=& \int P_+(\bm{\alpha},\bm{\beta})\, \re{\alpha_k\beta_k} \,d^{2N}\!\bm{\alpha}\,d^{2N}\!\bm{\beta} \nonumber\\
&=& \int P_{+k}(\alpha_k,\beta_k)\,\re{\alpha_k\beta_k} d^2\alpha_k d^2\,\beta_k \nonumber\\
&\propto& \int e^{-|\alpha_k|^2/2\sigma_k^2}(\,\re{\alpha_k}^2-\im{\alpha_k}^2)\, d^2\alpha_k \nonumber\\
&=& \int e^{-\re{\alpha_k}^2/2\sigma_k^2}\ e^{-\im{\alpha_k}^2/2\sigma_k^2}\, (\,\re{\alpha_k}^2-\im{\alpha_k}^2) \,d\re{\alpha_k}\, d\im{\alpha_k} \nonumber\\
&=& 0\
,}
the last line following from $\re{\alpha_k}\leftrightarrow\im{\alpha_k}$ symmetry.
In practice, however, it is not irrelevant which distribution \eqref{vacpp} is used, because the low-$\bm{\sigma}$ distributions are much more compact and thus give much better accuracy when calculating observable estimates from a finite sample. 

Generalizing, some  consequences of the non-uniqueness of distributions based on overcomplete basis sets are expected to be:
\ITEM{
\item The variables in the sampled set $C$ will not necessarily correspond to any physical observables. Only when they appear in the observable combinations allowed by \eqref{observables} is there a physical meaning. Because of this, such distributions are then often said to be in a ``non-classical'' phase space. (Also because of the number of variables (at least 4) specifying a local off-diagonal kernel (classically only two - e.g. position and momentum)\,). 
\item The same state may be represented by many different distributions of the variables. However, some of these will give more precise estimates of observables than others\footnote{Given the same number of samples.}.
\item Some distributions representing the same state may be narrower than others in certain directions of phase space, while broader in other directions.  This means that a distribution may be better than another for estimating one observable, but worse for a different one. 
\item Such differences in practical efficiency open the way for various optimizations, which here will be called stochastic gauges. See Chapters~\ref{CH4} and~\ref{CH7}.
\item It is generally more difficult to obtain an explicit expression for $P(C)$ given an initial state. If the kernel is orthogonal in a basis $\ket{C'}$, thus $\op{\Lambda}(C=\{C',\wt{C}'\})=\ket{C'}\langle\wt{C}'|$, obtaining the unique $P(\op{\rho})$ is easy: 
$P(C)=\bra{C'}\op{\rho}\,|\wt{C}'\rangle$. In an overcomplete basis, one can obtain expressions such as \eqref{pprho}, but these are usually only one of many possible.
}

\section{Equations of motion}
\label{CH3Equations}

  To simulate the evolution of the state one must make correspondences to the quantum master equation\footnote{Or, possibly different non-Markovian kinds of quantum evolution equations. In this thesis, however, only master equations are considered.}, analogous to those made for the density operator in the previous sections. The master equation results in a differential equation\footnote{A partial differential equation if $C$ is continuous, or a set of coupled (possibly partial) differential equations for the various probabilities $P(C)$ if $C$ is discrete.} for $P(C)$. This can then often be made to correspond to stochastic equations for the configuration variables $C$, which specify the evolution of the $\mc{S}$ initial samples.

  Now, while some equation for $P(C)$ can always be obtained, in its raw form it is not very useful, because solving it directly for $P(C,t)$ may be as (or even more) computationally intensive as solving the master equation directly.
For example for continuous variables, one would try to evaluate the evolution of $P(C)$ on a lattice of (say) $M_{\rm latt}$ points per variable. Since the number of variables in  the set $C$ is linear in $N$, the number of lattice points on which to evaluate the evolution of $P(C)$ would scale as $M^N_{\rm latt}$ --- exponential in $N$, just like a brute force density matrix element approach.

   One needs to make a correspondence between the evolution of $P(C)$ and stochastic evolution of $C$, its samples.
This is not in general a trivial (or perhaps even a feasible) exercise, but when $P(C)$ obeys one of the broad class of Fokker-Planck second order partial differential equations (FPE), there is a well-known ``textbook'' method\cite{Gardiner83,GardinerQN} to obtain 
stochastic equations for $C$. This is the approach that will be henceforth considered here. 

It is worth noting that there  may  also be ways of obtaining stochastic equations for the configuration variables $C$ 
for different kinds of equations. One direction that has been recently investigated by Olsen\etal\cite{Olsen-02} are differential equations for $P(C)$ involving third order partial derivative terms. The resulting equations have stochastic terms with different statistics than the Wiener increments that will be used here.  The results are promising but there appear to be serious numerical stability problems.

\subsection{Master equation to Fokker-Planck equation}
\label{CH3EquationsFPE}

Let us investigate in more detail what exactly is required to obtain an FPE for $P$. These are of the general form 
\EQN{\label{FPE}
\dada{}{t}P(C) = -\sum_j\dada{}{C_j}\left[\,A_j(C) P(C)\,\right] +\Half\sum_{jk}\frac{\partial^2}{\partial C_j \partial C_k} \left[\,D_{jk}(C) P(C)\,\right]
.}
A point on notation: here, the indices $j, k$ will label all system configuration variables $C_j$ consecutively, {\it not} implying any relationship between $C_j$ and a $j$th subsystem.

The first (rather trivial, but significant) comment is that since an FPE has terms of the differential form $\partial P(C)/\partial C_j$, the system variables $C_j$ should all be continuous so that partial derivatives can be defined. 

Additionally, not all combinations of master equations and operator kernels can lead to Fokker-Planck equations. Let us investigate what the exact requirements are. 
All master equations consist of terms $\mc{T}_l$ linear in the density operator, and can be written as
\EQN{\label{mastertermsum}
\dada{\op{\rho}}{t} = \sum_l \mc{T}_l\left[\,\op{\rho}\,\right]
,}
with the $\mc{T}_l$ composed of combinations of system operators $\op{A}_j$ and linear in\footnote{Or, equally well, linear in an un-normalized $\op{\rho}_u$.} $\op{\rho}$. 
When the kernel depends on continuous parameters, such system operators and the kernel can often be found  to satisfy mutual second-order differential identities of the forms
\SEQN{\label{generaloperatoridentity}}{
  \op{A}_j\op{\Lambda}(C) &=& \left[A_j^{(0)}(C) + \sum_k A_{jk}^{(1)}(C) \dada{}{C_k} + \sum_{kl}A^{(2)}_{jkl}(C) \dada{}{C_k}\dada{}{C_l}\right]\op{\Lambda}(C)\qquad\\
  \op{\Lambda}(C)\op{A}_j &=& \left[\wt{A}_j^{(0)}(C) + \sum_k \wt{A}_{jk}^{(1)}(C) \dada{}{C_k} + \sum_{kl}\wt{A}^{(2)}_{jkl}(C) \dada{}{C_k}\dada{}{C_l}\right]\op{\Lambda}(C)\qquad
} 
where all differential operators are henceforth defined to act on the right.
(a well-known example is in \eqref{ppoperatoridentities} for the positive P representation.)
If, however, higher than second order differential terms must also be included, no FPE can be obtained. 

Assuming \eqref{generaloperatoridentity} hold, by substitution of them into the forms of the $\mc{T}_l(\{\op{A}_j\},\op{\rho})$, differential expressions for 
$\mc{T}_l[\,\op{\Lambda}\,]$ can be found (note that now $\mc{T}_l$ acts on the kernel rather than the density matrix itself). Provided that the master equation does not contain terms of too high order in operators for which the differential terms are nonzero, then the operation of each term on the kernel will be able to (after some algebra) be written in a form:
\EQN{\label{termoperatoridentity}
  \mc{T}_l[\,\op{\Lambda}\,] &=& \left[T_l^{(0)}(C) + \sum_j T_{lj}^{(1)}(C) \dada{}{C_j} + \Half\sum_{jk}T^{(2)}_{ljk}(C) \dada{}{C_j}\dada{}{C_k}\right]\op{\Lambda}
.}
The factor $\half$ in second order terms is introduced for later convenience.

It is crucial that this expansion in partial derivatives terminates at second order for all terms $\mc{T}_l$, otherwise the final partial differential equation for $P$ will not be of the FPE form \eqref{FPE}. This usually creates a restriction on the complexity of processes in the model that can be considered. For example in the gauge P representation used in later parts of this thesis, one-particle and two-particle interactions are fine, but three-particle collisions cannot be treated. This restriction can often be surmounted if one increases the complexity of the kernel, at the cost of additional complexity in the equations.

Having confirmed that the identities \eqref{termoperatoridentity} hold with no 3rd or higher derivative terms, the FPE for $P$ can be derived as follows. Firstly, note that only kernels that are not explicitly time-dependent are being considered here, although time-dependent kernels can also be treatable by a similar, but more voluminous approach.
Having said this, the master equation expanded as \eqref{mastertermsum} can be written using \eqref{termoperatoridentity} as
\EQN{\label{masterinoperators}
\int \op{\Lambda}(C) \dada{P(C)}{t}\,dC &=& \sum_l\int P(C) \\
&&\times\left[T_l^{(0)}(C) + \sum_j T_{lj}^{(1)}(C) \dada{}{C_j} + \Half\sum_{jk}T^{(2)}_{ljk}(C) \dada{}{C_j}\dada{}{C_k}\right]\op{\Lambda}(C)\, dC\nonumber
.}
We now integrate the right hand side by parts, which gives 
\EQN{\label{operatorfpe}
\int \op{\Lambda}(C) \dada{P(C)}{t}\, dC &=& \sum_l\int \op{\Lambda}(C)\\
&&\times\left[T_l^{(0)}(C) - \sum_j \dada{}{C_j} T_{lj}^{(1)}(C) + \Half\sum_{jk}\dada{}{C_k}\dada{}{C_j} T^{(2)}_{ljk}(C) \right]P(C)\,dC\nonumber
}
{\it provided that boundary terms can be discarded}. This last is unfortunately not always true, most typically when the distribution $P(C)$ has power-law tails as the $|C_j|$ head to infinity. Chapter~\ref{CH6} is devoted to this issue and ways to ensure that these boundary terms are forced to zero using the stochastic gauge technique. Section~\ref{CH6FirstOrigin} derives the exact form of the boundary terms \eqref{btexpression}, which must be zero for the derivation to succeed. Also, there have been developed  numerical tests that allow one to check if nonzero boundary terms may be a problem\cite{Gilchrist-97}. 

Now, there may be many $P(C)$ that satisfy \eqref{operatorfpe}, but {\it one} certainly is 
\EQN{\label{almostfpe}
\dada{P(C)}{t}  = \sum_l\left[T_l^{(0)}(C) - \sum_j \dada{}{C_j} T_{lj}^{(1)}(C) + \Half\sum_{jk}\dada{}{C_k}\dada{}{C_j} T^{(2)}_{ljk}(C) \right]P(C) 
.}
This is almost an FPE, except for the possible terms constant in $P$. These typically appear when the normalization of $\op{\rho}$ is not conserved. It turns out that in many cases the constant terms are zero, and one can directly obtain a Fokker-Planck equation of the form \eqref{FPE}, noting the correspondences for the drift vector and diffusion matrices
\SEQN{\label{fpead}}{
A_j(C) &=& \sum_l T_{lj}^{(1)}(C) \label{ajequals}\\
D_{jk}(C) &=& \sum_l T_{ljk}^{(2)}(C)
.}
Nonzero constant terms can be easily treated using a gauge method described in Section~\ref{CH4Weighting}.

\subsection{Fokker-Planck equation to stochastic equations}
\label{CH3EquationsLangevin}

An appropriate set of diffusive stochastic Langevin equations for the variables $C_j$ is known to be equivalent to the FPE for the distribution $P(\{C_j\})$ in the limit of infinitely many samples of the variables $C_j$. This textbook topic is described in detail e.g. in Gardiner\cite{Gardiner83,GardinerQN}. An important qualification is that this correspondence only holds if the diffusion matrix $D_{jk}(C)$ when written for real variables $C_j$ is {\it positive semidefinite}\footnote{i.e. has no eigenvalues with negative real part.}.  

If this is the case, then Langevin equations for the (real) $C_j$, equivalent to the FPE \eqref{FPE},  are 
\EQN{\label{langevin}
 dC_j(t) = A_j(C,t)\, dt + \sum_k B_{jk}(C,t)\, dW_k(t)
}
in the Ito calculus.
Here, the {\it noise matrices} $B_{jk}$ are any matrices that satisfy the diffusion matrix decomposition
\EQN{\label{dbbt}
  D = B B^T
,}
i.e. in terms of matrix elements
\EQN{
  D_{jk} = \sum_l B_{jl} B_{kl}
.}
  The $dW_k(t)$ are real stochastic Wiener increments (i.e quantities satisfying the large sample ($\mc{S}\to\infty$) averages)
\SEQN{\label{Wiener}}{
  \average{ dW_j(t) } &=& 0\\
  \average{ dW_j(t) dW_k(t') } &=& \delta_{jk} \delta(t-t') dt^2
.}
For the small but finite time steps $\Delta t\to dt$ in a calculation, $dW_k(t)$ are usually implemented by real Gaussian noises\footnote{Other noise distributions are also possible, simply provided the conditions \eqref{Wiener} are satisfied. Since by the CLT the sum of several infinitesimal noises will always approach an infinitesimal Gaussian-distributed noise anyway, it is usually more efficient to start with a Gaussian noise right away, rather then make one the hard way by summing several non-Gaussian noises.} independent for each $k$ and each time step, having mean zero, and variance $\Delta t$. 
Some more detail of the numerical implementation of stochastic equations with Wiener increments can be found in Appendix~\ref{APPB}.
Lastly, it was chosen to write the equations immediately in difference form, as this is the form in which actual computer calculations are made. 

Now, clearly all the coefficients $A_j$ and $B_{jk}$ must be real, since the equations are for real variables $C_j$.
However, the decomposition \eqref{dbbt} in terms of real $B$ is only possible if the symmetric diffusion matrix $D(C)$ is positive semidefinite. This explains the reason for the positive semidefinite condition on $D$.

Finally, strictly speaking, a more general set of stochastic equations that remains equivalent to the FPE \eqref{FPE} is 
possible, and will be of use in later parts of this thesis.
Consider that only the means and variances \eqref{Wiener} of the Wiener increments are specified. If one writes \eqref{langevin} as
\EQN{\label{langevinx}
 dC_j(t) = A_j(C,t)\, dt + dX_j(C,t)
}
then by the properties of Ito stochastic calculus, the only binding relationships for the stochastic terms $dX_j$ are that 
\SEQN{\label{2cond}}{
  \average{dX_j(t)} &=& 0,\label{meancond}\\
  \average{dX_j(t)\,dX_k(t')} &=& \average{D_{jk}} \delta(t-t') dt^2\label{varcond}
.}
Indeed, $dX_j(t)$ need not even be Markovian (i.e. independent of quantities at times $t'\neq t$), provided the conditions \eqref{2cond} are satisfied. The relationship between the Fokker-Planck equation \eqref{FPE} and stochastic equations \eqref{langevinx} is considered in a more rigorous fashion and covered exhaustively in the texts by Gardiner\cite{Gardiner83,GardinerQN}, but equations \eqref{langevin} through to \eqref{2cond} will be sufficient for the considerations of this thesis.

\subsection{Ensuring positivity of diffusion for analytic kernels}
\label{CH3EquationsAnalytic}
   In the important case of a kernel that can be written as analytically dependent on complex parameters $z_j$ (rather than on real $C_j$)\footnote{%
i.e. $\op{\Lambda}$ depends on $z_j$ but not on any $z_j^*$.}, then we can use the freedom of interpretation of derivatives with respect to complex arguments to choose such a Fokker-Planck equation that the diffusion matrix $D$ is always positive semidefinite, and Langevin stochastic equations \eqref{langevin} can always be derived. 
The procedure described below is simply an application of the same method used for the special case of a positive P distribution  by Drummond and Gardiner\cite{DrummondGardiner80}.

This freedom of derivative choice arises from the property of all analytic functions $f(z_j)$ that 
\EQN{\label{complexderivatives}
  \dada{f(z_j)}{z_j} = \dada{f(z_j)}{z'_j} = -i\dada{f(z_j)}{z''_j}
.}
(Where real and imaginary parts of the variable have been denoted for brevity by the shorthand $z_j=z'_j+iz''_j$.)

The derivation in Section~\ref{CH3EquationsFPE}  follows through formally without change if the real variables $C_j$ are replaced with the complex $z_j$, the set of parameters now being $C=\{z_j\}$. 
Using \eqref{complexderivatives} the terms in first order derivatives of the kernel in \eqref{termoperatoridentity} can now be equated to 
\EQN{\label{forderterms}
 T_{lj}^{(1)}(C)\dada{}{z_j}\op{\Lambda}(C) = \left[\re{T_{lj}^{(1)}(C)}\dada{}{z'_j} + \im{T_{lj}^{(1)}(C)}\dada{}{z''_j}\right]\op{\Lambda}(C)
.}
To deal with the second order terms, one can decompose $\sum_lT_{ljk}^{(2)} = D_{jk} =  \sum_{p}B_{jp}B_{kp}$ into the noise matrix forms (which will in general be complex), use the shorthand \mbox{$B=B'+iB''$}, and express them (all together) using \eqref{complexderivatives} as 
\EQN{\label{sndorderterms}
\lefteqn{ \sum_l T_{ljk}^{(2)}(C)\dada{}{z_j}\dada{}{z_k}\op{\Lambda}(C) = \sum_p B_{jp} B_{kp} \dada{}{z_j}\dada{}{z_k}\op{\Lambda}(C)}&\nonumber\\
& = \sum_p \left[
B'_{jp} B'_{kp} \dada{}{z'_j}\dada{}{z'_k}
+ B'_{jp} B''_{kp} \dada{}{z'_j}\dada{}{z''_k} 
+B''_{jp} B'_{kp} \dada{}{z''_j}\dada{}{z'_k}
+ B'_{jp} B''_{kp} \dada{}{z''_j}\dada{}{z''_k}
\right]\op{\Lambda}(C)
.\nonumber\\}

Let us see what has happened to the diffusion matrix for real variables, as it is {\it this} that must be positive semidefinite for Langevin equations \eqref{langevin} to be obtained. 
While the indices $j,k,p = 1,\dots,N_z$ have referred to complex variables (of number $N_z$), let us also define real variables $v_{j'}$ (with primed indices $j',k',p'=1,\dots,2N_z$) by $z'_j=v_j$ and $z''_j$ = $v_{(j+N_z)}$. Denoting the diffusion and noise matrices of the real variables $v_{j'}$ as $D^{(v)}$ and $B^{(v)}$, respectively, 
we have $D^{(v)}_{j',k'}=\sum_{p'=1}^{2N_z}B^{(v)}_{j',p'}B^{(v)}_{k',p'}$, and from \eqref{sndorderterms}:
\SEQN{\label{canonicalnoisematrix}}{
  B^{(v)}_{j,p} &=& \re{B_{jp}}, \\
  B^{(v)}_{j+N_z,p} &=& \im{B_{jp}}, \\
  B^{(v)}_{j,p+N_z} &=& B^{(v)}_{j+N_z,p+N_z} = 0
,}
which are all explicitly real. Thus, the diffusion matrix is explicitly positive semidefinite, and Langevin equations for all the $v_{j'}$ can be obtained. 

From \eqref{langevin}, \eqref{forderterms}, \eqref{ajequals}, and \eqref{canonicalnoisematrix} the resulting stochastic equations for the real variables then are
\SEQN{}{
dv_j(t) &=& \re{A_j(C,t)}\,dt + \sum_k B^{(v)}_{j,k}(C,t)\,dW_k(t)\\
dv_{j+N_z}(t) &=& \im{A_j(C,t)}\,dt + \sum_k B^{(v)}_{j+N_z,k}(C,t)\,dW_k(t),
}
which can be written in shorthand form for the complex variables $z_j$ as 
\EQN{\label{langevinc}
 dz_j(t) = A_j(C,t)\, dt + \sum_k B_{jk}(C,t)\, dW_k(t)
.}
Note that the Wiener increments $dW_k$ remain real.

The only requirement (apart from obtaining an FPE) to carry out this procedure, which ensures a stochastic equation interpretation, is that the kernel be written as an analytic function of complex variables. 

Finally, the procedure described in this section can be carried out on only a part of the total set of variables $C$, if the kernel can be made analytic in only this part. Provided the resulting combined noise matrix $B^{(v)}$ for all the real variables of the system (real and imaginary parts of the complex variables, and the remaining real variables) is real, the diffusion matrix is positive semidefinite, and Langevin equations for all the variables can still be obtained in the same way as here.

%

\subsection{Positive P example:  dilute lattice Bose gas equations}
\label{CH3EquationsPP}

Continuing with the positive P representation example, let us obtain the stochastic equations corresponding to the quantum dynamics of the lattice interacting Bose gas Hamiltonian \eqref{deltaH} with single-particle losses to a zero temperature heat bath. The master equation is \eqref{dynamixmaster} with Linblad operators  \eqref{zerotemplosses}, and can be written in terms of $\op{\rho}$ and the creation and destruction operators $\op{a}_{\bo{n}}$,\,$\dagop{a}_{\bo{n}}$, which can be identified with the $\op{A}_j$ of Section~\ref{CH3EquationsFPE}. 

These can be easily verified (via \eqref{pplambda} and \eqref{ppaket}, and the mutual commutation of local kernels $\op{\Lambda}_{\bo{n}}$) to obey the identities 
\SEQN{\label{ppoperatoridentities}}{
\op{a}_{\bo{n}}\op{\Lambda} &=& \alpha_{\bo{n}}\op{\Lambda}, \\
\dagop{a}_{\bo{n}}\op{\Lambda} &=& \left(\beta_{\bo{n}}+\dada{}{\alpha_{\bo{n}}}\right)\op{\Lambda}, \\
\op{\Lambda}\dagop{a}_{\bo{n}} &=& \beta_{\bo{n}}\op{\Lambda}, \\
\op{\Lambda}\op{a}_{\bo{n}} &=& \left(\alpha_{\bo{n}}+\dada{}{\beta_{\bo{n}}}\right)\op{\Lambda}
}
for all $N$ lattice points $\bo{n}$.
 	It can be seen that to obtain an FPE there can be up to two operations of the kind $\dagop{a}\op{\rho}$ or $\op{\rho}\op{a}$ per term in the master equation, and any  number of $\op{a}\op{\rho}$ or $\op{\rho}\dagop{a}$. This 
allows any one- or two-particle processes but rules out three-particle processes, which contain terms of
third order in $\op{a}$ or $\dagop{a}$ in the Hamiltonian.

The terms in the master equation can be sorted (according to processes) as
\SEQN{}{
\mc{T}_1[\,\op{\rho}\,] &=& -i\sum_{\bo{nm}}\omega_{\bo{nm}}\left[ \dagop{a}_{\bo{n}}\op{a}_{\bo{m}}\op{\rho} - \op{\rho}\dagop{a}_{\bo{n}}\op{a}_{\bo{m}}\right],\\
\mc{T}_2[\,\op{\rho}\,] &=& -i\chi\sum_{\bo{n}} \left[ \dagop{a}_{\bo{n}}{}^2\op{a}_{\bo{n}}^2\op{\rho} - \op{\rho}\dagop{a}_{\bo{n}}{}^2\op{a}_{\bo{n}}^2\right],\\
\mc{T}_3[\,\op{\rho}\,] &=& \sum_{\bo{n}}\gamma_{\bo{n}}\left[ \op{a}_{\bo{n}}\op{\rho}\dagop{a}_{\bo{n}} -\Half\dagop{a}_{\bo{n}}\op{a}_{\bo{n}}\op{\rho} -\Half\op{\rho}\dagop{a}_{\bo{n}}\op{a}_{\bo{n}}\right].
}
Which, using \eqref{ppoperatoridentities}, gives 
\SEQN{}{
\mc{T}_1[\,\op{\Lambda}\,] &=& -i\sum_{\bo{nm}}\omega_{\bo{nm}}\left(
\alpha_{\bo{m}}\dada{}{\alpha_{\bo{n}}} - \beta_{\bo{n}}\dada{}{\beta_{\bo{m}}}\right)\op{\Lambda},\\
\mc{T}_2[\,\op{\Lambda}\,] &=& -i\chi\sum_{\bo{n}} \left(
2\alpha_{\bo{n}}^2\beta_{\bo{n}}\dada{}{\alpha_{\bo{n}}} - 2\alpha_{\bo{n}}\beta_{\bo{n}}^2\dada{}{\beta_{\bo{n}}}
+ \alpha_{\bo{n}}^2\frac{\partial^2}{\partial\alpha_{\bo{n}}^2} - \beta_{\bo{n}}^2\frac{\partial^2}{\partial\beta_{\bo{n}}^2}\right)\op{\Lambda},\\
\mc{T}_3[\,\op{\Lambda}\,] &=& -\Half\sum_{\bo{n}}\gamma_{\bo{n}}\left(
\alpha_{\bo{n}}\dada{}{\alpha_{\bo{n}}} + \beta_{\bo{n}}\dada{}{\beta_{\bo{n}}}\right)\op{\Lambda}
.}

Proceeding in the same manner as in Section~\ref{CH3EquationsAnalytic} (remember that the $\alpha_{\bo{n}}$ and $\beta_{\bo{n}}$ variables are complex), we obtain the (Ito) stochastic equations
\SEQN{}{
d\alpha_{\bo{n}} &=& -i\sum_{\bo{m}}\omega_{\bo{nm}}\alpha_{\bo{m}} dt -2i\chi\alpha_{\bo{n}}^2\beta_{\bo{n}}dt -\frac{\gamma_{\bo{n}}}{2}\alpha_{\bo{n}}dt + i\alpha_{\bo{n}}\sqrt{2i\chi} dW_{\bo{n}}(t),\\
d\beta_{\bo{n}} &=& i\sum_{\bo{m}}\omega_{\bo{mn}}\beta_{\bo{m}} dt + 2i\chi\alpha_{\bo{n}}\beta_{\bo{n}}^2dt -\frac{\gamma_{\bo{n}}}{2}\beta_{\bo{n}}dt + \beta_{\bo{n}}\sqrt{2i\chi} d\wt{W}_{\bo{n}}(t)
.}
Here the $2N$ real Wiener increments $dW_{\bo{n}}(t)$ and $d\wt{W}_{\bo{n}}(t)$ are all independent at each time step.
We have used the ``simplest'' diagonal noise matrix decomposition \mbox{$B_{\bo{nm}} = B_{\bo{nn}}\delta_{\bo{nm}}$} here. More on the possibilities with these decompositions in Chapter~\ref{CH4}.

\section{Convenience for parallel computation}
\label{CH3Parallel}
There is no cross-talk between separate realizations of the stochastic process in the  simulation scheme described in this chapter. This is a highly desirable property for numerical realization  because only one trajectory at a time need be stored in memory. In fact, many independent trajectories could be run at the same time on different computers in a cluster, or even on entirely separate computers, and then collected together at the end. The same applies to calculating observable averages \eqref{observables}, where all one needs is to keep track of the running sum of averaged quantities and their number\footnote{And possibly the sum of the squares of the averaged properties for assessment of uncertainty in the estimate as per \eqref{subensembleuncertainty}.}.

In combination with the linear scaling of variable number this makes for a potentially powerful and efficient combination for many-body simulations.

An exception to this convenient scaling with trajectory number are non-observable quantities such as fidelity, estimated by methods like in Section~\ref{CH3StochasticNonstatic}. In that kind of calculation  we have to keep in memory the variables for all $s$ samples in a subensemble to use all pairs of them in the calculation \eqref{fidelityestimate}. Memory requirements increase by a factor of $s$ to hold the entire subensemble. The actual calculation of estimates like \eqref{fidelityestimate} also takes a time $\propto s^2$ as opposed to $\propto s$ for static observable estimations \eqref{observables}.
Since for reasonable subensemble averages one typically needs \order{100} or more samples to a subensemble, this requires a increase of two orders of magnitude in computing resources --- highly nontrivial. Fortunately, at least the linear scaling $\propto N$ of the number of variables remains untouched. 

\section{Summary of representation requirements}
\label{CH3Requirements}
This chapter has shown or strongly indicated that for a representation to be useful for quantum simulations of many-body systems, it must satisfy the requirements listed below:
\ENUM{
\item\textbf{Non-negative real distribution:} To interpret $P(C)$ as a probability distribution over kernel operators $\op{\Lambda}(C)$, it must be real, normalizable to unity, and non-negative.
\item\textbf{Non-singular distribution:} Needed for finite probabilities and un-biased dynamic sampling, although delta functions in the initial state can be tolerated.
\item\textbf{Complete:} The representation must be able to describe the initial states, and any states reachable by the subsequent evolution, while satisfying the above conditions. In practice this usually implies that the kernel must form a complete or overcomplete basis, and that the kernel be off-diagonal. See Section~\ref{CH3RepresentationDual}. 
\item\textbf{Linear scaling:} The number of system variables in $C$ must scale linearly with the system size $N$  (Typically number of modes, particles, orbitals,\dots). In practice this requires the kernel to be a tensor product over local kernels $\op{\Lambda}_k$ for each subsystem, or a linear combination of a small ($\ll N$) number of such tensor product terms. See Section~\ref{CH3RepresentationProduct}.
\item\textbf{Most samples significant:} The majority of samples (i.e. $\order{\mc{S}}$) should contribute significantly to all observable estimates of interest, particularly observables local to the subsystems.  This appears to require the use of an overcomplete basis, and kernels normalized locally (i.e. $\tr{\op{\Lambda}_k}=1$), perhaps apart from some global multiplying functions. These issues are discussed in 
Section~\ref{CH3StochasticOvercomplete}.
\item\textbf{Finite kernel trace:} All configurations $C$ must lead to a finite kernel trace, otherwise observable averages will be undefined for $\mc{S}\to\infty$ and biased for finite $\mc{S}$ due to divergence of the denominator in \eqref{observables}.
}
If the stochastic interpretation of the quantum evolution is obtained via a Fokker-Planck equation as in Section~\ref{CH3Equations} (presently the only approach that has led to nontrivial simulations), then the following requirements must also be satisfied:
\ENUM{\setcounter{enumi}{6}
\item\textbf{Differential equation:} The configuration parameters $C$ should be continuous to allow a differential equation for $P(C)$.
\item\textbf{Fokker Planck equation:} The mapping $\dada{\op{\rho}}{t}\to\dada{P(C)}{t}$ must generate only first and second order partial derivatives of $P(C)$. See Section~\ref{CH3EquationsFPE}.
\item\textbf{Positive propagator:} The short time diffusion in phase space must be non-negative (i.e. the diffusion matrix for real variables $D^{(v)}$ has no eigenvalues with negative real part), to allow a stochastic interpretation. This can be guaranteed if the kernel is an analytic function of complex variables (see Section~\ref{CH3EquationsAnalytic}), although 
in some situations non-analytic kernels may also lead to satisfactory diffusion.
}
Furthermore, there are several additional requirements that appear generically in simulations, and must be satisfied to avoid bias and/or unmanageable noise in the observable estimates.
\ENUM{\setcounter{enumi}{9}
\item\textbf{Stable trajectories:} If trajectories in phase space diverge rapidly, so does the width of the distribution $P(C)$, and all useful precision in the estimates of observables is lost. 
\item\textbf{Unbiased:} Pathological cases are common (particularly in nonlinear underdamped systems), where the distribution $P(C)$ is too broad to allow unbiased sampling of some or all observable moments --- typically when $P(C)$ develops power-law tails. These ``boundary term'' errors are considered in detail in Chapter~\ref{CH6}, and usually go in tandem with unstable 
trajectories.
\item\textbf{UV convergent:} Numerical simulations almost always require one to approximate space by a finite lattice. This does not change the physical predictions of the model provided the lattice spacing is much finer than any other significant length scale in the model. This, in turn, requires that the effect of vacuum fluctuations on the equations of motion and observable estimates abates at large momentum (i.e. at small length scales). A mode-based lattice calculation will typically be non-UV-convergent if there are noise terms in the equations  that do not disappear at zero mode occupation.
}

\chapter{Stochastic gauges}
\label{CH4}
  The non-uniqueness of representations $P(C)$ leads to many potential sets of stochastic equations representing the same single master equation. In this chapter, the stochastic gauge formalism is introduced, which systematically characterizes the freedoms available in the stochastic equations after the physical model has been completely specified.
The term ``stochastic gauges'' is used here because of an analogy with electromagnetic gauges: In both cases the equations contain arbitrary (in principle) functions, which have no effect on physical observables, but can have a great influence on the ease with which the calculation proceeds. In the case of stochastic gauges,  the choice of gauge has no effect\footnote{In principle. In practice there is also the issue of boundary term errors, which may be present or not, depending on the gauge chosen. This is discussed in detail in Chapter~\ref{CH6}.} on observable quantum expectation values \eqref{observables} calculated in the limit of an infinite number of samples. For a finite number of samples, however, different gauges affect how rapidly this large sample limit is approached. It will be shown that an appropriate gauge choice can improve simulation efficiency by many orders of magnitude, and also correct biases caused by pathological distributions. 

During the derivation of stochastic Langevin equations \eqref{langevin} for the variables $C_j$ from a master equation for the density matrix, there are two distinct places where stochastic gauges can be introduced. Firstly, null differential identities on the kernel $\op{\Lambda}$, multiplied by arbitrary functions $\mc{F}$ and integrated over $P(C)$  can be introduced into the master equation at the point \eqref{masterinoperators}, which leads to $\mc{F}$ finding its way into the FPE, and into the final stochastic equations. These ``kernel stochastic gauges'' are described in detail in Sections~\ref{CH4Kernel} to \ref{CH4Drift}. Secondly, the diffusion matrix of an FPE does not uniquely specify the noise behavior of the resulting stochastic equations. These degrees of freedom can also be used to introduce arbitrary functions $g$ into the noise matrices $B$, as described in detail in Section~\ref{CH4Diffusion}.  A partly restricted set of kernel and diffusion gauges, which will be here dubbed the ``standard'' gauges, and are sufficient for most purposes, are summarized in Section~\ref{CH4Central}. 

As in the previous Chapter~\ref{CH3}, the derivation is kept as general as possible with the aim of sifting out those elements that are truly necessary for stochastic gauges to be present and to be useful. Several authors have recently (all starting in 2001) proposed phase-space distribution methods that make use of the freedoms that are here dubbed stochastic gauges. This includes the ``noise optimization'' of Plimak, Olsen, and Collett\cite{Plimak-01}, the ``stochastic wavefunction'' method of Carusotto, Castin, and Dalibard\cite{Carusotto-01,CarusottoCastin01}, and the ``stochastic gauges'' of Deuar, Drummond, and Kheruntsyan\cite{DeuarDrummond01,DeuarDrummond02,DrummondDeuar03,Drummond-04}.
All these are unified within the formalism presented in this chapter, which is itself a generalization of some ideas to be found in the articles by Deuar and Drummond\cite{DeuarDrummond01} and \cite{DrummondDeuar03}.

\section{Generalized kernel stochastic gauges}
\label{CH4Kernel}

  Whenever there are null differential identities on the kernel of the general form 
\EQN{\label{generalidentity}
  \left\{ J^{(0)}(C) + \sum_j J^{(1)}_j(C)\dada{}{C_j} + \Half\sum_{jk} J^{(2)}_{jk}(C)\frac{\partial^2}{\partial C_j\partial C_k}\right\}\op{\Lambda}(C) = \mc{J}\left[\op{\Lambda}(C)\right] = 0
,}
a stochastic gauge can be introduced in the following manner:

Since \eqref{generalidentity} is zero, so is its integral when multiplied by non-divergent functions. In particular,
\EQN{\label{zeromasterterm}
\int P(C,t) \mc{F}(C,t) \mc{J}\left[\op{\Lambda}(C)\right]\,dC=0
}
with the function $\mc{F}(C,t)$ otherwise unspecified. Now one adds zero (in the form \eqref{zeromasterterm}) to the right hand side of the master equation \eqref{masterinoperators} in the derivation of the Fokker-Planck equation of Section~\ref{CH3EquationsFPE}. 
(Actually, to be fully rigorous, one must make sure that the full integral $\int P (\cdot )\op{\Lambda}$ over the entire master equation \eqref{masterinoperators} plus all ``zero'' terms \eqref{zeromasterterm} is convergent, otherwise boundary term errors may result.  This issue is considered in detail in Chapter~\ref{CH6}, so for now let us defer this and assume that no boundary term errors are present.)

Adding \eqref{zeromasterterm} introduces an additional master equation term that can be written in the notation of Section~\ref{CH3EquationsFPE} as
\EQN{\label{jequalsft}
\mc{T}_{\mc{J}}\left[ \op{\Lambda} \right] = \mc{F}\mc{J}\left[\op{\Lambda}\right]
.}
 Following now the procedure of Sections~\ref{CH3EquationsFPE} and~\eqref{CH3EquationsLangevin}, 
the result is that the coefficients in the Langevin equations \eqref{langevin} become modified to 
\SEQN{\label{newcoeff}}{
  A_j(C) \to& A^{(\mc{F})}_j(C) =& A_j(C) + \mc{F}(C)J^{(1)}_j(C) 	\\
  D_{jk}(C) \to& D^{(\mc{F})}_{jk}(C) =& D_{jk}(C) + \mc{F}(C)J^{(2)}_{jk}(C)
.}
In this manner an arbitrary function $\mc{F}$ has been introduced into the stochastic equations. The number of these is unlimited provided we have at least one identity of the form \eqref{generalidentity}, although there will be at most as many gauges with distinct effects on the equations as we have distinct identities.  

The whole procedure is based on properties \eqref{generalidentity} of the kernel, and so the kernel gauges that can be used in a particular simulation will depend on the local subsystem bases used to expand the density  matrix, and on how the kernel is constructed out of these.

Some (though certainly not all) broad classes of gauges that are possible with various kernels include:
\ITEM{
\item {\scshape Analytic:} The procedure used in Section~\ref{CH3EquationsAnalytic} to ensure a positive diffusive propagator for representations based on kernels analytic in complex variables $z_j$ can be interpreted as using a gauge
of the form 
\EQN{
\left(\dada{}{\re{z_j}} + i\dada{}{\im{z_j}}\right)\op{\Lambda}(z_j) = 0.
}
This applies for any kernel analytic in $z_j$.
\item {\scshape Weighting:} If constant terms are present during the derivation of an FPE as in \eqref{almostfpe} in Section~\ref{CH3EquationsFPE}, they can be converted to deterministic evolution of global weights, provided the kernel contains a global factor $\Omega$, using an identity of the form
\EQN{\label{weightingidentity}
\left(\Omega\dada{}{\Omega} - 1\right)\op{\Lambda} = 0
.}
See Section~\ref{CH4Weighting}, below.
\item {\scshape Drift:} One can modify the deterministic part of the stochastic equations \eqref{langevin} in principle at will, provided compensation is made in the form of appropriate global trajectory weights. Building on the weighting gauge identity \eqref{weightingidentity} for kernels with global weights $\Omega$, the identities
\EQN{\label{generaldriftidentity}
\left(\Omega\dada{}{\Omega}-1\right)\dada{}{C_j}\op{\Lambda} = 0.
}
follow for any variable $C_j$, and can be used for this purpose. See Section~\ref{CH4Drift}.
\item {\scshape Reduction:} For some kernels, identities can be found that reduce second order partial derivatives of the kernel to first order. These can be used at the level of expressions \eqref{termoperatoridentity} for each part $\mc{T}_l$ of the master equation to reduce partial derivative terms obtained with the operator identities \eqref{generaloperatoridentity} to lower order. This will result in a change from diffusion behavior in the FPE to deterministic drift, or even better, in transformation of third- or higher-order partial derivative terms (which prevent one from obtaining any FPE) to lower-order terms, which can form part of an FPE. The identities required are of the general form 
\EQN{
\left(\dada{}{C_j}\dada{}{C_k} - f(C)\dada{}{C_p}\right)\op{\Lambda} = 0.
}
for some variables $C_j$, $C_k$, and $C_p$, and some function $f(C)$.
An example can be found in the recent work of Drummond and Corney on boson and fermion phase-space distributions\cite{CorneyDrummond03,Drummond04}
}

\section{Weighting stochastic gauges}
\label{CH4Weighting}

Consider the situation where during the derivation of the FPE in Section~\ref{CH3EquationsFPE}, one encounters nonzero constant terms $\sum_lT_l^{(0)}(C)P(C)$. Most typically, this situation occurs with un-normalized density matrices e.g. in the thermodynamic equation \eqref{thermomaster}.  As it stands, such an equation of the form \eqref{almostfpe} does not lead immediately to stochastic equations for the variables $C$. However, if the kernel contains a global weight factor $\Omega$, or by a simple modification is given one, this constant term can be converted to a deterministic evolution of this weight. 

What is needed is a kernel of the form
\EQN{\label{weightkernel}
\op{\Lambda}(C) = \Omega \ul{\op{\Lambda}}(\ul{C})
.}
where the ``base'' kernel $\ul{\op{\Lambda}}(\ul{C})$, containing all the local basis operators for subsystems does not depend on the global weight $\Omega$. 
For notational convenience let us define the logarithm of the weight by
\EQN{
\Omega = e^{C_0}
,}
then the full parameter set is $C=\{C_0, \ul{C}\}$. 
Directly from \eqref{weightkernel}, one can see that the gauge identity 
\EQN{\label{dadaco}
\left(\dada{}{C_0} - 1\right)\op{\Lambda}(C) = 0.
}
applies. If, now, one adds ``zero'' defined as 
\EQN{\label{weightinggaugemasterterm}
\sum_l\int P(C) T_l^{(0)}(\ul{C})\left(\dada{}{C_0} - 1\right)\op{\Lambda}(C) dC = 0
}
to the master equation in the form \eqref{masterinoperators}, the original terms in $T_l^{(0)}$ vanish, to be replaced by 
first order differential terms
\EQN{
\sum_l\int P(C) T_l^{(0)}(\ul{C})\dada{}{C_0}\op{\Lambda}(C) dC
.}
This can be incorporated into the formalism of the derivation in Section~\ref{CH3EquationsFPE} by including labels $j,\,k=0$ in the sums $\sum_j,\,\sum_k$, and noting that now
\SEQN{}{
T_{l0}^{(1)}(\ul{C}) &=& T_l^{(0)}(\ul{C})\\
T_{l0k}^{(2)}&=&T_{lj0}^{(2)} = T_{l00}^{(2)} = 0
.}
This then follows through to additional drift and diffusion coefficients \eqref{fpead} in the FPE \eqref{FPE}
\EQN{
A_0(\ul{C}) &=& \sum_l T_l^{(0)}(\ul{C})	\\
D_{j0} &=& D_{0j} = D_{00} = 0 			
.}
The new equation for $C_0=\log{\Omega}$ becomes
\EQN{
dC_0(t) = A_0(\ul{C},t)\,dt
}
(with no noise), while the stochastic equations for all the remaining variables $C_j$ remain unchanged.

Lastly, the global weight can be made complex with no formal change to the above, if one needs to deal with complex constant terms $T^{(0)}_l$.

\section{Drift stochastic gauges}
\label{CH4Drift}

\subsection{Mechanism}
\label{CH4DriftMechanism}

Consider again a kernel with global weight $\Omega=e^{C_0}$, defined as \eqref{weightkernel} in the previous section.
The weighting gauge identity \eqref{dadaco} implies further gauge identities involving each of the variables in $C$ (including $C_0$ itself):
\SEQN{\label{dadacocj}}{
0 &=& \left(\dada{}{C_0} - 1\right)\dada{}{C_j}\op{\Lambda}(C) \\
 &=& \left(\Half\dada{}{C_0}\dada{}{C_j} + \Half\dada{}{C_j}\dada{}{C_0} - \dada{}{C_j}\right)\op{\Lambda}(C) \\
 &=& \mc{J}_j\left[\op{\Lambda}(C)\right]
.}
While \eqref{dadaco} was used to convert constant terms in the ``FPE-like'' expression \eqref{almostfpe} to first order derivative terms (and hence deterministic evolution in the weight), the new identities can be used to convert first to second order terms. This allows conversion of deterministic evolution in a variable $C_j$ to stochastic changes in the weight. Such a conversion can be highly desirable as will be shown in Chapters~\ref{CH6} and later.

The procedure is similar to that employed in Section~\ref{CH4Weighting}, however some modifications are made to make the final expression for the stochastic equations in $C_j$ more revealing. 
If one  adds ``zero'', defined this time as
\EQN{\label{zerof}
\sum_j\int P(C) \mc{F}_j(C)\mc{J}_j\left[\op{\Lambda}(C)\right] dC = 0
,}
with arbitrary gauges $\mc{F}_j$ for each variable $C_j$, then in the formalism of Section~\ref{CH3EquationsFPE} 
additional terms  $\int P(C) \mc{T}_0\left[\op{\Lambda}(C)\right]dC=0$ are introduced into the master equation, with the superoperator $\mc{T}_0$ containing coefficients
\SEQN{}{
T_{0j}^{(1)}(C) &=& -\mc{F}_j(C)\\
T_{0\ul{j}0}^{(2)}(C)&=&T_{00\ul{j}}^{(2)}(C) = \mc{F}_{\ul{j}}(C)\\
T_{000}^{(2)}(C)&=&= 2\mc{F}_0(C)
,}
where underlined indices have been used to indicate labels for the ``base'' variables such that $\ul{j}=1,2,\dots$, etc.
Following the procedure through, the modified FPE coefficients 
\SEQN{\label{fcoef}}{
A_j &=& \ul{A}_j -\mc{F}_j \\
D_{\ul{j}\ul{k}} &=& \ul{D}_{\ul{j}\ul{k}} \\
D_{\ul{j}0} &=& \ul{D}_{\ul{j}0} +\mc{F}_{\ul{j}} = \ul{D}_{0\ul{j}}\\
D_{00} &=& 2\mc{F}_0
}
are obtained (the original coefficients when $\mc{F}_j=0$ are the underlined).
The modifications to the deterministic evolution of variables are seen to be $-\mc{F}_j\,dt$. 

A difficulty with the forms \eqref{fcoef} is that in general the form of the diffusion matrix elements $B_{jk}$ that actually go into the stochastic equations will depend in some complicated manner on all the $\mc{F}_j$. If the aim is to modify only the deterministic (drift) evolution of the variables $C_{\ul{j}}$ (as will be the case in later chapters), one may end up with the side-effect of additional changes in the noises $B_{\ul{j}k}\,dW_k$. This can prevent a clear assessment of what practical effect the change in the equations due a particular gauge will have. 

A convenient form to have the noise matrices $B$ in would be to have no change in the base variables $C_{\ul{j}}$ apart from the modification of the drift, and define new (also arbitrary) gauge functions $\mc{G}_{\ul{j}}(\mc{F})$ that give the exact stochastic terms for the weight variable $C_0$, and do not introduce any more independent noises than were had in the original formulation. That is, if the prior weight evolution had no diffusion, one can try the ansatz
\SEQN{\label{Bansatz}}{
  B_{\ul{j}\ul{k}} =& \ul{B}_{\ul{j}\ul{k}}\\
  B_{0\ul{k}} =& \mc{G}_{\ul{k}}
.}
Using $D=B B^T$, the diffusion matrix elements that would result from \eqref{Bansatz} are
\SEQN{\label{gcoef}}{
  D_{\ul{j}\ul{k}} &=& \sum_{\ul{p}} \ul{B}_{\ul{j}\ul{p}} \ul{B}_{\ul{k}\ul{p}} = \ul{D}_{\ul{j}\ul{k}}\\
  D_{\ul{j}0} &=& \sum_{\ul{p}} \ul{B}_{\ul{j}\ul{p}} \mc{G}_{\ul{p}} = D_{0\ul{j}}\\
  D_{00} &=& \sum_{\ul{p}} \mc{G}_{\ul{p}}^2
.}
Comparing these with \eqref{fcoef}, one identifies
\SEQN{\label{ftog}}{
  \mc{F}_0 &=& \Half\sum_{\ul{p}} \mc{G}_{\ul{p}}^2\label{ftogo}\\
  \mc{F}_{\ul{j}} &=& \sum_{\ul{p}} \ul{B}_{\ul{j}\ul{p}} \mc{G}_{\ul{p}}
.}
Note that there has to be a small restriction of the gauge freedoms from \eqref{fcoef} to \eqref{ftog} (one less arbitrary function due to \eqref{ftogo}\,) to achieve the convenient noise matrix $B$ given by the ansatz \eqref{Bansatz}.

The final stochastic Langevin equations with gauges $\mc{G}_{\ul{k}}$ then are (noting that $\ul{B}_{\ul{j}\ul{k}}=B_{\ul{j}\ul{k}}$)
\SEQN{\label{langevincg}}{
dC_{\ul{j}} &=& \ul{A}_{\ul{j}}\,dt + \sum_{\ul{k}} \ul{B}_{\ul{j}\ul{k}}\left( dW_{\ul{k}} - \mc{G}_{\ul{k}}\,dt\right),\\
dC_0 &=& \ul{A}_0\,dt + \sum_{\ul{k}} \mc{G}_{\ul{k}}\left(dW_{\ul{k}} -\Half\,\mc{G}_{\ul{k}}\,dt\right)
.}
The (Ito) equation for the actual weight $\Omega=e^{C_0}$ takes on a simpler form 
\EQN{\label{domega}
d\Omega = \Omega\left\{\ul{A}_o\,dt+   \sum_{\ul{k}} \mc{G}_{\ul{k}} dW_{\ul{k}}\right\}.
}
with no new drift terms. Recapping, several assumptions beyond the identities \eqref{dadacocj} have been made to obtain \eqref{langevincg}, and these are:
\ENUM{
\item The integral $\int P (\,\cdot\,)\op{\Lambda} dC$ for the master equation in form \eqref{masterinoperators}, including terms containing the $\mc{F}_j$, is convergent.
\item The function $\mc{F}_0$ is determined by \eqref{ftogo}, rather than being an independent arbitrary function.
\item The prior evolution of the weight contains no stochastic terms \mbox{($\ul{D}_{\ul{j}0}=\ul{D}_{00}=0$)}.
} 

Lastly, as with weighting gauges, if one is dealing with a kernel analytic in constant variables $z_j$, the above derivation goes through for $z_j$ instead of $C_j$ with no formal change, provided the weight variable $C_0=z_0=\log\Omega$ is now also complex. The gauges $\mc{G}_{\ul{k}}$ would then be arbitrary {\it complex} functions.

\subsection{Real drift gauges, and their conceptual basis}
\label{CH4DriftReal}
If one arbitrarily adjusts the drift behavior of the stochastic equations, there must be some kind of compensation to ensure continued correspondence to the original quantum master equation. 
When the gauges $\mc{G}_j$ are real, the modification of the weight can be understood in terms of compensation for the increments $\ul{B}_{jk}(dW_k-\mc{G}_k)$ no longer being described a Gaussian noise of mean zero (in the many infinitesimal timesteps limit when the CLT applies).  Let us first look at the simple case of one subsystem with one complex variable $z$, with the gauge-less weight is constant, and the drift gauge $\mc{G}$ real. 
The equations for the real and imaginary parts of the variables $z=z'+iz''$ and $z_0=\log{\Omega}=z'_0+iz''_0$ are
\SEQN{}{
  dz' &=& \ul{A}'dt +\ul{B}'dW -\ul{B}'\mc{G}dt\\
  dz'' &=& \ul{A}''dt + \ul{B}''dW -\ul{B}''\mc{G}dt\\
  dz'_0 &=& -\Half\,\mc{G}^2dt + \mc{G}dW\label{dzio}\\
  dz''_0 &=& 0\label{dziio}
,}
where we have used the shorthand $\ul{B}'=\re{\ul{B}}$ and $\ul{B}''=\im{\ul{B}}$ etc. for all quantities, and omitted writing subscripts $j,k=1$.

Consider a very small time step $dt$ in which the variables change from $z(t)$ to $z(t+dt)$ etc. This time step is small enough so that the stochastic equations for infinitesimal $dt$ apply, while large enough that the Wiener increment $dW$  is Gaussian distributed due to the CLT. Suppose also that we consider just one 
particular trajectory so that the initial distributions of $z$ and $z_0$ are delta functions around their initial values.
The probability distribution of the Gaussian noise is
\EQN{\label{pdw}
  {\rm Pr}(dW) = e^{-dW^2/2dt}
,}
and if there is no gauge ($\mc{G}=0$), the dependence of $z'(t+dt)$ on the realization of the noise $dW$ can be inverted to give
\EQN{\label{olddw}
  dW = \frac{z'(t+dt)-z'(t)-\ul{A}'dt}{\ul{B}'}
.}
Substituting this into \eqref{pdw}, we obtain an expression for the probability distribution\footnote{%
Note that we use only expressions involving $z'$ to characterize the probability distribution of the entire complex variable $z$. This is because $z'$ and $z''$ are completely correlated with each other, since only the one real noise $dW$ determines them both. The imaginary part, $z''$, could have been used just as well.} of $z(t+dt)$:
\EQN{\label{pvtpdt}
  P(z(t+dt)) = \exp\left[ -\frac{(z'(t+dt)-z'(t)-\ul{A}'dt)^2}{2(\ul{B}')^2dt}\right]
.}
If we now arbitrarily modify the drift with a nonzero gauge, \eqref{pdw} still applies, but the inverted relation $dW(z'(t+dt))$ now becomes 
\EQN{\label{newdw}
  dW = \frac{z'(t+dt)-z'(t)-\ul{A}'dt+\mc{G}\ul{B}'dt}{\ul{B}'}
,}
leading to a different probability distribution $P_{\mc{G}}(v(t+dt))$ than the ``correct'' gaugeless \eqref{pvtpdt}. Precisely:
\EQN{
  P_{\mc{G}}(z(t+dt)) &=& \exp\left[ -\frac{(z'(t+dt)-z'(t)-\ul{A}'dt+\mc{G}\ul{B}'dt)^2}{2(\ul{B}')^2dt}\right]\\
	&=&  P(z(t+dt))\exp\left[-\frac{\mc{G}}{2\ul{B}'}\left\{\mc{G}\ul{B}'dt + 2(z'(t+dt)-z'(t)-\ul{A}'dt)\right\}\right].\nonumber\\
\label{pgp}}
However, we can recover from this by introducing a global compensating weight $\Omega$ for the trajectory, which is always applied in all observable averages so that
\EQN{
\Omega P_{\mc{G}}(z(t+dt)) = P(z(t+dt))
.}
Using \eqref{newdw}\footnote{Not \eqref{olddw}, since that applied only for $\mc{G}=0$.} we can write this compensating weight from \eqref{pgp} as 
\EQN{
\Omega &=& \exp\left[\mc{G}\left(-\Half\mc{G}dt +dW\right)\right] 
.}
Identifying $\Omega$ from the definition of the kernel \eqref{weightkernel}
one sees that the drift gauge equations for $dz_0$ \eqref{dzio} and \eqref{dziio} produce exactly the right global weight to compensate for the arbitrary drift corrections introduced by nonzero $\mc{G}$.  

For the many-variable case with $N_z$ base variables $z_j$ plus the global complex log-weight $z_0$, the gauged stochastic equations \eqref{langevincg} can be written in vector form as
\SEQN{\label{vectorgaugelangevin}}{
  d\bm{z} &=& \ul{\bm{A}}\,dt +\ul{B}\bm{dW} - \ul{B}\bm{\mc{G}}dt\\
  dz_0 &=& -\Half\bm{\mc{G}}^T\bm{\mc{G}}dt +\bm{\mc{G}}^T\bm{dW}\label{z0equation}
,}
where all bold quantities denote column vectors of $N_z$ elements labeled by $j=1,\dots,N_z$ etc.  The probability distribution of a noise realization is 
\EQN{
{\rm Pr}(d\bm{W}) = \exp\left(-\frac{d\bm{W}^Td\bm{W}}{2dt}\right)
,}
since all noise elements $dW_j$ are independent. With no gauge, 
$d\bm{W} = (\ul{B}')^{-1}\left[ d\bm{z}' - \ul{\bm{A}}'dt\right]$,
leading to the probability distribution
\EQN{
P(\bm{z}(t+dt)) = \exp\left[ -\frac{1}{2dt}\left(d\bm{z}'-\ul{\bm{A}}'dt\right)^T [(\ul{B}')^{-1}]^T (\ul{B}')^{-1}\left(d\bm{z}'-\ul{\bm{A}}'dt\right)\right]
,}
while with a gauge the noise can be written
\EQN{\label{noisewgauge}
d\bm{W} = (\ul{B}')^{-1}\left[\, d\bm{z}' - \ul{\bm{A}}'dt+\ul{B}'\bm{\mc{G}}dt\right]
.}
This leads to a new probability distribution
\EQN{
P_{\mc{G}}(\bm{z}(t+dt)) &=& \exp\left[ -\frac{\left(d\bm{z}'-\ul{\bm{A}}'dt+\ul{B}'\bm{\mc{G}}dt\right)^T [(\ul{B}')^{-1}]^T (\ul{B}')^{-1}\left(d\bm{z}'-\ul{\bm{A}}'dt+\ul{B}'\bm{\mc{G}}dt\right)}{2dt}\right]\nonumber\\
&=& P_{\mc{G}}(\bm{z}(t+dt))\nonumber\\
&&\times	\exp\left[-\frac{
\bm{\mc{G}}^T(\ul{B}')^{-1}(d\bm{z}'-\ul{\bm{A}}'dt)
+(d\bm{z}'-\ul{\bm{A}}'dt)^T[(\ul{B}')^{-1}]^T\bm{\mc{G}}
+\bm{\mc{G}}^T\bm{\mc{G}}}{2}\right]\nonumber\\
&=& P_{\mc{G}}(\bm{z}(t+dt))\exp\left[-\bm{\mc{G}}^Td\bm{W}+\Half\bm{\mc{G}}^T\bm{\mc{G}}dt\right]
}
Where the last line was arrived at using \eqref{noisewgauge}. This  requires a compensating weight 
\EQN{
  \Omega  = \exp\left[\bm{\mc{G}}^T\left( -\Half\bm{\mc{G}}dt+d\bm{W}\right)\right]
,}
which is given by the gauged equations if $\Omega = e^{dz_0}$.

As corollaries of this:
\ENUM{
\item If there was no noise in the original equations ($\ul{B}=0$), or the noise matrix was singular, then there would be no way a global weight could compensate for an arbitrary change in the drift equations --- the new drift would just be plain wrong. 
For the case of very small noise in the $\bm{z}$ equations, even small changes with respect to the original  drift $\ul{\bm{A}}$ require very large weight compensation. These are the reasons why the gauges $\bm{\mc{G}}$ are multiplied by the coefficients $\ul{B}$ in the $d\bm{z}$ equations, but not in the weight equation for $dz_0$. Thus
\begin{quote}
If there is no random component to the evolution of a variable $z$, its drift cannot be modified using a drift gauge. If the noise matrix is singular, {\it no} variable drift can be modified using drift gauges.
\end{quote}
\item While the weight equation \eqref{domega} appears disordered and random due to the presence of noises $dW_k$, the evolution of $\Omega$ is actually strictly tied to the random walk of the other variables $z_j$, and acts to exactly compensate for any gauge modifications of the original drift. 
\item  If we consider the ($2N_z$--dimensional) phase space of all base variables in $\ul{C}=\{z_j\}$, then a single real 
drift gauge $\mc{G}_k$ changes the drift only in the phase-space direction specified by the $k$th column $\ul{\bm{B}}_k$ of the noise matrix\footnote{So that $\ul{B} = \left[ \ul{\bm{B}}_1\ \ul{\bm{B}}_2\ \dots\ \right]$.}, associated with the $k$th noise $dW_k$. 
}

\subsection{Complex gauges}
\label{CH4DriftComplex}
The change in the drift due to complex gauges $\bm{\mc{G}}=\bm{\mc{G}}'+i\bm{\mc{G}}''$ is always
\EQN{\label{complexzchange}
\bm{A} &=& \ul{\bm{A}}dt - \ul{B}\bm{\mc{G}}' dt - i\ul{B}\bm{\mc{G}}''dt
.}
Imaginary gauges $\mc{G}''_k$, in contrast to real ones, lead to changes in the drift only in a particular direction $i\ul{\bm{B}}_k$ orthogonal to that affected by a real gauge. Note however, that this is a {\it particular} direction, and not any of the infinitely many orthogonal to $\ul{\bm{B}}_k$.  
This kind of drift modification orthogonal to noise direction is not compensated for by changing the weight of the trajectory  $|\Omega|$, but by modification of its phase $e^{iz''_0}$, leading to an interference effect between trajectories. It appears harder to grasp intuitively that this fully compensates, but in simulations such gauges are also seen to preserve observable averages (see Part B).

Similarly to purely real gauges, no drift modification can be made if the noise matrix $\ul{B}$ is singular, as seen from \eqref{complexzchange}. Conversely, with non-singular $\ul{B}$ one has, in theory, full freedom to modify the deterministic evolution of the complex $z_j$ to any arbitrary functional form. The weight $\Omega$ will keep exactly track of these modifications by virtue of using the same noises $d\bm{W}$. In fact, the weight evolution can be written as a deterministic function of the evolution of the remaining variables:
\EQN{\label{dz0deterministic}
  dz_0 = \bm{\mc{G}}^T\left\{ \ul{B}^{-1}\left(d\bm{z}-\ul{\bm{A}}\,dt\right) +\Half\mc{G}\,dt\right\}
.}
A summary of drift gauge freedoms can be found in Table~\ref{TABLEDriftGauge}.

\begin{table}
\caption[Tally of drift gauge freedoms]{\label{TABLEDriftGauge} \footnotesize
Tally of drift gauge freedoms and comparison to phase space size.
\normalsize }\vspace*{3pt}
\begin{minipage}{\textwidth}\begin{tabular}{|l||c|c|}
\hline							& Number	& Kind		\\\hline
\hline
Size of base phase-space $\ul{C}=\{z_{\ul{j}}\}$ (degrees of freedom)
							& $2N_z$	& real		\\\hline
Size of full phase-space $C=\{z_0, \ul{C}\}$		& $2N_z+2$	& real		\\\hline
Number of standard drift gauges $\mc{G}_k$ 		& $N_z$		& complex 	\\\hline
Number of noises $dW_k$ in standard drift gauge scheme	& $N_z$		& real		\\\hline
\hline
Number of broadening drift gauges $\breve{\mc{G}}_j$	& $N_z$		& complex	\\\hline	
Number of noises in broadening drift gauge scheme \eqref{langevinbroadened}
							& $3N_z$	& real		\\\hline
\end{tabular}\end{minipage}
\end{table}

\subsection{Weight spread}
\label{CH4DriftWeightspread}
While the global weight $e^{z_0}$ is completely determined by the evolution of the base variables $z_j$, in observable calculations \eqref{observables} it appears (particularly in the denominator) as effectively a random variable. Other things being equal, it is certainly desirable to make the spread of the weights small during a simulation.  

\subsubsection{Variances to consider}
In general there are several properties of the weights that may be of relevance. $\re{\Omega}$ appears directly in the denominator average of observable estimates \eqref{observables} via $\tr{\op{\Lambda}}\propto\Omega$. On the other hand, for general observables, both the phase and magnitude of the weight may be important to calculate the numerator average in an obsrvable estimate \eqref{observables} --- the phase may be correlated in important ways with the other variables in the average. 

For some observables and kernels, the numerator expression of \eqref{observables} depends mostly only on $\re{\Omega}$. This is particularly so at short times when starting from a state well described as a classical (and separable) mixture of the kernels. 
In this case, both the numerator and denominator of the observable estimator \eqref{observables} depend mostly on $\re{\Omega}$, and the variance of $\re{\Omega}$ is the most relevant to consider. In more general cases, the variance of $|\Omega|$ will be more relevant.

\subsubsection{Log-weight spread estimate}
A common hindrance when choosing the gauge is the difficulty of accurately assessing the size of the weight spread that will be produced by a given gauge. Here approximate expressions for the variance of $z_0$ at small times \eqref{logweightestimate} will be derived.

Assuming there is no non-gauge weight drift ($\ul{A}_0=0$),
  the evolution in the variance of $z'_0=\re{z_0}$ is given by
\EQN{
  d\vari{z'_0} &=& d\average{(z'_0)^2} -2\average{z'_0}d\average{z'_0}
.}
Using properties of the Ito calculus, one can evaluate the time increments of these quantities from the stochastic equation \eqref{z0equation} as
\SEQN{}{
 d\average{z'_0} &=& \average{\re{A_0}}dt = \Half\sum_k\average{(\mc{G}''_k)^2 - (\mc{G}'_k)^2} \\
 d\average{(z'_0)^2} &=& \average{2z'_0\re{A_0}}dt + \sum_k \average{(B'_{0k})^2 dW_k}\\
		&=& \sum_k\average{z'_0(\mc{G}''_k)^2 - z'_0(\mc{G}'_k)^2 +(\mc{G}'_k)^2}
,}
and so in terms of covariances 
\begin{subequations}\label{dvari}\EQN{
d\vari{z'_0} = \sum_k \average{ (\mc{G}'_k)^2 + \text{covar}\left[ (\mc{G}''_k)^2, z'_0\right] - \text{covar}\left[(\mc{G}'_k)^2, z'_0\right] }
.}
Similarly one finds
\EQN{
d\vari{z''_0} = \sum_k \average{ (\mc{G}''_k)^2 -2\, \text{covar}\left[ \mc{G}'_k\mc{G}''_k, z''_0\right]  }
.}\end{subequations}

A practical approximation to the variance at short times can be gained by assuming lack of correlations between $z_0$ and $\mc{G}$ (i.e. that the covariances are negligible). These approximations then would be 
\SEQN{\label{logweightestimate}}{
  \vari{z'_0(t)} \approx V'_0(t) &=& \vari{z'_0(0)}+\int_0^t \sum_k\average{\mc{G}'_k(t')^2} dt' \label{Viz}\\
  \vari{z''_0(t)} \approx V''_0(t) &=& \vari{z''_0(0)}+\int_0^t \sum_k\average{\mc{G}''_k(t')^2} dt' \label{Viiz}
.}
Of course this is never exact because $z_0$ {\it does} depend on the gauges due to its evolution, however for small times, equal starting weights ($\vari{z'_0}=\vari{z''_0}=0$), and {\it autonomous} gauges (i.e. $\bm{\mc{G}}$ does not depend explicitly on $z_0$) the expressions \eqref{logweightestimate} are good approximations. They can be used to assess the amount of statistical noise that will be introduced by a particular gauge.  

\subsubsection{Small time regime for log-weights}
How small is ``small time'' in this context? Clearly times when the covariances are much less than the $\sum_k(\mc{G}_k)^2$ terms. From \eqref{dvari}, this will occur for autonomous gauges and equal starting weights  if 
\begin{subequations}\EQN{\label{smcovaricond}
z'_0(t) \sum_k |\mc{G}_k|^2 \ll \sum_k (\mc{G}'_k)^2
,}
and 
\EQN{\label{smcovariicond}
2z''_0(t) \sum_k \mc{G}'_k\mc{G}''_k \ll \sum_k (\mc{G}''_k)^2
.}\end{subequations}
To obtain limits on these, one needs an approximation to $z_0$. 
Since $dW_k\approx\order{\sqrt{dt}}$, then $\bm{\mc{G}}^T\bm{dW}$ is the dominant term in the evolution of $z_0$ at short enough times. This implies
\EQN{\label{shorttimezo}
z'_0(t) &\approx&  \order{\sqrt{dt}\sqrt{\sum_k(\mc{G}'_k)^2}}\\
z''_0(t) &\approx&  \order{\sqrt{dt}\sqrt{\sum_k(\mc{G}''_k)^2}}
.}
The root of sum of squares  is due to independence of  noises $dW_k$. For the approximations to $z'_0$ to apply one needs
\EQN{
  \sqrt{ dt\sum_k(\mc{G}'_k)^2} &\gg& |\re{A_0}\,dt| = \frac{dt}{2}\left|\sum_k\left[(\mc{G}'_k)^2-(\mc{G}''_k)^2\right]\,\right|\nonumber\\
\sum_k(\mc{G}'_k)^2 &\gg& \frac{dt}{4}\left\{\sum_k\left[(\mc{G}'_k)^2-(\mc{G}''_k)^2\right]\right\}^2
		\le \frac{dt}{4}\left\{\sum_k|\mc{G}_k|^2\right\}^2  \nonumber\\
dt &\ll& \frac{4\sum_k(\mc{G}'_k)^2}{\left\{\sum_k|\mc{G}_k|^2\right\}^2}\label{dtapproxi}
.}
 Given this is the case,  $z'_0$ is given by \eqref{shorttimezo}, and so the condition \eqref{smcovaricond} for \eqref{Viz} to hold becomes
\EQN{
 \sum_k(\mc{G}'_k)^2 &\gg&  \sqrt{dt}\,\sqrt{\sum_k(\mc{G}'_k)^2} \sum_k|\mc{G}_k|^2 \nonumber\\
		   dt&\ll& \frac{\sum_k(\mc{G}'_k)^2}{\left\{\sum_k|\mc{G}_k|^2\right\}^2}
.}
Since this agrees with the initial assumption of noise term dominance \eqref{dtapproxi},
it is the short time condition under which \eqref{Viz} is accurate.

For $dz''_0$ to be given by \eqref{shorttimezo}, one needs
\EQN{
  \sqrt{ dt\sum_k(\mc{G}''_k)^2} &\gg& |\im{A_0}\,dt| = dt\left|\sum_k\mc{G}'_k\mc{G}''_k\right|\nonumber\\
\sum_k(\mc{G}''_k)^2 &\gg& dt\left\{\sum_k\mc{G}'_k\mc{G}''_k\right\}^2
		\le dt\sum_j(\mc{G}'_j)^2\sum_k(\mc{G}''_k)^2  \nonumber\\
dt &\ll& \frac{1}{\sum_k(\mc{G}'_k)^2}\label{dtapproxii}
.}
The condition for \eqref{Viiz} is then 
\EQN{
 \sum_k(\mc{G}''_k)^2 &\gg&  2\sqrt{dt}\,\sqrt{\sum_k(\mc{G}''_k)^2} \sum_k\mc{G}'_k\mc{G}''_k \nonumber\\
 \sum_k(\mc{G}''_k)^2 &\gg&  4\,dt\,\left\{\sum_k\mc{G}'_k\mc{G}''_k\right\}^2 \le 4\,dt\,\sum_j(\mc{G}'_j)^2\sum_k(\mc{G}''_k)^2  \nonumber\\
dt &\ll& \frac{1}{4\sum_k(\mc{G}'_k)^2}
.}
Again, this agrees with the initial assumption of noise term dominance \eqref{dtapproxii}, and
is the short time condition under which \eqref{Viiz} is accurate.

\subsubsection{Log-weight variance limits}
 As will be discussed in Section~\ref{CH7Gaussian} and Appendix~\ref{APPA}, the variance of $z'_0$ should be $\lesssim \order{10}$ for the simulation to give results with any useful precision. Basically when  $z'_0$ of a variance \order{10} or more, the high positive $z'_0$ tail of the distribution contains few samples, but much weight, leading to possible bias.
Also if  $z''_0$ has a standard deviation of $\order{\pi}$ or more, mutual cancelling of trajectories with opposite phases will dominate the average, concealing any average in noise for reasonable sample sizes $\order{\lesssim 10^5}$.

This imposes a limit on how long a gauged simulation can last, and so it is extremely desirable to keep the magnitude of the gauges as small as possible. 

As a corollary to this point, it is desirable to ensure that all drift gauges are zero at any attractors in phase space, to avoid accumulating unnecessary randomness in the weights when no significant evolution is occuring.

\subsubsection{Direct weight spread estimate}
One can also investigate the evoultion of the variance of the weight $\Omega$ directly, and analytic estimates can be obtained  in some situations. To better understand the effect of real or imaginary gauges on the weight, let us consider the case where gauges $\mc{G}_k$ and weight $\Omega$ are decorrelated, there is no base weight drift ($\ul{A}_0=0$), and $\Omega=1$ initially for all trajectories. Let 
$\Omega = \Omega'+i\Omega''$, then from \eqref{domega}
\SEQN{\label{domegaiii}}{
d\Omega' =& \sum_k(\Omega'\mc{G}'_k -\Omega''\mc{G}''_k)&dW_k\\
d\Omega'' =& \sum_k(\Omega'\mc{G}''_k +\Omega''\mc{G}'_k)&dW_k
.}
If we consider the evolution of the second order moments of the weights, then under the uncorrelated $\Omega$ and $\bm{\mc{G}}$ assumption, it can be written as a closed system of equations
\EQN{\label{omegasystem}
d\average{\matri{{[\Omega']^2}\\{[\Omega'']^2}\\\Omega'\Omega''}} = 
\matrixx{c_1&c_2&-2c_3\\c_2&c_1&2c_3\\c_3&-c_3&c_1-c_2}
\average{\matri{{[\Omega']^2}\\{[\Omega'']^2}\\\Omega'\Omega''}}\hspace{-2ex}dt
,}
where 
\SEQN{\label{c123}}{
c_1(t) &=& \sum_k\average{[\mc{G}'_k(t)]^2}\\
c_2(t) &=& \sum_k\average{[\mc{G}''_k(t)]^2}\\
c_3(t) &=& \sum_k\average{\mc{G}'_k(t)\mc{G}''_k(t)}
.}
This system can be solved, and remembering that $\average{\Omega(t)}=1$ (from \eqref{domegaiii}), and $\Omega(0)=1$, one obtains the solutions
\EQN{
  \vari{\Omega'} &=& \Half  e^{\int_0^t c_1(t'\,) dt'} \left[ e^{\int_0^tc_2(t')\,dt'} + e^{-\int_0^tc_2(t')\,dt'}\cos\left(2\int_0^tc_3(t')\,dt'\right)\right]-1\nonumber\\
  \vari{\Omega''} &=& \Half  e^{\int_0^t c_1(t')\,dt'} \left[ e^{\int_0^tc_2(t')\,dt'} - e^{-\int_0^tc_2(t')\,dt'}\cos\left(2\int_0^tc_3(t')\,dt'\right)\right]-1\nonumber\\
  \average{\Omega'\Omega''} &=& \Half e^{\int_0^t[c_1(t')-c_2(t')]\,dt'}\sin\left(2\int_0^tc_3(t')\,dt'\right) 
.}
If the gauge averages $c_j(t)$ are approximately constant with time, then 
at short times, when the condition that $\Omega$ and $\mc{G}_k$ are uncorrelated holds, the variance of the weight is 
\EQN{\label{varomega}
\vari{\Omega'} = t\sum_k\average{(\mc{G}'_k)^2} +\frac{\,t^2}{2}\sum_k\average{(\mc{G}'_k-\mc{G''}_k)^2} + \order{t^3}
.}

In similar manner, one can also obtain 
\EQN{\label{varmomega}
\vari{\,|\Omega|\,} &=& \exp\left[ \int_0^t\average{|\mc{G}_k(t')|^2}dt'\right]-1
.}

\subsubsection{Some more conslusions about drift gauge forms}
For most observables it is the variance of $\Omega'$ that is most relevant for uncertainties in the finite sample estimates, because it appears both in the numerator and denominator of \eqref{observables}. However, for some like the local quadratures $(\dagop{a}_k-\op{a}_k)/2i$ the modulus of the weight is more relevant in the numerator of the observable expression \eqref{observables}. 
One can see from \eqref{varomega} that at short times imaginary gauges lead to smaller real weight spreads because the variance grows only quadratically, most of the gauge noise going into $\Omega''$. This makes imaginary drift gauges more convenient for short time estimates of most observables

\section{Diffusion stochastic gauges}
\label{CH4Diffusion}

The Fokker-Planck equation specifies directly only the diffusion matrix $D(C)$, which is then decomposed via 
\EQN{
  D = B B^T
}
into noise matrices $B$, however these are not specified completely. 
This freedom in the choice of $B$, leads to a different kind of stochastic gauge than considered in the previous sections, which will be termed ``diffusion gauges'' here.  No weights are required, and only the noise terms are modified. 
As with kernel stochastic gauges, the diffusion gauges may in some cases be used to choose a set of equations with the most convenient stochastic properties. The non-uniqueness of $B$ has always been known,  but has usually been considered to simply relabel the noises without any useful consequences. It has, however,  been recently shown by  Plimak\etal\cite{Plimak-01} that using a different $B$ than the obvious ``square root'' form $B=B_0=D^{1/2}$ leads to impressive improvement in the efficiency of positive P simulations of the Kerr oscillator in quantum optics. (This has a similar form of the Hamiltonian to the nonlinear term in \eqref{deltaH}.)
This somewhat surprising result leads us to try to quantify the amount of freedom of choice available in the noise matrices 
In the process, it will also become apparent why some extra properties of the kernel beyond the most general case are needed for diffusion gauges to be useful for simulations.

\subsection{Noise matrix freedoms with general and complex analytic kernels}
\label{CH4DiffusionFreedoms}

At first glance there appears to be a great deal of freedom in the choice of noise matrix $B$. Consider that the relationship $D=BB^T$ for $N_v$ real variables can be satisfied by a $N_v\times N_W$ noise matrix, subject to the $N_v(N_v+1)/2$ real constraints $D_{jk} = \sum_pB_{jp}B_{kp}$. The number ($N_W$) of independent real Wiener increments is formally unconstrained. It turns out, however, that much (if not all) of this freedom is simply freedom of labeling and splitting up a single Wiener increment into formally separate parts having no new statistical properties. More on this in Section~\ref{CH4DiffusionReal}.

Recaling Section~\ref{CH3EquationsLangevin}, the Ito Langevin equations \eqref{langevin}, \eqref{langevinc}, \eqref{langevincg}, or \eqref{vectorgaugelangevin} can also be written (Whether the $C_j$ are real or complex) as
\EQN{
  dC_j(t) = A_j(C,t) dt + dX_j(C,t)
,}
where the $\sum_kB_{jk}dW_k$ terms have been coalesced into a single stochastic increment $dX_j$. 
Only the means and variances of the Wiener increments $dW_k$ are specified, and so by the properties of Ito stochastic calculus, the only binding relationships for the stochastic terms $dX_j$ are \eqref{2cond} --- mutual variances specified by the stochastic average of the diffusion matrix, and zero means.

Let us ask the question ``do the relationships \eqref{2cond} completely specify the statistical properties of the 
$dX_j$?''. 

Consider firstly that because all the increments are taken to be ``practically'' infinitesimal in any simulation, then over any significant timescale the noise due to the Wiener increments $dW_k$ will be effectively Gaussian, whatever the actual distribution of $dW_k$ used\footnote{Provided that the variance of $dW_k$ is finite.}. This is due to the central limit theorem. 
Also, \eqref{meancond} is guaranteed by the independence of the $B_{jk}$ and $dW_k$ in the Ito calculus, combined with $\average{dW_k(t)}=0$.
These constraints then imply that to check if there is any freedom in the statistical properties of the $dX_j$ it suffices to compare the number of covariance  conditions \eqref{varcond} with the number of covariance relations for the $dX_j$. 

Seemingly this is trivial: for $N_v$ real variables $C_j$, there are $N_v(N_v+1)/2$ conditions \eqref{varcond}, and the same number of possible covariance relationships $\average{dX_j\,dX_k}$. This means that 
\begin{quote}
  In the general case of a kernel with no additional symmetry properties, 
all the (formally different) possible choices of the noise matrix $B$ have exactly the same effect on the statistical behavior of the Langevin equations.  
\end{quote}
  This is despite the noise matrix having formally many free parameters. These then are largely freedoms to relabel and split up the Wiener increments without introducing any new statistical behavior, as will be discussed in Section~\ref{CH4Diffusion}.

  This simple comparison of conditions and relationships also explains why trying various forms of the noise matrices $B$ was for a long time thought not to have any useful effect --- under general conditions it doesn't.

It has been found, however, that a positive P simulation can be optimized by particular choices of $B$\cite{Plimak-01}. Let us consider the case when the kernel $\op{\Lambda}$ can be written as an analytic function of $N_z$ complex variables  $C=\{z_j\}$. Now, there are  $N_z(N_z+1)/2$ complex conditions \eqref{varcond}.
As for statistical properties of the $dX_j$, we have to consider independently both the real and imaginary parts, $dX'_j$ and $dX''_j$ respectively. Their covariance relations are $\average{dX'_j\,dX'_k}$, $\average{dX''_j\,dX''_k}$, and $\average{dX'_j\,dX''_k}$, numbering $N_z(2N_z+1)$ real relations in all. Since there are two real conditions per one complex, this means that there are left over $N_z^2$ degrees of freedom in the covariance relations between elements of the complex quantities $dX_j$. These might be used to tailor these $N_z^2$ free variances or covariances to our needs.

This begs the obvious question of ``why do the analytic complex variable kernels seem allow more freedom than a general real variable kernel'', since they can also be written in terms of real variables -- we just split each complex variable into real and imaginary parts. 

The key lies in the fact that kernels analytic in complex variables have special symmetry properties that give us freedom to choose the complex derivatives of the kernel as \eqref{complexderivatives}, or indeed as
\EQN{\label{analyticgauge}
  \dada{\op{\Lambda}(C)}{z_j} = \mc{F} \dada{\op{\Lambda}(C)}{\re{z_j}} -i(1-\mc{F}) \dada{\op{\Lambda}(C)}{\im{z_j}}
}
using an analytic kernel stochastic gauge $\mc{F}$. This ``analytic'' symmetry then enters into the Fokker-Planck equation, and finds its way into the Langevin equations as freedom of choice of noise increments $dX_j$. 
As with all stochastic gauges, another way of looking at this is that because the kernel has certain symmetry properties, there is a whole family of distributions $P(C)$ of such kernels corresponding to the same quantum density matrix $\op{\rho}$. Choosing gauges chooses between these different distributions. 

In the following subsections, several specific types of freedoms (or ``stochastic gauges'') available in the noise matrix $B$ will be considered, and a tally will be made at the end in Table~\ref{TableDiffusionGauge}.

\subsection{Standard form of diffusion gauges for analytic kernels}
\label{CH4DiffusionCanonical}

  Let us consider the analytic kernel case, where it was seen above that there may be useful noise matrix freedoms. Such kernels tend also to be the most convenient generally because a positive propagator, and so a stochastic interpretation, is always guaranteed by the method in Section~\ref{CH3EquationsAnalytic}.  

Since $D=D^T$ is square and can be made symmetric\footnote{Since  $\partial^2/\partial v_1\partial v_2=\partial^2/\partial v_2\partial v_1$ for any variables $v_1$ and $v_2$ can be used in the FPE to achieve $D_{jk}=D_{kj}$.}, its matrix square root is also symmetric $\sqrt{D}=\sqrt{D}^T$, and can be used 
as a noise matrix $B_0$
\EQN{
 B_0 &=& \sqrt{D} \\
 D &=& B_0 B_0^T
.}
This square root form can be considered as the ``obvious'' choice of decomposition, unique apart from the $N_z$ signs of the diagonal terms\footnote{One for each complex variable.}. However, for any complex orthogonal $O$ such that $OO^T=I$, if $B_0$ is a valid decomposition of $D$, then so is the more general matrix $B=B_0O$. So, any matrix in the whole orthogonal family 
\EQN{\label{bo0}
B=B_0\, O
}
is a valid decomposition.
A general complex orthogonal matrix can be written explicitly using an antisymmetric matrix basis  $\sigma^{(jk)}$, ($j\neq k=1,\dots,N_z$) having matrix elements
\EQN{
  \sigma_{lp}^{(jk)} = \delta_{jl}\delta_{kp}-\delta_{jp}\delta_{kl}
.}
With these $(N_z-1)N_z/2$ independent $N_z\times N_z$ matrices $\sigma^{(jk)}$ the general form is
\EQN{\label{diffusiongaugeexplicit}
  O = \exp\left( \sum_{j<k} g_{jk}(C,C^*,t)\,\sigma^{(jk)}\right)
.}
The $g_{jk}$ are the $N_z(N_z-1)/2$ complex diffusion gauge functions, which can in principle be {\it completely arbitrary}, including dependence on all variables in $C$ (not necessarily analytic), and the time variable $t$, without affecting the correspondence between the Langevin stochastic equations, and the FPE.  

As an example, in the case of two complex variables, there is one complex gauge function $g_{12}$, and the resulting transformation is 
\EQN{\label{u12expr}
O &=& \exp\left(\,g_{12}\sigma^{(12)}\,\right)\nonumber\\
  &=& \cos(g_{12}) + \sigma^{(12)}\sin(g_{12})
,}
where the anti-symmetric matrix $\sigma^{(12)}$ is proportional to a Pauli matrix:
\EQN{
\sigma^{(12)} = \matrix{0&1\\-1&0}
.}
Hence, e.g. if the diffusion matrix is diagonal, $B_0$ is also, and the transformed  (but equivalent) noise matrix becomes:
\EQN{\label{diag2vardiffusion}
  B = \matrix{\sqrt{D_{11}}\cos(g_{12}) & \sqrt{D_{11}}\sin(g_{12})\\-\sqrt{D_{22}}\sin(g_{12})&\sqrt{D_{22}}\cos(g_{12})}
.}
The square form of $B$ \eqref{bo0} with $O$ given by  \eqref{diffusiongaugeexplicit} will be termed here the ``standard'' diffusion gauge, to distinguish it from some other forms that will be discussed in later subsections.

\subsection{Real standard diffusion gauges and noise mixing}
\label{CH4DiffusionReal}
  Not all the canonical diffusion gauges $g_{jk}$ are useful. One class of useless gauges to avoid are those that seemingly change $B$, but effectively only swap around the linear combinations of noises $dW_k$, without affecting $dX_j$ or hence any statistical properties of the equations. Their main  effect is usually  to complicate the math.

Consider, for example, the two-complex variable case given by \eqref{u12expr}. For a purely real diffusion gauge $g_{12}=g'_{12}$, the effect of the noise terms in the Langevin equations becomes  
\EQN{
\bm{dX} = B(g'_{12})\,\bm{dW} = B_0\matrix{\cos g'_{12}&\sin g'_{12}\\-\sin g'_{12} & \cos g'_{12}}\matri{dW_1\\dW_2} = B_0\bm{dW}'
.}
This is just a rotation of the noises $\bm{dW}$, leading to $\bm{dW}'$ being a linear combination of the noises $dW_1$ and $dW_2$. The new noises $dW'_j$ have the same statistical properties as the old $dW_j$. Thus $\bm{dX}(g'_{12})=B(g'_{12})\,\bm{dW}$ has exactly the same effect in the equations as the un-gauged $\bm{dX}(g'_{12}=0)$, regardless of any complicated form of $g'_{12}$ we choose to try. 

More generally, with $N_z$ complex variables, there are at least $N_z(N_z-1)/2$ such useless noise rotations available, one for each pair of variables. Looking at the standard orthogonal matrix form \eqref{diffusiongaugeexplicit}, one sees that the terms in the exponential proportional to real parts of gauges $g_{jk}$ represent simply all these useless rotations.

In a formalism with only real variables $C_j$, and no extra symmetries, diffusion gauges arise formally in the same manner as in Section~\ref{CH4DiffusionCanonical}, although all the matrices $D,B,B_0$, and $O$ must now be real, making $g_{jk}$ also only real.  This means that in such a case {\it all} the resulting real gauge functions in $O(g_{jk})$ are useless noise mixers, and no useful modifications of the stochastic equations can be achieved by choosing them. This explains why the potential of noise matrix choice to improve the stochastic equations went unnoticed for many years.

This subsection can be summarized as
\begin{quote}
  Only the imaginary parts of the standard diffusion gauges \eqref{diffusiongaugeexplicit} can affect the statistical properties of the stochastic equations.
\end{quote}

Lastly, it may be worth pointing out that real gauges can achieve the same useless noise rotations in non-square noise matrices as well. 
For example, for any arbitrary real function $f$ the noise matrix
\EQN{
  B^{(f)} = \matrix{ B\cos f & B\sin f}
}
satisfies $B^{(f)}[B^{(f)}]^T=D$ just as well as the original $B$, whatever the form of the arbitrary complex function $f$. 
The stochastic increments are (for $N_z$ complex variables)  
\EQN{
dX_j&=&\sum_{k=1}^{2N_z}B^{(f)}_{jk}dW_k \nonumber\\
&=& \sum_{k=1}^{N_z} B_{jk} (\cos f dW_k + \sin f dW_{N_z+k})\nonumber\\
&=& \sum_{k=1}^{N_z} B_{jk} dW'_k
,}
with the combined noises $dW'_k$ having the same statistical properties irrespective of the choice of $f$. 

\subsection{Imaginary standard diffusion gauges and statistics}
\label{CH4DiffusionImaginary}
The imaginary part of standard diffusion gauges $g''_{jk}=\im{g_{jk}}$ {\it do} affect the statistics. Again consider the two-variable case of diagonal $D$ of expression \eqref{u12expr}, this time with imaginary gauge $g_{12}=ig''_{12}$. 
\EQN{
\bm{dX} = B_0\matrix{\cosh g''_{12}&i\sinh g''_{12}\\-i\sinh g''_{12} & \cosh g''_{12}}\matri{dW_1\\dW_2}
.}
For example, in the simplest case of diagonal positive real $D$, one has (using the notation $dX_j=dX'_j+idX''_j$) 
\SEQN{\label{diagDdX}}{
dX'_1 &=& \sqrt{D_{11}}\cosh g''_{12}dW_1\\
dX'_2 &=& \sqrt{D_{22}}\cosh g''_{12}dW_2\\
dX''_1 &=& \sqrt{D_{11}}\sinh g''_{12}dW_2\\
dX''_2 &=& -\sqrt{D_{22}}\sinh g''_{12}dW_1 
.}
While the covariance requirements \eqref{varcond} are satisfied, the variances of the $dX_j$ components may vary depending on $g''_{12}$. The nonzero correlations are
\SEQN{}{
\average{(dX'_1)^2} &=& \cosh^2g''_{12}\average{D_{11}}\,dt\\
\average{(dX'_2)^2} &=& \cosh^2g''_{12}\average{D_{22}}\,dt\\
\average{(dX''_1)^2} &=& \sinh^2g''_{12}\average{D_{11}}\,dt\\
\average{(dX''_2)^2} &=& \sinh^2g''_{12}\average{D_{22}}\,dt\\
\average{dX'_1\,dX''_2} &=& -\Half\sinh 2g''_{12}\average{\sqrt{D_{11}D_{22}}}\,dt\\
\average{dX'_2\,dX''_1} &=& \Half\sinh 2g''_{12}\average{\sqrt{D_{11}D_{22}}}\,dt
.}

Such imaginary gauges only apply when the model is expressed in complex variables $z_j$ with an analytic kernel.  A glance at the canonical gauge form \eqref{diffusiongaugeexplicit}, indicates that there is then one such imaginary gauge $g''_{jk}=\im{g_{jk}}$ for each possible pair of variables. 

\subsection{Non-standard diffusion gauges and further freedoms}
\label{CH4DiffusionNoncanonical}

  In Section~\ref{CH4DiffusionFreedoms} it was seen that there are $N_z^2$ degrees of freedom\footnote{The case of a kernel analytic in complex variables is being considered all the time.} for the statistics of the $dX_j$, but later, in Sections~\ref{CH4DiffusionReal} and ~\ref{CH4DiffusionImaginary} it was shown that there are at most $N_z(N_z-1)/2$ useful (imaginary) standard gauges $g_{jk}=ig''_{jk}$. Conclusion: there are some more freedoms not included in the standard square noise matrix expression \eqref{bo0} and \eqref{diffusiongaugeexplicit}. The simplest example occurs when there is only one complex kernel parameter $z_1$: there is one degree of freedom in $dX_1$, but zero standard gauges. 

To see how the extra non-square gauges can enter the picture, let us consider the simplest $N_z=1$ case. Here there will be 
three covariance relations for $dX'_1$ and $dX''_1$, but two conditions \eqref{2cond} on these. 
Firstly, note that while the standard gauged noise matrix $B=B_0O$ is square in the complex variables $z_j$, in the real variables $z'_j$ and $z''_j$ it is a $2N_z\times N_z$ --- in this case a $2\times 1$. This leaves no room for a third degree of freedom in $dX_1=B\bm{dW}$.
There is, however, no particular limit on the number of columns of $B$ or the number of noises in $\bm{dW}$, provided $BB^T=D$, so let us instead consider  a $2\times 2$ noise matrix, 
\EQN{
B = \matrix{b_{\rm rr}&b_{\rm ri}\\b_{\rm ir}&b_{\rm ii}}
,}
where subscripts $\rm r$ and $\rm i$ denote elements relating to the real and imaginary parts of the complex variable $z_1$. 
Since there are only three degrees of freedom , we can set one of the elements to zero with no restrictions on the
statistical properties of $dX_1$ (say $b_{\rm ri}=0$). The remaining elements must satisfy the conditions (from \eqref{2cond})
\SEQN{\label{1varcond}}{
\average{\re{D}}\,dt =& \average{(dX'_1)^2-(dX''_1)^2} =& dt\average{ b_{\rm rr}^2 -b_{\rm ir}^2-b_{\rm ii}^2 }\\
\average{\im{D}}\,dt =& 2\average{ dX'_1dX''_1}	=& dt\average{ b_{\rm rr}b_{\rm ir} }
,}
leaving room for one arbitrary function (i.e. gauge) out of the three nonzero elements $b_{\rm rr}$, $b_{\rm ri}$, or $b_{\rm ii}$. 
In the standard noise matrix form from \eqref{canonicalnoisematrix} one had $b_{\rm ii}=0$, so all the noise matrix elements were then completely specified by \eqref{1varcond}, with no room for a gauge.

\subsection{Distribution broadening gauges}
\label{CH4DiffusionBroadening}
  Consider the case of no diffusion $D=0$.
When the kernel is analytic in complex variables, there can actually be nonzero noise matrices, leading to nonzero noise despite no diffusion in the FPE.
All that is required is that the complex $B$ obeys $BB^T=0$, or in terms of stochastic increments $\bm{dX}=B\bm{dW}$ that 
\SEQN{\label{incrementszero}}{
\average{\bm{dX}\cdot\bm{dX}^T} &=& D = 0 \\
\average{\bm{dX}} &=& 0
.}
A viable noise matrix is, for example,  
\EQN{\label{dummyB}
\breve{B} = \matrix{ \breve{g} & i \breve{g}}
,}
where $\breve{g}$ is a $N_z\times N_W$ complex matrix, all of whose elements can be any arbitrary functions (gauges) we like without affecting $\breve{B}\breve{B}^T=D=0$. There is no limit on the width $N_W$. 
For nonzero diffusion, the same kind of gauges can be attached to an existing noise matrix ($BB^T=D$) at one's leisure via
\EQN{\label{adddummy}
\wt{B} = \matrix{B& \breve{B}}
.}
The new noise matrix also obeys $\wt{B}\wt{B}^T=D$.
Nonzero noises corresponding to zero diffusion in the FPE are possible because the correspondence between stochastic equations and the FPE is exact only in the limit of infinite samples. Thus the conditions \eqref{incrementszero} apply only in the limit of infinitely many trajectories, and $B=0$ is just one special case in which these conditions are satisfied for every trajectory on its own.

The diffusion gauges $\breve{g}$ will be termed ``distribution broadening'' here because their effect is to make the distribution of the complex variables $z_j$ broader, while preserving their complex means and mutual correlations. 
Generally such broadening gauges are not of much use, simply making everything more noisy and reducing precision of observable averages, but there are some situations when this is advantageous. 

One example occurs when the noise matrix is singular, or there is no native diffusion in the FPE (i.e. $dX_j=0$) for a variable $z_j$. In this 
situation drift gauges are unable to make modifications to the deterministic evolution.
If one, however, uses a broadening gauge to force a nonzero noise matrix, then the usual drift gauge formalism can be used and arbitrary modifications to the deterministic evolution made. The idea of creating additional drift gauges in this manner was first proposed (in a $N_z=2$ case) by Dowling\cite{Dowling03}.  
Let us see how this proceeds in a general case: 

For clarity in the resultant equations it is best to define a diagonal broadening gauge matrix $\breve{g}$ with 
elements
\EQN{
\breve{g}_{jk} = \breve{g}_j\delta_{jk}
.}
Adding the broadening gauge to a pre-existing noise matrix $B$ as in \eqref{dummyB} and \eqref{adddummy}, one obtains a 
$N_z\times3N_z$ noise matrix and $3N_z$ independent real noises. Each of these noises can now have a drift kernel gauge attached to it in the manner described in Section~\ref{CH4Drift}.  For the purpose of arbitrary drift manipulation it suffices to introduce $N_z$ complex drift gauges $\breve{\mc{G}}_j$ on just $N_z$ of these noises --- say on the noise matrix elements $\breve{g}_j$. This leads then to the stochastic equations
\SEQN{\label{langevinbroadened}}{
dz_j &=& A_j\,dt + \sum_k B_{jk} dW_k + \breve{g}_j( d\breve{W}_j + id\breve{\wt{W}}_j - \breve{\mc{G}}_j\,dt)\\
dz_0 &=& A_0\,dt -\Half\sum_j \breve{\mc{G}}_j^2\,dt +\sum_j \breve{\mc{G}}_j d\breve{W}_j 
.}

The  ``broadening noises'' $d\breve{W}_j$ and $d\breve{\wt{W}}_j$ are independent real Wiener increments just like the $dW_k$.

The arbitrary gauge modifications to the drift of a variable $z_j$ are 
\EQN{
-\breve{g}_j\breve{\mc{G}}_j\,dt
.}

A related situation occurs when the native stochastic increment $dX_j$ for a variable $z_j$ is small. In such a situation,  making a significant change in the drift of $z_j$ would require large compensation in the log-weight $z_0$, leading to rapidly increasing statistical uncertainty or even bias from widely varying trajectory weights. However, if one introduces a broadening gauge such that 
\EQN{
\vari{\re{\breve{g}_j} d\breve{W}_j}  &\gg& \vari{\re{dX_j}}\nonumber\\
\average{\re{\breve{g}_j}^2}  &\gg& \sum_k\average{\re{B_{jk}}^2}
,}
then the drift of $\re{z_j}$ can be modified at smaller weighting cost then if one had used standard drift gauges.
Analogously for $\im{z_j}$. In fact, there appears to be a tradeoff here between the amount of noise introduced into the log-weight $z_0$ (proportional to $\breve{\mc{G}}_j$), and the amount of noise introduced directly into the variable whose drift is being modified (proportional to $\breve{g}_j$).

\subsection{Diffusion from different physical processes}
\label{CH4DiffusionMixing}
A commonly occurring situation is that several physically distinct processes give separate contributions to the diffusion matrix, e.g. 
\EQN{
D = \sum_l D^{(l)}
.}
Calculating the square-root noise matrix $B_0$ may, in some cases, give a very complicated expression in this situation. If this is a hindrance in calculations, a much more transparent noise matrix decomposition is possible. One decomposes each diffusion contribution 
$D^{(l)}$ separately into its own noise matrix as 
\EQN{
D^{(l)} = B^{(l)}(B^{(l)})^T
.}
and then combines them as 
\EQN{
B = \matrixx{B^{(1)}&B^{(2)}&\cdots}
.}
This results in separate noise processes for each diffusion contribution $D^{(l)}$, and formally separate drift gauges (one complex drift gauge per real noise). The benefit of doing this is that the resulting stochastic equations have stochastic terms of a relatively simple form. On the other hand, a possible benefit of doing things the hard way with $B_0$ a direct square root of the full diffusion matrix $D$ is that  diffusion contributions from different processes $l$ may partly cancel, leading to a less noisy simulation.



\section{Summary of standard gauges}
\label{CH4Central}
The  standard gauge choices that will be used in subsequent chapters can be summarized in vector form by the equations 
\SEQN{\label{langevinstd}}{
  d\bm{z} &=& \ul{\bm{A}}\,dt + \ul{B}_0 O(\{g\})\left( \bm{dW} - \bm{\mc{G}}\,dt\right),\\
  dz_0 &=& \ul{A}_0\,dt +\bm{\mc{G}}^T\left( \bm{dW} - \Half\,\bm{\mc{G}}\,dt\right)
.}
This uses a kernel of the form \eqref{weightkernel} proportional to a global weight $\Omega=e^{z_0}$, and analytic in $N_z+1$ complex variables $z_j,\,z_0$. A weighting kernel gauge, standard drift kernel gauges, and standard imaginary diffusion gauges have been used ($g_{jk}=ig''_{jk}$ forming the set $\{g\}$ of arbitrary real-valued gauge functions). Underlined quantities are those obtained from the FPE before introduction of drift gauges, while the bold quantities denote column vectors with an element per base configuration variable $z_{j\neq0}$. And so, $\ul{\bm{A}}$ and $\ul{A}_0$ are drift coefficients from the un-gauged FPE, $\bm{dW}$ are real Wiener increments (with each element $dW_j$ statistically independent for each $j$ and each timestep, of zero mean and variance $dt$), and $\bm{\mc{G}}$ are arbitrary complex drift gauge functions. 
$\ul{B}_0=\sqrt{\ul{D}}$ is the square-root noise matrix form, $O$ is the $N_z\times N_z$ orthogonal matrix given by \eqref{diffusiongaugeexplicit}, but in this case dependent only on imaginary gauges $ig''_{jk}$. 
The gauge freedoms in this standard formulation are summarized in Table~\ref{TABLEStandardGauge}.

\begin{table}[ht]
\caption[Tally of standard gauge freedoms]{\label{TABLEStandardGauge}\footnotesize 
\textbf{Tally of diffusion and noise matrix freedoms in the standard formulation \eqref{langevinstd}}: Kernel analytic in complex phase-space variables (Section~\ref{CH3EquationsAnalytic}), standard imaginary diffusion gauges (Section~\ref{CH4DiffusionCanonical}), standard drift gauges (Section~\ref{CH4Drift}).
\normalsize}
\begin{center}\begin{tabular}{|l||c|c|}
\hline
							& Number	& Kind			\\\hline
\hline\hline
Base phase-space variables $\ul{C}=\{z_j\}$		& $N_z$		& complex		\\\hline
All variables $C=\{z_0,\ul{C}\}$			& $N_z+1$	& complex		\\\hline
Noises (Wiener increments $dW_k$)			& $N_z$		& real			\\\hline
Drift gauges $\mc{G}_k$					& $N_z$		& complex		\\
Imaginary diffusion gauges $g_{jk}=ig''_{jk}$		& $N_z(N_z-1)/2$& imaginary		\\\hline
\end{tabular}\end{center}
\end{table}

\begin{table}[p]
\caption[Tally of diffusion gauge freedoms]{\label{TableDiffusionGauge}\footnotesize
 \textbf{Tally of diffusion gauge and noise matrix freedoms.} The degrees of freedom counted are always real, not complex.
\normalsize}\vspace*{3pt}
\hspace{-0.5cm}
\begin{tabular}{|l||c|c|}
\hline
						& Kernel analytic in 		& General kernel 		\\
Number of:						& $N_z$ complex variables 	& in $N_v$ real variables	\\
\hline\hline
Real variables					& $2N_z$			& $N_v$				\\\hline
Covariance relations between $dX_j$		& $N_z(2N_z+1)$			& $N_v(N_v+1)/2$		\\\hline
Covariance constraints on $dX_j$		& $N_z(N_z+1)$			& $N_v(N_v+1)/2$		\\\hline
Potentially useful				&&\\
statistical freedoms between $dX_j$		& $N_z^2$			& $0$				\\\hline
maximum Wiener increments $dW_k$		& $\infty$			& $\infty$			\\\hline
standard Wiener increments $dW_k$		& $N_z$				& $N_v$				\\\hline
Elements in standard square root  		&				& 				\\
noise matrix $B_0=\sqrt{D}$			& $2N_z^2$			& $N_v^2$			\\\hline
Elements in extended noise matrix 		&&\\
with $2N_z$ real columns			& $4N_z^2$			& $N_v^2$			\\\hline
Canonical gauges $g_{jk}$ in \eqref{diffusiongaugeexplicit}	& $N_z(N_z-1)$			& $N_v(N_v-1)/2$		\\
\hspace{1cm}--- useless real gauges $g'_{jk}$	& $N_z(N_z-1)/2$		& $N_v(N_v-1)/2$		\\
\hspace{1cm}--- useful imaginary gauges $g''_{jk}$& $N_z(N_z-1)/2$		& $0$				\\\hline
Potentially useful non-standard gauges 		& $N_z(N_z+1)/2$		& $0$				\\\hline
Useless potential noise spawning gauges		& $\infty$			& $\infty$			\\\hline

Potential broadening gauges in $\breve{g}$	& $\infty$			& not applicable		\\\hline
Broadening gauges $\breve{g}_j$ in		&				&				\\
 drift gauge scheme \eqref{langevinbroadened}. 	& $N_z$				& not applicable		\\\hline
\end{tabular}
\end{table}


\chapter{The gauge P representation}
\label{CH5}

The gauge P representation, which will be used in subsequent chapters, is explained and its properties investigated here. It is based on the positive P representation of quantum optics, which uses an off-diagonal coherent state kernel, but due to the inclusion of a global phase can be used to introduce weighting and drift gauges. The original distribution concept is due to P. D. Drummond, and its basic derivation is given in published work by Drummond and Deuar\cite{DeuarDrummond02,DrummondDeuar03,Drummond-04}.

The positive P distribution is a promising starting point because it has already been successfully used in some many-body problems in quantum optics (e.g. squeezing in optical solitons \cite{Carter-87,Drummond-93}) and with Bose atoms as well (evaporative cooling\cite{Steel-98,DrummondCorney99,Corney99}). The coherent-state mode-based approach gives simple equations similar to the mean-field Gross-Pitaevskii (GP) equations, and automatically applies to open systems. 
However, stability and systematic error problems can occur\cite{SmithGardiner89,SchackSchenzle91,Gilchrist-97,DeuarDrummond02} with these simulations, which hinders effective use of the method.
The gauge P representation inherits all the useful features of the positive P but allows drift gauges, which can be used to tailor the equations so that stability problems and systematic (boundary term) errors are removed.
Weighting gauges allow one to also perform thermodynamic simulations of grand canonical ensembles by including an evolving trajectory weight. 

Stochastic equations 
for the interacting Bose gas model are found in Sections~\ref{CH5Equations}, and \ref{CH5Thermo}.
Extension to non-local interparticle interactions is derived in Section~\ref{CH5Extended}.

\section{Properties of the representation}
\label{CH5Representation}
The representation uses an 
un-normalized (Bargmann) coherent state basis. On a subsystem $j$ this basis is
\EQN{\label{bargmannket}
||\alpha_j\rangle_j = \exp\left(\alpha_j\dagop{a}_j\right)\ket{0} = \exp\left(\frac{|\alpha_j|^2}{2}\right)|\alpha_j\rangle_j
,}
with the complex amplitude $\alpha_j$, and mean particle occupation $\langle\op{n}_j\rangle=|\alpha_j|^2$. 
 As usual, the boson annihilation operators at subsystem $j$ are $\op{a}_j$, and obey the commutation relations
\EQN{\label{commutationa}
\left[\, \op{a}_j\,,\,\dagop{a}_k\,\right] = \delta_{jk}
.}
The basis states are mutually non-orthogonal
\EQN{\label{bargmannproduct}
\langle\beta^*_j||_j||\alpha_k\rangle_k = \delta_{jk} e^{\alpha_j\beta_j}
,}
(from \eqref{cohorthog}) and overcomplete (from \eqref{cohovercomp}). 

The kernel (on $N$ separable subsystems) is chosen to be
\EQNa{\label{gaugekernel}
\op{\Lambda} &=& e^{z_0}\otimes_{j=1}^N\op{\Lambda}_j,\\
\op{\Lambda}_j &=& ||\alpha_j\rangle_j\langle\beta^*_j||_j\,\exp\left(-\alpha_j\beta_j\right)
.}
If one defines coherent amplitude vectors $\bm{\alpha}$ and $\bm{\beta}$ to contain all $N$  elements $\alpha_j$ and $\beta_j$, respectively,  then the full variable set is $C=\{z_0,\bm{\alpha},\bm{\beta}\}$, containing $2N+1$ complex variables.
The non-orthogonality of the basis coherent states allows normalization $\tr{\op{\Lambda}}=e^{z_0}$ apart from the complex global weight $\Omega=e^{z_0}$. The kernel is also seen to be analytic in all complex variables $\alpha_j$, $\beta_j$, and $z_0$, so the procedure of Section~\ref{CH3EquationsAnalytic} can be used to ensure a stochastic realization for any FPE. 

As per expression \eqref{basicform}, the density matrix (possibly un-normalized) is expanded  as
\EQN{\label{gaugeprho}
\op{\rho}_u = \int P_G(C) \op{\Lambda}(C)\,d^{2N}\!\bm{\alpha}\,d^{2N}\!\bm{\beta}\,d^2z_0
}
with the gauge P distribution $P_G(C)$.

Now, it can be verified by expansion that
\SEQN{\label{aopidentities}}{
\op{a}_k||\alpha_j\rangle_j &=& \delta_{jk}\alpha_j||\alpha_j\rangle_j\\
\dagop{a}_k||\alpha_j\rangle_j &=& \dada{}{\alpha_k}||\alpha_j\rangle_j
.}
This then leads to the basic kernel operator correspondences 
\SEQN{\label{correspondencesg}}{
\op{a}_j\op{\Lambda} &=& \alpha_j\op{\Lambda},\\
\dagop{a}_j\op{\Lambda} &=& \left(\beta_j+\dada{}{\alpha_j}\right)\op{\Lambda},\\
\op{\Lambda}\op{a}_j &=& \left(\alpha_j+\dada{}{\beta_j}\right)\op{\Lambda},\\
\op{\Lambda}\dagop{a}_j &=& \beta_j\op{\Lambda}\label{correspondenceLad}
}
in the form \eqref{generaloperatoridentity}
Together with 
\EQN{
\dada{}{z_0}\op{\Lambda} = \op{\Lambda}
,}
which is seen to apply by inspection of \eqref{gaugekernel}, these operator equations can be used to obtain observable moment  estimates and stochastic equations using the procedures of Sections~\ref{CH3StochasticMoments}, \ref{CH3EquationsFPE}, \ref{CH3EquationsAnalytic}.

The similarity of the gauge P to the positive P representation allows one to adopt some exact results obtained for the latter.
Any state that has the positive P representation $P_+(\bm{\alpha},\bm{\beta})$ can be represented by the gauge P representation 
\EQN{\label{Pgaugepp}
  P_G(z_0,\bm{\alpha},\bm{\beta}) = \delta^2(z_0)\,P_+(\bm{\alpha},\bm{\beta})
,}
by inspection of the kernels \eqref{gaugekernel} and \eqref{ppkernel}.
(This is not a unique correspondence, but the simplest of many.)

Since it has been shown that any quantum state $\op{\rho}$ has a positive P representation\cite{DrummondGardiner80}, it follows from \eqref{Pgaugepp} that all quantum states must also have gauge P representations. 
The constructive expression \eqref{pprho} for a positive P representation of arbitrary $\op{\rho}$ can be substituted into \eqref{Pgaugepp} to obtain a similar expression for the gauge P.

\section{Observables}
\label{CH5Observables}
Calculation of observables proceeds in a very similar manner to the positive P representation in Section~\ref{CH3StochasticPP}.
All operators 
on supported states can be written as linear combinations of the moments of the local annihilation and creation operators 
$\op{a}_j,\,\dagop{a}_j$. Thus, to evaluate any observable it suffices to know how to evaluate an expectation value of a Hermitian operator with two adjoint separable  terms of the form 
\EQN{\label{qdef}
\op{Q}(\{L_j,L'_k\},\theta) = \frac{e^{i\theta}}{2} \otimes_j \dagop{a}_{L_j}\otimes_{k} \op{a}_{L'_{k}} + \frac{e^{-i\theta}}{2} \otimes_k \dagop{a}_{L'_k}\otimes_{j} \op{a}_{L_j}
.}
$\theta$ is a phase, the $L_j$ and $L'_k$ are subsystem labels (not necessarily unique), while the $j$ and $k$ are ``subsystem label counters''. For example the particle number operator for subsystem $p$ has $j=k=\{1\},\, L_1=L'_1=p,\, \theta=0$; while a quadrature operator $\op{q}(\theta)=\half(\dagop{a}e^{i\theta}+\op{a}e^{-i\theta})$ on subsystem $p$ has $j=\{1\},\, k=\{\,\},\, L_1=p$. 

Using \eqref{correspondencesg} and $\tr{\op{\Lambda}}=e^{z_0}$, one then obtains 
\EQN{
\tr{\op{Q}\op{\Lambda}} &=& \frac{e^{i\theta}}{2} \prod_{k} \alpha_{L'_{k}}\prod_j \left(\beta_{L_j}+\dada{}{\alpha_{L_j}}\right)\tr{\op{\Lambda}} + \frac{e^{-i\theta}}{2} \prod_{j} \alpha_{L_j}\prod_k \left(\beta_{L'_k}+\dada{}{\alpha_{L'_k}}\right)\tr{\op{\Lambda}}\nonumber\\
&=& \frac{e^{i\theta+z_0}}{2} \prod_j \beta_{L_j}\prod_{k} \alpha_{L'_{k}} + \frac{e^{-i\theta+z_0}}{2} \prod_k \beta_{L'_k}\prod_{j} \alpha_{L_j}
.}
Similarly, 
\EQN{
\tr{\op{Q}\dagop{\Lambda}} &=& \frac{e^{i\theta+z^*_0}}{2} \prod_j \alpha^*_{L_j}\prod_{k} \beta^*_{L'_{k}} + \frac{e^{-i\theta+z^*_0}}{2} \prod_k \alpha^*_{L'_k}\prod_{j} \beta^*_{L_j}
.}
These can be entered directly into the general observable estimate expression \eqref{observables} giving 
\EQNa{\label{qest}
\bar{Q}(\{L_j,L'_k\},\theta) &=& \frac{\average{\re{\Half e^{z_0} \left(e^{i\theta}\prod_{jk}\beta_{L_j}\alpha_{L'_{k}} +e^{-i\theta}\prod_{jk}\beta_{L'_k}\alpha_{L_j}\right)}}}{\average{\re{e^{z_0}}}}\qquad\\
\langle\op{Q}(\{L_j,L'_k\},\theta)\rangle &=& \lim_{\mc{S}\to\infty} \bar{Q}(\{L_j,L'_k\},\theta)
.}

Comparing \eqref{qdef} and \eqref{qest}, one sees that the same form appears in both, and so an algorithm to  determine the finite-sample estimate of the expectation value of an arbitrary observable $\langle \op{O}\rangle$ can be given:
\ENUM{
\item Normally-order $\op{O}$, using \eqref{commutationa} by placing all annihilation operators to the right.
\item The numerator of the stochastic estimate $\bar{O}$ is formed by replacing $\op{a}_k$, $\dagop{a}_j$, and $\otimes$ in the  normally-ordered expression for $\op{O}$ by  $\alpha_k$, $\beta_j$, and $\prod$, respectively, and finally multiplying by the global weight $e^{z_0}=\Omega$.
\item Take the average of the real part of the numerator terms, and divide by the average of the real part of the weight $\average{\re{e^{z_0}}}$. 
}

In cases where the normalization of $\op{\rho}$ is known to be conserved (e.g. dynamical master equations \eqref{dynamixmaster}, but not thermodynamic \eqref{thermomaster}), the average in the denominator can be dropped, because it is known to always be equal to one in the limit of many trajectories. Its variations about unity are an indication of the sampling error in the calculation, but are not necessary to obtain expectation values.

The above algorithm applies also when $\op{O}$ is an infinite sum of moments, such as for example the parity operator at subsystem $k$, which can be written as
\EQN{\label{paritydef}
\op{\pi}_k = \sum_{n=0}^{\infty} (-1)^n\ket{n}_k\bra{n}_k
}
in terms of Fock number states $\ket{n}_k$ with occupation $n$ at  subsystem $k$. 
Since $\ket{n}_k = (1/\sqrt{n!})(\dagop{a}_k)^n\ket{0}$, the one obtains
\EQN{\label{parityestimate}
\langle\op{\pi}_k\rangle = \lim_{\mc{S}\to\infty}\frac{\average{\re{\exp\left(z_0-2\alpha_k\beta_k\right)}}}{\average{\re{e^{z_0}}}}
.}
  With such operators of infinite order in annihilation and creation operators, one should, however, be wary of boundary term errors (of the second kind), as explained in Section~\ref{CH6SecondMechanism}.

Finally, for estimates of fidelity using expression \eqref{fidelityestimate}, one requires the trace of kernel products. Using \eqref{gaugekernel} and the properties of Bargmann states \eqref{bargmannproduct}, one obtains
\SEQN{}{
\tr{\op{\Lambda}(\bm{\alpha}_1,\bm{\beta}_1,\Omega_1)\op{\Lambda}(\bm{\alpha}_2,\bm{\beta}_2,\Omega_2)}
&=& \Omega_1\Omega_2\exp\left[-(\bm{\alpha}_1-\bm{\alpha}_2)\cdot(\bm{\beta}_1-\bm{\beta}_2)\right]\qquad\\
\tr{\op{\Lambda}(\bm{\alpha}_1,\bm{\beta}_1,\Omega_1)\dagop{\Lambda}(\bm{\alpha}^*_2,\bm{\beta}^*_2,\Omega^*_2)}
&=& \Omega_1\Omega^*_2\exp\left[-(\bm{\alpha}_1-\bm{\beta}^*_2)\cdot(\bm{\beta}_1-\bm{\alpha}^*_2)\right]\qquad
.}

\section{Dynamics of locally-interacting Bose gas}
\label{CH5Equations}

  The master equation for the dynamics of a locally interacting Bose gas on a lattice is \eqref{dynamixmaster} with the Hamiltonian given by \eqref{deltaH}.  Using the operator correspondences \eqref{correspondencesg}, and the methods of Sections~\ref{CH3EquationsFPE}, \ref{CH3EquationsAnalytic}, one can obtain an FPE for this system. Subsequently using the methods of Section~\ref{CH3EquationsLangevin}, the square root form of the noise matrix $\ul{B}=\ul{B}_0=\sqrt{\ul{D}}$, and standard drift gauges, the stochastic equations given below are obtained from \eqref{langevincg} and \eqref{domega} (or more directly, from \eqref{langevinstd}, setting $g_{jk}=0$).  Diffusion gauges have been omitted at this stage (so, $O(g_{jk})=I$) for clarity. 
Note that the $M$ spatial modes are now labeled by the lattice labels $\bo{n}$, or $\bo{m}$ as defined in Section~\ref{CH2Lattice}. Each mode is  a ``subsystem''.

It was chosen to separate the noise contributions from each process, as discussed in Section~\ref{CH4DiffusionMixing}. This is done so that the noise terms take on a simple form. 

With no environment interactions, the purely Hamiltonian evolution is found to lead to the Ito stochastic equations 
\SEQN{\label{itoH}}{
  d\alpha_{\bo{n}} &=&  -i\sum_{\bo{m}}\omega_{\bo{nm}}\alpha_{\bo{m}}dt - 2i\chi\alpha_{\bo{n}}^2\beta_{\bo{n}}dt + i\alpha_{\bo{n}}\sqrt{2i\chi}(dW_{\bo{n}}-\mc{G}_{\bo{n}}),\\
  d\beta_{\bo{n}} &=&  i\sum_{\bo{m}}\omega_{\bo{mn}}\beta_{\bo{m}}dt +2i\chi\alpha_{\bo{n}}\beta_{\bo{n}}^2dt + \beta_{\bo{n}}\sqrt{2i\chi}(d\wt{W}_{\bo{n}}-\mc{\wt{G}}_{\bo{n}}),\\
d\Omega &=& \Omega \sum_{\bo{n}} \left[ \mc{G}_{\bo{n}}dW_{\bo{n}} + \mc{\wt{G}}_{\bo{n}}d\wt{W}_{\bo{n}}\right]
,}
with $2M$ independent real Wiener increments $dW_{\bo{n}}$ and $d\wt{W}_{\bo{n}}$. The corresponding complex drift gauges are $\mc{G}_{\bo{n}}$ and $\mc{\wt{G}}_{\bo{n}}$. 

Addition of a heat bath at temperature $T$ results in the following additions:
\SEQN{\label{itoheatbathT}}{
  d\alpha_{\bo{n}} &=& \dots - \frac{\gamma_{\bo{n}}}{2}\alpha_{\bo{n}} dt +\sqrt{\gamma_{\bo{n}}\bar{n}_{\rm bath}(T)}(d\eta_{\bo{n}} - \mc{G}^{(1)}_{\bo{n}})\\
  d\beta_{\bo{n}} &=& \dots - \frac{\gamma_{\bo{n}}}{2}\beta_{\bo{n}} dt +\sqrt{\gamma_{\bo{n}}\bar{n}_{\rm bath}(T)}(d\eta^*_{\bo{n}}-\wt{\mc{G}}^{(1)}_{\bo{n}})\\
d\Omega &=& \dots + 
\Omega\sum_{\bo{n}}\left[d\eta_{\bo{n}}\wt{\mc{G}}^{(1)}_{\bo{n}}+d\eta^*\mc{G}^{(1)}_{\bo{n}}\right]
,}
where $\bar{n}_{\rm bath}$ is given by the Bose-Einstein distribution \eqref{bedistribution}. 
Here the {\it complex} noises $d\eta_{\bo{n}}$ are independent, and satisfy 
\SEQN{}{
\average{d\eta_{\bo{n}}d\eta^*_{\bo{m}}} &=& \delta_{\bo{nm}}dt,\\
\average{d\eta_{\bo{n}}d\eta_{\bo{m}}} &=& 0,\\
\average{d\eta_{\bo{n}}} &=& 0
.}
The drift gauges $\mc{G}^{(1)}_{\bo{n}}$ and $\mc{\wt{G}}^{(1)}_{\bo{n}}$ are also complex.

Two-particle losses to a zero temperature heat bath result in the following additions to the dynamical equations:
\SEQN{\label{itotwoparticlelosses}}{
  d\alpha_{\bo{n}} &=& \dots -\gamma^{(2)}_{\bo{n}}\alpha_{\bo{n}}^2\beta_{\bo{n}} dt+i\alpha_{\bo{n}}\sqrt{\gamma^{(2)}_{\bo{n}}}\left(dW^{(2)}_{\bo{n}}-\mc{G}^{(2)}_{\bo{n}}\right), \\
  d\beta_{\bo{n}} &=& \dots -\gamma^{(2)}_{\bo{n}}\alpha_{\bo{n}}\beta_{\bo{n}}^2 dt +i\beta_{\bo{n}}\sqrt{\gamma^{(2)}_{\bo{n}}}\left(d\wt{W}^{(2)}_{\bo{n}}-\wt{\mc{G}}^{(2)}_{\bo{n}}\right), \\
d\Omega &=& \dots + \Omega \sum_{\bo{n}} \left[ \mc{G}^{(2)}_{\bo{n}}dW^{(2)}_{\bo{n}} + \mc{\wt{G}}^{(2)}_{\bo{n}}d\wt{W}^{(2)}_{\bo{n}}\right]
.}
Here, again, the noises $dW_{\bo{n}}^{(2)}$ and $d\wt{W}_{\bo{n}}^{(2)}$ are independent Wiener increments, and the drift gauges $\mc{G}^{(2)}_{\bo{n}}$ and $\mc{\wt{G}}^{(2)}_{\bo{n}}$ are complex.

Lastly, A coherent driving field leads to the deterministic corrections
\SEQN{\label{itoepsilon}}{
  d\alpha_{\bo{n}} &=& \dots + \varepsilon dt, \\
  d\beta_{\bo{n}} &=& \dots + \varepsilon^* dt
.}

The numerical simulation strategy is (briefly) \begin{enumerate}
\item Sample a trajectory according to the known initial condition $P_G(0)=P_G(\,\op{\rho}(0)\,)$ 
\item Evolve according to the stochastic equations, calculating moments of interest, and accumulating appropriate sums of them.
\item Repeat for $\mc{S}\gg1$ independent trajectories.
\end{enumerate}

One could equally well remain with $2M\times 2M$ complex noise matrix, which would use only $2M$ real noises and $2M$ drift gauges, instead of the $6M$ real noises and $6M$ complex drift gauges above. In general, however, the stochastic terms would then have a complicated dependence on the parameters $\chi, \bar{n}_{\rm bath}, \gamma_{\bo{n}},$ and $\gamma^{(2)}_{\bo{n}}$ --- possibly undesirable. On the other hand, the amount of noise in the simulation might be reduced, because the effects of one process may partly cancel the effects of another. 
This may be particularly so if both scattering $\propto\chi$ and two-boson heat bath interactions $\propto\gamma^{(2)}_{\bo{n}}$ are present. These two processes lead to similar terms in the equations (apart from factors of $i$, etc.), and may be expected to cancel some noise without causing excessively complicated equations.

Some brief comments about the behavior of the above equations:
\ITEM{
\item The (un-gauged) equations for $d\beta_{\bo{n}}$ are simply the complex conjugates of $d\alpha_{\bo{n}}$, apart from 
possessing independent noises.
\item The Hamiltonian evolution \eqref{itoH} leads to a nonlinear equation in the coherent amplitudes, with obligatory noise.
\item All interactions with a heat bath (irrespective of the temperature) cause a deterministic exponential loss of particles.
\item Finite temperature thermal interactions cause a directionless randomization of the coherent amplitudes $\bm{\alpha}$ and $\bm{\beta}$, leading also to a mean growth of boson numbers.
\item Two-particle losses also lead to nonlinear equations and noise. 
\item The noise from scattering and two-particle losses is directional in phase space, as opposed to thermal noise.
\item Coherent gain causes no noise in the equations.
}

\section{Comparison to Gross-Pitaevskii semiclassical equations}
\label{CH5GP}

The lossless equations \eqref{itoH} are similar in form to the Gross-Pitaevskii (GP) semiclassical equations ubiquitous in calculations on Bose-Einstein Condensates at temperatures well below condensation. (For details of these gases and results that are obtained with the GP equations see e.g. the review by Dalfovo\etal\cite{Dalfovo-99}.) Derivation of the GP equations basically assumes the particles all coherently occupy a single orbital, and can be described by its wavefunction. 

In fact, if one
\ENUM{
\item Ignores stochastic and drift gauge terms in \eqref{itoH}.
\item Makes the assumption that the field is coherent --- i.e. the local kernels are coherent state projectors $\op{\Lambda}_{\bo{n}}\to\ket{\alpha_{\bo{n}}}_{\bo{n}}\bra{\alpha_{\bo{n}}}_{\bo{n}}$, implying 
\EQN{
\bm{\beta}\to\bm{\alpha}^*
.}
Note that this is consistent with the first assumption of no stochastic terms, since $d\beta_{\bo{n}}$ differs from $d\alpha^*_{\bo{n}}$ only by having independent noises.
\item Makes a variable change
\EQN{
\psi_{\bo{n}} = \psi(\bo{x}_{\bo{n}}) = \frac{\alpha_{\bo{n}}}{\sqrt{\prod_d\Delta x_d}} = \frac{\beta^*_{\bo{n}}}{\sqrt{\prod_d\Delta x_d}} 
,}
such that $\psi$ is the order parameter (i.e. the wavefunction normalized to $\bar{N}=\int \psi(\bo{x})\, d^{\mc{D}}\bo{x}$, where $\bar{N}$ is the mean particle number): 
}
then precisely the (lattice) GP equations for $\psi(\bo{x}_{\bo{n}})$ are obtained.
From this, it can  be surmised that in regimes where the GP equations are a reasonably good approximation, the noise will be relatively small and calculations precise.

It is quite remarkable that just by the addition of simple stochastic terms, full first-principles quantum evolution is recovered from a mean field theory. This convenient property suggests that full quantum simulations using this method have the potential to remain numerically tractable.

\section{Extended interparticle interactions}
\label{CH5Extended}
  Equations for the case of extended interparticle interactions as in the Hamiltonian \eqref{latticeH} will be derived here.  
 Proceeding as in Section~\ref{CH5Equations}, the complex diffusion matrix (before introducing any drift gauges) in the FPE can now be written
\EQN{
  \ul{D} = \matrix{D^{(\alpha)}&0\\0&D^{(\beta)}}
,}
where $D^{(\alpha)}$ appears in the FPE as $\raisebox{2pt}{\half} D^{(\alpha)}_{\bo{nm}}\partial^2/\partial\alpha_{\bo{n}}\partial\alpha_{\bo{m}}$, and $D^{(\beta)}$ as $\raisebox{2pt}{\half} D^{(\beta)}_{\bo{nm}}\partial^2/\partial\beta_{\bo{n}}\partial\beta_{\bo{m}}$.
Their matrix elements are given by
\SEQN{\label{Dznm}}{
  D^{(\alpha)}_{\bo{nm}} &=& -i\frac{u_{\bo{nm}}}{\hbar}\alpha_{\bo{n}}\alpha_{\bo{m}},\\
  D^{(\beta)}_{\bo{nm}} &=& i\frac{u_{\bo{nm}}}{\hbar}\beta_{\bo{n}}\beta_{\bo{m}}
.}
Since the interaction potential $u_{\bo{nm}}$ is symmetric, then it could be orthogonally decomposed by its matrix square root 
$\upsilon$, which satisfies
\EQN{
  \upsilon\upsilon = u = \upsilon\upsilon^T
.}
The matrix $\upsilon$ could be calculated at the beginning of a simulation and subsequently  used in noise matrices satisfying $B^{(z)}[B^{(z)}]^T=D^{(z)}$ (with $z$ taking on the labels $\alpha$ or $\beta$), where  
\SEQN{}{
B^{(\alpha)}_{\bo{nm}} &=& \sqrt{\frac{-i}{\hbar}}\alpha_{\bo{n}}\,\upsilon_{\bo{nm}}\\
B^{(\beta)}_{\bo{nm}} &=& \sqrt{\frac{i}{\hbar}}\beta_{\bo{n}}\,\upsilon_{\bo{nm}}
.}
Unfortunately finding the matrix square root would usually need to be done numerically, which would require storing $\order{M^2}$ matrix elements $\upsilon_{\bo{nm}}$ and take a time $\order{M^4}$ to compute --- not efficient for large lattices.  

A much more efficient, though involved, approach is possible. One would like to obtain some orthogonal decomposition of $\ul{D}$ dependent directly on $U_{\bo{n}}$ that depends only on the interparticle spacing. This potential has $M$ elements, rather than the $M\times M$ of $u_{\bo{nm}}$.
Writing the Langevin equations in terms of stochastic increments $dX^{(z)}_{\bo{n}}$  (See Section~\ref{CH3EquationsLangevin}) as
\EQN{
dz_{\bo{n}} = A^{(z)}_{\bo{n}}\,dt + dX^{(z)}_{\bo{n}}
,}
the stochastic increments must obey 
\SEQN{}{
  \average{dX^{(z)}_{\bo{n}}} &=& 0\\
  \average{dX^{(z)}_{\bo{n}}dX^{(z)}_{\bo{m}}} &=& D^{(z)}_{\bo{nm}}\,dt\\
  \average{dX^{(\alpha)}_{\bo{n}}dX^{(\beta)}_{\bo{m}}} &=& 0
.}

Firstly, we can define new stochastic increments $dY^{(\alpha)}_{\bo{n}}$ and $dY^{(\beta)}_{\bo{n}}$, such that 
\SEQN{}{
  dX^{(\alpha)}_{\bo{n}} &=& \alpha_{\bo{n}}\sqrt{\frac{-i}{\hbar}}\,dY^{(\alpha)}_{\bo{n}}\\
  dX^{(\beta)}_{\bo{n}} &=& \beta_{\bo{n}}\sqrt{\frac{i}{\hbar}}\,dY^{(\beta)}_{\bo{n}}
,} and these must obey relationships (Remembering from \eqref{uuu} that $u_{\bo{nm}}=u_{\bo{mn}}=U_{|\bo{n}-\bo{m}|}$), 
\SEQN{\label{dycond}}{
  \average{dY^{(z)}_{\bo{n}}} &=& 0\\
  \average{dY^{(z)}_{\bo{n}}dY^{(z)}_{\bo{m}}} &=& U_{|\bo{n}-\bo{m}|}\,dt  \label{dycond2},
}
which now depend only on the {\it a-priori} potential $U_{\bo{n}}$, not on the dynamically evolving mode amplitudes. 
Note that while the variances of the $dY^{(z)}_{\bo{n}}$ are the same for both choices of $z$, the actual realizations of the increments must be independent.

It is useful to now consider the $\mc{D}$-dimensional discrete Fourier transform of the interparticle potential:
\EQN{\label{Utilde}
\wt{U}_{\wt{\bo{n}}} =  \wt{U}'_{\wt{\bo{n}}} +i\wt{U}''_{\wt{\bo{n}}}  =\frac{1}{C_{\rm norm}}\sum_{\bo{n}} U_{\bo{n}} e^{-i\bo{k}_{\wt{\bo{n}}}\cdot\bo{x}_{\bo{n}}}
,}
with the normalization constant $C_{\rm norm}=(2\pi)^{\mc{D}/2}/\prod_d \Delta x_d$.
All the lattice notation used from here on has been defined in  Section~\ref{CH2Lattice}.
Note now that since $U_{\bo{n}}$ is real, and $\bo{k}_{-\wt{\bo{n}}}=-\bo{k}_{\wt{\bo{n}}}$, then \eqref{Utilde} implies
\SEQN{\label{utildeconj}}{
\wt{U}'_{\wt{\bo{n}}} &=& \wt{U}'_{-\wt{\bo{n}}}\\
\wt{U}''_{\wt{\bo{n}}} &=& -\wt{U}''_{-\wt{\bo{n}}}
.}
The inverse transform is 
\EQN{\label{iUtilde}
U_{\bo{n}} = \frac{C_{\rm norm}}{M}\sum_{\wt{\bo{n}}} \wt{U}_{\wt{\bo{n}}} e^{i\bo{k}_{\wt{\bo{n}}}\cdot\bo{x}_{\bo{n}}}
.}
Expanding out the elements of lattice coordinate vectors, one can write
$|\bo{x}_{\bo{n}}-\bo{x}_{\bo{m}}|=\{\vartheta_1\Delta x_1(n_1-m_1),\dots,\vartheta_{\mc{D}}\Delta x_{\mc{D}}(n_{\mc{D}}-m_{\mc{D}})\}$, where the quantities $\vartheta_d$ can take on the values $+1$ or $-1$, depending on what is required to take the modulus. Recalling the symmetry property of the interparticle potential that was assumed in Section~\ref{CH2Lattice}
 ($U_{\{n_1,\dots,n_d,\dots\}} = U_{\{n_1,\dots,M_d-n_d,\dots\}}$ for any dimension $d$), one obtains 
\EQN{
U_{\{n_1,\dots,n_d,\dots\}} &=& \frac{C_{\rm norm}}{M}\sum_{\wt{\bo{n}}} \wt{U}_{\wt{\bo{n}}} \prod_d e^{ik_d(\wt{n}_d)x_d(n_d)}\nonumber\\
&=& \frac{C_{\rm norm}}{M}\sum_{\wt{\bo{n}}} \wt{U}_{\wt{\bo{n}}} e^{-ik_d(\wt{n}_d)x_d(n_d)}
\prod_{d'\neq d} e^{ik_{d'}(\wt{n}_{d'})x_{d'}(n_{d'})}
,}
 since $\exp[ik_dM_d\Delta x_d]=1$ for any $k_d$. So, the phase for dimension $d$ can have either sign. This result can then be applied to each dimension where $\vartheta_d=-1$ to obtain (using also $\bo{x}_{\bo{n}-\bo{m}}=\bo{x}_{\bo{n}}-\bo{x}_{\bo{m}}$) the expression (subtly different from \eqref{iUtilde})
\EQN{\label{imUtilde}
U_{|\bo{n}-\bo{m}|} = \frac{C_{\rm norm}}{M}\sum_{\wt{\bo{n}}} \wt{U}_{\wt{\bo{n}}} e^{i\bo{k}_{\wt{\bo{n}}}\cdot(\bo{x}_{\bo{n}}-\bo{x}_{\bo{m}})}
.}

If one introduces new stochastic increments $dZ^{(v)}_{\bo{n}}$, then using \eqref{imUtilde},  condition \eqref{dycond2} is equivalent to 
\EQN{\label{dydy2}
\average{dY^{(z)}_{\bo{n}}dY^{(z)}_{\bo{m}}} &=&\frac{C_{\rm norm}}{M}\sum_{\wt{\bo{n}}\wt{\bo{m}}}e^{i\bo{k}_{\wt{\bo{n}}}\cdot\bo{x}_{\bo{n}}}e^{-i\bo{k}_{\wt{\bo{m}}}\cdot\bo{x}_{\bo{m}}}
\sqrt{\wt{U}_{\wt{\bo{n}}}\wt{U}_{\wt{\bo{m}}}}\average{dZ^{(v)}_{\wt{\bo{n}}}dZ^{(v)}_{\wt{\bo{m}}}}	
}
provided that 
\SEQN{}{
  \average{dZ^{(z)}_{\wt{\bo{n}}}} &=&  0\\
  \average{dZ^{(z)}_{\wt{\bo{n}}}dZ^{(z)}_{\wt{\bo{m}}}} &=& \delta_{\wt{\bo{n}}\wt{\bo{m}}}\,dt
.}
This can be checked by substitution. This is not yet quite what one wants to be able to decompose into $dY^{(z)}_{\bo{n}}$ because the second phase in \eqref{dydy2} has the wrong sign. Again using $-\bo{k}_{\wt{\bo{n}}}=\bo{k}_{-\wt{\bo{n}}}$, to relabel $\wt{\bo{m}}\to-\wt{\bo{m}}$, and applying \eqref{utildeconj} one has 
\EQN{
\average{dY^{(z)}_{\bo{n}}dY^{(z)}_{\bo{m}}} &=&\frac{C_{\rm norm}}{M}\sum_{\wt{\bo{n}}\wt{\bo{m}}}e^{i\bo{k}_{\wt{\bo{n}}}\cdot\bo{x}_{\bo{n}}}e^{i\bo{k}_{\wt{\bo{m}}}\cdot\bo{x}_{\bo{m}}}
\sqrt{\wt{U}_{\wt{\bo{n}}}\wt{U}^*_{\wt{\bo{m}}}}\ \delta_{\bo{n},-\bo{m}}dt
.}
The phase factor is now fine, but the complex conjugate $U^*_{\wt{\bo{m}}}$ spoils the potential decomposition. 
What is needed are some 
stochastic increments $d\wt{Z}^{(z)}_{\wt{\bo{n}}}$ that will satisfy 
\SEQN{}{
\average{d\wt{Z}^{(z)}_{\wt{\bo{n}}}}&=& 0\\
\average{d\wt{Z}^{(z)}_{\wt{\bo{n}}}d\wt{Z}^{(z)}_{\wt{\bo{m}}}} &=& \delta_{\wt{\bo{n}},-\wt{\bo{m}}}\wt{U}_{\wt{\bo{n}}}\,dt
,}
to allow then a decomposition 
\EQN{
dY^{(z)}_{\bo{n}} = \sqrt{\frac{C_{\rm norm}}{M}}\sum_{\wt{\bo{n}}} e^{i\bo{k}_{\wt{\bo{n}}}\cdot\bo{x}_{\bo{n}}}\,d\wt{Z}^{(z)}_{\wt{\bo{n}}}
.}
This can be achieved by separating out the real and imaginary parts of $\wt{U}$ as
\EQN{
d\wt{Z}^{(z)}_{\wt{\bo{n}}} = \sqrt{\wt{U}'_{\wt{\bo{n}}}}d\zeta^{(z)}_{\wt{\bo{n}}} + \sqrt{\wt{U}''_{\wt{\bo{n}}}}d\wt{\zeta}^{(z)}_{\wt{\bo{n}}}
,}
with the new stochastic increments
\SEQN{}{
\average{d\zeta^{(z)}_{\wt{\bo{n}}}}&=& \average{d\wt{\zeta}^{(z)}_{\wt{\bo{n}}}} = 0\\
\average{d\zeta^{(z)}_{\wt{\bo{n}}}d\zeta^{(z)}_{\wt{\bo{m}}}} &=& \delta_{\wt{\bo{n}},-\wt{\bo{m}}}\,dt\\
\average{d\wt{\zeta}^{(z)}_{\wt{\bo{n}}}d\wt{\zeta}^{(z)}_{\wt{\bo{m}}}} &=& \delta_{\wt{\bo{n}},-\wt{\bo{m}}}\,dt\\
\average{d\zeta^{(z)}_{\wt{\bo{n}}}d\wt{\zeta}^{(z)}_{\wt{\bo{m}}}} &=& 0
.}
Note that $\sqrt{\wt{U}''_{\wt{\bo{n}}}\wt{U}''_{-\wt{\bo{n}}}} = i\wt{U}''_{\wt{\bo{n}}}$ by \eqref{utildeconj}.
Let us divide the $\wt{\bo{n}}\neq0$ momentum mode space into two symmetric halves $\mc{R}$ and $\wt{\mc{R}}$, such that e.g. when $\mc{D}=3$:
\EQN{
\wt{\bo{n}}\in \mc{R}\text{ if }\left\{\begin{array}{clcl}&\wt{n}_1>0&&\\
		 \text{or,}&\wt{n_1}=0&\text{and}& \wt{n_2}>0\\
		 \text{or,}&\wt{n}_1=\wt{n}_2=0&\text{and}&\wt{n}_3>0
\end{array}\right.
}
while $\wt{\bo{n}}\in\wt{\mc{R}}$ if $\wt{\bo{n}}\not\in\{\mc{R},\bo{0}\}$. 
The $d\zeta^{(z)}_{\wt{\bo{n}}}$ noises can now be realized by the construction 
\EQN{\label{zetaxi}
d\zeta^{(z)}_{\wt{\bo{n}}} = \left\{\begin{array}{cl}
(dW_{\wt{\bo{n}},1}+idW_{\wt{\bo{n}},2})/\sqrt{2}    & \text{ if } \wt{\bo{n}}\in\mc{R}\\
dW_{\bo{0},1}		       	& \text{ if } \wt{\bo{n}}=\bo{0}\\
(dW_{-\wt{\bo{n}},1}-idW_{-\wt{\bo{n}},2})/\sqrt{2}  & \text{ if } \wt{\bo{n}}\in\wt{\mc{R}}.
\end{array}\right.
}
in terms of real independent Wiener increments $dW_{\wt{\bo{n}},j}$, for all $\wt{\bo{n}}\in\{\mc{R},\bo{0}\}$, numbering $M$ in total\footnote{ $M+1$ if $M$ is even.}. $d\wt{\zeta}^{(z)}_{\wt{\bo{n}}}$ requires separate $M$ independent noises $dW_{\wt{\bo{n}},3}$ and $dW_{\wt{\bo{n}},4}$.

Collecting all this together, one obtains 
\SEQN{\label{dXexpr}}{
dX^{(\alpha)}_{\bo{n}} &=& \alpha_{\bo{n}}\sqrt{\frac{-iC_{\rm norm}}{\hbar M}}\sum_{\wt{\bo{n}}}e^{i\bo{k}_{\wt{\bo{n}}}\cdot\bo{x}_{\bo{n}}}
\left\{ \sqrt{\wt{U}'_{\wt{\bo{n}}}}d\zeta^{(\alpha)}_{\wt{\bo{n}}} + \sqrt{\wt{U}''_{\wt{\bo{n}}}}d\wt{\zeta}^{(\alpha)}_{\wt{\bo{n}}}
\right\}\\
dX^{(\beta)}_{\bo{n}} &=& \beta_{\bo{n}}\sqrt{\frac{iC_{\rm norm}}{\hbar M}}\sum_{\wt{\bo{n}}}e^{i\bo{k}_{\wt{\bo{n}}}\cdot\bo{x}_{\bo{n}}}
\left\{ \sqrt{\wt{U}'_{\wt{\bo{n}}}}d\zeta^{(\beta)}_{\wt{\bo{n}}} + \sqrt{\wt{U}''_{\wt{\bo{n}}}}d\wt{\zeta}^{(\beta)}_{\wt{\bo{n}}}
\right\}
.}
All four $\zeta^{(z)}_{\wt{\bo{n}}}$ and $\wt{\zeta}^{(z)}_{\wt{\bo{n}}}$ complex noises per (Fourier space) lattice point are independent of each other, and of the noises at all other (momentum) lattice points and times. Explicitly they have the form \eqref{zetaxi} in terms of the $4M$ real Wiener increments\footnote{Or $4(M+1)$ if $M$ is even.} $dW_{\wt{\bo{n}},j}$. 
Storage space for $M$ complex variables $\wt{U}_{\wt{\bo{n}}}$ is required, and calculation of these\footnote{Using the  ``fast Fourier transform'' algorithm.} takes a time $\propto M\log M$ --- much more tractable than the calculations of $v_{\bo{nm}}$. The $dX^{(z)}_{\bo{n}}$ reduce to the noise terms of \eqref{itoH} under local interactions $U_{\bo{n}}=2\hbar\chi\delta_{\bo{n},\bo{0}}$.

With no drift gauges, the two-particle interaction terms of the Ito stochastic equations become (the terms due to other processes are unchanged)
\SEQN{\label{nodriftUeq}}{
d\alpha_{\bo{n}} &=& \dots -i\sum_{\bo{m}}\frac{U_{|\bo{n}-\bo{m}|}}{\hbar}\alpha_{\bo{n}}\alpha_{\bo{m}}\beta_{\bo{m}} +dX^{(\alpha)}_{\bo{n}},\\
d\beta_{\bo{n}} &=& \dots +i\sum_{\bo{m}}\frac{U_{|\bo{n}-\bo{m}|}}{\hbar}\beta_{\bo{n}}\alpha_{\bo{m}}\beta_{\bo{m}} +dX^{(\beta)}_{\bo{n}}
.}
Drift gauges can be introduced by making the replacements
\EQN{
dW_{\bo{n},j}\to dW_{\bo{n},j}-\mc{G}_{\bo{n},j}\,dt
,}
in \eqref{zetaxi} (this follows straight from the standard form \eqref{langevinstd}) and 
\EQN{
d\Omega = \dots + \Omega\sum_{\bo{n}}\sum_j\mc{G}_{\bo{n},j}dW_{\bo{n},j} 
.}

Simulations of such models pose no fundamental problem (see e.g. Sections~\ref{CH10Extended} and~\ref{CH10Trap}), however the required computer time scales more steeply with $M$. There are now $2M$ complex terms to calculate in the equation for each variable rather than the $2$ for a local interaction model \eqref{itoH}. 

\REM{
\section{Multi-species simulations}
\label{CH5Species}
An extension to models that involve several distinct boson species is straightforward. At each lattice point, a tensor product local kernel is introduced that has separate independent coherent state amplitudes for each species $\varsigma$:
\EQN{
\op{\Lambda}_k = \otimes_{\varsigma} ||\alpha^{(\varsigma)}_k\rangle_{k,\varsigma}\langle\beta^{(\varsigma)}_k||_{k,\varsigma}
.}
For Hamiltonians of a similar form to \eqref{hamiltonian} that involve terms up to second order in boson creation operators (i.e. contain up to two-particle processes), stochastic equations can be derived with the same procedure as for the one-species case. There will be separate terms in the stochastic equations for each combination of interacting species. 
}

\section{Thermodynamics of interacting Bose gas}
\label{CH5Thermo}

The grand canonical thermodynamics of a system with Hamiltonian $\op{H}$ in thermal and diffusive contact with a reservoir at temperature $T$ and chemical potential $\mu$ can be simulated using the master equation \eqref{thermomaster}. 

With the locally-interacting Hamiltonian \eqref{deltaH}, using the operator correspondences \eqref{correspondencesg}, and the methods of Sections~\ref{CH3EquationsFPE}, \ref{CH3EquationsAnalytic}, one can obtain the FPE, and then directly by \eqref{langevinstd} (not using diffusion gauges here, so $g_{jk}=0$) the stochastic equations. Using the same notation as in Section~\ref{CH5Equations}, these can be written
\SEQN{\label{gaugepthermo}}{
d\alpha_{\bo{n}} &=& -\hbar\sum_{\bo{m}}\omega_{\bo{nm}}\alpha_{\bo{m}}\,d\tau 
+\left(\mu_e-2\hbar\chi\alpha_{\bo{n}}\beta_{\bo{n}}\right)\alpha_{\bo{n}}\,d\tau +i\alpha_{\bo{n}}\sqrt{2\hbar\chi}\left(dW_{\bo{n}}-\mc{G}_{\bo{n}}\right),\qquad\\
d\beta_{\bo{n}} &=& 0,\\
d\Omega &=& \Omega\left[ -\hbar\sum_{\bo{nm}}\omega_{\bo{nm}}\alpha_{\bo{m}}\beta_{\bo{n}}\,d\tau 
+\sum_{\bo{n}}\left(\mu_e-\hbar\chi\alpha_{\bo{n}}\beta_{\bo{n}}\right)\alpha_{\bo{n}}\beta_{\bo{n}}\,d\tau  + \sum_{\bo{n}}\mc{G}_{\bo{n}}dW_{\bo{n}}\right]\qquad
.}
There are $M$ real Wiener increments $dW_{\bo{n}}$, and hence the same number of complex drift gauges $\mc{G}_{\bo{n}}$. 

The asymmetric form (no $\beta_{\bo{n}}$ evolution) arises because the Kamiltonian acts only from the left on the density matrix in \eqref{thermomasterstub}. The initial $\beta_{\bo{n}}$ takes on a range of random initial values (see \eqref{inidist} below), which then remain constant. A symmetric set of equations is also possible by starting from the middle (anticommutator) term  of \eqref{thermomasterstub}, however this appears to serve no useful purpose but needs more noises (another set of $M$ for the $\beta_{\bo{n}}$ evolution).  

The initial condition \eqref{rhouo} is $\op{\rho}_u(0) = \exp\left[-\lambda_n\op{N}\right] = \otimes_{\bo{n}}\op{\rho}_{\bo{n}}(0)$, where $\lambda_n$ is given by \eqref{lambdandef}, and with 
\EQN{\label{rhouo5}
\op{\rho}_{\bo{n}}(0) = \exp\left[-\lambda_n\dagop{a}_{\bo{n}}\op{a}_{\bo{n}}\right]
.}
One could use \eqref{Pgaugepp} and \eqref{pprho} to obtain a gauge P distribution corresponding to this initial state, but a more compact distribution can be found as follows:

Since $\op{\rho}_u(0)$ is separable, let us just consider the initial conditions in a single mode $\bo{n}$ (all modes are in this same state initially). In a local Fock number state complete orthogonal basis $\ket{n}$, the initial state can be written 
\EQN{
\op{\rho}_{\bo{n}}(0)=
\exp\left[-\lambda_n\op{n}_{\bo{n}}\right]\op{I}_{\bo{n}} &=& \sum_p\frac{(-\lambda_n)^p}{p!}\op{n}_{\bo{n}}^p\sum_n\ket{n}\bra{n}\nonumber\\
&=& \sum_n e^{-\lambda_n n}\ket{n}\bra{n}\label{inifock}
.}
The local kernel \eqref{gaugekernel} , on the other hand, expanding the Bargmann states, is
\EQN{
\op{\Lambda}_{\bo{n}} &=& e^{\alpha_{\bo{n}}\dagop{a}_{\bo{n}}}\ket{0}\bra{0}e^{\beta_{\bo{n}}\op{a}_{\bo{n}}} e^{-\alpha_{\bo{n}}\beta_{\bo{n}}}\nonumber\\
&=& \sum_{n\wt{n}} \frac{\alpha_{\bo{n}}^n\beta_{\bo{n}}^{\wt{n}}}{\sqrt{n!\wt{n}!}}\ket{n}\bra{\wt{n}}e^{-\alpha_{\bo{n}}\beta_{\bo{n}}}\label{localkerneln}
.}
Since the high temperature state should be classical let us try a gauge P distribution over just diagonal coherent states and so postulate $\bm{\alpha}=\bm{\beta}^*$. 
Also, at high temperature the modes should be separable, so let us try a distribution where the amplitudes at each mode are independent. From \eqref{rhouo5}, the state of each mode should be identical, and lastly, for simplicity, let us choose the initial weight of each trajectory to be equal:  $\Omega(0)=e^{z_0(0)}=1$
Let us start, then, with the Gaussian ansatz
\EQN{
P^{\rm try}_G(\bm{\alpha},\bm{\beta},z_0) = \delta^2(z_0)\delta^{2M}(\bm{\beta}-\bm{\alpha}^*) \prod_{\bo{n}} \frac{1}{2\pi\sigma^2}\exp\left(\frac{-|\alpha_{\bo{n}}|^2}{2\sigma^2}\right)
.}
To see whether this is sufficient to represent the initial state, and to find the  value of $\sigma$, let us substitute into \eqref{gaugeprho} and see if \eqref{inifock} can be satisfied. 
Integrating over $\Omega$ and $\bm{\beta}$, and separating modes, one has (for each mode $\bo{n}$):
\EQN{
  \op{\rho}^{\rm\;try}_{\bo{n}}(0) &=& \int \frac{1}{2\pi\sigma^2}\exp\left(\frac{-|\alpha_{\bo{n}}|^2}{2\sigma^2}\right)
\sum_{n\wt{n}} \frac{\alpha_{\bo{n}}^n(\alpha^*_{\bo{n}})^{\wt{n}}}{\sqrt{n!\wt{n}!}}\ket{n}\bra{\wt{n}}e^{-|\alpha_{\bo{n}}|^2} d^2\alpha_{\bo{n}}
.}
Writing $\alpha_{\bo{n}}=re^{i\theta}$, one has
\EQN{
\op{\rho}^{\rm\;try}_{\bo{n}}(0) &=& \frac{1}{2\pi\sigma^2}\sum_{n\wt{n}}\frac{1}{\sqrt{n!\wt{n}!}}\int_{-\pi}^{\pi} \!\!\!d\theta\,e^{i\theta(n-\wt{n})}\int_0^{\infty}\!\!\!dr\,r^{n+\wt{n}+1}\exp\left(-\frac{r^2}{1+1/2\sigma^2}\right)\ket{n}\bra{\wt{n}}\nonumber\\
&=& \frac{1}{\sigma^2}\sum_{n}\frac{1}{n!}\int_0^{\infty}\!\!\!dr\,r^{2n+1}\exp\left(-\frac{r^2}{1+1/2\sigma^2}\right)\ket{n}\bra{n}
,}
and using\cite{GradshteynRyzhik65a} $\int_0^{\infty}r^{2n+1}e^{-fr^2} dr = n!/2f^{n+1}$, 
\EQN{
\op{\rho}^{\rm\;try}_{\bo{n}}(0) = \frac{1}{1+2\sigma^2}\sum_n\left(\frac{\sigma^2}{1+2\sigma^2}\right)^n\ket{n}\bra{n}
.}
One wants to have $\op{\rho}^{\rm\;try}_{\bo{n}}(0) = \op{\rho}_{\bo{n}}(0)/\tr{\,\op{\rho}_{\bo{n}}(0)}$. Since $\tr{\op{\rho}_{\bo{n}}(0)}=1/[1-e^{-\lambda_n}]$, this implies
\EQN{
\sigma = \frac{1}{\sqrt{2\left(e^{\lambda_n}-1\right)}} = \sqrt{\frac{\bar{n}_0}{2}}
,}
($\bar{n}_0=1/[e^{\lambda_n}-1]$ is the mean occupation per mode). So then, it has been verified that a gauge P distribution 
for $\op{\rho}_u(0)$ is just a Gaussian in $\bm{\alpha}$.
\EQN{\label{inidist}
P_G(\bm{\alpha},\bm{\beta},z_0) = \delta^2(z_0)\delta^{2M}(\bm{\beta}-\bm{\alpha}^*) \prod_{\bo{n}} \frac{1}{\pi \bar{n}_0}\exp\left(\frac{-|\alpha_{\bo{n}}|^2}{\bar{n}_0}\right)
.}
This is easily sampled to obtain initial values of $\alpha_{\bo{n}}=\beta^*_{\bo{n}}$, and $z_0=0$ for each trajectory.

For the case of extended interparticle interactions $u_{\bo{nm}}=U_{|\bo{n}-\bo{m}|}$, a similar procedure can be followed as was done for the dynamics in Section~\ref{CH5Extended}.  The diffusion matrix in the FPE is now
\SEQN{}{
  D^{(\alpha)}_{\bo{nm}} &=& -u_{\bo{nm}}\alpha_{\bo{n}}\alpha_{\bo{m}},\\
  D^{(\beta)}_{\bo{nm}} &=& 0
,}
rather than \eqref{Dznm}. With no drift gauges, the Ito stochastic equations are
\SEQN{}{
d\alpha_{\bo{n}} &=& -\hbar\sum_{\bo{m}}\omega_{\bo{nm}}\alpha_{\bo{m}}\,d\tau 
+\mu_e\alpha_{\bo{n}}\,d\tau -\sum_{\bo{m}}U_{|\bo{n}-\bo{m}|}\alpha_{\bo{n}}\alpha_{\bo{m}}\beta_{\bo{m}}\,d\tau + dX^{(\alpha)}_{\bo{n}}\\
d\beta_{\bo{n}} &=& 0,\\
d\Omega &=& \Omega\left[ -\hbar\sum_{\bo{nm}}\omega_{\bo{nm}}\alpha_{\bo{m}}\beta_{\bo{n}}
+\sum_{\bo{n}}\left(\mu_e-\sum_{\bo{m}}\frac{U_{|\bo{n}-\bo{m}|}}{2}\alpha_{\bo{m}}\beta_{\bo{m}}\right)\alpha_{\bo{n}}\beta_{\bo{n}} \right]\,d\tau\qquad
,} 
With the stochastic increments given by 
\EQN{
dX^{(\alpha)}_{\bo{n}} &=& i\alpha_{\bo{n}}\sqrt{\frac{C_{\rm norm}}{M}}\sum_{\wt{\bo{n}}}e^{i\bo{k}_{\wt{\bo{n}}}\cdot\bo{x}_{\bo{n}}}
\left\{ \sqrt{\wt{U}'_{\wt{\bo{n}}}}d\zeta^{(\alpha)}_{\wt{\bo{n}}} + \sqrt{\wt{U}''_{\wt{\bo{n}}}}d\wt{\zeta}^{(\alpha)}_{\wt{\bo{n}}}
\right\}
.}
Here the noises $d\zeta^{(\alpha)}_{\bo{n}}$ and $d\wt{\zeta}^{(\alpha)}_{\wt{\bo{n}}}$ are given again by \eqref{zetaxi}, but the $2M$ real Wiener increments $dW_{\wt{\bo{n}},j}$ now have variance $d\tau$ instead of $dt$. 
Drift gauges can again be introduced by making the replacements
\EQN{
dW_{\bo{n},j}\to dW_{\bo{n},j}-\mc{G}_{\bo{n},j}\,d\tau
,}
in \eqref{zetaxi} and 
\EQN{
d\Omega = \dots + \Omega\sum_{\bo{n}}\sum_j\mc{G}_{\bo{n},j}dW_{\bo{n},j} 
.}

\section{Comparison with historical distributions}
\label{CH5Comparison}

To put the gauge P representation in perspective, let us compare to the more commonly used phase-space representations from the field of quantum optics where phase-space distributions have arguably been most used. 
This section is based on Section~2 of the published article by Drummond and Deuar\cite{DrummondDeuar03}. 
The concepts and general layout of this section are due to P. D. Drummond.

To understand the reasons for development of the various distributions, it is useful to peruse the requirements listed in Section~\ref{CH3Requirements} that they have to satisfy to result in stochastic simulations of many-body models. 

Phase-space mappings were first introduced by Wigner as the famous Wigner representation\cite{Wigner32}. Historically, phase-space distributions have developed in three stages:
\ITEM{
\item \textbf{Stage one (classical-like phase space):} A classical-like phase space was used in which the number of real configuration variables in $C$ was the same as the number of classical degrees of freedom. Typically the kernel is of the form of a diagonal projector, and not all quantum states can be represented by positive nonsingular $P$. This usually manifests itself as either a non positive-definite propagator or 3rd order terms in the FPE in master equations involving several-body processes. Either way, a quantum-equivalent stochastic process is not recovered. 
The Wigner\cite{Wigner32}, Husimi Q\cite{Husimi40}, Glauber-Sudarshan P\cite{Glauber63,CahillGlauber69,Sudarshan63}, and the Poisson\cite{GardinerChaturvedi77,GardinerChaturvedi78,GardinerQN} representations, among others, all fall into this category.
\item \textbf{Stage two (doubled phase space):} By the use of off-diagonal kernels, representations in a higher-dimensional phase space were developed, for which a non-singular distribution $P$ exists for all quantum states. These typically have at least two  real configuration variables for each classical degree of freedom (one for the ket and one for the bra in the kernel). Examples of such representations are the Glauber R\cite{Glauber63,CahillGlauber69}  and the positive P\cite{Chaturvedi-77,DrummondGardiner80} representations. While these work very well for highly damped systems, in models with several-body processes and low damping, they typically develop unstable trajectories. This leads to large sampling uncertainties or even systematic ``boundary term errors''\cite{SmithGardiner89,SchackSchenzle91}.
\item \textbf{Stage three (global amplitude and stochastic gauges):} Addition of a global weight to the kernel allows the introduction of drift stochastic gauges as in Section~\ref{CH4Drift}, and (by appropriate gauge choice) modification of the stochastic equations to remove the instabilities. The gauge P representation introduced above is of this type, as is the stochastic wavefunction method of Carusotto\etal\cite{Carusotto-01,CarusottoCastin01}. This work of Carusotto\etal\ is in some ways complementary to that presented in this thesis, and has been developed approximately in parallel. One fundamental difference between the two representations is that the gauge P representation allows the particle number to vary, allowing open behavior such as lasing or evaporative cooling, whereas the stochastic wavefunction method is hardwired to a constant particle number $N$. 
}

\REM{
The kernels used in the abovementioned distributions are shown in Table~\ref{TABLEKernels}.
}

How do these various representations compare? A check of their applicability to interacting Bose gas simulations with Hamiltonians \eqref{deltaH} or \eqref{latticeH} is shown in Table~\ref{TABLEDistributions}. One can see that 
the early distributions were often hindered in obtaining a many-body simulation by a whole variety of problems.

For completeness, it should be mentioned that there are other distributions of the general form \eqref{basicform} that are more suited to non-stochastic calculations. For example the symplectic tomography scheme of Mancini\etal\cite{Mancini-96,Mancini-97},
which expresses the quantum state as a probability distribution of a quadrature observable depending on a range of lab parameters. This
has been used to investigate quantum entanglement and failure of local realism, but apparently has not led to many-mode
quantum simulations, presumably due to the lack of a positive propagator in nonlinear evolution. The complex P representation\cite{DrummondGardiner80}
allows one to derive exact results for certain problems, but does not lead to stochastic equations, since the distribution
is neither real nor positive.

\REM{
\begin{table}
\caption[Commonly used representation kernels]{\label{TABLEKernels} Kernels in the more commonly used phase-space representations. Subsystems are labeled by $i$ and are taken to be spatial lattice points, unless stated otherwise.}
\newlength{\thirdcolwidth}\setlength{\thirdcolwidth}{10cm}
\newcommand{\boxit}[1]{\parbox{\thirdcolwidth}{#1}}
\newcommand{\dispit}[1]{\ensuremath{\displaystyle{#1}}}
\begin{center}\begin{minipage}{\textwidth}\begin{tabular}{|l|c|l|}
\hline
Distribution		& Kernel	& Variables\footnote{Complex.}\\
\hline\hline
\multicolumn{3}{|l|}{\scshape First Stage }\\\hline
			&&\\\cline{2-3}
\raisebox{1ex}{Wigner}	&\multicolumn{2}{|l|}{}\\\hline

			&
			& $\alpha_i$	\\\cline{2-3}
\raisebox{1ex}{Husimi Q}&\multicolumn{2}{|l|}{$\ket{\alpha_i}_i$ is a coherent state at $i$th lattice point.}\\\hline

			& \dispit{ \otimes_i \left(\ket{\alpha_i}_i\bra{\alpha_i}_i\right) }
			& $\alpha_i$	\\\cline{2-3}
\raisebox{1ex}{Glauber-Sudarshan P}
	&\multicolumn{2}{|l|}{$\ket{\alpha_i}_i$ is a coherent state at $i$th lattice point.}\\\hline

			& \dispit{ \otimes_i \left( \sum_{n=0}^{\infty} e^{-\lambda_i}\frac{\lambda_i^n}{n!}\ket{n}_i\bra{n}_i\right)}
			& $\lambda_i$	\\\cline{2-3}
\raisebox{1ex}{Poisson}
	&\multicolumn{2}{|l|}{$\ket{n_i}_i$ is a Fock number state at $i$th lattice point.}\\\hline
\hline

\multicolumn{3}{|l|}{\scshape Second Stage }\\\hline
			& \dispit{ \otimes_i \left(\frac{1}{\pi^2}\ket{\alpha_i}_i\bra{\beta_i^*}_i e^{-\{|\alpha_i|^2+|\beta_i|^2\}/2}\right) } 
			& $\alpha_i,\,\beta_i$	\\\cline{2-3}
\raisebox{1ex}{Glauber R}
	&\multicolumn{2}{|l|}{$\ket{\alpha_i}_i$ is a coherent state at $i$th lattice point.}\\\hline

			& \dispit{ \otimes_i \left(\frac{\ket{\alpha_i}_i\bra{\beta_i^*}_i}{\braket{\beta_i^*}{\alpha_i}_i}\right) } 
			& $\alpha_i,\,\beta_i$	\\\cline{2-3}
\raisebox{1ex}{positive P}
	&\multicolumn{2}{|l|}{$\ket{\alpha_i}_i$ is a coherent state at $i$th lattice point.}\\\hline

stochastic wavefunction	& \dispit{ \otimes_j \left\{\left(\sum_\bo{x}\phi_{\bo{x}}\ket{\bo{x}}_j\right)\otimes\left(\sum_{\bo{y}}\wt{\phi}_{\bo{y}}\bra{\bo{y}}_j\right)\right\} }
			& $\phi_{\bo{x}},\,\wt{\phi}_{\bo{y}}$ \\\cline{2-3}
(Fock state) 		&\multicolumn{2}{|l|}{$\ket{\bo{x}}_j$ is a position eigenstate of the $j$th particle at position $\bo{x}$.}\\\hline
\hline

\multicolumn{3}{|l|}{\scshape Third Stage }\\\hline
			& \dispit{ e^{z_0}\otimes_i \left(\frac{\ket{\alpha_i}_i\bra{\beta_i^*}_i}{\braket{\beta_i^*}{\alpha_i}_i}\right) } 
			& $\alpha_i,\,\beta_i,\,z_0$	\\\cline{2-3}
\raisebox{1ex}{gauge P}
	&\multicolumn{2}{|l|}{$\ket{\alpha_i}_i$ is a coherent state at $i$th lattice point.}\\\hline

stochastic wavefunction	& \dispit{ \Pi \otimes_i \ket{c\phi_i}_i 
					\otimes_j \langle c\wt{\phi}_j|_j }
			& $\phi_i,\,\wt{\phi}_j,\,\Pi$ \\\cline{2-3}
(Coherent state) 		&\multicolumn{2}{|l|}{\boxit{$\ket{c\phi_i}_i$ is a coherent state at the $i$th lattice point, $c$ is a real normalizing constant, dependent on mean particle number $\bar{N}$.}}\\\hline

\hline
\end{tabular}\end{minipage}\end{center}
\end{table}
}

\begin{table}
\caption[Properties of several representations for the interacting Bose gas]{\label{TABLEDistributions} \footnotesize
Check list of required representation properties for the more commonly used phase-space representations, when applied to a lattice interacting Bose gas Hamiltonianof the forms \eqref{deltaH} or \eqref{latticeH},  with no external coupling. 
\normalsize}\vspace*{3pt}
\hspace{-1cm}
\begin{minipage}{\textwidth}
\begin{tabular}{|l||c|c|c|c|c|c|c|}
\hline
distribution 	& positive& complete&      &positive& stable & UV 	& open		\\
type		& real	& non-singular	& FPE\footnote{i.e. only 1st and 2nd partial derivative terms in $\partial P/\partial t$.}
					 & propagator& unbiased& convergent & systems \\\hline
\hline
Wigner		& no	& yes	& no	& varies\footnote{There are no second order terms in the FPE for dynamics, while for thermodynamics the propagator may or may not be positive semi-definite, depending on occupation of modes.}
						& --	& no	& yes	\\\hline
Q		& yes	& yes	& yes	& no	& --	& no	& yes	\\\hline
P		& yes	& no	& yes	& no	& --	& yes	& yes	\\\hline
R		& no	& yes	& yes	& no	& --	& yes	& yes	\\\hline
positive P	& yes	& yes	& yes	& yes	& no	& yes	& yes	\\\hline
sw.\footnote{Stochastic wavefunction} Fock
		& yes	& yes	& yes	& yes	& yes\footnote{Given an appropriate choice of gauge --- see Chapter~\ref{CH7}}
							 	& yes	& no	\\\hline
sw.${}^b$ coherent
		& yes	& yes	& yes	& yes	& yes${}^c$& yes& no	\\\hline
gauge P		& yes	& yes	& yes	& yes	& yes${}^c$& yes& yes	\\\hline
\end{tabular}
\end{minipage}
\end{table}


\chapter{Boundary term errors and their removal}
\label{CH6}

  As has been mentioned previously, when one attempts to simulate non-separable physical models using stochastic equations derived via phase-space distributions, instabilities in the equations are common. These can  give rise to systematic biases in the observable estimates \eqref{observables} using a finite number of samples. Sometimes these biases abate at large sample number (Such a situation is considered in detail in Appendix~\ref{APPA}), while in other cases the biases remain even in the limit $\mc{S}\to\infty$. 

The latter have been called {\it boundary term errors} because the root of the biasing lies in discarding nonzero boundary terms in integrations involving the distribution $P(C)$ (and some factors) over $C$. 
Rigorous results regarding how and when these biases arise in practical simulations are few, and many researchers while aware of phase-space distribution methods, are either 1) unaware of the potential for biases, or 2) know that biases exist and because of this avoid using phase-space distribution methods altogether. With the aim of reducing this confusion, in Sections~\ref{CH6First} and~\ref{CH6Second}, the mechanisms by which these biases come about are investigated and classified for general phase-space distributions. To clarify the conditions under which boundary term errors may be a problem, 
some symptoms that can be checked for by inspection of the stochastic equations and observable estimators are pointed out.

It will be seen in Chapters~\ref{CH7} and ~\ref{CH9} that simulations of interacting Bose gases can contain boundary term errors (at least when using coherent state kernels), and a method to overcome this problem is developed in Section~\ref{CH6Removal}. 
Subsequently, this un-biasing procedure is demonstrated on the two rather different examples from the literature where 
boundary term errors  have been seen when using a positive P representation: Two-boson absorption (Section~\ref{CH6Absorber}), and a single-mode laser model (Section~\ref{CH6Laser}). Removal of possible or confirmed  biases in the interacting Bose gas is carried out in Sections~\ref{CH7Drift} and~\ref{CH9Gauge}, respectively.

\section{Boundary term errors of the first kind}
\label{CH6First}
\subsection{Origin and general form}
\label{CH6FirstOrigin}
  During the derivation of the Fokker-Planck equation (from the original master equation), there is a point between \eqref{masterinoperators} and \eqref{operatorfpe} where integration by parts is performed, and boundary terms discarded. 
The correspondence between full quantum evolution and FPE (and hence the final stochastic equations) occurs {\it only} if these discarded boundary terms are zero. Otherwise, there is a discrepancy between quantum mechanics and what is calculated by the stochastic equations in the limit of $\mc{S}\to\infty$. This has been termed a boundary term error. Let us see what exact form these discarded boundary terms take.

The partial integration is made on a master equation that already contains all kernel gauge expressions \eqref{zerof}, \eqref{weightinggaugemasterterm} and/or \eqref{zeromasterterm} as well as the basic nonzero terms arising from directly applying the operator correspondences \eqref{generaloperatoridentity} to the master equation. In any case, all these terms can be incorporated into the formalism of \eqref{masterinoperators}. With the real variable set $C=\{C_j\}$, when going from \eqref{masterinoperators} to \eqref{operatorfpe} the discarded boundary terms are found to be (after some algebra)
\begin{subequations}\label{btexpression}\EQN{
\op{\mc{B}}(C) = \sum_{lj}\int \bigg[ \op{\mc{B}}_j^{\rm int}(C) \bigg]_{\min[C_j]}^{\max[C_j]} \prod_{p\neq j} dC_p
\label{btexpressiona}}
where
\EQN{\lefteqn{
\op{\mc{B}}_j^{\rm int}(C) = 
P(C)T_{lj}^{(1)}(C)\op{\Lambda}(C) 
\displaystyle+\Half P(C)\sum_k T_{ljk}^{(2)}\dada{\op{\Lambda}(C)}{C_k}
\displaystyle-\Half\op{\Lambda}(C)\sum_k\dada{}{C_k}\left(P(C)T_{lkj}^{(2)}(C)\right).}&&\nonumber\\
&&\hspace*{14.32cm}\mbox{}\label{btexpressionb}}\end{subequations}
 The usual definite  evaluation notation \mbox{$\left[f(v)\right]_a^b = \lim_{v\to b}f(v)-\lim_{v\to a}f(v)$} has been used. 
Note that the kernel is still an operator, and hence so is the boundary term expression $\op{\mc{B}}$. Every matrix element of the boundary term operator should be zero for the correspondence between quantum mechanics and the FPE to be exact. Such matrix elements could in principle be evaluated by inserting the elements of $\op{\Lambda}$ and $\partial\op{\Lambda}/\partial C_k$, which can usually be worked out exactly, into \eqref{btexpression}. This will be attempted for the gauge P representation in Section~\ref{CH6FirstGaugeP}, and a toy example of the calculation of $\op{\mc{B}}$ is given in Section~\ref{CH6SecondBargmann}.

While \eqref{btexpression} is nontrivial, there are generally a variety of ways in which it could turn out to be zero.
\ENUM{
\item Especially for an open phase space, one expects that $P(C)$ will tend towards zero at the boundary, and if this rate is faster than the growth of the multiplying factors ($T^{(\bullet)}(C)$ times the matrix elements of $\op{\Lambda}$), then the terms in \eqref{btexpression} will be zero.   
\item Even if individually the terms in \eqref{btexpression} do not tend to zero, but the distribution $P(C)$ times multiplying factors is an even function of $C_j$, then the values evaluated at $\max[C_j]$ and $\min[C_j]$ may cancel each other. A typical example of this is when a particular $C_j=\theta$ is the phase of a complex variable ---  any function of $\theta$ evaluated at the limits of its domain $[-\pi,\pi]$ is the same at both limits, and any boundary terms in $\theta$ sum to zero.
\item Also, there may be a cancellation between separate terms in the expression \eqref{btexpression}.
}
Conversely, the boundary terms can easily turn out to be nonzero in an open phase-space if $P(C)$ drops off too slowly as $|C_k|$ grows (typically --- if the tails of $P(C)$ obey a power law), or, in a partly or fully closed phase-space, when $P(C)\neq0$ at the boundary.

\subsection{Symptoms indicating boundary term errors}
\label{CH6FirstSymptoms}
Unfortunately, it is not generally possible to directly evaluate the boundary terms \eqref{btexpression} and check correctness, because this involves knowing at least the far tail behavior (in an open phase space) of $P(C)$ as a function of time. In a fully or partly closed phase-space one needs to know the exact behavior near the boundary.  These require detailed knowledge about aspects of the full solution to the master equation --- generally not achievable in non-trivial systems, although power law tails of $P(C)$ can be considered a bad sign.

What is really needed are some symptoms of likely boundary term errors that could be looked for in  the stochastic equations before simulation. Since boundary term errors arise when the distribution falls off too slowly as the (possibly open) boundaries of phase space are approached, at least four (there may be more) broad classes of features come to mind that indicate the probable presence of boundary term errors of the first kind.

\begin{enumerate}
\item
The first such symptom, fairly well known,  is the presence of so-called
{\scshape moving singularities.} In a set of deterministic equations, these manifest themselves as solutions (usually a set of measure zero) that diverge in a finite time. Previous studies \cite{Gilchrist-97,DeuarDrummond02} have shown that when moving singularities are present in the deterministic parts of stochastic equations, boundary term errors usually result. 
In particular, it was seen that
\ITEM{
\item In a number of studied systems, moving singularities in equations derived using the positive P representation occur whenever systematic errors are seen in observables calculated with estimators polynomial in the system variables. 
\item In these systems, when the moving singularities are removed with the aid of stochastic gauges (as explained in Section~\ref{CH6Removal}), the systematic errors are removed as well\cite{DeuarDrummond02}.
}

 The presence of pathological behaviour when moving singularities are present can be grasped intuitively: Due to the stochasticity of the equations, and the continuous nature of the noise, one expects that after small times, the distribution of trajectories (at least the far tails of it) will extend over all of phase space. In particular into those phase-space regions where the solutions that diverge in a finite time occur.
While the divergent trajectories are usually a set of measure zero, 
an infinitesimal proportion of divergent trajectories may have a finite effect on the observable means, on the general basis that $0\times\infty$ does not necessarily equal zero. However, because the number of such trajectories is infinitesimal in comparison with the rest, this effect can never be estimated by a finite sample of them, hence the systematic errors.  

Moving singularities may arise either in a dynamic fashion (e.g. in an equation $dv=v^2\,dt$, $v(t) = v(0)/[1-tv(0)]$, and a divergence occurs at $t=1/v(0)$), or due to divergent drift terms for finite phase-space variables (e.g. $dv=dt/(1-v)$ at $v=1$, or $dv=dt/(1-t)$ at $t=1$). In the latter situation the divergent trajectories may even be a set of nonzero measure.

\item
Secondly, one should also be wary of {\scshape noise divergences} ---  i.e. instabilities in the noise matrices $B$. For some small set of trajectories the Wiener increments $dW(t)$ will conspire to act approximately as a constant $(\Delta W) \approx \order{\sqrt{\Delta t}}$ over a time interval $\Delta t\gg dt$. This results in approximately an effective drift term proportional to $B$, and if such ``drift terms'' give rise to what effectively are moving singularities, then systematic errors  will also result by an analogous mechanism as for instabilities of the drift terms.

\item 
Thirdly, Another indicator of possible boundary term errors is a {\scshape discontinuous drift and/or noise matrix $B$.} This may cause an effect that can not be simulated stochastically, as the discontinuity is only ever sampled by a set of trajectories of measure zero. In particular, there may be a region near such a boundary that contains only an infinitesimal portion of total trajectories, but those trajectories have large impact on the observable estimates.

\item
Fourthly, if the {\scshape initial distribution is too broad,} boundary term errors may be present from the outset. An example would be a Lorentzian initial distribution of variable $v(0)$ if the Fokker-Planck equation contains terms quadratic in $v$. Another example is the single-mode laser model of Section~\ref{CH6Laser} with non-delta-function initial conditions.
\end{enumerate}

If any of the above symptoms is present, boundary term errors may be suspected. Conversely, if the symptoms are absent, it {\it appears likely} that no boundary term errors occur, but unfortunately rigorous results for general distributions are hard to come by. No boundary term errors were seen for any simulation that did not suffer from one or more of these symptoms.

There have also been developed  several numerical indicators (described in detail in \cite{Gilchrist-97} ) that one can use to check for the presence of boundary term errors from a finite set of trajectories.

\subsection{Moving singularities in a complex phase-space}
\label{CH6FirstPractice}

For a set of stochastic equations (e.g. such as \eqref{langevin} or \eqref{langevinstd}) it is relatively easy to rule out 
boundary term symptoms 3 and 4 from the previous section (by inspection). Moving singularities and noise divergences (symptoms 1 and 2) are a little more slippery, but practical conditions for their absence can be stated. Simulations in this thesis are with complex variable sets on an open phase space, so here a practical condition for absence of moving singularities will be derived only for this particular (though commonly occurring) kind of phase-space.

In an open  complex phase space, 
barring symptoms 3 and 4,
moving singularities cannot occur if the both the drift and noise matrices of the stochastic equations contain no radial components that lead to super-exponential growth.  This is because then all variables remain finite at finite times, and so at finite times never reach the ``boundary'' at modulus infinity.  Since $z=ze^{2i\pi}$ for any complex $z$, tangential evolution does not lead to divergences, because boundary terms evaluated at the boundaries of the polar variable ($\angle z = 0$ or $2\pi$) are equal, and thus cancel in \eqref{btexpressiona}.

More precisely, for a set of stochastic equations 
\begin{equation}
dz_j = A_j(C)\,dt + \sum_k B_{jk}(C) dW_k(t), 
\end{equation}
in complex variables $z_j$\ $(C=\{z_j\}$, including, possibly $z_0$), this gives the \textit{Condition:}

\begin{itemize}\item
Moving singularities or noise divergences will not occur provided that
the limits
\begin{equation}\label{mvsingcondition}
\lim_{|z_j|\to\infty} \frac{A_j}{|z_j|} \qquad\ \text{and}\ \qquad
\lim_{|z_j|\to\infty} \frac{B_{jk}}{|z_j|} 
\end{equation}
converge for all $z_j\in C$ and all $k$, and that symptoms 3 and 4 of Section~\ref{CH6FirstSymptoms} are ruled out (by inspection).
\end{itemize}

\subsection{Generic presence in many-mode simulations}
\label{CH6FirstManymode}

Off-diagonal kernels appear necessary to allow all quantum states to be described by a positive real distribution that can be sampled stochastically (see Section~\ref{CH3RepresentationDual}). In such a case, the independent ``ket'' and ``bra'' parameters of the kernel will (if complex) almost always enter the kernel in an analytic way.  Then, to ensure a positive propagator, requires in turn that the complex derivatives  be chosen as in Section~\ref{CH3EquationsAnalytic}, preserving the analytic nature of the formulation into the stochastic equations.

The presence of moving singularities in (deterministic) dynamical equations for an analytically continued phase-space is a well-studied problem for both ordinary and partial differential nonlinear evolution equations. 
According
to the Painleve conjecture (see, e.g. \cite{Ablowitz-78, Bountis-82,Weiss-83}), this is equivalent to non-integrability - and is therefore generic to many (in fact, all but a set of measure zero) complex dynamical equations
with large numbers of degrees of freedom. Just the sort one obtains when simulating many-body systems.

Kernel gauges are, however,  not restricted by the above mechanisms to be analytic. This is because they arise as arbitrary (hence, possibly non-analytic) multipliers to operators on the kernel, rather than as a result of applying analytic operator identities \eqref{generaloperatoridentity} to it.  Because of this freedom of gauge choice, it has been conjectured by P. D. Drummond\cite{DrummondDeuar03} that 
breaking the analyticity of the complex equations using stochastic gauges is a necessary (though not sufficient just by itself) condition to remove boundary term errors. 
This agrees with previous work\cite{DeuarDrummond02,Carusotto-01}, and also with the gauges used to remove boundary term errors in this thesis. 

Diffusion gauges, while possibly non-analytic, do not appear useful for removal of boundary term errors of the first kind because they appear only at the FPE-Langevin stage, after any nonzero boundary term operators have been already discarded.

\subsection{Boundary terms in the gauge P representation}
\label{CH6FirstGaugeP}

Let us consider the gauge P representation. The kernel \eqref{gaugekernel}, expanded in the complete Fock number state basis for each subsystem is $\op{\Lambda} = \Omega\otimes_k\op{\Lambda}_k$, with 
\EQN{\label{localkernelk}
\op{\Lambda}_k = \sum_{n\wt{n}} \frac{\alpha_k^n\beta_k^{\wt{n}}}{\sqrt{n!\wt{n}!}}\ket{n}_k\bra{\wt{n}}_ke^{-\alpha_k\beta_k}
,}
as in \eqref{localkerneln}. A master equation leads via \eqref{correspondencesg}, without invoking gauges, to terms $T^{(1)}_{lj}$ and $T^{(2)}_{ljk}$ polynomial (of low order) in $\alpha_k$ and $\beta_k$. Use of a weighting gauge will lead to some terms $T^{(\bullet)}\propto\Omega$. Drift gauges could generally lead to any form of the corresponding coefficients $T^{(\bullet)}$, but let us {\it assume} that the drift gauges $\mc{G}_k$, etc.  are polynomial in $\alpha_k,\,\beta_k,\,\alpha_k^*,\,\beta^*_k$, and autonomous (i.e. not dependent on $\Omega$). Under this assumption, the terms $T^{(\bullet)}$ due to drift gauges will be polynomial in all the variables, and at most $\propto \Omega^2$ (Diffusion coefficient in $\Omega$ due to standard drift gauge introduction is $\propto\Omega^2\mc{G}_k^2$).  

Combining this together, the (matrix elements of the) multipliers of $P(C)$  appearing in the boundary term expression \eqref{btexpression} will at most scale as :
\EQN{\label{proptobt}
\le\ \propto \Omega^3 \prod_k \alpha_k^{a_{k}} \beta_k^{b_{k}}e^{-\alpha_k\beta_k}
,}
where the coefficients take on all values $a_{k}\ge0$ and $b_{k}\ge0$ for various matrix elements. 

In an open phase space of complex variables $z_j$ (as here), the boundary lies at $|z_j|\to\infty$. The multipliers of P(C) in \eqref{btexpression} scale in these far tails as
\EQN{
\le\ \propto\quad |\Omega^3| \prod_k  |\alpha_k|^{a_k} |\beta_k|^{b_k} e^{c_k|\alpha_k||\beta_k|}
,}
where $c_k=-e^{i\angle\alpha_k}e^{i\angle\beta_k}$, can have positive real part. This implies that if one wants to be sure that no boundary term errors are present, (assuming any drift gauges $\mc{G}_k$ are polynomial in $\alpha_k$, $\beta_k$, or their conjugates, and independent of $\Omega$), then: 
$P(C)$ should decay in the far tails no slower than 
\SEQN{\label{gaugebtcond}}{
 P(C) &<\ \propto& 1/|\Omega^{4+\wt{b}_k}|,\\
 P(C) &<\ \propto& \exp\left(-\wt{c}_k |\alpha_k|^{\wt{a}_k}\right),\\
 P(C) &<\ \propto& \exp\left(-\wt{c}_k |\beta_k|^{\wt{a}_k}\right),
}
where $\wt{b}_k>0$, $\wt{c}_k>0$ and $\wt{a}_k>1$. (And there should be no time-dependent discontinuities in the drift or noise equations).

This does not preclude that by some canceling of terms, a less favorable scaling of the tails may also become acceptable, but certainly if $P$ scales like \eqref{gaugebtcond} or better in the far tails, boundary term errors (of the first kind) will be absent in a gauge P representation simulation. Note in particular that from \eqref{gaugebtcond}:
\ENUM{
\item Power law tails of $P(C)$ in the $\bm{\alpha},\bm{\beta}$ part of phase-space will invariably lead to boundary term errors. For some matrix elements of $\op{\mc{B}}$ for which the coefficient of $\ket{n}_k\bra{\wt{n}}_k$ in \eqref{localkernelk} is of high enough order, the boundary terms will become nonzero. This is consistent with the conclusions drawn by Gilchrist\etal\cite{Gilchrist-97} that power law tails in a positive P distribution are an indicator of boundary term errors.
\item Exponential decay of $P(C)$ with $|\alpha_k|$ or $|\beta_k|$ is also not good enough in general. This is because the coefficient of (say) $|\alpha_k|$ in the exponents of \eqref{gaugebtcond} is not constant, but $\propto|\beta_k|$ --- which may be any arbitrarily large value. Only a faster-than-exponential decay will be able to overcome this when $|\beta_k|$ is large. However, Gilchrist\etal\cite{Gilchrist-97} found no boundary term errors in a single-mode anharmonic oscillator \eqref{bargmannh} simulation for which the far tail behaviour of $P_+$ appeared to be exponential, and no indications of boundary term errors in this system are seen here in Chapter~\ref{CH7} either. This may be because the single-mode system is a special case --- certainly in Chapter~\ref{CH8} it is seen that when acting on several coupled modes, a similar Hamiltonian {\it can} lead to boundary term errors.
}

\section{Boundary term errors of the second kind}
\label{CH6Second}

Biases that do not abate for $\mc{S}\to\infty$ may also arise via a different mechanism when the estimator functions to be averaged to obtain the expectation value of a given observable grow too rapidly in phase space.

\subsection{Mechanism and overview}
\label{CH6SecondMechanism}
  Working at the level of the distribution $P(C)$, expectation values of observables are given by the real part of \eqref{obsintegral}. When one takes a number $\mc{S}$ of samples of this distribution, the estimator \eqref{observables} is used, but if the real parts of the integrals in \eqref{obsintegral} do not converge then this finite sample estimate does not approach the quantum mechanical value even in the limit $\mc{S}\to\infty$. Systematic errors arising from this source will be called here boundary term errors of the second kind, and have very different properties to those of the first kind.   
Explicitly, from \eqref{obsintegral} and by imposing Hermiticity, the integrals that must converge are 
\SEQN{\label{secondint}}{
\int P(C) &\re{ \tr{ \op{O}\Big(\op{\Lambda}(C)+\dagop{\Lambda}(C^*)\Big) } }& dC\label{numeratorint}\\
\int P(C) &\re{ \tr{ \op{\Lambda}(C) }}& dC \label{denominatorint}
,}
for any particular observable $\op{O}$. 

One can see that even when \eqref{denominatorint} is convergent, errors may occur for some observables for which only \eqref{numeratorint} diverges. Given a particular distribution, divergences may come about due to the matrix elements of either the kernel $\op{\Lambda}$, the operator $\op{O}$, or both growing too rapidly as the boundaries of phase space are approached. 
Observables that involve an infinite sum of moments polynomial in the system variables (such as the parity in the gauge P representation, whose estimate  is given by the sum \eqref{parityestimate}) may be particularly prone to these boundary term errors, because the dependence of $\tr{\op{O}\op{\Lambda}}$ on the phase-space variables will grow particularly rapidly.

Boundary term errors of the second kind depend on the equations of motion only by their effect on the distribution of samples $P(C)$, and so do not show up as moving singularities or noise divergences in the equations. 

A sobering point is that, in theory, one could always come up with some observables that grow fast enough as the boundary of phase space is approached to overcome whatever far tail behaviour occurs in $\op{\Lambda}(C)P(C)$. Such an observable could be defined as an infinite sum of polynomial moments whose terms grow appropriately rapidly. Thus, boundary term errors of the second kind are likely to occur for {\it some} observable estimates for any phase-space distribution simulation method, for any model. In practice, fortunately, it suffices to make the estimators of only all observables {\it of interest} convergent. 

\REM{
The converse is not always true: One cannot always think up some observable that would decay fast enough as the boundaries of phase space are approached so that whatever divergent behaviour occurs in $\op{\Lambda}(C)P(C)$ is overcome. This is because \eqref{denominatorint} does not depend on what observable we choose, only on the kernel. 
For norm-preserving master equations, however, such as e.g. purely Hamiltonian evolution of a previously normalized density matrix, the denominator term in \eqref{observables} need not be calculated at all.  One {\it can} then always find some observable that decays fast enough as the boundaries of phase space are approached so that the normalized expectation estimate \eqref{numeratorint} converges while \eqref{denominatorint} is ignored. 
}

\subsection{Example: un-normalized positive P representation}
\label{CH6SecondBargmann}
A good example of boundary term errors of the second kind occurs for the ``anharmonic'' oscillator single-mode Hamiltonian, when using an un-normalized coherent state kernel. This Hamiltonian\footnote{In some appropriate time units.} is 
\EQN{\label{bargmannh}
  \op{H} = \hbar\, \dagop{a}{}^2\op{a}^2
,}
the single-mode analogue of a repulsive two-body potential in the interaction picture\footnote{That is, with terms linear in $\op{a}$ and $\dagop{a}$ transformed away. Compare with \eqref{deltaH}.}. Let us consider the kernel 
\EQN{\label{bargmannkernel}
\op{\Lambda} = ||\alpha\rangle\langle\beta^*||
,}
off-diagonal in the un-normalized Bargmann coherent states \eqref{bargmannket}. It is convenient to change to logarithmic variables $n_L=\log(\alpha\beta)$, and $m_L=\log(\alpha/\beta)$, then
the basic operator correspondences are, from \eqref{aopidentities},
\SEQN{}{
\op{a}\op{\Lambda} =& \alpha\op{\Lambda} =& e^{(n_L+m_L)/2}\op{\Lambda}\\
\dagop{a}\op{\Lambda} =& \dada{}{\alpha}\op{\Lambda} =& e^{-(n_L+m_L)/2}\left(\dada{}{n_L}+\dada{}{m_L}\right)\op{\Lambda}\\
\op{\Lambda}\dagop{a} =& \beta\op{\Lambda} =& e^{(n_L-m_L)/2}\op{\Lambda}\\
\op{\Lambda}\op{a} =& \dada{}{\beta}\op{\Lambda} =& e^{(m_L-n_L)/2}\left(\dada{}{n_L}-\dada{}{m_L}\right)\op{\Lambda}
.}
It is also convenient to split the variables into real and imaginary parts: $n_L=n'_L+in''_L$, and $m_L=m'_L+im''_L$.
For a  dynamical evolution model with no coupling to the environment, utilizing the analytic nature of the kernel as explained in Section~\ref{CH3EquationsAnalytic}, one obtains the FPE (provided no boundary term errors of the first kind are present) 
\EQN{\label{fpebargmann}
\dada{P}{t} &=& \left\{ 2\dada{}{m''_L} + \frac{\partial^2}{\partial n'_L{}^2} + \frac{\partial^2}{\partial n''_L{}^2}
+ \frac{\partial^2}{\partial m'_L{}^2} + \frac{\partial^2}{\partial m''_L{}^2} \right. \nonumber\\
&&- \left. \frac{\partial^2}{\partial n'_L\partial m''_L} -\frac{\partial^2}{\partial m''_L\partial n'_L}
- \frac{\partial^2}{\partial n''_L\partial m'_L} -\frac{\partial^2}{\partial m'_L\partial n''_L}
\right\} P
.}
This gives the Ito stochastic equations (with standard square root form of $B_0=\sqrt{D}$)
\SEQN{}{
dn_L =& i\sqrt{2i}\,[dW(t)-id\wt{W}(t)]  	&= 2d\eta(t)\\
dm_L =& i\sqrt{2i}\,[dW(t)+id\wt{W}(t)] -2i\,dt	&= 2i[d\eta^*(t) - dt]
.}
$dW(t)$ and $d\wt{W}(t)$ are the usual independent Wiener increments, and the complex noise $d\eta(t)$ defined here obeys
\SEQN{}{
\average{d\eta(t)} &=& 0\\
\average{d\eta^*(t)d\eta(t')} &=& \delta(t-t')\,dt^2\\
\average{d\eta(t)d\eta(t')} &=& 0
.}
The FPE (or stochastic equations) can be solved, and in the case of coherent state initial conditions (with initial amplitude $\alpha_0$)
\EQN{\label{bargmanninitial}
 P(n_L,m_L,0) = \delta^2(m_L-2i\angle\alpha_0)\ \delta^2(n_L-2\log|\alpha_0|)
,}
the distribution at time $t$ is
\EQN{\label{bargmannsolution}
  P(n_L,m_L,t) &=& \delta^2\Big(m_L-\Big[i(n_L^*-2\log|\alpha_0|)-2it+2i\angle\alpha_0\Big]\,\Big)\nonumber\\
&&\qquad\times\ \frac{1}{2\pi t} \exp\left(-\frac{|n_L-2\log|\alpha_0|\,|^2}{4t}\right)
.}

In this toy model, one can actually check if there really were no boundary terms of the first kind, by  substituting \eqref{bargmannsolution}, and \eqref{bargmannkernel} into the expression for $\op{\mc{B}}$. 
The  boundary term operator $\op{\mc{B}}$ is given by the integrals \eqref{btexpression} of operators $\op{\mc{B}}_j^{\rm int}$ composed of  $P$,  $\op{\Lambda}$ and coefficients $T^{(\bullet)}$, with individual terms of $\op{\mc{B}}$ evaluated at the boundaries of phase space. In our case these boundaries are $C_j\to-\infty$ and $C_j\to\infty$ for all $j=1,\dots,4$, and one requires all the matrix elements of $\op{\mc{B}}$ to be zero.
By comparison of  \eqref{fpebargmann} to \eqref{almostfpe}, and labeling variables as $n'_L=C_1$, $n''_L=C_2$, $m'_L=C_3$,  and $m''_L=C_4$, and also setting the term counter $l=1$, 
the nonzero $T^{(\bullet)}$ coefficients in the $\op{B}_j^{\rm int}$ are 
$T^{(1)}_{14} = -2$, $T^{(2)}_{1jj}=2$ for all $j$, and $T^{(2)}_{114}=T^{(2)}_{141}=T^{(2)}_{123}=T^{(2)}_{132}=-2$.
 Also, the matrix elements of $\op{\Lambda}$ (in a Fock basis $\ket{n}$) can be written (from the definition \eqref{bargmannket} of $||\alpha\rangle$) as
\EQN{\label{bargmannmelements}
\bra{\wt{n}}\op{\Lambda}\ket{n} = \frac{1}{\sqrt{n!\wt{n}!}}\exp\Big[\ n_L(n+\wt{n})/2+ m_L(n-\wt{n})/2\ \Big]
.}
So:
\ENUM{
\item The $T^{(\bullet)}$ coefficients are constants.
\item The matrix elements of $\op{\Lambda}$ and  $\partial\op{\Lambda}/\partial z$  are exponential in both variables $z=n_L$ or $m_L$, with some constant positive or negative coefficient.
\item The distribution $P$ is Gaussian in real and imaginary parts of both variables ($n_L$ around mean $2\log|\alpha_0|$), $m_L$ around mean $2i(\angle\alpha_0-t)$).
}
One can see that all matrix elements of all the integrands ($\bra{\wt{n}}\mc{B}_j^{\rm int}\ket{n}$)  will be dominated by the Gaussian decay of $P(C)$ in the far tails for all $n$ and $\wt{n}$. For $j=1, 2$,
after integration over $m_L$ and a (real or imaginary) component of $n_L$, the expression to be evaluated at large $n_L$ will be Gaussian, hence zero in the far tail limit. Also, for the terms ($j=3,4$) in which one integrates over $n_L$ and a component of $m_L$, the expression to be evaluated will be a Dirac delta function --- also zero in the far tail limit.  Hence, 
\EQN{
\op{\mc{B}}=0
.}
Thus it has been confirmed that there are no boundary term errors of the first kind. 

So far it seems like this might be a very good kernel to try even for many-body locally interacting Bose gases, which have the same kind of interparticle potential term. The advantages would be that  there are no boundary term errors of the first kind, no moving singularities (compare to Section~\ref{CH6FirstManymode}) and even no nonlinear terms in the stochastic equations (compare to \eqref{itoH}). 

\REM{
However, when one actually tries to calculate an observable, the estimates can be wrong. Consider e.g. the expectation value of the mean occupation of the mode: the expectation value  $\langle\dagop{a}\op{a}\rangle$, which is conserved in this system since $\op{H}=\hbar\dagop{a}\op{a}(\dagop{a}\op{a}-1)$. The estimator for this observable is (via \eqref{observables}) 
\EQN{
\average{\re{\alpha\beta^{\alpha\beta}}}/\average{\re{e^{\alpha\beta}}}
.}
Results of a calculation are shown in Figure~\ref{FCH6BARGMANNN}.
}

Having the luxury of knowing $P(C)$ for this toy problem, let us see what happens when one tries to calculate an observable. Directly from \eqref{bargmannkernel} and \eqref{bargmannproduct}:
\EQN{\label{bmantrl}
\tr{\op{\Lambda}}= \exp(\alpha\beta) = \exp\left[e^{n_L}\right]
.}
The essential denominator term \eqref{denominatorint} in any observable calculation, 
is then (using \eqref{bargmannsolution} and defining $n_L=n'_L+in''_L$),
\EQN{
&& \int \re{\exp\left[e^{n_L}\right]}\frac{1}{4\pi t} e^{-|n_L-2\log|\alpha_0||^2/4t} d^2n_L \nonumber\\
&\propto& \int \exp\left[e^{n'_L}\cos n''_L\right] \cos\left(e^{n'_L}\sin n''_L\right) e^{-(n'_L-2\log|\alpha_0|)^2/4t} e^{-(n''_L)^2/4t}\, d^2n_L\qquad\qquad
.}
For $n_L$ with large positive real part, the factor from $\tr{\op{\Lambda}}$ dominates (it's an exponential of an exponential!), and since the integrand is symmetric in $n''_L$ and nonzero, it must become unbounded, and the integral diverge.  

\begin{figure}[t]
\center{\includegraphics[width=11cm]{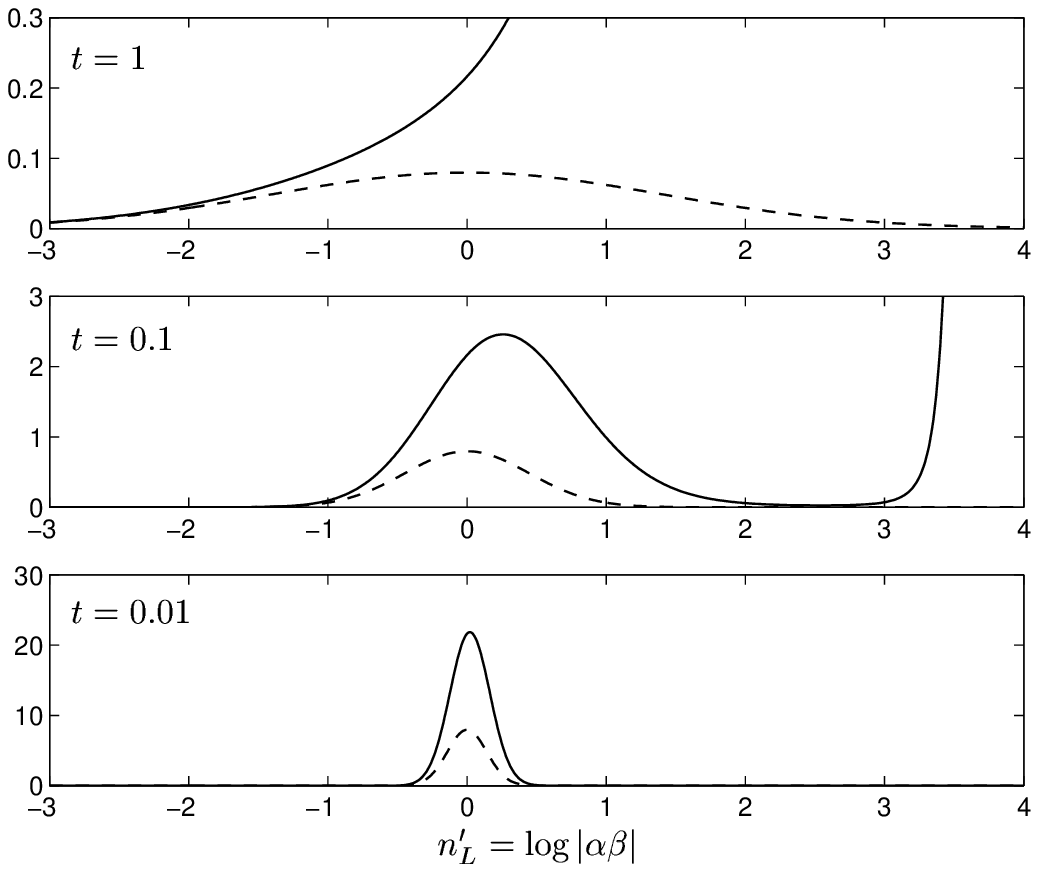}}\vspace{-8pt}\par
\caption[Unbounded integrands in an un-normalized P representation]{\label{FIGUREBargmann}\footnotesize
 \textbf{Integrands appearing in the denominator} of observable estimates for anharmonic oscillator dynamics governed by the Hamiltonian \eqref{bargmannh}. Values plotted are ({\scshape Solid line}): $\int \tr{\op{\Lambda}(C)}P(C)\,d^2m_L$, evaluated (using \eqref{bmantrl} and \eqref{bargmannsolution}) at $n''_L=0$. Coherent state initial conditions \eqref{bargmanninitial} were used with mean occupation number unity ($|\alpha_0|=1$). 
To show what parts of the integral \eqref{denominatorint} would be sampled by a stochastic simulation, the probability distribution of $n_L$ evaluated at $n''_L=0$ (i.e. $\int P(C) d^2m_L$ evaluated at $n''_L=0$) is also plotted as the {\scshape Dashed line.} Note the unbounded behaviour of the integrand at large positive $n'_L$ values (which are not sampled). 
\normalsize}
\end{figure}

A plot of the integrand 
$P\tr{\op{\Lambda}}$ for several time values is shown in Figure~\ref{FIGUREBargmann}. The unbounded behaviour at high $n'_L$ values indicates that \eqref{denominatorint} will diverge and boundary term errors (of the second kind) will be present.  A comparison to the probability distribution of $n'_L$ is also made there to indicate which contributions to the integral \eqref{denominatorint} are sampled by a stochastic simulation. 

This means that: {\it All observable estimates using \eqref{observables} are suspect with this representation!}

\subsection{In the gauge P representation}
\label{CH6SecondGaugeP}

For the gauge P representation, $\tr{\op{\Lambda}}=e^{z_0}$, and so the denominator integral \eqref{denominatorint} becomes simply
\EQN{
\int P(C)\, e^{z'_0}\cos z''_0\, dC 
}
in terms of $z_0=z'_0+iz''_0$.
This will be convergent provided the far tails of the marginal probability distribution of $z'_0$ decay as 
$e^{-z'_0}$, or faster in the limit $|z'_0|\to\infty$.

From the  estimator expression \eqref{qest} for an arbitrary observable that is a polynomial function of  $\op{a}$ and $\dagop{a}$, one can see that the numerator integral \eqref{numeratorint} will be of the form 
\EQN{
\int P(C) \sum_j f_j(C)\, dC
,} 
where the functions for each term $f_j$ are 
\EQN{
  f_j(C) = \Half\, e^{z'_0}\left(\prod_k |\alpha_k|^{a_k}|\beta_k|^{b_k}\right) \cos\left( z''_0\pm\theta+\sum_k [a_k\angle\alpha_k+b_k\angle\beta_k]\right)
,}
with some non-negative integer coefficients $a_k$ and $b_k$, which take on different values for different $f_j$. 
The behaviour of each such term as one approaches the boundaries of phase space, where the modulus of at least one variable tends to $+\infty$, is exponential in $z'_0$ (with unity coefficient), and polynomial in $|\alpha_k|$ and $|\beta_k|$. 

This means that 
\ITEM{
\item Boundary term errors  can be expected for observables polynomial in $\op{a}_k$ and/or $\dagop{a}_k$ if the tails of $P(C)$ decay as any power law in the $|\alpha_k|$, or $|\beta_k|$, or as $e^{-z'_0}$ or slower in $z'_0$. 
}
Conversely, if the tails of $P(C)$ decay faster than power law in all $|\alpha_k|$ and $|\beta_k|$ and faster than $e^{-z'_0}$ in the modulus of the weight, then boundary term errors {\it(at least of the second kind)} will not occur for any observable  polynomial in the annihilation and creation operators on subsystems. That is, any observable given by \eqref{qdef}, or a finite linear combination of such expressions.

Power law tails of $P(C)$ in $|\alpha_k|$ or $|\beta_k|$ do not immediately exclude all observables, only those whose estimators grow too rapidly. By inspection of \eqref{qdef}, \eqref{qest}, one can see that an observable that is an $a_k$th polynomial  of $\op{a}_k$ will lead to a term $f_j\propto e^{z'_0}|\alpha_k|^{a_k}$. 
If $P(C)\propto e^{-cz'_0}|\alpha_k|^{-\wt{a}_k}$ in the far tails, then the indefinite integral $\int P(C) f_j(C) dC$ will be $\propto e^{(1-c)z'_0}|\alpha_k|^{a_k-\wt{a}_k+1}$ (if $a_k-\wt{a}_k+1\neq0$), or $\propto e^{(1-c)z'_0}\log|\alpha_k|$ (otherwise) in this far tail region.  When this is evaluated in the far tail limit, the integral will convergent if the decay of $P(C)$ is fast enough. Similarly for the relationship between $\dagop{a}_k$ and $\beta_k$. Hence, one can state the conditions for expectation value estimates of a particular operator $\op{A}$ to not suffer from boundary term errors of the second kind: When $\epsilon_j$ are infinitesimal positive real constants, one requires 
\begin{subequations}\label{gaugeppowercond}\EQN{
\op{A} \propto \op{a}_k^{a_k} \implies \lim_{|\alpha_k|\to\infty} P(C)\  &<\ \propto&\ \frac{1}{|\alpha_k|^{-(1+a_k+\epsilon_1)}}\\
\op{A} \propto \dagop{a}_k{}^{b_k} \implies \lim_{|\beta_k|\to\infty} P(C)\  &<\ \propto&\ \frac{1}{|\beta_k|^{-(1+b_k+\epsilon_2)}}
}
as well as 
\EQN{
 \lim_{z'_0\to\infty} P(C)\  &<\ \propto&\ e^{-(1+\epsilon_3)z'_0}
.}\end{subequations}

Non-polynomial observables defined as infinite sums (e.g. the parity \eqref{paritydef}) can be more troublesome, because the functions $\tr{\op{\Lambda}\op{O}}$ etc. that are to be sampled by \eqref{numeratorint} can grow faster than polynomially as $|\alpha|_k, |\beta_k| \to\infty$. Conditions of the kind \eqref{gaugeppowercond} can also be derived, but will not scale as a power law. For example since the parity estimator \eqref{parityestimate} has a numerator expression that scales $ \propto e^{-2\re{\alpha_k\beta_k}}$, one requires $P(C)$ to decay faster than $e^{2\re{\alpha_k\beta_k}}$  in the far tails for an unbiased estimate of parity expectation values (remember that $\alpha_k\beta_k$ may have negative real part in highly non-classical parts of phase-space).

\subsection{Locally-interacting Bose gas calculations}
\label{CH6SecondGas}
Here the likelihood of boundary term errors of the second kind is considered for a many-mode interacting Bose gas model using the gauge P representation as described in Sections~\ref{CH5Equations} and~\ref{CH5Thermo}.
By inspection of the dynamics and thermodynamics gauge P equations for the locally interacting Bose gas \eqref{itoH}--\eqref{itoepsilon}, and \eqref{gaugepthermo}, one sees that the stochastic terms (assuming no gauges)
take the forms 
\SEQN{\label{dalphaprop}}{
d\alpha_{\bo{n}}&\dots +& c\,\alpha_{\bo{n}}\,dW_{\bo{n}}\\
d\beta_{\bo{n}}&\dots +& \wt{c}\,\beta_{\bo{n}}\,d\wt{W}_{\bo{n}}
,}
with some complex constants $c$ and $\wt{c}$, apart from possibly additional terms caused by finite temperature heat bath interactions 
\eqref{itoheatbathT}
\SEQN{\label{dalphapropbath}}{
d\alpha_{\bo{n}}&\dots +& \sqrt{\gamma_{\bo{n}}\bar{n}_{\rm bath}(T)}d\eta_{\bo{n}}\\
d\beta_{\bo{n}}&\dots +& \sqrt{\gamma_{\bo{n}}\bar{n}_{\rm bath}(T)}d\eta^*_{\bo{n}}
.}
In this Subsection, the far-tail regime of phase space where $|\alpha_{\bo{n}}|$ and/or $|\beta_{\bo{n}}|$ are large will be of interest, and there the terms \eqref{dalphapropbath} are negligible, and hence are ignored here. 

Upon change of variables to $z_{\bo{n}}=\log\alpha_{\bo{n}}$ and $\wt{z}_{\bo{n}}=\log\beta^*_{\bo{n}}$, one finds by the rules of Ito calculus that these logarithmic variables will have stochastic evolution
\SEQN{}{
dz_{\bo{n}} &\dots +&  c\,dW_{\bo{n}}\\
d\wt{z}_{\bo{n}} &\dots +& \wt{c}\,d\wt{W}_{\bo{n}}
.}
This indicates that the {\it short time} evolution of $z_{\bo{n}}$ and $\wt{z}_{\bo{n}}$ is dominated by Gaussian noise. 
Thus at short times, (defining $z_{\bo{n}}=z'_{\bo{n}}+iz''_{\bo{n}}$, and $\wt{z}_{\bo{n}}=\wt{z}'_{\bo{n}}+i\wt{z}''_{\bo{n}}$)
\EQN{
P(C,t)\prod_{\bo{n}}d^2z_{\bo{n}}d^2\wt{z}_{\bo{n}}
\propto P(C,0) \prod_{\bo{n}} \exp\left( -\frac{[z'_{\bo{n}}(t)-z'_{\bo{n}}(0)]^2}{2\re{c}^2t} -\frac{[\wt{z}'_{\bo{n}}(t)-\wt{z}'_{\bo{n}}(0)]^2}{2\re{\wt{c}^2}t} \right)\quad
.}
Changing back to the coherent state amplitude variables, 
\EQN{\label{pcz}
P(C,t)\prod_{\bo{n}}d^2\alpha_{\bo{n}}d^2\beta_{\bo{n}}
\propto P(C,0) \prod_{\bo{n}} \frac{1}{|\alpha_{\bo{n}}\beta_{\bo{n}}|}\exp\left\{ -\frac{\left(\log\left|\frac{\alpha_{\bo{n}}(t)}{\alpha_{\bo{n}}(0)}\right|\right)^2}{2\re{c}^2t} -\frac{\left(\log\left|\frac{\beta_{\bo{n}}(t)}{\beta_{\bo{n}}(0)}\right|\right)^2}{2\re{\wt{c}}^2t} \right\}\quad
.}
The far tail decay as a Gaussian of a logarithm of $|\alpha_{\bo{n}}|$ or $|\beta_{\bo{n}}|$ is strong enough to overcome any polynomial of these variables if one moves far enough into the tails. (polynomials will be only exponential in the logarithm). So, there is no immediate reason to suspect boundary term errors of the second kind for polynomial observables here. (They cannot be ruled out by the above discussion, however, because at longer times the evolution of $P(C)$ is much more complicated and no proof for such a regime has been given). 

On the other hand, \eqref{pcz} does not decay strongly enough to make an average over a parity estimator converge, since these latter can grow as fast as  $\exp(2e^{z'_{\bo{n}}+\wt{z}'_{\bo{n}}})$ (when $\alpha_{\bo{n}}\beta_{\bo{n}}$ is negative real), which grows much faster than \eqref{pcz} decays. One thus suspects that parity expectation values will not be estimated by \eqref{parityestimate} in an unbiased way for a gauge P representation of the locally interacting Bose gas model.

\subsection{Unresolved issues}
\label{CH6SecondIssues}

Lest one think that boundary term errors are fully understood, 
a variety of issues are still unclear (to the best of the author's knowledge). Some of these have to do with tradeoffs between the two kinds of error. For example:
\ITEM{
\item \textbf{The role of normalization:}
In Section~\ref{CH6SecondBargmann} it was seen that for a  (single-mode) interacting Bose gas, the un-normalized positive P representation  did not suffer boundary term errors of the first kind, but had severe problems with errors of the second kind when $\re{\alpha\beta}\to+\infty$. On the other hand, the normalized positive P representation or the gauge P do not suffer from the second kind of error for most observables, but (as will be seen in Chapter~\ref{CH8}) does have problems with moving singularities, and hence presumably the first kind of boundary term error, when $\re{\alpha\beta}\to-\infty$.  

  There appears to be a tradeoff between the two processes causing boundary term errors, which can involve the normalization of the kernel. Nevertheless, the normalization does not parameterize the tradeoff in a straightforward way, since the two kinds of boundary term error occur in different regions of phase-space.

\item \textbf{The role of moving singularities:}
Only the first kind of boundary term error  appears during derivation of the FPE and stochastic equations while on the other hand,  errors of the second kind can occur simply on a static initial distribution, which does not evolve in any way. For these reasons it appears that moving singularities in the deterministic part of the stochastic equations should be related to boundary term errors of the first kind. 

\enlargethispage{1cm}
On the other hand, moving singularities do lead to pathological tail behaviour in $P(C)$ after a certain time $t_{\rm sing}$ (even without any noise terms, if the initial distribution is nonzero on any point of a trajectory that escapes to infinity at finite time.) Because the far tail behaviour of $P(C)$ changes in this qualitative way after $t_{\rm sing}$, one should be able (at least in some cases) to find some observables that would be estimated in an unbiased way for $t<t_{\rm sing}$, but suffer from boundary term errors of the second kind for $t>t_{\rm sing}$. So it appears that moving singularities can also influence the second kind of boundary term error in some way, presumably by causing slowly decaying distribution tails. 
} 
A better understanding of the relationship between the two kinds of error might lead to a way to make an advantageous tradeoff that avoids both, at least in some cases.

\section{Removal of the errors using kernel gauges}
\label{CH6Removal}
\subsection{Conceptual justification}
\label{CH6RemovalConceptual}

If one introduces a kernel gauge $\mc{F}$ (e.g. a drift gauge $\mc{G}_k$) using a gauge identity $\mc{J}\left[\op{\Lambda}\right]=0$, defined by \eqref{generalidentity}  as in Section~\ref{CH4Kernel}, then zero is added to the master equation in the form $\int P(C)\mc{F}(C)\op{J}\left[\op{\Lambda}\right]dC$, as in \eqref{zeromasterterm}. 
  Introducing such a kernel gauge can have an effect on the presence of boundary term errors in two ways:
\ENUM{
\item \textbf{Modification of the distribution:} 
  A change in the FPE due to addition of extra gauge terms will cause a direct change to its solution $P(C,t)$. A good choice of gauge can lead to faster decaying tails, and hence directly to $\op{\mc{B}}=0$ from evaluation of expression \eqref{btexpression}, and no boundary term errors of the first kind. This is the easiest approach in practice, and will be considered in some detail in the next Subsection~\ref{CH6RemovalHeuristic}.
  
Less directly, this approach also has the potential to remove boundary term errors of the {\it second} kind. This is again because $P(C)$ changes, and if the far tail behaviour is made to decay faster, the observable expressions \eqref{numeratorint} and \eqref{denominatorint} may become convergent.

\item \textbf{Renormalization of boundary terms:} 
  The expression \eqref{btexpressionb} gains extra gauge terms, which can have an effect of the final value of $\op{\mc{B}}$.
During the derivation of the FPE in Section~\ref{CH3EquationsFPE}, these extra added terms to the master equation can be written in the $T^{(\bullet)}$ notation of \eqref{masterinoperators} and also \eqref{btexpression} for $\op{\mc{B}}$ as 
\SEQN{}{
T^{(0)}_{0}&=&\mc{F}J^{(0)}\\
T^{(1)}_{0j}&=&\mc{F}J^{(1)}_j\\
T^{(2)}_{0jk}&=&\mc{F}J^{(2)}_{jk}
.}
One finds that the additional boundary terms introduced by this kernel gauge are (using \eqref{btexpression})
are of the same general form as \eqref{btexpressiona}, but with integration  over some additional terms $\op{\mc{B}}^{\rm int}_{\mc{J},j}$ on top of the usual $\op{\mc{B}}_j^{\rm int}$.
Using \eqref{btexpressionb}, these new terms are
\EQN{\label{btexpressiong}
\lefteqn{\op{\mc{B}}_{\mc{J},j}^{\rm int}(C) =} &&\\
&&\mc{F}(C)\left\{
P(C)J_{j}^{(1)}(C)
\displaystyle+\Half P(C)\sum_k T_{jk}^{(2)}\dada{}{C_k}
\displaystyle-\Half\sum_k\dada{}{C_k}\left(P(C)T_{kj}^{(2)}(C)\right)
\right\}\op{\Lambda}(C)\nonumber
.}
The point to be made is that while the identity $\mc{J}\left[\op{\Lambda}\right]$, which can be expressed in differential form as \eqref{generalidentity}, is zero, the integrands $\op{\mc{B}}_{\mc{J},j}^{\rm int}$ for the boundary term expression 
$\mc{\op{B}}$ have a different form, and are not necessarily zero at all.  This then may open the possibility that if originally (with no gauges), the boundary terms were nonzero, an appropriate choice of gauge can cancel these, again leading to $\op{\mc{B}}=0$, and no boundary term errors of the first kind.
}

\subsection{Practical implementation using drift gauges}
\label{CH6RemovalHeuristic}

Let us take the direct approach of changing $P(C,t)$ to make the boundary terms $\op{\mc{B}}=0$.  A difficulty is that the solution of $P(C,t)$ is not known for non-trivial cases, so one cannot work with it directly. However, as pointed out in Section~\ref{CH6FirstSymptoms}, moving singularities in the noiseless equations are very closely tied to boundary term errors of the first kind. Certainly in the sense that their presence indicates weakly decaying distribution tails, and hence boundary term errors. The drift part of stochastic equations can be modified practically at will using drift kernel gauges $\mc{G}_k$ (see Section~\ref{CH4Drift}), and in particular, the moving singularities can be removed in this way. As will be shown in examples of Sections~\ref{CH6Absorber} and~\ref{CH6Laser},  removal of these moving singularities  does indeed appear to result in disappearance of the boundary term errors. 

  Given a kernel amenable to drift gauges (i.e. possessing a global weight $\Omega$), and keeping in mind the symptoms of boundary term errors of the first kind outlined in Section~\ref{CH6FirstSymptoms} one wishes to choose drift gauges according to the following heuristic criteria:
\ENUM{
\item Moving singularities that were present in the original (ungauged) stochastic equations are removed.
This can be assessed by analysis of the deterministic parts of the stochastic equations. See examples in subsequent sections.
\item No new moving singularities are introduced. 
\item No noise divergences are introduced (in the $d\Omega = \dots + \Omega\sum_k\mc{G}_kdW_k$ evolution).
\item No discontinuities in the drift or diffusion of any variables are introduced.
\item Given the above criteria 1--4 are satisfied (which can be assessed by inspection or analysis of the gauged stochastic equations), one wants to minimize the weight spread in the resulting simulation, to optimize efficiency. 

 A broadly applicable rule of thumb is to \textbf{keep the gauges as small as possible while achieving the goal of removing boundary term errors.}
This can be aimed for using several complementary approaches:
\ENUM{
  \item Make the gauge functions zero in regions of phase space where no correction to the un-gauged drift is necessary.
  \item Minimize a relevant variance of the weights. This weight spread has been considered in detail in Section~\ref{CH4DriftWeightspread}. Depending on the situation, the quantity  $\sum_k|\mc{G}_k|^2$ (more generally) or $\sum_k\re{\mc{G}}_k^2$ (for short time calculations of some observables) can be minimized, since the short-time growth of the variance of $|\Omega|$ or $\re{\Omega}$ (respectively) has been shown in Section~\ref{CH4DriftWeightspread} to be proportional to these quantities.
  \item Ensure that drift gauges are zero at deterministic attractors in phase-space, so that no weight spread is accumulated when no significant evolution is taking place.
  \item Minimize the variances of estimators of observables of interest. That is, the variance of quantities $\tr{\op{O}\op{\Lambda}}$ and $\tr{\op{O}\dagop{\Lambda}}$ appearing in the numerator of observable estimates \eqref{observables}. In this manner, one can optimize the simulation for the estimation of a particular observable $\op{O}$, rather than more generally as in point (a) above.
} 
\item One should also avoid gauges with the following features:
\ENUM{
\item Gauges that are nonzero only in a rarely visited region of phase space.  The problem is that good sampling of the gauged region is needed to make the correct change to observable estimates --- otherwise, if the gauged region is not visited by any trajectories, one has practically the same simulation as without gauges! In a less extreme case where only a tiny (but nonzero) fraction of trajectories visit the gauged region, the correction from biased-ungauged to  unbiased-gauged observable estimates is badly resolved, and the simulation results become very noisy. Furthermore, if the bias to be corrected is significant, then the change in the few gauged trajectories needs to be large to compensate for the bias, and hence the weight correction on those trajectories will become large as well. This leads to a large weight spread. It is much better to make small changes in many trajectories contribute to removal of the bias.
\item Gauges rapidly changing in phase-space. These typically lead to stiff equations, and a gauged region of phase space that is badly sampled by reasonable-sized ensembles. This bad sampling arises when no single trajectory stays in the gauged region long, to such a degree that at a given time few (or, on average, possibly less than one) trajectories are sampling the gauged region, i.e. sampling the resulting observable estimate corrections. This results in similar problems as in the previous point (a).
\item Non autonomous gauges (i.e. depending directly on the weight $\Omega$). The problem with these is that they are usually much more difficult to analyze to check whether or not new moving singularities or noise divergences have been introduced. 
\item Gauges analytic in complex variables if the kernel is also analytic in these. As discussed in Section~\ref{CH6FirstManymode}, it is desirable to break the analyticity of the stochastic equations to avoid moving singularities.
}
  Note that an explicit variational minimization according to 5(b) or 5(d) could be in conflict with 
the recomemndations 6(a) or 6(b), in which case it is probably best to try the simulations and see which works better.
}

This technique appears to be broadly applicable, and only requires the recognition of what instabilities in the stochastic equations could lead to problems. It does not require detailed knowledge of what the boundary terms $\op{\mc{B}}$ are, nor of the behaviour of the far tails of $P(C)$, provided instabilities are removed.

\subsection{Ito vs. Stratonovich moving singularities}
\label{CH6RemovalIS}

There is a subtle ambiguity that can arise when using the method just presented: Should one remove moving singularities from the Ito or the Stratonovich form of the stochastic equations? 

As indicated in Appendix~\ref{APPB}, when one numerically integrates the stochastic differential equations using a finite but small time step $\Delta t$, the exact form of the drift that is integrated can depend on details of the integration algorithm chosen. For example, if one evaluates the derivatives using the variable values at the beginning of a time step (i.e. simply those calculated using the previous step), then one integrates the standard Ito form of the equations. However, with a semi-implicit method, where one first estimates the variables in the middle of the time step, before using {\it these} to calculate the derivative and actually advance the simulation, one must integrate the {\it Stratonovich} form of the stochastic differential equations. The difference between these is given by the Stratonovich correction to the drift \eqref{stratcorrection}. In practice,  multiplicative\footnote{i.e. equations in which the noise terms are proportional to the variables, and hence of non-constant variance.}   stochastic equations, such as those considered in this thesis, are more stable (and hence, efficient) using the semi-implicit method\cite{DrummondMortimer91}. 

In some situations it may occur that moving singularities occur in the deterministic parts of the Stratonovich equations but not the Ito equations (or vice-versa). 
To arrive at practical guidelines on this issue, one can argue in the following way:  
\ITEM{
\item Since both the explicit (Ito) and semi-implicit (Stratonovich) algorithm simulate the same underlying equations, then 
to be sure that no bias is being introduced, moving singularities should not occur in {\it either} algorithm. In fact, a whole family of hybrid Ito-Stratonovich algorithms are possible where derivatives are evaluated at arbitrary time values during the time step $\Delta t$, and the difference in the drift terms between two such algorithms is a multiple $\in[0,2]$ of the Stratonovich correction \eqref{stratcorrection} --- See Appendix~\ref{APPB}. {\it None} of these algorithms should contain moving singularities, otherwise some kind of bias may be suspected.  
\item The Stratonovich correction for variable $C_j$ \eqref{stratcorrection} is highly dependent on the form of the noise matrix elements ($\propto \sum_{kl} B_{kl} \partial B_{jl}/\partial C_k$). If one finds that moving singularities are present for one algorithm but not for another, this indicates that the Stratonovich correction  is growing very rapidly as one heads for the boundaries of phase space. This in turn indicates that the noise matrix elements must be growing in some appropriately rapid fashion as well --- possibly rapidly enough to cause noise divergences.   
}
Taking the above points into consideration, it is conjectured here that:
\ENUM{
\item In a simulation that does not suffer from boundary term errors, there must be no moving singularities in the deterministic part of either the Ito, Stratonovich, or any of the hybrid Ito-Stratonovich forms of the stochastic equations. 
\item If a simulation has moving singularities for one of the family of algorithms (e.g. Ito), but not for some other member of this family, then noise divergences may be suspected as an additional cause of boundary term errors. 
\item To eliminate biases caused by moving singularities using a drift gauge, one must remove moving singularities from {\it all} hybrid Ito-Stratonovich algorithms with the same gauge choice. 
}

\section{Removal example 1: Two-particle absorber}
\label{CH6Absorber}

Two examples will now be worked through --- those systems in which boundary term errors in positive P simulations have been reported in published work. Removal of boundary term errors using the guidelines of Section~\ref{CH6RemovalHeuristic} will be demonstrated, and the procedure considered in some detail. 
The first example will be a single-mode two-boson absorber (in this section), while the second example of a single-mode laser will be treated in Section~\ref{CH6Laser}. These two examples demonstrate somewhat different aspects of boundary term errors and their removal.

This section is closely based on Section IV of the published article by Deuar and Drummond\cite{DeuarDrummond02}. 

\subsection{The single-mode model}
\label{CH6AbsorberModel}

Consider a single optical cavity mode driven by coherent radiation, and damped by
 a zero temperature bath that causes both one- and two-photon losses.
For convenience, units of time are chosen so that the rate of two-photon loss is unity.
 The scaled one-photon loss rate
is $\gamma $, and $\varepsilon$ is the scaled (complex) driving field amplitude.  Following standard techniques, the system can be described with the master
equation 
\begin{eqnarray}
\dada{\op{\rho}}{t} &=&  \left[\, \varepsilon \dagop{a}-\varepsilon ^{*}\op{a}\,,\,\op{\rho}\,\right] +\frac{\gamma }{2}(2\,\op{a}\op{\rho}\,\dagop{a}-\dagop{a}\op{a}\op{\rho}-\op{\rho}\,\dagop{a}\op{a}) \nonumber\\
 &&   + \frac{1}{2}(2\,\op{a}^{2}\op{\rho}\,\op{a}^{\dagger 2}-\op{a}^{\dagger 2}\op{a}^{2}\op{\rho}-\op{\rho}\,\op{a}^{\dagger 2}\op{a}^{2})\, \, .\label{12pdmaster} 
\end{eqnarray}
This is also the master equation of two-particle absorption in an interaction picture where kinetic processes have has been relegated to the Heisenberg picture evolution of the operators.

Let us use the gauge P representation to convert this master equation into a set of stochastic equations. 
The  single-mode kernel is written as
\EQN{\label{gaugekernelch6}
\op{\Lambda}(\ C=\{\alpha,\beta,\Omega\}\ ) = \Omega||\alpha\rangle\langle\beta^*||e^{-\alpha\beta}
.}
Following the treatment of Sections~\ref{CH3Equations}, \ref{CH4Drift}, and~\ref{CH5Equations}, and converting to Stratonovich form using \eqref{stratcorrection}, one arrives at the 
Stratonovich stochastic equations
\SEQN{}{
d\alpha  & = & [\,\varepsilon ^{\, }-\alpha (\alpha \beta +i\mc{G}+\{\gamma-1\}/2)]\,dt+i\alpha dW\, \, \\
d\beta  & = & [\,\varepsilon ^{*}-\beta (\alpha \beta +i\mc{\wt{G}}+\{\gamma-1\}/2)]\,dt+i\beta d\wt{W}\, \,  \\
d\Omega  & = & S_{\Omega}\,dt+\Omega \left[ \mc{G}dW+\wt{\mc{G}}d\wt{W}\right] \, \, \, \, .\label{baseeq} 
}
 Here $S_{\Omega }dt$ is the appropriate Stratonovich correction 
term given by \eqref{stratcorrection}, which depends on the functional form of the particular gauges chosen.

With no gauge ($\mc{G}=\wt{\mc{G}}=0$), the positive P Stratonovich
equations are recovered:
\SEQN{}{
d\alpha  & = & [\,\varepsilon \, \, \, -\alpha (\alpha \beta +\{\gamma -1\}/2)]\,dt+i\alpha dW\, \, \label{12pdposp} \\
d\beta  & = & [\,\varepsilon ^{*}-\beta (\alpha \beta +\{\gamma -1\}/2)]\,dt+i\beta d\wt{W}\, \, .
}

This two-particle loss model can give either correct or incorrect results when treated with
the usual positive P representation methods. Generally, problems only arise when any accompanying single-particle losses
 have small rates or when the number of bosons per
mode is small (see Figure~\ref{FIGURE2ban}).
It is a well-studied case, and a detailed treatment of the boundary term errors that appear can be found in \cite{Gilchrist-97}. It also has
the merit that exact solutions can be readily found using other means.
Analyzing this example gives a good check that the moving singularity removal procedure of Section~\ref{CH6RemovalHeuristic}
 does in fact eliminate boundary
terms and leads to correct results.
Several  simplifications of this model will now be concentrated on,
which correspond to existing literature.

\subsection{Two-boson absorber}
\label{CH6AbsorberTwo}

\label{2PD} In the simplest form of this model, corresponding to $\gamma =\varepsilon =0$,
only two-boson absorption takes place. One expects that for some state
$\left| \psi \right\rangle =\sum _{n=0}^{\infty} c_{n}|n\rangle $ in a Fock number state basis, all even
boson number components will decay to vacuum, and all odd-numbered
components will decay to $\left| 1\right\rangle $, leaving a
mixture of vacuum and one-boson states at long times.

The positive P representation has been found to give erroneous results \cite{SmithGardiner89,CraigMcNeil89,McNeilCraig90,Gardiner-93}
accompanied by moving singularities \cite{Gilchrist-97}. There are
 power-law tails in the distribution that lead to boundary
term errors --- consistently with the conditions \eqref{gaugebtcond}. The observable usually concentrated on in this system is the occupation number
 $\op{n}=\dagop{a}\op{a}$, which is estimated by $\bar{n}=\average{\alpha\beta=\breve{n}}$ in the positive P representation. 
The variable $\breve{n}=\alpha\beta$ has a convenient closed equation (Stratonovich),
\begin{equation}
\label{2PDeq}
d\breve{n}=-2\breve{n}(\breve{n}+i\mc{G}^+-1/2)\,dt +i\breve{n}\,dW^{+}
\end{equation}
 with $dW^{+}=(dW+d\wt{W})$, and $\mc{G}^+=(\mc{G}+\wt{\mc{G}})/2$.

Let us examine the behavior of the above equation when $\mc{G}^+=0$,
i.e., in the standard, un-gauged formulation. The deterministic part
of the evolution has a repellor at $\breve{n}=0$, and an attractor at
$\breve{n}=1/2$. The noise is real, finite, and of standard deviation
$\sqrt{dt/2}$ at the attractor. The deterministic
part of the evolution has a single trajectory of measure zero that
can escape to infinity along the negative real axis, \begin{equation}
\alpha(t) =-\beta(t) =\frac{1}{\sqrt{2(t_{\rm sing}-t)}},
\end{equation}
 where $t_{\rm sing}=1/2\alpha (0)^{2}=-1/2\breve{n}(0)$. This moving singularity
is a symptom of likely boundary term errors.

Indeed, in the steady-state limit, all
trajectories in a simulation will head toward $\breve{n}=1/2$,
with some remnant noise around this value being kept up by the stochastic term.
This noise does not affect the average, however, and the estimator of $\langle\op{n}\rangle$ becomes  $\lim_{t\rightarrow\infty} \bar{n} = \lim_{t\to\infty}\average{ \breve{n} } =1/2$ in the many trajectories limit.
Quantum mechanics, however, predicts that if we start from a state
$\op{\rho}(0)$, the steady state will be \begin{equation}
\lim _{t\rightarrow \infty }\langle \op{n}\rangle =\sum _{j=0}^{\infty }\left\langle 1+2j\right| \op{\rho}(0)\left| 1+2j\right\rangle \, \, ,
\end{equation}
since all even occupations decay to vacuum, and all odd to $\ket{1}$.
 For a coherent state $\left| \alpha _{0}\right\rangle $ input,
say, this will be \begin{equation}
\label{exss}
\lim _{t\rightarrow \infty }\langle \op{n}\rangle =\frac{1}{2}\left( 1-e^{-2|\alpha_{0} |^{2}}\right) \, \, .
\end{equation}
 Thus one expects that the positive P simulation will give correct
results only when $e^{|\alpha_{0} |^{2}}\gg 1$.

To correct the problem one has to change the phase-space topology
in some way to prevent the occurrence of such moving singularities. 
It was found that a good gauge for a two-boson absorber nonlinearity
in general is \begin{equation}
\label{goodgauge}
\mc{G}=\wt{\mc{G}}=\mc{G}^+=i(\breve{n}-|\breve{n}|)\, \, .
\end{equation}
 This will be dubbed the ``circular'' gauge due to the creation of an attractor at $|\breve{n}|=1/2$.

This gauge has the effect of replacing the $-2\breve{n}^{2}$ term in \eqref{2PDeq} with $-2\breve{n}|\breve{n}|$, which is always a restoring
force, and so never leads to any escape to infinity in finite time. 
With the gauge (\ref{goodgauge}), the Stratonovich equations become
\SEQN{\label{2pdeqs}}{
d\breve{n} & = & \breve{n}(1-2|\breve{n}|)\,dt +i\breve{n}dW^{+}\, \, ,\\
d\alpha &=& -\alpha(|\breve{n}|-1/2)\,dt +i\alpha dW\,\,,\\
d\Omega  & = & \Omega \left\{\ [\breve{n}+(\breve{n}-|\breve{n}|)^2]\,dt+i(\breve{n}-|\breve{n}|)\,dW^{+}\right\} \, \, .
}

How do these new equations measure up to the four symptoms of boundary term errors of the first kind outlined in Section~\ref{CH6FirstSymptoms}?
  In logarithmic variables $\alpha_L=\log\alpha$, $n_L=\log \breve{n}$, and $z_0=\log\Omega$, the Stratonovich system equations are
\SEQN{\label{logeqns2pd}}{
dn_L &=& (1-2|\breve{n}|)\,dt+idW^+\\
d\alpha_L&=& -(|\breve{n}|-1/2)\,dt +idW\\
dz_0 &=& [\breve{n}+(\breve{n}-|\breve{n}|)^2]\,dt + i(\breve{n}-|\breve{n}|)\,dW^+.
}
 One now has an attractor on the circle
$|\breve{n}|=1/2$, and a repellor at $\breve{n}=0$, with free phase
diffusion of $n$ in the tangential direction. Once trajectories reach the
attractor, only phase diffusion occurs. 
So:
\ENUM{
\item {\it Moving singularities:} There are no moving singularities in $n_L$, since $|\breve{n}|$ always remains bounded. Also, since the deterministic evolution of the remaining independent variables $\alpha_L$ and $z_0$ depends only on $\breve{n}$ or $|\breve{n}|$ in such a way that $d\alpha_L$ and $dz_0$ remain finite for finite $\breve{n}$, then these variables never escape to infinity in finite time (being just a time integral of finite values of $\breve{n}$). Thus, no moving singularities occur. 
\item {\it Noise divergences:} Similarly because the noise coefficient of $n_L$ is constant, while the noise coefficients of $\alpha_L$ and $z_0$ depend only on $\breve{n}$, and in such a way that for finite $\breve{n}$ the noise magnitude remains finite --- then one concludes that noise divergences are not present either.
\item {\it Discontinuous drift or noise:} Not present, by inspection.
\item {\it Initial distribution:} Initial coherent state has \mbox{$P(0) = \delta^2(\alpha-\alpha_0)\delta^2(\beta-\alpha^*)\delta^2(\Omega-1)
.$} There are no boundary terms at the outset with such an infinitely-peaked distribution.
}
How does this change in the Ito formulation? Using \eqref{stratcorrection} on \eqref{logeqns2pd}, one finds that the Ito and Stratonovich 
equations are identical for $n_L$ and $\alpha_L$, while the Ito weight equation is $dz_0 = (\breve{n}-|\breve{n}|)^2dt + i(\breve{n}-|\breve{n}|)\,dW^+$. The above arguments apply without change to Ito, and to hybrid Ito-Stratonovich algorithms.
So --- no symptoms of boundary term errors remain in any formulation.

\begin{figure}[tp]
\center{\includegraphics[width=9cm]{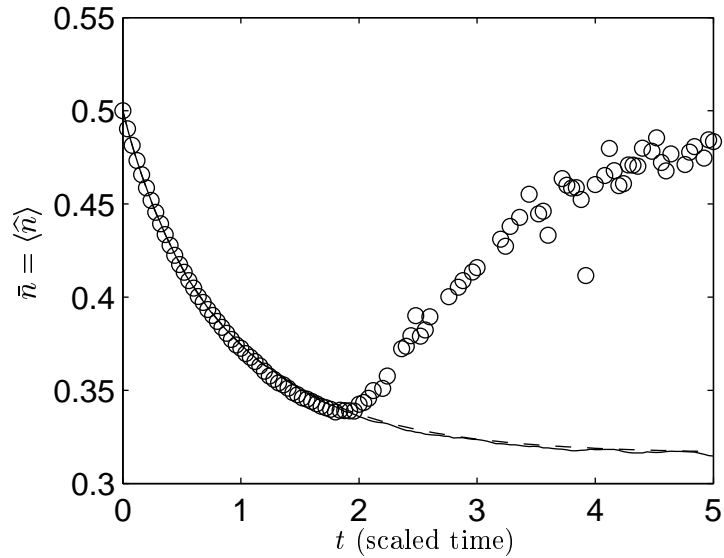}}\vspace{-8pt}\par
\caption[Comparison of two-boson absorption simulations]{\label{FIGURE2bat} \footnotesize
\textbf{Comparison of two-boson absorption simulations:} time-varying mode occupation. {\scshape circles}:
positive P simulation; {\scshape solid line}: gauge simulation using circular gauge \eqref{goodgauge};
{\scshape dashed line}: exact calculation (truncated number-state basis).
Simulation parameters: $\mc{S}=4\times10^4$ trajectories; initial
coherent state.
\normalsize}
\end{figure}

\begin{figure}[tp]
\center{\includegraphics[width=9cm]{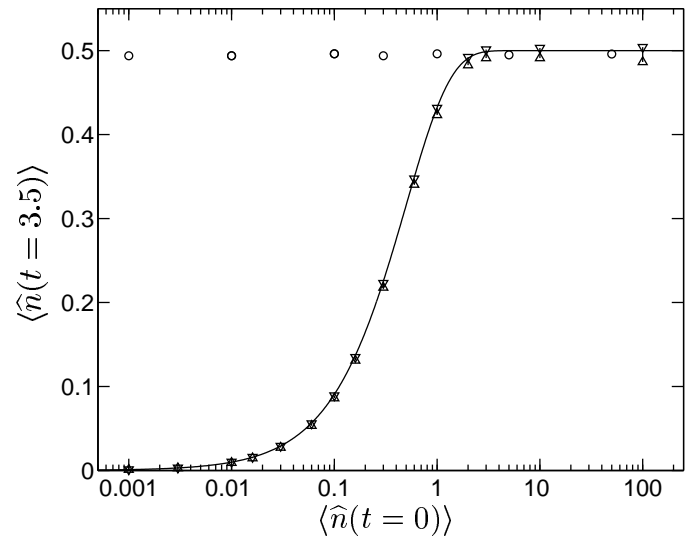}}\vspace{-8pt}\par
\caption[Steady state expectation values of boson number]{\label{FIGURE2ban} \footnotesize
\textbf{Steady state expectation values of boson number} \protect$\langle\op{n}\rangle\protect$
obtained by gauge simulations ({\scshape double triangles}) compared
to exact analytic results  (\ref{exss})\ ({\scshape solid line})
and positive P simulations ({\scshape circles}) for a wide range of
initial coherent states. Size of uncertainty in gauge results due
to finite sample size is indicated by vertical extent of `double-triangle'
symbol. Steady state was observed to have been reached in all simulations
by \protect$t =3.5\protect$ or earlier (compare with Figure~\ref{FIGURE2bat}
and \ref{FIGURE12bd}), hence this is the time for which the simulation
data is plotted. $\mc{S}=10^5$ trajectories 
\normalsize}
\end{figure}

Let us see how the gauge P simulation compares with the positive P in practice --- i.e. are boundary term errors {\it really} removed along with the moving singularities?
Figure \ref{FIGURE2bat} compares results for an (exact) truncated
number-state basis calculation, a positive P calculation, and a 
gauge  calculation using the circular gauge \eqref{goodgauge} for an initial coherent state
of $\alpha _{0}=1/\sqrt{2}$. Figure~\ref{FIGURE2ban} compares steady-state
 values for exact, positive P, and gauge calculations for a wide range of 
initial coherent states. It is seen that the gauge
calculation is correct to within the small errors due to finite sample
size, whereas the positive P calculation leads to severe errors at low occupations.

\subsection{One- and two-boson absorber}
\label{CH6AbsorberOneTwo}

If one now considers single-boson decay as well, but still with no
driving, one expects that all states will decay to the vacuum on a time scale $1/\gamma$ 
on top of the two-particle losses, which produce decay of particle numbers on a timescale of $1$ (although not necessarily right down to the vacuum.) 
When $\gamma\gg 1$, the single-particle decay dominates and nothing interesting is seen beyond an exponential decay of particle number. However, if $\gamma \lesssim 1$,
one should first see a rapid decay to a mixture of vacuum and one-boson
states due to the two-boson process, and then a slow decay of the
one-boson state to the vacuum on a time scale of $t \approx 1/\gamma $.

In this case the positive P (Stratonovich) equations display different behavior depending
on whether $\gamma $ is above or below the threshold $\gamma =1$.
Below threshold, there is an  attractor at $\breve{n}=(1-\gamma )/2$, and
a repellor at $\breve{n}=0$, while above threshold, the attractor is
at $\breve{n}=0$, and the repellor at $\breve{n}=-(\gamma -1)/2$. In either
case, there is a divergent trajectory along the negative real axis,
which is again a moving singularity. It turns out that the steady
state calculated with the positive P is erroneous while $\gamma <1$, and
that there are transient boundary term errors while $\gamma <2$ \cite{SmithGardiner89}.
The false steady state below threshold lies at the location of the
attractor: $(1-\gamma )/2$.

Let us try to fix this problem using the same circular gauge (\ref{goodgauge})
as before. The gauged equation for $n_L$ is now \begin{equation}
dn_L=(1-\gamma -2|\breve{n}|)\,dt +idW^+ \, \, ,
\end{equation}
 while the $\alpha $ and $\Omega $ evolution is unchanged.
So, above threshold one is left with only an attractor at $\breve{n}=0$,
while below threshold one has a repellor at $\breve{n}=0$ surrounded
by an attracting circle at $|\breve{n}|=(1-\gamma )/2$. This phase space
again has no moving singularities or noise divergences.

\begin{figure}[t]
\center{\includegraphics[width=9cm]{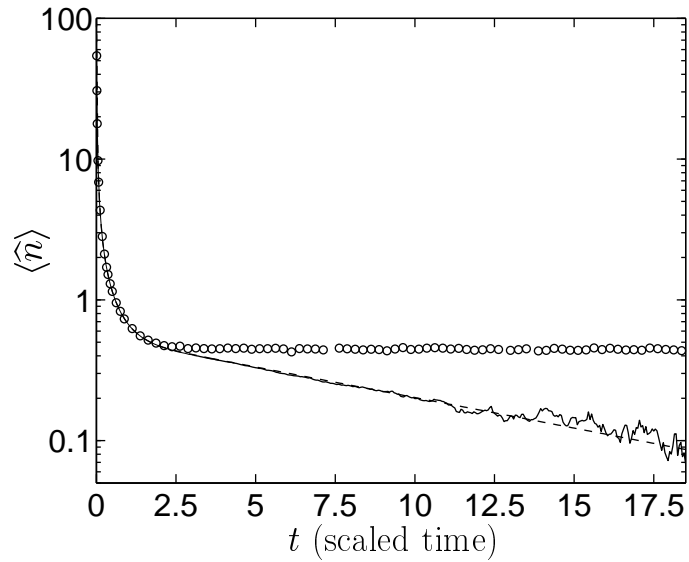}}\vspace{-8pt}\par
\caption[Simulation with both one- and two-particle absorption]{\label{FIGURE12bd} \footnotesize
\textbf{Comparison of simulations for system with both one- and 
two-boson damping.} Relative single-boson damping strength \protect$\gamma =0.1\protect$;
{\scshape Circles}: positive P simulation. {\scshape solid line}: 
simulation using the circular gauge \eqref{goodgauge}; {\scshape dashed line}: exact calculation (truncated
number-state basis). Gauge simulation parameters: \mbox{100 000}
trajectories;  initial coherent state \protect$\left| 10\right\rangle \protect$
with \protect$\langle \op{n}\rangle =100\protect$ bosons. 
\normalsize}
\end{figure}

The results of simulations for the parameter $\gamma =0.1$ are
shown in Figure~\ref{FIGURE12bd} ($\gamma \ll 1$ was chosen so that a system with
two widely differing time scales is tested.) The gauge simulation tracks the exact
results and avoids the false results of the positive P simulation.  Note also that the
gauge simulation remains efficient for a wide range of occupation
numbers --- from $\langle \op{n}\rangle \approx 100\gg 1$, where
the positive P is also accurate, to $\langle \op{n}\rangle \approx 0.1\ll 1$
where it is totally incorrect.

\subsection{Driven two-boson absorber}
\label{CH6AbsorberDriven}

\begin{figure}[t]
\center{\includegraphics[width=9cm]{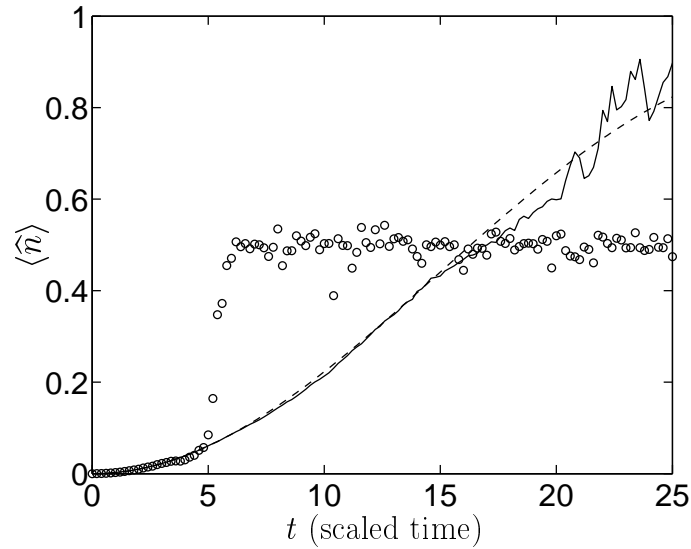}}\vspace{-8pt}\par
\caption[Driven two-boson absorber simulations]{\label{FIGURE2bdrivent} \footnotesize
\textbf{Driven two-boson absorber} with \protect$\varepsilon =0.05\protect$.
{\scshape Circles}: positive P simulation (\protect$1000\protect$
trajectories); {\scshape solid line}: circular gauge \eqref{goodgauge} simulation (\mbox{100 000}
trajectories); {\scshape dashed line}: exact calculation (truncated
number-state basis). 
Initial vacuum state. 
\normalsize}
\end{figure}

The other type of situation to consider is when there is a driving
field as well as two-boson damping. In this Subsection
the coherent driving $\varepsilon$ is nonzero, but the single-particle loss rate is assumed negligible ($\gamma =0$), since this
process never causes any of the simulation problems anyway, but leaving
it out simplifies analysis. Failure of the positive P representation
method has been found in this limit as well  \cite{SchackSchenzle91}, and is
evident in Figure~\ref{FIGURE2bdrivent}. The equation for $\breve{n}$ is no longer
stand-alone in this case, and all three complex variables
must be simulated as in \eqref{baseeq}, the $\Omega$ equation being the same as in the 
undriven case \eqref{2pdeqs}.

A treatment of the singular trajectory problem with the same circular
gauge~(\ref{goodgauge}) leads again to correct results, as seen
in Figure~\ref{FIGURE2bdrivent}. 

\subsection{Relevance to interacting Bose gas}
\label{CH6AbosrberRelevance}

The master equation \eqref{12pdmaster} was first considered in the context of quantum optics. There,  single-photon losses are usually large relative to two-photon processes, and mode occupations are high,  so the boundary term errors have  not been a practical problem in their original quantum optics regime. However, for other physical systems such
as interacting Bose gases (which this thesis is most interested in), two-particle losses can be stronger, one expects a large number of modes to have low or practically zero occupation, and so the  boundary term errors may be significant. 
For example, collisional many-body processes  that remove particles in pairs from the modeled field $\op{\Psi}$, will lead to similar two-particle loss  terms \eqref{twoparticlelosses} in the master equation.

\section{Removal example 2: Single-mode laser}
\label{CH6Laser}

 Let us now consider the second quantum system for which
systematic errors have been seen with the positive P representation.
The mechanism by which boundary term errors occur here is somewhat different than in the two-boson absorber. For two-boson damping, boundary term
errors occur even when one chooses an optimal (i.e., compact) initial
distribution to represent the starting state $\ket{\alpha_0}\bra{\alpha_0}$, whereas for this laser model the systematic
errors occur only if one starts with  unusually broad initial distributions. Nevertheless, such broader distributions may arise during time-evolution from some other state, so consideration of what occurs for significant distribution broadness is also relevant to practical simulations.

This section is closely based on Section IV of the published article by Deuar and Drummond\cite{DeuarDrummond02}.

\subsection{The laser model}
\label{CH6LaserModel}

For a simple model of a photonic or atomic laser,
using the single-mode positive P representation, 
\EQN{\label{ppkernelch6}
\op{\Lambda}(\ C=\{\alpha,\beta\}\ ) = ||\alpha\rangle\langle\beta^*||e^{-\alpha\beta}
,}
one can derive\cite{SchackSchenzle91,Gilchrist-97} the 
the Fokker-Planck equation 
\begin{equation}
\dada{P}{t} = \left\{ 
\dada{}{\alpha}\left[ \frac{\alpha\beta}{\mc{N}}-G\right]\alpha 
+\dada{}{\beta}\left[ \frac{\alpha\beta}{\mc{N}}-G\right]\beta + 2Q\mc{N}\frac{\partial ^{2}}{\partial \alpha \partial \beta }\right\} P
,\end{equation}
where the average number of photons in the mode is $\bar{n}=\average{\alpha\beta}$, and $\mathcal{N}$ is a scaling parameter that equals the number of atoms in the gain medium. Both $G$, the gain parameter, and $Q\ge G/\mathcal{N}$, the
noise parameter, are real and positive.
One usually considers the scaled variables 
\SEQN{}{
\wt{\alpha } & = & \alpha /\sqrt{\mathcal{N}}\\
\wt{\beta } & = & \beta /\sqrt{\mathcal{N}}
,}
 The stochastic Ito equations are then\cite{SchackSchenzle91,Gilchrist-97}
\SEQN{\label{sseqns}}{
d\wt{\alpha} &=& (G-\wt{\alpha}\wt{\beta})\,\wt{\alpha}\,dt + \sqrt{2Q}\,d\eta\\
d\wt{\beta} &=& (G-\wt{\alpha}\wt{\beta})\,\wt{\beta}\,dt + \sqrt{2Q}\,d\eta^*
,}
where $d\eta$ is a complex Wiener increment obeying $\average{d\eta^*d\eta}=dt$, $\average{d\eta}=\average{d\eta^2}=0$.

One is again usually interested in the (scaled) occupation of the lasing mode
\EQN{
\bar{n_{\rm sc}} = \average{n_{\rm sc}=\wt{\alpha}\wt{\beta}} = \langle\dagop{a}\op{a}\rangle/\mathcal{N}
,}
i.e. the number of bosons in the lasing mode per each atom in the gain medium.
Changing variables to $n_{\rm sc}$ and an auxiliary ``phase-like'' variable $z_2=\wt{\alpha}/\wt{\beta}$, the FPE becomes
\EQN{
\dada{P}{t} = \left\{2\dada{}{n_{\rm sc}}\left[Q+n_{\rm sc}(n_{\rm sc}-G)\right] + 2Q\dada{}{z_2}\frac{n_{\rm sc}}{z_2} + 2Q\frac{\partial^2}{\partial n_{\rm sc}^2} n_{\rm sc} -2Q\frac{\partial^2}{\partial z_2}\frac{z_2^2}{n_{\rm sc}}\right\}P.\nonumber\\
}
The diffusion matrix is now decoupled 
\EQN{
4Q\matrix{n_{\rm sc}&0\\0&-z_2^2/n_{\rm sc}}
,}
and so the two complex variables can evolve under independent noises.  In fact the entire $n_{\rm sc}$ evolution decouples from $z_2$, and can be written as the closed Ito equation
 \begin{equation}\label{eqnnsc}
dn_{\rm sc}=-2(n_{\rm sc}-a)(n_{\rm sc}-b)dt +2\sqrt{Qn_{\rm sc}}dW\, \, ,
\end{equation}
 where now the \textit{real} Wiener increments (implemented as Gaussian noises) have variance $\average{dW^2} =dt$,
and the deterministic stationary points 
are 
\SEQN{}{
a  = & \frac{1}{2}\left( G+\sqrt{G^{2}+4Q}\right) &\ge 0\\
b  = & \frac{1}{2}\left( G-\sqrt{G^{2}+4Q}\right) &\le 0
.}
 (In the Stratonovich calculus, $\left[\,{}^a_b\right]=\half(G\pm\sqrt{G^2+2Q})\,$).
  One finds that the stationary point at $a$ is an attractor, and
at $b$ there is  a repellor. Defining $n_{\rm neg} =b-n_{\rm sc}$, as the distance (in the negative direction) from the repellor, one obtains 
\begin{equation}\label{nneg}
\dd{n_{\rm neg}}{t} =2n_{\rm neg} (n_{\rm neg} +\sqrt{G^{2}+4Q})+\text{noise}\, \, ,
\end{equation}
 which shows that there is a singular trajectory escaping to
infinity in finite time along the negative real axis $n_{\rm sc}<b$ --- i.e. a moving singularity.

\subsection{Initial conditions}
\label{CH6LaserInitial}

Let us consider the case of vacuum initial conditions. A vacuum
can be represented by the positive P representation \begin{equation}
\label{deltap}
P_+(\wt{\alpha },\wt{\beta })=\delta^2 (\wt{\alpha })\,\delta^2 (\wt{\beta })\, \, ,
\end{equation}
 but also by Gaussian distributions of any variance $\sigma _{0}^{2}$,
around the above, \begin{equation}
\label{varini}
P^{(+)}(\wt{\alpha },\wt{\beta })=\frac{1}{4\pi ^{2}\sigma _{0}^{4}}\exp \left\{ -\frac{|\wt{\alpha }|^{2}+|\wt{\beta }|^{2}}{2\sigma _{0}^{2}}\right\} \, \, .
\end{equation}
 It can be checked that this satisfies the vacuum positive P representation condition \eqref{vacuumpp}, giving another family of equivalent vacuum distributions different to \eqref{vacpp}, which was found in Section~\ref{CH3StochasticNonunique}.
 Note: the distribution of $n_{\rm sc}$ is non-Gaussian, but has
a standard deviation of $\sigma _{n_{\rm sc}}\approx \sqrt{2}\,\sigma _{0}^{2}$
in both the real and imaginary directions.

It has been found by Schack and Schenzle \cite{SchackSchenzle91} that for this
single-mode laser model, a positive P simulation of pumping from a
vacuum will give correct answers if the usual $\delta$-function initial
condition (\ref{deltap}) is used, but will have systematic errors
if an initial condition \eqref{varini} has a sufficiently large variance
(see Figure~\ref{FIGUREsml}). 

This can be understood by inspection of the stochastic equation \eqref{eqnnsc}. If one has a sufficiently broad initial
distribution, that some trajectories start with $\re{n_{\rm sc}}<b$, then the singular
trajectory will be explored by the distribution. From this it appears that for all $\sigma_0>0$, some boundary term errors may be present. In practice, no errors were seen by Schack and Schenzle for small enough $\sigma_0$, presumably because their magnitude was below the signal-to-noise resolution obtained in those simulations --- an interesting effect in itself. 
Additionally, even if initially
no trajectories fall on the singular trajectory (e.g. some initial distribution that is exactly zero for $\re{n_{\rm sc}}<b$),  the region $\re{n_{\rm sc}}<b$ may
be subsequently explored due to the presence of the noise terms.

Apart from the obvious $\delta$-function initial condition, one might
want to try the canonical distribution \eqref{pprho} applicable generally to any density matrix.
For optical laser models, this will not cause practical problems as then the variance is $\sigma _{0}^{2}=1/{\mathcal{N}}$,
which for any realistic case will be very small (i.e., $\sigma _{n_{\rm sc}}\ll |b|$). Hence, boundary term errors 
would be negligible for a wide range of simulation times, based on the evidence that they were not seen  \cite{SchackSchenzle91} in this regime. Nevertheless, broader distributions may arise during time-evolution of some other state {\it to} a vacuum, so consideration of what occurs for significant $\sigma_0$ is also relevant to practical simulations.

Incidentally, the anomalous results were discovered in \cite{SchackSchenzle91} when  $\sigma _{0}^{2}=1$ was chosen 
by the  erroneous procedure of scaling the equations while not
simultaneously scaling the canonical initial condition in $\alpha$.

\subsection{Drift gauged equations}
\label{CH6LaserDrift}

Let us introduce the single-mode gauge P kernel \eqref{gaugekernel} (compare with \eqref{ppkernelch6}, which differs only by lack of the global weight factor $e^{z_0}$).
Proceeding to introduce drift gauges as in Section~\ref{CH4Drift} (for $n_{\rm sc}$ but not for $z_2$, as averages of this variable will not figure in what follows), one obtains the Ito stochastic equations
\SEQN{}{
  dn_{\rm sc} &=& 2n_{\rm sc}( G - n_{\rm sc})\,dt +2Q\,dt +2\sqrt{Q n_{\rm sc}\,}(dW-\mc{G}\,dt) \\
  dz_0 &=& -\Half\mc{G}^2\,dt+\mc{G}dW 
,}
as per the standard gauge formulation \eqref{langevinstd}.
 It is convenient to define a transformed gauge function 
\EQN{\mc{G}_{(n)} = \mc{G}\sqrt{\frac{Q}{n_{\rm sc}}} },
which is also arbitrary and enters into the simulation as
\SEQN{}{
  dn_{\rm sc} &=& 2n_{\rm sc}( G - n_{\rm sc}-\mc{G}_{(n)})\,dt +2Q\,dt +2\sqrt{Q n_{\rm sc}\,}dW \\
  dz_0 &=& -\frac{n_{\rm sc}\mc{G}_{(n)}^2}{2Q}\,dt+\mc{G}_{(n)}\sqrt{\frac{n_{\rm sc}}{Q}}dW 
.}
 
\subsection{Correcting for the moving singularity}
\label{CH6LaserCorrection}

Consider the (Ito) deterministic evolution of, $n'_{\rm sc}$, the real part
of $n_{\rm sc}=n'_{\rm sc}+in''_{\rm sc}$, \begin{equation}
\dd{n'_{\rm sc}}{t}=-2n'_{\rm sc}{}^{2}+2Gn'_{\rm sc}+2Q+2n''_{\rm sc}{}^{2}-2\re{n_{\rm sc}\mc{G}_{(n)}}\, \, .
\end{equation}
 The moving singularity is due to the $-2n'_{\rm sc}{}^{2}$ leading
term for negative values of $n'_{\rm sc}$. 
The drift gauge is now chosen according to the criteria below (as in Section~\ref{CH6RemovalHeuristic}, but in a different order):
\ENUM{
\item  As pointed out in Sections~\ref{CH4DriftWeightspread} and~\ref{CH6RemovalHeuristic}, it is desirable to keep the gauge terms to a minimum because whenever
they act the weights of trajectories become more randomized. Thus, let us restrict ourselves to functions
$\mc{G}_{(n)}$ that are only nonzero for $n'_{\rm sc}<0$.
\item 
  There should not be any discontinuities in the drift equations, as this might lead to bad sampling of some parts of the distribution and thus possibly to systematic biases. A further problem with discontinuous gauges is that the Stratonovich correction term for $dz_0$ cannot be calculated at the discontinuity, preventing use of the more stable semi-implicit numerical algorithm (See Appendix~\ref{APPB}).

Hence, in particular,  in light of the previous point 1., one should have $\lim _{n'_{\rm sc}\rightarrow0 }(\ \mc{G}_{(n)}\ )=0$.
For ease of analysis, let us start with a simple form for the gauge,
$\mc{G}_{(n)}=c-\lambda n'_{\rm sc}+\lambda _{y}n''_{\rm sc}$.
Continuity  immediately implies $c=\lambda _{y}=0$, hence\begin{equation}
\mc{G}_{(n)}=\left\{ \begin{array}{cl}
-\lambda n'_{\rm sc} & \text {if } \ n'_{\rm sc}<0\\
0 & \text {if } \  n'_{\rm sc}\geq 0
\end{array}\, \, ,\right. 
\label{lasergauge}\end{equation}
\item 
 The next necessary (and most important) condition, to remove moving singularities, is that the $-2n'_{\rm sc}{}^{2}$
term is canceled, hence: \begin{equation}
\lambda \geq 1\, .
\end{equation}
  Using \eqref{stratcorrection}, the Stratonovich correction for $dn_{\rm sc}$ is $-Q\,dt$, and so the family of hybrid Ito-Stratonovich equations differ only in having this term  a multiple of $Q\,dt$ in the range $[0,2Q\,dt]$. This has no effect on the removal or not of the moving singularity. 
\item
 Now, if $\lambda =1$ there are no systematic errors, but the
sampling error very quickly obscures any observable estimates because $n'_{\rm sc}$
still heads to $-\infty $ exponentially fast due to the $2Gn'_{\rm sc}$
term. This takes it into regions of ever increasing $|\mc{G}_{(n)}|$,
and weights quickly become randomized. For slightly larger parameters
$\lambda $, the $n'_{\rm sc}$ evolution takes trajectories
to a point lying far into the negative $n'_{\rm sc}$ region where the
two leading terms balance. Here the trajectories sit, and rapidly
accumulate weight noise. It is clear that for an optimum simulation
all stationary points of $n'_{\rm sc}$ in the nonzero gauge
region must be removed. In this system the  condition for this is \begin{equation}
\lambda >1+\frac{G^{2}}{2Q},
\end{equation}
for the Stratonovich calculus, and $\lambda>1+G^2/4Q$ for Ito. 
\item Are any new moving singularities or noise divergences introduced? 
The gauged equation for the imaginary part of $n_{\rm sc}$ is
\EQN{
dn''_{\rm sc} = 2n''_{\rm sc}[G+n'_{\rm sc}(\lambda-2)]\,dt+ 2\sqrt{Q|n_{\rm sc}|}\sin\left(\half\angle n_{\rm sc}\right)\,dW
,}
which is attractive towards $n''_{\rm sc}=0$ when $\lambda>2$ in the gauged ($n'_{\rm sc}<0$) region, and exponentially 
growing otherwise. In either case, no super-exponential escape occurs, and no moving singularities. The evolution of $z_0$ depends only on finite expressions of finite $n_{\rm sc}$, is thus an integration over the history of a trajectory, and so does not diverge in finite time.  
The Stratonovich correction for $dz_0$ is $\lambda\left(n'_{\rm sc}+n_{\rm sc}+|n_{\rm sc}|\right)/2$ in the gauged region, zero otherwise. Introduction of such a term or its multiple does not lead to any moving singularities for $z_0$ evolution either.
One concludes that no new moving singularities or noise divergences are introduced by the gauge \eqref{lasergauge}.
}

\begin{figure}[t]
\center{\includegraphics[width=8cm]{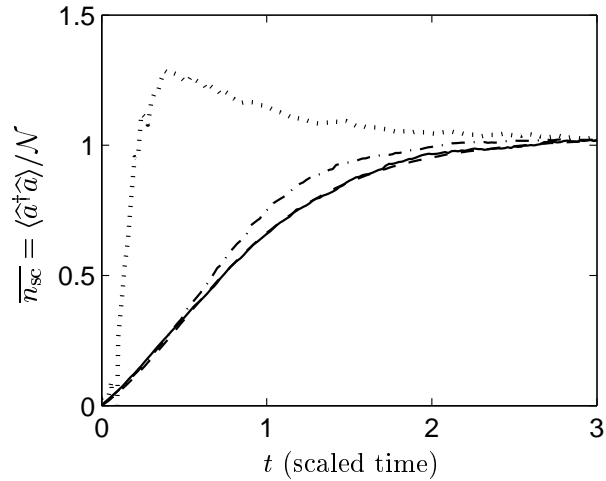}}\vspace{-8pt}\par
\caption[Single-mode laser simulation]{\label{FIGUREsml} \footnotesize
One-mode laser \protect$G=1\protect$, \protect$Q=0.25\protect$.
{\scshape Dashed line}: (correct) positive P simulation with $\delta$-function
initial conditions (\ref{deltap}) \protect$\sigma _{0}^{2}=0\protect$,
and \protect$100 000\protect$ trajectories. {\scshape Dotted-dashed
line}: erroneous positive P simulation with Gaussian initial conditions
(\ref{varini}) \protect$\sigma _{0}^{2}=0.1\protect$ initially,
and \protect$100 000\protect$ trajectories. {\scshape Dotted line}:
erroneous positive P simulation with \protect$\sigma _{0}^{2}=1\protect$,
and \protect$10 000\protect$ trajectories. {\scshape Solid line}:
gauge calculation for \protect$\sigma _{0}^{2}=0.1\protect$ with
\protect$\lambda =4\protect$, which corrects the systematic error
of the positive P. Only \protect$4000\protect$ trajectories,
so as not to obscure other data. 
\normalsize}
\end{figure}

The results for an example have been plotted in Figure~\ref{FIGUREsml}. The parameters there were 
$G=1$, $Q=0.25$ (leading to $a\approx 1.1124$ and $b\approx -0.1124$ in Stratonovich formulation
). The initial condition in the example was $\sigma _{0}^{2}=0.1$,
which is already much larger than the canonical variance for physically
likely parameters in a photonic laser. Typical values of $|n_{\rm sc}|$ initially will
be of order $\sigma _{n_{\rm sc}}\approx 0.14\gtrsim |b|$ here.
A good choice of gauge was found to be $\lambda =4$. The use of this gauge
restores the correct results.

\subsection{Non-optimal initial conditions}
\label{CH6LaserNonoptimal}

As one increases the spread of the initial distribution beyond $\sigma _{n_{\rm sc}}\approx |b|$,
it becomes increasingly difficult to find a gauge that will give reasonable
simulations. For example a wide variety of what seemed
like promising gauges for $\sigma_0^{2}=0.3$ have been tried (with the same values
of parameters $Q$ and $G$ as in Figure~\ref{FIGUREsml}), and none have come close to success.
The problem is that while systematic errors are removed, large random
noise appears and obscures whatever one is trying to calculate.

Trajectories that start off at a value of $n'_{\rm sc}$ lying
significantly beyond $b$ require a lot of modification to their
subsequent evolution to (1) stop them from escaping to $-\infty$
and (2) move them out of the gauged region of phase space so that they
do not accumulate excessive weight noise. If there are many of these,
the trade-off between the gauge size and length of time spent in the gauged
region does not give much benefit anymore. Some consolation is gained in knowing that results will at worst be noisy and
unusable, rather than being systematically incorrect.

\section{Summary of main boundary term results}
\label{CH6Summary}

The two mechanisms that lead to boundary term errors have been specified and investigated for general phase-space distribution methods: 1) Errors of the first kind, which are caused by discarding nonzero boundary terms in the partial integration step between the master and Fokker-Planck equations, and 2) Errors of the second kind, which can arise if the estimator of a particular observable for a given kernel grows too rapidly as the boundaries of phase-space $C$ are approached, and may be present or absent depending on the moment chosen.

The exact expression for the first kind of boundary terms has been found: \eqref{btexpression}, and the integrals that must converge for moment calculations to avoid the second kind are given by \eqref{secondint}. While these cannot in general be explicitly evaluated for non-toy models because the exact form of the distribution $P(C)$ at the boundaries of phase space is required, a simple example has been solved for instructive purposes in Section~\ref{CH6SecondBargmann}, and these boundary term expressions explicitly evaluated.  For the gauge P representation, conditions on $P(C)$  \eqref{gaugebtcond} and \eqref{gaugeppowercond} are found, which are sufficient to avoid boundary term errors of the first and second kind (respectively).  In particular, for the gauge P representation: 
\ENUM{
\item The simulation will be free of boundary terms of the first kind when: A) Kernel drift gauges are polynomial in the system variables or their conjugates, and B) the far tails of $P(C)$ decay faster than exponential in the coherent amplitudes  and faster than power law in $\Omega$. 
\item Estimators of observables polynomial in the $\op{a}_k$ and $\dagop{a}_k$ operators are found to be free of boundary term errors of the second kind
provided that the far tails of $P(C)$ decay faster than a power law.  
}
This is consistent with the results found by Gilchrist\etal\cite{Gilchrist-97}. No indication that such biases will occur for interacting Bose gas simulations is found in Section~\ref{CH6SecondGas}. 

  In Section~\ref{CH6FirstSymptoms}, four symptoms of boundary term errors of the first kind are pointed out. These are moving singularities, noise divergences, discontinuities, and excessively broad initial conditions. All four can be checked for by inspection and some analysis of the stochastic equations and initial condition, before any simulation. Explicit conditions \eqref{mvsingcondition} have been found for simulations on open complex phase-spaces, which are sufficient to ensure that neither moving singularities nor noise divergences occur. It is also noted that by the Painleve conjecture, boundary term errors of the first kind are likely to be generically present in ungauged many-mode simulations based on analytic kernels.

All the above results have been collected here with the aim of systematizing and reducing the confusion on the topic of boundary term errors.  Subsequently, Section~\ref{CH6RemovalHeuristic} introduces a broadly-applicable heuristic method by which the symptoms of boundary term errors of the first kind can be removed using drift kernel gauges (and hence, presumably, also the boundary term errors as such). This removal method is applied to the two cases in the literature where systematic bias has been reported when using the positive P representation: two-boson absorption, and a single-mode laser model. It is found that 
the systematic bias is indeed removed from simulation results by this method for these examples. 

\part{Development of gauges for interacting Bose gases \mbox{}\hspace*{10cm}\mbox{}}
\chapter{Gauges for single-mode interacting Bose gas dynamics}
\label{CH7}

\section{Motivation}
\label{CH7Motivation}

Simulations of the dynamics of multi-mode (locally) interacting Bose gases (with lattice Hamiltonian \eqref{deltaH}) using the positive P representation suffer two major technical problems caused by instabilities in the stochastic equations: 
1) Exponential growth of distribution broadness leading to rapid loss of any useful accuracy.  2) Probable moving singularities  when two or more modes are coupled (See Section~\ref{CH7ModelCoupling}).
These issues prevent simulation of all but short time behaviour. Gases with non-local interactions (Hamiltonian \eqref{latticeH}) are expected to suffer from similar problems on the basis that the locally-interacting lattice model is a special case of these more general models. 

The instabilities arise from the two-body interaction terms in the Hamiltonian, and so for a locally-interacting model 
the unstable processes decouple and are local to each spatial lattice mode of the model. Hence, if the instabilities are brought under control for each mode on its own, then simulations of the full many-mode model should benefit as well.

In this chapter, a single-boson mode with two-body interactions is considered, and gauges (kernel drift and diffusion) are introduced to make improvements. A drift gauge can remove the offending instability, while the diffusion gauge makes a tradeoff between noise in the phase-space (i.e. $\alpha$ and $\beta$) equations and noise in the global weight $z_0$, and will be optimized to improve simulation times. 
In Chapter~\ref{CH8}, the performance of the method developed here  will be investigated in a coupled two-mode system as a prelude to multi-mode simulations.

\section{The single-mode interacting Bose gas}
\label{CH7Model}

\subsection{Physical model}
\label{CH7ModelModel}

Consider a single mode extracted from the open multi-mode locally-interacting lattice model of Section~\ref{CH2Dynamics}.
Annihilation and creation operators for the mode are $\op{a}$ and $\dagop{a}$, and obey $[\op{a},\dagop{a}]=1$.
From \eqref{deltaH}, and with possible coherent gain added as in \eqref{coherentgain}, the Hamiltonian is 
\EQN{
\op{H} = \hbar\op{n}\left( \omega+\chi\left[\op{n}-1\right]\,\right) + i\hbar(\varepsilon\dagop{a}-\varepsilon^*\op{a})
,}
 The master equation of this system interacting with an environment is given by the usual Linblad form \eqref{dynamixmaster}, with Linblad operators 
\SEQN{}{
\op{L}_1 &=& \op{a}\sqrt{\gamma(1+\bar{n}_{\rm bath})}\\
\op{L}_2 &=& \dagop{a}\sqrt{\gamma\bar{n}_{\rm bath}}
}
modeling single-particle interactions with  a standard boson heat bath with a mean number of particles $\bar{n}_{\rm bath}(T)$ per bath mode, as in Section~\ref{CH2Dynamics}. 

Physically this model approximates a single mode of interest $\bo{n}$ of a multi-mode system with two-body scattering, where evolution of the other modes $\bo{m}\neq\bo{n}$ is assumed negligible. Roughly, the linear self-energy term becomes  
$\omega=\omega_{\bo{nn}}$, highly occupied coherent modes $\bo{m}$ can be collected into $\varepsilon=-i\sum_{\bo{m}}\omega_{\bo{nm}}\langle\op{a}_{\bo{m}}\rangle$, while the remainder of modes can become the heat bath. 
The single-mode model can also sometimes represent an approximation to a single orbital of a more complex system, if the constants $\omega$,$\varepsilon$, $\bar{n}_{\rm bath}$ and $\chi$ are chosen appropriately.

\subsection{Stochastic gauge P equations}
\label{CH7ModelGauge}
A single-mode gauge P kernel \eqref{gaugekernelch6} is used, where the complex configuration variables are coherent state amplitudes $\alpha$ and $\beta$, and a global weight $\Omega=e^{z_0}$.
From the multi-mode equations \eqref{itoH}, \eqref{itoheatbathT}, and \eqref{itoepsilon} of Section~\ref{CH5Equations}, the gauge P Ito equations for this model are
\SEQN{\label{dynamixdab}}{
d\alpha &=& -i\omega\alpha\,dt -2i\chi\alpha^2\beta\,dt -\frac{\gamma}{2}\alpha\,dt+ \epsilon\,dt 
+ \sum_k \ul{B}_{1k}(dW_k-\mc{G}_k\,dt)\\
d\beta &=& i\omega\beta\,dt +2i\chi\beta^2\alpha\,dt -\frac{\gamma}{2}\beta\,dt+\epsilon^*\,dt
+ \sum_k \ul{B}_{2k}(dW_k-\mc{G}_k\,dt)\\
d\Omega &=& \Omega\sum_k \mc{G}_k dW_k
,}
where the pre-drift-gauge noise matrices $\ul{B}_{jk}$ defined as in Section~\ref{CH4DriftMechanism} and drift gauges $\mc{G}_k$ have not been specified yet. The $dW_k$ are independent Wiener increments, which can be implemented by Gaussian noises of variance $dt$.
The elements of the  $2\times N_W$ complex noise matrices $\ul{B}$ satisfy
\SEQN{}{
\ul{D}_{11}=&\sum_k \ul{B}_{1k}^2 &= -2i\chi\alpha^2\\
\ul{D}_{22}=&\sum_k \ul{B}_{2k}^2 &= 2i\chi\beta^2\\
\ul{D}_{12}=&\sum_k \ul{B}_{1k}\ul{B}_{2k} &= \gamma\bar{n}_{\rm bath}(T)
.}
The $N_W$ complex drift gauge functions $\mc{G}_k$ are arbitrary in principle. 

The standard gauge formulation \eqref{langevinstd} will be used, with the proviso that noises related to the bath interaction and the interparticle interactions  are chosen to be independent to simplify the equations as outlined in Section~\ref{CH4DiffusionMixing}. That is, using diffusion matrix $\ul{D}=\ul{B}\ul{B}^T=\sum_l\ul{D}^{(l)}$ with
\EQN{
\ul{D}^{(1)} = 2i\chi\matrix{-\alpha^2&0\\0&\beta^2} \qquad;\qquad \ul{D}^{(2)}=\gamma\bar{n}_{\rm bath}(T)\matrix{0&1\\1&0}
,}
one obtains the square root noise matrix forms 
\EQN{
\ul{B}_0^{(1)} = \sqrt{2i\chi}\matrix{i\alpha&0\\0&\beta} \qquad;\qquad \ul{B}_0^{(2)}=\sqrt{\frac{\gamma\bar{n}_{\rm bath}(T)}{2}}\matrix{1&i\\1&-i}
.}

In Section~\ref{CH7ModelPp} it will be shown that the process responsible for instabilities is the  two-particle collisions (parameterized by $\chi$), and so improvements should be searched for by gauging this process, not the bath interactions (parameterized by $\gamma$). Thus, the standard gauges are applied as: 
\ENUM{
\item Imaginary diffusion  gauges $g''_{jk}$ as in Section~\ref{CH4Central}. 
Recall from Sections~\ref{CH4DiffusionReal} and~\ref{CH4DiffusionImaginary} that only imaginary standard diffusion gauges can alter the stochastic behaviour of the simulation, so only these will be considered. 
One obtains 
\EQN{
\ul{B} = \matrix{\ul{B}_0^{(1)}O(g''_{12}) & \ul{B}_0^{(2)}}
,}
with 
an orthogonal matrix $O$ of the form \eqref{diffusiongaugeexplicit} dependent on a single imaginary diffusion gauge $g_{12}=ig''$ with real $g''$. 
\item Complex drift gauges $\mc{G}_1$ and $\mc{G}_2$, with the remaining $\mc{G}_{k>2}=0$.
}
The two-variable diagonal diffusion gauge expression \eqref{diag2vardiffusion} applies here, and the Ito stochastic equations become 
\SEQN{\label{ahofull}}{
d\alpha &=& -i\omega\alpha\,dt -2i\chi\alpha^2\beta\,dt -\frac{\gamma}{2}\alpha\,dt+ \epsilon\,dt+\sqrt{\gamma\bar{n}_{\rm bath}}d\eta_{\rm bath} \nonumber\\
&&+ i\alpha\sqrt{2i\chi}\left[(dW_1-\mc{G}_1\,dt)\cosh g'' +i(dW_2-\mc{G}_2\,dt)\sinh g''\right] \\
d\beta &=& i\omega\beta\,dt +2i\chi\beta^2\alpha\,dt -\frac{\gamma}{2}\beta\,dt+\epsilon^*\,dt+\sqrt{\gamma\bar{n}_{\rm bath}}d\eta_{\rm bath}^*\nonumber\\
&&+ \beta\sqrt{2i\chi}\left[-i(dW_1-\mc{G}_1\,dt)\sinh g'' +(dW_2-\mc{G}_2\,dt)\cosh g''\right] \\
d\Omega &=& \Omega\left\{\mc{G}_1dW_1+\mc{G}_2dW_2\right\}
}
in terms of the real Wiener increments $dW_1$, $dW_2$ implemented by Gaussian noises of variance $dt$, and the complex stochastic Wiener-like increment $d\eta_{\rm bath}$ 
obeying $\average{d\eta_{\rm bath}d\eta_{\rm bath}^*}=dt$ and $\average{d\eta_{\rm bath}^2}=0$.

\subsection{Anharmonic oscillator}
\label{CH7ModelAnharmonic}

Consider the gain-less system ($\bar{n}_{\rm bath}(T)=\varepsilon=0$), which contains all the essential features of the two-body interactions. This system is known in the literature as the damped anharmonic oscillator. 

For this toy system, the observables can be solved analytically, and  comparing simulations with these  will be of great use here to optimize diffusion gauge choices.
In particular, consider  coherent state projector initial conditions $|\alpha_0\rangle\langle\beta_0^*|$, 
which correspond to the configuration of a single trajectory out of the $\mc{S}$ system samples that together can represent any arbitrary quantum state.
Let us concentrate on the mean particle number 
\EQN{\label{obsn}
\langle\op{n}\rangle=\langle\op{a}^{\dagger}\op{a}\rangle
,}
 and the first-order time correlation function\footnote{This first order correlation function is usually written in the Heisenberg picture as $G^{(1)}(0,t) = \langle\dagop{a}(0)\op{a}(t)\rangle$. Moving to the Schr\"odinger picture in which the gauge P representation is defined, one has $G^{(1)}=\tr{e^{i\op{H}t/\hbar}\op{a}e^{-i\op{H}t/\hbar}\op{\rho}(0)\dagop{a}}$. 
Because of the coherent state conditions used here, $\op{\rho}(0)=\op{\Lambda}(\alpha_0,\beta_0,\Omega=1)$. Thus, $\op{\rho}(0)\dagop{a}=\beta_0\op{\rho}(0)$ using \eqref{correspondenceLad}. This gives $G^{(1)}=\beta_0\tr{\op{a}e^{-i\op{H}t/\hbar}\op{\rho}(0)\,e^{i\op{H}t/\hbar}}$, and identifying the Schr\"odinger picture density matrix as $\op{\rho}(t)=e^{-i\op{H}t/\hbar}\op{\rho}(0)\,e^{i\op{H}t/\hbar}$, this gives the desired expression $G^{(1)}(0,t)=\beta_0\langle\op{a}\rangle$. To obtain the right-hand side expression in \eqref{G1tdef}, note that $G^{(1)}$ is a complex quantity composed of two observables (see \eqref{G1inq}) --- one for the real, and one for the imaginary part. Taking the adjoint, $G^{(1)}(0,t)^*=\langle\dagop{a}(t)\op{a}(0)\rangle$ in the Heisenberg picture, and following the same procedure as above, one obtains $G^{(1)}(0,t)^*=\alpha_0\langle\dagop{a}\rangle$.}
 \begin{equation}\label{G1tdef}
G^{(1)}(0,t)=\beta_0\langle\op{a}\rangle = \alpha_0^*\langle\dagop{a}\rangle^*
 ,\end{equation}
 which contains phase coherence information.  Normalizing by $\langle\alpha_0|\beta_0^*\rangle=\tr{\,\ket{\alpha_0}\bra{\beta_0^*}\,}$, their expectation values are found to be
\begin{subequations}\label{1obs}\begin{eqnarray}
\langle\op{n}\rangle  &=& n_0\,e^{-\gamma t},\\
G^{(1)}(0,t) &=& n_0\,e^{-\gamma t/2}\, e^{-i\omega t}
 \exp{\left\{\frac{n_0}{1-i\gamma/2\chi}\left(e^{-2i\chi t-\gamma t}-1\right)\right\}},\label{G1exact}
\end{eqnarray}\end{subequations}
where $n_0=\alpha_0\beta_0$. 

  When the damping is negligible, $n_0$ real, and the number of particles is $n_0\gg 1$, one sees that the initial phase oscillation period (ignoring $\omega$) is 
\EQN{
t_{\text{osc}}=\frac{1}{2\chi}\sin^{-1}\left(\frac{2\pi}{n_0}\right) \approx \frac{\pi}{\chi n_0}
,}
 and the phase coherence time, over which $|G^{(1)}(0,t)|$ 
decays is 
\EQN{\label{tcoh}
t_{\text{coh}}=\frac{1}{2\chi}\cos^{-1}\left(1-\frac{1}{2n_0}\right) \approx \frac{1}{2\chi\sqrt{n_0}}
.} 
The first quantum revival occurs at 
\EQN{
t_{\text{revival}} = \frac{\pi}{\chi}
.}

The stochastic equations for this gain-less system are more convenient in terms of the logarithmic number-phase variables
\SEQN{\label{mndef}}{
n_L&=&\log(\alpha\beta)\\
m_L&=&\log\left(\frac{\alpha}{\beta}\right)
.}
For coherent states where $\beta=\alpha^*$, $n_L$ is the logarithm of the mean number of particles $|\alpha|^2$, while $m_L=2i\angle\alpha$, and characterizes the phase of the coherent amplitude.
The Ito equations in the new variables are
\SEQN{\label{ahologequations}}{
dn_L &=& -\gamma\,dt + 2i\sqrt{i\chi}e^{-g''}(d\eta-\mc{G}_{(n)}\,dt)\\
dm_L &=& -2i(\omega-\chi+2\chi e^{n_L})\,dt +2i\sqrt{i\chi}e^{g''}(d\eta^*-\mc{G}_{(m)}\,dt)\\
dz_0 &=& \mc{G}_{(m)}d\eta +\mc{G}_{(n)}d\eta^* -\mc{G}_{(n)}\mc{G}_{(m)}\,dt = d(\log\Omega)
,}
where the complex Wiener-like increment $d\eta=(dW_1-idW_2)/\sqrt{2}$ has variances
\SEQN{}{
\average{d\eta d\eta^*}&=&dt\\
\average{d\eta^2}&=&0
,}
and the independent transformed complex drift gauge functions are 
\SEQN{\label{loggauges}}{
\mc{G}_{(n)} = \frac{\mc{G}_1-i\mc{G}_2}{\sqrt{2}}\qquad;\qquad
\mc{G}_{(m)} = \frac{\mc{G}_1+i\mc{G}_2}{\sqrt{2}}
.}
Note: these are not necessarily complex conjugates since the original $\mc{G}_1$ and $\mc{G}_2$ are in general complex.

\subsection{Behaviour of the positive P simulation}
\label{CH7ModelPp}

\subsubsection{Lack of moving singularities}
Consider the anharmonic oscillator equations \eqref{ahologequations} in the positive P representation, where $\mc{G}_k=0$, and $g''=0$. 
   By comparison with the condition \eqref{mvsingcondition}, one can check by inspection that moving singularities do not occur : $dn_L$ does not depend on any variables and so is stable, while  $dm_L$ depends only on the other variable $n_L$. Thus the deterministic behaviour of $m_L$ is just integration of a finite function of $n_L$ (which itself remains finite). The noise in $m_L$ is of constant magnitude. $z_0$ does not evolve. All variables remain finite, no moving singularities or noise divergences occur. Therefore, none of the symptoms of boundary term errors from Section~\ref{CH6FirstSymptoms} are present in the equations. 

Indeed, the investigations of Gilchrist\etal\cite{Gilchrist-97} found that boundary term errors do not occur for the special case of the single-mode anharmonic oscillator.

\subsubsection{Rapid growth of statistical error}
    If the observable estimators converge to the quantum mechanical expectation values in the limit of many trajectories so that \textit{accuracy} of the simulation is not a problem, there still remains the issue of how large a sample of trajectories has to be to give a \textit{precise} result. 

  While the evolution of $n_L$ is well-behaved, being simply constant isotropic diffusion and possibly some decay, the deterministic evolution of $m_L$ can be very sensitive to $n_L$ due to the exponential drift term. 

Particularly  $\re{m_L}$ has a great influence on observable estimates of phase-dependent observables, such as $G^{(1)}(0,t)$.
Consider the quadrature observables
\EQN{
\op{q}(\theta) = \frac{\op{a}e^{i\theta}+\dagop{a}e^{-i\theta}}{2}
,}
which are closely related to the first-order correlation function
\EQN{\label{G1inq}
G(0,t) = \beta_0\langle\op{q}(0)-i\op{q}(\pi/2)\rangle = \left(\alpha_0\langle\op{q}(0)+i\op{q}(\pi/2)\rangle\right)^*
.}
Comparing to \eqref{qdef} and \eqref{qest}, the estimator for $\langle\op{q}\rangle$ is 
\EQN{
\bar{q}(\theta) &=& \frac{\average{\re{\frac{\Omega}{2}\left(\alpha e^{i\theta}+\beta e^{-i\theta}\right)}}}{\average{\re{\Omega}}}\\
&=& \frac{\average{\re{\exp\left(z_0+\frac{n_L}{2}\right)\cosh\left(i\theta+\frac{m_L}{2}\right)}}}{\average{\re{e^{z_0}}}}
.}
We can see that $\bar{q}$ is very sensitive to $\re{m_L}$ due to the $\cosh$ factor. 
This real part of $m_L$ evolves as 
\EQN{
d\re{m_L}=4\chi e^{\re{n_L}}\sin(\im{n_L})\,dt+ \dots
,}
which can be a very rapid growth even for moderate values of $\re{n_L}$. Worse, even a moderate spread in $\re{n_L}$ leads rapidly to a very wide {\it spread} of $\re{m_L}$, not to mention the spread in the factor $\cosh(m_L/2)$ that appears in the observable estimate. The behaviour of this spread will be considered in detail in Section~\ref{CH7BothOptimization}, but for now it suffices to point out that there arises a characteristic time scale $t_{\rm sim}$ beyond which the uncertainty in phase-dependent observable estimates grows faster than exponentially with time, and any simulations become effectively useless no matter how many trajectories are calculated. 

Numerical investigations (See Figure~\ref{FIGUREsimtime} and Table~\ref{TABLEsimtime}) found that a positive P simulation lasts at most for times 
\EQN{\label{trmsim}
t_{\rm sim} \approx \frac{(1.27 \pm 0.08)}{\chi n_0^{2/3}}
.}
For large mode occupation $n_0\gg1$, this is not enough time for significant phase decoherence to occur (compare to $t_{\rm coh}$ in \eqref{tcoh}).  This unfavorable scaling is known to be a major unresolved stumbling block for  positive P simulations of the interacting Bose gas model \eqref{hamiltonian} in many physical regimes. For example, evaporative cooling simulations encounter such sampling problems upon the onset of Bose condensation\cite{DrummondCorney99}. 
The expression \eqref{trmsim} indicates that the simulation time for a many-mode system is likely to be limited by the simulation time of the most highly occupied mode (this is confirmed by the simulations in Chapter~\ref{CH10}), which makes it especially worthwile to improve simulation times at large mode occupations.

\subsubsection{Greater stability when damped}
\label{CH7ModelPpDamping}
Consider the evolution of the log-occupation $n_L$ in \eqref{ahologequations}. In a positive P simulation (with $g''=0$), the noise term produces a spread in both the real and imaginary parts of $n_L$ of standard deviation $\sigma=\sqrt{2\chi t}$. To a good approximation, all trajectories in a reasonable-sized sample will lie within about $4\sigma$ of the mean value (which is $\log(n_0)-\gamma t$). If the values of $\re{n_L}$ for all trajectories are $\ll -\log{2\chi}$, then the nonlinear term in the $dm_L$ evolution equation becomes negligible, and stability problems abate. This will occur if the damping $\gamma$ is large enough in comparison with the two-particle collisions $\chi$. 
In Section~\ref{CH7BothCases}, this behaviour will be found to depend on the ``damping strength'' parameter \eqref{qparamdef}, which will quantify what is meant by ``$\gamma$ big enough''. The increased stability at higher damping is a well-known feature of positive P simulations of the anharmonic oscillator, and/or an interacting Bose gas\cite{Corney99}.

\subsection{Coupling to other modes and moving singularities}
\label{CH7ModelCoupling}

Non phase-dependent observables such as $\op{n}=\dagop{a}\op{a}$ can still be calculated at long times for the anharmonic oscillator with the positive P method, because the evolution of their estimators (e.g. $\average{e^{n_L}}$) does not depend on the unstable $m_L$. This is no consolation, however,  because we are {\it ultimately} interested in simulating many-mode systems. (Singe-mode toy problems will fall even to a brute force truncated number-state basis calculation, so all these involved stochastic schemes are only ultimately justified for many-mode models). When many modes are present, the convenient separation of $n_L$ and $m_L$ evolution seen in \eqref{ahologequations} is no longer present due to mode-mixing terms (from kinetics and external potentials). This occurs even for a single-mode model with nonzero pumping $\varepsilon$, or a finite-temperature heat bath $\bar{n}_{\rm bath}>0$, which also model underlying mode-coupling processes.
If the $n_L$ and $m_L$ equations are coupled, the growth of the spread of $\re{m_L}$ will also feed the growth of the spread of $n_L$, making {\it all} moment calculations intractable after some relatively short time.

Furthermore, moving singularities may appear. Reverting back to considering the $\alpha$ and $\beta$ equations \eqref{ahofull}, the Ito positive P evolution of $\alpha$ will be of the form 
\EQN{
d\alpha = -2i\chi\alpha^2\beta\,dt +\epsilon\,dt + \dots
,}
and in particular
\EQN{\label{dralpha}
d|\alpha| = 2\chi|\alpha|^2|\beta|\sin(\angle\alpha+\angle\beta)\,dt + \dots 
.}
When $|\beta|\sin(\angle\alpha+\angle\beta)$ is positive, this violates the no-moving-singularities conditions \eqref{mvsingcondition}, $|\alpha|$ grows faster then exponentially, and moving singularities may be possible, leading to boundary term errors.  This did not occur for the closed anharmonic oscillator only because of its special symmetry properties that allowed the convenient decoupled logarithmic form of the equations \eqref{ahologequations}.

Thus it can be seen that when mode mixing occurs, there are two potential problems that can arise after short times:
\ENUM{
  \item Rapid appearance of massive statistical errors, masking observable estimates in noise.
  \item Possible systematic biases caused by moving singularities in the equations.  
}
Ways to deal with these issues will be investigated in this chapter for a single-mode system, and for coupled-mode systems in Chapter~\ref{CH8}.

\section{Drift gauges: Removal of instability}
\label{CH7Drift}
As was seen in \eqref{dralpha}, the instabilities in the equations arise from the nonlinear two-body drift terms, and in particular from the part that leads to super exponential growth of $|\alpha|$ or $|\beta|$. Defining the real and imaginary parts of the number variable 
\EQN{\label{brevendef}
\breve{n}=\alpha\beta=e^{n_L}=n'+in''
}
 for convenience,
the offending terms are:
\SEQN{\label{badtermsaho}}{
d\alpha &=& 2\chi\alpha n''\,dt + \dots\\
d\beta &=& -2\chi\beta n''\,dt + \dots
}
whereas the terms 
\SEQN{}{
d\alpha &=& -2i\chi\alpha n'\,dt + \dots\\
d\beta &=& 2i\chi\beta n'\,dt + \dots
}
affect only the phase of $\alpha$ or $\beta$, and are harmless. 
Furthermore we can see that any nonzero value of $n''=|\alpha|\im{\beta e^{i\angle\alpha}}$ can lead to moving singularities (in $\alpha$ evolution if $n''>0$, or in $\beta$ evolution if $n''<0$). Note that nonzero $n''$ are characteristic of ``non-classical'' states that cannot be represented by non-singular Glauber P distributions --- i.e. as a mixture of coherent states. This is because for all coherent state trajectories $\beta=\alpha^*$, and so $\breve{n}=n'$ only. 

Let us follow the heuristic procedure of Section~\ref{CH6RemovalHeuristic} to remove the instabilities. Considering the points there, in order:
\ENUM{
\item {\scshape Remove instability:} Drift gauges $\mc{G}_k$ are chosen so that the terms causing the instability \eqref{badtermsaho} are canceled by the drift correction. Assuming no other changes to the drift takes place, then from \eqref{ahofull}, after some algebra, one obtains the required gauges:
\SEQN{\label{ahogauge}}{
\mc{G}_1 &=& -\sqrt{2i\chi}\,n'' e^{-g''}\\
\mc{G}_2 &=& i\sqrt{2i\chi}\,n'' e^{-g''} = -i\mc{G}_1
}
These effectively cause the replacement in the stochastic equations
\SEQN{\label{newtermsaho}}{
-2i\chi\alpha\breve{n}\,dt &\to& -2i\chi\alpha n'\,dt \\
2i\chi\beta\breve{n}\,dt &\to& 2i\chi\beta n'\,dt
.}
In terms of the logarithmic variables, the scaled gauges \eqref{loggauges} are
\SEQN{}{
\mc{G}_{(n)} &=& 0\\
\mc{G}_{(m)} &=& -2\sqrt{i\chi}\,n''e^{-g''} = \mc{G}_1\sqrt{2}
.}
For the anharmonic oscillator, $dn_L$ is unchanged from the positive P simulation, and the full equations are
\SEQN{\label{ahologeq}}{
dn_L &=& -\gamma\,dt +2i\sqrt{i\chi}e^{-g''}d\eta\\
dm_L &=& -2i(\omega-\chi+2\chi\re{e^{n_L}})\,dt +2i\sqrt{i\chi}e^{g''}d\eta^*\\
dz_0 &=& -2\sqrt{i\chi}\,\im{e^{n_L}}e^{-g''}\,d\eta \label{dzoaho}
.}
\item {\scshape Check again for moving singularities:}
(Ignoring, for the time being, any dependence on $g''$ since the diffusion gauge has not been chosen yet)

 \ITEM{
\item
  Comparing to conditions \eqref{mvsingcondition}, it is seen that the linear drift terms in parameters $\varepsilon$, $\omega$, $\gamma$, and $\bar{n}_{\rm bath}$ in \eqref{ahofull} will not lead to moving singularities. 
\item The new nonlinear drift terms \eqref{newtermsaho} have lost their radial component due to the drift gauges, and so do not lead to any moving singularities either. 
\item From \eqref{stratcorrection}, the Stratonovich corrections for $d\alpha$ and $d\beta$ are $i\chi\alpha\,dt$ and $-i\chi\beta\,dt$, respectively. These extra terms satisfy \eqref{mvsingcondition}, and so do not contribute any moving singularities in any of the Ito-Stratonovich hybrid family of algorithms. 
\item That leaves the new $dz_0$ evolution. 
The weight deterministically tracks the $n_L$ evolution, and from \eqref{ahologeq}
\EQN{
dz_0 &=& i\,\im{e^{n_L}}(dn_L+\gamma\,dt)
,}
and so if $n_L=\log(\alpha\beta)$ remains finite for all trajectories (as is seen from the lack of moving singularities in $\alpha$ and $\beta$ evolution), then so will $z_0$.
}
Thus it is seen that no new moving singularities are introduced provided $g''$ is well-enough behaved. The dependence on $g''$ is considered in Section~\ref{CH7BothBoundaryterm}.
\item The noise terms in $d\alpha$ and $d\beta$ are exponential in $\alpha$, and so also satisfy \eqref{mvsingcondition} together with the noise terms of $dz_0$.  thus noise divergences are absent, given a  well behaved $g''$.
\item As desired, there are no discontinuities in equations, by inspection (provided $g''$ is well behaved).

\item {\scshape Gauge efficiency}. Comparing to corresponding sub-points from Section~\ref{CH6RemovalHeuristic}:
\ENUM{
\item Corrections to the drift are necessary in all of phase space apart from the subspaces $n''=0$ of measure zero, so generally $\mc{G}_k\neq 0$ apart from this special region. 
\item For general purpose calculations, one wishes to minimize the quantity $\sum_k|\mc{G}_k|^2$, given that the instabilities are removed. This is achieved by the gauges \eqref{ahogauge} as they are only just large enough to remove the instabilities, and do not introduce any other modifications to the $d\alpha$, or $d\beta$ equations. 
\item Attractors? 
For the single-mode anharmonic oscillator when $\varepsilon=0$, there are deterministic attractors in the phase space at 
vacuum ($n=0$), and $n=(\chi-\omega)/2\chi$ (Stratonovich) or $n=-\omega/2\chi$ (Ito). At all these, $n''=0$, so the gauge is also zero. This is the desired situation.

In the more complicated case of nonzero $\varepsilon$, there will be some stationary points elsewhere in phase space. One could try to construct some gauges that would behave as \eqref{ahogauge} in the far tails of phase space i.e. as $|\alpha|,|\beta|\to\infty$, but would be zero at these stationary points. In the broad picture, however, there seems little point to do this for the single-mode case, because in a many-mode simulation all modes will be coupled together by the kinetic interaction $\omega_{\bo{n\neq m}}$, and $\varepsilon$ will be different or absent. 
\item Rather than tailoring the simulation for a single observable $\op{O}$, the quantity $\sum_k|\mc{G}_k|^2$ was minimized here to keep the gauge applicable for a general case. 
}
\item None of the features to be avoided occur.\ENUM{
\item The gauge is nonzero over a wide range of phase-space, including regions often visited.
\item The gauge does not change in a particularly rapid fashion in phase-space
\item The gauge is autonomous.
\item The drift gauges break the analyticity of the equations, as suggested by the conjecture of Section~\ref{CH6FirstManymode}.
}
}

\section{Exponentials of Gaussian random variables}
\label{CH7Gaussian}
From the single-mode equations for the locally-interacting Bose gas \eqref{ahofull} (or indeed from the many-mode equations \eqref{itoH} or \eqref{gaugepthermo}), it is seen that noise terms due to the interparticle interactions are generally of the multiplicative form 
\EQN{
d\alpha \propto \alpha\sqrt{\chi}\,dW_k+\dots\qquad;\qquad d\beta\propto \beta\sqrt{\chi}\,dW_k + \dots
.}
This leads to Brownian-like motion in the {\it logarithmic} variables $n_L$ or $m_L$, as e.g. in \eqref{ahologequations}. On the other hand, the observable estimates (e.g. $\langle\op{n}\rangle$ or $G^{(1)}(0,t)$) typically involve quantities such as  $\alpha$ or $\alpha\beta$ --- in the original non-logarithmic variables. Clearly there will be a lot of averaging over random variables that are of similar form to the exponential of a Gaussian. 
Let us investigate the behaviour of such random variables.

Let $\xi$ be a Gaussian random variable of mean zero and variance unity, thus its distribution is 
\EQN{\label{gausdist}
  \text{Pr}(\xi) = \frac{1}{\sqrt{2\pi}}\exp\left(-\frac{\xi^2}{2}\right)
.}
The moments of $\xi$ can be found by integration of \eqref{gausdist} to be 
\begin{equation}\label{gaumom}
  \average{ \xi^k} = \left\{\begin{array}{ll}
  \displaystyle\frac{k!}{2^{k/2}(k/2)!}& \text{if $k>0$ is even}\\
  0 & \text{if $k>0$ is odd}
\end{array}\right.
\end{equation}
Now let us define the exponential random variable 
\EQN{
v_{\sigma} = v_0e^{\sigma\xi} = e^{v_L}
}
with positive real $\sigma$ and $v_0$.
Using \eqref{gaumom}, the exponential variable's mean is 
\begin{eqnarray}\label{mexp}
\average{v_{\sigma}} = \bar{v}_{\sigma}= \sum_{k=0}^{\infty} \frac{\sigma^{k}\average{ \xi^k }}{k!} =
 v_0\exp\left( \frac{\sigma^2}{2}\right).
\end{eqnarray}
and so, also 
\EQN{\label{varexp}
\vari{v_{\sigma}} = \sigma_v^2 = 
 v_0^2e^{\sigma^2}\left(e^{\sigma^2}-1\right) = 
\left[\bar{v}_{\sigma}\sqrt{\left(\frac{\bar{v}_{\sigma}}{v_0}\right)^2-1}\right]^2
.}

\begin{figure}[t]
\center{\includegraphics[width=\textwidth]{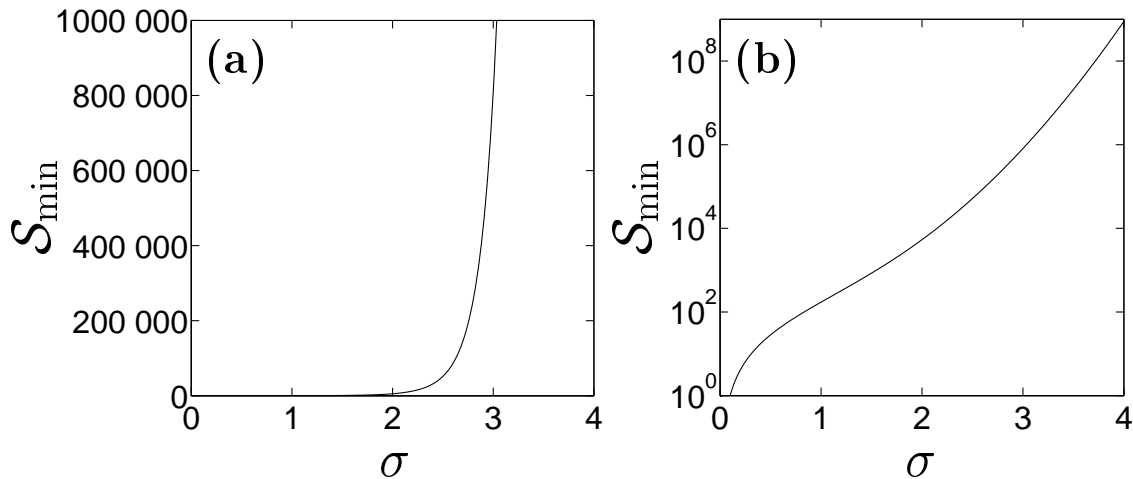}}\vspace{-8pt}\par	 
\caption[Sample size growth for exponentials of a Gaussian random variable]{\label{FIGUREvargaus} \footnotesize
\textbf{Number of samples} $\mc{S}_{\rm min}$ required to obtain a single significant digit of precision in the mean of a random variable $v_{\sigma}$ that is an exponential of a Gaussian random variable with standard deviation $\sigma$. 
Scales: {\bf(a)} Linear, {\bf (b)} logarithmic.
\normalsize}
\end{figure}

If one is interested in estimating quantities such as $\bar{v}_{\sigma}$ using $\mc{S}$ samples, then by the Central Limit Theorem, the relative uncertainty in such a finite-sample estimate will be 
\EQN{\label{deltabarvsigma}
\Delta\bar{v}_{\sigma} = \frac{\sigma_v}{\bar{v}_{\sigma}\sqrt{\mc{S}}}
.}
To obtain only a single significant digit of accuracy, the number of trajectories needed is then
\EQN{
\mc{S}_{\rm min} = \frac{100\sigma_v}{\bar{v}_{\sigma}} = 100\sqrt{e^{\sigma^2}-1}
.}
This minimum number of required samples (independent of $v_0$) is plotted in Figure~\ref{FIGUREvargaus}.
It can be seen there that for $\sigma\gtrsim 3$ the number of samples required becomes intractable, and this sets a  
practical limit on the variance of the logarithmic variable $v_L=\sigma\xi$:
\EQN{\label{sdlimit}
\sigma ^2 \lesssim \order{10}
,}
 which will be referred to numerous times in this and following chapters. 

At small standard deviations $\sigma\lesssim 1$, on the other hand, 
$\sigma_v\approx\sigma$, and the variance of the logarithmic and exponential variables ($\vari{v_L}$ and $\sigma_v^2$, respectively)
is approximately the same --- a result that will also be useful.

Lastly, this sampling problem can also lead to systematic biases (but not boundary term errors) in averages of $v$ with finite sample number $\mc{S}$ once $\sigma\gtrsim3$, when \eqref{sdlimit} is violated. This is considered in detail in Appendix~\ref{APPA}.
\enlargethispage{0.5cm}

\subsubsection{Means of some combinations of Gaussian noises} 
Some expressions that will be used in Sections~\ref{CH7Both}, and~\ref{CH7DiffusionOptimization} are given below.
Since they contain only odd powers of $\xi$,
\EQN{
\average{\sin(\sigma\xi)} &=& 0\label{msin}\\
\average{\xi\cos(\sigma\xi)} &=& 0\label{mxicos}
.}
And, by the same series expansion approach as for \eqref{mexp}
\EQN{
\average{\xi e^{\sigma\xi}} &=& \sigma\exp\left(\frac{\sigma^2}{2}\right)\label{mxiexp}\\
\average{\cos(\sigma\xi)} &=& \exp\left(-\frac{\sigma^2}{2}\right)\label{mcos}\\
\average{\xi\sin(\sigma\xi)} &=& \sigma\exp\left(-\frac{\sigma^2}{2}\right)\label{mxisin}
.}

\section{Optimization of diffusion gauges}
\label{CH7Both}

\subsection{Aims}
\label{CH7BothAims}
  Upon choice of the drift gauges \eqref{ahogauge} to remove the offending moving singularities, diffusion gauges can be chosen to vastly (as it turns out) improve the sampling behaviour of the simulation.
  The aim here will be to arrive at a diffusion gauge $g''$ for the one-mode system that satisfies the following:
\ENUM{
\item Improves useful simulation times in the low-damped high boson occupation regime (this is the regime where simulations without diffusion gauges give very unsatisfactory results -- see Figure~\ref{FIGUREsimtime} and Section~\ref{CH7ModelPpDamping}). 

\item Is easily generalized to many-mode situations. This means, in particular, that the gauge choice should allow for the possibility of mode occupations changing dynamically due to inter-mode coupling (although this is not actually seen in the one-mode toy system).

\item Is expressed as an exact expression, or does not require excessively involved calculations to evaluate. This is important, since in a simulation the gauge should be evaluated at each time step if it is to adapt to dynamically changing mode occupations.

\item Applies on timescales of the order of the coherence time (and shorter timescales as well). This is a relevant timescale only when mode occupations are larger than order unity, and is then smaller than the quantum revival time $t_{\rm revival}=\pi/\chi$. At lower mean occupations (order unity or smaller), we require that the simulation remains stable for a time $\order{1/\chi}$, which should be sufficient not to prematurely destabilize any coupled highly-occupied modes in multi-mode systems.

\item Depends only on variables in the current trajectory --- i.e. on parameters of the current coherent-state projector $|\alpha\rangle\langle\beta^*|$ component of the full state $\op{\rho}$. This is necessary if we want to be able to parallelize the calculation, which enormously improves calculation times.
Also, if the evolution of all possible coherent-state projectors can be stabilized to restrict distribution size and statistical error, then so will the evolution of all possible states since these are always expressed as a distribution, and estimated as an appropriate sample, of such projectors.
  
\item We are especially interested in the case of low or absent damping $\gamma \ll 2\chi$, since this is the regime where  quantum effects are strongest, and also where the simulation is most unstable (see e.g. Sections~\ref{CH7ModelPpDamping} and~\ref{CH7BothCases}). The highly-damped regime is a lesser priority, since there a mean-field, or other approximate simulation would be sufficient for most purposes.
}
This may seem like a lot of conditions for one quantity, but it helps to remember that $g''$ is in principle an {\it arbitrary} function.

\subsection{Variables to be optimized}
\label{CH7BothVariables}
Before proceeding directly to searching for advantageous values of the diffusion gauge, there remain several more issues to address:
\subsubsection{Which moment to optimize?}
   While, strictly speaking, this depends on which moment we might be interested in, in general we should concentrate on 
occupation number $\langle\op{n}\rangle$ or phase correlations $G^{(1)}$. If these basic low-order observables are badly calculated then higher-order observables will not do any better, because they are more sensitive to the distribution broadness.

Consider that estimators for these two observables are (by comparing their definitions \eqref{obsn} and \eqref{G1tdef} to observable estimators in the gauge P representation \eqref{qdef} and \eqref{qest}) are
\EQN{
\langle\op{n}\rangle &\propto& \re{\average{\Omega\breve{n}}}\\
G^{(1)}(0,t) &\propto& \beta_0\average{\Omega\alpha}\qquad\text{or}\qquad \propto\alpha_0^*\average{\Omega^*\beta^*}\label{G1est}
}
(the normalization by denominators $\re{\average{\Omega}}$ has been omitted). 
One finds from \eqref{ahologequations} or \eqref{ahofull} that both $d\breve{n}$ and $d\Omega$ are proportional to $e^{-g''}$,  so varying the diffusion gauge $g''$ acts to scale noise in the number observable $\langle \op{n} \rangle \sim \average{\Omega\breve{n}}$ smoothly from very large at negative $g''$ to very small at positive $g''$. This does not suggest any ``optimal'' values of $g''$. On the other hand, the evolution of phase-dependent expectation values such as $G^{(1)}(0,t)$  displays 
high noise at both large negative and positive values of $g''$ but via different processes. (Large noise in $\Omega=e^{z_0}$ at negative $g''$, and large noise directly in $\alpha$ or $\beta$ at positive values). This suggests that $g''$ parameterizes some tradeoff between phase-space and weight noise, and that 
there is some intermediate  value of $g''$ at which the resulting uncertainty in phase observables is  minimized. Thus optimization should be based on such phase variables.

\subsubsection{Logarithmic variables}
 Ideally one would like to optimize for the variance of variables like $(\alpha\Omega)$, since these appear directly in the calculation of the observable $G^{(1)}$. Unfortunately, exact expressions for such variances are difficult to obtain in a closed form\footnote{The reason for this becomes clear after proceeding to calculate some logarithmic variances in \eqref{lgvars}. These  require the evaluation of quantities such as $\average{n''(t')n''(t'')}$ given in \eqref{niinii}. To evaluate $\vari{\Omega\alpha}=\vari{e^{G_L}}$, on the other hand, requires calculating  averages of the form $\average{n''(t_1)n''(t_2)n''(t_3),\cdots}$ to all orders, which becomes increasingly involved as the number of factors grow.}, and would probably be very complicated if obtained. A complicated form makes it difficult to arrive at an expression for the optimum $g''$ that would satisfy condition 3. in Section~\ref{CH7BothAims} (i.e. simple to evaluate during each simulation step).

It turns out, however, that exact expressions for the variances of the logarithms of  phase-dependent observables such as $\vari{\log(\alpha\Omega)}=\vari{z_0+n_L/2+m_L/2}$ can be found  with (relative) convenience and ease, and we will look  for values of  $g''$ to minimize this logarithmic variance instead of $\vari{\alpha\Omega}$. 

As further justification of this choice, one can make an analogy between the behaviour of $\log\alpha$ and $\alpha$, etc. 
versus the behaviour of the Gaussian random variable $v_L$ of Section~\ref{CH7Gaussian} and $v_{\sigma}=e^{v_L}$. In both cases the logarithmic variable is generated by Brownian motion, although in the $\log\alpha$ case, there is additional drift. One sees from \eqref{varexp} that for $\sigma\lesssim\order{1}$, the variances of both the logarithmic and ``normal'' variables are the same. This corresponds to short time evolution of the anharmonic oscillator in the above analogy. Furthermore, because of the rapid rise of variance with time, the range of $\sigma$ 
for which  $\sigma\not\approx\sigma_v$ but the simulation gives any useful accuracy is fairly narrow and occurs in the more noisy part of the simulation. 

In summary, it can be expected that a optimization of $g''$ in  logarithmic variances will still give good results. This is borne out by the massive improvement in simulation times seen in Figure~\ref{FIGUREsimtime}. Certainly, however, some further improvement could  be obtained by considering non-logarithmic variances (particularly since in the full simulation, $\log\alpha\Omega$ etc. are not exactly Gaussian distributed due to the effect of the drift terms), although it is not clear whether the difference would be significant or not.

\subsubsection{Which phase-dependent variable to optimize}
The first-order correlation function $G^{(1)}(0,t)$ can be estimated in two ways, as seen in \eqref{G1est}. Hence, it is best to optimize for the average variance of the logarithm of the two random variables $\Omega\alpha\beta_0$ and $(\Omega\alpha_0\beta)^*$ corresponding to the $G^{(1)}$ estimate.
  Also, since general coherent state initial conditions to be optimized will have arbitrary phases, we should optimize a variance of variables related to $|G^{(1)}(0,t)|$ rather than the complex $G^{(1)}$.

\subsection{Optimization of $g''$}
\label{CH7BothOptimization}
  $|G^{(1)}(0,t)|$ is estimated by either $e^{G_L}$ or $e^{\wt{G}_L}$, where 
\begin{subequations}\label{Lg}\begin{eqnarray}
G_L(t) &=& \re{\log\beta_0 +\log\alpha(t)+z_0(t)}  \\
\wt{G}_L(t) &=& \re{\log\alpha_0 +\log\beta(t)+z_0(t)} 
,\end{eqnarray}\end{subequations}
while the uncertainty in the estimate will be proportional to $\vari{e^{G_L}}$ or $\vari{e^{\wt{G}_L}}$.
  Taking the considerations of the previous Section~\ref{CH7BothVariables} into account, let us look for such $g''=g''_{\rm opt}$ that the 
mean variance of $G_L$ and $\wt{G}_L$
is minimized for the anharmonic oscillator system. 
This minimum occurs when 
\begin{equation}\label{mincond}
\frac{\partial\left(\vari{G_L(t_{\text{opt}})} + \vari{\widetilde{G}_L(t_{\text{opt}})}\right)}{\partial g''} = 0.	
\end{equation}
Note that the optimal value $g''_{\rm opt}$ will in general depend upon the target time $t_{\text{opt}}=t$ at which the variances are considered.

The formal solution of the gauged anharmonic oscillator equations \eqref{ahologeq}, assuming uniform initial weight $\Omega(0)=1$, is straightforward to find:
\SEQN{\label{formalsoln}}{
  n_L(t) &=& \log(n_0) -\gamma t + \sqrt{\chi}e^{-g''}\left[i\zeta^+(t)-\zeta^-(t)\right]\\
  m_L(t) &=& \log(\alpha_0/\beta_0) -2i(\omega-\chi)t -4i\chi\int_0^t n'(t')dt'\nonumber\\
&&+ \sqrt{\chi}e^{g''}\left[i\zeta^-(t)-\zeta^+(t)\right]\label{formalsolnml}\\
  z_0(t) &=& -\sqrt{\chi}e^{-g''}\int_0^tn''(t')\left[\eta^+(t')+i\eta^-(t')\right] dt',
}
where using the definition \eqref{brevendef}.
\EQN{\label{etadef}
\eta^{\pm}(t) = \dd{W_1(t)}{t}\pm\dd{W_2(t)}{t}
}
are independent real Gaussian noises obeying
\SEQN{\label{etaprop}}{
\average{\eta^{\pm}(t)}&=&0\\
\average{\eta^{\pm}(t)\eta{\pm}(t')}&=& 2\delta(t-t')\label{etaeta}\\
\average{\eta^{\pm}(t)\eta^{\mp}(t')}=& 0
,}
and so $d\eta = \sqrt{i}(\eta^--\eta^+)\,dt/2$.
Also,
\EQN{\label{zetadef}
\zeta^{\pm}(t) = \int_0^t \eta^{\pm}(t') dt'
.}
These $\zeta^{\pm}$ are time-correlated Gaussian random variables. Using \eqref{etaprop}, they obey the relationships
\SEQN{\label{zetaprop}}{
\average{\zeta^{\pm}(t)}&=&0\\
\average{\zeta^{\pm}(t)\zeta{\pm}(t')}&=& 2\,\text{min}\left[t,t'\right]\label{zetazeta}\\
\average{\zeta^{\pm}(t)\zeta^{\mp}(t')}=& 0
.}

The variables to optimize are 
\EQN{\label{Gsoln}
\matri{G_L\\\wt{G}_L} &=& \log|n_0|-\frac{\gamma}{2}\,t -\frac{\sqrt{\chi}}{2}\left(e^{-g''}\zeta^-(t)\matri{+\\-}e^{g''}\zeta^+(t)\right)-\sqrt{\chi}e^{-g''}\int_0^tn''(t')\eta^+(t')dt'
.\nonumber\\}
In the Ito calculus the noises at $t'$ are independent of any variables at $t'$, so one finds
\EQN{\label{lg1}
\matri{\average{G_L}\\\langle\wt{G}_L\rangle_{\rm stoch}} = \log|n_0|-\frac{\gamma}{2}\,t,
}
and using \eqref{zetazeta}
\EQN{\label{lg2}
\matri{\average{G_L^2}\\\langle\wt{G}_L^2\rangle_{\rm stoch}} &=& \matri{\average{G_L}^2\\\langle\wt{G}_L\rangle_{\rm stoch}^2}
+\chi t \cosh 2g'' \matri{+\\-} \chi\int_0^t\average{n''(t')\eta^+(t')\zeta^+(t)}dt' \nonumber\\
&&+\chi e^{-2g''}\int_0^tdt'\int_0^tdt''\average{n''(t')n''(t'')}\average{\eta^+(t')\eta^+(t'')}
,}
where use has also been made of the independence of $\eta^{\pm}(t')$ and $\zeta^{\mp}(t)$.

To further evaluate (\ref{lg2}), one needs to calculate the averages containing $n''$. Define $c=\sqrt{\chi}e^{-g''}$, and $n_0=n'_0+in''_0$.
Firstly, 
\begin{eqnarray}\label{nii}
\average{ n''(t) } &=& e^{-\gamma t} \average{e^{-c\zeta^-}}
 \left[ n'_0 \average{\sin c\,\zeta^+} + n''_0 \average{\cos c\,\zeta^+}\right].\nonumber\\
&=& n''_0 e^{-\gamma t}
\end{eqnarray}
where we have first used the independence of the $\zeta^{\pm}$ to separate the stochastic averages, and then applied expressions (\ref{mexp}), (\ref{msin}), and(\ref{mcos}) to evaluate them. 

Now, since noises at times $t>t'$ are uncorrelated with those at $t'$ or earlier,
\begin{eqnarray}\label{netaxi}
\average{n''(t') \eta^+(t') \zeta^+(t)} &=& \average{ n''(t') \eta^+(t') \zeta^+(t') } \nonumber \\
&=& \int_0^{t'} \average{ n''(t') \eta^+(t') \eta^+(t'') } dt''\nonumber\\
&=& \int_0^{t'} \average{ n''(t') }  \average{ \eta^+(t')\eta^+(t'')} dt'' \nonumber\\
&=& 2\average{ n''(t') },
\end{eqnarray}
using (\ref{etaeta}).

Secondly, if $t''>t'$,
noting that 
\begin{equation*}
  \zeta^{\pm}(t'') = \zeta^{\pm}(t') + \widetilde{\zeta}^{\pm}(t''-t'),
\end{equation*}
  where the $\widetilde{\zeta}^{\pm}(t)$ are independent of the $\zeta^{\pm}(t)$, but have the same properties \eqref{zetaprop}, one finds that 
\EQN{
\lefteqn{\average{ n''(t') n''(t'') } }&&\\
&=&e^{-\gamma(t'+t'')} \average{ e^{-2c\,\zeta^-(t')}} \average{ e^{-c\,\widetilde{\zeta}^-(t''-t')}} 
  \average{ \cos c\,\widetilde{\zeta}^+(t''-t')} \nonumber\\
&&\times \left\{ [n'_0]^2 \average{\sin^2 c\,\zeta^+(t')}
[n''_0]^2 \average{\cos^2c\,\zeta^+(t')}
+n'_0 n''_0 \average{\sin 2c\,\zeta^+(t')}
\right\}\nonumber
.}
Terms containing $\average{\sin c\,\widetilde{\zeta}}=0$ have already been discarded. Evaluating the averages using (\ref{mexp}) and (\ref{mcos}), one obtains
\begin{equation}\label{niinii}
\average{ n''(t') n''(t'') } = \frac{e^{-\gamma(t'+t'')}}{2}
\left\{ |n_0|^2e^{4t' c^2} -(n'_0)^2+(n''_0)^2   \right\}. 
\end{equation}
Due to symmetry
\begin{equation*}
\int_0^t\! dt' \int_0^t\! dt'' n''(t') n''(t'') = 2\int_0^t\! dt'' \int_0^{t''}\! dt'\ n''(t') n''(t''),
\end{equation*}
 and substituting  (\ref{niinii}), (\ref{netaxi}), and (\ref{nii}) into (\ref{lg2}) and (\ref{lg1}), then integrating, one obtains
\begin{eqnarray}\label{lgvars}
\matri{\vari{G_L(t)}\\\text{var}[\wt{G}_L(t)]} &=&  \chi t \cosh(2 g'') \matri{+\\-} 2\chi n''_0\left(\frac{1-e^{-\gamma t}}{\gamma}\right) \\&&+ 
\chi e^{-2g''}|n_0|^2\left(\frac{e^{4\chi e^{-2g''}t} e^{-2\gamma t} - 1}{4\chi e^{-2g''}-2\gamma}\right) -\chi e^{-2g''}[(n'_0)^2-(n''_0)^2]\left(\frac{1-e^{-2\gamma t}}{2\gamma}\right).\nonumber
\end{eqnarray}
The optimum $g''=g''_{\rm opt}$ can now in theory be calculated by imposing (\ref{mincond}).

Since this would involve the solution of a transcendental equation, it would be cumbersome to use in a numerical simulation, going against requirement 3 of  Section~\ref{CH7BothAims}, because one would have to execute a time-consuming algorithm at each time step. By considering, below,  some important special cases, an approximation to the optimum $g''_{\rm opt}$ applicable under the broad conditions aimed for in Section~\ref{CH7BothAims} will be found, which can then be easily used in actual simulations.

\subsection{Important special cases}
\label{CH7BothCases}
\ENUM{
\item Perhaps the most important special case is when damping is absent ($\gamma=0$). This is also the worst case in terms of simulation stability.  At relatively short times one has
\begin{equation}\label{shtcond} 4\chi t_{\text{opt}}e^{-2g''_{\text{opt}}} \ll 1,\end{equation}
 and the optimum is given (from \eqref{mincond} and \eqref{lgvars}) by the roots of the cubic in $V_g=e^{-2g''_{\rm opt}}$:
\begin{equation}\label{vcubic}
8\chi t_{\text{opt}}|n_0|^2 V_g^3 +[4(n''_0)^2+1]V_g^2 -1 = 0.
\end{equation}
In the usual case of simulations with sizeable occupation numbers in modes, and times up to coherence time (which then is much shorter than the quantum revival time $\pi/\chi$) this short time condition is satisfied. 
A very useful expression 
\begin{equation}\label{approxgiiopt1}
  g''_{\text{opt}} \approx \frac{1}{3}\log\left(|n_0|\sqrt{8\chi t_{\text{opt}}}\right).
\end{equation}
applies when the $V_g^2$ term is negligible. This occurs at long enough times when $n_0$ is large enough and mostly real: i.e. when $1+4(n''_0)^2 \ll (8\chi t_{\text{opt}} |n_0|^2)^{2/3}$. So $\chi t_{\text{opt}}$ must be at least $\gg 1/8|n_0|^2$  (a higher limit applies if $n''_0\neq0$). 

The opposite case when $n_0$ is either too small, too imaginary, or the time is too short has the $V_g^3$ term negligible and leads to 
\begin{equation}\label{approxgiiopt2}
  g''_{\text{opt}} \approx \frac{1}{4}\log\left[1+4(n''_0)^2\right].
\end{equation}

\item It can be seen that the long time behaviour of the variances depends on the ``damping strength'' parameter
\begin{equation}\label{qparamdef}
q = 2\gamma-4\chi e^{-2g''}.
\end{equation}
When $q$ is positive, the long time (i.e. $qt\gg 1$) behaviour of (\ref{lgvars}) is asymptotic to a linear increase
\EQN{\label{longtimeqt}
\matri{\vari{G_L}\\\text{var}[\wt{G}_L]}\to \chi t \cosh 2g''  - b
}
 where  $b=\chi e^{-2g''}\{[(n'_0)^2-(n''_0)^2]/2\gamma-|n_0|^2/q\}\mp 2\chi n''_0/\gamma$ is a constant. This means that the sampling uncertainty grows relatively slowly, and long simulation times are possible. When $q$ is negative, on the other hand, the long time behaviour is $\chi e^{-2g''}|n_0|^2e^{|q|t}/|q|$, and we expect a rapid 
appearance of intractable sampling error after some time $|q|t\gg1$. 

The parameter $q$ depends on the relative strengths of the nonlinearity and the damping, and determines the long time behavior of the statistical error. Note though that if damping is present, then for large enough diffusion gauge $g''$, $q$ can always be made positive, and the linear (in $t$)  variance regime can be reached with a choice $g'' \ge \log\sqrt{2\chi/\gamma}$.

\item An immediate extension to nonzero damping of expressions (\ref{vcubic}), (\ref{approxgiiopt1}) and (\ref{approxgiiopt2}) can be derived for times $|qt_{\text{opt}}|\ll 1$ to give the cubic
\begin{subequations}\label{vacubic}\begin{equation}
  8\chi t_{\text{opt}} |n_0|^2 V_g^3  + a_2 V_g^2 - 1 = 0
\end{equation}
where
\begin{eqnarray}
a_2(n_0,\gamma t_{\rm opt}) &=& 1 + 4(n''_0)^2\left(\frac{1-e^{-2\gamma t_{\text{opt}}}}{2\gamma t_{\text{opt}}}\right) \label{a2def}\nonumber\\
&& - 2|n_0|^2\left(\frac{1-2\gamma t_{\text{opt}} + 2(\gamma t_{\text{opt}})^2-e^{-2\gamma t_{\text{opt}}}}{\gamma t_{\text{opt}}}\right).
\end{eqnarray}\end{subequations}
The expression (\ref{approxgiiopt1}) still applies at large enough and real enough $n_0$: i.e. when $a_2 \ll (8\chi t_{\text{opt}} |n_0|^2)^{2/3}$. In the opposite case of $a_2 \gg (8\chi t_{\text{opt}} |n_0|^2)^{2/3}$, one has
\begin{equation}\label{approxgiiopt2gam}
g''_{\text{opt}} \approx \frac{1}{4}\log(a_2).
\end{equation}
}

\subsection{Suggested approximate form of diffusion gauge}
\label{CH7BothForm}
Guided by conditions 2, 4, 5,  and 6, in Section~\ref{CH7BothAims}, we can see that the choice of $g''_{\text{approx}}$ for the anharmonic oscillator might depend on the four real parameters $n'_0$, $n''_0$, $\chi t_{\text{opt}}$ and $\gamma$. Because of conditions  4 and 5, we are most interested in the regime of  small $\gamma$, and either $\chi t_{\text{opt}} \lesssim \order{1/2\sqrt{|n_0|}}$ when $|n_0|\gtrsim \order{1}$, or 
$\chi t_{\text{opt}} \lesssim \order{1}$ when $n_0\lesssim \order{1}$. These cases are covered by expression (\ref{vacubic}).

   This can be most easily seen in the two limits $|n_0|\gg1$ and $|n_0|\ll1$ where for small $\gamma$, we have
$qt_{\text{opt}}\approx -2(\chi t_{\text{opt}}/|n_0|)^{2/3}$ and $\approx -4\chi t_{\text{opt}}/\sqrt{1+4(n''_0)^2}$ respectively. Hence the condition $|q|t_{\text{opt}}\ll 1$ applies for target times $\chi t_{\text{opt}} \ll |n_0|/2\sqrt{2}$ and 
$\ll\, \approx(1+2|n''_0|)/4$ respectively, which is roughly sufficient for low occupations, and more than sufficient for high occupations. 

To obtain an explicit estimate for $g''_{\rm opt}$, one
can either evaluate the roots of the polynomials \eqref{vcubic} or \eqref{vacubic}  by standard expressions, which can be still quite complicated although reasonably rapid, or use the approximation 
\begin{equation}\label{approxgiiopto}
g''_{\text{approx}} = \frac{1}{6}\log\left\{ 8|n_0|^2\chi t_{\text{opt}} + a_2^{3/2}\right\}, 
\end{equation}
which reduces to (\ref{approxgiiopt1}) and (\ref{approxgiiopt2gam}) in their limits of applicability, and works very well in practice (see figures \ref{FIGUREsimtime}, \ref{FIGUREG}, and \ref{FIGUREvarGt}).

\begin{figure}[t]
\center{\includegraphics[width=10cm]{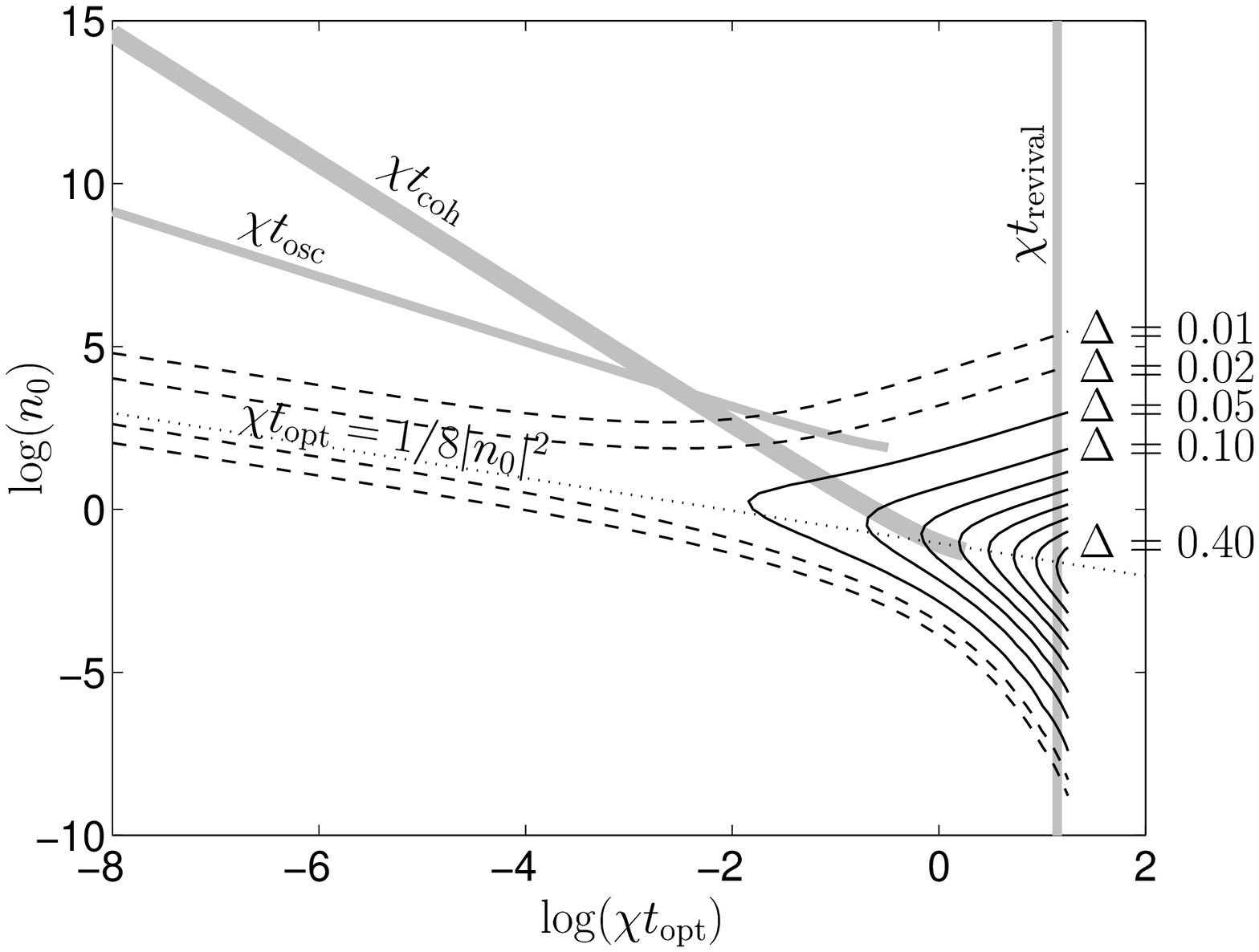}}\vspace{-8pt}\par
\caption[Discrepancy between exact $g''_{\rm opt}$ and its approximation]{\label{FIGUREgiidiscrepancy} \footnotesize
\textbf{Discrepancy} $\Delta=g''_{\text{opt}}-g''_{\text{approx}}$ between $g''_{\rm opt}$ (the exact optimization of $g''$ by solving \eqref{mincond} using \eqref{lgvars}), and the approximate expression (\ref{approxgiiopto}). Displayed is the case of no damping ($\gamma=0$) and classical initial occupation ($n_0=n'_0$), shown as a function of $t_{\text{opt}}$ and $n_0$. Discrepancy values $\Delta$  are shown as {\scshape solid contours} with spacing $0.05$. Additional {\scshape dashed contours} shown at very low discrepancy. {\scshape Dotted line} approximates region of greatest discrepancy $\chi t_{\rm opt}\approx 1/8n'_0{}^2$.
For comparison, several physical timescales are also shown in {\scshape grey}: time of first quantum revival $t_{\text{revival}}$, phase coherence time $t_{\text{coh}}$ and phase oscillation period $t_{\text{osc}}$. 
\normalsize}
\end{figure}

The discrepancy $\Delta$ between (\ref{approxgiiopto}) and the exact optimization obtained by solving \eqref{mincond} with \eqref{lgvars} is 
shown for real $n_0$ for a wide range of parameters in Figure~\ref{FIGUREgiidiscrepancy}. Note that the ubiquitous $e^{-2g''}$ factor is, for small discrepancy, $e^{-2g''_{\rm opt}}\approx(1-2\Delta)e^{-2g''_{\rm approx}}$. It can be seen that for occupations $\gtrsim \order{10}$ and/or for times shorter than, or of the order of, singly-occupied coherence time the approximation is very good.

The diffusion gauge choice (\ref{approxgiiopto}) will be used from here on.

\subsection{Relationship between target time and mode occupation}
\label{1NT}
  The expression (\ref{approxgiiopto}) was worked out under the conditions that the mean occupation 
of the mode is conserved. In coupled-mode simulations this is no longer the case, and could be adapted for by replacing 
$n_0$ by $\breve{n}(t)$, which explicitly assumes independence of trajectories, and a Markovian process. Although the mean over all trajectories $\langle \Omega(t)\breve{n}(t)\rangle$ would be a better estimator of the mean boson occupation, this would be in conflict with aim 5 in Section~\ref{CH7BothAims}, and might lead to biases due to complicated feedback mechanisms between trajectories. 
   
 Another assumption used to arrive at (\ref{approxgiiopto}) was that $g''$ would be constant in time, which is now no longer the case. This raises the question of how to include the target time in $g''_{\rm opt}$. Two ways that quickly come to mind 
is  either to calculate $g''_{\rm opt}(t)$  
\ENUM{
\item Always optimizing for a time $t_{\text{opt}}$ forward from the present $t$ (choosing then, explicitly, $g''_{\rm opt}(t_{\text{opt}})$), or:

\item Optimizing for only the remaining time to the absolute target time $t_{\text{opt}}\ge t$ (choosing, then, $g''_{\rm opt}(\text{max}[t_{\text{opt}}-t,0])$).
}

In combination, this gives rise to the four possible gauge forms (in the \eqref{approxgiiopto} approximation, using \eqref{a2def})
\begin{subequations}\label{giiforms}\begin{eqnarray}
g''_{\rm approx}(t) &=& \frac{1}{6}\log\left\{ 8|n_0|^2\chi t_{\text{opt}} + a_2(n_0,\gamma t_{\text{opt}})^{3/2}\right\},\label{giinoto}\\
g''_{\rm approx}(t) &=& \frac{1}{6}\log\left\{ 8|n_0|^2\chi t_{\rm rem} + a_2(n_0,\gamma t_{\rm rem})^{3/2}\right\},\label{giinot}\\
g''_{\rm approx}(t) &=& \frac{1}{6}\log\left\{ 8|\breve{n}(t)|^2\chi t_{\text{opt}} + a_2(\breve{n}(t),\gamma t_{\text{opt}})^{3/2}\right\},\label{giinto}\\
g''_{\rm approx} &=& \frac{1}{6}\log\left\{ 8|\breve{n}(t)|^2\chi t_{\rm rem} + a_2(\breve{n}(t),\gamma t_{\rm rem})^{3/2}\right\}(t),\label{giint} 
\end{eqnarray}\end{subequations}
where the ``remaining time to target'' is 
\EQN{\label{tremdef}
t_{\rm rem}=\text{max}[t_{\text{opt}}-t,0]
.}

These strategies have been numerically investigated for the undamped ($\gamma=0$) case with real $n_0$ starting conditions. It turns out that form (\ref{giint}) has some advantage over the others, particularly at high occupations. Details are discussed in Section~\ref{CH7NumericalProcedure} and shown in Table~\ref{TABLEsimtime}.

\subsection{Boundary term issues}
\label{CH7BothBoundaryterm}
When considering the use of an adaptive $g''(t,\breve{n}(t))$, one should be careful that the $g''$ dependent terms do not introduce new noise divergences  that were not present when the standard $g''=0$ square root noise matrix $B_0$ was used\footnote{Incidentally, any boundary term errors introduced by such noise divergences would have to be of the second kind, since the choice of diffusion gauge is made at the level of the FPE-stochastic equation correspondence, long after any  boundary terms $\op{\mc{B}}$ has been discarded.}.
The condition to avoid noise divergence symptoms (and so, presumably, boundary term errors) is as always \eqref{mvsingcondition}: The stochastic equations should contain no radial component that grows faster than exponentially as large radial values of the phase-space variables are reached.

    Since the suggested forms \eqref{giiforms} depend directly only on the complex occupation variable $\breve{n}=e^{n_L}$, and not on either of the other independent variables ($m_L$, and $z_0$), then it suffices to only check that the $\breve{n}$ evolution equation contains no radial super-exponential growth. This is because the $g''$ dependent terms in the $m_L$ and $z_0$ evolutions simply accumulate integrals of functions of $n_L$, and if $\breve{n}$ remains finite, then so will the other variables.

  The $\breve{n}$ evolution is now 
\begin{eqnarray}
d\breve{n}&=&2i\breve{n}\sqrt{i\chi}e^{-g''}d\eta +\gamma(\bar{n}_{\rm bath}-\breve{n})\,dt \nonumber\\
&&+(\varepsilon\,dt+\sqrt{\gamma\bar{n}_{\rm bath}}\,d\eta_{\rm bath})\beta 
+(\varepsilon\,dt+\sqrt{\gamma\bar{n}_{\rm bath}}\,d\eta_{\rm bath})^*\alpha .\label{ahodn}
\end{eqnarray}
 Provided the time dependence of parameters $\gamma,\bar{n}_{\rm bath},\varepsilon$, or $\chi$ is not pathological, the condition \eqref{mvsingcondition} means $g''$ must obey 
\begin{equation}\label{1btcond}
 \lim_{|\breve{n}(t)|\to\infty} e^{-g''} \propto |n(t)|^a \text{ where $a\le 0$}.
\end{equation}
for no super-exponential growth to occur. That is, $e^{g''}$ must grow as a non-negative power law (or faster) as $|\breve{n}|$ becomes large. This is seen to be satisfied by the suggested forms \eqref{giiforms}. 
Finally, non $g''$-dependent terms were considered in Section~\ref{CH7Drift}, and found not to lead to any moving singularities. 

It is concluded, then,  that the diffusion gauge form (\ref{approxgiiopto}) does not lead to any new noise divergences or moving singularities, and hence none of the usual boundary term error symptoms have been reintroduced.

\subsection{Particle gain}
\label{CH7BothGain}

External particle gain (rather than loss due to a zero-temperature heat bath) complicates the behaviour. Looking at \eqref{ahodn}, the bath interactions tend to equilibrate the mode occupation to the bath value $\bar{n}_{\rm bath}$. 
If $\bar{n}_{\rm bath}\ll\langle\op{n}\rangle$, then one can expect that the optimum gauge calculated with nonzero $\gamma$ will be largely unchanged. 
For highly occupied bath modes, or strong coherent particle gain $\varepsilon$, a new analysis of $g''$ optimization may give better results that \eqref{giiforms}, however this appears to be a very involved procedure. 
An  obvious first try is to just see what happens with the gauge form \eqref{giiforms}.

\section{Estimates of simulation times}
\label{CH7Times}

The limit \eqref{sdlimit} on the variance of exponentials of Gaussian random variables can be used to estimate 
times of useful simulation $t_{\rm sim}$ from expressions for $\vari{G_L}$ --- since $G_L$ behaves much like a Gaussian random variable in analogy to $\sigma\xi$, and observables are estimated by $e^{G_L}$ in analogy to $v_{\sigma}\propto e^{\sigma\xi}$. 
Imposing 
\EQN{\label{timeeqn}
(\vari{G_L(t_{\rm sim})}+\text{var}[\wt{G}_L(t_{\rm sim})])/2\approx 10
,}
 one can solve for $t_{\rm sim}$ to obtain at estimate at least of the scaling of $t_{\rm sim}$ with system parameters.

Consider first the \textbf{gauged} simulation. For
 coherent state initial conditions $\beta_0=\alpha_0^*$ at
 small damping $\gamma t\ll1$,
 and short enough times $|q|t\ll 1$, 
\EQN{\label{varglboth}
\vari{G_L}=\frac{\chi t}{2}\left(V_g+\frac{1}{V_g}\right) +2(\chi V_g t n'_0)^2
.}
For large particle number $n'_0\gg1$, $g''_{\rm opt}$ takes the form (\ref{approxgiiopt1}), so 
$V_g\approx 1/2(n'{}^2_0\chi t)^{1/3}$, the $\chi t V_g/2$ term is negligible, and 
 $\vari{G_L}\approx 3(n'_0\chi^2 t^2)^{2/3}/2$, and \eqref{timeeqn} leads to a useful simulation time 
\begin{equation}\label{bignsimtime}
t_{\text{sim}} \approx \frac{(20/3)^{3/4}}{\chi\sqrt{n'_0}} \approx \order{10\,t_{\rm coh}}
.\end{equation}
Since in this regime $|q|t\approx 4\chi t V_g \approx 2(\chi t/n'_0)^{2/3}$, then $|q|t_{\text{sim}}\approx \order{5}/n'_0\ll1$, and so the expression (\ref{bignsimtime}) is consistent with the original short time assumption $|q|t\ll1$ when $n'_0\gg\order{5}$.

At small occupations $n'_0\ll 1$ on the other hand, $g''_{\rm opt}\to 0$, and at long times, $\vari{G_L} \approx \frac{1}{4}n'_0{}^2e^{4\chi t}$. Thus 
useful simulation time scales very slowly with $n'_0$, and is 
\begin{equation}\label{smallnsimtime}
t_{\text{sim}}\approx \frac{\order{1}-\frac{1}{2}\log n'_0}{\chi}. 
\end{equation}
The ``long time'' condition holds while $\frac{1}{4}n'_0{}^2e^{4\chi t}$ is much greater than the lower order term $\chi t$. That is, 
$n'_0\gg \order{10^{-8}}$. At even smaller $n'_0$, the $\chi t$ term dominates \eqref{varglboth}, and 
\EQN{\label{tinynsimtime}
t_{\rm sim}\approx \order{10}/\chi
.}

For comparison, consider the \textbf{positive P} behaviour, which can be estimated from the logarithmic variances \eqref{varlgnopt} for the case with no drift gauge, which are worked out in Section~\ref{CH7DiffusionOptimization}. For coherent state initial conditions $n_0=n'_0$ and small damping $\gamma t\ll 1$, 
\EQN{
\vari{G_L}\approx \chi t +2n'_0(\chi t)^2 +2\chi^2n'_0{}^2\left\{ \frac{e^{4\chi t}-1}{8\chi^2}-\frac{t}{2\chi}-t^2\right\}
.}
At short times $4\chi t\ll 1$ the leading term in the $\{\bullet\}$ expression is $\frac{8}{3}(\chi t)^3n'_0{}^2$. In such a case, for large mode occupation $n'_0\gg\order{1}/\chi t$ this term dominates, and using \eqref{timeeqn}, one obtains 
\EQN{\label{tsimpp}
t_{\rm sim} \approx \frac{\order{1}}{\chi n'_0{}^{2/3}}
.}
Checking back, $1/\chi t\approx \order{1}n'_0{}^{2/3}$, so \eqref{tsimpp} is consistent with the short time assumption for $n'_0\gg 1$.
At long times, one again has \eqref{smallnsimtime} and \eqref{tinynsimtime}.
The expression \eqref{tsimpp} is also useful in estimating simulation times for many-mode models, as will be seen in Section~\ref{CH10LatticePp}.

Curiously, the ``stable'' drift-gauged simulation with \textbf{no diffusion gauge} does even worse than the positive P. 
In this case, $V_g=1$, and using \eqref{varglboth}, at large $n'_0$
\EQN{
t_{\rm sim} \approx \frac{\order{2}}{\chi n'_0} \approx t_{\rm osc}
,}
which is consistent when $n'_0\gg\order{\frac{1}{4}}$.

The numerical results in Table~\ref{TABLEsimtime} and Figure~\ref{FIGUREsimtime}, are seen to agree with these estimates to within constant factors of $\order{1}$.

\section{Numerical investigation of improvement}
\label{CH7Numerical}
  To unambiguously determine the improvements in simulation time that are achieved by the use of the proposed gauges, simulations of an undamped anharmonic oscillator were carried out for a wide range of mode occupations and a variety of gauges, testing various target times $t_{\rm opt}$ where appropriate. 
(The undamped system was chosen because this is the worst case.)

Figure~\ref{FIGUREsimtime} compares $t_{\rm sim}$, the maximum time achieved with various methods at which useful precision in the phase correlations $G^{(1)}(0,t)$ can be obtained. Note the logarithmic scale. Results at high boson occupation are tabulated in Table~\ref{TABLEsimtime}, which includes data for a larger set of gauge choices.
Figure~\ref{FIGUREG} gives examples of calculated values $|G^{(1)}(0,t)|$ along with error estimates. Table~\ref{TABLEtimesfit} gives some empirical fitting parameters to expression \eqref{timefit} for the useful simulation time $t_{\rm sim}$. These may be useful to assess simulation times when the particle occupation is of $\order{10}$ or smaller, and the expressions \eqref{bignsimtime} or \eqref{tsimpp} are not accurate.

Details of the sampling error behaviour are shown in Figure~\ref{FIGUREvarGt} for a range of gauge choices, while 
Figure~\ref{FIGUREtopttprec} shows the dependence of useful simulation times on the target time parameter $t_{\text{opt}}$
for a variety of gauge forms for two example mode occupations: $n_0=1$ and $n_0=10^4$.

\subsection{Procedure}
\label{CH7NumericalProcedure}

Simulations were carried out with coherent state initial conditions for a wide range of 
mean occupation number 
\EQN{\label{ahon0}
n_0 = \{ 10^{-5}, 10^{-4}, 10^{-3}, 0.01, 0.1, 1, 10, 100, 1000, 10^4, 10^5, 10^6, 10^8, 10^{10}\}.
}
The following gauges, and their varieties as seen in Table~\ref{TABLEsimtime}, were tried for each $n_0$ value:

i) No gauge: $\mc{G}_k=g''=0$. This is the standard positive P distribution technique.

ii) Drift gauge only: $\mc{G}_k$ given by \eqref{ahogauge}, and $g''=0$.

iii) Both drift gauge (\ref{ahogauge}) and diffusion gauges  of the four related forms \eqref{giiforms}.

iv) Diffusion gauge (\ref{plimakgauge}) only, as described by Plimak\etal\cite{Plimak-01} (See Section~\ref{CH7ComparisonPlimak}). 

v) Drift gauge (\ref{CCDgauge}) only, as described by Carusotto\etal\cite{Carusotto-01} (See Section~\ref{CH7ComparisonCCD}) 

vi) Adaptive diffusion gauges \eqref{thenoptgauge} only (of the four forms \eqref{giiformsnopt}), and $\mc{G}_k=0$.

\begin{figure}[tp]
\center{\includegraphics[width=\textwidth]{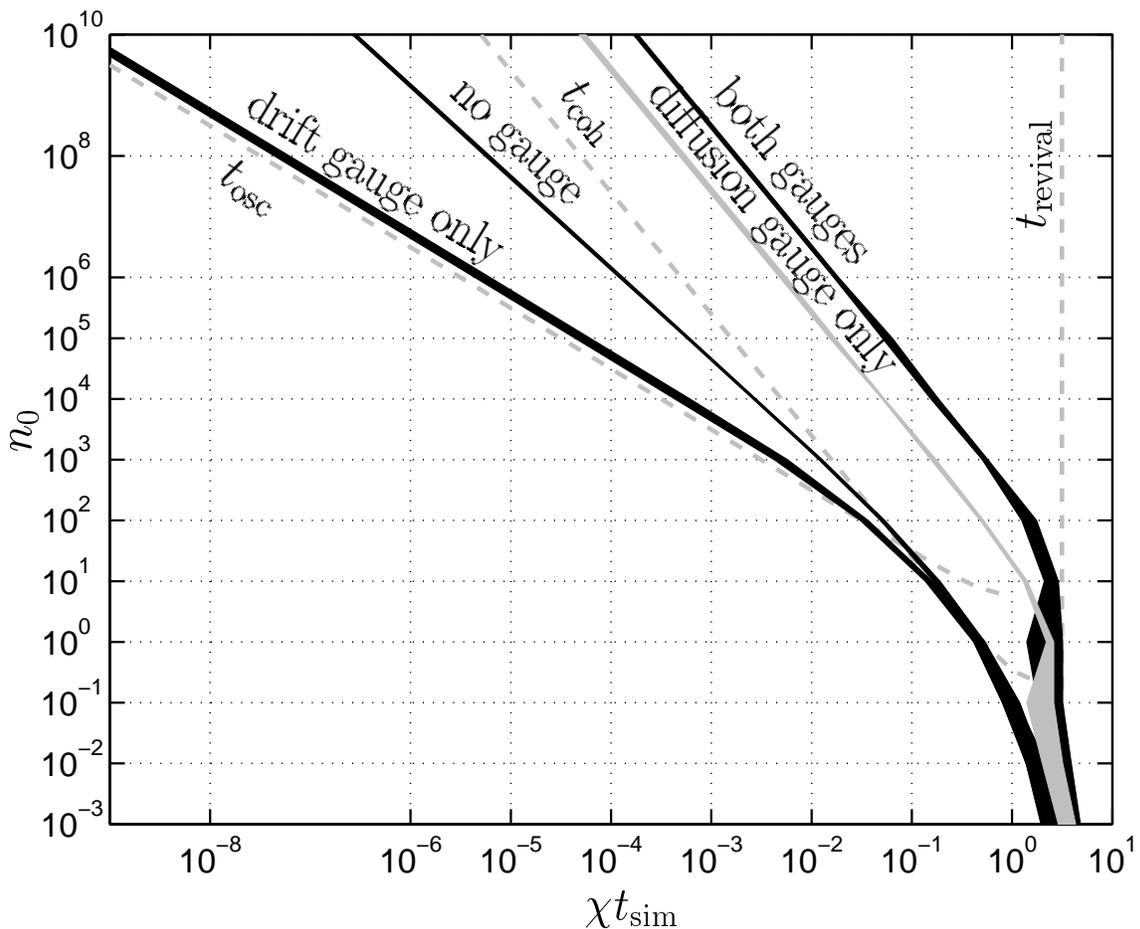}}\vspace{-8pt}\par
\caption[Maximum useful simulation time]{\label{FIGUREsimtime} \footnotesize
\textbf{Maximum useful simulation time} $t_{\rm sim}$,  of the one-mode undamped anharmonic oscillator with various gauge choices. Initial coherent state mean mode occupations $n_0=n'_0$. 
Width of plotted lines shows estimated uncertainty in the values shown.
The {\scshape drift gauge} is (\ref{ahogauge}), while the {\scshape diffusion gauge} is \eqref{plimakgauge} of Plimak\etal\cite{Plimak-01} when on its own, or \eqref{giint} when with the drift gauge \eqref{ahogauge}. 
For comparison, several timescales from Section~\ref{CH7ModelAnharmonic} are also shown as broken lines: time of first quantum revival $t_{\text{revival}}$, phase coherence time $t_{\text{coh}}$ and phase oscillation period $t_{\text{osc}}$. 
\normalsize}
\end{figure}

\begin{figure}[tp]
\center{\includegraphics[width=\textwidth]{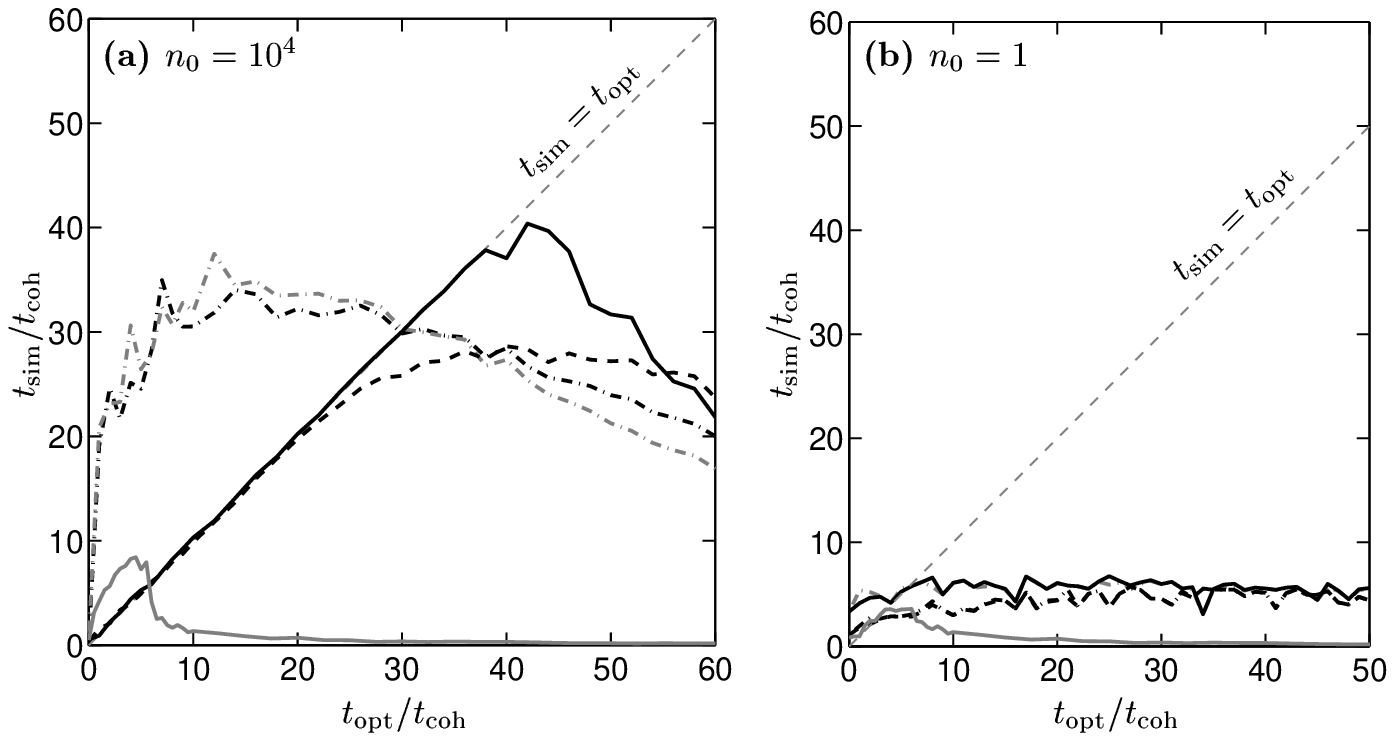}}\vspace{-8pt}\par
\caption[Comparison of target time and actual simulation time]{\label{FIGUREtopttprec}\footnotesize
\textbf{Comparison of {\it a priori} target time $t_{\rm opt}$  with actual useful simulation time $t_{\rm sim}$} for a variety of diffusion gauges: 
Results for drift gauges \eqref{ahogauge} with the four $g''$ forms \eqref{giiforms} are shown as: (\ref{giint}) -- {\scshape solid dark}; (\ref{giinto}) -- {\scshape dash-dotted light}; (\ref{giinot}) -- {\scshape dashed}; (\ref{giinoto}) -- {\scshape dash-dotted dark}.
Relationship obtained using the diffusion gauge (\ref{plimakgauge}) of Plimak\etal\cite{Plimak-01} (but no drift gauge)
 is also shown ({\scshape solid light}). The dashed line in the background shows, for reference $t_{\text{opt}}=t_{\text{sim}}$.
Subplot {\bf(a)}: mean particle number $n_0=10^4$, {\bf (b)}: $n_0=1$. 
\normalsize}
\end{figure}

\begin{figure}[tp]
\center{\includegraphics[width=\textwidth]{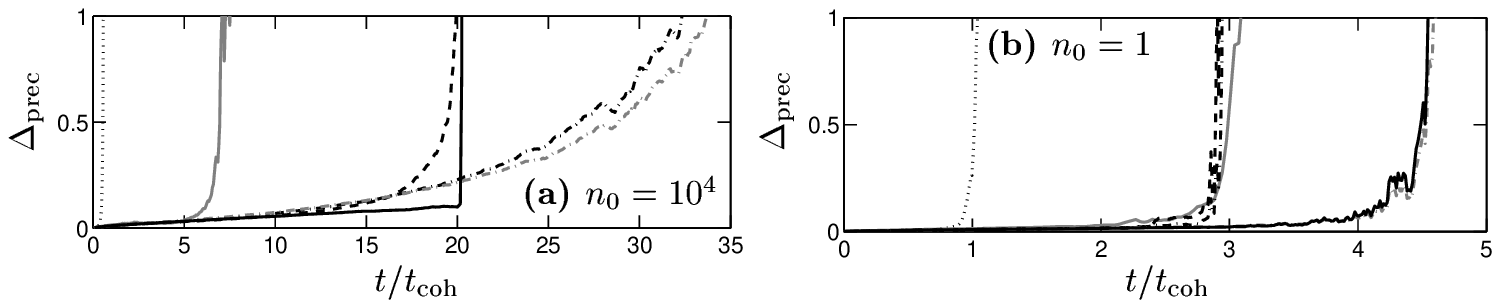}}\vspace{-8pt}\par
\caption[Uncertainty in $G^{(1)}(0,t)$ for various gauges]{\label{FIGUREvarGt} \footnotesize
\textbf{Uncertainty in $|G^{(1)}(0,t)|$}, as a function of time. The quantity plotted is $\Delta_{\text{prec}} = 
\sqrt{\mc{S}/10^6}(10\ \Delta|G^{(1)}(0,t)|\,/\,|G^{(1)}(0,0)|)$, so that $\Delta_{\text{prec}}\le1$ corresponds to 
useful precision as defined in Section~\ref{CH7NumericalProcedure}. Results are plotted for combined drift (\ref{ahogauge}) and diffusion (\ref{giiforms}) gauges, where the four forms are shown as : (\ref{giint}) -- {\scshape solid dark}; (\ref{giinto}) -- {\scshape dash-dotted light}; (\ref{giinot}) -- {\scshape dashed}; (\ref{giinoto}) -- {\scshape dash-dotted dark}. Data for the diffusion gauge (\ref{plimakgauge}) of Plimak\etal\cite{Plimak-01}  with $t_{\rm opt}=3t_{\text{coh}}$ as used therein is shown as a {\scshape light solid} line, while the ungauged positive P calculation is shown as a {\scshape dotted} line. Simulations were carried out with $\mc{S}=10^4$ trajectories, starting with initial coherent state conditions: \textbf{(a)} $n_0=10^4$ \textbf{(b)} $n_0=1$. The combined drift-and-diffusion-gauge plots were calculated with target times of: \textbf{(a)} $t_{\text{opt}}=20t_{\text{coh}}$ \textbf{(b)} $t_{\text{opt}}=4t_{\text{coh}}$.
\normalsize}
\end{figure}

\begin{figure}[t]
\center{\includegraphics[width=\textwidth]{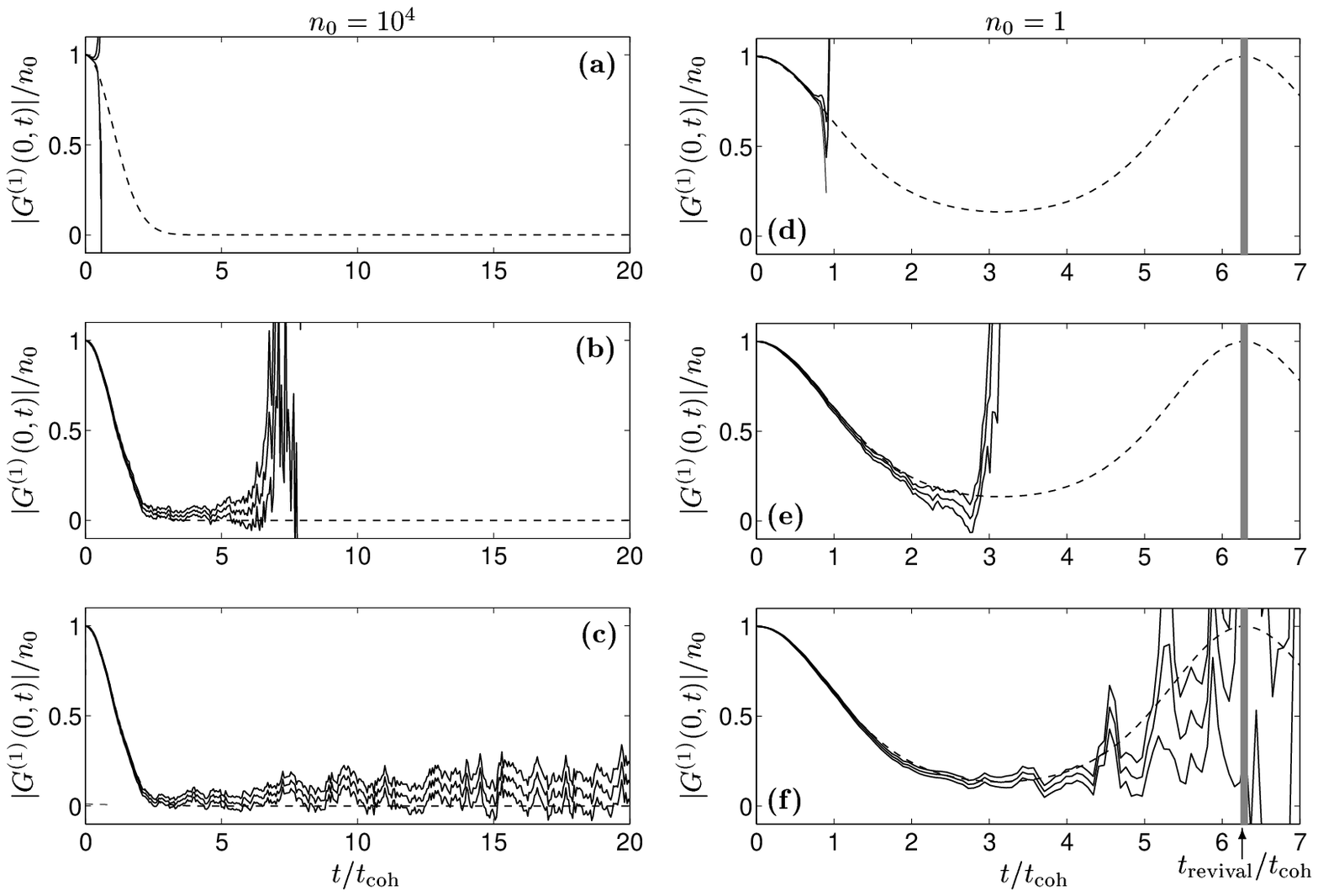}}\vspace{-8pt}\par
\caption[Phase correlations using various gauges]{\label{FIGUREG} \footnotesize
\textbf{Modulus of the phase correlation function $G^{(1)}(0,t)$}. Comparison of calculations with different gauges:
Subplots {\bf (a)} and {\bf(d)}: Ungauged positive P.
Subplots {\bf (b)} and {\bf(e)}: Diffusion gauge \eqref{plimakgauge} of Plimak\etal\cite{Plimak-01} with $t_{\rm opt}=3t_{\rm coh}$.
Subplots {\bf (c)} and {\bf(f)}: Combined drift and diffusion gauges (\ref{ahogauge}) and (\ref{giint}) with the choice $t_{\text{opt}}=20\,t_{\text{coh}}$ 
The initial conditions were a coherent state with $\langle\op{n}\rangle=n_0$, where in subplots {\bf (a)-(c)}: $n_0=10^4$ particles, and {\bf (d)-(f)}: $n_0=1$.  
Triple {\scshape Solid} lines indicate $G^{(1)}$ estimate with error bars. Exact values are also shown (single {\scshape dashed line}). The quantum revival time is shown {\scshape shaded} for the $n_0=1$ plots. 
$\mc{S}=10^4$ trajectories in all cases
\normalsize}
\end{figure}

\noindent For the diffusion gauges of iii), iv), and vi), which depend on an {\it a priori} target time parameter $t_{\rm opt}$, a wide variety of target times were tried to investigate the dependence between $t_{\rm sim}$ and $t_{\rm opt}$, and ascertain what are the longest simulation times achievable. The dependence of $t_{\rm sim}$ on $t_{\rm opt}$ is plotted in Figure~\ref{FIGUREtopttprec}, while the best times are tabulated in Tables~\ref{TABLEsimtime},~\ref{TABLEtimesfit}, and Figure~\ref{FIGUREsimtime}.

The term \textit{useful} precision has been taken to indicate a situation where the estimate $\bar{O}$ of the expectation value of some observable $\op{O}$ is known to be precise to at least one significant digit 
when $\mc{S}=10^6$ trajectories are used. That is, $\Delta\bar{O}(\mc{S}=10^6)\le|\bar{O}|/10$. 
Since for many-mode systems,  the calculation of even one trajectory is reasonably time consuming, it is clear that $\mc{S}=10^6$ trajectories is very unlikely to be exceeded for most non-trivial problems. Hence, meaningful results for non-trivial problems are at best likely to be obtained only in the parameter region where the above-defined ``useful'' precision condition is satisfied. If one calculates some lesser (than $10^6$) number of trajectories, yet $\mc{S}\gg1$, then the Central Limit theorem can be used to extrapolate $\Delta\bar{O}$ to the situation when $\mc{S}=10^6$, since $\Delta\bar{O}\propto1/\sqrt{\mc{S}}$. This gives finally, that the precision is taken to be ``useful'' when 
\EQN{\label{usefulprecision}
  \Delta\bar{O}(\mc{S})\sqrt{\frac{\mc{S}}{10^6}} \le \frac{1}{10} |\bar{O}|
.}

In the current simulations, the observable in question is $|G^{(1)}(0,t)|$, and 
the ``simulation time'' $t_{\rm sim}$ that will be referred to here is the maximum time at which  useful precision in $|G^{(1)}(0,t)/G^{(1)}(0,0)|$ is retained. 
 Each actual simulation was done with $\mc{S}=10^4$ trajectories.
  
Uncertainties in the calculated useful simulated  times arise because the $\Delta|G^{(1)}|$ are themselves estimated from the finite ensemble of $\mc{S}$ trajectories. The uncertainty in $\Delta|G^{(1)}|$ was estimated by inspection of several (usually 10) independent runs with identical parameters. The range of values of $t_{\rm sim}$ seen was taken to be twice the uncertainty in $t_{\rm sim}$.

The scalings of simulation time with particle number for $n'_0\gg1$ and $n'_0\ll 1$ have been worked out in Section~\ref{CH7Times}. Taking these results into account, Table~\ref{TABLEtimesfit} empirically fits the simulation time $t_{\text{sim}}$ to 
\begin{equation}\label{timefit}
  t_{\text{est}} = \frac{1}{\chi}\left\{ \left[ {c_1} n'_0{}^{-{c_0}} \right]^{-{c_2}} + \left[ \log\left(\frac{e^{c_3}}{n'_0{}^{c_4}}+1\right)\right]^{-{c_2}}\right\}^{-1/{c_2}}
\end{equation}
for intermediate times, as this may be useful for evaluation of simulation strategies in a many-mode case.
$c_0$ characterizes the power-law scaling of $t_{\rm sim}$ at high occupation, $c_1$ the pre-factor for high $n'_0$, $c_3$ a constant residual $t_{\rm sim}$ at near vacuum, $c_4$ characterizes the curvature at small $n'_0$, while $c_2$ is related to the stiffness of the transition.
 The expression 
(\ref{timefit}) reduces to ${c_1}n'_0{}^{-{c_0}}/\chi$ and $({c_3}-c_4\log n'_0)/\chi$, when $n'_0\gg1$ and $n'_0\ll1$, respectively, in agreement with the limiting expressions (\ref{bignsimtime}) and (\ref{smallnsimtime}). 
Uncertainty $\Delta c_j$ in parameters $c_j$ was worked out by requiring $\sum_{n_0}\{[t_{\text{est}}(c_j\pm\Delta c_j,n_0)-t_{\text{sim}}(n_0)]/\Delta t_{\text{sim}}\}^2 = \sum_{n_0}\{1+([t_{\text{est}}(c_j,n_0)-t_{\text{sim}}(n_0)]/\Delta t_{\text{sim}})^2\}$. In the range checked ($n_0\in[10^{-5},10^{10}]$), the fit is good --- i.e. there are no 
outlier data that would lie significantly beyond the range of $t_{\rm est}$ specified by parameters $c_j\pm\Delta c_j$.

\begin{table}[tp]
\caption[Maximum useful simulation time]{\label{TABLEsimtime}\footnotesize
\textbf{Maximum simulation time}, at useful precision, of the one-mode system in the limit $(n_0=n'_0)\gg1$, achievable with various gauge choices. 
Calculations are for the undamped ($\gamma=0$) system, which is the worst case in terms of sampling error.
 More details are  given in Section~\ref{CH7NumericalProcedure}.
For comparison, several timescales from Section~\ref{CH7ModelAnharmonic} are also given: time of first quantum revival $t_{\text{revival}}$, phase coherence time $t_{\text{coh}}$ and phase oscillation period $t_{\text{osc}}$. 
\normalsize}\vspace*{3pt}
\hspace{-24pt}
\begin{minipage}{\textwidth}
\hfilll\begin{tabular}{|c|c|r@{\,}l|c|}\hline
Drift gauge $\mc{G}_k$& Diffusion gauge $g''$&\multicolumn{2}{c|}{Useful simulation time}&Maximum $n_0$\\
&&\multicolumn{2}{c|}{$t_{\rm sim}$ when $n_0=n'_0\gg1$}&for which $\chi t_{\rm sim}\ge1$\\
\hline\hline
\multicolumn{2}{|r@{\ =\ }}{$t_{\rm osc}$}&\multicolumn{3}{l|}{$\pi/\chi n'_0$}\\
\hline
\eqref{CCDgauge}	& 0			& $(1.06\pm0.16)$ 		& $t_{\text{osc}}$		
												&$0.014\ \mm{+\ 0.016}{-\ 0.008}$\\
\eqref{ahogauge}	& 0			& $(1.7\pm0.4)$ 		& $t_{\text{osc}}$	
												&$0.08\ \mm{+\ 0.07}{-\ 0.05}$\\
\bf 0			& \bf 0			&$\mathbf{(1.27\pm0.08)}$	&$/\chi n'_0{}^{2/3}$	
												&$0.11\pm 0.06$\\
\hline
\multicolumn{2}{|r@{\ =\ }}{$t_{\rm coh}$}&\multicolumn{3}{l|}{$1/2\chi \sqrt{n'_0}$}\\
\hline
0			&\eqref{plimakgauge} 
	or \eqref{giinopt} or \eqref{giinnopt}	& $(8.2\pm0.4)$			& $t_{\text{coh}}$ 	&$12\pm3$\\
0			&\eqref{giitnopt}
			   or \eqref{giintnopt}	& $(10.4\pm0.7)$		& $t_{\text{coh}}$ 	&$19\pm4$\\
\eqref{ahogauge} 	&\eqref{giinot}		& $(25.6\pm1.0)$		& $t_{\text{coh}}$ 	&$120\pm30$\\
\eqref{ahogauge} 	&\eqref{giinoto}	& $(30\pm3)$			& $t_{\text{coh}}$ 	&$150\pm40$\\
\eqref{ahogauge} 	&\eqref{giinto}		& $(32\pm3)$			& $t_{\text{coh}}$ 	&$190\pm15$\\
\textbf{\eqref{ahogauge}}
			&\textbf{\eqref{giint}}	& $\mathbf{(35\pm4)}$	& $t_{\text{coh}}$ 		&$240\pm70$\\
\hline
\multicolumn{2}{|r@{\ =\ }}{$t_{\rm revival}$}&\multicolumn{3}{l|}{$\pi/\chi$}\\
\hline
\end{tabular}\hfilll
\end{minipage}
\end{table}

\begin{table}[tp]
\caption[Empirical fitting parameters for simulation time]{ \label{TABLEtimesfit}\footnotesize
\textbf{Empirical fitting parameters} for maximum useful simulation time $t_{\text{sim}}$ with several gauge choices. The fit is to expression (\ref{timefit}).
\normalsize}\vspace*{3pt}
\begin{minipage}{\textwidth}
\hfilll\begin{tabular}{|l|r@{$\ \pm\ $}l|r@{$\ \pm\ $}l|r@{$\ \pm\ $}l|r@{$\ \pm\ $}l|}\hline
	& \multicolumn{2}{l|}{Positive P}&\multicolumn{2}{l|}{Both gauges} &\multicolumn{2}{l|}{Drift gauge}&\multicolumn{2}{l|}{Diffusion gauge} \\
\hline
Drift Gauge 	& \multicolumn{2}{c|}{0} 	& \multicolumn{2}{c|}{(\ref{ahogauge})}
	& \multicolumn{2}{c|}{(\ref{ahogauge})} 	&\multicolumn{2}{c|}{0}\\
Diffusion Gauge	& \multicolumn{2}{c|}{0}	& \multicolumn{2}{c|}{(\ref{giint})} 
	&\multicolumn{2}{c|}{0} 		&\multicolumn{2}{c|}{\eqref{giintnopt}}\\
${c_0}$		& \multicolumn{2}{c|}{$2/3$}	&\multicolumn{2}{c|}{$1/2$}	
	&\multicolumn{2}{c|}{$1$}		&\multicolumn{2}{c|}{$1/2$}\\
\hline\hline
${c_1}$	& $1.27$	& $0.08$	& $17.6$	& $1.7$		& $5.5$		&$1.2$ 				
												&\hspace{3ex}$5.2$
													&$0.4$	\\
${c_2}$	& $3.2$		&$\mm{\infty}{1.2}$
					& $3.6$	& $\mm{\infty}{2.3}$	& $1.4$		& $\mm{\infty}{0.4}$
												 &\hspace{3ex}$2.7$ 
													&$\mm{\infty}{1.0}$\\
${c_3}$	& $-0.5$	& $0.3$		& $2.8$		& $0.9$		& $-0.5$	& $0.3$ 
												&\hspace{3ex}$-2.4$	
													&$0.6$	\\
${c_4}$	& $0.45$	& $0.07$	& $0.23$	& $0.13$	& $0.49$	& $0.08$		
												&\hspace{3ex}$0.23$	
													&$0.13$	\\
\hline
\end{tabular}\hfilll
\end{minipage}
\end{table}

\subsection{Features seen}
\label{CH7NumericalFeatures}

\begin{itemize}
\item Combining drift and diffusion gauges gives the longest useful simulation times. 
Such simulations give good precision well beyond the point at which all coherence has decayed away for highly-occupied modes - potentially up to about 35 coherence times. 
\item Diffusion-gauge-only simulations also give quite good statistical behaviour (although useful simulation times are about 4 times shorter at high occupation than with both gauges). 
\item Despite the efficient behaviour of combined gauge simulations, using only a drift-gauge gives even worse statistical error than no gauge at all (although, this does eliminate potential boundary term systematics in a multi-mode system, which are a problem with no gauge). Such simulations are restricted in time to about one phase oscillation.
\item Plain positive P simulations at high occupation last for about $n'_0{}^{1/3}$ phase oscillation periods, which is much less that the time required for significant coherence to be lost if $n_0$ is large. 
\item The diffusion gauge forms (\ref{giinot}) and (\ref{giint}) that optimize for the ``remaining time to target''  $(t_{\text{opt}}-t)$ show 
statistical errors that are well controlled by the choice of the target time $t_{\text{opt}}$. Basically statistical error can be reliably expected to remain small up to the explicit target time $t_{\text{opt}}$, provided that this is within the useful simulation range given in Table~\ref{TABLEsimtime} --- see Figure~\ref{FIGUREvarGt}. Thus, a heuristic approach to get the maximum time out of a simulation would be to
\begin{enumerate}
\item Pick $t_{\rm opt}$ to be some time that one wants to be able to simulate to.
\item Run a simulation, and see how long useful observable estimates occur: $t_{\rm sim}(t_{\rm opt})$. 
\item Choose:
\begin{enumerate}
\item If $t_{\rm sim}(t_{\rm opt}) < t_{\rm opt}$ then reduce $t_{\rm opt}$ to some value between $t_{\rm sim}(t_{\rm opt})$ and the present $t_{\rm opt}$. Using this new value should give a new better simulation time also in that range. Iterate back to step 2. Or,
\item if $t_{\rm sim}(t_{\rm opt}) \approx t_{\rm opt}$ then either keep the simulation if one is happy with it, or one can try to increase $t_{\rm opt}$ to perhaps obtain more simulation time, going back to step 2.
\end{enumerate}
\end{enumerate}
The rest of the diffusion gauge forms \eqref{giiforms} do not follow such a simple dependence and appear to require tedious parameter searching to find the best gauge parameter choice for given initial conditions.
\item Simulations showing quantum revivals have not been achieved. Generally it appears that useful simulation times are proportional to coherence times at high occupations. In this context, however, it is worth mentioning that quantum revivals {\it have} been seen in this system by Dowling\cite{Dowling03} using the gauge P representation, by imposing externally-induced time reversal in the mode at a time $t_{\rm reversal}$, where  $t_{\rm coh} < t_{\rm reversal} \ll t_{\rm revival}$. The phase coherence was seen by Dowling\cite{Dowling03} to decay and then revive at $t=2t_{\rm reversal}$. This indicates that the potential for quantum revivals in these stochastic simulations is there, just not realized in the free evolution analyzed in this chapter.
\item At low occupation, i.e. of the order of one boson or less, combined-gauge methods still give the best results, but the magnitude of their advantage becomes smaller. 
\item Of the four gauge forms (\ref{giiforms}), the form (\ref{giint}) gives the longest useful simulations.
\item The diffusion gauge forms (\ref{giint}) and (\ref{giinto}) optimizing depending on the occupation of the current trajectory do better than those optimizing on the basis of only the initial condition. This is not surprising, since this approach is simply better tailored to the predicted subsequent evolution of each trajectory. 
\item At low occupations, a broad range of $t_{\text{opt}}$ choices give much the same statistical behaviour. This is most likely because the forms (\ref{giiforms}) are not always the best optimizations of (\ref{lgvars}) in this regime for two reasons: 1) Because the ``small time'' condition $|q|t\ll 1$ is not always satisfied at the long time end of the simulation, and 2)  because as seen in Figure~\ref{FIGUREgiidiscrepancy}, the form (\ref{approxgiiopto}) is not an accurate root of the cubic (\ref{vacubic}).
\item The simulation times with diffusion gauges (whether accompanied by drift gauge \eqref{ahogauge} or not) 
not only have better scaling with $n_0$ when $n_0$ is large, but this power-law decay of simulation time starts much later, as seen in Figure~\ref{FIGUREsimtime} and the right hand column of Table~\ref{TABLEsimtime}.
\end{itemize}

\section{Diffusion gauges on their own}
\label{CH7Diffusion}
\subsection{Motivation}
\label{CH7DiffusionMotivation}
  The previous Sections~\ref{CH7Both} to~\ref{CH7Numerical} considered optimization of the diffusion gauge given that 
instabilities in the equations have been removed using the drift gauge \eqref{ahogauge}. An approach that could be an alternative is to only use a diffusion gauge $g''$ to make a tradeoff between noise in the number and phase variables as before, but without introducing any drift gauge. 

Such an approach could have the advantage that no weight spread is introduced since $dz_0=0$. The weight spread is not a major issue in a single-mode system, and clearly from Table~\ref{TABLEsimtime} the drift-gauged simulations last longer, but 
with many modes there may be a problem: {\it All} the gauges introduce modifications to the {\it same} single weight variable.
  A vivid example occurs if we have $M$ identical but independent modes. Then, $dz_0=\sum_{k=1}^{2M}\mc{G}_kdW_k$, and 
the standard deviation of $dz_0$ is $\sqrt{M}$ times greater than for a single mode but with no new physics.  Since the log-weight $z_0$ enters into the observable estimates in exponential fashion, the situation is again analogous to the average of an exponential of a Gaussian random variable, and there is the non-scalable useful simulation limit $\vari{z_0}\lesssim\order{10}$ by condition\eqref{sdlimit}. Since $\vari{z_0(t)} \approx\, \propto Mt$ at short times, this translates to 
\EQN{
t_{\rm sim}\propto \frac{1}{M}
}
in a worst case. 

The major disadvantage of the $\mc{G}=0$ approach, is that boundary term errors cannot be ruled out, as in the plain positive P method. One can, however, try to monitor for their appearance using the indicators developed by Gilchrist\etal\cite{Gilchrist-97}:
Spiking in moment calculations, rapid onset of statistical error after some time, or power-law distribution tails.
A caveat is that these indicators tend to emerge only for fairly large ensemble sizes $\mc{S}$ (Typically  what happens is that the indicators appear with a certain delay time if $\mc{S}$ small).

This $\mc{G}=0$ method variant, although less desirable due to the need for careful monitoring of the simulation, was found to be much more successful than the drift-gauged method (in its present form) at increasing simulation times for the preliminary example many-mode models simulated in Chapter~\ref{CH10}.

\subsection{Optimization of $g''$ and comments}
\label{CH7DiffusionOptimization}
Proceeding as before in Section~\ref{CH7BothOptimization}, the formal solution of the anharmonic oscillator equations 
with $\Omega(0)=1$ are as \eqref{formalsoln}, but with the changes
\SEQN{\label{formalnoptsoln}}{
m_L &=& \text{\eqref{formalsolnml}} + 4\chi\int_0^tn''(t')\,dt\\
z_0(t) &=& 0
.}
Following the same procedure as before, one obtains
\EQN{\label{varlgnopt}
\matri{\vari{G_L}\\\text{var}[\wt{G}_L]} &=&
\chi t \cosh(2g'') -4\chi^2\left(n'_0\matri{-\\+}n''_0e^{-2g''}\right)\left\{\frac{1-e^{-\gamma t}(1+\gamma t)}{\gamma^2}\right\}\nonumber\\
&& +4\chi^2|n_0|^2\left\{\frac{e^{-qt}-1}{q(q-\gamma)}-\frac{1-e^{-\gamma t}}{\gamma(\gamma-q)}-\Half\left(\frac{1-e^{-\gamma t}}{\gamma}\right)^2\right\}
,}
where the ``damping strength'' $q$ is given by \eqref{qparamdef}. The variational condition \eqref{mincond} can now be used to optimize $g''$.

Considering the same special cases as in Section~\ref{CH7BothCases}, 
\ENUM{
\item With damping absent, and at relatively short times $|q|t\ll1$, the optimized gauge is
\EQN{
g''_{\rm opt} \approx g_{\rm approx} = \frac{1}{4}\log\left[\frac{1}{3}(4\chi t_{\rm opt}|n_0|)^2+1\right]
.}
(no cubic polynomial this time.) 
\item The long time behaviour when $|q|t\gg1$: When $q>0$, the increase in the variances of $G_L$ and $\wt{G}_L$ is still linear just like in \eqref{longtimeqt}, but with the constant $b=4\chi^2|n_0|^2(q/2\gamma-1)/q \gamma-4\chi^2[n'_0\mp n''_0e^{-2g''}]/\gamma^2$. When $q<0$ the long time behaviour is again exponential: $G_L\approx4\chi^2|n_0|^2e^{|q|t}(q/2\gamma-1)/\gamma q$.
\item For nonzero damping in the $|qt_{\rm opt}|\ll1$ regime, the optimized diffusion gauge is 
\begin{subequations}\label{thenoptgauge}\EQN{
  g''_{\rm opt} \approx g_{\rm approx}= \frac{1}{4}\log\left[\frac{(4\chi t_{\rm opt}|n_0|)^2}{3}c_2(\gamma t_{\rm opt})+1\right]
,}
where the coefficient is 
\EQN{
  c_2(v) = \frac{3}{2}\left(\frac{e^{-2v}(3+2v)+1-4e^{-v}}{v^3}\right)
,}\end{subequations}
which reduces to $c_2(0)=1$ in the undamped case.
}


\begin{figure}[t]
\center{\includegraphics[width=10cm]{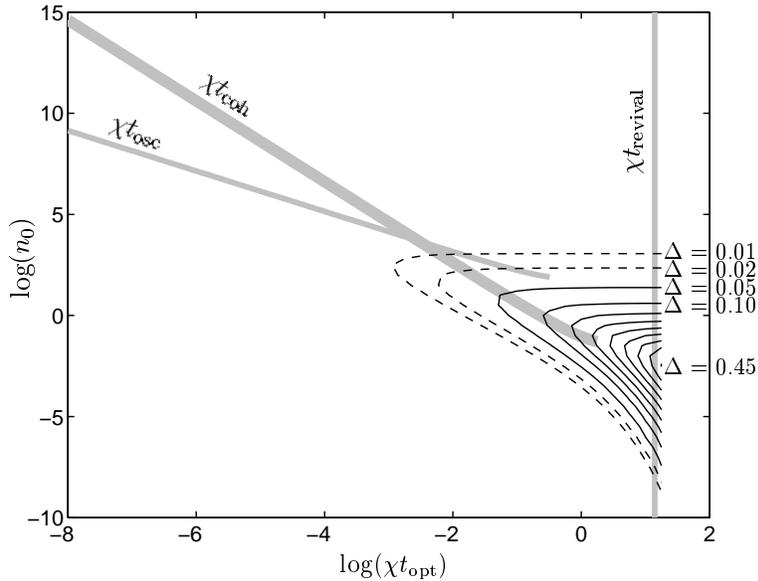}}\vspace{-8pt}\par
\caption[Discrepancy between $g''_{\rm opt}$ and its approximation for $\mc{G}=0$ scheme]{\label{FIGUREgiidiscrepancynopt} \footnotesize
\textbf{Discrepancy} $\Delta=g''_{\text{opt}}-g''_{\text{approx}}$ between $g''_{\rm opt}$ (the exact optimization of $g''$ obtained by solving \eqref{mincond} using \eqref{varlgnopt}) and the approximate expression (\ref{thenoptgauge}). Displayed is the case of no damping ($\gamma=0$) and classical initial occupation ($n_0=n'_0$), shown as a function of $t_{\text{opt}}$ and $n_0$. Discrepancy values $\Delta$  are shown as {\scshape solid contours} with spacing $0.05$. Additional {\scshape dashed contours} shown at very low discrepancy. 
For comparison, several physical timescales are also shown in {\scshape grey}: time of first quantum revival $t_{\text{revival}}$, phase coherence time $t_{\text{coh}}$ and phase oscillation period $t_{\text{osc}}$. 
\normalsize}
\end{figure}

The discrepancy between (\ref{thenoptgauge}) and the exact optimization obtained by solving \eqref{mincond} with \eqref{varlgnopt} is 
shown for real $n_0$ for a wide range of parameters in Figure~\ref{FIGUREgiidiscrepancynopt}. It can be seen that for occupations $\gtrsim \order{10}$ and/or for times shorter than singly-occupied coherence time $1/2\chi$,  the approximation is good. 
Compare with the analogous case with drift gauge shown in Figure~\ref{FIGUREgiidiscrepancy}.


For zero damping, and coherent state initial conditions $n_0=n'_0$, and relatively short times $4\chi te^{-2g''}\ll1$ one finds
\EQN{\label{varinopt}
\vari{G_L(t)} = \text{var}[\wt{G}_L(t)] = \chi t\cosh 2g''-2(\chi t)^2n'_0 +\frac{8}{3}\chi t (n'_0)^2 e^{-2g''}
.}
At large occupation $n'_0$, the gauge
\eqref{thenoptgauge} gives  $e^{g''}\approx (4\chi t_{\rm opt} n'_0)^2/3$, and one finds that all three terms in \eqref{varinopt} are of similar size. Using the condition \eqref{sdlimit} the simulation time is  
\EQN{
t_{\rm sim} \approx \order{10}t_{\rm coh}
.}
The $4\chi t e^{-2g''_{\rm approx}}\ll 1$ condition for $t<t_{\rm opt}$ implies large occupation $n'_0\gg\order{1}$. 
At low occupation on the other hand, $e^{g''_{\rm approx}}\to1$, the variances are $\approx e^{4\chi t}n'_0{}^2/4$, and one has again \eqref{smallnsimtime}

\subsection{Numerical investigation of performance}
\label{CH7DiffusionNumerical}

\begin{figure}[p]
\center{\includegraphics[width=\textwidth]{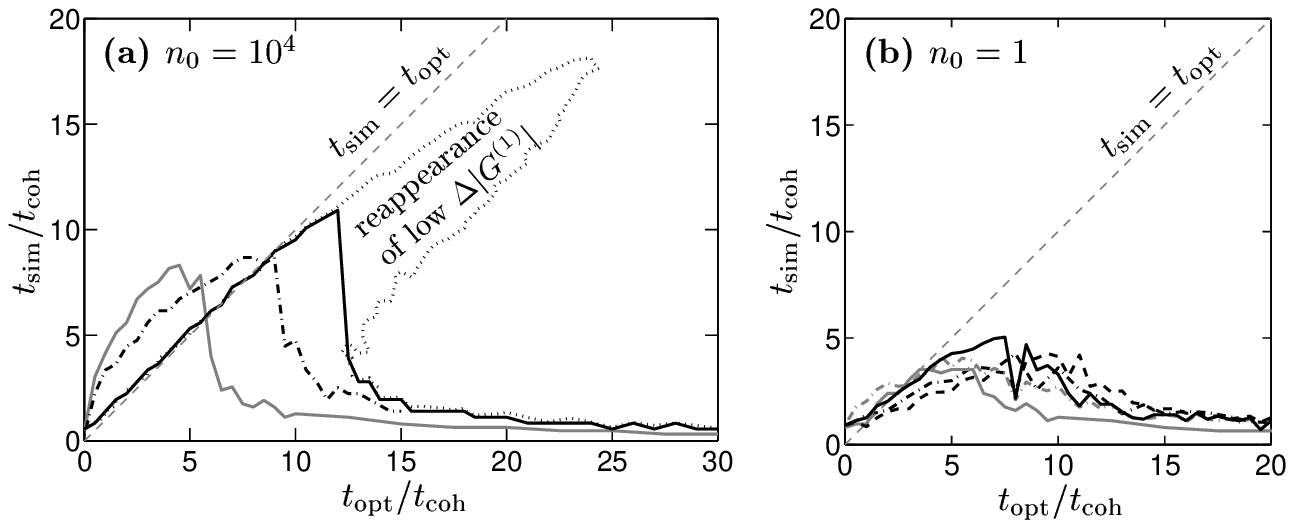}}\vspace{-8pt}\par
\caption[Comparison of target time and actual simulation time for $\mc{G}=0$ schemes]{\label{FIGUREtoptnopt}\footnotesize
\textbf{Comparison of {\it a priori} target time $t_{\rm opt}$  with actual useful simulation time $t_{\rm sim}$} for a variety of diffusion gauges in the $\mc{G}=0$ schemes: 
Results for the four forms \eqref{giiformsnopt} are shown as: (\ref{giintnopt}) -- {\scshape solid dark}; (\ref{giinnopt}) -- {\scshape dash-dotted light}; (\ref{giitnopt}) -- {\scshape dashed}; (\ref{giinopt}) -- {\scshape dash-dotted dark}.
Relationship obtained using the diffusion gauge (\ref{plimakgauge}) of Plimak\etal\cite{Plimak-01} (but no drift gauge)
 is also shown ({\scshape solid light}). The dashed line in the background shows, for reference $t_{\text{opt}}=t_{\text{sim}}$. For the gauge \eqref{giintnopt}, the whole region where useful precision occurs is shown by the {\scshape dotted} line.
Subplot {\bf(a)}: mean particle number $n_0=10^4$, {\bf (b)}: $n_0=1$. 
\normalsize}
\end{figure}

\begin{figure}[p]
\center{\includegraphics[width=\textwidth]{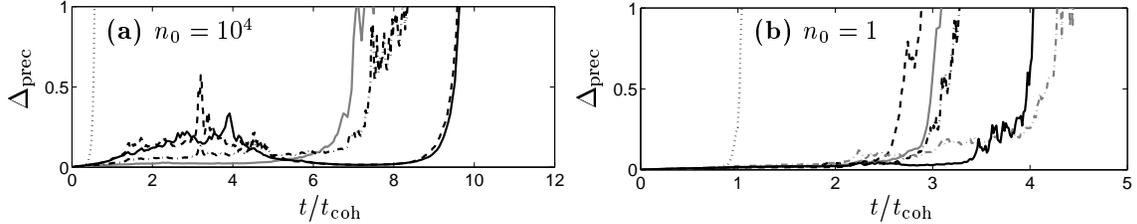}}\vspace{-8pt}\par
\caption[Uncertainty in $G^{(1)}(0,t)$ for various gauges in $\mc{G}=0$ schemes]{\label{FIGUREtusenopt}\footnotesize
\textbf{Uncertainty in $|G^{(1)}(0,t)|$}, as a function of time for various $\mc{G}=0$ schemes. The quantity plotted is $\Delta_{\text{prec}} = 
\sqrt{\mc{S}/10^6}(10\ \Delta|G^{(1)}(0,t)|\,/\,|G^{(1)}(0,0)|)$, so that $\Delta_{\text{prec}}\le1$ corresponds to 
useful precision as defined in Section~\ref{CH7NumericalProcedure}. Results are plotted for the four  diffusion gauge forms (\ref{giiformsnopt}), shown as : (\ref{giintnopt}) -- {\scshape solid dark}; (\ref{giinnopt}) -- {\scshape dash-dotted light}; (\ref{giitnopt}) -- {\scshape dashed}; (\ref{giinopt}) -- {\scshape dash-dotted dark}. Data for the diffusion gauge (\ref{plimakgauge}) of Plimak\etal\cite{Plimak-01}  with $t_{\rm opt}=3t_{\text{coh}}$ (as used therein) is shown as a {\scshape light solid} line, while the ungauged positive P calculation is shown as a {\scshape dotted} line. Simulations were carried out with $\mc{S}=10^4$ trajectories, starting with initial coherent state conditions: \textbf{(a)} $n_0=10^4$ \textbf{(b)} $n_0=1$. The diffusion-gauge plots in \textbf{(a)} were calculated with target times of $t_{\text{opt}}=10t_{\text{coh}}$ for forms \eqref{giintnopt} and \eqref{giitnopt}, and $t_{\rm opt}=7t_{\rm coh}$ for forms \eqref{giinnopt} and \eqref{giinopt}. In subplot \textbf{(b)}, the four forms \eqref{giiformsnopt} were simulated with  $t_{\text{opt}}=5t_{\text{coh}}$.
\normalsize}
\end{figure}

The results of numerical simulations using such diffusion gauge only schemes are shown in Figures~\ref{FIGUREtoptnopt} and ~\ref{FIGUREtusenopt}, in analogy with Figures~\ref{FIGUREtopttprec} and~\ref{FIGUREvarGt} for the drift gauged schemes. Some data is also given in Tables~\ref{TABLEsimtime} and~\ref{TABLEtimesfit}. As before, simulations were made with $\mc{S}=10^4$ for the mode occupations \eqref{ahon0}, and a wide range of target times $t_{\rm opt}$. 

The four gauges simulated were all of the general form \eqref{thenoptgauge}, but with the four adaptive varieties
forms
\SEQN{\label{giiformsnopt}}{
g''(t) &=& g''_{\rm approx}(n_0,t_{\rm opt})\label{giinopt}\\
g''(t) &=& g''_{\rm approx}(n_0,t_{\rm rem})\label{giitnopt}\\
g''(t) &=& g''_{\rm approx}(\breve{n}(t),t_{\rm opt})\label{giinnopt}\\
g''(t) &=& g''_{\rm approx}(\breve{n}(t),t_{\rm rem})\label{giintnopt}
}
in analogy with \eqref{giiforms}, where $t_{\rm rem}$ is given by \eqref{tremdef}.
Otherwise, the procedure was the same as described in Section~\ref{CH7Numerical}.
Additional features beyond what is mentioned there include:
\ITEM{
\item The diffusion-gauge-only simulations give improvement over the positive P, but give simulation times $\order{4}$ times shorter than when combined with drift gauge \eqref{ahogauge}. For example, compare Figure~\ref{FIGUREtoptnopt} to Figure~\ref{FIGUREtopttprec}). $t_{\rm sim}$ is still $\gg t_{\rm coh}$ at large $n_0$, however. 
\item The full adaptive gauge form \eqref{giintnopt} gives the most predictable results, in analogy to the drift-gauged case. That is, $t_{\rm sim}\approx t_{\rm opt}$ for times up till those given in Table~\ref{TABLEsimtime}.
\item At low $n'_0$ the simulation time appears more sensitive to the choice of $t_{\rm opt}$ than for the drift-gauged simulations.
\item At high occupations $n'_0\gg1$, the pairs of adaptive gauges \eqref{giintnopt}\&\eqref{giitnopt} and \eqref{giinopt}\&\eqref{giinnopt} behave identically, and are thus not shown in all plots. The simulation appears to be insensitive to whether one uses a gauge choice dependent on $\breve{n}(t)$ or $n_0$.  This is probably a feature peculiar to the particle number conserving single-mode anharmonic oscillator model, because here $\breve{n}(t)\approx n_0$ while useful precision is seen.
\item For $n'_0\gg1$, the time-adaptive gauge forms \eqref{giitnopt} and \eqref{giintnopt} lead to a peculiar effect if 
the optimization time $t_{\rm opt}$ is chosen larger than the usual maximum $t_{\rm sim}$ given in Table~\ref{TABLEsimtime}. 
The 
statistical error in the $G^{(1)}$  estimate first rises rapidly, then falls again, and finally grows definitively. This is seen in Figure~\ref{FIGUREtusenopt}\textbf{(a)}, and the parameter region for which this occurs is shown in Figure~\ref{FIGUREtoptnopt}\textbf{(a)}.
In effect one has two time intervals when the simulation gives useful results: at short times, and later in a time interval around $t\approx\order{10t_{\rm coh}}$. 
}

\section{Comparison to recent related work}
\label{CH7Comparison}

	Improvements to the basic positive P simulation method for specific cases of interacting Bose gas systems have been tried with some success in several recent publications\cite{Carusotto-01, Plimak-01,DeuarDrummond01,DrummondDeuar03}. Here we tie these together with the stochastic gauge formalism, and make some comparison to the results and analysis in the present chapter.

\subsection{The work \cite{Carusotto-01} of Carusotto, Castin, and Dalibard}
\label{CH7ComparisonCCD}
  The article \textit{$N$-boson time-dependent problem: A reformulation with stochastic wavefunctions} considered an isolated (i.e. particle-conserving) system of $\bar{N}$ interacting bosons. The ``coherent state simple scheme'' described in Section III B 2 therein can be identified as using drift gauges of the form 
\SEQN{\label{CCDgaugem}}{
\mc{G}_{\bo{n}} &=& i\sqrt{2i\chi}\left(\alpha_{\bo{n}}\beta_{\bo{n}}-|\alpha_{\bo{n}}|^2\right)\\
\wt{\mc{G}}_{\bo{n}} &=& \sqrt{2i\chi}\left(\alpha_{\bo{n}}\beta_{\bo{n}}-|\beta_{\bo{n}}|^2\right)
}
when re-written in the notation of this thesis for a many-mode system as in equations \eqref{itoH}, with mode labels $\bo{n}$. For the single-mode system  considered in this chapter, this corresponds to
\SEQN{\label{CCDgauge}}{
\mc{G}_1 &=& i\sqrt{2i\chi}\left(\breve{n}-|\alpha|^2\right)\\
\mc{G}_2 &=& \sqrt{2i\chi}\left(\breve{n}-|\beta|^2\right)
.}
This gauge causes  a full decoupling of the complementary $\alpha$ and $\beta$ equations. Like \eqref{ahogauge} it is also successful in removing moving singularities, since the nonlinear terms in the radial equations for $d|\alpha|$ and $d|\beta|$ are removed. 

There are two major differences between the coherent state wavefunction method and the gauge P method:
Firstly, the former 
is hardwired to models that conserve particle number. This hardwiring to $\bar{N}$ particles has both a major benefit and a disadvantage, as compared to the gauge P method considered in this thesis:
\ITEM{
\item {\scshape Benefit:} 
coherent state amplitudes are not free to explore the entire phase space, are bounded from above, and cannot form moving singularities. 
\item {\scshape Disadvantage:} Losses or gains from external reservoirs are unable to be simulated, and so it  is not applicable in its present form to simulations of e.g.  evaporative cooling or a continuously loaded system such as an atom laser. 
}
Another major difference is that diffusion gauges were not considered by Carusotto\etal, and hence simulation times with this method for a single mode were very short ($\approx t_{\rm osc}$).

\begin{figure}[t]
\center{\includegraphics[width=\textwidth]{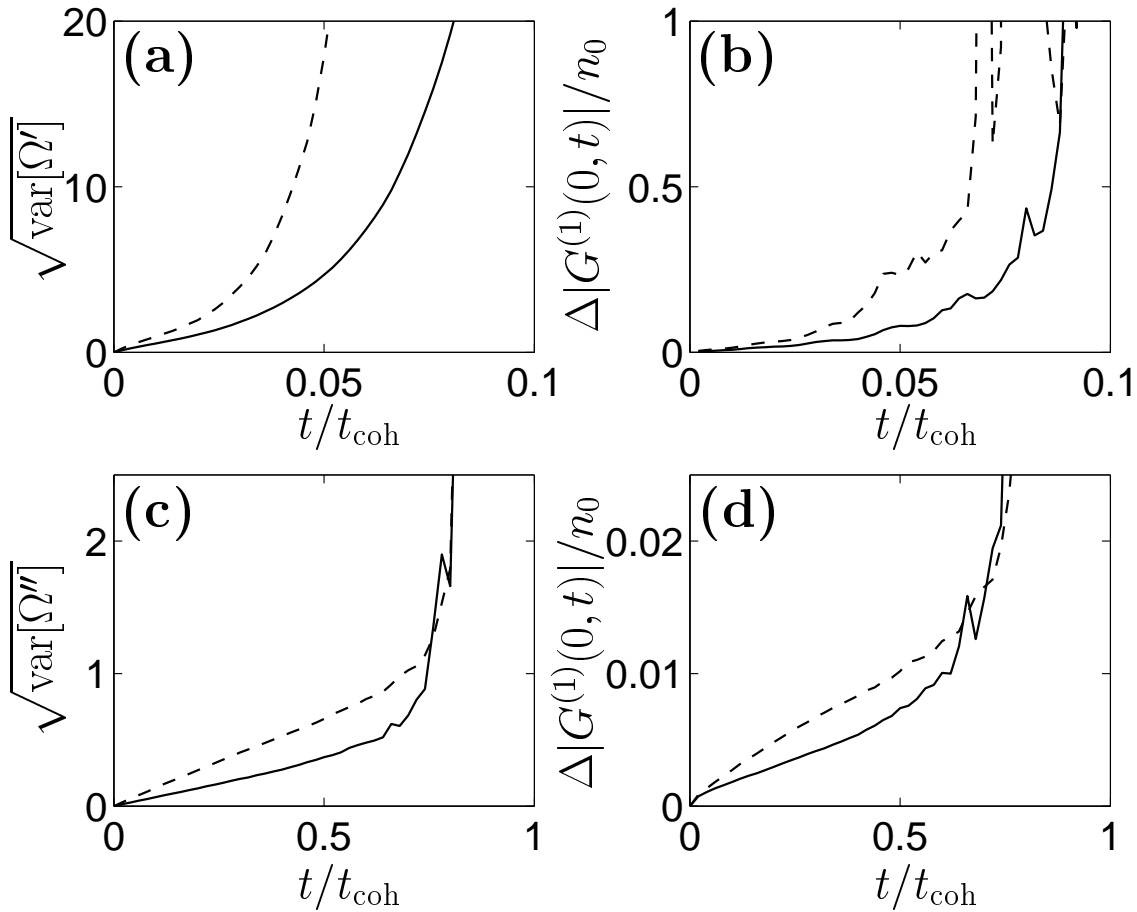}}\vspace{-8pt}\par
\caption[Comparison of drift gauge to that in Carusotto\etal\cite{Carusotto-01}]{\label{FIGUREccdga}\footnotesize
\textbf{Spread in trajectory weights}
 (\textbf{(a)} and \textbf{(c)}) \textit{and phase correlation function uncertainties} (\textbf{(b)} and \textbf{(d)}), compared for the drift (only) gauges (\ref{CCDgauge}) -- {\scshape dashed}, and (\ref{ahogauge}) (with no diffusion gauge: $g''=0$)-- {\scshape solid}, in a one-mode, undamped, gainless system. Coherent state initial conditions with initial mean occupations \textbf{(a)} and \textbf{(b)}: $n_0=10^4$, while in \textbf{(c)} and \textbf{(d)}: $n_0=1$. All calculations are with $10^4$ trajectories.
\normalsize}
\end{figure}

The gauges \eqref{CCDgauge} could also be applied to the present gauge P method, and their efficiency can be compared with the efficiency of gauges developed here: \eqref{ahogauge} when $g''=0$. 
 The gauge (\ref{CCDgauge}) mediates the replacements  $\breve{n}\to|\alpha|^2\text{\, or\,}|\beta|^2$ in the $d\alpha$ and $d\beta$ equations (respectively), as opposed to \eqref{ahogauge}, which only replaces $\breve{n}\to \re{\breve{n}}$ and does not decouple $d\beta$ from $d\alpha$.   So, since the magnitude of the spread in $\Omega$ behaves approximately proportional to $\sqrt{\sum_k|\mc{G}_k|^2}$, the gauge \eqref{CCDgauge}
 is expected to produce a broader distribution of weights. Numerical simulations were carried out, and the results are shown in Figure~\ref{FIGUREccdga} for both high and low mode occupations, , and Table~\ref{TABLEsimtime} for high occupation. 
Useful simulation time with \eqref{CCDgauge} is somewhat smaller than seen with (\ref{ahogauge}).

In the same published work\cite{Carusotto-01}, a stochastic wavefunction method was developed based on Fock number states as well. This method was shown to give good results and not be prone to boundary term errors but  unfortunately does not appear extensible to open systems in any straightforward fashion because it is very strongly hardwired to $\bar{N}$ total particles.

The stochastic wavefunction method was also adapted in \cite{Carusotto-01} to some cases of non-local interactions, arriving at equations that use effectively a similar noise expansion to that in \eqref{dXexpr}, but were only applicable to potentials having exactly real Fourier transforms $\wt{U}_{\wt{\bo{n}}}=\wt{U}'_{\wt{\bo{n}}}$.

\subsection{The work \cite{Plimak-01} of Plimak, Olsen, and Collett}
\label{CH7ComparisonPlimak}

The article \textit{Optimization of the positive-$P$ representation for the anharmonic oscillator} considered a single-mode 
undamped, gainless system at high Bose occupation with coherent state initial conditions. The ``noise optimization'' scheme applied therein to greatly improve simulation times can be identified as an imaginary diffusion gauge of the form (rewritten in the present notation)
\begin{equation}\label{plimakgauge}
g_{12} = \frac{i}{2}\cosh^{-1}\left[2n_0\chi t_{\rm opt}\right] = ig''_A
\end{equation}
defined at high occupation or long times (i.e. while $2n_0\chi t_{\rm opt}\ge 1$). 
This is dependent on a target time $t_{\rm opt}$ (which was taken to be $t_{\rm opt}=3t_{\text{coh}}$ in the calculations of Ref.~\cite{Plimak-01}), and the initial Bose occupation $n_0=n'_0=|\alpha_0|^2$. The value of the gauge was held constant throughout the calculation, without allowance for a changing mode occupation.

The useful simulation times obtainable with this method are also shown in Figures~\ref{FIGUREsimtime} and~\ref{FIGUREvarGt}, and Table~\ref{TABLEsimtime}. Their precise dependence on the target time parameter $t_{\rm  opt}$ has been calculated here,  and is shown in Figure~\ref{FIGUREtopttprec}.

Comparing \eqref{plimakgauge} with the optimized diffusion gauge \eqref{giinopt} that is constant with time, one finds that in the $n_0\gg1$ limit the two gauges differ by a constant:
\EQN{
g''_A(n_0,t_{\rm opt})\approx g''_{\rm approx}(n_0,t_{\rm opt}) +\frac{\log 3}{4}
.}
While the simulation times achievable using $g''_A$ are comparable with those obtained with the adaptive gauge $g''_{\rm approx}(\breve{n}(t),t_{\rm opt})$ of \eqref{giinopt}, the target time $t_{\rm opt}$ is no longer a good indicator, and has a complicated relationship with $t_{\rm sim}$, as seen in Figure~\ref{FIGUREtoptnopt}. The simulation time is also much shorter than with the drift gauged simulation.

Extensions to multi-mode systems or low mode occupations were not considered, however
since the system under consideration was only a single mode, the lack of drift gauges did not lead to any boundary term errors in this particular case.

\subsection{The work \cite{DeuarDrummond01} of Deuar and Drummond}
\label{CH7ComparisonCCP2000}

Some initial research by Drummond and Deuar into gauges for the single-mode undamped anharmonic oscillator system was reported in\cite{DeuarDrummond01}. Here the state was written as an expansion over a normalized, Hermitian, coherent state projector kernel
\begin{equation}\label{ccp2kkernel}
\op{\Lambda}(\alpha,\beta,\theta) = \frac{e^{i\theta}||\alpha\rangle\langle\beta^*||+e^{-i\theta}||\beta^*\rangle\langle\alpha||}{2e^{n'}\cos(\theta+n'')},
\end{equation}
where the phase variable $\theta$ was real. This is equivalent to imposing Hermiticity on the positive P kernel before normalizing it.

Drift gauges of the form
\SEQN{}{
\mc{G}_1 &=-&\lambda\sqrt{\chi}\left[n''-n'+|\alpha|^2\right]\,[\tan(\theta+n'')+i]\\
\mc{G}_2 &= &\lambda\sqrt{\chi}\left[n''+n'-|\beta|^2\right]\,[\tan(\theta+n'')+i]
}
were used, where $\lambda$ is a (real) strength parameter chosen \textit{a priori}.

Simulation precision was greatly improved, but evident boundary term errors were present. This may have been caused by several factors, which are pointed out here because they are good examples of what should be avoided when choosing kernels and gauges:
\ENUM{
\item The radial super-exponential behavior of $d\alpha$ and $d\beta$ is not removed in full, hence one can expect that moving singularities will still be present --- as is ultimately seen as boundary term errors.

\item The magnitude of the kernel (\ref{ccp2kkernel}) diverges when $\theta\to-n''$. Hence, this parameter region might be unable to be sampled in an unbiased way. In particular, matrix elements of $\op{\Lambda}$ appearing in the expression \eqref{btexpression} for $\op{\mc{B}}$ diverge when $\theta=\pm\pi/2$, and may lead to nonzero boundary terms once integrated. 

\item The gauges were dependent on the trajectory weight $e^{i\theta}$, and thus non-autonomous. This causes feedback behavior between the phase-space variables $\alpha$ and $\beta$, and the weight, which is difficult to analyze to definitively determine whether moving singularities are present or not. 
(Compare to point 6(c) in Section~\ref{CH6RemovalHeuristic}).

\item The gauges diverge at $\theta\to\pm\pi/2$, against  recommendation 6(b) of Section~\ref{CH6RemovalHeuristic}.
}

\section{Gauge recommendation}
\label{CH7Gauge}
Collecting together the analysis, and results of numerical calculations reported in this chapter, the following gauge choices appear to be advantageous for a single mode (labeled $\bo{n}$) of a two-body locally interacting Bose gas that is free to interact with its environment:

Defining 
\EQN{\label{nexpr}
\breve{n}_{\bo{n}}=\alpha_{\bo{n}}\beta_{\bo{n}}
,}
then
\EQN{\label{ahodriftgauge}
\mc{G}_{\bo{n}} =-i\wt{\mc{G}}_{\bo{n}} = -\sqrt{2i\chi}\,\im{\breve{n}_{\bo{n}}} e^{-g''_{\bo{n}}}
}
The local diffusion gauge dependent on the single ``target time'' parameter $t_{\rm opt}$, or more precisely on the 
``remaining time to target''
\EQN{\label{trem}
 t_{\rm rem} = \left\{\begin{array}{c@{\text{\ if\ }}l}t_{\rm opt}-t & t< t_{\rm opt}\\ 0 & t\ge t_{\rm opt}\end{array}\right.
.}
is then given by
\EQN{\label{ahodiffusiongauge}
g''_{\bo{n}} = \frac{1}{6}\log\left\{ 8|\breve{n}_{\bo{n}}(t)|^2\chi t_{\rm rem} + a_2^{3/2}(\breve{n}_{\bo{n}}(t),\gamma_{\bo{n}} t_{\rm rem})\right\}
,}
i.e. \eqref{approxgiiopto}. The coefficient $a_2$ is
\EQN{\label{a2coef}
a_2(\breve{n},v) = 1 + 4\im{\breve{n}}^2\left(\frac{1-e^{-2v}}{2v}\right) - 2|\breve{n}|^2\left(\frac{1-2v+2v^2-e^{-2v}}{v}\right)
.}

\section{Summary}
\label{CH7Summary}

The principal aim of this chapter has been to develop a ``black box'' form  of drift and diffusion gauges. A form that can be expected to remove boundary term errors and extend simulation times beyond what is possible with the positive P for 
a single mode of a locally-interacting Bose gas with two-particle collisions that is open to the environment.
The recommended form is given by \eqref{nexpr}-\eqref{a2coef}.
These gauges:
\ENUM{
\item Extend simulation time for the anharmonic oscillator beyond what was possible with the positive P representation (i.e. when $g''_{jk}=0$ and $\mc{G}_{\bo{n}}=\wt{\mc{G}}_{\bo{n}}$). For high mode occupations, the simulation time is extended by a factor of $\order{15n_0^{1/6}}$ (see Table~\ref{TABLEsimtime}) when starting from a coherent initial state, representative of a single trajectory. This simulation time is  $\order{40}$ coherence times at high mode occupation, and all decoherence behavior is simulated.  The improvement in numerical performance is best summarized by Figures~\ref{FIGUREsimtime},~\ref{FIGUREG}, and in Table~\ref{TABLEsimtime}.
\item Remove the instability responsible for moving singularities.
\item Apply for dynamically  changing mode occupation.
\item Apply for simulations of  open systems.
}
For the aim of using these gauges in many-mode simulations of open interacting Bose gases, it is crucial that the 
gauge choice  be freely adaptable to any changes caused to the mode occupation (or other properties). Such  changes will  be caused by interactions with the rest of the modes, and an external environment.

\REM{Some properties of single-mode simulations using the recommended gauges have been checked in detail. e.g. simulation time at both high and low occupation, the variance of the estimator of the phase-dependent variable $G^{(1)}(0,t)$, the dependence of stochastic error on the damping rate, and the best way to adapt the optimized gauge to changing mode occupation.}

\REM{Performance of the recommended gauge has been investigated numerically, and found to be an improvement of up to several orders of magnitude in simulation time as compared with the plain positive P simulation method.}

A second possibly alternative method where only diffusion gauges are used has been considered, and appropriate optimized diffusion gauges derived. This method does not guarantee removal of systematic biases, but appears to  have advantages in efficiency for many-mode calculations provided the simulation can be successfully monitored using the indicators of Gilchrist\etal\cite{Gilchrist-97} to catch any boundary term errors if these form.

Lastly, the method developed has been tied in with and compared to some previous related work in the field\cite{Carusotto-01,Plimak-01,DeuarDrummond01}.

\chapter{Dynamics of two coupled Bose gas modes}
\label{CH8}

The next step up towards a many-mode simulation is to see 
how the statistical behavior of the gauge methods is affected once mode  mixing occurs.
In this chapter, the effect of the combined drift and diffusion gauges suggested in Section~\ref{CH7Gauge} on dynamical simulations of a two-mode system will be considered (undamped, gain-less, with  two modes coupled by Rabi oscillations). This system is simple enough that there are only a few parameters, yet many of the changes caused by mode mixing are visible.

\section{The model}
\label{CH8Model}

The system consists of two orthogonal modes labeled $1$ and $2$ with interparticle interactions in each mode, and Rabi coupling between them. 
  The coupling  frequency will be restricted to be real for simplicity (complex values can always be considered in similar vein with no significant alterations to the following discussion).  
Time units are chosen so that the nonlinear interaction frequency is $\chi=1$, and a transformation is made to an interaction picture in which the linear mode self-energies $\hbar\omega_{jj}\op{n}_j$ are moved into the Heisenberg evolution of  operators. The interaction picture  Hamiltonian then is
\EQN{\label{2H}
\op{H} = \hbar\omega_{12}\left[\dagop{a}_1\op{a}_2 + \dagop{a}_2\op{a}_1\right] + \hbar\sum_{j=1}^2\dagop{a}_j{}^2\op{a}_j^2
.}
The Rabi frequency is $\omega_{12}$ (in scaled time units).

Physically this model can represent, for example, a simplified model of two internal boson states coupled by an EM field, or two trapped condensates spatially separated by a barrier. This approximation has been used by a multitude of authors to investigate the quantum behavior of BECs in two-state systems. For references to some of the work in this field see Leggett~\cite{Leggett01}, Part VII.

Comparing to the lattice Hamiltonian \eqref{deltaH} of Chapter~\ref{CH2}, the modes $1$ and $2$ can be identified with some modes $\bo{n}$ and $\bo{m}$, respectively. Then, $\omega_{12}=\omega_{\bo{nm}}/\chi$, while $\chi$, $\omega_{\bo{nn}}$, and $\omega_{\bo{mm}}$ have been scaled out or moved into the Heisenberg evolution of the $\op{a}_j$. It has been assumed that $\omega_{\bo{nm}}$ is real, and that all $\gamma_{\bo{n}}$ and $\varepsilon_{\bo{n}}=0$.

From the gauge P equations \eqref{itoH} of Chapter~\ref{CH5}, and not yet specifying the exact form of the noise matrices, one obtains gauged Ito stochastic equations
\SEQN{\label{2modeeqn}}{
d\alpha_j &=& -i\omega_{12}\alpha_{\neg j}\,dt -2i\alpha_j^2\beta_j\,dt+\sum_k B^{(\alpha)}_{jk} (dW_k -\mc{G}_k\,dt)\\
d\beta_j &=& i\omega_{12}\beta_{\neg j}\,dt +2i\alpha_j\beta_j^2\,dt +\sum_k B^{(\beta)}_{jk} (dW_k-\mc{G}_k\,dt)\\
d\Omega &=& \sum_k \mc{G}_k\,dW_k
,}
where $\neg j= 1+\delta_{1j}$, and $dW_k$ are the usual Wiener increments. The noise matrices obey 
\SEQN{}{
\sum_l B^{(\alpha)}_{jl} B^{(\alpha)}_{jk} &=& -2i\delta_{jk}\alpha_j^2\\
\sum_l B^{(\beta)}_{jl} B^{(\beta)}_{jk} &=& 2i\delta_{jk}\beta_j^2\\
\sum_l B^{(\alpha)}_{jl} B^{(\alpha)}_{jk} &=& 0
.}

Applying the drift and diffusion gauges \eqref{ahodriftgauge} and \eqref{ahodiffusiongauge} worked out in Chapter~\ref{CH7} to each mode separately, these equations become
\SEQN{\label{gauged2mode}}{
d\alpha_j &=& -i\omega_{12}\alpha_{\neg j}\,dt -2i\alpha_j\re{\breve{n}_j}\,dt+
i\alpha_j\sqrt{2i}\left[\cosh g''_j dW_{j1} +i\sinh g''_j dW_{j2}\right]\qquad\quad\\
d\beta_j &=& i\omega_{12}\beta_{\neg j}\,dt +2i\beta_j\re{\breve{n}_j}\,dt+
\beta_j\sqrt{2i}\left[-i\sinh g''_j dW_{j1} +\cosh g''_j dW_{j2}\right]\\
d\Omega &=&  -\sqrt{2i}\Omega\sum_j \im{\breve{n}_j}e^{-g''_j}(dW_{j1}-idW_{j2})
,}
where $\breve{n}_j=\alpha_j\beta_j$, and 
\EQN{\label{giig2mode}
g''_j = \frac{1}{6}\log\left\{8|\breve{n}_j|^2t_{\rm opt}+(1+4\,\im{\breve{n}_j}^2)^{3/2}\right\}
.}
The four independent real Wiener increments $dW_{jk}$ can be implemented by independent Gaussian random variables of variance $dt$ at each time step.

Proceeding to check the no-moving-singularities condition \eqref{mvsingcondition} as for the single mode in Section~\ref{CH7Drift}, there is no super-exponential growth of any $|\alpha_j|$, $|\beta_j|$ or $|\Omega|$. Moving singularities or noise divergences will not occur.

Two kinds of initial conditions were numerically investigated in detail:

\section{Case 1: Coupling to a vacuum mode}
\label{CH8Case1}

\subsection{Description}
\label{CH8Case1Description}
The system starts initially with  a coherent state of mean particle number $n_0$ in mode 1, and vacuum in mode 2. 
In all simulations of this case, the inter-mode coupling strength was taken to be $\omega_{12}=5$, but the mean particle number $n_0$ was varied. This mean particle number $\bar{N}=n_0=\sum_j\langle\op{n}_j\rangle=\sum_j\langle\dagop{a}_j\op{a}_j\rangle$ is conserved during each simulation.

At low particle number, the Rabi oscillations dominate the Hamiltonian, and particles oscillate between the modes, without much decoherence.
At high particle number $n_0$, on the other hand, phase decoherence dominates mode 1, suppressing also the coherent transfer of particles to mode 2.

The particular values chosen to simulate were 
\EQN{
n_0=\{1, 17, 200, 1500, 10^4\}
.}

The two observables considered were: The fraction of particles in the (initially empty) mode 2:
\EQN{
  p_2 = \frac{\langle\op{n}_2\rangle}{\bar{N}}
,}
and the local normalized second order correlation functions of modes $j$:
\EQN{\label{g2def}
  g^{(2)}_j(t,t) = \frac{\langle:\op{n}_j(t):\rangle}{\langle\op{n}_j(t)\rangle} = \frac{\langle\op{n}_j^2\rangle}{\langle\op{n}_j\rangle^2}-\frac{1}{\langle\op{n}_j\rangle}
.}
The latter quantify the amount of (instantaneous) bunching/antibunching in the boson field, and for example, are unity for coherent states, two for thermal fields, and $1-1/n$ for Fock number states of $n$ particles. Large values $g^{(2)} > 2$  occur  e.g. for quantum superpositions of vacuum and Fock number states with two or more particles where the average particle number is small. For example in the state $\ket{\psi}=\sin\theta\ket{0}+\cos\theta\ket{n}$, $g^{(2)} = (1-\frac{1}{n})/\cos^2\theta$.

\subsection{Procedure}
\label{CH8Case1Procedure}

Three gauge choices were compared for this case. 
\ENUM{
\item \textit{ No gauge}. i.e. positive P.
\item \textit{Drift and diffusion gauges} as in equations \eqref{gauged2mode} with \eqref{giig2mode}.The target time $t_{\rm opt}$ was varied to obtain the longest useful simulation time $t_{\rm sim}$.
\item
\textit{Drift gauge} \eqref{CCDgauge} of Carusotto\etal\cite{Carusotto-01}. As pointed out before, a subtle difference from Ref. \cite{Carusotto-01} is that here particle number conservation is not explicitly hardwired into the simulation, but happens to arise because no environment interactions are included in this particular simulation. This freedom is left unconstrained to make conclusions reached on the basis of the two modes relevant to a larger open many-mode  system.
}

In each run, $\mc{S}=2\times10^5$ trajectories were used, and useful simulation precision was defined as in \eqref{usefulprecision}, and simulation time $t_{\rm sim}$ accordingly.

\begin{figure}[p]
\center{\includegraphics[width=\textwidth]{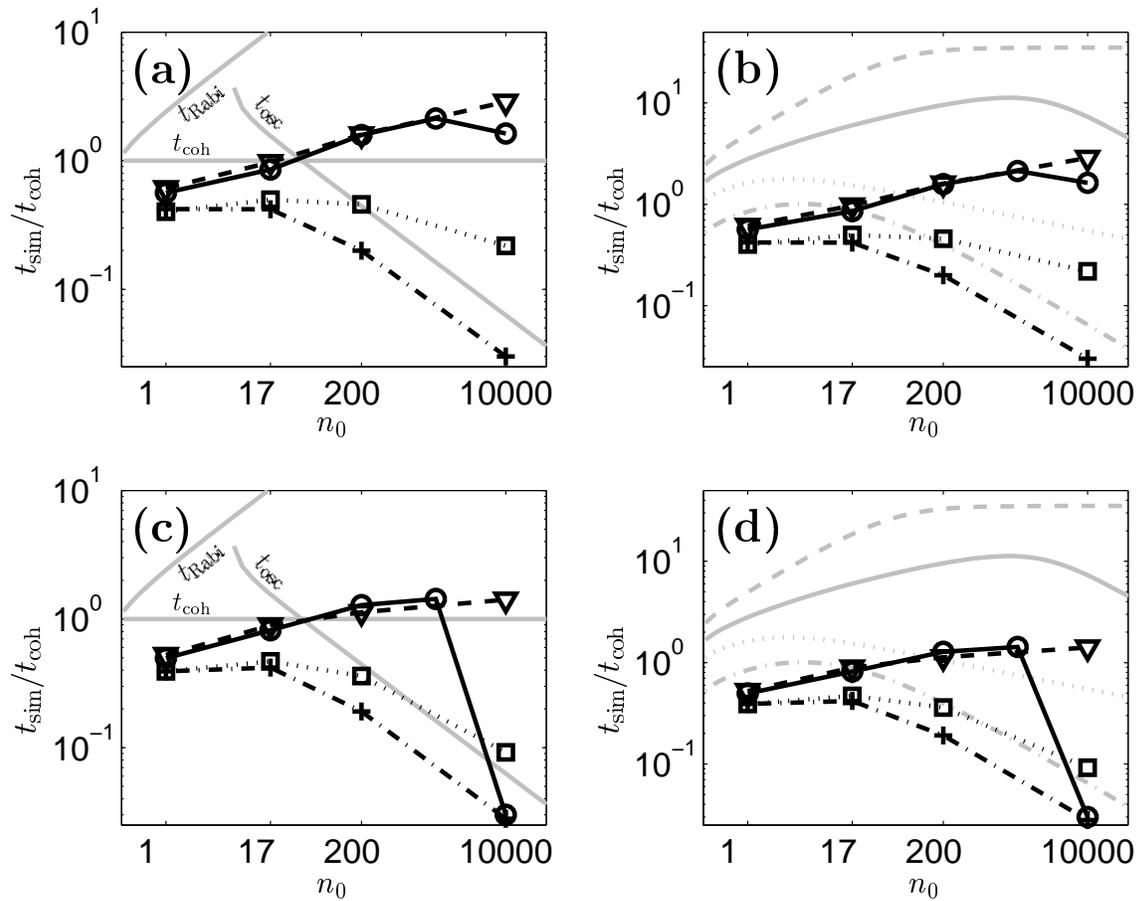}}\vspace{-8pt}\par
\caption[Simulation times for coupling to vacuum mode]{\label{FIGUREcase1st}\footnotesize
\textbf{Useful simulation times} $t_{\rm sim}$ for a mode coupled to vacuum as in Section~\ref{CH8Case1}.
Calculated simulation times are shown as data points, with the symbols denoting gauge used. ``$\Box$'': positive P; ``$\bigcirc$'': drift gauge \eqref{ahodriftgauge} and diffusion gauge \eqref{giig2mode} with $t_{\rm opt}=0$; ``$+$'' drift gauge \eqref{CCDgauge} of Carusotto\etal\cite{Carusotto-01}; ``$\bigtriangledown$'': drift and diffusion gauges \eqref{ahodriftgauge} and \eqref{giig2mode} with best target time parameters: $\omega_{12}t_{\rm opt} = \{2.5,0.5,0.5,0.075\}$ for $n_0=\{1,17,200,10^4\}$, respectively.
 Subplots \textbf{(a)} and \textbf{(b)} show simulation times based on estimates of the observable $p_2$, while \textbf{(c)} and \textbf{(d)} times based on $g^{(2)}_2$. 
Subplots \textbf{(a)} and \textbf{(c)} compare to physical time scales, including Rabi oscillation period $t_{\rm Rabi}=2\pi/\omega_{12}$, while subplots \textbf{(b)} and \textbf{(d)} compare to expected simulation times for a single mode using the empirical fits of Table~\ref{TABLEtimesfit} and \eqref{fitccd}--\eqref{fitga0}. The expected $t_{\rm sim}$ are plotted as light lines: {\scshape dotted}: positive P; {\scshape solid}: drift gauge \eqref{ahodriftgauge} and diffusion gauge \eqref{giig2mode} with $t_{\rm opt}=0$; {\scshape dashed}: with optimum $t_{\rm opt}$ choice; {\scshape dash-dotted}: drift gauge \eqref{CCDgauge} of Carusotto\etal\cite{Carusotto-01}.
\normalsize}
\end{figure}

In Figure~\ref{FIGUREcase1st} actual simulation times are compared to physical timescales and expected  simulation times based on the single-mode analysis of Chapter~\ref{CH7}. 
 Note that fitting parameters to \eqref{timefit} for the Carusotto\etal\cite{Carusotto-01}  gauge \eqref{CCDgauge} were calculated from single-mode simulations to be 
\EQN{\label{fitccd}
\{ c_0,\dots,c_4\}=\{1,3.3\pm0.5,0.9\,{}^{+\infty}_{-0.2} ,-0.3\pm0.4,0.31\pm0.09\}
,}
 and for \eqref{ahodriftgauge} with \eqref{giig2mode} and $t_{\rm opt}=0$ were 
\EQN{\label{fitga0}
\{c_0,\dots,c_4\}=\{1,400\pm130,3.6\,{}^{+\infty}_{-2.3},1.2\pm0.2,0.42\pm0.03\}
.}

\begin{figure}[t]
\center{\includegraphics[width=\textwidth]{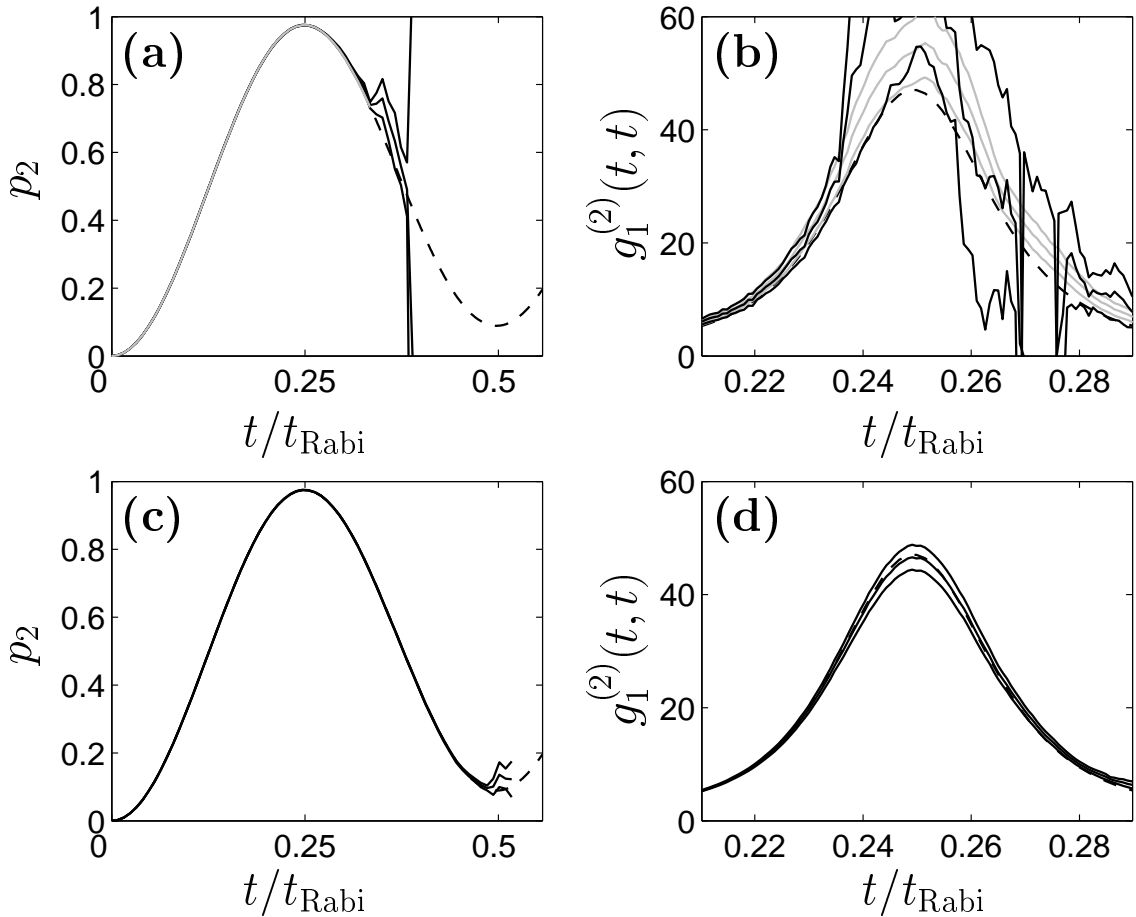}}\vspace{-8pt}\par
\caption[Coupling to vacuum mode: $n_0=1$]{\label{FIGURE2moden1}\footnotesize
Coupling to vacuum mode: \textbf{System with  few particles on average}
($n_0=1$). Time scaled to Rabi period $t_{\rm Rabi}=2\pi/\omega_{12}$. Subfigures \textbf{(a)} and \textbf{(c)} show mode $2$ particle fraction, while  \textbf{(b)} and \textbf{(d)} show instantaneous particle number correlations in mode $2$.  Subfigures \textbf{(a)} and \textbf{(b)} show results with positive P {\scshape light solid} and drift gauge (\ref{CCDgauge}) {\scshape dark solid}, whereas \textbf{(c)} and \textbf{(d)} show results with combined drift gauge (\ref{ahodriftgauge}) and diffusion gauge (\ref{ahodiffusiongauge}) using target time $t_{\text{opt}}=2.5/\omega_{12}\approx 0.4t_{\rm Rabi}$. Exact results also shown (dashed). Triple lines indicate mean and error bars.
\normalsize}
\end{figure}

\begin{figure}[t]
\center{\includegraphics[width=\textwidth]{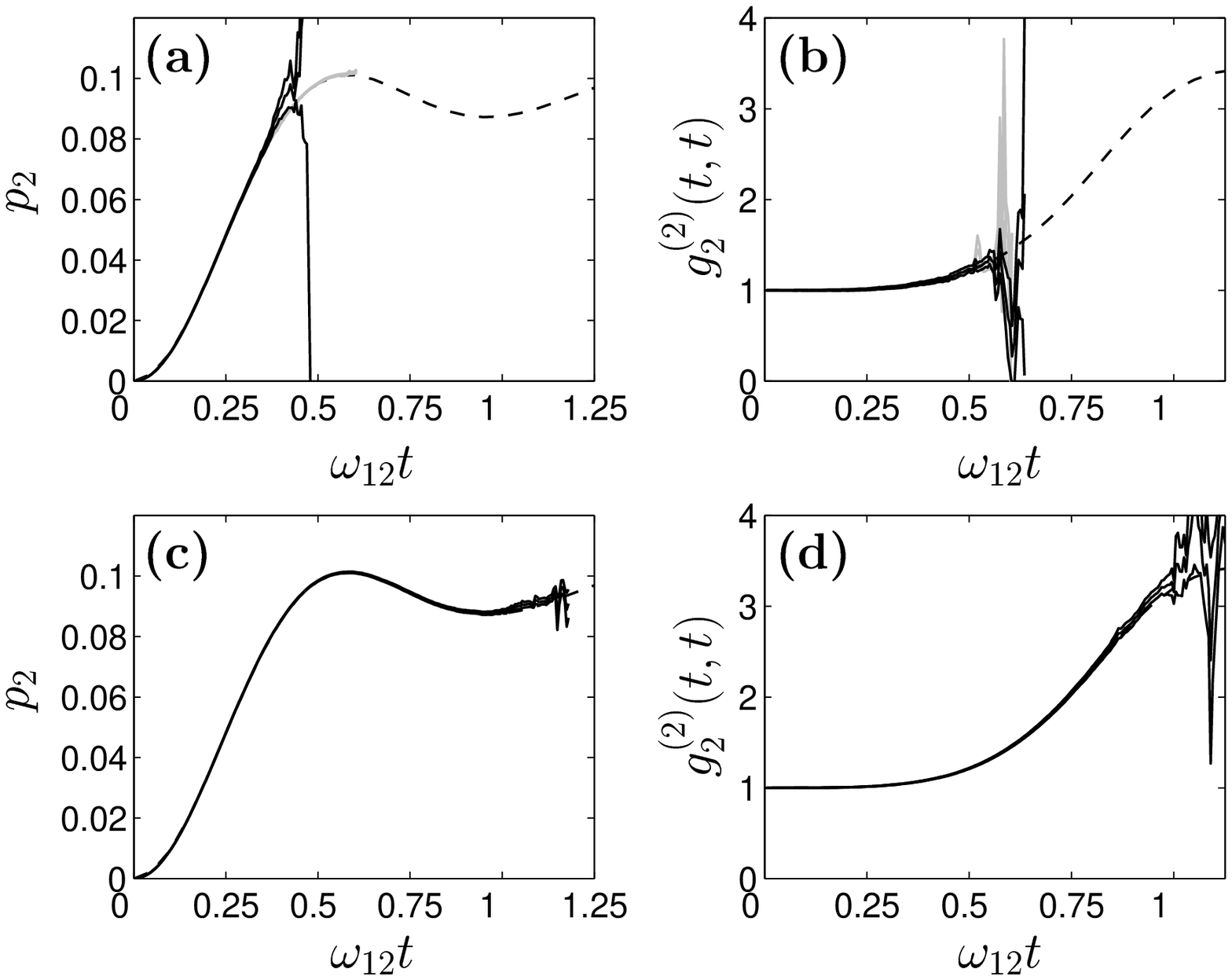}}\vspace{-8pt}\par
\caption[Coupling to vacuum mode: $n_0=17$]{\label{FIGURE2moden17}\footnotesize
Coupling to vacuum mode:\textbf{System with  intermediate number of particles}
($n_0=17$). Subfigures \textbf{(a)} and \textbf{(c)} show mode $2$ particle fraction, while  \textbf{(b)} and \textbf{(d)} show instantaneous particle number correlations in mode $2$.  Subfigures \textbf{(a)} and \textbf{(b)} show results with positive P {\scshape light solid} and drift gauge (\ref{CCDgauge}) {\scshape dark solid}, whereas \textbf{(c)} and \textbf{(d)} show results with combined drift gauge (\ref{ahodriftgauge}) and diffusion gauge (\ref{ahodiffusiongauge}) using target time $\omega_{12}t_{\text{opt}}=0.5$. Exact results also shown (dashed).
Triple lines indicate mean and error bars.
\normalsize}
\end{figure}

\begin{figure}[t]
\center{\includegraphics[width=\textwidth]{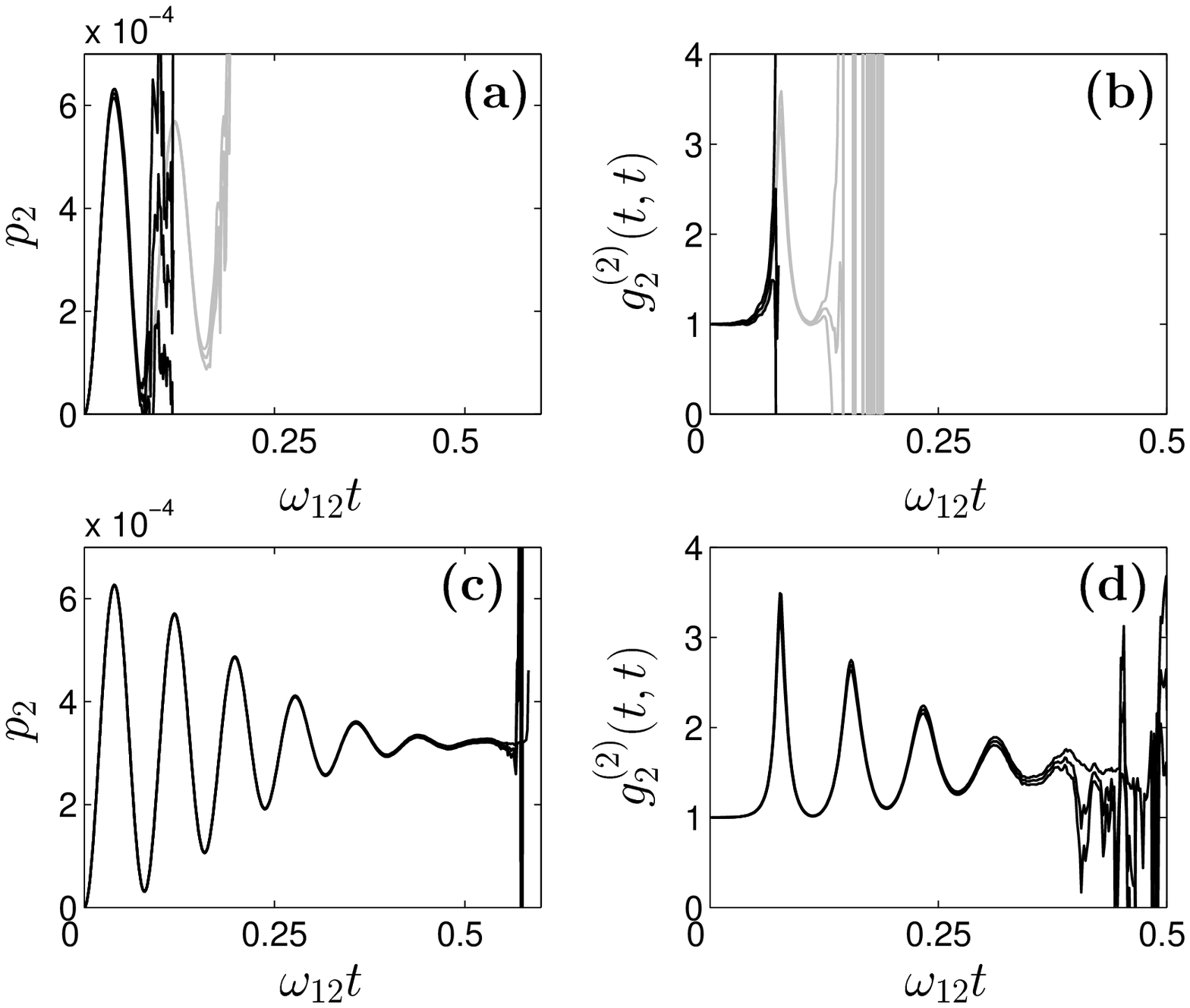}}\vspace{-8pt}\par
\caption[Coupling to vacuum mode: $n_0=200$]{\label{FIGURE2moden200}\footnotesize
Coupling to vacuum mode: \textbf{System with  large number of particles}
($n_0=200$). Subfigures \textbf{(a)} and \textbf{(c)} show mode $2$ particle fraction, while  \textbf{(b)} and \textbf{(d)} show instantaneous particle number correlations in mode $2$.  Subfigures \textbf{(a)} and \textbf{(b)} show results with positive P {\scshape light solid} and drift gauge (\ref{CCDgauge}) {\scshape dark solid}, whereas \textbf{(c)} and \textbf{(d)} show results with combined drift gauge (\ref{ahodriftgauge}) and diffusion gauge (\ref{ahodiffusiongauge}) using target time $\omega_{12}t_{\text{opt}}=0.25$. 
Triple lines indicate mean and error bars.
\normalsize}
\end{figure}

\begin{figure}[t]
\center{\includegraphics[width=\textwidth]{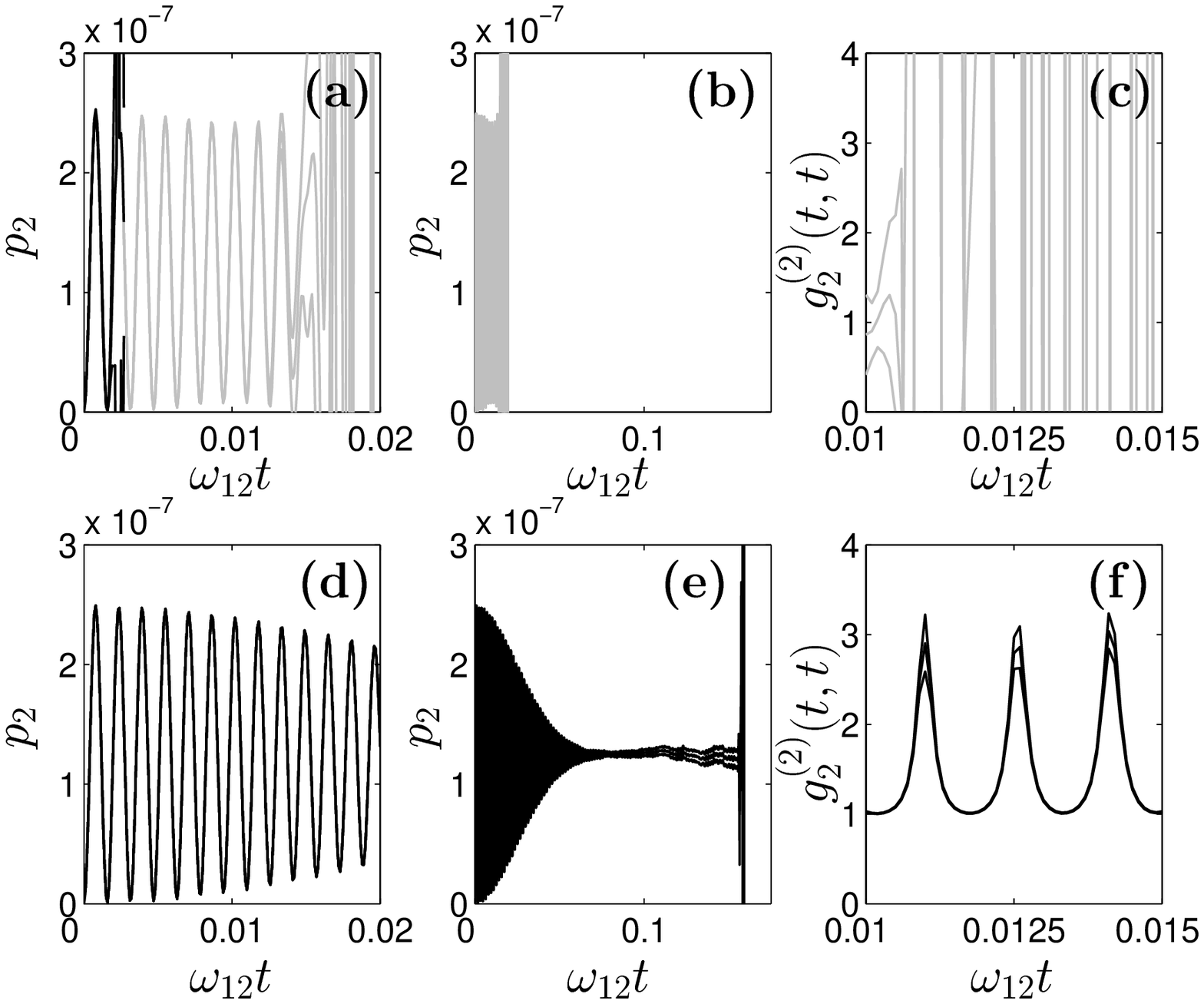}}\vspace{-8pt}\par
\caption[Coupling to vacuum mode: $n_0=10^4$]{\label{FIGURE2moden1e4}\footnotesize
Coupling to vacuum mode: \textbf{System with  very large number of particles}
($n_0=10^4$). Subfigures \textbf{(a)}, \textbf{(b)}, \textbf{(d)} and \textbf{(e)} show mode $2$ particle fraction, while  \textbf{(c)} and \textbf{(f)} show instantaneous particle number correlations in mode $2$.  Subfigures \textbf{(a)}--\textbf{(c)}  show results with positive P {\scshape light solid} and drift gauge (\ref{CCDgauge}) {\scshape dark solid}, whereas \textbf{(d)}-- \textbf{(f)} show results with combined drift gauge (\ref{ahodriftgauge}) and diffusion gauge (\ref{ahodiffusiongauge}) using target time $\omega_{12}t_{\text{opt}}=0.1$. 
Triple lines indicate mean and error bars.
\normalsize}
\end{figure}

\begin{figure}[t]
\center{\includegraphics[width=\textwidth]{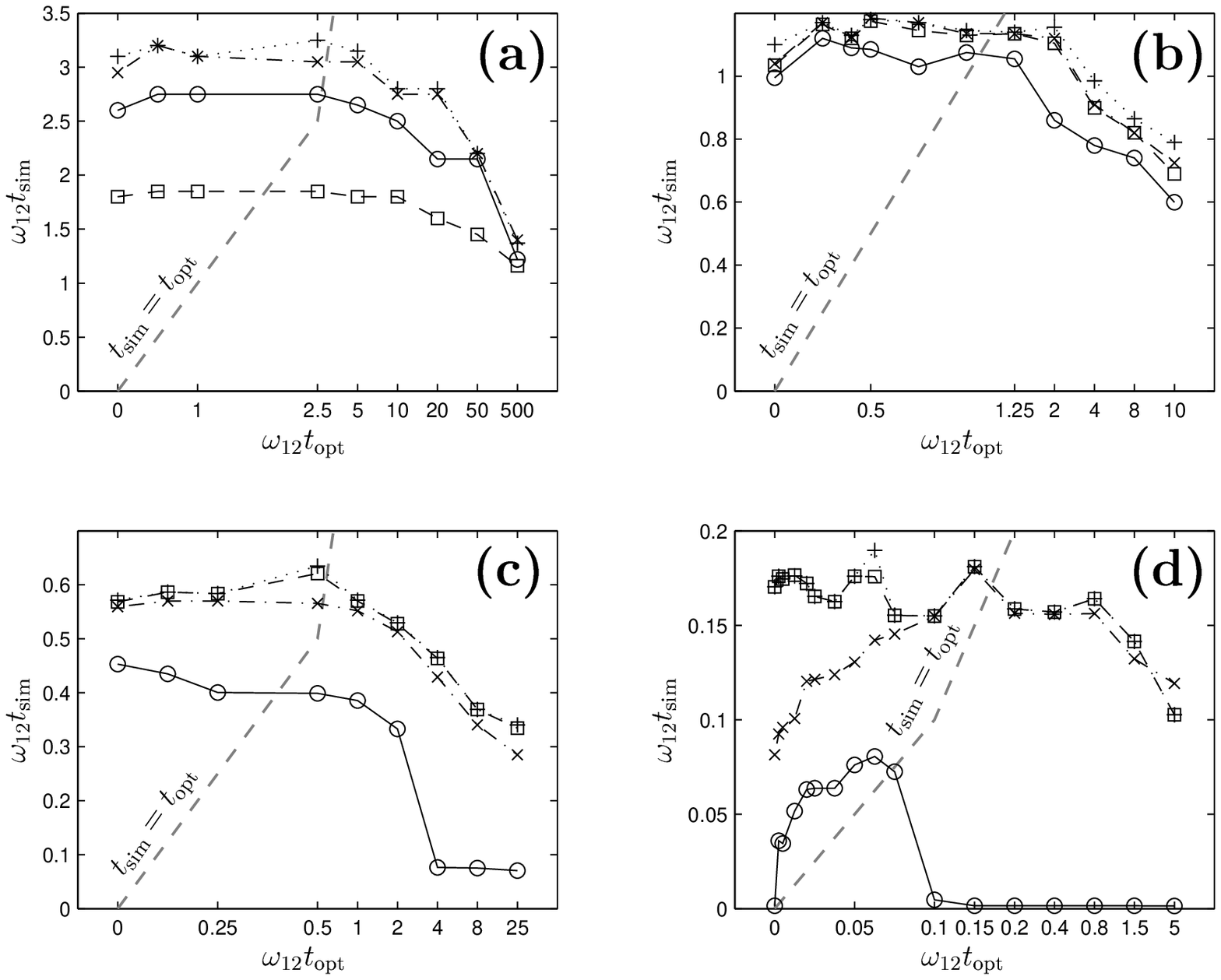}}\vspace{-8pt}\par
\caption[Coupling to vacuum mode $t_{\rm sim}$ vs. $t_{\rm opt}$]{\label{FIGURE2modett}\footnotesize
Dependence of \textbf{useful simulation time} $t_{\rm sim}$ on the target time parameter $t_{\rm opt}$ for simulations of Section~\ref{CH8Case1}
``$+$'': useful simulation of $\langle\op{n}_1\rangle$, 
``$\times$'': of $\langle\op{n}_2\rangle$,
``$\Box$'': of $g^{(2)}_1$,
``$\bigcirc$'': of $g^{(2)}_2$.
Nonlinear but monotonic $t_{\rm opt}$ scale used to avoid overlap of data points, while showing trend.
Subplot \textbf{(a)}: $n_0=1$; \textbf{(b)}: $n_0=17$; \textbf{(c)}: $n_0=200$; \textbf{(d)}: $n_0=10^4$.
$t_{\rm opt}=t_{\rm sim}$ shown for reference as {\scshape light dashed} line.
\normalsize}
\end{figure}

Example simulations are shown in Figures~\ref{FIGURE2moden1},~\ref{FIGURE2moden17},~\ref{FIGURE2moden200}, and~\ref{FIGURE2moden1e4}, and dependence of $t_{\rm sim}$ on the target time parameter $t_{\rm opt}$ are shown in Figure~\ref{FIGURE2modett}.

 Any data from positive P simulations after the onset of spiking ( numerical signature of possible boundary term errors) were discarded. The simulations up to such a point, showed no sign of any systematic biases\footnote{The onset of spiking naturally moves to slightly  earlier times as larger ensembles of trajectories are simulated. 
This may cause a beneficial ``bias-concealing'' effect in the following sense: As precision increases due to larger $\mc{S}$, some small boundary term biases that were originally concealed by noise might emerge, but simultaneously some rare trajectories with earlier spikes become included in the larger ensemble. These then warn one to discard data from an earlier time than one would have with the original smaller ensemble.}.

\subsection{Features seen}
\label{CH8Case1Features}
\ITEM{
\item With the single-mode gauge choices \eqref{ahodriftgauge} and \eqref{giig2mode} applied directly to the twin-mode system, improvement over the positive P and Carusotto\etal gauge \eqref{CCDgauge}\cite{Carusotto-01} is seen for all $n_0$
\item This improvement becomes more and more significant at high mode occupations,  like in the single-mode system. Generally speaking, the proportions between the simulation times obtained with the different gauge methods are qualitatively similar to the single-mode situation.
\item At large mode occupations, one can still simulate well beyond coherence times, so most interesting phenomena should not be missed (apart from quantum revivals, of course). (See Figures~\ref{FIGURE2moden200}--\ref{FIGURE2moden1e4})
\item However, simulation times with all methods are significantly smaller than those predicted (by using expected mode occupation $n_0$) for single modes from Tables~\ref{TABLEsimtime} or~\ref{TABLEtimesfit}, or \eqref{fitccd}--\eqref{fitga0}.
\item No biases from the exact solution are seen for the cases ($n_0=1,17$) where this was compared.
\item The $n_0=17$ case was also simulated by Carusotto\etal\cite{Carusotto-01} with (solely) the drift gauge \eqref{CCDgauge}, and a simulation time of $\approx0.55/\omega_{12}$ was seen. This is about $20\%$ longer than with the same gauge here. This effect is most likely because there total particle number conservation was {\it a priori} imposed on each trajectory rather than only in the ensemble mean as here. The extra freedom in the present simulations is necessary to allow straightforward extension to open systems, but the price paid appears to be some extra inefficiency.
\item Precision in the highly-occupied mode appears largely insensitive to the choice of target time $t_{\rm opt}$, as long as it is within reasonable values (See Figure~\ref{FIGURE2modett}). A choice of $t_{\rm opt}=0$ achieves very large improvement over positive P and drift-gauge only simulations for these observables. Note that $t_{\rm opt}=0$ is not the same as no diffusion gauge, because \eqref{giig2mode} becomes $g''_j =\frac{1}{4}\log(1+4\im{\breve{n}_j}^2)$, and this appears to have an important effect. The same applies to mode 2 when $n_0$ is not large.
\item At high $n_0$ the observables for the nearly-vacuum mode 2 become sensitive to target time choice (See Figure~\ref{FIGURE2modett}). This is particularly evident for $g^{(2)}_2$ when $n_0=10^4$. The optimum target time $t_{\rm opt}$ is then sharply peaked, much as in the single-mode case, and $t_{\rm sim}\approx \ge t_{\rm opt}$ in the optimum region where $t_{\rm opt}$ is of the order of coherence time for the highly-occupied mode $1$.
\item One point to note is that ``useful simulation time'' $t_{\rm sim}$ was based on the moment when relative error in a quantity was {\it first} found to be too large. In calculations of $g^{(2)}$, uncertainties are much greater when $g^{(2)}(0,t)$ peaks --- see e.g. Figure~\ref{FIGURE2moden1e4}\textbf{(f)}. Good accuracy can often be obtained between peaks for much longer times than shown in Figure~\ref{FIGURE2modett}, up to about the same simulation time as worked out based on the occupation number. This is particularly evident in the $n_0=10^4$ case, where the sudden drop-off in $t_{\rm sim}$ is due to this effect (see solid line of Figure~\ref{FIGURE2modett}\textbf{(d)}). 
\item Best simulation times at high $n_0$ occur for $t_{\rm opt}\le\approx t_{\rm sim}$.
\item Positive P simulations do better than those based on the drift gauge \eqref{CCDgauge} for all $n_0$.
\item Best simulation times (based on $\langle\op{n}_2\rangle$) with both gauges appear to scale roughly as 
\EQN{
\omega_{12}t_{\rm sim}\approx3/n_0^{1/3}
}
}
Collective conclusions are left till the end of the chapter.

\subsection{Comparison to Fock state stochastic wavefunctions}
\label{2MODECCD} 
The simulation of a seemingly similar $17$-particle two-mode system was carried out by Carusotto\etal\cite{Carusotto-01} using  the ``simple Fock state scheme'' with stochastic wavefunctions. This method gave a useful simulation time of 
$\approx3/\omega_{12}$ with $\mc{S}=2\times10^5$ trajectories. This is longer than seen with either the ``simple coherent state scheme'' therein ($\approx0.55/\omega_{12}$), or the simulation in Figure~\ref{FIGURE2moden17} here ($\approx1.18/\omega_{12}$). 
This was, however, for number-state initial conditions, so the two types of simulation cannot be easily directly compared, and the observable expectation values calculated are different.

 Coherent states with mean occupation $n_0$, used as starting points here, contain Fock components with much larger numbers of particles than $n_0$. Since the spread in the norm, and hence the uncertainty of observables, has been found to rise exponentially depending on the quantity $\bar{N}t=n_0 t$ (from equation (78) in \cite{Carusotto-01}), one expects that useful simulation time of the simple Fock state scheme will be $\approx\propto1/\,\bar{N}=1/n_0$. This is a sharp decrease compared to the $1/n_0^{1/3}$ dependence with the combined gauges discussed here, and one expects that for $n_0\gg17$, the Fock state wavefunction method will give much shorter simulation times.

One can attempt to compare schemes for $n_0=17$ in spite of the differences in the  physical system simulated. 
It is not clear how best to sample a coherent state (which includes Fock state components with different particle numbers)
with the Fock scheme, but one can estimate that for accuracy one should consider components with occupation at least three standard deviations away from the mean giving $\bar{N}\lesssim 30$ for such Fock state components. A trajectory corresponding to such a component  might be expected to not diverge for times $\omega_{12}t\lesssim 1.7$.

This is about $30\%$ longer than seen in Figure~\ref{FIGURE2moden17}, a factor similar to that noticed between simulations using the gauge \eqref{CCDgauge} in Ref.~\cite{Carusotto-01} using the ``simple coherent scheme'' and here.
It is likely the price paid for having the formalism left open to simulating gains and losses.

\section{Case 2: Coherent mixing of two identical modes}
\label{CH8Case2}
\subsection{Description}
\label{CH8Case2Description}
  The system starts initially with identical coherent states of mean particle number $n_0$ in each mode. The total mean particle number is $\bar{N}=2\,n_0$. During time evolution, Rabi coupling of the modes occurs along with some decoherence from inter-atom collisions in each mode.  
  This time simulations were carried out with constant particle number $n_0=100$, but the coupling frequency $\omega_{12}$ was varied. 

At low frequency $\omega_{12}\ll n_0=100$, decoherence local to each mode dominates, and phase oscillations in each mode occur with period $t_{\rm osc}=\pi/n_0$, while at high frequency $\omega_{12}\gg n_0=100$, the inter-mode coupling dominates and  phase oscillations for each mode occur with period $2\pi/\omega_{12}$. One expects that for low coupling the two modes should behave largely as two independent single modes of Chapter~\ref{CH7}.

  The particular values chosen to simulate were 
\EQN{
\omega_{12} = \{ 5000, 500, 50, 5, 0.5, 0.05, 0.005, 0.0005\}
.}
The observable of most interest here was $G^{(1)}(0,t)$ as in the single-mode case. Due to the symmetry of the system, all single-mode observables such as this have identical expectation values for both modes, so either mode can be considered.

The procedure was the same here as outlined in Section~\ref{CH8Case1Procedure}, with $\mc{S}=10^4$ trajectories per simulation.

\begin{figure}[t]
\center{\includegraphics[width=\textwidth]{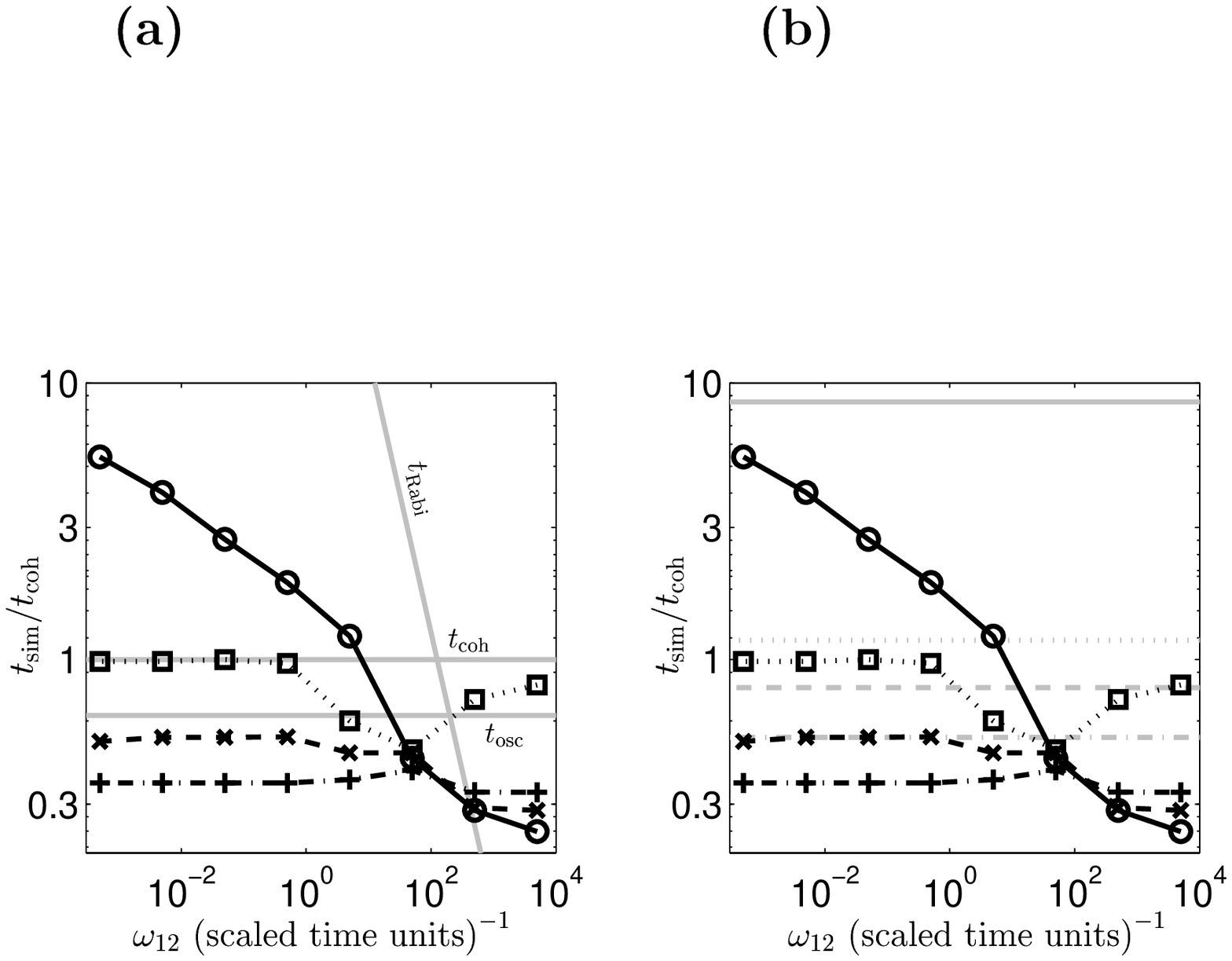}}\vspace{-8pt}\par
\caption[Simulation times for mixing of two identical modes]{\label{FIGUREcase2st}\footnotesize
\textbf{Useful simulation times} $t_{\rm sim}$ for a two identical modes undergoing coherent mixing as in Section~\ref{CH8Case2}.
Calculated simulation times are shown as data points, with the symbols denoting gauge used. ``$\Box$'': positive P; ``$\bigcirc$'': drift gauge \eqref{ahodriftgauge} and diffusion gauge \eqref{giig2mode} with $t_{\rm opt}=0$; ``$+$'' drift gauge \eqref{CCDgauge} of Carusotto\etal\cite{Carusotto-01}; ``$\times$'': drift gauge \eqref{ahodriftgauge} only.
 Subplot \textbf{(a)} compares to physical time scales, including Rabi oscillation period $t_{\rm Rabi}=2\pi/\omega_{12}$, while subplot \textbf{(b)}  compares to expected simulation times for a single mode using the empirical fits of Table~\ref{TABLEtimesfit} and \eqref{fitccd}--\eqref{fitga0}. The expected $t_{\rm sim}$ are plotted as light lines: {\scshape dotted}: positive P; {\scshape solid}: drift gauge \eqref{ahodriftgauge} and diffusion gauge \eqref{giig2mode} with $t_{\rm opt}=0$; {\scshape dashed}: with drift gauge \eqref{ahodriftgauge} only; {\scshape dash-dotted}: drift gauge \eqref{CCDgauge} of Carusotto\etal\cite{Carusotto-01}.
\normalsize}
\end{figure}

In Figure~\ref{FIGUREcase2st} actual simulation times are compared to physical timescales and expected  simulation times based on the single-mode analysis of Chapter~\ref{CH7}. Simulation times $t_{\rm sim}$ were based on the condition 
\eqref{usefulprecision} for the observable $|G^{(1)}(0,t)|$.

\begin{figure}[t]
\center{\includegraphics[width=\textwidth]{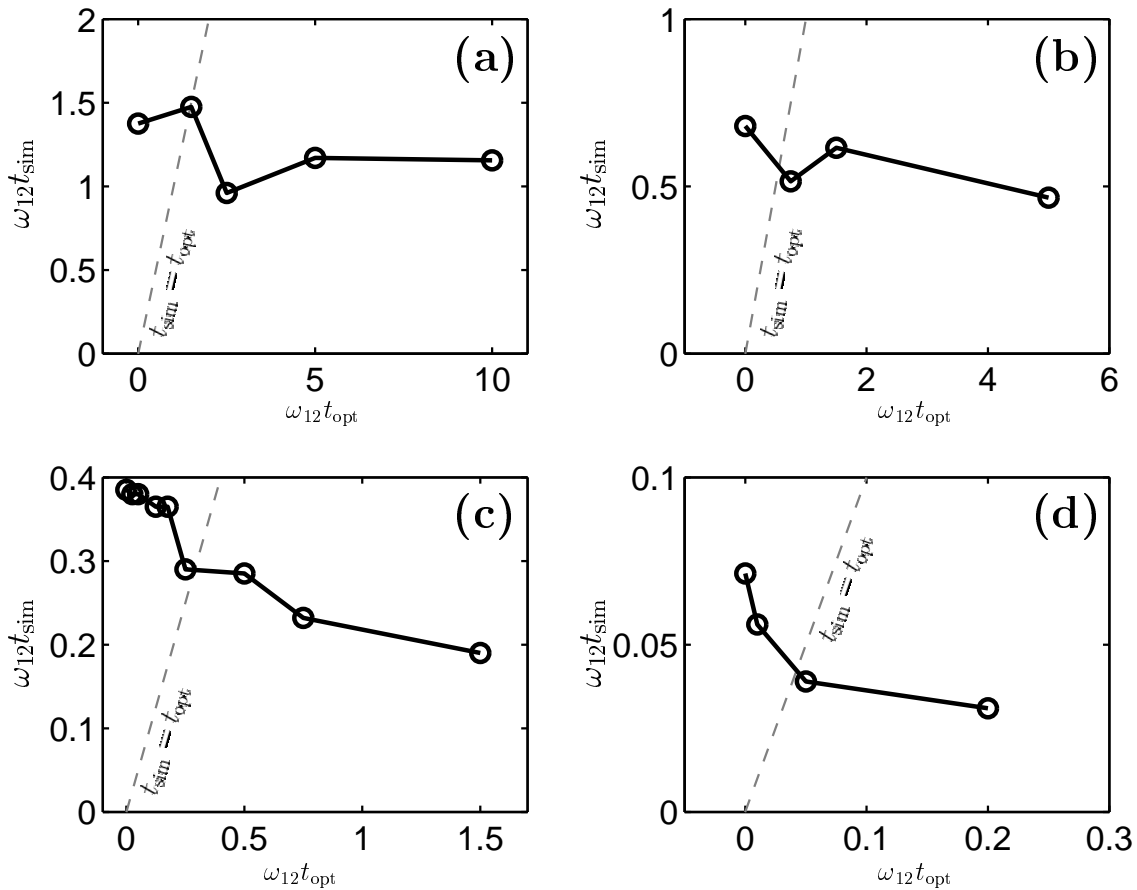}}\vspace{-8pt}\par
\caption[Mixing of two identical modes $t_{\rm sim}$ vs. $t_{\rm opt}$]{\label{FIGUREcase2tots}\footnotesize
Mixing of two identical modes:
Dependence of \textbf{useful simulation time} $t_{\rm sim}$ on the target time parameter $t_{\rm opt}$ for simulations of Section~\ref{CH8Case2}. Subplots \textbf{(a)} to \textbf{(d)} are for coupling strengths $\omega_{12}=\{0.0005, 0.05, 5, 500\}$, respectively.
Data are from single simulations with $\mc{S}=10^4$.
\normalsize}
\end{figure}

\begin{figure}[htb]
\center{\includegraphics[width=\textwidth]{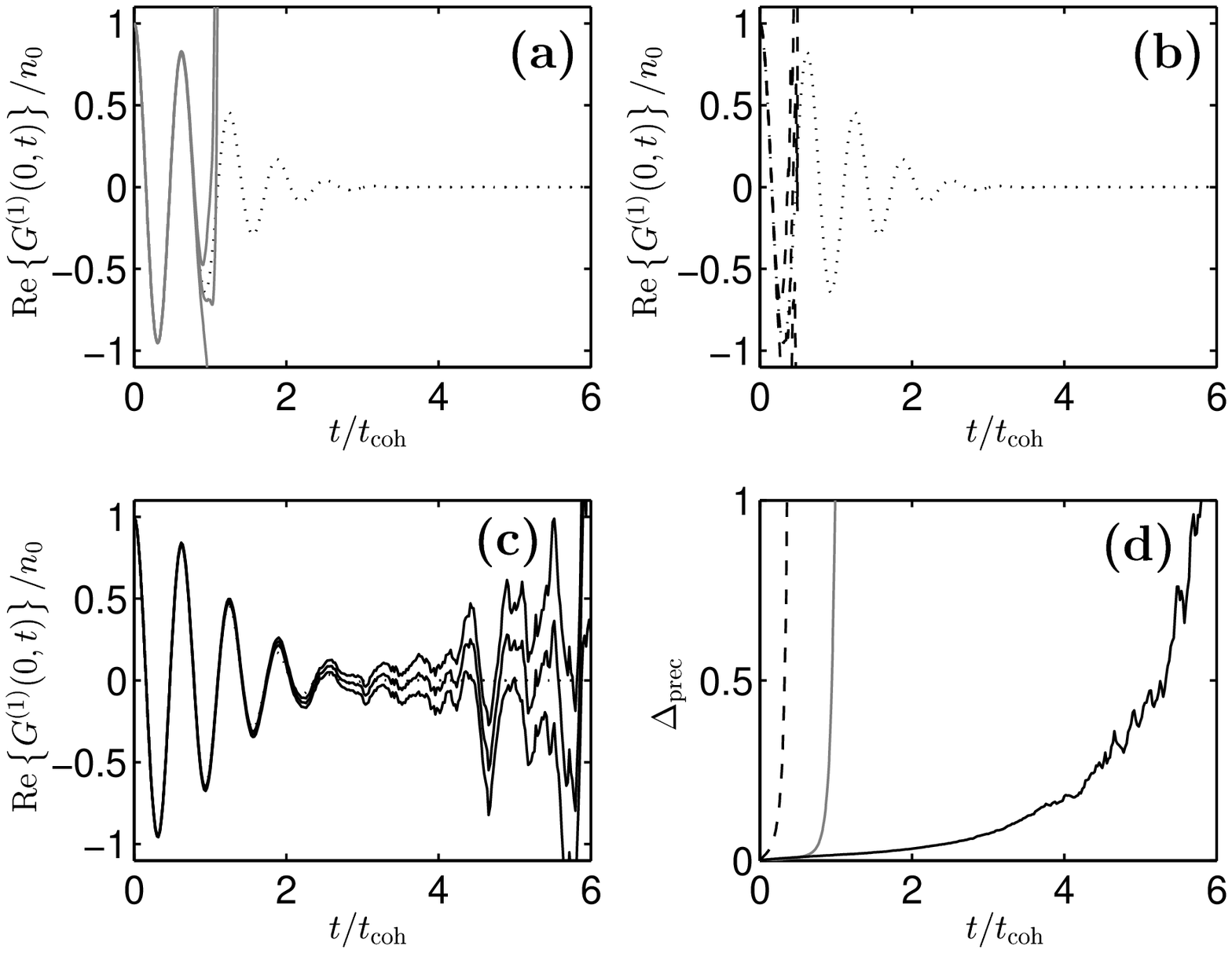}}\vspace{-8pt}\par
\caption[Mixing of two identical modes $\omega_{12}=0.0005$]{\label{FIGUREcase2c1e3}\footnotesize
Mixing of two identical modes:
\textbf{Weak coupling:} ($\omega_{12}=0.0005$). Subplots \textbf{(a)}--\textbf{(c)} show, respectively,  simulations with positive P ({\scshape light line}); with Carusotto\etal\cite{Carusotto-01} drift gauge \eqref{CCDgauge} ({\scshape Dashed line}); and with both drift gauge \eqref{ahodriftgauge} and diffusion gauge \eqref{giig2mode} ({\scshape Solid line}). The  exact solution \eqref{G1exact} for  no coupling ($\omega_{12}=0$) is shown as a {\scshape dotted line}.
 Subplot \textbf{(d)} shows relative sampling error in the estimate of $|G^{(1)}(0,t)|$, such that $\Delta_{\rm prec} = 10\sqrt{\mc{S}/10^6}\Delta|G^{(1)}(0,t)|/|G^{(1)}(0,t)$ is less then unity while ``useful'' precision is present, when defined as in \eqref{usefulprecision}. 
\normalsize}
\end{figure}

\begin{figure}[htb]
\center{\includegraphics[width=\textwidth]{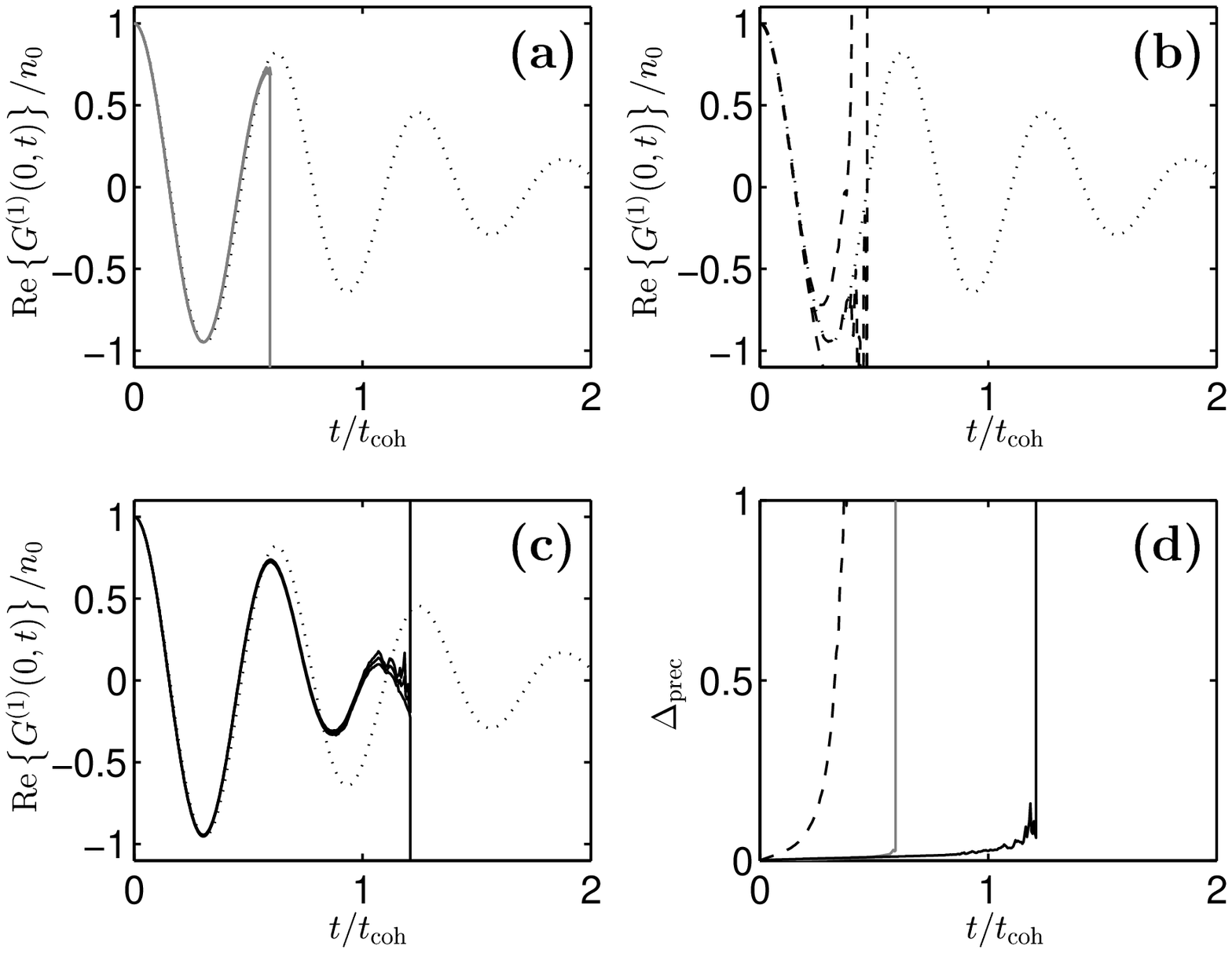}}\vspace{-8pt}\par
\caption[Mixing of two identical modes $\omega_{12}=5$]{\label{FIGUREcase2c01}\footnotesize
Mixing of two identical modes:
\textbf{Intermediate coupling:} ($\omega_{12}=5$). Subplots \textbf{(a)}--\textbf{(c)} show, respectively,  simulations with positive P ({\scshape Light line}); with Carusotto\etal\cite{Carusotto-01} drift gauge \eqref{CCDgauge} ({\scshape Dashed line}); and with both drift gauge \eqref{ahodriftgauge} and diffusion gauge \eqref{giig2mode} ({\scshape Solid line}). The  exact solution \eqref{G1exact}  for  no coupling ($\omega_{12}=0$) is shown as a {\scshape dotted line}.
 Subplot \textbf{(d)} shows relative sampling error in the estimate of $|G^{(1)}(0,t)|$, such that $\Delta_{\rm prec} = 10\sqrt{\mc{S}/10^6}\Delta|G^{(1)}(0,t)|/|G^{(1)}(0,t)$ is less then unity while ``useful'' precision is present, when defined as in \eqref{usefulprecision}. 
\normalsize}
\end{figure}

\begin{figure}[htb]
\center{\includegraphics[width=\textwidth]{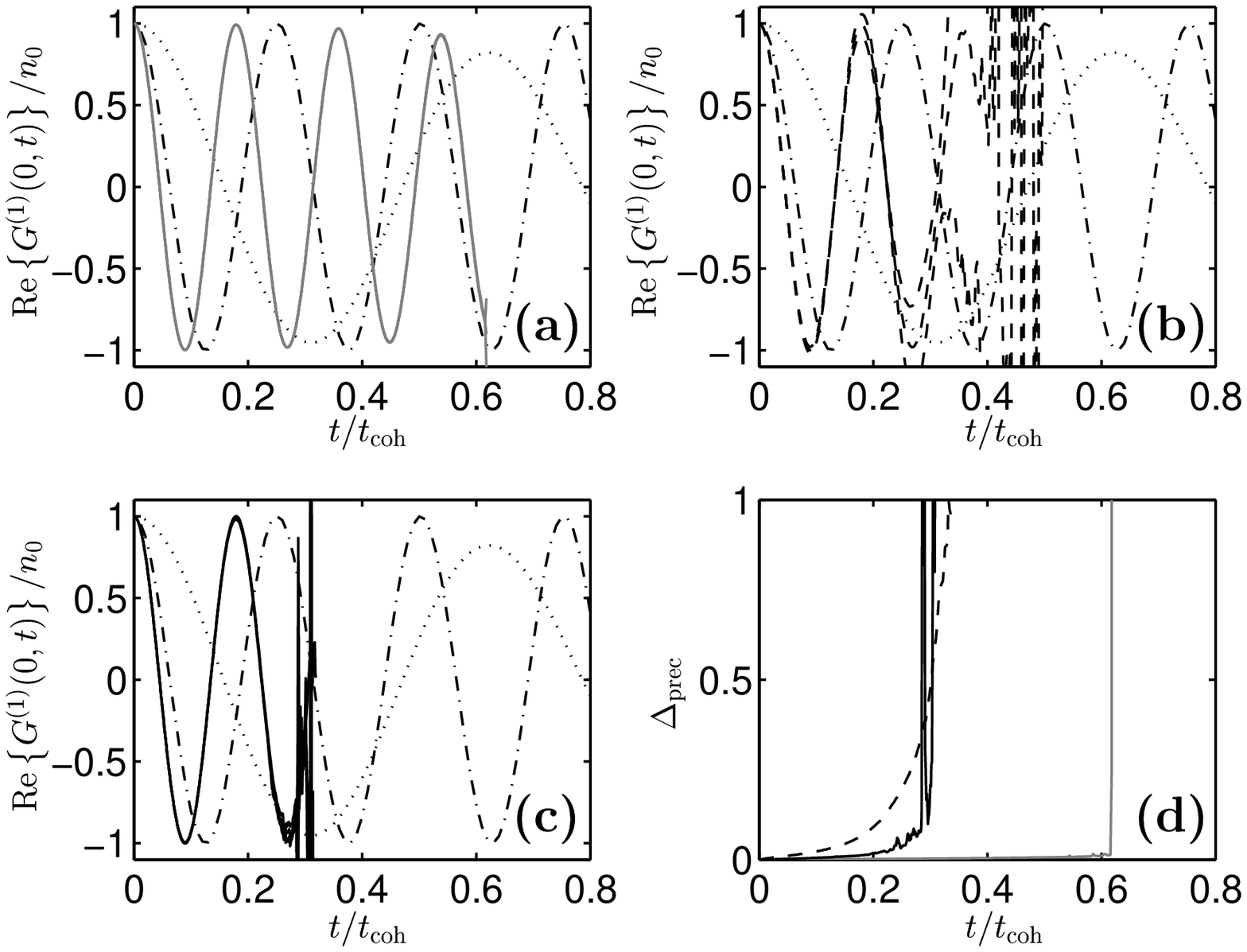}}\vspace{-8pt}\par
\caption[Mixing of two identical modes $\omega_{12}=500$]{\label{FIGUREcase2c1em3}\footnotesize
Mixing of two identical modes:
\textbf{Strong coupling:} ($\omega_{12}=500$). Subplots \textbf{(a)}--\textbf{(c)} show, respectively,  simulations with positive P ({\scshape Light line}); with Carusotto\etal\cite{Carusotto-01} drift gauge \eqref{CCDgauge} ({\scshape Dashed line}); and with both drift gauge \eqref{ahodriftgauge} and diffusion gauge \eqref{giig2mode} ({\scshape Solid line}). The  exact solution for  \eqref{G1exact}  no coupling ($\omega_{12}=0$) is shown as a {\scshape dotted line}, and for no collisions ($\omega_{12}\to\infty$) as a {\scshape dot-dashed line}.
 Subplot \textbf{(d)} shows relative sampling error in the estimate of $|G^{(1)}(0,t)|$, such that $\Delta_{\rm prec} = 10\sqrt{\mc{S}/10^6}\Delta|G^{(1)}(0,t)|/|G^{(1)}(0,t)$ is less then unity while ``useful'' precision is present, when defined as in \eqref{usefulprecision}. 
\normalsize}
\end{figure}

Figure~\ref{FIGUREcase2tots} shows data from a quick search for optimum target times $t_{\rm opt}$, while some
  example simulations are shown in Figures~\ref{FIGUREcase2c1e3},~\ref{FIGUREcase2c01}, and~\ref{FIGUREcase2c1em3}.

\subsection{Features seen}
\label{CH8Case2Features}
\ITEM{
\item At weak coupling $\omega_{12}\ll n_0$, the simulation times with various gauge choices behave qualitatively 
similar to the single mode of Chapter~\ref{CH7}, and vacuum coupled modes of Section~\ref{CH8Case1}. That is, the dual gauge combination \eqref{ahodriftgauge} and \eqref{giig2mode} allows simulation until the system has decohered, and give much longer simulation times than the positive P, drift gauge \eqref{ahodriftgauge} only, or Carusotto\etal gauge \eqref{CCDgauge} (in that order). The drift-gauge-only methods simulate for less time than a single phase oscillation period.
\item	Simulation times are shorter for all methods than those given in Table~\ref{TABLEsimtime} for the relevant single-mode case.
\item 	At strong coupling, $\omega_{12}\gg n_0$, the plain positive P method appears to give the best results out of the gauges tried. The times achievable with the gauged methods are much shorter. Relatively longest with the Carusotto\etal gauge \eqref{CCDgauge}, then the drift gauge \eqref{ahodriftgauge}, and worst performance is obtained from the diffusion-gauged simulation. 
\item At strong coupling, none of the methods tried reach coherence time $t_{\rm coh}$, but many Rabi oscillations can be simulated by all methods, since these are now much shorter than $t_{\rm osc}$.
\item  Choosing various target times $t_{\rm opt}$ appears not to have no effect apart from a worsening of the simulation at large $t_{\rm opt}$ values. Note that this was also the situation for the vacuum-coupled simulations in Section~\ref{CH8Case1} with the same total particle number $\bar{N}=200$, and so this lack of improvement with $t_{\rm opt}$ may be typical of lower mode occupations when any significant coupling $\omega_{12}$ is present.
\item At strong coupling, the Carusotto\etal gauge \eqref{CCDgauge} allowed somewhat longer simulation times, while the signal to noise ratio  in the estimate of $G^{(1)}(0,t)$ was lower with the gauge \eqref{ahodriftgauge} at short and intermediate times  --- see Figure~\ref{FIGUREcase2c1em3}\textbf{(d)}.
}

\section{Analysis and Conclusions}
\label{CH8Analysis}
   The two kinds of initial conditions considered above  represent the two kinds of situations generically occurring in all many-mode simulations. Coupling between modes of widely differing occupation will lead to behavior similar to that seen in Section~\ref{CH8Case1} (Case 1). Coupling between modes of similar occupation will display features similar to those seen in Section~\ref{CH8Case2} (Case 2). Adjacent spatial or momentum modes typically behave like Case 2, since if a field model is well resolved by the lattice, then physical properties (e.g. density, and hence mode occupation) should not change much over the distance between neighboring lattice points.

Still, this chapter has not really been a comprehensive assessment of gauge performance for all coupled mode cases, since 
  not all parameter regimes have been explored.
For the most general two modes starting in (off-diagonal) coherent state initial conditions one could be looking at the 5 complex parameters $\breve{n}_j,\,\alpha_j,\,\omega_{12}$ --- no mean feat to investigate.
Rather, this chapter has been a study of several commonly occurring cases to get an indication of whether the simulation improvements found for a single mode survive when inter-mode coupling is present, and to what degree.

Several aspects of the situation have emerged:
\ENUM{
\item When scattering within a mode dominates over the coupling between modes,  the response of the system to the gauges 
is similar to what was seen for a single mode. That is, 
\ENUM{
\item When mode occupation is high $\breve{n}_j\gg1$, use of the combined drift \eqref{ahodriftgauge} and diffusion \eqref{giig2mode} gauges developed in Chapter~\ref{CH7} gives dramatic improvement in useful simulation times and observable precision over the positive P method. In this regime, the drift-gauged only  (e.g. \eqref{ahodriftgauge} or \eqref{CCDgauge}) simulations reduce simulation time from even the positive P case, although boundary terms are removed.
Using both gauges \eqref{ahodriftgauge} and \eqref{giig2mode} allows simulation to times by which  full decoherence 
has occurred.
\item When mode occupation is low $n_0\lesssim \order{1}$, the improvement gained with the gauges is smaller although some improvement is always seen. The absolute simulation time is longer than for $n_0\gg1$, however.
}
Still, it turns out that in this strong scattering regime, one doesn't need much inter-mode coupling $\omega_{12}$ to reduce the simulation time in absolute terms by a factor $\order{2-10}$, although this reduction mainly affects the time after full decoherence has occurred, and not much is physically happening.

\item The beneficial effect of choosing $t_{\rm opt}>0$ appears to be strongly suppressed when the coupling $\omega_{12}$ is weak but not yet {\it very} weak, or the mode occupation is large but not yet {\it very} large. The $t_{\rm opt}=0$ diffusion gauge
\EQN{\label{giig0gauge}
g''_j(t_{\rm opt}=0)=\frac{1}{4}\log(1+4\im{\breve{n}_j}^2)
,}
 does however have a marked beneficial effect whenever two body scattering is significant,  and simulations using this gauge show very significant improvement over positive P or $g''_j=0$ simulations for a wide range of parameters. This  parameter range appears to be  $\omega_{12} \ll \breve{n}\chi$, and $\order{1}\lesssim n_0\lesssim\order{10^3}$. Simulation times obtained are smaller by about a factor of $\order{2}$ than those given for a single mode with the same gauge. 
At higher occupations, the benefit gained with  \eqref{giig0gauge} abates but nonzero $t_{\rm opt}$ values appear to become useful again, and continue to provide strong improvements over positive P simulations. (See, e.g. Figure~\ref{FIGUREcase1st}\textbf{(d)}.) 
The gauge \eqref{giig0gauge} is convenient also because there is no {\it a priori}  parameter $t_{\rm opt}$. 

\item When the inter-mode coupling dominates, the gauges developed in Chapter~\ref{CH7} or by Carusotto\etal\cite{Carusotto-01} do not appear to be very useful. They actually reduce simulation time as compared to the plain positive P method, although several Rabi oscillation periods can always be simulated.  

The transition between the strong and weak coupling  behavior appears to be at around 
\EQN{
\omega_{12}\approx \chi\breve{n}
,}
which is the point at which the expectation values of the coupling  and two-body scattering energies in the Hamiltonian are approximately equal.
}

At the level of the stochastic equations \eqref{2modeeqn}, the evolution of (say) $d\alpha_j$ can gather noise from three sources
\ENUM{
\item Directly from the local noise term $\propto \alpha_j \sqrt{\chi}\,dW_j$
\item Indirectly from the local nonlinear term $\propto \chi\alpha_j\breve{n}_j$, which can amplify variation in the local noise term.
\item From the other mode through the coupling term $\propto \omega_{12}\alpha_{\neg j}$.
}
The drift gauges  \eqref{ahodriftgauge} developed here or \eqref{CCDgauge} developed by Carusotto\etal\cite{Carusotto-01} 
neutralize source 2. The diffusion gauges \eqref{giig2mode} or \eqref{giintnopt} suppresses the direct noise source 1.
No local gauge can suppress the third source of randomness, however, because the randomness in $\alpha_{\neg j}$ is largely independent of any processes occurring in mode $j$. What happens is that even small randomness in one mode feeds into the other, can become amplified, and fed back again.   

The part of the diffusion gauge that is retained at $t_{\rm opt}=0$ appears more robust to this mixing than 
the additional optimization obtained with $t_{\rm opt}>0$.

Based on this, and the behavior $t_{\rm sim}$ vs. $t_{\rm opt}$ shown e.g. in Figures~\ref{FIGUREtopttprec} and~\ref{FIGURE2modett}, a way to proceed with a full many-mode simulation would be to 
\ENUM{
\item Run a simulation with local drift gauge \eqref{ahodriftgauge}, and $t_{\rm opt}=0$ diffusion gauge \eqref{giig0gauge}.
	Call the simulation time obtained $t_{\rm sim}^{(0)}$.
\item Run a simulation with $t_{\rm opt}=t_{\rm sim}^{(0)}$, and obtain a simulation time $t_{\rm sim}^{(1)}$.
\item If $t_{\rm sim}^{(1)}>t_{\rm sim}^{(0)}$ then nonzero $t_{\rm opt}$ is be beneficial, one expects $t_{\rm sim}\gtrsim t_{\rm opt}$, and one can make a search for a reasonable value by now running a simulation with $t_{\rm opt}=t_{\rm sim}^{(1)}$, and iterating in this manner until no more significant improvement is gained. 

If, on the other hand, $t_{\rm sim}^{(1)}\lesssim t_{\rm sim}^{(0)}$, then it appears that one is in a regime where the $t_{\rm opt}$ parameter does not affect the dominant noise source, and $t_{\rm sim}^{(0)}$ is the best simulation time one can obtain with the \eqref{ahodriftgauge}\&\eqref{giig2mode} dual gauge method.
\item If simulation time is too short, it may be worth trying a plain positive P simulation, or even a diffusion-gauge-only simulation using \eqref{giintnopt}, while carefully monitoring for the presence of spiking, and discarding any data beyond the first spike time. (The simulations carried out in this chapter with this approach did not show any bias when compared with exact results --- see e.g. Figures~\ref{FIGURE2moden1} and~\ref{FIGURE2moden17}).
}

In conclusion, it has been found in this chapter that simulation time and precision improvement can still be gained with coupled modes by using the local drift and diffusion gauges \eqref{ahodriftgauge} and \eqref{giig2mode}, although marked improvement will mostly be seen only when two-particle scattering dominates the inter-mode coupling. Other non-local gauges may improve simulations in a regime where inter-mode coupling dominates but development of such is for future research.

\REM{
\begin{table}[htb]
\caption[Coupling to vacuum mode: useful simulation times]{\label{TABLE2modecase1}\footnotesize
\textit{Useful simulation times} using various gauges, for the two mode system when coupling to a vacuum mode.
Values of $t_{\rm sim}$ taken from a single simulation, hence no error estimate given. Simulation times are calculated independently for the observables $\langle\op{n}\rangle$ and $g^{(2)}_j(t,t)$ by requiring single digit precision at $\mc{S}=10^6$, as in \eqref{usefulprecision}.
\normalsize}
\begin{minipage}{\textwidth}
\hfilll\begin{tabular}{|c|c|c||c|c|c|c||c|c|}
\hline
	&\multicolumn{2}{c|}{Gauge}	&\multicolumn{4}{c|}{$\omega_{12}t_{\rm sim}$}
									&	&		\\
\cline{2-7}
$n_0$	& 		&		&\multicolumn{2}{c|}{based on $\langle\op{n}_j\rangle$} 
							&\multicolumn{2}{c|}{based on $g^{(2)}_j(t,t)$}
									& $\omega_{12}t_{\rm opt}$
										&$\mc{S}$	\\
\cline{4-7}
	& Drift		& Diffusion	&mode 1	&mode 2	&mode 1	&mode 2	&	&		\\
\hline
\hline
1	& 0		& 0		& 2.1	& 2.1	& 2.0 	& 2.0 	& 	& $2\times10^5$	\\
	&\eqref{CCDgauge}& 0		& 2.1 	& 2.2	& 2.1	& 2.0 	&	& $2\times10^5$	\\
	&\eqref{ahodriftgauge}&\eqref{ahodiffusiongauge}	& 3.1	& 2.9 	& 1.8	& 2.6	& 0	& $2\times10^5$	\\
	&\eqref{ahodriftgauge}&\eqref{ahodiffusiongauge}	& 3.2 	& 3.0 	& 1.8 	& 2.7 	& 2.5	& $2\times10^5$	\\
	&\eqref{ahodriftgauge}&\eqref{ahodiffusiongauge}	& 2.2	& 2.2	& 1.4 	& 2.1 	& 50	& $1.9\times10^5$\\
\hline
17	& 0		& 0		& 0.60 	& 0.60 	& 0.60 	& 0.57	& 	& $2\times10^5$	\\
	&\eqref{CCDgauge}& 0		& 0.50	& 0.51	& 0.46	& 0.51	&	& $2\times10^5$	\\
	&\eqref{ahodriftgauge}&\eqref{ahodiffusiongauge}	& 1.10	& 1.04	& 1.03 	& 0.99 	& 0	& $2\times10^5$	\\
	&\eqref{ahodriftgauge}&\eqref{ahodiffusiongauge}	& 1.18	& 1.18 	& 1.17 	& 1.08  & 0.5	& $2\times10^5$	\\
	&\eqref{ahodriftgauge}&\eqref{ahodiffusiongauge}	& 0.85	& 0.85	& 0.81	& 0.73 	& 8	& $0.8\times10^5$\\
\hline
200	& 0		& 0		& 0.192	& 0.162	& 0.190 & 0.127 & 	& $2\times10^5$	\\
	&\eqref{CCDgauge}& 0		& 0.076 & 0.070 & 0.069	& 0.067 &	& $2\times10^5$	\\
	&\eqref{ahodriftgauge}&\eqref{ahodiffusiongauge}	& 0.574 & 0.562 & 0.567	& 0.441	& 0	& $0.6\times10^5$	\\
	&\eqref{ahodriftgauge}&\eqref{ahodiffusiongauge}	& 0.586 & 0.565 & 0.586 & 0.459	& 0.25	& $2\times10^5$	\\
	&\eqref{ahodriftgauge}&\eqref{ahodiffusiongauge}	& 0.369	& 0.369	& 0.369	& 0.073 & 8	& $0.5\times10^5$\\
\hline
$10^4$	& 0		& 0		&0.0197	&0.0109	&0.0196	&0.0046	& 	& $2\times10^5$	\\
	&\eqref{CCDgauge}& 0		&0.0016	&0.0015	&0.0014	&0.0014	&	& $2\times10^5$	\\
	&\eqref{ahodriftgauge}&\eqref{ahodiffusiongauge}	&0.1703	&0.0815	&0.1703	&0.0015	& 0	& $2\times10^5$	\\
	&\eqref{ahodriftgauge}&\eqref{ahodiffusiongauge}	&0.1635	&0.1299	&0.1635	&0.0791	& 0.05	& $2\times10^5$	\\
	&\eqref{ahodriftgauge}&\eqref{ahodiffusiongauge}	&0.1550	&0.1548	&0.1549	&0.0046	& 0.1	& $2\times10^5$\\
	&\eqref{ahodriftgauge}&\eqref{ahodiffusiongauge}	&0.1414	&0.1322	&0.1414	&0.0015	& 0.15	& $0.5\times10^5$\\
\hline
\end{tabular}\hfilll
\end{minipage}
\end{table}
}

\chapter{Single-mode interacting Bose gas thermodynamics}
\label{CH9}

  In Section~\ref{CH5Thermo} it was shown that the gauge P representation can also be used
to calculate properties of the grand canonical ensemble of  an interacting Bose gas described by the lattice Hamiltonian \eqref{latticeH}. A convenient feature of calculations made with this method (rather than, say path integral Monte Carlo, or variational Monte Carlo methods) is that a single simulation can estimate any observables, and gives results for a wide range of temperatures $T\ge 1/k_B\text{max}[\tau]$. 

  Before attempting full many-mode simulations (see Chapter~\ref{CH11}), it is pertinent to make sure that the simplest special case of a single-mode is correctly simulated. This is a good test case because it can easily be solved exactly for a unambiguous comparison, and because it is simple enough that a broad analysis of the statistical behavior of the simulation is possible.

The single-mode model is particularly relevant to a locally-interacting many-mode Hamiltonian \eqref{deltaH}. Here the stochastic equations to simulate are given by \eqref{gaugepthermo}, and it can be seen that the nonlinearity and noise in the evolution of local amplitudes $\alpha_{\bo{n}}$ for each mode $\bo{n}$ depend only on the local variables $\alpha_{\bo{n}}$ and $\beta_{\bo{n}}$. 
In such a situation the nonlinear and stochastic features of the many-mode behavior should appear in their entirety in the single-mode toy model.

\section{The single-mode model}
\label{CH9Model}
\subsection{Quantum description}
\label{CH9ModelQuantum}
Isolating a single mode from the description of Section~\ref{CH2Thermodynamics},
the master equation for the evolution of the (un-normalized) density matrix is 
\EQN{\label{canmaster}
\dada{\op{\rho}_u}{\tau} = \left[\mu_e(\tau)\op{n} - \op{H}\right]\op{\rho}_u
}
with $\op{n}=\dagop{a}\op{a}$ (the number operator) defined in the usual way in terms of boson creation and annihilation operators.
The ``imaginary time'' parameter is $\tau=1/k_BT$, and the ``effective chemical potential'' $\mu_e$ is defined in \eqref{muedef}.
That expression can be integrated to give 
\EQN{\label{muexpr}
\mu(\tau) = \frac{1}{\tau}\left[-\lambda_n+ \int_0^{\tau}\mu_e(t)dt\right]
}
with constant $\lambda_n$, and it is then seen that  at low temperatures $\mu(T)\to\mu_e(T)$.
The single-mode analogue of the interacting Bose gas  Hamiltonian is
\EQN{
\op{H} = \hbar\chi\op{n}\left(\op{n}-1\right)
,}
where any linear contribution $\propto\op{n}$ has been amalgamated into the chemical potential. 
The initial state is (as per \eqref{rhouo})
\EQN{
\op{\rho}_u(0) = \exp\left[-\lambda_n\op{n}\right]
.}
Physically, this single-mode model is an approximation of e.g. a boson orbital in a heat bath at temperature $T$, and chemical potential $\mu$, where this chemical potential includes any kinetic and external potential effects.

\subsection{Gauge P stochastic equations}
\label{CH9ModelEquations}

Defining 
\EQN{
\breve{n}=\alpha\beta=n'+in''
} 
as in previous chapters,  the Ito stochastic equations in a gauge P representation are obtained from \eqref{gaugepthermo}: 
\SEQN{\label{canequations}}{
  d\alpha &=&	\left(\mu_e-2\hbar\chi\breve{n}\right)\alpha\,d\tau + i\alpha\sqrt{2\hbar\chi}\left(\,dW-\mc{G}\,d\tau\right)\\
  d\beta &=& 0\\
  d\Omega &=& \Omega\left[ \left(\mu_e-\breve{n}\right)n\,d\tau +\mc{G}\,dW\right]
.}
 The Wiener increments $dW$ can be implemented as independent real Gaussian noises of mean zero, and variance $d\tau$ at each time step of the simulation. The single complex drift gauge is $\mc{G}$. 

Because only a single non-weight variable $\alpha$ experiences diffusion, then the un-gauged diffusion matrix is $\ul{D}=-2\hbar\chi\alpha^2$, a single complex function. There are then no standard diffusion gauges (as described in Section~\ref{CH4DiffusionCanonical}) because the orthogonal $1\times 1$ matrix \eqref{diffusiongaugeexplicit} is just $O=1$, apart from a phase factor.

Initial conditions on the variables $\alpha$ and $\beta$ are (from \eqref{inidist})
\EQN{\label{pginit}
  P_G(\alpha,\beta,\Omega) = \delta^2(\Omega-1)\delta^2(\beta-\alpha^*)\frac{1}{\pi\bar{n}_0}\exp\left(\frac{-|\alpha|^2}{\bar{n_0}}	\right)
,}
where 
\EQN{
\bar{n}_0 = \frac{1}{e^{\lambda_n}-1}
,}
is the mean particle number at high temperature. $\alpha$ is easily sampled using Gaussian random variables.

The mean number of particles is $\langle\op{n}\rangle = \tr{\op{\rho}_u\op{n}}/\tr{\op{\rho}_u}$, and is estimated by
\EQN{
\bar{n}=\frac{\average{\re{\breve{n}\Omega}}}{\average{\re{\Omega}}}
.}

These stochastic equations exhibit completely different behavior to the dynamics of this system \eqref{dynamixdab}. Rather they are more similar to the two-boson absorption of Section~\ref{CH6Absorber}.

\subsection{Exact solution}
\label{CH9ModelExact}
  The quantum evolution of this model is easily evaluated exactly in a Fock number state basis $\ket{n}$. This is useful to make a definitive check of the correctness of the stochastic simulations. Using $\op{\rho_u} = \exp\left[\mu\tau\op{n}-\op{H}\tau\right]$, the density matrix elements in this basis 
$\rho_{n\wt{n}}=\bra{n}\op{\rho}_u\ket{\wt{n}}$
 are found to be:
\EQN{\label{exactrho}
\rho_{n\wt{n}} &=& \delta_{n\wt{n}}\exp\left\{\left[\mu-\hbar\chi(n-1)\right]n\tau\right\}
.}

A long times $\tau\to\infty$, i.e. low temperatures, the density matrix will be dominated by the populations for which the exponent is largest.  This maximum of the exponent  occurs when
$n_{\infty}=[\lim_{\tau\to\infty}\mu(\tau)+\hbar\chi]/2\hbar\chi$. Since $n$ takes on only integer values, then the dominant mode occupation will be at the nearest integer to $n_{\infty}$ (or the two nearest components if $n_{\infty}$ takes on half-integer values). Using \eqref{muexpr}, 
\EQN{\label{nextrdef}
n_{\infty}  = \frac{1}{2}+\frac{1}{2\hbar\chi}\lim_{\tau\to\infty}\left[\frac{1}{\tau}\int_0^{\tau}\mu_e(t)dt\right]
,}
and the $T\to0$ state $\op{\rho}(\tau\to\infty)$ is:
\ITEM{
\item A vacuum if $n_{\infty}<0.5$.
\item Otherwise, if $n_{\infty}$ is a  half integer, then $\op{\rho}(\tau\to\infty)$ is an equal mixture of the $\ket{n_{\infty}\pm\half}$ states.
\item Otherwise, $\op{\rho}(\tau\to\infty)$ is the Fock number state with the nearest whole occupation number to $n_{\infty}$.
}

\section{Moving singularities and removal with gauges}
\label{CH9Gauge}

\subsection{Moving singularity}
\label{CH9GaugeMvsing}
\begin{figure}[t]
\center{\includegraphics[width=\textwidth]{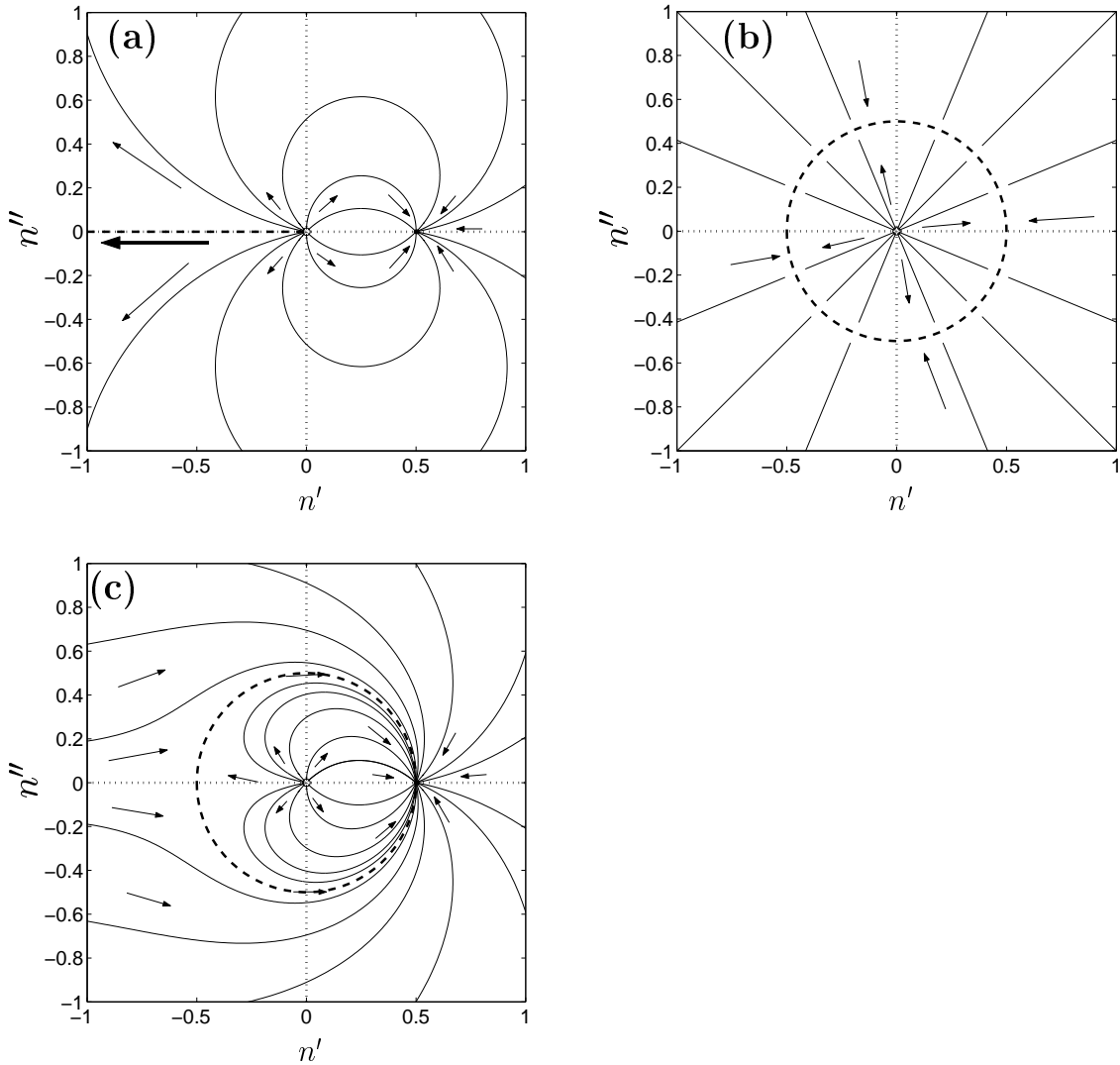}}\vspace{-8pt}\par
\caption[Deterministic phase space for single-mode thermodynamics]{\label{FIGUREcanphase}\footnotesize
\textbf{Deterministic phase space} for Stratonovich form of the $d\breve{n}$ equation.  The case of $n_{\infty}=0.5$ is shown. 
\textbf{(a)}: positive-P-like simulation $\mc{G}=0$; 
\textbf{(b)}: radial gauge  \eqref{radialgauge};
\textbf{(c)}: minimal gauge \eqref{minimalgauge}.
The moving singularity in \textbf{(a)} is shown with large arrow, the attractors at $|\breve{n}|=n_{\infty}$ with a thick dashed line.
\normalsize}
\end{figure}

Consider, for now, the case of constant effective chemical potential $\mu_e$.
From \eqref{canequations}, and using \eqref{stratcorrection}, the Stratonovich stochastic equation for the complex occupation variable $\breve{n}$ is 
\EQN{\label{dbreven}
  d\breve{n} &=&	\breve{n}\left[\mu_e\,d\tau-\hbar\chi(2\breve{n}-1)\,d\tau+i\sqrt{2\hbar\chi}\left(\,dW-\mc{G}\,d\tau\right)\right]
.}
In a positive-P-like  simulation\footnote{But with a weighting gauge to allow for complex weight evolution $d\Omega$.}, $\mc{G}=0$, and the deterministic part of the $d\breve{n}$ equation has the phase-space structure shown in Figure~\ref{FIGUREcanphase}\textbf{(a)}. There are stationary points at vacuum and at
\EQN{
\breve{n}=n_{\infty}=\Half+\frac{\mu_e}{2\hbar\chi}
,}
with the more positive stationary point being an attractor, and the more negative a repellor.
The deterministic evolution is easily solved, and with initial condition $\breve{n}(0)=n_0$ gives
\EQN{\label{candetsol}
\breve{n}(\tau) = \frac{n_{\infty}n_0}{n_0+(n_{\infty}-n_0)e^{-2\hbar\chi n_{\infty}\tau}}
.}
A moving singularity appears along the negative real axis, and if one starts with a negative real $n_0=-|n'_0|$, then 
$\breve{n}\to \infty$ at time 
\EQN{
\tau_{\rm sing} = \frac{1}{2\hbar\chi n_{\infty}}\log\left(1-\frac{n_{\infty}}{n_0}\right)
.}

Since initial conditions \eqref{pginit} lead to samples with  $n_0\in(0,\infty)$, then some low-$n_0$ trajectories can be expected to rapidly diffuse into the negative real part of phase space where the super-exponential growth occurs. 
This will tend to cause either misleading systematic boundary term errors or uncontrolled spiking, which renders the simulation useless after a short time. As is seen in Figure~\ref{FIGUREcanppgp} in this case it is spiking.

\subsection{Minimal drift gauge}
\label{CH9GaugeMinimal}

To correct the problem one has to change the phase-space topology
in some way to prevent the occurrence of such moving singularities. 
The radial evolution $d|\breve{n}|=-2\hbar\chi|\breve{n}|n'\,d\tau +\dots $ is at fault, and the offending term can be removed with the gauge
\EQN{\label{minimalgauge}
\mc{G} = i\sqrt{2\hbar\chi}\left(\re{n}-|\breve{n}|\right)
.}
This gauge has the effect of replacing the $-2\hbar\chi|\breve{n}|\re{n}\,d\tau$ term in $d|\breve{n}|$  with 
$-2\hbar\chi|\breve{n}|^2\,d\tau$, which always attracts trajectories into the phase space in the vicinity of the origin. Polar phase evolution is unchanged, and 
the resulting deterministic phase space is shown in Figure~\ref{FIGUREcanphase}\textbf{(c)}.
If $n_{\infty}>0$ the attractor and repellor remain, while for $n_{\infty}\le0$, there is just an attractor at vacuum.
For near-classical trajectories having real $\breve{n}$, phase-space evolution is unchanged, and the gauge is zero.

How do the modified equations measure up to the gauge choice criteria of Section~\ref{CH6RemovalHeuristic}? In order:
\ENUM{
\item The moving singularity in $d\breve{n}$ has been removed (at high $|\breve{n}|$ the deterministic behavior is restorative towards the origin), and no other moving singularities are present. Note that if $\breve{n}$ remains finite, then so does $\Omega$, which just experiences exponential growth with an exponent that is always finite, since it depends only on $\breve{n}$.
\item No new moving singularities have been introduced.
\item No noise divergences (new or old) are present since all noise terms satisfy \eqref{mvsingcondition}.
\item No discontinuities in the drift or diffusion coefficients are present provided the time-dependence of $\mu_e(\tau)$ is not singular.
\item Is the weight spread minimized?
\ENUM{
\item Gauge is zero when $\breve{n}$ is positive real, and small in its neighborhood, where gauge corrections are unnecessary.
\item Explicit variational minimization of $\mc{G}$ has not been carried out, but rather criteria 6 have been applied.
\item Gauge is zero at deterministic attractor $\breve{n}=n_{\infty}$.
}
\item No unwanted gauge behaviors are present:
\ENUM{
\item Gauge is nonzero in a large part of phase space, so most trajectories contribute to removal of bias.
\item Gauge changes smoothly in phase space.
\item Gauge is autonomous.
\item Gauge breaks the analyticity of the equations.
}
}

\subsection{Radial drift gauge}
\label{CH9GaugeRadial}

The gauge \eqref{minimalgauge} looks good, but it was found that a more severe phase-space modification usually gives smaller statistical uncertainties. This better gauge form is
\EQN{\label{radialgauge}
\mc{G} = i\sqrt{2\hbar\chi}\left(\breve{n}-|\breve{n}|\right)
,}
which will be called here the ``radial'' gauge, due to elimination of any deterministic polar-angle evolution as shown in Figure~\ref{FIGUREcanphase}\textbf{(b)}. The deterministic attractor is the entire $|\breve{n}|=n_{\infty}$ circle (or the vacuum if $n_{\infty}\le0$), with some phase diffusion in the phase of $\breve{n}$.

\begin{figure}[t]
\center{\includegraphics[width=12cm]{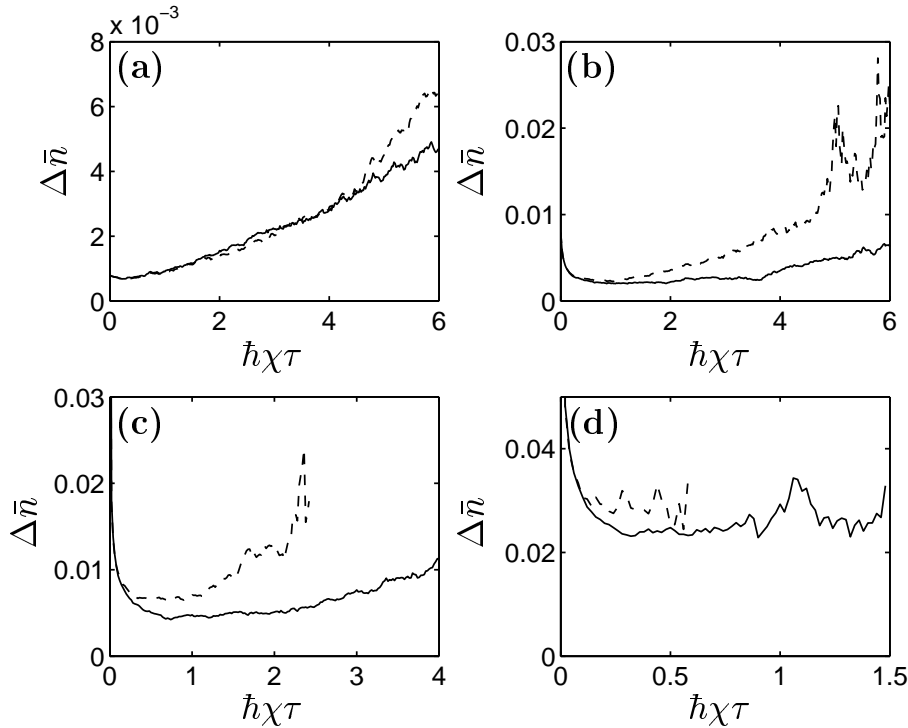}}\vspace{-8pt}\par
\caption[Efficiency of two gauges]{\label{FIGUREgagr}\footnotesize
\textbf{Uncertainty in $\bar{n}$}, the estimate of $\langle\op{n}\rangle$ for the $\mu_e=0$ system. 
\textbf{(a)}: $\bar{n}_0=0.1$;
\textbf{(b)}: $\bar{n}_0=1$;
\textbf{(c)}: $\bar{n}_0=10$;
\textbf{(d)}: $\bar{n}_0=100$;
{\scshape Solid line}: radial gauge \eqref{radialgauge}; {\scshape Dashed line}: gauge \eqref{minimalgauge}.
Simulations were carried out with $\mc{S}=2\times10^4$ trajectories. 
\normalsize}
\end{figure}

The efficiency of the two gauges is compared in Figure~\ref{FIGUREgagr} for  $\mu_e=0$ and a variety of initial $\bar{n}_0$.
It is not immediately clear why the radial gauge is better, but the improvement is significant whenever $\bar{n}_0\ge\order{1}$.
Similar behavior was seen for other $\mu_e$ values. In particular also for $\mu_e=\hbar\chi$, which corresponds to a different low temperature state (i.e. a pure Fock number state --- see Figure~\ref{FIGUREcanng}).

Comparison to the heuristic gauge choice criteria of Section~\ref{CH6RemovalHeuristic} follows through identically as for the previous gauge \eqref{minimalgauge}, apart from the gauge not being zero on the entire $|\breve{n}|=n_{\infty}$ attractor but only at $\breve{n}=n_{\infty}$.

\section{Numerical simulations}
\label{CH9Numerical}

\begin{figure}[p]
\center{\includegraphics[width=\textwidth]{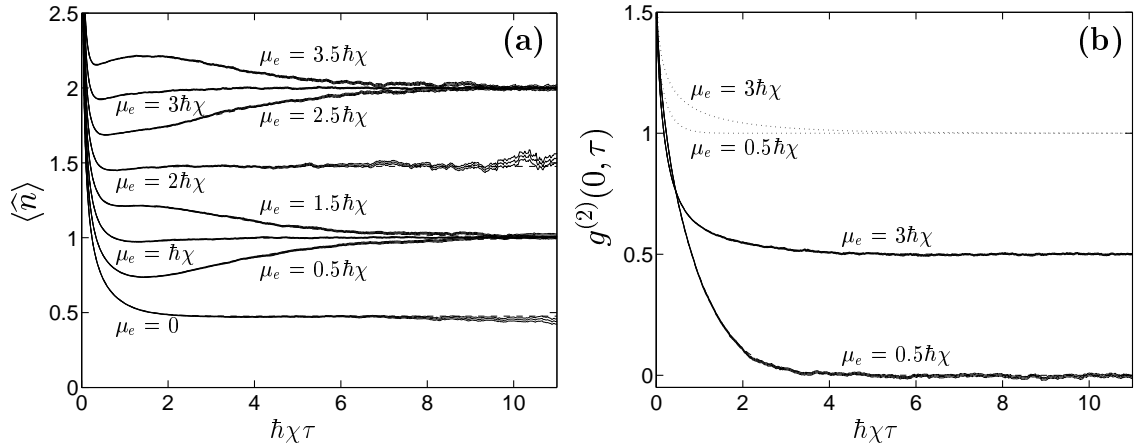}}\vspace{-8pt}\par
\caption[Mean occupation and two-body correlations]{\label{FIGUREcanng}\footnotesize
\textbf{Temperature dependent observables} given constant $\mu_e$ and $\bar{n}_0=10$ at high temperature. 
{\scshape solid lines}: simulation with $\mc{S}=2\times10^5$ trajectories. Triple lines indicate error bars (often not resolved at this scale).
{\scshape dashed lines}: exact values (mostly obscured by simulation results)
{\scshape dotted lines}: mean field calculation (subplot \textbf{(b)}).
The second order correlation function $g^{(2)}(0,\tau)$ is given by \eqref{g2def}.
\normalsize}
\end{figure}

\begin{figure}[p]
\center{\includegraphics[width=9cm]{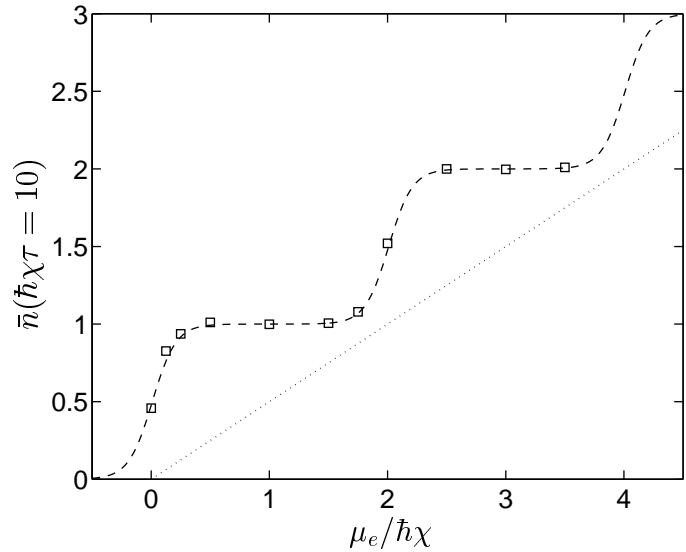}}\vspace{-8pt}\par
\caption[Mode occupations at low temperature]{\label{FIGUREnstep}\footnotesize
\textbf{Mean mode occupations at low temperature} $T=\hbar\chi/10k_B$ as a function of constant $\mu_e$.
{\scshape dashed}: exact result --- note the quantization, {\scshape dotted}: mean field result, {\scshape squares}: simulation results with radial gauge \eqref{radialgauge}. Uncertainty was of symbol size or smaller.
\normalsize}
\end{figure}

\begin{figure}[t]
\center{\includegraphics[width=7.5cm]{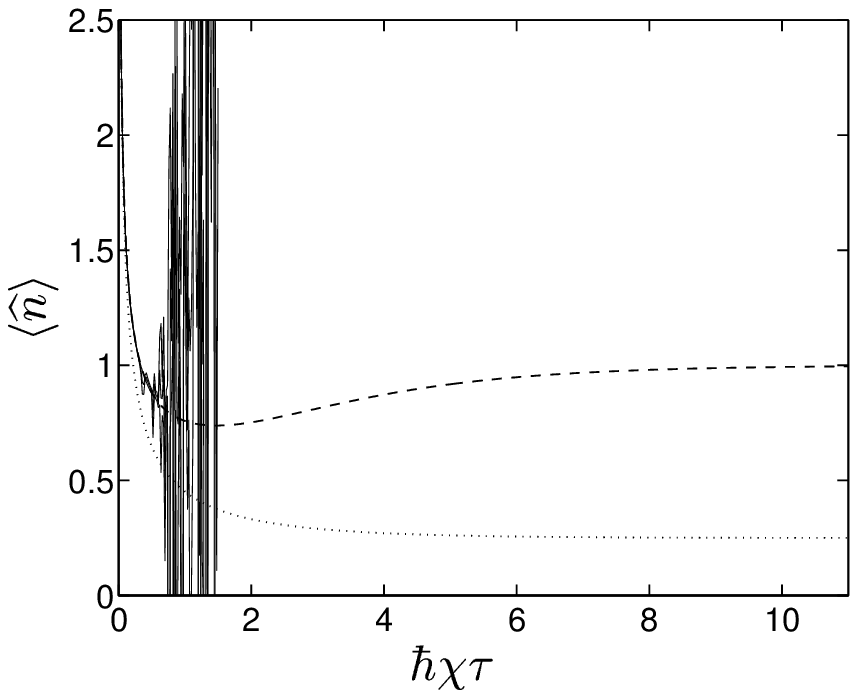}}\vspace{-8pt}\par
\caption[Performance of un-gauged and mean field calculations]{\label{FIGUREcanppgp}\footnotesize
Performance of the mean field and un-gauged simulations for comparison with Figure~\ref{FIGUREcanng}\textbf{(a)}.
\textbf{Mean mode occupation} with $\mu_e=0.5\hbar\chi$ and $\bar{n}_0=10$.
{\scshape dashed}: exact result, {\scshape dotted}: mean field result, {\scshape solid}: un-gauged simulation.
\normalsize}
\end{figure}

The results of some example simulations using the radial gauge \eqref{radialgauge} are shown in Figures~\ref{FIGUREcanng} and~\ref{FIGUREnstep}. These are  compared to exact results obtained from \eqref{exactrho} and also to mean field semiclassical calculations. As outlined in Section~\ref{CH5GP}, mean field calculations such as solution of Gross-Pitaevskii equations
are equivalent to simulation of only the deterministic part of the Ito gauge P equations.

Additionally, a positive-P-like simulation with no gauge ($\mc{G}=0$) is shown in Figure~\ref{FIGUREcanppgp} for comparison with the gauged technique.

A subtlety to keep in mind in these thermodynamics simulations is that (in contrast to dynamics) the normalization $\langle\re{\Omega}\rangle$, which appears in the denominator of observable estimates \eqref{observables}, is not unity in the $\mc{S}\to\infty$ limit. 
This then requires a sufficient number of trajectories per subensemble  so that the denominator is always positive and not too close to zero for any of these, as discussed in detail in Appendix~\ref{APPC}.

Features seen in these figures include:
\ITEM{
\item Convergence of the gauged simulation is excellent, and precisely reproduces the exact quantum behavior.
\item At low temperatures $\tau\gtrsim 1/\hbar\chi$ the semiclassical approximation gives completely wrong results for $g^{(2)}$ and 
is out by $\order{1}$ in $\langle\op{n}\rangle$, which is also completely wrong if mode occupation is $\le\order{1}$.
\item The gauge P method, on the other hand, reproduces the exact quantum behavior precisely despite using a semiclassical coherent-state basis.
\item The un-gauged simulation breaks down while still at high temperatures due to boundary term induced spiking.
\item Low temperature observable estimates are consistent with the Fock state low temperature ground state described in Section~\ref{CH9ModelExact}. ($\langle\op{n}\rangle$ approaches integer values\footnote{Apart from when $\mu_e$ is an even integer multiple of $\hbar\chi$, since then $\langle\op{n}\rangle$ approaches half-integer values.} and the two-particle correlation $g^{(2)}$ approaches $1-1/\langle\op{n}\rangle$.)
\item Mid-temperature behavior is also simulated precisely.
\item At low temperature (long times) uncertainty in the observable estimates is greatest for $\mu_e$ even integer multiples of $\hbar\chi$ (and hence even integer multiple chemical potential $\mu$ at low temperatures), and lowest for $\mu_e$ odd integer multiples of $\hbar\chi$. These cases correspond to the low temperature state being a mixture of two Fock number states, or just a single pure Fock state, respectively.
}

\section{Chemical potential as a free gauge parameter}
\label{CH9Chempot}

Suppose one is interested in obtaining properties at a given temperature $T$ and chemical potential $\mu$.
This determines the final simulation target time $\tau_T$ and target chemical potential $\mu(\tau_T)$, but 
the chemical potential at intermediate times $\tau<\tau_T$ is not specified. The only conditions are 1) that the quantity $\mu(\tau)\tau$ be (piecewise) differentiable so that $\mu_e(\tau)$ can be calculated, and 2) that $\lambda_n=-\lim_{\tau\to0}(\mu\tau)$ be positive so that the initial distribution \eqref{pginit} is a normalizable probability. Apart from these, $\mu(\tau)$ is formally an arbitrary function (and $\lambda_n$ an arbitrary positive number). 

    This is all in the limit of many trajectories, however. The form of $\mu(\tau)$ at intermediate $\tau$ has no effect on observable estimate means in the limit $\mc{S}\to\infty$, but can have a strong effect on the broadness of the trajectory distribution. And, hence --- on the precision of a finite sample estimate. In this sense then, $\mu_e(\tau)$ (given by \eqref{muedef}), acts as effectively an additional gauge function.

\begin{figure}[tp]
\center{\includegraphics[width=8.5cm]{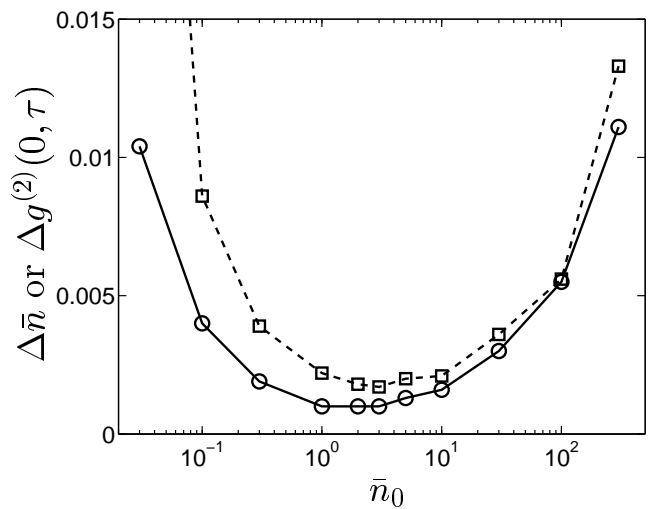}}\vspace{-8pt}\par
\caption[Dependence of efficiency on starting occupation $\bar{n}_0$]{\label{FIGUREnodn}\footnotesize
\textbf{Dependence of observable estimate uncertainties on the starting occupation $\bar{n}_0$}. 
Shown are calculated uncertainties  in 
{\scshape Circles}: $\bar{n}=\langle\op{n}\rangle$
{\scshape Squares}: $g^{(2)}(0,\tau)$,
for  $\hbar\chi\tau=1$ and $\mu=\hbar\chi$.
Calculations were made with the radial gauge \eqref{radialgauge} varying the starting occupation $\bar{n}_0$ and choosing a constant $\mu_e$ according to \eqref{muexpr}. Each calculation was with $2\times10^4$ trajectories.
\normalsize}
\end{figure}

\begin{figure}[p]
\center{\includegraphics[width=13cm]{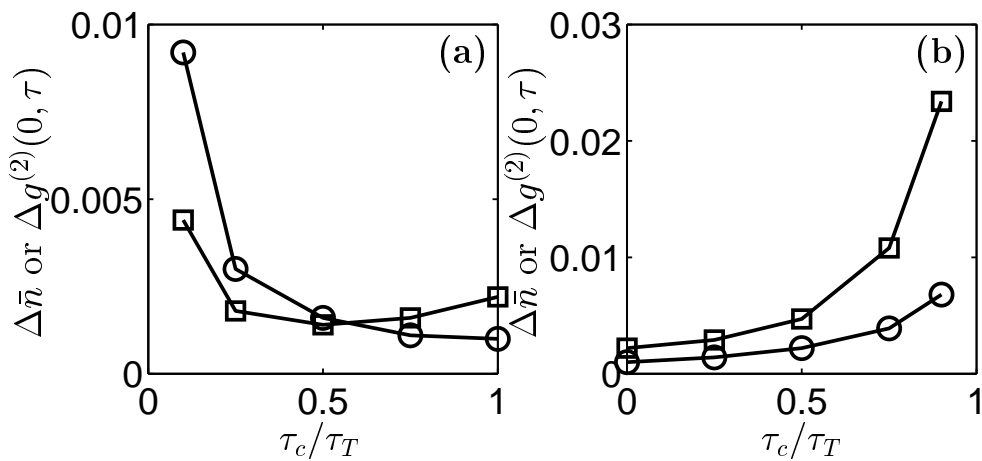}}\vspace{-8pt}\par
\caption[Dependence of efficiency on form of $\mu_e$]{\label{FIGUREt12dn}\footnotesize
\textbf{Dependence of observable estimate uncertainties on the form of $\mu_e$}. Shown are calculated uncertainties  in 
{\scshape Circles}: $\bar{n}=\langle\op{n}\rangle$
{\scshape Squares}: $g^{(2)}(0,\tau)$,
for  $\hbar\chi\tau=1$ and $\mu=\hbar\chi$.
Calculations were made with the radial gauge \eqref{radialgauge}, starting with $\bar{n}_0=1$ initially, but 
varying the form of $\mu_e$. In \textbf{(a)}, this was taken to be a nonzero constant for $\tau\le\tau_c$ and zero for $\tau>\tau_c$. In \textbf{(b)}, $\mu_e$ was zero for $\tau<\tau_c$, and nonzero constant for $\tau\ge\tau_c$. Each calculation was with $2\times10^4$ trajectories.
\normalsize}
\end{figure}

  An indication of the characteristics of an efficient form of $\mu_e(\tau)$ can be gained by comparing calculated observable uncertainties for several basic cases. Let us look at the situation when $\tau_T=1/\hbar\chi$, and $\mu(\tau_T)=\hbar\chi$. (This gives 
mean occupation $\bar{n}(\tau_T)=0.5628$).
\ITEM{
\item \textbf{Case 1:} Variation in $\bar{n}_0(\lambda_n)$, i.e. in the high-temperature starting mode occupation, while $\mu_e(\tau)$ is a constant chosen appropriately using \eqref{muexpr}. Calculated uncertainties in $\langle\op{n}\rangle$ and $g^{(2)}(0,\tau)$ are compared in Figure~\ref{FIGUREnodn}. It is seen that initial occupation $\order{1}-\order{10}$ gives the best performance, while too large or too small initial occupation lead to excessive uncertainty in the observables. 
\item \textbf{Case 2a:} Time-varying $\mu_e$, while $\bar{n}_0=1$ is held constant. $\mu_e$ is chosen to be a nonzero constant up to a time $\tau_c$, and zero for times $\tau>\tau_c\le\tau_T$. Note that as $\tau_c\to 0$, the size of $\mu_e$ grows to make up for the shorter time over which it acts.
Uncertainties are compared in Figure~\ref{FIGUREt12dn}\textbf{(a)}. It is seen that if $\mu_e$ is varied too strongly, an inefficient simulation results, although some relatively small variation may be beneficial for the $g^{(2)}$ calculation.
\item \textbf{Case 2b:} Again time-varying $\mu_e$, while $\bar{n}_0=1$ is held constant, but this time $\mu_e$ is held zero at high temperatures (low times $\tau<\tau_c$), and makes up for this by becoming a nonzero constant for $\tau_c<\tau<\tau_T$. As $\tau_c\to\tau_T$, the size of the required $\mu_e$ grows.
Uncertainties are compared in Figure~\ref{FIGUREt12dn}\textbf{(b)}. Again, it is seen that strong variation in $\mu_e$ leads to an inefficient simulation.
}
From the simulation data it appears that, other things being equal, 
\ENUM{
\item $\bar{n}_0\approx\order{\bar{n}(\tau_T)}$ or slightly greater appears to give the best performance
\item A constant (or nearly so) $\mu_e(\tau)$, which keeps the phase-space behavior time independent appears to give better performance than simulations for which $\mu_e$ strongly varies with $\tau$.
}

\section{Summary}
\label{CH9Summary}

The drift gauges \eqref{minimalgauge} and \eqref{radialgauge} have been developed to overcome shortcomings of the un-gauged positive-P-like technique. Using these gauges it has been shown that a stochastic simulation can precisely simulate the full quantum grand canonical thermodynamics of a two-body interacting Bose gas mode or orbital. Features include
\ENUM{
\item Simulation with good precision down to the lowest temperatures $\hbar\chi\tau\gg1$ where discretization of mode occupation takes place (See e.g. Figures~\ref{FIGUREcanng} and~\ref{FIGUREnstep}).
\item Simulation of a range of temperatures down to a minimum $1/k_B\tau_T$ in one run.
\item Calculation of any desired set of observables from one run.
}
The uncontrollable spiking (which was present in an un-gauged simulation) is removed. 

Numerical simulations indicate that the radial gauge \eqref{radialgauge} gives superior efficiency for a wide range of parameters. Also, it has been found that if one is aiming for a particular target temperature and chemical potential, 
it appears advantageous to choose the intermediate-time chemical potentials in the simulation such that: 1) $\mu_e$ is slowly varying with $\tau$ (or constant), and 2) The starting occupation $\bar{n}_0(\lambda_n)$ is of the same order as the mean occupation at the target time. 

The gauges developed should carry over in a straightforward manner to multi-mode simulations when the interparticle scattering is local to the lattice sites as described by the Hamiltonian \eqref{deltaH}. This is implemented in Chapter~\ref{CH11}.

\part{Examples of many-mode simulations}
\chapter{Many-mode dynamics}
\label{CH10}

Part C of this thesis investigates the performance of the gauge P representation (including the special case of the positive P with $\mc{G}_{\bo{n}}=0$) for many-mode calculations. Specifically, the interacting Bose gas. 
Chapter~\ref{CH10} will consider simulations of many-mode dynamics, while Chapter~\ref{CH11} concerns 
calculations of grand canonical ensembles.

Regarding dynamics, 
it will be seen that an un-gauged simulation (equivalent to positive P) can already give many useful results, 
while under some conditions the diffusion gauges of Chapter~\ref{CH7} can be applied directly (i.e. without any additional gauge improvements for kinetic coupling) to lengthen the simulation time and improve precision.

The particular example systems simulated in this chapter do not have particle losses, which is a ``worst case'' in terms of simulation time and stability, as has been discussed in Section~\ref{CH7ModelPpDamping}. This is also the case in which nonclassical phenomena are most pronounced. It is worth pointing out, however, that the gauge P representation method allows the addition of damping and/or losses in a completely straightforward way by adding some linear terms of the forms \eqref{itoheatbathT}, or \eqref{itoepsilon} into the stochastic equations. Chapter~\ref{CH11}, on the other hand,  considers simulations where the particle number strongly varies during the simulation.

The stochastic wavefunction method of Carusotto\etal\cite{Carusotto-01} and gauge P representation methods share many common features, but the former is applicable only to explicitly particle-conserving systems. That method has been used to calculate the evolution of the width of an atomic cloud  with extended interactions in a similar breathing trap arrangement to the example calculation of Section~\ref{CH10Trap} here.

\section{Simulation procedure}
\label{CH10Simulation}
   To simulate dynamics with the gauge P representation, one starts from the equations \eqref{itoH} to \eqref{itoepsilon}.
For an $M$-mode system, there are then $2M+1$ complex stochastic differential equations. 
The Wiener stochastic increments were implemented using Gaussian noises of variance $\Delta t$, independent at each time step of length  $\Delta t$. 

For efficiency reasons, it is convenient to use a split-step algorithm to evolve the kinetic energy part of the  terms $\omega_{\bo{nm}}$ in Fourier space, while doing the rest of the evolution in position space. For notational convenience
let us define the dual coherent state amplitudes (of the bra vector in the kernel \eqref{gaugekernel}) as	
\EQN{
\alpha'_{\bo{n}} = \beta^*_{\bo{n}}
.}
If the label $z$ is allowed to stand for either $\alpha$ or $\alpha'$, then 
the (discrete) Fourier space variables are 
\EQN{\label{fouriervardef}
\wt{z}_{\wt{\bo{n}}} =  \frac{(2\pi)^{\mc{D}/2}}{V}\sum_{\bo{n}}e^{-i\bo{k}_{\bo{\wt{n}}}\cdot\bo{x}_{\bo{n}}}z_{\bo{n}}
,}
where $V$ is the volume spanned by the spatial lattice.
Expanding $\omega_{\bo{nm}}$ as per \eqref{omegadef} and \eqref{k2def}, the kinetic evolution in momentum space takes the form
\SEQN{\label{kineticincrement}}{
d\wt{\alpha}_{\bo{\wt{n}}} &=& -\frac{i}{2}|\bo{k}_{\wt{\bo{n}}}|^2\wt{\alpha}_{\bo{n}}\,dt + \dots \\
d\wt{\alpha}'_{\bo{\wt{n}}} &=& -\frac{i}{2}|\bo{k}_{\wt{\bo{n}}}|^2\wt{\alpha}'_{\bo{n}}\,dt + \dots
}
after some algebra. (The remainder of the terms denoted by ``$\dots$'' will be evaluated in position space.) The split-step simulation algorithm is then,
for each trajectory out of $\mc{S}$:
\ENUM{
\item Initialize variables choosing randomly either from explicitly known initial conditions, or by passing variables  from a trajectory in a previous calculation (e.g. a thermal grand canonical ensemble as will be calculated in Chapter~\ref{CH11}).
\item For each time step $\Delta t$,
\ENUM{
\item Transform variables to $\wt{\alpha}_{\bo{\wt{n}}}$, $\wt{\alpha}'_{\bo{\wt{n}}}$ in Fourier space $\{\bo{\wt{n}}\}$.
\item Evolve variables forward by a time step $\Delta t$ by applying the change \eqref{kineticincrement} due to kinetic terms.
\item Accumulate any moments of variables in momentum space required for observable estimates.
\item Transform variables to $\alpha_{\bo{n}}$, $\alpha'_{\bo{n}}$ in normal space $\{\bo{n}\}$.
\item Evolve variables (including weight $\Omega$) forward by a time step $\Delta t$ by applying the rest of the evolution due to interparticle collisions, particle gains or losses, and external potential. This last appears as diagonal linear frequency terms, which now have the form $\omega_{\bo{nn}} = V_{\bo{n}}^{\rm ext}/\hbar$ after kinetic processes have been moved to Fourier space.
\item Accumulate any moments of variables in normal space required for observable estimates.
}
}

In the case when there is no particle gain from the environment (but losses to a zero temperature heat bath can be tolerated), 
the differential equations for all the variables $\alpha_{\bo{n}},\,\alpha'_{\bo{n}},\,\Omega$,  take on an exponential and local form $dz_j \propto z_j$ ($j$ is here a generic variable label). It is then convenient to change to logarithmic variables 
$\log(\alpha_{\bo{n}})$, $\log(\alpha'_{\bo{n}})$, and $z_0=\log\Omega$. The radial and tangential evolution are then simulated separately, which is found to lead to superior numerical stability. This allows one to use larger time steps. Efficiency is improved despite the need to change back to the non-logarithmic amplitude variables $\alpha_{\bo{n}}$, $\alpha'_{\bo{n}}$ to make the Fourier transform.

  The integration algorithm used  was a semi-implicit half-step iterative method, ($\kappa=\half$ in the notation of  \eqref{tkappa} in Appendix~\ref{APPB}).
The variable increments used ($d\alpha_{\bo{n}}$, $dz_0$ etc.) are in the Stratonovich calculus.
 This algorithm has been found by Drummond and Mortimer\cite{DrummondMortimer91} to have much superior stability to a plain first order method (although it is still first order in $\Delta t$).

  During a simulation, several indicators were monitored to make sure no sampling biases occur. These indicators were
\ENUM{
\item \textbf{Excessive variance in logarithmic variables}. All or most of the evolution for the interacting Bose gas occurs in an exponential fashion
as per $dz_j \propto z_j$, when written in terms of variables that are averaged to obtain observable estimates. As described in Appendix~\ref{APPA}, bias may occur if the variance of $\log|z_j|$ exceeds $\order{10}$.
\item \textbf{Sudden spiking}, which may be an indicator of boundary term errors in the simulations with no drift gauge.  This is described by Gilchrist\etal\cite{Gilchrist-97}. Some care had to be taken not to confuse this spiking with the far more benign kind caused by too-small ensemble sizes $\mc{S}$, which occurs when evaluating quotients of variable moments (as described in Appendix~\ref{APPC}). A characteristic distinguishing feature between these is what happens to the time of first spiking when the ensemble size is changed.
\ITEM{
  \item Boundary term related spiking becomes more severe and/or first occurs at slightly earlier times as the number of trajectories $\mc{S}$ is increased. This is because as  $\mc{S}$ increases, trajectories closer to the divergent moving singularity (and hence spiking sooner) become included in the ensemble.
\item Subensemble-size related  spiking abates and/or first occurs at later times as the number of trajectories in an ensemble $\mc{S}$ is increased, due to better precision in evaluating the denominator in the quotient.
}
\item \textbf{Change in mean with ensemble size}. A robust and widely applicable indicator of possible bias is when means of variables or of their functions  change in a systematic manner (with statistical significance) as the ensemble size $\mc{S}$ is increased. 
}
On the whole, reasonably precise observable estimates were not brought into question by any of the above indicators, apart from some unusual situations. (An example of such a pathological simulation were thermodynamic calculations  of ideal gases, where no noise occurs during the evolution. See Section~\ref{CH11Preweighting}.) Typically, worrying symptoms were seen only at times when noise had already obscured any observable estimates.

\section{Lattice size and simulation performance}
\label{CH10Lattice}
  A lattice simulation will be equivalent to the continuum system provided that 1) the lattice encompasses the entire system (i.e. the volume spanned by the lattice is large enough) and 2) all relevant features are resolved (i.e. the lattice is fine enough). Once in this limit, all simulations  will converge to the same physical predictions in the many trajectories limit $\mc{S}\to\infty$ (provided there are no boundary term errors). However, the \textit{rate} of convergence turns out to be affected by the size of the lattice spacing. This affects the useful simulation time.

  Before considering some example simulations in Sections~\ref{CH10Uniform} to~\ref{CH10Scattering}, the effect of lattice size on the simulation time is investigated.

\subsection{Scaling due to interparticle scattering}
\label{CH10LatticePp}
  To get a rough feel for the issue, let us see what happens in an un-gauged (positive P) simulation when the kinetic coupling is negligible, and the scattering process dominates.  Consider the case of  local rather than extended ($U_{\bo{n}}$) interactions. 
Each lattice point represents an effective volume $\Delta V=\prod_d\Delta x_d$,  and the mean  occupation is $\bar{n}_{\bo{n}}\approx \rho(\bo{x}_{\bo{n}})\Delta V$ for a system with density $\rho(\bo{x})$. The self-interaction strength is $\chi = g/2\hbar\Delta V$ from \eqref{gdef}.

At large mean lattice point occupations $\bar{n}$, the simulation time for a single mode using the positive P is given from Table~\ref{TABLEsimtime} by $t_{\rm sim} \approx 1.27 /\chi\bar{n}^{2/3}$,  based on numerical calculations. This is also in agreement with the analytic result \eqref{tsimpp}. At low occupations the empirical fits of Table~\ref{TABLEtimesfit} can be used instead.
 Since single-mode simulation time decreases in absolute terms with mode occupation, then the expected simulation time is limited by the most highly occupied mode:
\EQN{\label{ppsimtime}
t_{\rm sim} \lesssim \frac{2.5 \hbar (\Delta V)^{1/3}}{g (\max[\rho(\bo{x})])^{2/3}}
.}
Kinetic processes and external potentials have been ignored for the time being, although at least the kinetic mode coupling will reduce the useful simulation time below this level, as was seen in the calculations of Chapter~\ref{CH8}. The expression \eqref{ppsimtime} (or its extensions to include the empirical fits \eqref{timefit}at low mode occupation)  is qualitatively borne out in the simulations presented later in this chapter.

Summarizing, \textbf{finer lattices lead to shorter simulations}, scaling as $(\Delta V)^{1/3}$ in the strong interaction limit.

\subsection{Effect of kinetic interactions}
\label{CH10LatticeKinetic}
   Let us now consider the effect of the kinetic interactions. The simplest case is a uniform coherent gas with density $\rho$, scattering strength $g$, and lattice point volume $\Delta V$.  Also, no environment interactions, no external potential, and a constant standard diffusion gauge $g''$ applied to each mode separately.

When this is implemented as described in Chapter~\ref{CH7} and Section~\ref{CH4DiffusionCanonical} (e.g. as in \eqref{gauged2mode}), the evolution of the amplitudes of a single spatial mode $\bo{n}$  is given (via \eqref{itoH}, \eqref{omegadef} and $\breve{n}_{\bo{n}}=\alpha_{\bo{n}}{\alpha'}^*_{\bo{n}}$) by
\SEQN{\label{kineticeqs}}{
d\alpha_{\bo{n}} &=& -i\mc{E}_{\bo{n}}^{(\alpha)}\,dt -iV^{\rm ext}_{\bo{n}}\alpha_{\bo{n}}\,dt/\hbar - 2i\chi\alpha_{\bo{n}}\breve{n}_{\bo{n}}\,dt 
+dX^{(\alpha)}_{\bo{n}}\\
d\alpha'_{\bo{n}} &=& -i\mc{E}_{\bo{n}}^{(\alpha')}\,dt -iV^{\rm ext}_{\bo{n}}\alpha'_{\bo{n}}\,dt/\hbar- 2i\chi\alpha'_{\bo{n}}\breve{n}^*_{\bo{n}}\,dt +dX^{(\alpha')}_{\bo{n}}
.}
The quantities $\mc{E}_{\bo{n}}^{(z)}$ (with $z$ representing either $\alpha$ or $\alpha'$) contain the kinetic interactions, and 
 from \eqref{itoH}, \eqref{omegadef}, and \eqref{k2def},
\EQN{\label{mcEdef}
\mc{E}_{\bo{n}}^{(z)} = \frac{\hbar\Delta V}{2mV}\sum_{\bo{m},\bo{\wt{n}}} z_{\bo{m}} |\bo{k}_{\bo{\wt{n}}}|^2 e^{i\bo{k}_{\bo{\wt{n}}}\cdot(\bo{x}_{\bo{n}}-\bo{x}_{\bo{m}})}
,}
where $V$ is the entire lattice volume.
 The  direct noise terms are explicitly
\SEQN{\label{dXzdef}}{
dX_{\bo{n}}^{(\alpha)} &=& i\sqrt{2i\chi}\alpha_{\bo{n}}\left[\cosh g''\,dW_{\bo{n}} + i\sinh g''\, d\wt{W}_{\bo{n}}\right]\\
dX_{\bo{n}}^{(\alpha')} &=& i\sqrt{2i\chi}\alpha'_{\bo{n}}\left[-i\sinh g''\,dW_{\bo{n}} - \cosh g''\, d\wt{W}_{\bo{n}}\right]
.}

Let us write the $\mc{E}^{(z)}_{\bo{n}}$  in terms of a mean  $\bar{\mc{E}}^{(z)}$ and a fluctuating part $\delta\mc{E}_{\bo{n}}^{(z)}$ of mean zero:
\EQN{\label{delEdef}
\mc{E}_{\bo{n}}^{(z)}(t)= \bar{\mc{E}}^{(z)}(t) + \delta\mc{E}_{\bo{n}}^{(z)}(t) 
.}
Similarly the solution of \eqref{kineticeqs} will have a constant and fluctuating part:
\EQN{
z_{\bo{n}}(t) = \bar{z}(t) + \delta z_{\bo{n}}(t)
.}
Since the initial conditions are uniform and coherent then both $\bar{z}$ and $\bar{\mc{E}}^{(z)}$ are the same for all modes and trajectories. Also, because there are no external forces on this system, 
\EQN{\label{barzexpr}
\bar{z}(t) = \sqrt{\bar{n}}e^{i\theta(t)} = \sqrt{\rho\Delta V}e^{i\theta(t)}
}
with some real phase $\theta(t)$.

If (temporarily) the influence of the kinetic terms was ignored, then the analysis of the logarithmic variances of a single mode from Sections~\ref{CH7BothOptimization} or ~\ref{CH7DiffusionOptimization} would apply, and the variance of $\log|z_{\bo{n}}|$ would be given by  \eqref{lgvars} or \eqref{varlgnopt}.
(Compare $z_{\bo{n}}$ to the definition of $G_L$ in \eqref{Lg}). In particular, at short times the dominant fluctuating contribution would be due directly to the noise terms rather than amplification of $\breve{n}_{\bo{n}}$ fluctuations by the nonlinear drift term. From  \eqref{formalsoln}  and \eqref{mndef}, then
\EQN{\label{redalfa}
\matri{\log\alpha_{\bo{n}}(t)\\
\log\alpha'_{\bo{n}}(t)}
 &\approx& \Half\log\bar{n} -2i\chi\bar{n}t +\frac{\sqrt{\chi}}{2}\left\{  e^{-g''}[i\zeta_{\bo{n}}^+(t)-\zeta_{\bo{n}}^-(t)] \matri{+\\-}  e^{g''}[i\zeta^-_{\bo{n}}(t)-\zeta_{\bo{n}}^+(t)] \right\}
,\nonumber\\&&}
where the $\zeta_{\bo{n}}^{\pm}(t)$ are time-correlated Gaussian random variables defined as in \eqref{zetadef} and \eqref{zetaprop}, and are independent for each mode $\bo{n}$.
At short enough times that $\vari{\re{\log z_{\bo{n}}(t)}}\lesssim 1$ , one would have 
\EQNa{\label{ddirectdef}
z_{\bo{n}}(t) &\approx& \bar{z}(t)\left[1+\log z_{\bo{n}}(t)-i\theta(t)\right]\\
&=& \bar{z} + \delta^{\rm direct}z_{\bo{n}}(t)
,}
where $\theta(t) = -2\chi\bar{n}t$, and $z_{\bo{n}}$ represents either $\alpha_{\bo{n}}$ or $\alpha'_{\bo{n}}$.

In reality there is also, a fluctuating contribution from the other modes mediated by the kinetic interactions so that the total short time fluctuations are 
\EQN{
\delta z_{\bo{n}}(t) \approx \delta^{\rm direct} z_{\bo{n}}(t) + \delta^{\rm kinetic} z_{\bo{n}}(t)
.}
In the Ito calculus, where Wiener increments are uncorrelated with variables at the same time step, the 
direct and kinetic fluctuations at $t$ are un-correlated, and so the variance of fluctuations will be 
\EQN{
\vari{\,|\delta z_{\bo{n}}(t)|\,} = \vari{\,|\delta^{\rm direct} z_{\bo{n}}(t)|\,} + \vari{\,|\delta^{\rm kinetic} z_{\bo{n}}(t)|\,}
.}

One expects that as long as the first term (due to the direct noise terms $dX^{(z)}_{\bo{n}}$) dominates the fluctuations, then  the single-mode analysis of Chapter~\ref{CH7} is accurate. In such a regime:
\ITEM{
\item The diffusion gauge \eqref{ahodiffusiongauge} (or  \eqref{thenoptgauge} for the case of $\mc{G}_{\bo{n}}=0$) will be well optimized.
\item The simulation times of Table~\ref{TABLEsimtime} and ~\ref{TABLEtimesfit} are accurate (at least for $\mc{G}_{\bo{n}}=0$ simulations --- see Section~\ref{CH10LatticeDrift}).
\item The diffusion gauges \eqref{ahodiffusiongauge} or \eqref{thenoptgauge} will give improvements of simulation time while the mean mode occupation is high (i.e.  $\bar{n}=\rho\Delta V\gg 1$ ).  
}

Let us estimate the relative size of these two fluctuation contributions at short times.
Using \eqref{ddirectdef} and \eqref{redalfa}, the properties \eqref{zetaprop} of the $\zeta^{\pm}_{\bo{n}}$, one finds 
\EQN{\label{varidirect}
\vari{\,|\delta^{\rm direct} z_{\bo{n}}(t)|\,} \approx \frac{\rho gt\cosh 2g''}{\hbar}
.}
Also, directly from the equations \eqref{kineticeqs} and \eqref{delEdef} 
\EQN{
\delta^{\rm kinetic} z_{\bo{n}}(t) = -i\int_0^t\delta \mc{E}^{(z)}_{\bo{n}}(t')\, dt'
.}
Using \eqref{mcEdef}, and substituting in for $z_{\bo{m}}$ with the approximate (direct noise only) short time expression 
$z_{\bo{m}}\approx\bar{z}+\delta^{\rm direct} z_{\bo{m}}$, one obtains
\EQN{
\vari{\,|\delta^{\rm kinetic} z_{\bo{n}}(t)|\,}
 &=& \left(\frac{\hbar\Delta V}{2mV}\right)^2
\int_0^tdt' \int_0^t dt'' \sum_{\bo{m},\bo{m}',\bo{\wt{n}},\bo{\wt{n}}'}
\average{\delta^{\rm direct} z^*_{\bo{m}'}(t'')\delta^{\rm direct} z_{\bo{m}}(t')}\nonumber\\
&&\hspace*{70pt}\quad\times |\bo{k}_{\bo{\wt{n}}}|^2  |\bo{k}_{\bo{\wt{n}}'}|^2 
e^{-i\bo{k}_{\bo{\wt{n}}'}\cdot(\bo{x}_{\bo{n}}-\bo{x}_{\bo{m}'})}
e^{i\bo{k}_{\bo{\wt{n}}}\cdot(\bo{x}_{\bo{n}}-\bo{x}_{\bo{m}})}
.}
Since the direct noise at each lattice point in the locally-interacting model is independent, then
\EQN{
\average{\delta^{\rm direct} z^*_{\bo{m}'}(t'')\delta^{\rm direct} z_{\bo{m}}(t')}
= \delta_{\bo{m},\bo{m}'} \frac{\rho g \cosh 2g''\text{min}[t'',t']}{\hbar}
,}
similarly to \eqref{varidirect}, and the same for all modes.
After performing the integrations over $t'$ and $t''$, and simplifying the Fourier transforms, one obtains
\EQN{
\vari{\,|\delta^{\rm kinetic} z_{\bo{n}}(t)|\,}
= \vari{\,|\delta^{\rm direct} z_{\bo{n}}(t)|\,} \frac{\hbar^2t^2\Delta V}{12m^2V}\sum_{\bo{\wt{n}}}|\bo{k}_{\wt{\bo{n}}}|^4
.}
For a $\mc{D}$-dimensional system with many modes, one can approximate 
\EQN{
\sum_{\bo{\wt{n}}}|\bo{k}_{\wt{\bo{n}}}|^4 &\approx& \frac{V}{(2\pi)^{\mc{D}}}\int  |\bo{k}|^4\,d^{\mc{D}}\bo{k}\nonumber\\
				&=& \frac{V}{(2\pi)^{\mc{D}}} c_1(\bo{k}^{\rm max})\left(\frac{(2\pi)^{\mc{D}}}{5\Delta V}\right)\left(\frac{\pi}{(\Delta V)^{1/\mc{D}}}\right)^4\nonumber\\
&=& c_1(\bo{k}^{\rm max})\frac{V\pi^4}{5(\Delta V)^{1+4/\mc{D}}}\label{k4int}
,}
where $c_1(\bo{k}^{\rm max})$ is a shape factor $\order{1}$ that depends on the ratios between momentum cutoffs $k^{\rm max}_d$ for  the various lattice dimensions. For example in $1D$, $c_1 = 1$, while for $2D$ and $3D$ when the momentum cutoffs in each dimension are equal, one has $c_1=3\frac{1}{9}$, and $c_1=6\frac{1}{3}$. Using \eqref{k4int}, one obtains the final estimate 
\EQN{\label{varikd}
\vari{\,|\delta^{\rm kinetic} z_{\bo{n}}(t)|\,}
\approx \vari{\,|\delta^{\rm direct} z_{\bo{n}}(t)|\,} \frac{\hbar^2t^2\pi^4c_1}{60m^2(\Delta V)^{4/\mc{D}}} 
.}

\textbf{It can be seen that:} the fluctuations due to kinetic terms
\ITEM{
\item become relatively more important with time, and
\item are more dominant for fine lattices (i.e. when $\Delta V$ is small). 
}

\subsection{Lattice size and diffusion gauge usefulness}
\label{CH10LatticeHealing}

If the kinetic coupling remains weak (in relative terms) for the duration of a 
un-gauged or diffusion-gauged simulation, then one expects that the local diffusion gauges \eqref{thenoptgauge} are fairly well optimized, and will give simulation time improvements. Let us investigate under what conditions this is expected to occur based on \eqref{varikd}.

Firstly, simulation times were found in Chapter~\ref{CH7} to be significantly improved when the mean particle occupation per mode is $\bar{n}\gtrsim 1$. The expected single-mode simulation time was found to be 
$t_{\rm sim} \approx \order{10}/\chi\sqrt{\bar{n}}$ --- see Table~\ref{TABLEsimtime}. It will be convenient to write the lattice interaction strength $\chi=g/2\hbar\Delta V$ in terms of the  \textbf{healing length} 
\EQN{\label{heallength}
\xi^{\rm heal} = \frac{\hbar}{\sqrt{2m\rho g}}
.}
 This is the minimum length scale 
over which a local density inhomogeneity  in a Bose condensate wavefunction can be in balance with the quantum pressure due to kinetic effects (e.g. it is the typical size of a quantized vortex in a BEC\cite{Gross61,Pitaevskii61},  or the typical size of a density correlation disturbance, as will be seen here). The healing length is discussed in more detail e.g. in Dalfovo\etal\cite{Dalfovo-99}, p. 481.

In terms of $\xi^{\rm heal}$, the diffusion gauged simulation time is 
$t_{\rm sim}\approx \order{40}(\xi^{\rm heal})^2m\sqrt{\bar{n}}/\hbar$. 
From \eqref{varikd}, the kinetic fluctuations are expected to be weak  ($\vari{\,|\delta^{\rm kinetic} z_{\bo{n}}(t)|\,} \ll 
 \vari{\,|\delta^{\rm direct} z_{\bo{n}}(t)|\,}$) up to this time provided that
\EQN{
\frac{80}{3}c_1\bar{n}\left(\frac{\pi\xi^{\rm heal}}{(\Delta V)^{1/\mc{D}}}\right)^4 \ll 1
.}
  The quantity $(\Delta V)^{1/\mc{D}}=\bar{\Delta x}$ is the geometric mean of the lattice spacing. One sees that 
simulation time improvement using a local gauge diffusion gauge occurs only when 
\EQN{
  \bar{\Delta x} &\gtrsim& \xi^{\rm heal} \bar{n}^{1/4} c_3\nonumber\\
		&\gtrsim& \order{\xi^{\rm heal}}
}
with $c_3=\pi(3/80c_1)^{1/4}\approx\order{1}$ a weakly lattice shape dependent constant.
(The second line follows since improvements occur only for $\bar{n}\gtrsim 1$.)

That is, \textbf{the local diffusion gauges \eqref{ahodiffusiongauge} or \eqref{thenoptgauge} can be expected to give simulation time improvements only if the lattice spacing is of the order of the healing length or greater}.

\subsection{Drift gauges and many-mode dynamics}
\label{CH10LatticeDrift}

As has been noted in Section~\ref{CH7DiffusionMotivation}, drift gauged simulations additionally encounter a scaling problem in many-mode systems because the single log-weight variable $z_0$ accumulates fluctuations from all modes. 

 One has the Ito equation for the log-weight:
\EQN{
dz_0 = \sum_{\bo{n}}\left\{\mc{G}_{\bo{n}}\left[dW_{\bo{n}}-\Half\mc{G}_{\bo{n}}\,dt\right]  
+\wt{\mc{G}}_{\bo{n}}\left[d\wt{W}_{\bo{n}}-\Half\wt{\mc{G}}_{\bo{n}}\,dt\right]\right\}
,}
with independent contributions from each mode.
In a uniform gas on $M=V/\Delta V$ modes each with occupation $\bar{n}$, the contribution from each mode is identical on average, and the variance of real and imaginary parts of $dz_0$ will scale as 
$\vari{z_0} \propto M$. Because $z_0$ appears as $\Omega=e^{z_0}$ in observable estimates, there is a limit \eqref{sdlimit} to how large the variance of $\re{z_0}$ can be if any precision is to be retained in the simulation (see Section~\ref{CH7Gaussian}). At short times $\vari{\re{z_0}}\propto tM$, which leads to a reduction of simulation time $t_{\rm sim} \propto 1/M$ in this case.

In the simulations with drift gauge \eqref{ahodriftgauge} the diffusion gauge $g''$ acts to trade-off fluctuations in the amplitudes $\propto e^{g''}$ with fluctuations in the log-weight $\propto\sqrt{M}e^{-g''}$. This indicates that when many modes are present, a more advantageous tradeoff  might be achieved by increasing $g''$ relative to the single-mode expression than \eqref{ahodiffusiongauge}. 

Let us investigate this. With $M$ identical modes in a uniform gas, the changes in the optimization  of Section~\ref{CH7BothOptimization} can be easily tracked. The last term of \eqref{lg2} for each mode is $M$ times greater, and so the $\propto|n_0|^2$ term in \eqref{lgvars} also acquires a  factor $M$. This factor also appears in the third-order (in $V_g$) term of the cubics \eqref{vcubic} and \eqref{vacubic}. In the end, the new optimization gives 
\EQN{\label{ahodiffusiongaugeM}
g''_{\bo{n}} = \frac{1}{6}\log\left\{ 8M|\breve{n}_{\bo{n}}(t)|^2\chi t_{\rm rem} + a_2^{3/2}(\breve{n}_{\bo{n}}(t),\gamma_{\bo{n}} t_{\rm rem})\right\}
}
instead  of the single-mode expression \eqref{ahodiffusiongauge}. Simulation time can be estimated as in Section~\ref{CH7Times}. At short times 
\EQN{
\vari{G_L}=\frac{\chi t}{2}\left(V_g+\frac{1}{V_g}\right) +2M(\chi V_g t n'_0)^2
}
(compare to \eqref{varglboth}), while for large particle number $V_g\approx1/2(n'{}^2_0\chi t M)^{1/3}$. 
One obtains 
\begin{equation}\label{bignsimtimeM}
t_{\text{sim}} \approx \frac{(20/3)^{3/2}}{M^{1/4}\chi\sqrt{n'_0}} \approx \order{\frac{40\,t_{\rm coh}}{M^{1/4}}}
.\end{equation}
so simulation appears to be reduced by a factor of $M^{-1/4}$ as compared to the single-mode case (Much better than a $1/M$ reduction with the single-mode expression \eqref{ahodiffusiongauge}).

Unfortunately there is some further noise process that limits simulation times with drift gauged dynamics simulations on  large lattices. A preliminary trial with 
 a uniform 1D gas on a 50 point lattice with  $\Delta x_1=\Delta V=10\,\xi^{\rm heal}$, and $\bar{n}=1000$ bosons per lattice point was tried, but  numerical simulations with the gauge \eqref{ahodiffusiongaugeM}  did not show improvement to simulation times over the $\mc{G}_{\bo{n}}=0$ case. (Nor did a combination of drift gauge \eqref{ahogauge} with  \eqref{ahodiffusiongauge} or $g''_{\bo{n}}=0$.) Only some improvement of precision at very small times $\ll t_{\rm sim}$ was seen. Presumably  kinetic effects lead to a large increase in $z_0$ fluctuations in this regime in comparison with what is expected for $M$ uncoupled modes, and neither the original optimization not the above re-optimization  of $g''_{\bo{n}}$ gives improvement. 

A combination of local drift and diffusion gauges {\it was} seen to give improvement under some conditions in Chapter~\ref{CH8}. There it was concluded that the kinetic coupling needs to be relatvely weak for this to occur, but
details remain to be investigated. Simulations shown in the remainder of this chapter were made with zero drift gauge, and diffusion gauges only.

\section{Correlation functions}
\label{CH10Correlation}

   In most experimentally realized Bose gas systems, low order local observables such as density or energy are often well described by approximate theories such as Gross-Pitaevskii (GP) equations (for a BEC),  or statistical approaches in a high temperature gas. Multi-particle correlations are not well described by these  theories, however, and so are of more interest for first-principles calculations. In terms of local lattice annihilation/creation operators those correlation functions that will be considered here are:
\ENUM{
\item \textbf{First order:} 
\EQN{
g^{(1)}(\bo{x}_{\bo{n}},\bo{x}_{\bo{m}}) = \frac{\langle\dagop{a}_{\bo{n}}\op{a}_{\bo{m}}\rangle}{\sqrt{\langle\dagop{a}_{\bo{n}}\op{a}_{\bo{n}}\rangle\langle\dagop{a}_{\bo{m}}\op{a}_{\bo{m}}\rangle}}.
}
This is a phase-dependent correlation function. Its magnitude $|g^{(1)}|$ tells one the degree of first-order spatial coherence, while its phase gives the relative phase of the wavefunction at spacing $\bo{x}_{\bo{n}}-\bo{x}_{\bo{m}}$.
\item \textbf{Second order:}
\EQN{
g^{(2)}(\bo{x}_{\bo{n}},\bo{x}_{\bo{m}}) = \frac{\langle\dagop{a}_{\bo{n}}\dagop{a}_{\bo{m}}\op{a}_{\bo{n}}\op{a}_{\bo{m}}\rangle}{\langle\dagop{a}_{\bo{n}}\op{a}_{\bo{n}}\rangle\langle\dagop{a}_{\bo{m}}\op{a}_{\bo{m}}\rangle}.
}
  This two-particle number correlation function is not phase dependent, and always real positive. It describes the likelihood of finding two particles at a spacing $\bo{x}_{\bo{m}}-\bo{x}_{\bo{m}}$ from each other, relative to what is expected of a coherent field. For a bunched field, $g^{(2)}(\bo{x},\bo{x})>1$ (e.g. a thermal state has $g^{(2)}(\bo{x},\bo{x})=2$), while antibunching is evidenced by $g^{(2)}(\bo{x},\bo{x})<1$.
\item \textbf{Third order:}
\EQN{
g^{(3)}(\bo{x}_{\bo{n}},\bo{x}_{\bo{m}},\bo{x}_{\bo{m}'}) = \frac{\langle\dagop{a}_{\bo{n}}\dagop{a}_{\bo{m}}\dagop{a}_{\bo{m}'}\op{a}_{\bo{n}}\op{a}_{\bo{m}}\op{a}_{\bo{m}'}\rangle}{\langle\dagop{a}_{\bo{n}}\op{a}_{\bo{n}}\rangle\langle\dagop{a}_{\bo{m}}\op{a}_{\bo{m}}\rangle\langle\dagop{a}_{\bo{m}'}\op{a}_{\bo{m}'}\rangle}.
}
  This three-particle correlation function describes the likelihood of particles at $\bo{x}_{\bo{n}}$, $\bo{x}_{\bo{m}}$, and $\bo{x}_{\bo{m}'}$ (ralative to a coherent field). The rate of three-body processes is proportional to $g^{(3)}(\bo{x},\bo{x},\bo{x})$ (or, if the effective radius of particles is $r$, three body processes will depend on  values of 
$g^{(3)}(\bo{x},\bo{x}+\epsilon_1,\bo{x}+\epsilon_2)$, where $|\epsilon_j|\lesssim r$.) For example in a BEC, the rate of three-body recombination, which can limit the condensate's lifetime, is proportional to $g^{(3)}(\bo{x},\bo{x},\bo{x})$\cite{Kagan-88}.
}
For the uniform gas, average values of these (over all space of $M$ lattice points) are of relevance:
\SEQN{\label{bargdef}}{
\bar{g}^{(j)}(\bo{x}) &=& \frac{1}{M}\sum_{\bo{m}}g^{(j)}(\bo{x}_{\bo{m}},\bo{x}_{\bo{m}}+\bo{x})\\
\bar{g}^{(3)}(\bo{x},\bo{y}) &=& \frac{1}{M}\sum_{\bo{m}}g^{(3)}(\bo{x}_{\bo{m}},\bo{x}_{\bo{m}}+\bo{x},\bo{x}_{\bo{m}}+\bo{y})
.}

\section[Example 1: Correlation waves in a uniform gas]{Example 1: \\Correlation waves in a uniform gas}
\label{CH10Uniform}

\subsection{The system}
\label{CH10UniformSystem}
  This system consists of a uniform gas of bosons with density $\rho$ and interparticle \mbox{$s$-wave} scattering length $a_s$. The lattice is chosen with a spacing $\Delta x_d\gg a_s$ so that interparticle interactions are effectively local at each lattice point with effective field interaction strength  $g$ given by \eqref{gcoup} or \eqref{gdef}, depending on the dimensionality of the system. Coupling strength between lattice points, $\chi$, is then given by \eqref{chidef}. Periodic boundary conditions are assumed. 
  The results obtained from a simulation will be invariant of lattice size provided the lattice spacing is fine enough to resolve all occupied momenta, and the lattice volume $V$ is large enough to encompass all phenomena.

The initial state is taken to be a coherent wavefunction, which is a stationary state of the ideal gas with no interparticle interactions (i.e. $g=0$). Subsequent evolution is with $g>0$, so that there is a disturbance at $t=0$ when interparticle interactions are rapidly turned on.
Physically, this kind of disturbance can be created in a BEC by rapidly increasing the scattering length at $t\approx 0$ by e.g. tuning the external magnetic field near a Feshbach resonance.

The behavior of a uniform gas can tell us a lot about what goes on in more complicated inhomogeneous systems. If the density of a system is slowly varying on length scales of $L_{\rho}$ or less, then any uniform gas phenomena of spatial extent $L_{\rho}$ or smaller will also be present in the inhomogeneous system. In a trapped gas, for example, uniform gas phenomena that occur on length scales of the order of a fraction of the trap size or smaller.

\subsection{Correlation waves in a one-dimensional gas}
\label{CH10Uniform1D}

\begin{figure}[p]
\center{\includegraphics[width=\textwidth]{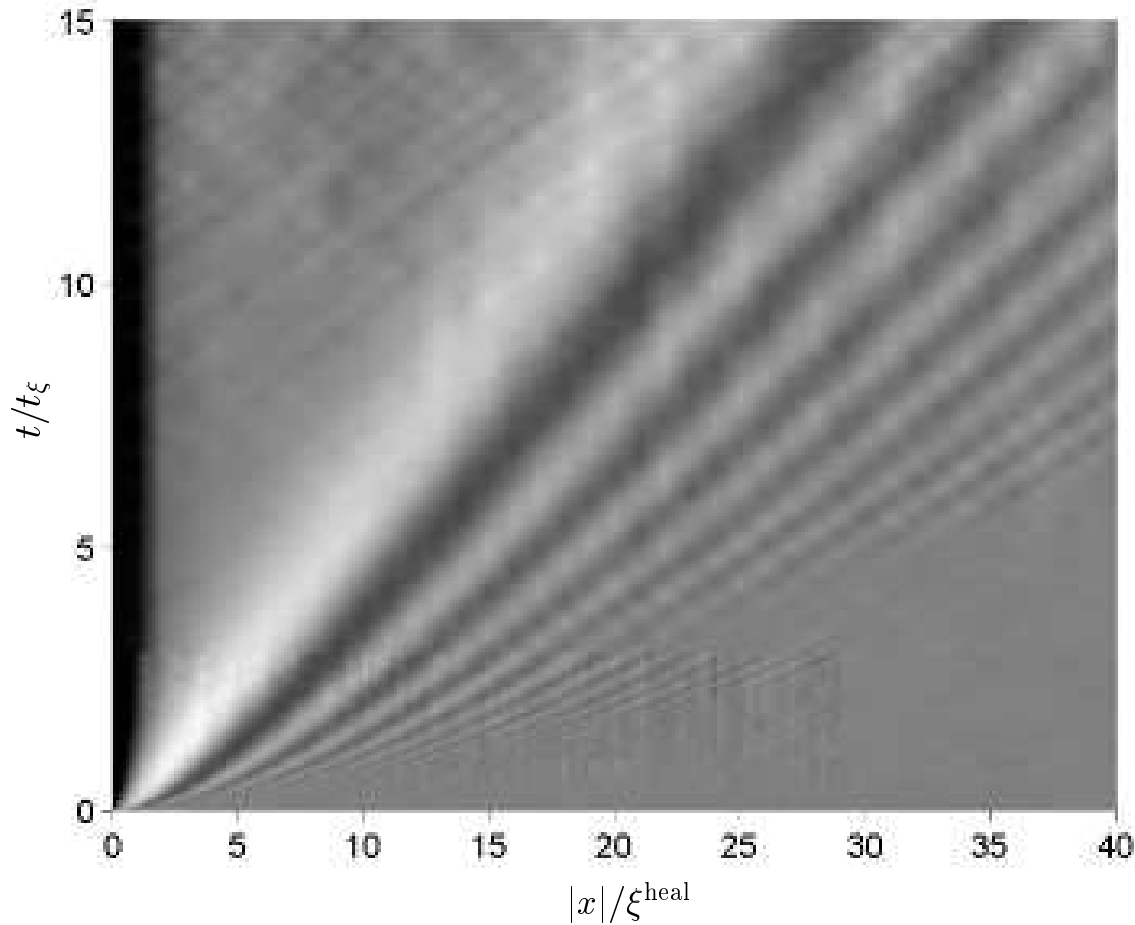}}\vspace{-8pt}\par
\caption[Evolution of second order correlations $g^{(2)}$ in a uniform 1D gas]{\label{FIGUREg2composite}\footnotesize
\textbf{Evolution of $\bar{g}^{(2)}(x)$ in a uniform 1D gas} with density $\rho=100/\xi^{\rm heal}$. Shading indicates relative magnitude of $\bar{g}^{(2)}(x)$, with light regions indicating high values $>1$, and dark indicating low values $<1$. The diagram is a composite of data from three simulations with different lattices: $\Delta x = 0.12\xi^{\rm heal}$, $500$ lattice points and $\mc{S}=3000$ for $t\in[0,t_{\xi}]$ and $|x|\in[0,29\xi^{\rm heal}]$. $\Delta x = 0.24\xi^{\rm heal}$, 250 points  and $\mc{S}=10^4$ for $t\in[t_{\xi},3t_{\xi}]$. $|x|\in[0,29\xi^{\rm heal}]$, and $\Delta x = \xi^{\rm heal}$, 200 points and $\mc{S}=10^4$ for the rest of the data. These changes in resolution are responsible for the apparent slight discontinuities in the data. The apparent weak cross-hashing superimposed on the plot at long times and/or large distances is due to finite sample size and/or lattice discretization effects and is not statistically significant.
 Magnitudes of $\bar{g}^{(2)}$ are shown in Figure~\ref{FIGUREp100plots}. [arXiv note: higher resultion available at Piotr Deuar's homepage, currently {\tt http://www.physics.uq.edu.au/people/deuar/thesis/}]
\normalsize}
\end{figure}

\begin{figure}[p]
\center{\includegraphics[width=11cm]{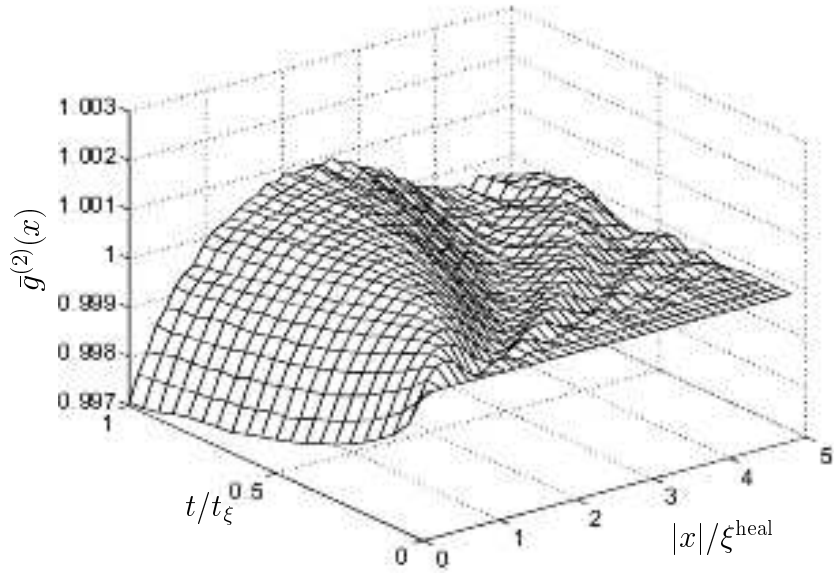}}\vspace{-8pt}\par
\caption[Second order correlations $g^{(2)}$ at short times in a uniform 1D gas]{\label{FIGUREg2short}\footnotesize
\textbf{Short time evolution of $\bar{g}^{(2)}(x)$ in a 1D gas} with density $\rho=100/\xi^{\rm heal}$. From a simulation 
with $\mc{S}=3000$ trajectories, 500 points and $\Delta x=0.12\xi^{\rm heal}$. [arXiv note: higher resultion available at Piotr Deuar's homepage, currently {\tt http://www.physics.uq.edu.au/people/deuar/thesis/}]\normalsize}
\end{figure}

\begin{figure}[p]
\center{\includegraphics[width=8cm]{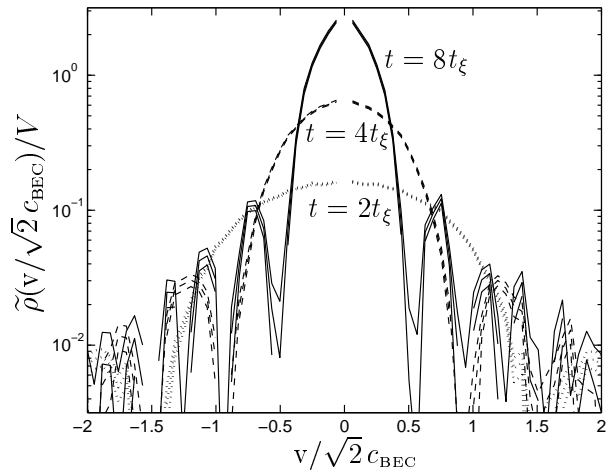}}\vspace{-8pt}\par
\caption[Evolution of velocity distribution in a uniform 1D gas]{\label{FIGUREp100pk}\footnotesize
\textbf{Distribution of velocity in the uniform 1D gas} of spatial density $\rho=100/\xi^{\rm heal}$ at several times $t$ after the initial disturbance. Plotted is the velocity density $\wt{\rho}$ per unit volume of the gas. Velocities ${\rm v}$ are given in units of $\sqrt{2}c_{\rm\scriptscriptstyle BEC}$, which is the  speed of the strongest correlation wave. The  background density of stationary atoms has been omitted. Triple lines indicate error bars  at one standard deviation significance.  $\mc{S}=10^4$ trajectories, $\Delta x=\xi^{\rm heal}$, 100 lattice points.
\normalsize}
\end{figure}

\begin{figure}[t]
\center{\includegraphics[width=\textwidth]{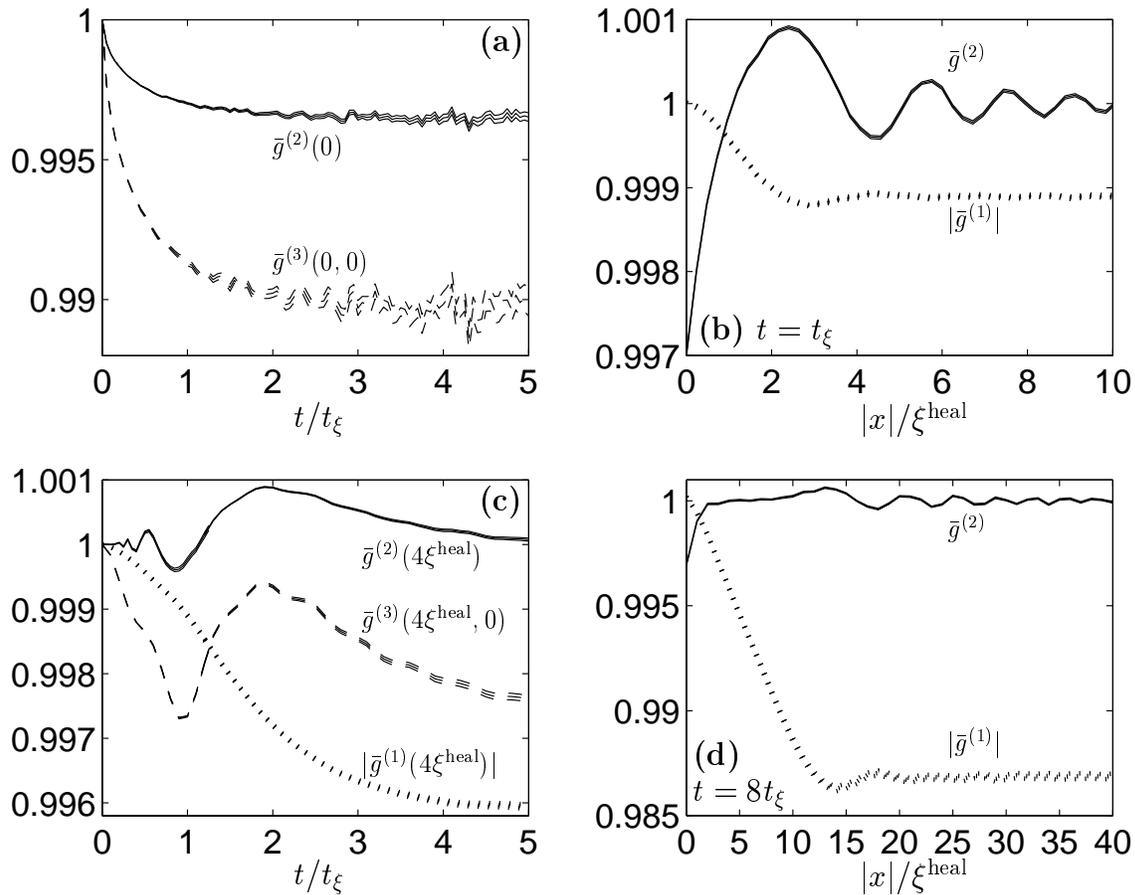}}\vspace{-8pt}\par
\caption[Correlation functions in a uniform 1D gas]{\label{FIGUREp100plots}\footnotesize
\textbf{Correlation functions in a uniform 1D gas} with density $\rho=100/\xi^{\rm heal}$. {\scshape solid} lines show $\bar{g}^{(2)}$, {\scshape dashed} lines show $\bar{g}^{(3)}$, {\scshape dotted} lines show $|\bar{g}^{(1)}|$. Triple lines indicate error bars at one standard deviation significance. Subplot \textbf{(a)}: Time dependence of local correlations. \textbf{(b)}: Correlations at time $t=t_{\xi}$. \textbf{(c)}: Time dependence of correlations at distance $x=4\xi^{\rm heal}$. \textbf{(d)}: Long time correlations at $t=8t_{\xi}$. Data in \textbf{(a)}, \textbf{(b)}, and $t<1.25t_{\xi}$ in \textbf{(c)} are from a simulation with $\Delta x=0.24\xi^{\rm heal}$ and 250 points, while the rest is from a $\Delta x=\xi^{\rm heal}$ simulation with 200 points. $\mc{S}=10^4$ trajectories in both cases.
\normalsize}
\end{figure}

The evolution of some of the quantum correlations that appear in a one-dimensional gas with density $\rho=100/\xi^{\rm heal}$ is shown in Figures~\ref{FIGUREg2composite}, \ref{FIGUREg2short}, and \ref{FIGUREp100plots}, based on $\mc{G}_{\bo{n}}=0$ simulations. Figure~\ref{FIGUREp100pk} also shows the momentum distribution of particles at times after being excited by the disturbance at $t=0$. 

Time is given in units of the timescale
\EQN{\label{txidef}
t_{\xi} = \frac{m(\xi^{\rm heal})^2}{\hbar} = \frac{\hbar}{2\rho g}
}
(the ``healing time''), which is approximately the time needed for the short-distance ($\order{\xi^{\rm heal}}$) inter-atomic correlations to
 equilibrate after the disturbance (see Figure~\ref{FIGUREp100plots}\textbf{(a)}).

Perhaps the most interesting feature noted in the simulations is 
the propagating wave train that appears in the \textit{two-particle correlations}. These waves are seen only in the two- (or more) particle correlations, while the density always remains uniform (by symmetry, since there are no gains or losses in this model), and shows no sign of any wave behavior.  This ``correlation wave'' is best seen in the 
 time-dependent behavior of $\bar{g}^{(2)}(x)$ in Figure~\ref{FIGUREg2composite}. 

Once some initial transient effects at times $t\lesssim\order{t_{\xi}}$ after the disturbance die out (these are shown in Figure~\ref{FIGUREg2short}), the long time behavior of first and second order correlations appears as follows:
\ITEM{
\item At distances $\order{\xi^{\rm heal}}$, antibunching occurs --- See e.g. Figures~\ref{FIGUREg2composite},~\ref{FIGUREp100plots}\textbf{(b)}, \ref{FIGUREp100plots}\textbf{(d)}.
\item Enhanced two-particle correlations occur at a quite well-defined interparticle spacing, which increases at a constant rate: $\approx \sqrt{2}\,\xi^{\rm heal}$ per $t_{\xi}$. This is a factor $\sqrt{2}$ faster than the low-momentum sound velocity in a BEC, obtained from the Bogoliubov dispersion law: 
\EQN{\label{cbecdef}
c_{\rm\scriptscriptstyle BEC}=\sqrt{\rho g/m}
.}
See e.g. Dalfovo\etal\cite{Dalfovo-99} p. 481. 
\item There are also many weaker correlations (and anti-correlations) at larger distances that move apart at increasingly  faster rates, as seen in Figure~\ref{FIGUREg2composite}.  These leading correlation wavelets are particularly well visible in Figure~\ref{FIGUREp100plots}\textbf{(b)}. The leading disturbances appear to move at $\approx 20 c_{\rm\scriptscriptstyle BEC}$ for the $\rho=100/\xi^{\rm heal}$ system, although this is only clearly seen in this simulation for times $\lesssim 1.5 t_{\xi}$ when transient effects may still be significant. 
\item Looking at the momentum spectrum of the particles (Figure~\ref{FIGUREp100pk}), there are peaks, but their position changes with time (towards lower momenta), and in particular there is no peak corresponding to any particles travelling at the speed of the main (trailing) correlation wave. 
\item The peak to peak width of these correlation waves is of the order of several healing lengths for the times simulated, although some slow spreading is seen.
\item At long times $\gtrsim \order{t_{\xi}}$ at distances shorter than the strongest trailing correlation wave with velocity $\sqrt{2}c_{\rm\scriptscriptstyle BEC}$, but longer than the healing length,  long range second order coherence ($\bar{g}^{(2)}(x)=1$) between particles reappears.
\item Long range phase coherence $|\bar{g}^{(1)}(x)|$ decays with time due to the scattering processes. (approximately linearly, at least while $|\bar{g}^{(1)}|\approx\order{1}$). See e.g. the shorter time behavior of Figure~\ref{FIGUREp100plots}\textbf{(c)}, and the long distance behavior of Figure~\ref{FIGUREp100plots}\textbf{(b)} and \textbf{(d)}.
\item However, in the (2nd order) coherent neighborhood that appears after the last correlation wave has passed,  phase coherence decays only up to a certain value (see Figure~\ref{FIGUREp100plots}\textbf{(c)} after the last $\bar{g}^{(2)}$ correlation wave has passed. This steady state phase coherence value drops off approximately linearly with distance, as seen especially in Figure~\ref{FIGUREp100plots}\textbf{(d)}.
\item The third-order correlation $\bar{g}^{(3)}$ function also displays some wave behavior at similar times and distances as $g^{(2)}(x)$. This is visible e.g. in Figure~\ref{FIGUREp100plots}\textbf{(c)}.
}

The uniform one-dimensional gas is peculiar in some respects, one of these being that as density increases, the system 
becomes more like an ideal gas\cite{LiebLiniger63,YangYang69}. 
From \eqref{ppsimtime} and \eqref{txidef} the expected simulation time in units of the healing time $t_{\xi}$
is 
\EQN{\label{tsimppxi}
t_{\rm sim} \approx 5\bar{n}^{1/3}t_{\xi}
,}
at high lattice point occupation $\bar{n}$.
To resolve the correlation waves (with width $\approx\xi^{\rm heal}$), one requires $\Delta x\approx\xi^{\rm heal}$, which gives the scaling
\EQN{
\frac{t_{\rm sim}}{t_{\xi}} \propto \left(\rho\,\xi^{\rm heal}\right)^{1/3}
.}
That is, at lower densities, it is more difficult to observe the long time correlation wave behavior that evolves on the  $t_{\xi}$ timescale. This is borne out in Figure~\ref{FIGUREscaling}\textbf{(b)}, and is the reason why the relatively weak correlations in the $\rho=100/\xi^{\rm heal}$ have been investigated here in most detail.

\begin{figure}[tb]
\center{\includegraphics[width=\textwidth]{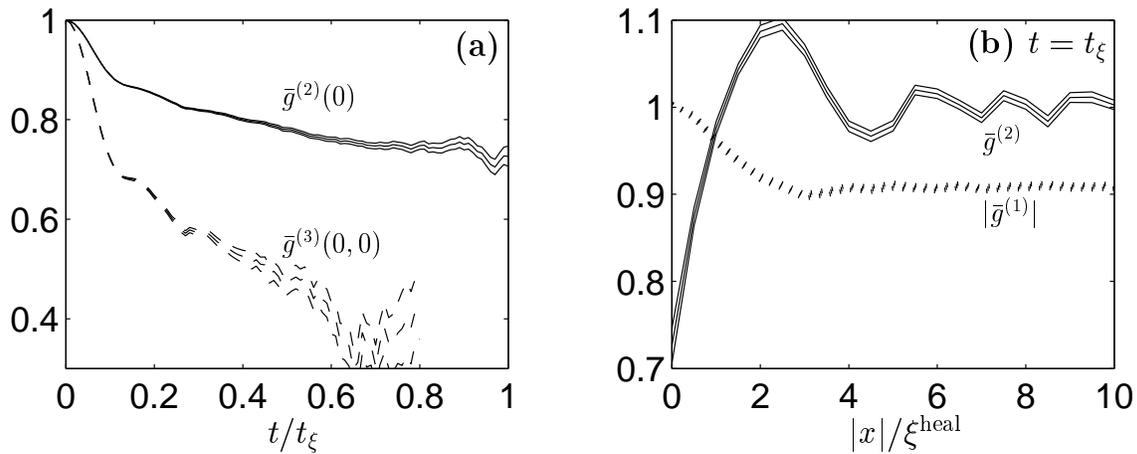}}\vspace{-8pt}\par
\caption[Correlation functions in a dense uniform 1D gas]{\label{FIGUREp1plots}\footnotesize
\textbf{Correlation functions in a uniform 1D gas} with density $\rho=1/\xi^{\rm heal}$. {\scshape solid} lines show $\bar{g}^{(2)}$, {\scshape dashed} lines show $\bar{g}^{(3)}$, {\scshape dotted} lines show $|\bar{g}^{(1)}|$. Triple lines indicate error bars at one standard deviation significance. Subplot \textbf{(a)}: Time dependence of local correlations. \textbf{(b)}: Correlations at time $t=t_{\xi}$. Simulations with $\Delta x = 0.5\xi^{\rm heal}$, 50 lattice points,  and $\mc{S}=10^4$ trajectories.
\normalsize}
\end{figure}

Lower density simulations were also carried out for $\rho=1/\xi^{\rm heal}$ and $\rho=10/\xi^{\rm heal}$. Qualitatively the same kind of correlation wave behavior was seen for these densities up to simulation time, but the correlations are much stronger. Some examples for $\rho=1/\xi^{\rm heal}$ are shown in Figure~\ref{FIGUREp1plots}.

  When the length and time scales are given in density-dependent dimensionless units $x/\xi^{\rm heal}$ and $t/t_{\xi}$, the scaling of correlations in the simulated parameter regimes was closely approximated by 
\EQN{
\bar{g}^{(j)} \propto \frac{\xi^{\rm heal}}{\rho}
.}

\subsection{Lattice dependence of simulation time}
\label{CH10UniformSimtime}
\begin{figure}[p]
\center{\includegraphics[width=\textwidth]{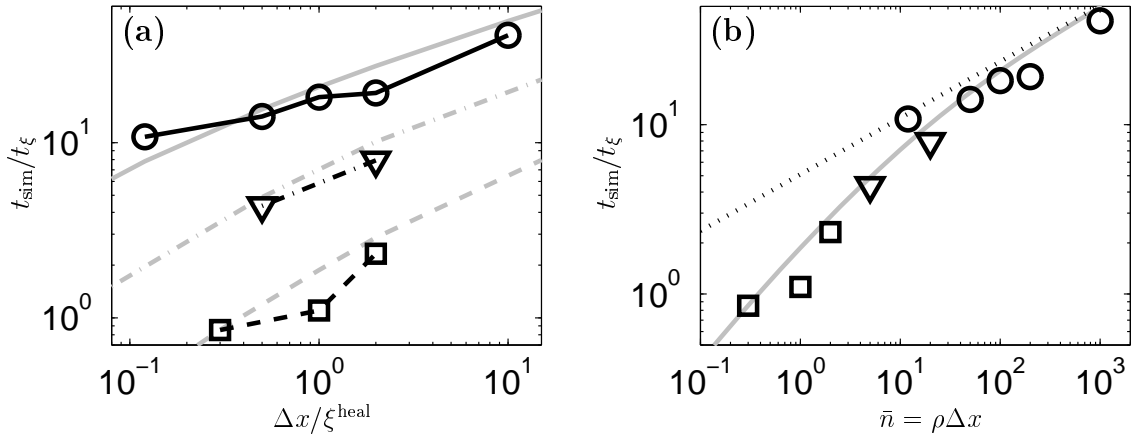}}\vspace{-8pt}\par
\caption[Simulation time vs. lattice spacing / occupation in uniform 1D gas]{\label{FIGUREscaling}\footnotesize
\textbf{Scaling of useful simulation time} with lattice spacing $\Delta x$ and mean lattice point occupation $\bar{n}=\rho\Delta x$.
Data points  for simulations of 1D gases of densities $100/\xi^{\rm heal}$, $10/\xi^{\rm heal}$, and $1/\xi^{\rm heal}$, are shown with {\scshape circles, triangles,} and {\scshape squares}, respectively. 
 $\mc{S}=10^4$ trajectories.
Also shown  for comparison as {\scshape light lines} are expected simulation times based on the empirical relationship \eqref{timefit} for the single-mode positive P. The {\scshape dotted line} in subplot \textbf{(b)} gives the high-occupation estimate \eqref{tsimppxi}.
\normalsize}
\end{figure}

  The observed scaling of simulation time with lattice spacing $\Delta x$ and lattice occupation $\bar{n}=\rho\Delta x$ is shown in Figure~\ref{FIGUREscaling}, and compared to the expected un-gauged times when kinetic effects are ignored. The estimated times shown in gray are based on the empirical relationship \eqref{timefit} with the positive P (i.e. un-gauged) fitting parameters given in Table~\ref{TABLEtimesfit}.
  It can be seen that for the physical regimes simulated, this empirical estimate of simulation time is very good also for  many-mode simulations. 	(Most of the residual variation in $t_{\rm sim}$ is due to statistical effects of finite sample size).

\begin{figure}[p]
\center{\includegraphics[width=\textwidth]{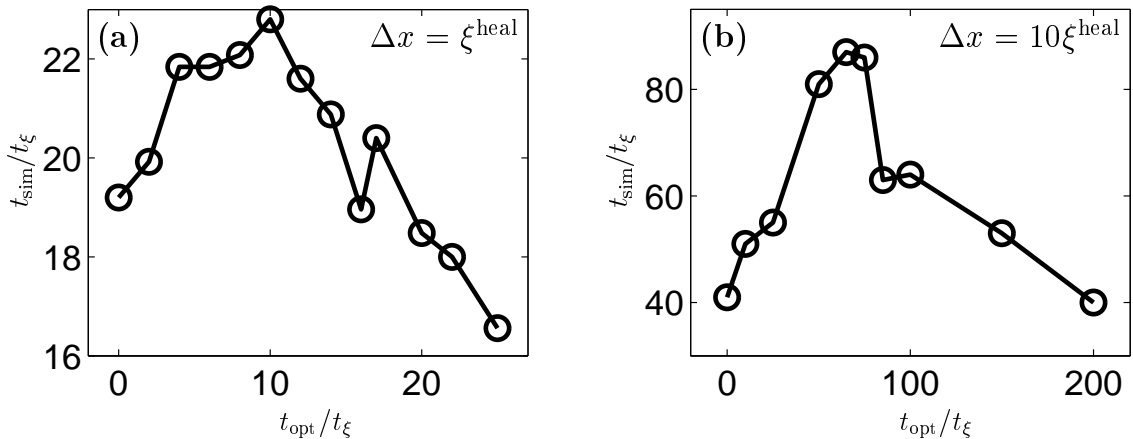}}\vspace{-8pt}\par
\caption[Improvement of uniform 1D gas simulation time with diffusion gauges]{\label{FIGURE1dtopt}\footnotesize
\textbf{Improvement of simulation time with diffusion gauges}
Simulation times as a function of the target time $t_{\rm opt}$ when using the diffusion (only) gauge \eqref{thenoptgauge}. 
All simulations are of a $\rho=100/\xi^{\rm heal}$ gas. In subplot \textbf{(a)}: $\Delta x=\xi^{\rm heal}$ with $\mc{S}=10^4$ trajectories, and in \textbf{(b)}, and $\Delta x=10\xi^{\rm heal}$ with $\order{500}$ trajectories.
\normalsize}
\end{figure}

\subsection{Diffusion gauge dependence of simulation time}
\label{CH10UniformGauge}
  From the analysis of Section~\ref{CH10LatticeHealing}, it is expected that diffusion gauges will give  significant improvements once $\Delta x \gtrsim \xi^{\rm heal}$. This is confirmed in Figure~\ref{FIGURE1dtopt}, which shows the dependence of simulation time on $t_{\rm opt}$ for some  simulations using the gauge \eqref{thenoptgauge}.
At large lattice spacings, significant gains can be made with the local diffusion gauge, while for spacing of the order of the healing length the improvement is small. Note the  similarity of 
Figure~\ref{FIGURE1dtopt}\textbf{(b)} to the single-mode case Figure~\ref{FIGUREtoptnopt}\textbf{(a)}. In both cases simulation time increases steadily while $t_{\rm opt}\le \text{max}[t_{\rm sim}]$,  but drops off sharply once the critical $t_{\rm opt}$ is reached ($t_{\rm opt}\approx 80t_{\xi}$ in this case).

\subsection{Two-dimensional gas}
\label{CH10Uniform2D}
\begin{figure}[tb]
\center{\includegraphics[width=\textwidth]{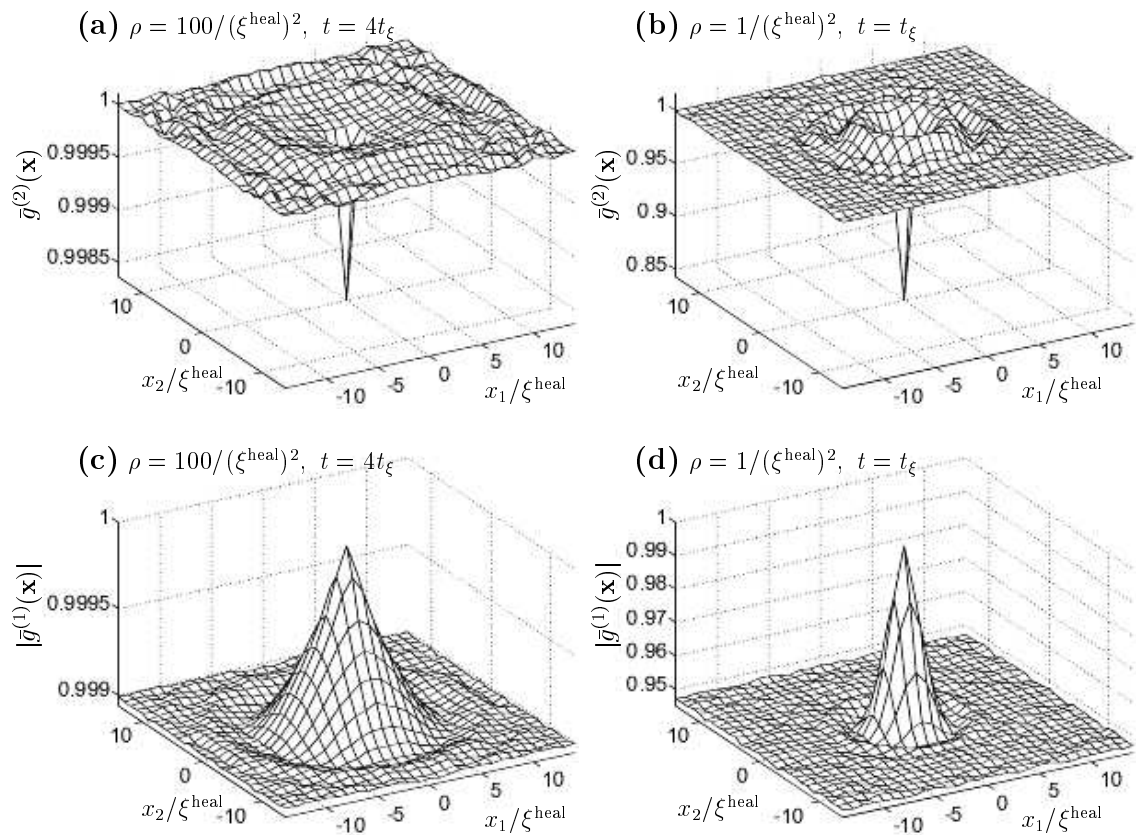}}\vspace{-8pt}\par
\caption[Correlation functions in 2D uniform gases]{\label{FIGURE2dmesh}\footnotesize
\textbf{Correlation functions in 2D uniform gases} at times $4t_{\xi}$ and $t_{\xi}$ for gases of density $100/(\xi^{\rm heal})^2$ and $1/(\xi^{\rm heal})^2$, respectively. Coordinates in dimension $d$ are denoted $x_d$. Simulations used a lattice spacing of $\Delta x_d=\xi^{\rm heal}$, $30\times30$ lattice points, and $\mc{S}=10^4$ trajectories.
[arXiv note: higher resultion available at Piotr Deuar's homepage, currently {\tt http://www.physics.uq.edu.au/people/deuar/thesis/}]\normalsize}
\end{figure}

Simulation of a two-dimensional gas is also straightforward using the equations \eqref{itoH}. Some examples of calculated correlation functions for 2D uniform gases with densities $\rho=100/(\xi^{\rm heal})^2$ and $\rho=1/(\xi^{\rm heal})^2$ are shown in Figure~\ref{FIGURE2dmesh}.
Correlation wave behavior similar to the 1D case is seen.

\vfill
\pagebreak

\section[Example 2: Extended interactions in a uniform gas]{Example 2: \\Extended interactions in a uniform gas}
\label{CH10Extended}
  The simulation of a gas with extended interparticle interactions is also possible with the gauge P representation, as described in Section~\ref{CH5Extended}.
One uses the equations \eqref{itoH} but with scattering effects modified to become the nonlinear terms \eqref{nodriftUeq} and noise terms given by \eqref{dXexpr} and \eqref{zetaxi}. 

  A Gaussian interparticle potential 
\EQN{\label{ugauss}
  U(x) = \frac{g}{\sigma_U\sqrt{2\pi}}\exp\left[\frac{-x^2}{2\sigma_U^2}\right]
.}
is assumed for the interparticle potential in this example. The bulk interaction energy density in a system with density varying much slower in $x$ than $U(x)$ is 
$\bar{u} = \frac{1}{2V}\int U(x-y) \dagop{\Psi}(x)\dagop{\Psi}(y)\op{\Psi}(x)\op{\Psi}(y) dx dy 
\approx \frac{\rho^2}{2}\int U(x) dx$. The potential \eqref{ugauss} has been normalized so that $\bar{u}$ is the same as for a local-only  ``delta-function'' interaction with strength $g$. 

\begin{figure}[p]
\center{\includegraphics[width=11cm]{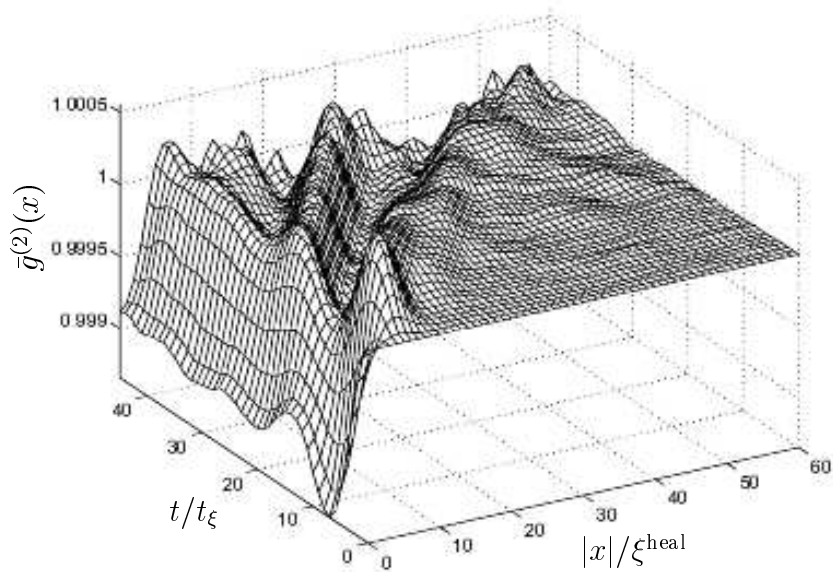}}\vspace{-8pt}\par
\caption[$\bar{g}^{(2)}(x)$ in a uniform 1D gas with extended interparticle interactions]{\label{FIGUREug2}\footnotesize
\textbf{Evolution of $\bar{g}^{(2)}(x)$} in a uniform 1D gas with extended interparticle interactions \eqref{ugauss}: $\sigma_U=3\xi^{\rm heal}$.
Gas of density $\rho=100/\xi^{\rm heal}$. Simulation was on a lattice with spacing $\Delta x=\xi^{\rm heal}$, 200 lattice points (thus with $20\,000$ total particles on average), and $\mc{S}=10^4$ trajectories. [arXiv note: higher resultion available at Piotr Deuar's homepage, currently {\tt http://www.physics.uq.edu.au/people/deuar/thesis/}]
\normalsize}
\end{figure}

\begin{figure}[p]
\center{\includegraphics[width=10.5cm]{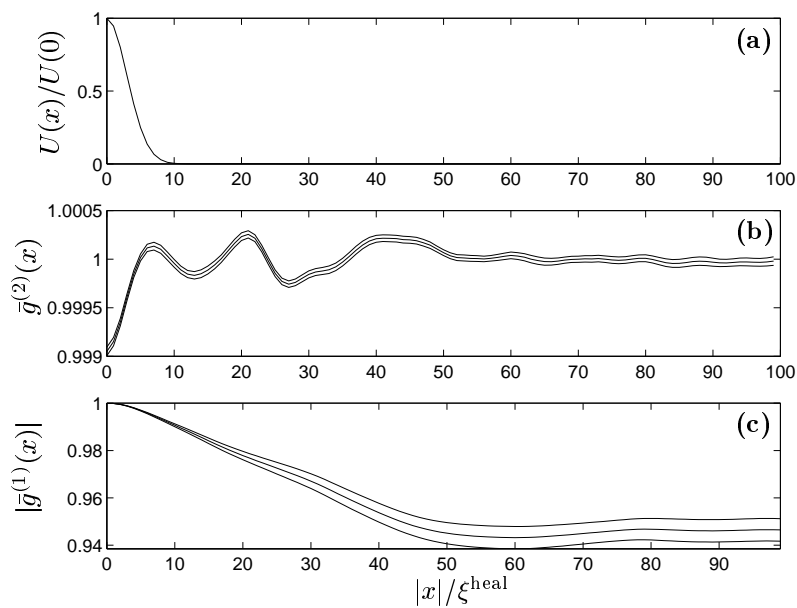}}\vspace{-8pt}\par
\caption[Uniform 1D gas with extended interparticle interactions at $t=35t_{\xi}$]{\label{FIGUREugt35}\footnotesize
\textbf{Correlations at $t=35t_{\xi}$} for the uniform 1D gas with extended interparticle interactions. Gas and simulation parameters as in Figure~\ref{FIGUREug2}. Triple lines indicate error bars at one standard deviation. The form of the interparticle potential is shown to scale in \textbf{(a)}.
\normalsize}
\end{figure}

The simulated second order correlation function is shown in Figures~\ref{FIGUREug2} and~\ref{FIGUREugt35} for a potential  of standard deviation $\sigma_U=3\xi^{\rm heal}$. 
Density was again chosen $\rho=100/\xi^{\rm heal}$, so that at large length scales $\gg \sigma_U$, the behavior of the gas approaches the behavior of the locally-interacting model shown in Section~\ref{CH10Uniform1D} and Figures~\ref{FIGUREg2composite},~\ref{FIGUREg2short}, and~\ref{FIGUREp100plots}.

Correlation wave phenomena are seen in this system as well, but with significant differences to the locally interacting gas of Section~\ref{CH10Uniform1D}:
\ITEM{
\item The antibunching at short distance scales occurs out to a range $\order{\sigma_U}$ rather than $\xi^{\rm heal}$, and takes on approximately a Gaussian form. See Figure~\ref{FIGUREugt35}\textbf{(a)} and \textbf{(b)}, and compare to the linear (in $x$) growth of  $\bar{g}^{(2)}(x)$ in Figure~\ref{FIGUREp100plots}\textbf{(a)}.
\item The strongest positive correlation wave moves at approximately the speed of sound $c_{\rm\scriptscriptstyle BEC}$, not at the faster rates seen in the locally-interacting gas. 
\item A  train of weaker correlation waves is created (at least two more wavefronts are seen  in Figure~\ref{FIGUREug2}), but these form \textit{behind} the strongest leading wavefront rather than in front of it, and these secondary  waves appear to also move at the sound velocity. 
\item Some decaying oscillations of $\bar{g}^{(2)}(0)$ occur.
}
Phase coherence (Figure~\ref{FIGUREugt35}\textbf{(c)}) decays in similar fashion to the locally-interacting gas.

The differences in the  long time behavior between this system and the locally interacting gas may be  reconciled despite the potential having the same coarse-grained interaction strength $g$. An assumption made to arrive at an effectively  locally-interacting gas was that the s-wave scattering length $a_s=mg/4\pi \hbar^2$ is much smaller than all relevant length scales -- in this case e.g. $\xi^{\rm heal}$. This is satisfied for the simulations in Section~\ref{CH10Uniform1D}, but the equivalent quantity here (which is $\sigma_U$) is {\it not} smaller than $\xi^{\rm heal}$.

As in the gas with only local interactions on the lattice, long time behavior is best observed with high density but small correlations as in Figure~\ref{FIGUREug2}. Much stronger correlations can be simulated, but for shorter times (scaling will again be $t_{\rm sim}\propto\bar{n}^{1/3}$).

Lastly, a convenient feature of these extended interaction simulations is that the useful simulation time is found to be significantly longer than for a system with local interactions of the same coarse-grained  strength $g$. For example in the $\rho=100/\xi^{\rm heal}$ system, 
$t_{\rm sim}\approx 46t_{\xi}$ for the $\sigma_U=3\xi^{\rm heal}$ gas, while the locally-interacting gas had $t_{\rm sim}\approx17t_{\xi}$ (Note that $g=\hbar/2\rho t_{\xi}$).

\section{Example 3: Correlations in a trap}
\label{CH10Trap}

\begin{figure}[tbp]
\center{\includegraphics[width=\textwidth]{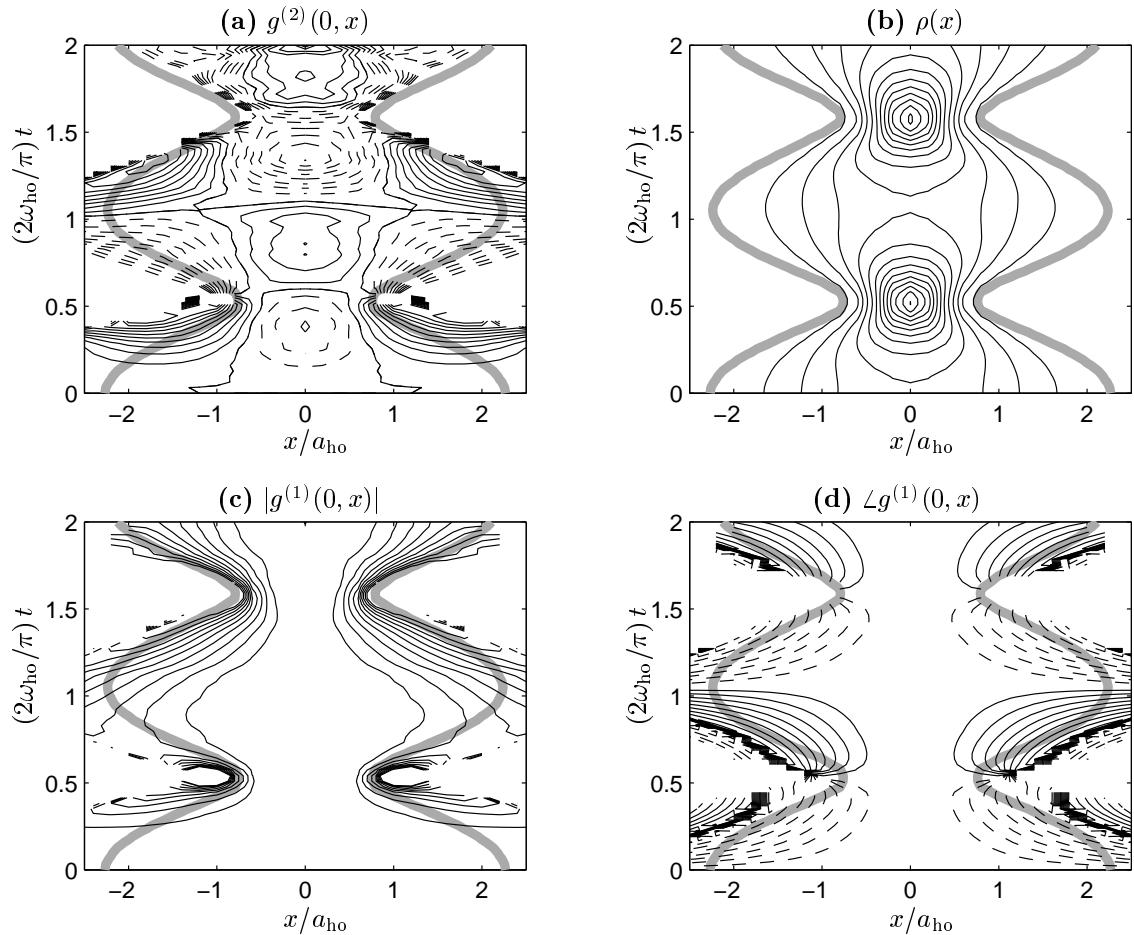}}\vspace{-8pt}\par
\caption[Correlation functions in a breathing trapped condensate]{\label{FIGUREtrapcontour}\footnotesize
\textbf{Contour plots of the evolution of correlations and boson density} in the trapped Bose gas described in Section~\ref{CH10Trap}. Correlations shown are between bosons in the center of the trap, and those a distance $x$ from the center. \textbf{(a)}: Contours of density correlations $g^{(2)}(0,x)$ with a spacing of $0.01$. {\scshape Solid} lines indicate $g^{(2)}>1$, {\scshape dashed} $g^{(2)}<1$. \textbf{(b)}: contours of mean density $\rho(x)$ with spacing of $1/a_{\rm ho}$. \textbf{(c)}: contours of phase coherence $|g^{(1)}(0,x)|\le1$ with spacing $0.01$ (note: $g^{(1)}(0,0)=1$). \textbf{(d)}: contours of the relative condensate phase $\angle g^{(1)}(0,x)$ with spacing $\pi/10$.  {\scshape Solid} lines indicate $0<\angle g^{(1)}<\pi$, {\scshape dashed} $-\pi<\angle g^{(1)}<0$. $\angle g^{(1)}=0$ contour omitted.  
{\scshape Light broad} lines are contours of density at $5\%$ of the central value, and indicate the approximate extent of the boson cloud. Simulation with $\mc{S}=10^4$ trajectories on a 60 point lattice with $\Delta x=0.2a_{\rm ho}$. Note that $a_{\rm ho}$ is the harmonic oscillator length of the initial cloud, while $a_{\rm ho}/2$ is the width of the narrower trap at $t>0$ with frequency $2\omega_{\rm ho}$.
\normalsize}
\end{figure}

\begin{figure}[tb]
\center{\includegraphics[width=8cm]{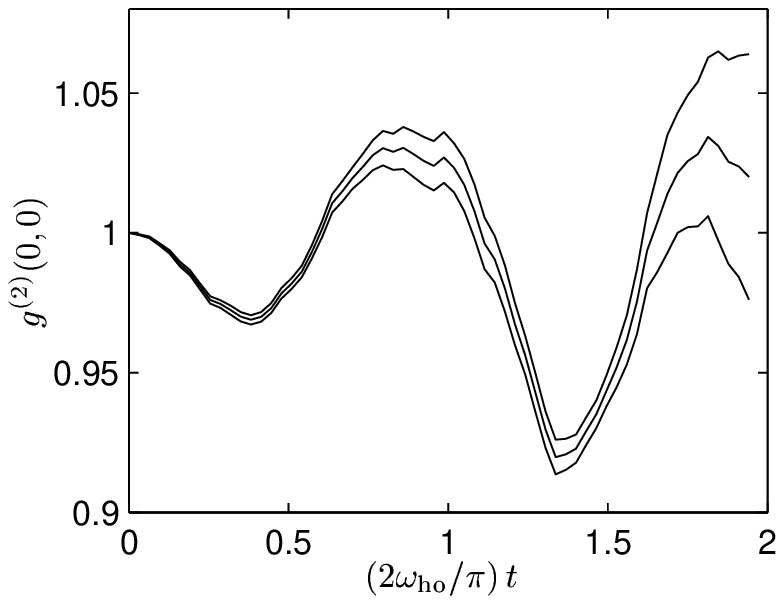}}\vspace{-8pt}\par
\caption[Local two-particle correlations in the breathing trapped condensate]{\label{FIGUREtrapg20}\footnotesize
\textbf{Local two-particle correlations in the center of the trap} for the breathing condensate described in Section~\ref{CH10Trap}. Triple lines indicate error bars at one standard deviation. Parameters as in Figure~\ref{FIGUREtrapcontour}.
\normalsize}
\end{figure}

Simulations in an external trapping potential $V^{\rm ext}$ pose no particular problem. 
For example, the following system was simulated: 
\ITEM{
\item Bosons are prepared in a harmonic trap with trapping potential 
\EQN{
V^{\rm ext}(x) = \Half m \omega_{\rm ho}^2 x^2
,}
which has a harmonic oscillator length $a_{\rm ho}=\sqrt{\hbar/m\omega_{\rm ho}}$.
Initially they are in the coherent zero temperature ground state obtained by solving the Gross-Pitaevskii mean field equations\footnote{i.e. the stationary state of \eqref{itoH} with noise terms removed. --- see Section~\ref{CH5GP} for more on the correspondence between the deterministic part of the gauge P equations and the Gross-Pitaevskii equations.}.  The mean number of atoms in the trap in this example is $\bar{N}=10$.
\item The bosons experience two-body interactions with an effective Gaussian interparticle potential \eqref{ugauss}
with radius $\sigma_U=a_{\rm ho}$, and a strength $g=0.4 \hbar a_{\rm ho}\omega_{\rm ho}$.
\item At $t=0$, breathing of the atomic cloud is induced by switching to a more confined harmonic potential with double the trapping frequency: i.e. $V^{\rm ext}\to\Half m(2\omega_{\rm ho})^2x^2$. 
}
Some data from the simulation are shown in Figures~\ref{FIGUREtrapcontour} and~\ref{FIGUREtrapg20}. 

This model is in a regime where qualitative results are hard to achieve using approximate methods because several length scales from different processes are of the same order: Trap width is $a_{\rm ho}$,  initial cloud width is $\approx 2a_{\rm ho}$, interparticle scattering range is also $a_{\rm ho}$.

Phenomena seen in this the simulation include:
\ITEM{
\item The two-particle density correlations $g^{(2)}(0,x)$ display different behaviors in the contracting and expanding phase. See Figure~\ref{FIGUREtrapcontour}\textbf{(a)}. 
\ITEM{
\item When the particle cloud is contracting, \textit{anti}bunching appears at the center of the trap, while during the later part of the contraction there is an enhanced likelihood of pairs of atoms with one in the outer region of the cloud and one in the center.
\item During expansion, on the other hand, the particles tend to bunch in the center of the trap (i.e. there is increased likelihood of two particles at small separation), while pairs of particles with one in the tails, one in the center are suppressed. 
}
\item This bunching during the expansion and antibunching during the contraction appear counter-intuitive. The reason for this appears to be a time lag before the dominant effect makes itself felt in the correlations. This results in the  bunching trailing the contraction by a significant part of the breathing period. Initially, the cloud is stationary, and during the contraction phase, there is a long period of time when the atoms are laregly in free fall, and  interparticle repulsion dominates the correlations causing antibunching despite the contraction. Eventually, the atoms become squashed together leading to bunching, but this occurs only at the end of the contraction phase. Bunching now remains for a large part of the expansion phase before it is finally overcome by the interparticle repulsion. 
\item The oscillations of $g^{(2)}(0,0)$ (i.e. bunching at the center of the trap) due to the breathing of the atomic cloud become more pronounced with time --- see Figure~\ref{FIGUREtrapg20}. This may indicate a resonance between the breathing and the repulsion, although it is also possible that this is a transient initial effect.
\item Coherence between the center of the trap and outlying regions of the cloud deteriorates as time proceeds -- compare the first and second contraction phase in Figure~\ref{FIGUREtrapcontour}\textbf{(c)}.
}


\section[Example 4:Bosonic enhancement of scattered atoms]{Example 4: \\
Bosonic enhancement of atoms scattered from colliding condensates}
\label{CH10Scattering}

\subsection{The Vogels\etal\ four wave-mixing experiment}
\label{CH10ScatteringExperiment}
In a recent experiment of Vogels\etal\cite{Vogels-02} at MIT, strong coherent four-wave mixing between components of a Bose-Einstein condensate at different velocities was observed.
The matter wave components were created by applying Bragg pulses to an initially stationary trapped condensate so as to impart a velocity $2\bo{v}^{\rm cm}$ to approximately half of the atoms. ($\bo{v}^{\rm cm}$ is their velocity with respect to the center of mass). A small seed  population ($\order{1\%}$ of all the atoms) moving at velocity $\bo{v}_s$ was also created. Bose-enhanced scattering of the atoms during the half-collision of the two main wavepackets led to $\order{10\times}$ coherent amplification  of the of the seed population and the creation of a fourth coherent population at velocity $\bo{v}_4=2\bo{v}^{\rm cm}-\bo{v}_s$. 
   
As the wavepackets move through each other, scattering of atom pairs occurs into velocities $\bo{v}$ and $2\bo{v}^{\rm cm}-\bo{v}$, with energy conservation favoring $|\bo{v}|\approx\bo{v}^{\rm cm}$. If the seed wavepacket is present at some $|\bo{v}_s|\approx\bo{v}^{\rm cm}$ then scattering of atoms into its modes is preferred due to Bose enhancement, analogously to stimulated emission in photonic systems. 

In the experiment the initially empty momentum modes also acquired sufficient density after some  time, so that Bose enhancement of scattering occurred into these non-seed modes as well. These momentum modes (which are much more numerous than the seed wave) eventually competed with the seed wavepacket and limited its growth.

A first-principles method could be desirable to quantitatively describe the effect of the initially empty modes in this system. Such a calculation is difficult with approximate methods as
both single-particle  effects (to occupy the empty modes in the first place), and subsequent amplification 
of the coherent many-particle wavefunction are involved. The first process can  be estimated with perturbative methods, and the second with mean field GP equations, but combining the two has been difficult.

Experimental parameters (relevant to the subsequent simulation described in the next subsection) were : $\approx 30\,000\,000 $ atoms of ${}^{23}$Na ($a_s=2.75$nm, $m=3.82\times10^{-26}$kg) in the initial wavepacket. This was created in an axially symmetric longitudinal trap with frequencies $\omega/2\pi$ of 20Hz in the longitudinal and 80Hz in the axial directions. This trap is subsequently turned off, and the two main wavepackets created with the Bragg pulse move at a velocity $\pm\bo{v}^{\rm cm}$ relative to the center of mass, with $\bo{v}^{\rm cm}\approx 10$mm/s.

\subsection{Simulating Bose-enhanced scattering into initially \\empty modes}
\label{CH10ScatteringSimulation}
To study the enhanced scattering into the initially empty modes, an un-gauged positive P representation simulation was made 
with  similar parameters as the Vogels\etal\ experiment. There were two differences:
\ITEM{
\item In the simulation there were $1.5\times10^5$ atoms (on average), while there were $\approx3\times10^7$ atoms in the experiment. This smaller atom number was needed to achieve a long enough simulation time to see significant Bose enhancement of the scattered modes.
\item In the simulation no seed wave was placed.
}
The simulation was carried out in the center-of-mass frame, and
the initial  coherent wavefunction was taken to be
\EQN{\label{psicos}
\psi(\bo{x}) =\psi_{\rm GP}(\bo{x})\sqrt{2}\cos\left(\frac{m\bo{v}^{\rm cm}\cdot\bo{x}}{\hbar}\right)
,}
where $\psi_{\rm GP}(\bo{x})$ is the coherent $T=0$ ground state solution of the Gross-Pitaevskii (GP) equation in a trap of the same dimensions as in the experiment, and the full number of atoms. The Bragg pulse used to impart velocity to the moving  wavepacket was of short duration $\approx 40\mu$s,  and so to a good approximation no significant evolution of the condensate took place during the pulse, hence \eqref{psicos} is a good approximation to the state of the system just after the pulse. 
The initial wavepackets move in the longitudinal direction $\pm\bo{v}^{\rm cm}=[\pm {\rm v}^{\rm cm}_1,0,0]$.
The simulation was on a $438\times48\times48$ lattice with $\mc{S}=200$ trajectories, and lattice spacing of $\Delta x_1=0.2209\mu$m in the longitudinal, and $\Delta x_{2,3}=0.4921\mu$m in the axial directions. Simulation time is $\order{6 {\rm days}}$  on a PC of 2002 vintage.

\begin{figure}[tb]
\center{\includegraphics[width=\textwidth]{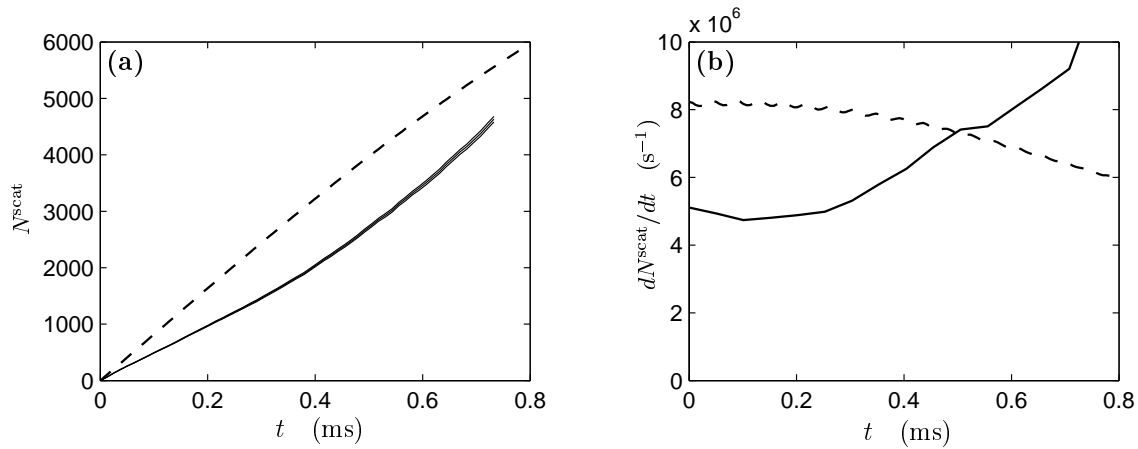}}\vspace{-8pt}\par
\caption[Scattered atoms in a condensate collision]{\label{FIGUREScat}\footnotesize
\textbf{}\textbf{(a)}: Total number of scattered atoms $N^{\rm scat}$, and \textbf{(b)}: rate of scattering in the colliding BEC system of Section~\ref{CH10Scattering}. {\scshape solid line}: scattered atoms calculated as described in Section~\ref{CH10ScatteringSimulation}, from a positive P simulation. Triple lines indicate uncertainty at one standard deviation. {\scshape dashed line}: Estimate obtained with the imaginary scattering length method\cite{Band-00}.
\normalsize}
\end{figure}

Figure~\ref{FIGUREScat} shows the change of scattering rate with time, and the total number of scattered atoms for this simulation. 
 Since the overlap between the two wavepackets travelling at $\pm\bo{v}^{\rm cm}$ decreases as they move apart, and
the rate of scattering into empty modes is  proportional to this overlap, then the increasing scattering rate in the simulation is \textbf{clear evidence of the beginning of Bose enhancement of the initially empty modes}.

An interesting technical difficulty occurs when estimating the number of scattered atoms. Just as in a real experiment, the atoms in the first-principles calculation are not labeled as ``scattered'' or ``non-scattered''. Some of them are scattered back into momentum modes already occupied by the initial wavepackets, and cannot be separated from the unscattered atoms by counting. To nevertheless make an estimate of scattering rate, a similar procedure was used to what would be needed for experimental data. It was aimed to count only those atoms with momenta beyond the coherent wavepackets. Explicitly, momentum modes with longitudinal velocities 
differing by less than 1mm/s from $\pm {\rm v}^{\rm cm}_1$, \textit{and} with radial velocities of less than 2.3mm/s in both orthogonal directions are excluded from the count. (For comparison, the rms velocity deviation of atoms from $\pm\bo{v}^{\rm cm}=[\pm {\rm v}_1^{\rm cm},0,0]$ in the condensates  is initially about 0.13mm/s  and 0.5mm/s in the longitudinal and radial directions, although some later spreading occurs). This gives a scattered atom count $N^{\rm scat}_1$.

 Furthermore, in a mean field GP equation calculation, there appear some atoms at momenta that would be counted as ``scattered'' with the above counting method. (These occur mainly in the far tails of the wavepacket momentum distribution due to spreading with time, and also at velocities $\pm3\bo{v}^{\rm cm}$ due to the stimulation of a weak scattering process $\bo{v}\, \& \bo{v}\to3\bo{v}\,\&-\bo{v}$ when $\bo{v}\approx\pm\bo{v}^{\rm cm}$ to a  state with short lifetime (due to lack of energy conservation). Initially their number is $\order{200}$ initially, growing to $\order{3000}$ at the end of the simulation).
These GP background atoms, which are not due to the spontaneous scattering  process of interest or its Bose enhancement, are subtracted away from  the $N^{\rm scat}_1$ count obtained previously to arrive at the final scattered atoms estimate $N^{\rm scat}$ shown in Figure~\ref{FIGUREScat}.

To check how well the expression \eqref{ppsimtime} assesses simulation time \textit{a priori}, let us see what estimate it gives in this more complex system. 
The peak density in the middle of the initial cloud in the Thomas-Fermi approximation is $\rho_0^{\rm TF}=\hbar\omega_{\rm ho}(15\bar{N}a_s/a_{\rm ho})^{2/5}/2g$. 
The maximum density is well approximated by twice this Thomas-Fermi peak density (the factor of two arises because the  initial condition \eqref{psicos} has local density peaks with twice the local average density). Using $\text{max}[\rho(\bo{x})]=2\rho_0^{\rm TF}$ in \eqref{ppsimtime}, one obtains  an expected simulation time of 0.6ms, which agrees fairly well with the observed simulation time of 0.78ms. The extra simulation time as compared to the estimate may be due to the decrease of  peak density with time as the wavepackets  move apart.

  Lastly, the local diffusion gauges \eqref{thenoptgauge} and \eqref{ahodiffusiongauge} of Chapter~\ref{CH7} were not expected (nor found) to give significant simulation time improvements for this calculation, because the 
healing length is larger than the lattice spacing. (Using $\rho_0^{\rm TF}$ again, one obtains $\xi^{\rm heal}\approx 0.6\mu$m. The small lattice size used is needed to resolve the phase oscillations in the moving condensate.

\subsection{Comparison to imaginary scattering length estimate}
\label{CH10ScatteringImag}

The first-principles scattering rate is also compared  to that obtained using the imaginary scattering length technique\cite{Band-00} (see Figure~\ref{FIGUREScat}). This approximate method applies to colliding BECs when (among other conditions) their relative velocity is much larger than the momentum spread in a single wavepacket. (Here, the packet velocity relative to the center of mass is $\approx10$mm/s in the longitudinal direction, while the initial velocity spread in this direction is $\approx0.13$mm/s rms). Scattering losses from the condensate wavefunction to empty modes are estimated from the GP equations 
by making the replacement
\EQN{
a_s\to a'_s = a_s(1 -i|\bo{k}|a_s)
,}
where $\bo{k}=m\bo{v}^{\rm cm}/\hbar$.
 $a'_s$ enters the GP equations as 
\EQN{
g = \frac{4\pi\hbar^2a'_s}{m}
}
in the (now complex) scattering strength, which leads to particle loss form the wavefunction.

In Figure~\ref{FIGUREScat}, one immediately sees that the scattering rate is significantly overestimated with this method.
This difference is due to suppression of scattering to momenta lying close to the  primary direction of motion $\pm\bo{v}^{\rm cm}$, which is not taken into account in  the imaginary scattering length approximation. 

\begin{figure}[htbp]
\center{\includegraphics[width=\textwidth]{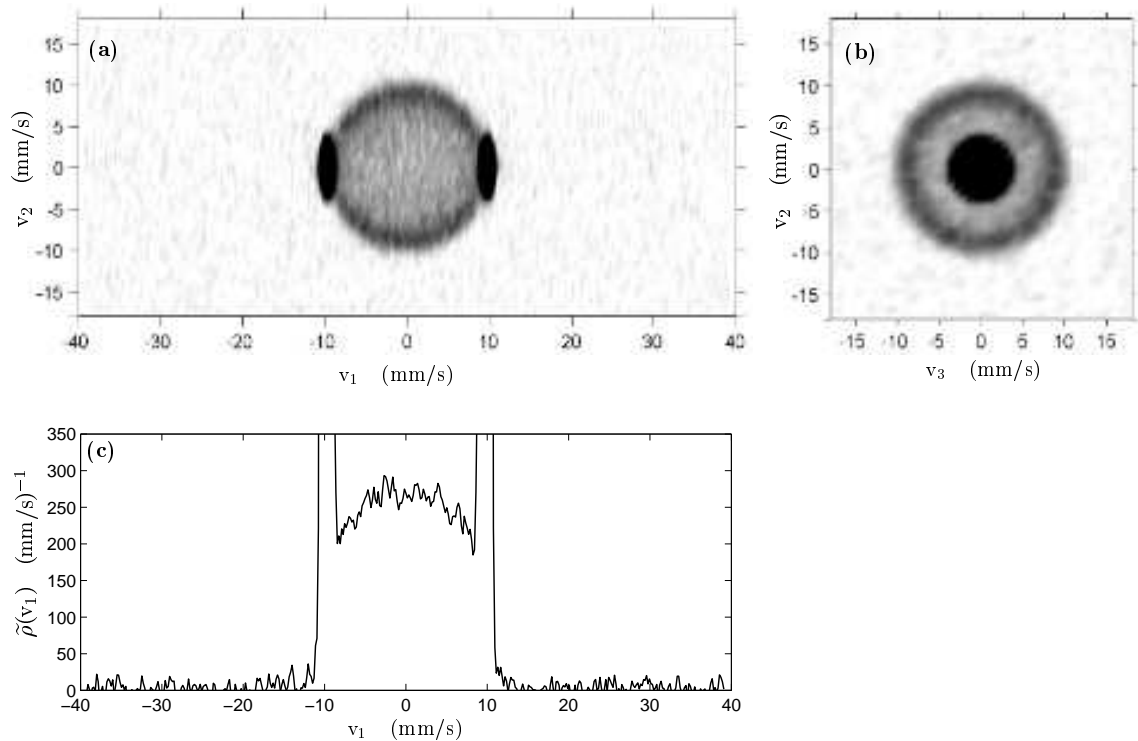}}\vspace{-8pt}\par
\caption[Velocity distribution of atoms during condensate collision]{\label{FIGUREscatdist}\footnotesize
\textbf{Velocity distribution} of atoms at a time $t=0.783$ms after the end of the Bragg pulse that separates the two condensate wavepackets. Subplot \textbf{(a)} shows the combined distribution of longitudinal velocity ${\rm v}_1$ and velocity in one axial direction ${\rm v}_2$, after summing over all ${\rm v}_3$ values. Density of atoms grows as shading darkens, with no shading corresponding to vacuum. The high spatial frequency noise is due finite sample ($\mc{S}=200$)  uncertainty rather than actual atoms. Subplot \textbf{(b)} shows the atom distribution in the axial directions, with summing over all longitudinal velocity values ${\rm v}_1$. 
 Subplot \textbf{(c)} shows the distribution of longitudinal atom velocity $\wt{\rho}({\rm v}_1)$. The noise is due to finite sample effects, and shows the degree of uncertainty in the calculated values.
Data are from the positive P simulation described in Section~\ref{CH10ScatteringSimulation}.
[arXiv note: higher resultion available at Piotr Deuar's homepage, currently {\tt http://www.physics.uq.edu.au/people/deuar/thesis/}]\normalsize}
\end{figure}

   In Figure~\ref{FIGUREscatdist}, marginals of the (positive P) simulated momentum distribution are plotted, and in particular, density of the velocity component $v_1$ in the longitudinal direction is shown in  subplot \textbf{(c)}. 
One sees that the density of scattered atoms decreases (slowly) as $|{\rm v}_1|\to {\rm v}_1^{\rm cm}$, whereas if the scattering was to an isotropic spherical shell of momenta, this linear density would rise towards ${\rm v}_1^{\rm cm}$.

This suppression of scattering has been predicted by Bach\etal\cite{Bach-02} when considering the scattering from two plane waves under the Bogoliubov approximation. They found that the suppression of scattering was dependent on the ratio $r_E$ of kinetic single-particle energy due to the plane wave motion with respect to the center of mass, and the interaction energy per particle. This ratio was 
\EQN{\label{redef}
r_E=m|\bo{v}^{\rm cm}|^2/2g\rho
,}
 where $\rho$ was the spatial density of each plane wave component. 
The suppression of scattering in the $\pm\bo{v}^{\rm cm}$ directions becomes less significant at large $r_E$, but is still strong when $r_E\approx\order{10}$. For the system simulated here, $r_E$ can be estimated by using $\rho_0^{\rm TF}$ as an estimate for $\rho$ in \eqref{redef}, and gives $r_E\approx 5$, well in the regime where scattering along the direction $\pm\bo{v}^{\rm cm}$ direction is suppressed. 

   A reduction of the magnitude of scattered particle momenta from the condensate values was also predicted in Ref.~\cite{Bach-02} and is also seen in the simulation here (Note how the circle of scattered atoms in Figure~\ref{FIGUREscatdist}\textbf{(c)} lies at a slightly smaller radius in momentum space than the condensates.)

\section{Summary}
\label{CH10Summary}

The above examples demonstrate that qualitative predictions of dynamics can be made with the gauge P representation (or the special case of the positive P representation) for a wide range of many-mode interacting Bose gas systems.  This includes predictions for spatial correlation functions (including non-local and/or high order) as well as local observables such as densities. Processes simulated included interaction with external trap potentials and two-body scattering under the influence of an extended interparticle potential, or of one acting only locally at each lattice point as in a Bose-Hubbard model.  Spatial and momentum densities and their fluctuations have also been previously calculated for a lossy system with local interactions by Corney and Drummond \cite{Corney99,DrummondCorney99} using the un-gauged method. There, the onset of condensation in an evaporative cooling simulation was seen. 

\enlargethispage{1cm}
The examples in this chapter show that such first-principles simulations can be tractable even with very large numbers of modes or particles in the system given the right conditions. This is evidenced by example 4 above, where there were \mbox{1009\,152} lattice points and on average \mbox{150\,000} atoms.  

Situations where there are several length scales of similar order, or processes of similar strength are of particular interest for first-principles simulations because it is  difficult to make accurate quantitative predictions otherwise. The examples of  Sections~\ref{CH10Trap} and ~\ref{CH10Scattering}  shown that predictions for systems in such regimes can  be made with the present method.

  Improvement of simulation times by use of local diffusion gauges was seen in Section~\ref{CH10Uniform}. This occurs when the lattice spacing is of the order of the (local) healing length or greater. This confirms what was expected from the analysis of Section~\ref{CH10LatticeHealing}.

In Section~\ref{CH10LatticePp}, an analytic estimate of useful simulation time with the un-gauged positive P method was derived. This was subsequently seen to be quite accurate both in the uniform gas with local interactions, and in the more complex colliding condensates system of Section~\ref{CH10Scattering}. 
 The simulation time is seen to scale as $(\Delta V)^{1/3}$ with the effective volume of a lattice point $\Delta V$, so that coarse-grained simulations last longer. There is a tradeoff between a fine enough lattice to include all relevant processes, and one coarse enough for the simulation to last long enough to see the desired phenomena. The analysis of Section~\ref{CH10LatticeKinetic} also shows that kinetic coupling between spatial modes (which can also reduce simulation time) is more dominant for fine lattices.
Simulations with Gaussian extended interparticle interactions \eqref{ugauss} were seen in Sections~\ref{CH10Extended} and~\ref{CH10Trap} to last at least several times longer than those with the same bulk interaction energy density.

 Some of the phenomena predicted by the example simulations include: 
\ITEM{
\item  The propagation of correlation disturbances (``correlation waves'') in a coherent gas without any corresponding density waves, and at a velocity $\approx\sqrt{2}$ times faster than the speed of sound. (Section~\ref{CH10Uniform})
\item The complex interaction between the breathing motion of a condensate in a trap and the interparticle correlations in the condensate. (Section~\ref{CH10Trap})
\item Bosonic enhancement of initially empty momentum modes during the collision of two condensates. (Section~\ref{CH10Scattering})
\item Suppression of spontaneous scattering processes  in these colliding condensates in the direction of motion. (Section~\ref{CH10ScatteringImag})
}

\chapter{Thermodynamics of a one-dimensional Bose gas}
\label{CH11}

The one-dimensional (1D) regime has been recently observed in trapped Bose gases\cite{Gorlitz-01,Schreck-01,Greiner-01}, giving  strong impetus to theoretical studies of such systems.
This chapter describes 
exact predictions of spatial correlation functions and momentum densities, which have been obtained for 1D uniform Bose gases in a grand canonical ensemble by using gauge P simulations.

A part of these results  have been recently published\cite{Drummond-04}.

\section{One-dimensional uniform Bose gas at finite temperature}
\label{CH1Gas}

\subsection{Exact solutions}
\label{CH11GasExact}

The  interacting uniform 1D Bose gas model described by the Hamiltonian 
\EQN{\label{1dHam}
\op{H} = \frac{\hbar^2}{2m}\int  \dada{\dagop{\Psi}(x)}{x}\dada{\op{\Psi}(x)}{x}\, dx + \frac{g}{2}\int \op{\Psi}^{\dagger 2}(x)\op{\Psi}^2(x)\, dx
}
was found to be exactly solvable several decades ago, and is one of the few exactly solvable nontrivial many-body problems\cite{Mattis94}. 
The solutions were found in the pioneering works of Girardeau\cite{Girardeau60,Girardeau65} at $g\to\infty$,  Lieb and Liniger\cite{LiebLiniger63,Lieb63} at $T=0$, and by Yang and Yang\cite{YangYang69} for the grand canonical ensemble at finite temperature $T>0$. See also \cite{Popov83,Thacker81,Korepin-93} for reviews.

  Explicltly, these solutions provide a numerical algorithm to calculate a density of states $\rho_k$ and holes $\rho_h$. These can then be used to obtain  some intensive physical quantities of the system such as density, energy density, and pressure. 
 Recently there have been some  further advances, especially in the zero temperature and strong interactions limit. These include $T=0$ expressions for the tails of the momentum distribution and short range $\bar{g}^{(1)}(x)$\cite{OlshaniiDunjko03}; $T=0$ expressions for $\bar{g}^{(2)}(0)$, $\bar{g}^{(3)}(0,0)$, and $\bar{g}^{(1)}(x)$\cite{GangardtShlyapnikov03}; and finite $T$ values of $\bar{g}^{(2)}(0)$\cite{Kheruntsyan-03}. 
These last finite temperature $\bar{g}^{(2)}(0)$ solutions indicate that there is a rich variety of  correlation phenomena in the $T>0$ regime. 

Still, exact results for the vast majority of observables remain hitherto inaccessible because they cannot be obtained from the density of states and holes (or perhaps, a way to obtain them has not yet been found). This includes such basic observables as spatial correlations and even momentum distributions for $T>0$.

The Hamiltonian \eqref{1dHam} is a special case of \eqref{hamiltonian}, and as such, its grand canonical thermodynamics can be simulated using the gauge P equations \eqref{gaugepthermo} in a wide range of physical regimes. This has allowed the calculation of exact spatial correlation functions and momentum distributions for this model here --- see Section~\ref{CH11Calc}.

Noteworthy in this context are also the stochastic wavefunction simulations of Carusotto and Castin\cite{CarusottoCastin01}, who calculated spatial correlation functions $\bar{g}^{(2)}(x)$ and $\bar{g}^{(1)}(x)$ of a different but related model: The canonical ensemble (i.e. with set particle number, as with all stochastic wavefunction calculations) of a uniform interacting gas in a system of small spatial extent where finite size effects were important.

\subsection{Correspondence with trapped gases}
\label{CH11GasTrapped}

Under present day experimental conditions, the Bose gas confinement is well approximated by a harmonic trap of transverse angular frequency $\omega_{\bot}$ and longitudinal frequency $\omega_0$. The correspondence between the uniform gas model, and a trapped gas experiment can be summarized in the following points:
\ITEM{

\item The atoms behave effectively as a 1D gas provided that the transverse zero point oscillations 
\EQN{\label{ltransv}
l_{\bot} = \sqrt{\frac{\hbar}{m\omega_{\bot}}}
}
are much smaller than other characteristic length scales of the system such as the healing length  \eqref{lheal} and the thermal de Broglie wavelength of excitations \eqref{ltherm}.  This regime has been reached in recent experiments\cite{Gorlitz-01,Schreck-01,Greiner-01}.

\item On length scales smaller than the trap size 
\EQN{
x \ll l_0 = \sqrt{\frac{\hbar}{m\omega_0}}
}
the density of the gas is slowly varying. On this and smaller scales the 
1D trapped gas and uniform gas will show the same physical behavior, and so the results obtained in Section~\ref{CH11Calc} will be directly applicable to a trapped gas while $x\ll l_0$. It is worth noting that if, instead,  details of the trap are important, a gauge P simulation with the external potential $V^{\rm ext}$ explicitly specified can be used instead. A dynamical example of this was given in Section~\ref{CH10Trap}.

\item The field interaction strength $g$ is related to the 3D scattering length $a_s$  by the expression \eqref{gdef}, where 
the effective trap cross-section in the transverse direction is\cite{Olshanii98} 
\EQN{
\lambda_0 = 2\pi l_{\bot}^2 = \frac{2\pi\hbar}{m\omega_{\bot}}
.}
}

\subsection{Uniform gas regimes}
\label{CH11GasRegimes}

The properties of a the finite temperature uniform 1D gas with the Hamiltonian \eqref{1dHam} depend on two dimensionless 
parameters: The first is the \textbf{coupling parameter}\footnote{Not to be confused with the unrelated single-boson loss rate $\gamma$ of Chapters~\ref{CH6} and~\ref{CH7}. $\gamma$ is used here in Chapter~\ref{CH11} for the quantity \eqref{gammadef} as this is the standard notation in the literature.}
\EQN{\label{gammadef}
\gamma = \frac{mg}{\hbar^2\rho}
,}
dependent on interaction strength $g$ and density $\rho$. The ideal Bose gas is reached as $\gamma\to0$, while the Tonks-Girardeau (TG) hard sphere gas occurs in the limit $\gamma\to\infty$. In this limit the particles undergo an effective ``fermionization''\cite{Girardeau60,Girardeau65,LiebLiniger63,Lieb63}.

The second dimensionless parameter is  the \textbf{reduced temperature}
\EQN{\label{taudef}
\wt{T} = \frac{T}{T_d} =  \frac{mk_BT}{2\pi\hbar^2\rho^2}
,}
where $T_d$ is the quantum degeneracy temperature. For $T<T_d$, the gas is denser than the quantum concentration $n_Q=\sqrt{mk_BT/2\pi\hbar^2}$ and the average spacing between particles $1/\rho$ is less than  the thermal de Broglie wavelength \eqref{ltherm}.

Several characteristic length scales in the uniform gas are:
\ITEM{
\item Mean interparticle spacing
\EQN{\label{lspac}
l_{\rho} = \frac{1}{\rho}
.}
\item Thermal de Broglie wavelength
\EQN{\label{ltherm}
\lambda_T = \sqrt{\frac{2\pi\hbar^2}{mk_BT}} = \frac{l_{\rho}}{\sqrt{\wt{T}}}
.}
Atoms with typical energy behave as particles on length scales $\gg\lambda_T$, as waves on $\ll\lambda_T$
\item Healing length
\EQN{\label{lheal}
\xi^{\rm heal} = \frac{\hbar}{\sqrt{2m\rho g}} = \frac{l_{\rho}}{\sqrt{2\gamma}}
,}
which is the length scale of typical spatial density fluctuations.
\item Degenerate phase coherence length
\EQN{\label{lphase}
l_{\phi} = \frac{\rho\lambda_T^2}{2\pi} = \frac{l_{\rho}}{2\pi\wt{T}}
,}
which applies when $\wt{T}\ll 1$ and $\gamma\ll 1$. When $\wt{T}\gg1$, the phase coherence length is $l_{\rho}=\lambda_T/\sqrt{2\pi}$.
\item 1D scattering length\cite{Olshanii98}
\EQN{\label{la1d}
a_{\rm 1D} = \frac{\hbar^2}{mg} = \frac{l_{\rho}}{\gamma}
}
is the characteristic distance over which the influence of a single particle is felt. In the fermionized large $\gamma$ regime, this is the effective size of the hard sphere due to a single particle.
}

The 1D gas behaves rather differently than the 2D and 3D cases, in that, for example, 
in a  strong coupling regime the density is \textit{low}. 
Also, a true BEC does not form, because long range order on length scales $\gg l_{\phi}$ is destroyed. This follows from the Bogoliubov $k^{-2}$ theorem, as explained in \cite{MerminWagner66,Hohenberg67}. Nevertheless, at degenerate temperatures 
$\wt{T}\lesssim 1$ and relatively weak couplings $l_{\phi}\ll\xi^{\rm heal}$ and $\gamma\ll 1$, a finite-range quasicondensate with phase coherence exists.

\section{Simulation details}
\label{CH11Details}

\subsection{Overview}
\label{CH11DetailsOverview}

Integration of the gauge P stochastic differential equations \eqref{gaugepthermo} proceeds in a similar fashion to the dynamics calculations of Chapter~\ref{CH10}. These are based on the master equation formulation of the grand canonical ensemble, as described in Section~\ref{CH2Thermodynamics}. A split-step algorithm is used, as described in Section~\ref{CH10Simulation} with integration using the semi-implicit method described in Appendix~\ref{APPB}. To exclude potential finite-sample bias of the kind described in Appendix~\ref{APPA}, the variance of the logarithms of quantities to be averaged for observable estimates was also tracked. 

To simulate the \textit{uniform} gas, care needed to be taken to avoid  finite size effects. This can usually be achieved for all practical purposes by  choosing a lattice spacing several times smaller than the smallest length scale of Section~\ref{CH11GasRegimes}, and a number of lattice points large enough so that the total spatial extent is several times larger than the largest length scale. Operationally, the lattice resolution was increased in  normal  and/or  momentum space until no changes with lattice resolution were seen in the observables of interest.

There are several further issues that arise only in the thermodynamics calculations and require more detailed discussion:

\subsection{Drift Gauge}
\label{CH11DetailsGauge}
   As was discussed in more detail in Section~\ref{CH9GaugeMvsing}, the un-gauged positive P equations are not suitable for simulation because of severe moving singularities as $\re{\alpha_{\bo{n}}\beta_{\bo{n}}}$ takes on negative values. This behavior appears in the many-mode case as well, and leads to rapid onset of spiking, as well as possible systematic errors after only a short simulation time. 
 
The radial drift gauge \eqref{radialgauge} acting locally on each mode was used to remove the moving singularities in the same manner as for the single-mode system of Chapter~\ref{CH9}.  Explicitly, when adapted to the many-mode situation this gauge is
\EQN{\label{radialgaugeM}
\mc{G}_{\bo{n}} = i\sqrt{2\hbar\chi}\left(\alpha_{\bo{n}}\beta_{\bo{n}}-|\alpha_{\bo{n}}\beta_{\bo{n}}|\right)
,}
where  $\chi$  is given by \eqref{chidef}, as usual. 

Spiking abates, and boundary term errors are not seen.

\subsection{Importance sampling}
\label{CH11Preweighting}

The simulated equations \eqref{gaugepthermo} include evolution of both  amplitudes  $\alpha_{\bo{n}}$ and weight $\Omega=e^{z_0}$. Deterministic evolution of the weight is a new feature compared to dynamics and can cause sampling problems at nonzero times $\tau>0$.

It is convenient to initialize in Fourier space. Here 
the variables $\wt{\alpha}_{\wt{\bo{n}}}(\tau)$ and $\wt{\alpha}'_{\wt{\bo{n}}}(\tau)$ are given by \eqref{fouriervardef}, 
 and bold quantities indicate vectors with $M$ elements --- one per mode. 
For an $M$-mode system of length $L$, the standard initial distribution \eqref{inidist} becomes in Fourier space 
\EQN{\label{inidistk}
P_G(\bm{\wt{\alpha}},\bm{\wt{\alpha}}',z_0) = \delta^2(z_0)\delta^{2M}(\bm{\wt{\alpha}}'-\bm{\wt{\alpha}}) \prod_{\bo{\wt{n}}} \frac{1}{\pi \wt{n}_0}\exp\left(\frac{-|\wt{\alpha}_{\bo{\wt{n}}}|^2}{\wt{n}_0}\right)
,}
where 
$\wt{n}_0 = L^2\bar{n}_0/2\pi M$
is the mean of all $|\wt{\alpha}_{\bo{\wt{n}}}|^2$ at $\tau=0$. The lattice factors arise because the number of particles per Fourier mode is $\propto\frac{L^2}{2\pi M}$. A sample 
is  generated by the simple procedure
\SEQN{}{
\wt{\alpha}_{\bo{\wt{n}}} &=& \sqrt{\wt{n}_0}\,\eta_{\bo{\wt{n}}} = \wt{\alpha}_{\bo{\wt{n}}}^{(0)}\\
\wt{\alpha}'_{\bo{\wt{n}}} &=& \wt{\alpha}^{(0)}_{\bo{\wt{n}}}\\
z_0 &=& 0
,}
where the $\eta_{\bo{\wt{n}}}$ are independent complex Gaussian noises of variance unity.

The problem is that this is a good sample of the $\tau=0$ distribution, but not necessarily so at later times. This is so even for an ideal gas when two-body interactions are absent. 
In the simplest case of the ideal gas with no external potential $V^{\rm ext}=0$, the equations of motion \eqref{gaugepthermo} can be solved exactly in Fourier space, which gives
\SEQN{\label{fourierideal}}{
\wt{\alpha}_{\bo{\wt{n}}}(\tau) &=&  \wt{\alpha}_{\bo{\wt{n}}}^{(0)}e^{h_{\bo{\wt{n}}}(\tau)\tau}\label{fourierideala}\\
\wt{\alpha'}_{\bo{\wt{n}}}(\tau) &=& \wt{\alpha}_{\bo{\wt{n}}}^{(0)}\\
z_0(\tau) &=& \frac{L^2}{2\pi M}\sum_{\bo{\wt{n}}}\left\{
|\wt{\alpha}_{\bo{\wt{n}}}(\tau)|^2-|\wt{\alpha}_{\bo{\wt{n}}}^{(0)}|^2\right\}\label{fourieridealz}
,} 
where 
\EQN{
h_{\bo{\wt{n}}}(\tau) = \mu(\tau) - \frac{\hbar^2k_{\bo{\wt{n}}}^2}{2m}
}
is the Gibbs factor exponent. 

For the ideal gas, the exact finite temperature solution consists of an independent thermal state with mean occupation $\bar{n}_{\bo{\wt{n}}}$ for each Fourier mode. An unbiased sample of this is
\SEQN{\label{idealgasd}}{
\wt{\alpha}_{\bo{\wt{n}}}(\tau) &=& L\sqrt{\frac{\bar{n}_{\bo{\wt{n}}}(\tau)}{2\pi M}}\,\eta_{\bo{\wt{n}}}\label{idealgasda}\\ 
\wt{\alpha}'_{\bo{\wt{n}}}(\tau) &=& \wt{\alpha}_{\bo{\wt{n}}}(\tau)\\
z_0 &=& 0
,}
where for the ideal Bose gas, the mode occupations obey the Bose-Einstein distribution
\EQN{\label{bedist}
\bar{n}_{\bo{\wt{n}}}(\tau) =\bar{n}_{\bo{\wt{n}}}^{\rm BE}(\tau)= \frac{1}{e^{-h_{\bo{\wt{n}}}(\tau)\tau}-1}
.}
The actual simulated $\wt{\alpha}_{\bo{\wt{n}}}(\tau)$ in \eqref{fourierideala} are not necessarily anywhere near \eqref{idealgasda} with Bose-Einstein  occupations \eqref{bedist}, so the nonzero variable weight \eqref{fourieridealz} is needed to compensate for this. 

This is fine in the limit of infinite samples $\mc{S}\to\infty$, but in practice the distribution of the weight $\Omega=e^{z_0}$ can be badly sampled once the variance in $z_0$ becomes too large. Since $|\wt{\alpha}_{\bo{\wt{n}}}^{(0)}|^2$ is a Gaussian, then $z_0$ also will have a similar nature and the usual limit \eqref{sdlimit} for exponentials of a Gaussian applies: The weights (and hence, observable estimators, which all average quantities $\propto e^{z_0}$) are well sampled only while 
\EQN{
\vari{\re{z_0}} \lesssim \order{10}
.}

A method known as  \textbf{importance sampling} offers a way to alleviate this weighting problem if one is primarily interested in the gas around a particular target time $\tau_T$ (i.e. temperature $1/k_B\tau$ and chemical potential $\mu(\tau_T)$). The importance sampling  approach is  widely used for weighted stochastic integration, and a comprehensive discussion of this for general situations can be found in \cite{NumericalRecipes}, Section 7.8. 
In the case considered here, the idea is as follows: 

In the gauge P representation, the density matrix is written as 
\EQN{\label{puzexpr}
\op{\rho}_u(\tau) = \int  P_G(\tau,\bm{z},z_0) \op{\Lambda}(\bm{z},z_0)\,d^{4M}\bm{z}\,d^2z_0
,}
 where $\bm{z}$ stands for all the phase-space variables $\bm{\wt{\alpha}}$, $\bm{\wt{\alpha}}'$. The simulation here uses standard drift gauges, so that the evolution of the weight $z_0$ is actually deterministically dependent on the evolution of the $\bm{z}$, as given in  \eqref{dz0deterministic}. In the initial distribution \eqref{inidistk}, $z_0=0$ is independent of $\bm{z}$. Together, these properties allow  the $\bm{z}$ and $z_0$ evolutions to be separated. 	
Explicitly, 
\EQN{
z_0(\tau,\bm{z}) = \int_0^{\tau} \dd{z_0(\tau',\bm{z}(\tau'))}{\tau'}\,d\tau'
,}
where $dz_0$ is given by \eqref{dz0deterministic}. After defining  $\op{\Lambda}(\bm{z},z_0) = e^{z_0}\op{\ul{\Lambda}}(\bm{z})$, this allows us to rewrite \eqref{puzexpr} as
\EQN{\label{pgsep}
\op{\rho}_u(\tau) = \int \ul{P}_G(\tau,\bm{z})\op{\ul{\Lambda}}(\bm{z})\, e^{z_0(\tau,\bm{z})}\ d^{4M}\bm{z}
,} 
where the underlined quantities depend only on the amplitude variables $\bm{z}$. 

The multiplying factor of the operator kernel $\op{\ul{\Lambda}}$ in \eqref{pgsep} is $\ul{P}_Ge^{z_0}$, and in fact there is a whole range of possible initial distributions provided that this factor remains the same. In particular we could start with a different initial distribution of the amplitude variables $\ul{P}_{G0}(0,\bm{z})$ provided that this is compensated for in the initial weights by
\EQN{\label{newz0}
z_0 =  \int_0^{\tau} \dd{z_0(\tau',\bm{z}(\tau'))}{\tau'}\,d\tau' +\log \ul{P}_G(0,\bm{z}) - \log\ul{P}_{G0}(0,\bm{z})
.}

Then (and this is the crucial point), if we could choose some initial distribution $\ul{P}_{G0}$ that leads to only a small spread in $z_0$ at target time $\tau_T$, then the simulation would be well sampled there. (It would also be sampled much worse at $\tau=0$, but this doesn't bother us).  This then is the sampling according to importance (i.e. according to the situation at $\tau_T$). 

For the purposes of the simulations reported in this chapter, a fairly crude yet effective importance sampling was applied:
At relatively weak coupling, a very rough but useful estimate of  the state 
is that the Fourier modes are uncoupled, and thermally distributed with mean occupations $\bar{n}_{\bo{\wt{n}}}(\tau_T)$. The equal-weight samples \eqref{idealgasd} 
correspond to the distribution 
\EQN{\label{PgGest}
\ul{P}_{G}^{\rm est}(\tau_T,\bm{z}) \propto \delta^{2M}(\bm{\wt{\alpha}}'-\bm{\wt{\alpha}}) \exp\left[ -\frac{2\pi M}{L^2}\sum_{\bo{\wt{n}}}\frac{|\wt{\alpha}_{\bo{\wt{n}}}|^2}{\bar{n}_{\bo{\wt{n}}}(\tau_T)}\right]
}
with appropriate normalization. This is not yet quite what is wanted because the desired sampling distribution $\ul{P}_{G0}$ is to be at $\tau=0$. An estimate of the initial distribution that leads to $\ul{P}_G^{\rm est}(\tau,\bm{z})$ can be obtained by evolving \eqref{PgGest} back in time with only kinetic interactions. This is again rather rough, since deterministic interaction terms $\propto g$ have been omitted (not to mention noise), but is simple to carry out, and was found to be sufficient for the purposes of the preliminary calculations presented here. One obtains the sampling distribution 
\EQN{\label{PG0}
\ul{P}_{G0}(0,\bm{z}) \propto \delta^{2M}(\bm{\wt{\alpha}}'-\bm{\wt{\alpha}}) \exp\left[ -\frac{2\pi M}{L^2}\sum_{\bo{\wt{n}}}\frac{|\wt{\alpha}_{\bo{\wt{n}}}|^2}{\bar{n}^{(0)}_{\bo{\wt{n}}}}\right]
,}
where 
\EQN{\label{barn0def}
\bar{n}^{(0)}_{\bo{\wt{n}}} = \bar{n}_{\bo{\wt{n}}}(\tau_T)\exp\left[-\lambda_n-h_{\bo{\wt{n}}}(\tau_T)\tau_T\right]
.}
(Note that $\lim_{\tau\to 0}[h_{\bo{\wt{n}}}(\tau)\tau] = -\lambda_n$).
Using \eqref{newz0}, the initial variables are then sampled according to 
\SEQN{}{
\wt{\alpha}_{\bo{\wt{n}}}(0) &=& L\sqrt{\frac{\bar{n}^{(0)}_{\bo{\wt{n}}}}{2\pi M}}\,\eta_{\bo{\wt{n}}}\\
\wt{\alpha}'_{\bo{\wt{n}}}(0) &=& \wt{\alpha}_{\bo{\wt{n}}}(0)\\
z_0(0) &=& \frac{L^2}{2\pi M}\sum_{\bo{\wt{n}}} |\wt{\alpha}_{\bo{\wt{n}}}(0)|^2\left(\frac{1}{\sqrt{\bar{n}^{(0)}_{\bo{\wt{n}}}}}-\frac{1}{\bar{n}_0}\right)\label{iniweightz0}
.}

For most of the simulations of Section~\ref{CH11Calc}, the target Fourier space distribution $\bar{n}_{\bo{\wt{n}}}(\tau_T)$  was just taken to be the plain ideal gas Bose-Einstein distribution \eqref{bedist}. 
However, once the interactions become significant, this is not satisfactory  because the actual interacting gas momentum distribution is far from the Bose-Einstein form (See e.g. Figure~\ref{FIGUREnk}). A better choice of $\bar{n}_{k_{\bo{\wt{n}}}}(\tau_T)$ is the density of states function $\rho_k$ of the exact Yang \& Yang solution\cite{YangYang69}.  In practice, to simplify the calculation, an estimate of $\rho_k$ was used instead. This was obtained by making a least squares (unweighted) fit of parameters $C_{\rm est}$ and $\sigma_{\rm est}$ in the expression
\EQN{
\bar{n}_{\bo{\wt{n}}} = \bar{n}_0e^{h_{\bo{\wt{n}}}(\tau_T)\tau_T+\lambda_n} + C_{\rm est} e^{-k_{\bo{\wt{n}}}^2\tau_T/2\sigma_{\rm est}^2}
}
to $\rho_k$.   This form was chosen because in the tails of the distribution in Fourier space when $k_{\bo{\wt{n}}}^2\tau_T\gg1$, $\bar{n}^{(0)}_{\bo{\wt{n}}}\approx\bar{n}_0$, and so the initial weights $z_0(0)$ in \eqref{iniweightz0} are not dependent on the momentum cutoff.

\subsection{Momentum cutoff}
\label{CH11DetailsKmax}

In a dynamical simulation, one requires that all occupied momentum modes are contained within the momentum cutoff $k^{\rm max}_d=\pi/\Delta x_d$. Strictly speaking, this cannot ever be satisfied in these thermodynamic simulations because the $T\to\infty$ ideal gas grand canonical ensemble is used as a starting condition. This has all momentum modes occupied up to infinite energy. 
This turns out not to be a practical problem  because the precision of calculated results is limited anyway by the finite sample size $\mc{S}$. It is then sufficient to increase the $k^{\rm max}_d$ until all observable estimates calculated are invariant with $k^{\rm max}_d$ to the precision achievable with a given ensemble size. This was carried out.

\subsection{Chemical potential at intermediate temperatures}
\label{CH11DetailsMu}

As was noted in Section~\ref{CH9Chempot}, if one is primarily interested in the behavior of the system at (or around) a given temperature $\tau_T$, and chemical potential $\mu$, then the values of $\mu$ at intermediate times $\tau<\tau_T$ can in principle be chosen at will. 

In practice, some choices lead to better precision per ensemble size than others. The analysis in  Section~\ref{CH9Chempot} indicated that the single-mode simulations led to the best precision when the ``effective'' chemical potential $\mu_e=\dada{\mu\tau}{\tau}$ was roughly constant during the simulation. 
It was assumed that a similar dependence occurs in the many-mode system, and so, accordingly a constant $\mu_e$ was chosen so that at $\tau_T$ the chemical potential becomes $\mu(\tau_T)$. Using \eqref{muexpr}, the constant $\mu_e$ assumption leads to
\EQN{\label{constmue}
\mu_e = \mu(\tau_T) + \frac{\lambda_n}{\tau_T}
.}

It remains to choose the initial density $\rho_0=M\bar{n}_0(\lambda_n)/L$  (with initial occupation $\bar{n}_0$ given by \eqref{barnexpr}). It was found that for target densities $\rho(\tau_T)$  a choice of initial density 
\EQN{
\rho_0= \order{\frac{\rho(\tau_T)}{10}}
}
gave the best precision at $\tau_T$ in most cases. Details varied depending on system parameters, although a clear tradeoff between two noise generating processes was seen:
\ENUM{
\item When $\rho_0$ was too large, a lot of added randomness is introduced into $z_0$ at early times, and precision is lost very rapidly once $\vari{\re{z_0}}\gtrsim\order{10}$, as discussed in Section~\ref{CH7Gaussian} and Appendix~\ref{APPA}.  This randomness arises largely independently of gauges or noise since the spread of $\alpha_{\bo{n}}$ in the initial thermal state is proportional to $\sqrt{\rho_0}$, and this then leads directly to a spread  in the deterministic evolution terms of $dz_0$. 
\item When $\rho_0$ is too small, excessive spread in the amplitude variables arises as well (which also then feeds into the log-weight $z_0$). 
}

\subsection{Scaling of weight variance}
\label{CH11DetailsScaling}

For uniform gas calculations it was found to be desirable to use the smallest system size that does not introduce finite size effects.  The reason is that (apart from gauge-dependent terms), the weight $\Omega(\tau)=\exp\int_0^\tau \dd{z_0}{\tau'}\,d\tau' $ has the form of a Gibbs factor $\exp((\mu \bar{N}-E)/k_BT)$. This can be verified by inspection of the equations \eqref{gaugepthermo}.  The magnitude but also the \textit{spread} in log-weights $z_0$ will thus grow with energy $E$ and particle number $\bar{N}$. Since $\vari{\re{z_0}}$ must be $\lesssim\order{10}$ for good sampling of the distribution, this sets an upper limit on how large a system can be simulated.

\section{Physical regimes simulated}
\label{CH11Regimes}

\subsection{Parameter targeting}
\label{CH11RegimesTargeting}

The procedure to calculate properties at given target gas parameters $\gamma$ and $\wt{T}$ was as follows:
\ENUM{
\item Three physical quantities specify units. For example, $m$, $\hbar$, and $k_BT=1/\tau_T$ can be chosen unity. For the purposes of investigating gases at a given $\gamma$ and $\wt{T}$, these can be chosen arbitrarily and properties of the whole family of physical gases with those $\gamma$ and $\wt{T}$ follow by scaling. 
\item The target time $\tau_T$ is determined by the choice of units above, and the remaining essential parameters required to simulate \eqref{gaugepthermo} are $g$ and $\mu(\tau_T)$. The coupling strength is determined by 
\EQN{
 g = \hbar \gamma \sqrt{\frac{k_BT}{2\pi m\wt{T}}}
,}
while the required target density is
\EQN{\label{rhoTexpr}
\rho(\tau_T) = \sqrt{\frac{mk_BT}{2\pi\hbar^2\wt{T}}}
.}
\item The target density is not a direct input parameter into the simulation, but it is has a one-to-one correspondence with $\mu(\tau_T)$ (while in dimensionless units). The Yang \& Yang exact solutions\cite{YangYang69} provide an algorithm to calculate $\rho(T,\mu,g)$, and the chemical potential needed to obtain the desired density \eqref{rhoTexpr} was calculated by numerically inverting this relation.
}

\subsection{Regimes attained}
\label{CH11RegimesRegimes}
\begin{figure}[t]
\center{\includegraphics[width=\textwidth]{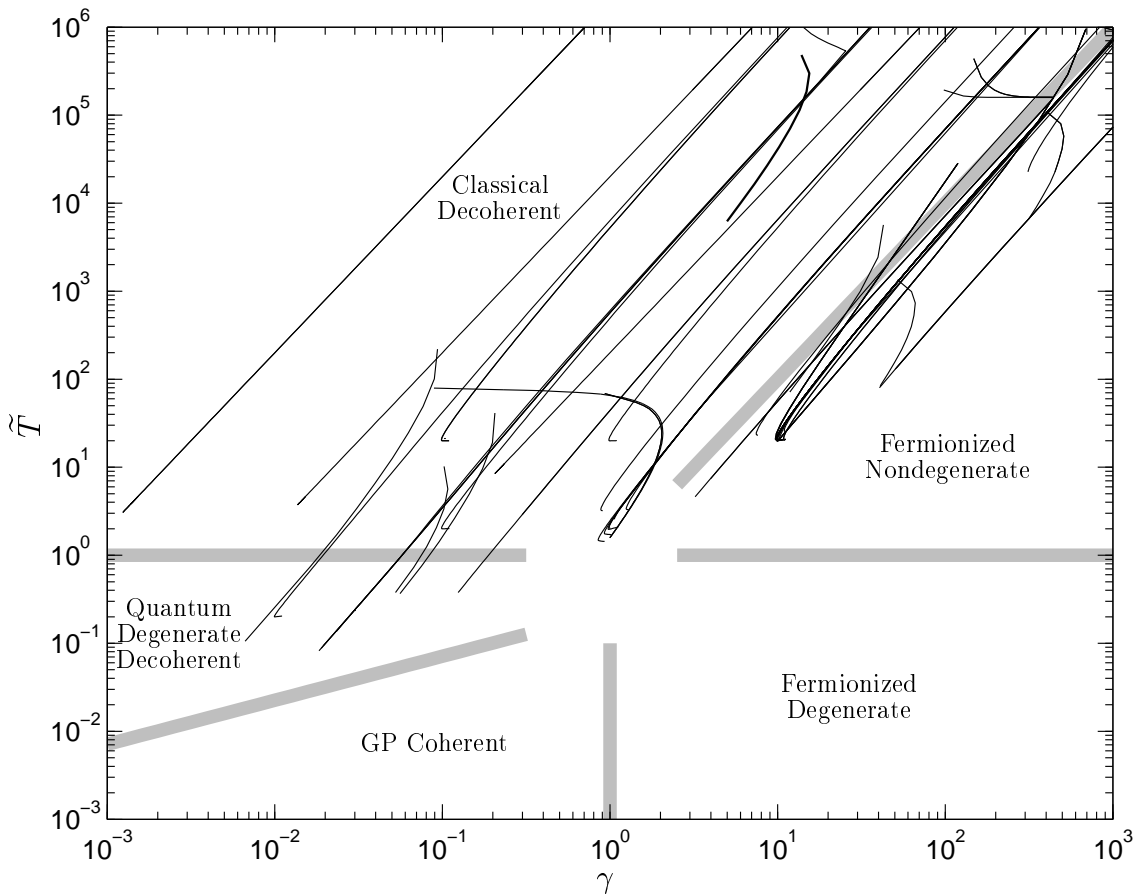}}\vspace{-8pt}\par
\caption[Physical regimes attained for the uniform interacting 1D Bose gas]{\label{FIGUREgamtau}\footnotesize
\textbf{Physical regimes attained} for the uniform interacting 1D Bose gas. {\scshape dark solid} lines show the paths in $\gamma$, $\wt{T}$ parameter space taken by simulations for the $\tau$ range where useful precision in $\bar{g}^{(2)}(x)$ was obtained. Paths shown include simulations with various forms of $\mu_e$: either given by \eqref{constmue}, or by $2e^{\mu\tau}=e^{-\lambda_n}+e^{\mu(\tau_T)\tau_T}+(e^{\mu(\tau_T)\tau_T}-e^{-\lambda_n})\cos[4\pi(\tau-\tau_T)/3\tau_T]$ described in \cite{Drummond-04}. The constant form \eqref{constmue} was found to be as good or superior at reaching low temperatures.
{\scshape Thick shaded} bars indicate approximate transition regions between physical regimes based on the classification by Kheruntsyan\etal\cite{Kheruntsyan-03} on the basis of $\bar{g}^{(2)}(0)$. 
\normalsize}
\end{figure}

A wide range of physical regimes were simulated directly with the procedure described in the previous sections. In terms of the two characteristic dimensionless gas parameters $\gamma$ and $\wt{T}$, the physical regions that have been explored are shown in Figure~\ref{FIGUREgamtau}.

In terms of the physical regimes classification of Kheruntsyan\etal\cite{Kheruntsyan-03} based on $\bar{g}^{(2)}(0)$ behavior (These are described in more detail in Section~\ref{CH11CalcCorr}), the accessible regions are both the \textbf{quantum degenerate decoherent} and \textbf{classical decoherent} regimes, as well as the \textbf{nondegenerate fermionized} (strong coupling) regime. Simulations also access \textbf{transition regions} between these. The interesting transition region $\gamma\approx\wt{T}\approx\order{1}$ where several length scales are of the same order is also accessible in part.
The lower limit of accessible $\wt{T}$ does not appear to be a hard limit, and more sophisticated importance sampling techniques and/or choices of $\mu_e(\tau)$ may be able to make some further inroads into lower temperature regions. 

   The basic limiting factor is growth of the variance of the real and/or imaginary parts of the log-weight $z_0$. As explained Section~\ref{CH11DetailsScaling}, there is a tradeoff between the variance of $z_0$ and the system size (e.g. lattice length $L$ if the density is set), but there is a limit on how small the system can be made if it is to correspond to the uniform gas model. $L$ must be significantly larger than the longest relevant length scale. 
\enlargethispage{0.5cm}
However, this limit is only applicable if an unconstrained uniform gas is required. 
The behavior of a gas in a finite box, torus, or trap is an easier simulation and can reach temperatures not accessible for uniform gas simulations if $L$ can be made small enough to control the growth of $\vari{z_0}$.
The previously mentioned calculations of by Carusotto and Castin\cite{CarusottoCastin01} for the different but related closed system model were made in such a ``low temperature with finite-size effects'' regime, with  $\gamma\approx\order{0.001}-\order{0.01}$, $\wt{T}\approx\order{0.001}-\order{0.005}$.

\section{Observable predictions}
\label{CH11Calc}

\subsection{Spatial correlation functions}
\label{CH11CalcCorr}

A classification of uniform gas behaviors based on the local second order correlation function $\bar{g}^{(2)}(0)$ has been determined  from the Yang \& Yang solution by Kheruntsyan\etal\cite{Kheruntsyan-03}. In this chapter, the finite temperature behavior of $\bar{g}^{(1)}(x)$, $\bar{g}^{(2)}(x)$, and $\bar{g}^{(3)}(x,y)$ has been calculated in several of these regimes and is shown in Figures~\ref{FIGUREgfig1} to~\ref{FIGUREg3fig}. 
Exact results are also compared to ideal gas values. 
The simulated  physical regimes were:

\subsubsection{Classical decoherent regime ($\wt{T}\gg\text{max}[1,\gamma^2]$)} 
\begin{figure}[tbp]
\center{\includegraphics[width=\textwidth]{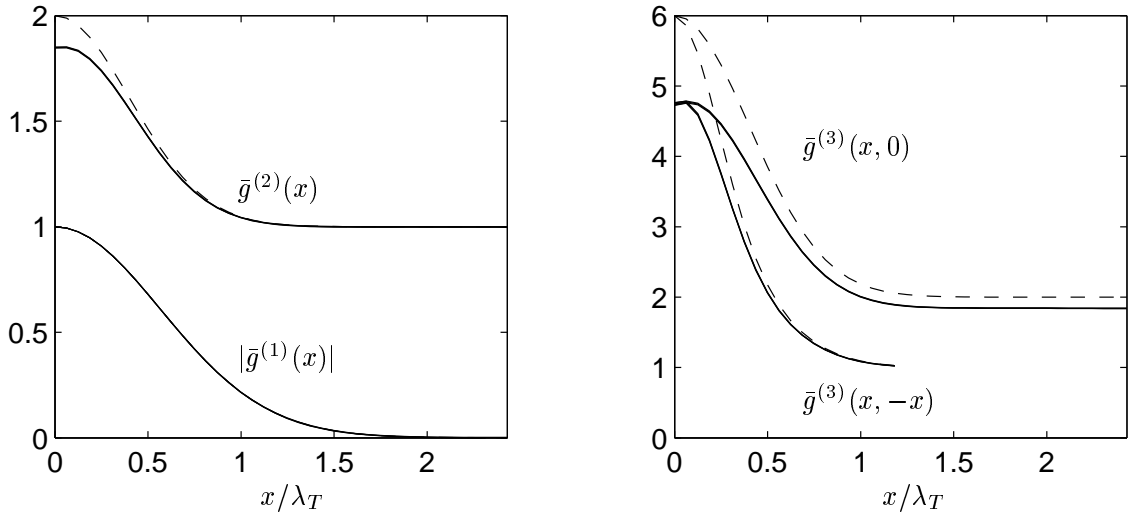}}\vspace{-8pt}\par
\caption[Correlations in a classical gas]{\label{FIGUREgfig1}\footnotesize
\textbf{Correlation functions in $\gamma=5$, $\wt{T}=1000$ decoherent classical gases.}
Exact results are shown as {\scshape solid} lines, with triple lines indicating uncertainty (mostly not visible at this scale). {\scshape dashed} lines show the ideal Bose gas values for comparison ($g^{(1)}(x)$ is indistinguishable). Here, $l_{\rho}=10\pi\lambda_T$.
$\mc{S}=10^5$, $M=80$, $L=5\lambda_T$.
\normalsize}
\end{figure}

This regime occurs while the thermal wavelength dominates the small scale behavior of the particles. In particular, while it is smaller than the interparticle spacing $l_{\rho}$ and the effective scattering length $a_{\rm 1D}$. Correlation functions in a fairly strongly interacting part of this regime are shown in Figure~\ref{FIGUREgfig1}. Correlations decay on the length scale of $\lambda_T$, with only a small reduction in close range correlation with respect to the ideal gas.

\subsubsection{Quantum degenerate decoherent regime ($\sqrt{\gamma/2\pi^2}\ll\wt{T}\ll1$)} 
\begin{figure}[tbp]
\center{\includegraphics[width=\textwidth]{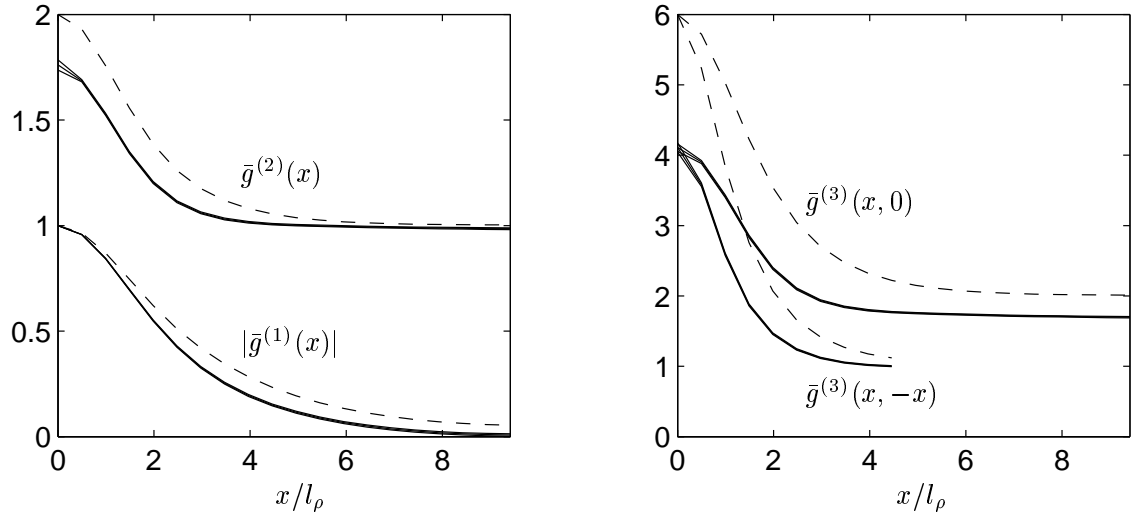}}\vspace{-8pt}\par
\caption[Correlations in a decoherent degenerate quantum gas]{\label{FIGUREgfig2}\footnotesize
\textbf{Correlation functions in $\gamma=0.03$, $\wt{T}=0.236$ decoherent quantum degenerate gases.}
Exact results are shown as {\scshape solid} lines, with triple lines indicating uncertainty (mostly not visible at this scale). {\scshape dashed} lines show the ideal Bose gas values for comparison. Here,  $\lambda_T=2.06l_{\rho}$.
$\mc{S}=10^5$, $M=40$, $L=5.14l_{\rho}$. 
\normalsize}
\end{figure}

Here the gas is degenerate, but phase coherence exists only on length scales $l_{\phi}$ shorter or of the same order as  the healing length $\xi^{\rm heal}$. This is not enough for any significant quasicondensate to form. Example correlations shown in Figure~\ref{FIGUREgfig2}. The short-range correlations are still largely thermal $g^{(2)}>1$,  as for the classical gas, but 
a systematic reduction of the range of all multi-particle correlations in comparison to the ideal gas is seen.

\subsubsection{Fermionized nondegenerate regime ($1\gg\wt{T}\ll\gamma^2$)} 
\begin{figure}[tbp]
\center{\includegraphics[width=\textwidth]{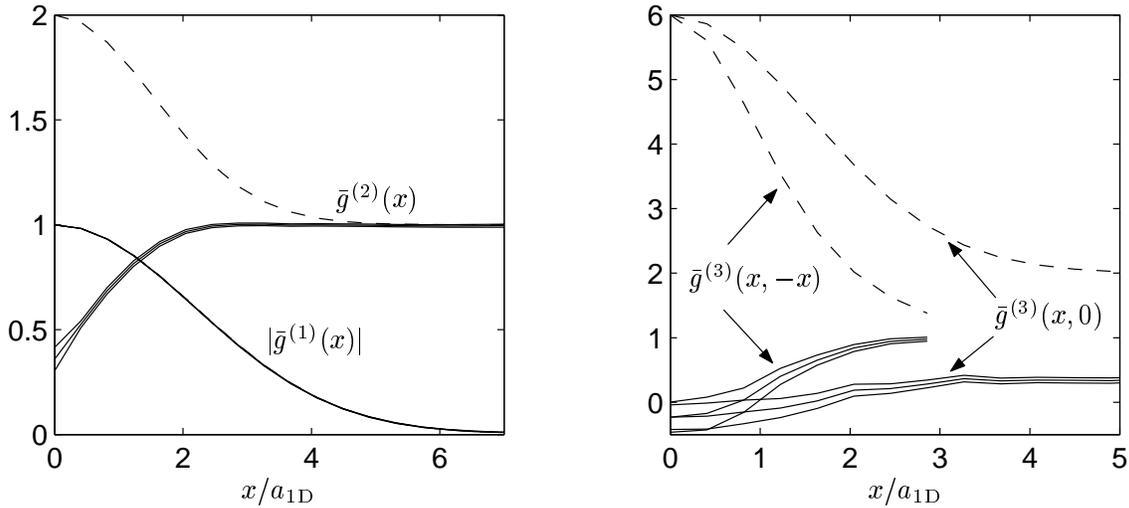}}\vspace{-8pt}\par
\caption[Correlations in a fermionized gas (1)]{\label{FIGUREgfig3}\footnotesize
\textbf{Correlation functions in $\gamma=300$, $\wt{T}=6000$ strongly fermionized nondegenerate gases.}
Exact results are shown as {\scshape solid} lines, with triple lines indicating uncertainty. {\scshape dashed} lines show the ideal Bose gas values for comparison ($g^{(1)}(x)$ is indistinguishable). Here,  $l_{\rho}=300a_{\rm 1D}$, $\lambda_T=3.873a_{\rm 1D}$, and $\xi^{\rm heal}=12.24a_{\rm 1D}$.
$\mc{S}=10^5$, $M=36$, $L=14.7a_{\rm 1D}$.
\normalsize}
\end{figure}

Here the fermion-like nature of strongly repulsive atoms dominates the small-scale behavior of the gas. The effective scattering length $a_{\rm 1D}$ is much smaller than both the thermal wavelength and the interparticle spacing $l_{\rho}$ leading to behavior similar to a hard-sphere (Tonks-Girardeau) model. 
In Figure~\ref{FIGUREgfig3}, one sees strong antibunching on length scales $a_{\rm 1D}$, several times smaller than the phase coherence length. Three-particle correlations are also very strongly reduced at small distances $\lesssim a_{\rm 1D}$, with $|\bar{g}^{(3)}(0,0)|\lesssim 0.2$ in the example at $\gamma=300$, $\wt{T}=6000$ shown in Figure~\ref{FIGUREgfig3}. This will lead to a strong reduction in three-particle inelastic scattering losses despite the very high temperature.

\subsubsection{Transition regime with atom pairing ($\lambda_T\approx a_{\rm 1D}\approx\order{1}$)}
Interesting new physical phenomena were seen in simulations of the transition regime where $\gamma\approx\wt{T}\approx\order{1}$. Here all the length scales of the system noted in Section~\ref{CH11GasRegimes} are of similar order, the exact relationship between them depending fairly sensitively on $\gamma$ and $\wt{T}$. Because of the presence several competing processes, approximate methods do not usually give accurate quantitative predictions.

\begin{figure}[tbp]
\center{\includegraphics[width=\textwidth]{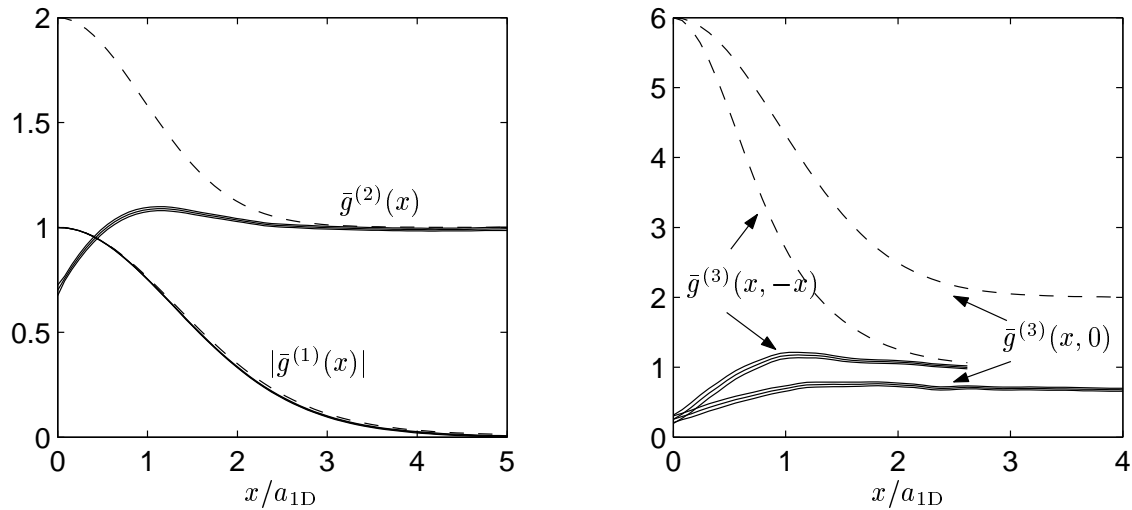}}\vspace{-8pt}\par
\caption[Correlations in a fermionized gas (2)]{\label{FIGUREgfig4}\footnotesize
\textbf{Correlation functions in $\gamma=10$, $\wt{T}=20$ fermionized nondegenerate gases.}
Exact results are shown as {\scshape solid} lines, with triple lines indicating uncertainty. {\scshape dashed} lines show the ideal Bose gas values for comparison. Here,  $l_{\rho}=10a_{\rm 1D}$, $\lambda_T=\xi^{\rm heal}=2.24a_{\rm 1D}$.
$\mc{S}=10^5$, $M=135$, $L=10.7a_{\rm 1D}$.
\normalsize}
\end{figure}

\begin{figure}[tbp]
\center{\includegraphics[width=\textwidth]{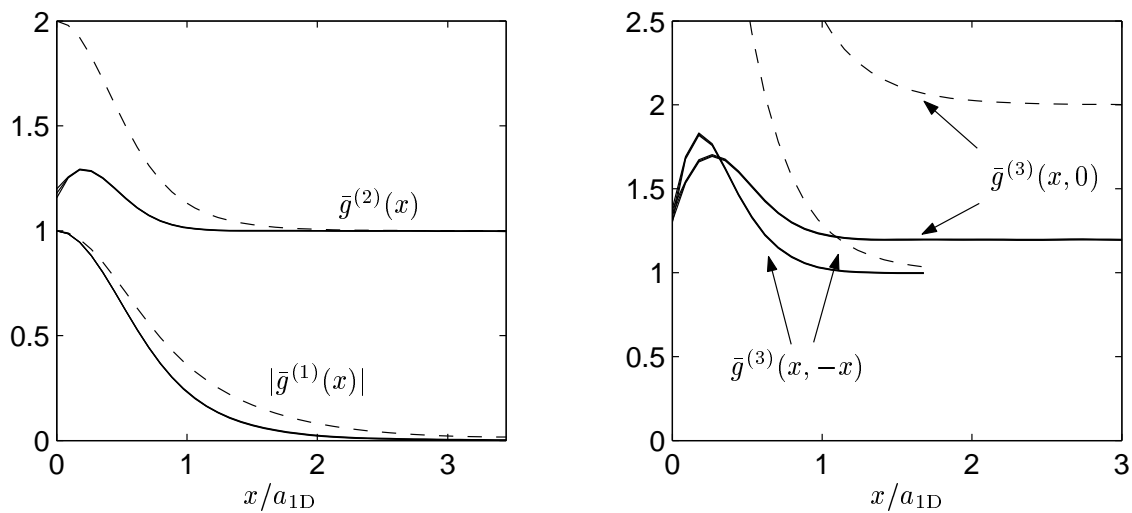}}\vspace{-8pt}\par
\caption[Correlations in a transition gas]{\label{FIGUREgfig5}\footnotesize
\textbf{Correlation functions in $\gamma=1$, $\wt{T}=1.6$ gases.}
Exact results are shown as {\scshape solid} lines, with triple lines indicating uncertainty (mostly not visible at this scale). {\scshape dashed} lines show the ideal Bose gas values for comparison. Here,  $l_{\rho}=a_{\rm 1D}$, $\lambda_T=0.793a_{\rm 1D}$, and $\xi^{\rm heal}=0.708a_{\rm 1D}$.
$\mc{S}=10^5$, $M=80$, $L=7.07a_{\rm 1D}$.
\normalsize}
\end{figure}

\begin{figure}[tbp]
\center{\includegraphics[width=300pt]{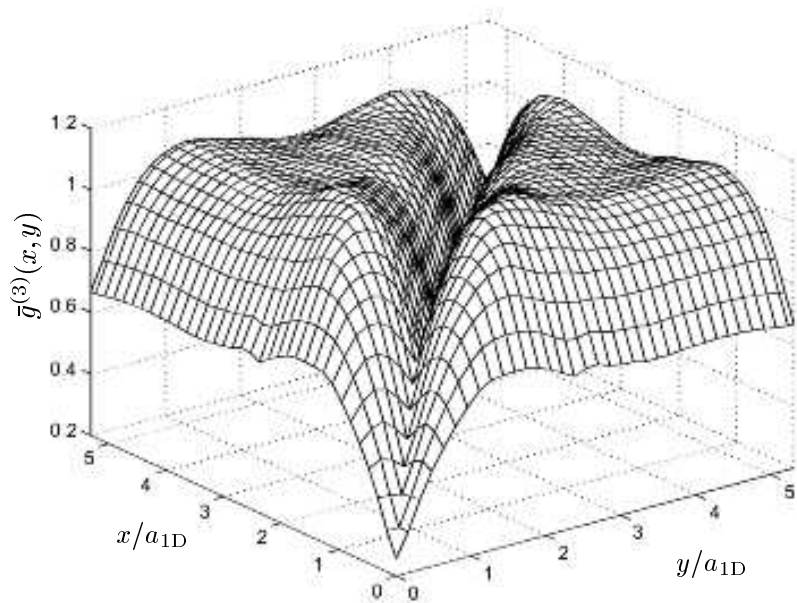}}\vspace{-8pt}\par
\caption[Three-particle correlations in a fermionized gas (2)]{\label{FIGUREg3fig}\footnotesize
\textbf{General three-particle correlation function} in $\gamma=10$, $\wt{T}=20$ fermionized nondegenerate gases.
Simulation details as in Figure~\ref{FIGUREgfig4}. [arXiv note: higher resultion available at Piotr Deuar's homepage, currently {\tt http://www.physics.uq.edu.au/people/deuar/thesis/}]
\normalsize}
\end{figure}

The transition between the nondegenerate fermionized gas and a decoherent quantum degenerate gas is shown in the sequence of Figures~\ref{FIGUREgfig3},~\ref{FIGUREgfig4},~\ref{FIGUREgfig5}, and~\ref{FIGUREgfig2}. Details of the three-particle correlations in the still-fermionized part of the transition are shown in Figure~\ref{FIGUREg3fig}.

There is a parameter regime where a peak in correlations is seen at finite separations, indicating the appearance of \textbf{pairing between atoms}. The pairing arises at interparticle distances of 
\EQN{
l_{\rm pair}\approx\frac{\lambda_T}{2}
}
when $\lambda_T$ and $a_{1D}$ are of the same order. It appears to be a consequence of  competition between repulsion, which promotes relative antibunching at short lengths, and the inherent bunching in a thermal field on length scales $\lesssim\lambda_T$. Enhanced pairing at distances $l_{\rm pair}$ is seen to occur both in situations where the point density correlations $\bar{g}^{(0)}$ indicate local bunching ($>1$ --- e.g. Figure~\ref{FIGUREgfig5}) or antibunching ($<1$ --- e.g. Figure~\ref{FIGUREgfig4}). Enhanced three-particle correlations at distances $\approx l_{\rm pair}$ were also seen.

\subsection{Momentum distributions}
\label{CH11CalcMomentum}
\begin{figure}[tb]
\center{\includegraphics[width=\textwidth]{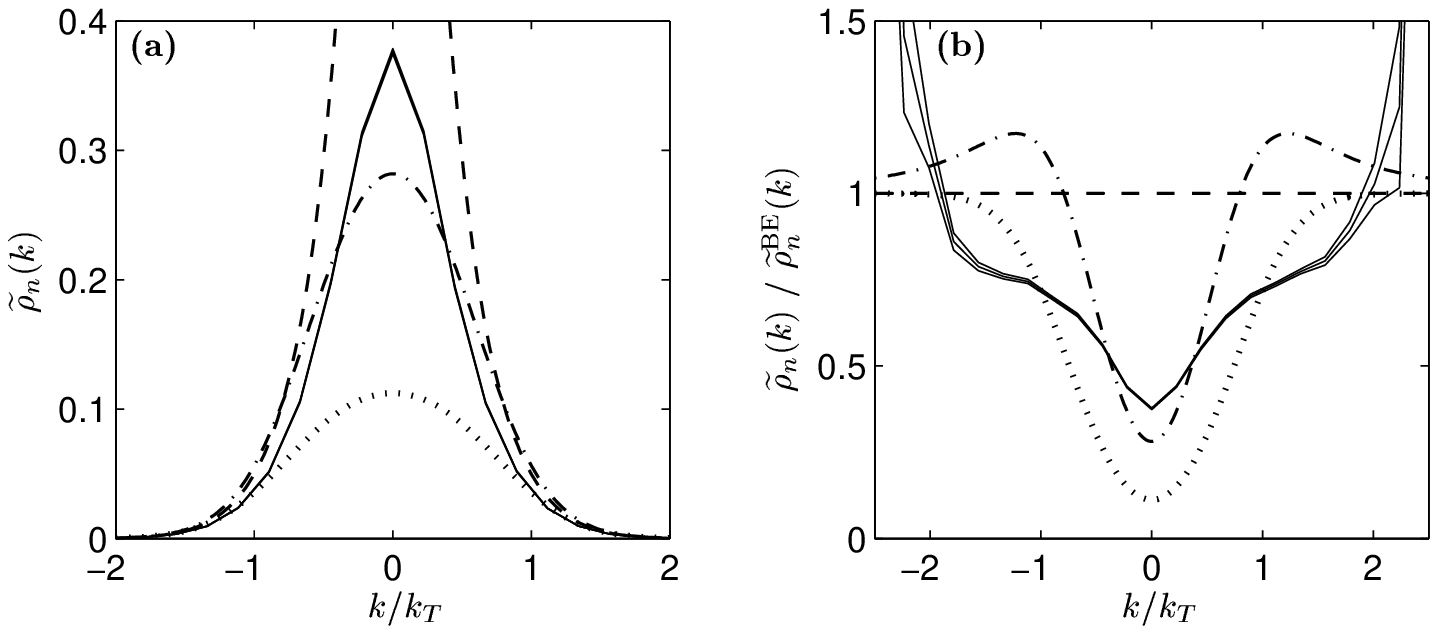}}\vspace{-8pt}\par
\caption[Momentum distribution in a transition gas]{\label{FIGUREnk}\footnotesize
\textbf{Momentum distribution in the $\gamma=1$, $\wt{T}=1.6$ gases.} {\scshape Solid lines} indicate the results (with error bars) of a calculation with parameters as in Figure~\ref{FIGUREgfig5}. Wave vectors $k$ are scaled with respect to the thermal momentum $k_T=\lambda_T/\sqrt{2\pi}$, while the momentum distribution is normalized so that $\int \wt{\rho}_n(k) dk = 1$. Also shown for comparison are the Bose-Einstein ideal gas distribution ({\scshape dashed}), Fermi-Dirac distribution ({\scshape dotted}), and density of states $\rho_k$ from the Yang \& Yang exact solution\cite{YangYang69} ({\scshape dash-dotted}), all calculated at the same $T$, $\mu$, and $g$. These are normalized with the same factor as $\wt{\rho}_n$, rather than to unity. Subplot \textbf{(b)} shows the ratio with the ideal gas distribution $\wt{\rho}_n^{\rm BE}$ to resolve the high $k$ behavior.
\normalsize}
\end{figure}

An example of a calculated momentum distribution in the transition region is shown in Figure~\ref{FIGUREnk}. At low momenta  $k\lesssim k_T=\lambda_T/\sqrt{2\pi}$, the distribution is seen to be intermediate between an the ideal gas Bose-Einstein distribution and the Fermi-Dirac distribution for the same $\mu$ and $T$. One also sees that the density of states function $\rho_k$ from the exact Yang \& Yang solution\cite{YangYang69} differs significantly from the actual distribution of momenta.

\begin{figure}[tb]
\center{\includegraphics[width=200pt]{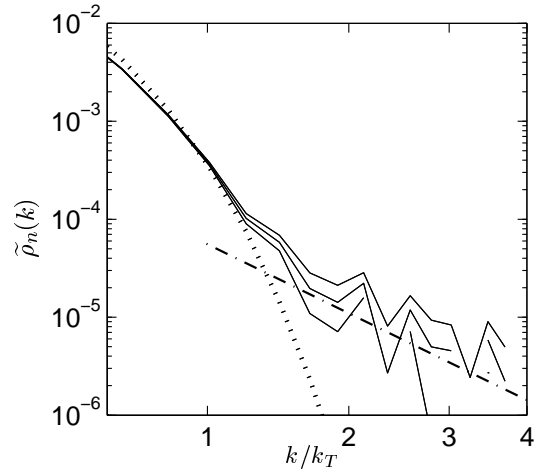}}\vspace{-8pt}\par
\caption[Momentum distribution in the far tails]{\label{FIGUREolsh}\footnotesize
\textbf{Far tails of the momentum distribution in the $\gamma=1$, $\wt{T}=1.6$ gases.} Parameters as in Figure~\ref{FIGUREnk}. {\scshape solid} triple lines indicate calculated normalized momentum distribution with error bars, {\scshape dotted} the Bose-Einstein ideal gas momentum distribution. The {\scshape Dot-dashed} line shows the $T=0$ high $k$ asymptotic behavior\cite{OlshaniiDunjko03}.
\normalsize}
\end{figure}

At intermediate momenta $k_T\lesssim k\lesssim 2k_T$, a depletion in comparison to the ideal gas is seen, while very high momentum particles $k\gtrsim2k_T$ are more common than in the ideal gas. This regime is shown in Figure~\ref{FIGUREolsh}, and one sees that the high momentum distribution approaches the $T=0$ power-law decay $\wt{\rho}(k)\propto k^{-4}$ found by Olshanii and Dunjko\cite{OlshaniiDunjko03} from the exact Lieb and Liniger solution\cite{LiebLiniger63}.

\subsection{Comparison to exact Yang \& Yang solution}
\label{CH11CalcYang}
\begin{figure}[tb]
\center{\includegraphics[width=\textwidth]{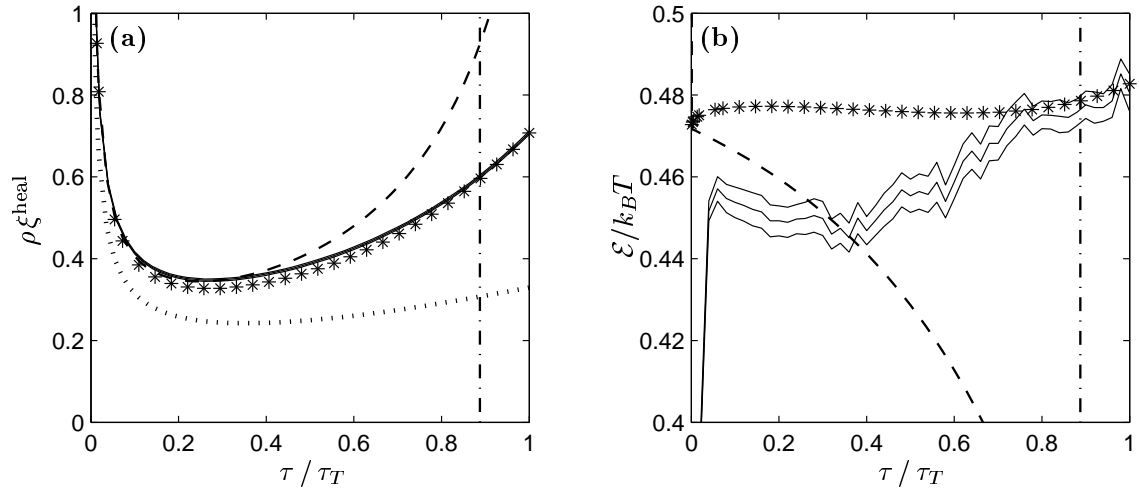}}\vspace{-8pt}\par
\caption[Comparison of density and energy to exact solution]{\label{FIGUREdenseper}\footnotesize
\textbf{Comparison of numerical calculation} ({\scshape solid lines} with error bars) to exact solution ({\scshape asterixes}) for the $\gamma=1$, $\wt{T}=1.6$ calculation with parameters given in Figure~\ref{FIGUREgfig5}. \textbf{(a)}: density $\rho$, \textbf{(b)}: energy per particle $\mc{E}$. Ideal Bose gas ({\scshape dashed}) and Fermi gas values also shown for comparison. The {\scshape dash-dotted} vertical line indicates the lower limit on times $\tau$ for which accuracy is expected with a finite sample on the basis of the $\vari{\re{z_0}}<10$ indicator (See Appendix~\ref{APPA}).  
\normalsize}
\end{figure}

As a test, these calculations have been compared with the density and energy per particle calculated from the Yang \& Yang 
exact solution\cite{YangYang69}. This is shown in Figure~\ref{FIGUREdenseper}. At and around the importance sampling target time $\tau_T$, the agreement is excellent. As expected, at $\tau\ll\tau_T$ the true distribution is badly sampled due to excessive spread in $z_0$. The indicator used to catch this problem is $\vari{\re{z_0}}$ (described in Section~\ref{CH10Simulation}), which should be $\lesssim\order{10}$ to be confident of accuracy. For the case shown in Figure~\ref{FIGUREdenseper}, this indicates accurate calculations at $\tau\ge 0.89\tau_T$, which is quite conservative.
A similar comparison with $\bar{g}^{(2)}(0)$ values calculated\cite{Kheruntsyan-03} from the exact solution also shows agreement.

\chapter{Conclusions}
\label{CH12}

\section{Overview of main results}
\label{CH12Overview}
  As with the thesis, the main results obtained can be divided into three parts: 
\ENUM{

\item \textbf{Theoretical background work regarding general phase-space distributions}. The first aim here has been to 
specify what freedoms are available to the broadly conceived phase-space distribution approach when it is to be used for first-principles simulations of mesoscopic quantum mechanics. The considerations of Chapter~\ref{CH3} led to a set of necessary  requirements for any useful simulation, summarized in Section~\ref{CH3Requirements}. Some fundamental expressions (e.g.  observable estimators) have been given for  general representations (i.e. choices of kernel).  The usefulness of a given representation (which is closely related to basis choice) is highly system dependent, and so it is hoped that the results and considerations of Chapter~\ref{CH3} will aid in matching representations to problems.
  
With a given representation, often there still  remains a wide range of stochastic equations that simulate the same physical problem, but strongly differ in efficiency. The stochastic gauge formalism developed in Chapter~\ref{CH4} describes the freedoms available in a systematic way. At the coarsest level there are two main kinds: Those that arise when making the correspondences from master to Fokker-Planck equations (kernel gauges), and from Fokker-Planck to stochastic equations (diffusion gauges).  Some analysis of these freedoms for general gauge choices has also been made. 
Uses have been found to include 
\ENUM{
\item Improvement of efficiency by tailoring the shape of the distribution of random variables to improve sampling.
\item Removal of biases due to a broad class of so-called boundary term errors, which are due to instabilities or divergences in the stochastic equations.
\item Allowance for calculations of grand canonical ensemble properties by the inclusion of a dynamically  varying (with simulation steps) weight.
}
Part of the freedoms (e.g. real diffusion gauges) have been determined to not be useful.
The stochastic gauge formalism includes in a systematic way several recent developments\cite{Plimak-01,Carusotto-01,CarusottoCastin01,DeuarDrummond01,DeuarDrummond02,DrummondDeuar03,Drummond-04} as special cases. A relatively restricted set of gauges that still allows the uses listed above has been identified in Section~\ref{CH4Central} as the ``standard gauges''.

The causes and symptoms of boundary term errors (sampling biases that do not abate in the infinite sample number limit), have been considered in considerable detail in Chapter~\ref{CH6}. These have been major obstacles for many phase-space simulations in the past. Two kinds of processes have been identified as causing these --- either arising when making the correspondence between master and Fokker-Planck equations (first kind), or when calculating observable estimates using the simulated random variables (second kind). Several warning symptoms of these errors, which can be identified by inspection of the stochastic equations prior to simulation, have been identified.
  
It has been found that appropriate choices of drift stochastic gauges (a class of Kernel gauge) can be used to remove the biases and instabilities in many (or possibly all) cases of boundary term errors of the first kind. 
This is demonstrated in several simple examples where boundary term errors have been known to occur, and also used in some subsequent many-mode calculations where required. Heuristic guidelines for appropriate gauge choice have been found and given in Section~\ref{CH6RemovalHeuristic}.

\item \textbf{Investigation of gauge P representation and development of gauges for use with interacting Bose gases}. This representation is an extension of the successful positive P representation based on coherent states. The added complex weight factor is shown to allow the use of the standard gauges to improve simulations, remove possible boundary term bias, and extend the range of physical models that can be simulated (thermodynamic calculations).
 
The emphasis here has been almost exclusively on simulations of open Bose gases with binary interactions. Equations for dynamics and thermodynamics calculations  have been obtained (including non-delta-function interparticle potentials). 
Part B searched for useful choices of gauges for dynamics and thermodynamics calculations, given the constraint that they be local (i.e. dependent only on variables at a single lattice point). This constraint is not optimal, but is a basic starting point for possible further investigation. Both diffusion and drift gauges that improve simulations under appropriate conditions were found. Diffusion gauges were able to significantly improve sampling and useful simulation times in dynamics calculations, while drift gauges  are essential for any accurate thermodynamic calculation, and were found to also  give sampling improvement in dynamics under some conditions (See Chapter~\ref{CH8}).

Analyses were made of the expected behavior of multi-mode simulations, and subsequently confirmed in the actual  calculations. In dynamics calculations, simulation time is found to be limited  by properties of the most highly occupied lattice point, and a robust estimate of this time for positive P simulations is given by \eqref{ppsimtime}. Relative to this positive P simulation time, it was found that simulation time improvement with diffusion gauges occurs when lattice spacing is greater than the healing length in the gas.  Local drift gauges were found to give even more improvement in simulation time but for a much narrower range of systems in which two-body collision effects dominate kinetic processes, as in some simulations in Chapter~\ref{CH8}. Thermodynamic simulations  were made possible with judicious use of a drift gauge. It was also found that simulation precision at a target temperature and chemical potential was dependent (sometimes strongly)  on the (otherwise free) choice of chemical potential at higher  temperatures.

\item \textbf{Simulations of many-mode interacting Bose gases} Several nontrivial mesoscopic systems of open interacting Bose gases have been simulated. These were dynamics of:
\ITEM{
\item Uniform 1D and 2D  gas with effectively local scattering. Wave behavior occurring in the two- or three-particle correlations (but not density) was found, and displays some interesting properties (e.g. movement of main wavefront at $\sqrt{2}$ the speed of sound).
\item Uniform 1D gas with extended nonlocal interparticle interactions. Correlation wave behavior was also seen, but with different properties than for the locally-interacting gas. Simulation time was found to increase significantly with respect to a locally-interacting gas with the same interaction energy density.
\item Trapped 1D gas with extended interactions on the length scale of the trap. Strong interplay between the scattering and breathing of the atom cloud in the trap was seen. 
\item Scattering of atoms from colliding BECs in 3D. Bosonic enhancement of initially empty modes was seen in a physical situation similar to the Vogels\etal\ four wave mixing experiment\cite{Vogels-02} (the difference was that there were less atoms in the simulation). Additionally, a significant suppression of the spontaneous scattering process compared  to the imaginary scattering length estimate is predicted.
}
Also:
\ITEM{
\item Calculations of grand canonical ensembles were made for the uniform 1D gas at temperature $T$ and chemical potential $\mu$. Physical regimes reached included the nondegenerate strongly interacting fermionized regime, and the quantum degenerate decoherent regime. Spatial correlation functions of first $\bar{g}^{(1)}(x)$, second $\bar{g}^{(s){(x}}$ and  third order $\bar{g}^{(3)}(x,y)$ have been calculated as well as momentum distributions. 
A transition regime where enhanced atom pairing occurs at a preferred interparticle distance has been found. 
This occurs when the de Broglie thermal wavelength $\lambda_T$ is of the same order as the 1D scattering length $a_{\rm 1D}$ (effectively the ``size'' of the atom for scattering processes). The pairing was found to occur at distances $\approx\lambda_T/2$ for the parameters simulated. 
}
}
These simulations can be tractable even with very large numbers of modes or particles in the system given the right conditions, as evidenced by Section~\ref{CH10Scattering}, where there were 1009 152 lattice points and 150 000 atoms on average. 

Situations where there are several length scales of similar order, or processes of similar strength are of particular interest for first-principles simulations because it is  difficult to make accurate quantitative predictions otherwise. The calculations in Sections~\ref{CH10Trap} and ~\ref{CH10Scattering} and the thermodynamics calculations were of this type, and were seen to be amenable to simulation using the gauge P method.
In fact, even in a perturbative regime (e.g. the correlation phenomena at low intensity can probably be treated by Bogoliubov approximation approach), the equations are convenient to use: One simply repeatedly simulates a GP equation with noise  and takes appropriate averages. All quantum effects are included to within statistical precision.  The caveat is  that simulation time is limited. Nevertheless, many phenomena can still be seen, and simulation time can be extended with appropriate gauges under some conditions.

\section{Future directions of improvement}
\label{CH12Future}

The local diffusion gauges developed here give significant improvements only when the lattice is relatively sparse, and/or scattering interactions are dominant. A way to overcome these limitations  might be found by using nonlocal diffusion gauges that would depend on neighboring, or preferably all, lattice amplitudes $\alpha_{\bo{n}}$, $\beta_{\bo{n}}$. A possible starting point would be to consider optimizing diffusion gauges in the opposite regime of strong kinetic interactions and weak scattering. The equations for such a model are probably best considered in Fourier space where kinetic processes take on a simpler form.

Another major issue is the reduction of simulation time with system size due to increasing weight or $z_0$ spread when drift gauges are used. (As was discussed in Section~\ref{CH10LatticeDrift}). A starting point here may be to determine why the lattice-dependent diffusion gauge \eqref{ahodiffusiongaugeM} did not give the expected simulation time improvements \eqref{bignsimtimeM}. If the expected $M^{-1/4}$ scaling could be achieved, many useful mesoscopic simulations should  become accessible with a drift gauge. Such a drift-gauged simulation would  be free of moving singularities, and possibly also extend simulation time  as was seen for the drift gauged one- and two-mode systems.

A completely different but potentially very promising approach is to use non-coherent state kernels for the representation. Certainly in many low temperature regimes this would be a more efficient approach if viable representations could be found because the low temperature ground and low excited states are often far from coherent states. Some preliminary attempts with squeezed state kernels give mixed results\cite{squeezed}.

For thermodynamics calculations of properties at and around a target temperature and chemical potential, an important outstanding issue is to characterize the influence of the choice of $\mu(T)$ at intermediate temperature values. It is conceivable that a good choice of $\mu(T)$ may give dramatic improvement in precision and significantly extend the physical region that can be simulated. Some precision and simulation time improvement should also be obtained by using a more sophisticated importance sampling technique than the rough method used in Chapter~\ref{CH11}. Some preliminary results indicate that using a Metropolis sampling algorithm\cite{Metropolis-53} extends the reachable $\gamma$, $\wt{T}$ regime\cite{metropolis}.


%
%


\bibliographystyle{unsrt}\rm	
\bibliography{phd}
\addcontentsline{toc}{chapter}{Bibliography}


\begin{appendix}
\chapter{Exponentials of Gaussian random variables}
\label{APPA}
  Gauge P simulations of interacting Bose gases (whether dynamic or thermodynamic) involve multiplicative noise terms of the form 
\EQN{\label{czdw}
dz = czdW(t) + \dots
,}
where $z$ is a complex variable ($\alpha_{\bo{n}}$, $\beta_{\bo{n}}$, $z_0$), and $c$ a constant. This leads to $\log z$ being distributed approximately as a Gaussian. This is exact at short times, but some modification of the distribution occur also as a consequence of the other ``$\dots$'' drift (or noise) terms in the equations. Observable estimates are usually obtained through averages of non-logarithmic combinations of $z$, however, and this requires some care. Let us consider the idealized situation of $z\approx v_{\sigma}=v_0e^{\sigma\xi}=e^{v_L}$, where $\xi$ is a Gaussian random variable of mean zero, variance unity. The notation of Section~\ref{CH7Gaussian} will be used. 

Using the distribution of $\xi$, \eqref{gausdist}, one obtains 
\EQN{
\text{Pr}(v_{\sigma}) = \frac{1}{\sigma v_{\sigma}\sqrt{2\pi}}\exp\left\{-\frac{(\log[v_{\sigma}/v_0])^2}{2\sigma^2}\right\}
.}
This distribution falls off slowly as $v_{\sigma}\to\infty$, and for a finite sample  may not be sampled correctly if $\xi$ is chosen according to its Gaussian distribution. 
With $\mc{S}$ samples, the greatest value of $\xi$ obtained can be expected to be $\xi_{\rm max}$, the solution of 
$\half[1-\text{erf}(\xi_{\rm max}/\sqrt{2})]\approx2/\mc{S}$, where $\text{erf}(x)=2\int_0^xe^{-t^2}dt$ is the error function. For $\mc{S}\approx\order{1000}\text{ to }\order{10^5}$, this gives $\xi_{\rm max}\approx 3\text{ to }4$. 
Some bias in $\average{v_{\sigma}}$ (underestimation) may be expected due to the lack of sampling of the far tails  once 
\EQN{
\int^{\infty}_{v_0\exp(\sigma\xi_{\rm max})} \text{Pr}(v_{\sigma})\ dv_{\sigma} \gg 1/\mc{S}
.} 
For typical sample sizes $\mc{S}\approx\order{1000}\text{ to }\order{10^5}$ this will usually occur around $\sigma^2\gtrsim10$. This is actually the same large $\sigma$ region given by \eqref{sdlimit} where all precision is lost anyway because of excessive distribution spread (as shown in Section~\ref{CH7Gaussian}). 

\begin{figure}[tb]
\center{\includegraphics[width=\textwidth]{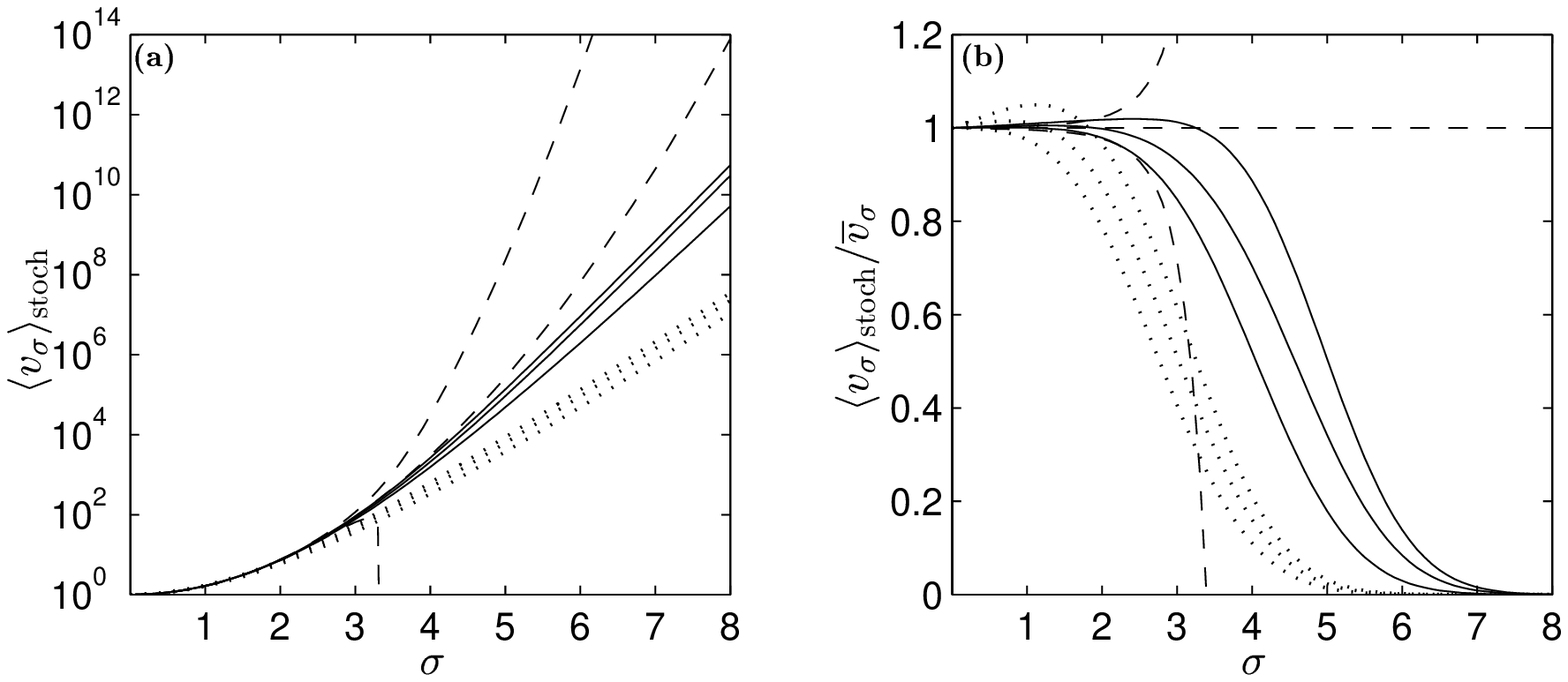}}\vspace{-8pt}\par
\caption[Finite sample estimates with multiplicative noise]{\label{FIGUREgausbias}\footnotesize
\textbf{Finite sample estimates} of means of exponentials of gaussian random variables $v_{\sigma}=e^{\sigma\xi}$. 
{\scshape solid} lines: $\mc{S}=10^5$ sample estimate of $\average{v_{\sigma}}$ and error bars, where $\xi$ was sampled from a Gaussian distribution.
{\scshape dotted} lines: Same but using $\mc{S}=1000$ samples.
{\scshape dashed} lines: Exact ($\mc{S}\to\infty$) value of the mean $\langle v_{\sigma}\rangle\to\bar{v}_{\sigma}$, along with error estimates for the $\mc{S}=10^5$ calculation based on the exact value of $\vari{v_{\sigma}}$, from \eqref{mexp} and \eqref{varexp}. This corresponds to what would be expected from an unbiased sample of $v_{\sigma}$.
All error bars were calculated according to \eqref{deltabarvsigma}.
Subplot \textbf{(b)} shows the same values, but scaled with respect to the $\mc{S}\to\infty$ mean $\bar{v}_{\sigma}$.
Values at all $\sigma$ were made with the same sample of $\xi$ to show the direct dependence on $\sigma$ only.
\normalsize}
\end{figure}

However, the problem is that \textit{both} the mean $\average{v_{\sigma}}$ and the CLT precision estimate $\sqrt{\vari{v_{\sigma}}/\mc{S}}$ are underestimated when $\sigma$ is large, but the error estimate is underestimated \textit{by an even greater amount} than the mean. As a result, badly sampled data may still appear to be significant for large $\sigma$ values. The situation is shown in Figure~\ref{FIGUREgausbias}.

The simplest practical solution is to simply discard any calculated means $\average{z}$ when
\EQN{\label{varilimit}
\vari{\log|z|} \gtrsim 10
.}
If an unbiased mean $\average{z}$ and precision estimate $\Delta z$ was obtained by some more sophisticated method, then it would not be significant anyway since one expects $\Delta z \gg \average{z}$ at these large $\sigma$ values.
The main point here is that quantities \eqref{varilimit} should be monitored when dealing with multiplicative noise of the form \eqref{czdw} in the stochastic equations.

Some subtleties can arise because in realistic simulations $\log z$ is not exactly Gaussian. If the large $\log z$ distribution tails fall off more rapidly than Gaussian, then an unbiased simulation can be obtained for larger $\vari{\log|z|}$, while the converse is true if these tails fall of less rapidly.  A \textbf{more robust} indicator of possible bias than \eqref{varilimit} is to compare the observable estimate obtained with two sample sizes $\mc{S}$, differing by at least an order of magnitude. If sampling bias is present, the two estimates of the average $\average{z}$ will usually differ by a statistically significant amount. An example of this for the idealized $v_{\sigma}$ model is shown in Figure~\ref{FIGUREgausbias}. This kind of sample-size-dependent behavior is always a strong warning sign.

\chapter{Some details of Stochastic Calculus}
\label{APPB}

Some results regarding random terms in differential equations relevant to the thesis are gathered here. Proofs of  these can be found in Gardiner\cite{Gardiner83}, with details on computer algorithms in \cite{DrummondMortimer91}.

In many cases, a set of stochastic differential equations (i.e. differential equations with random terms) can be written in the Langevin form 
\EQN{\label{langevinform}
dx_j(t)  =  A_j(\bo{x},t)\,dt + \sum_k B_{jk}(\bo{x},t)\,dW_k(t)
,}
where $\bo{x}$ contains all variables $x_j$. The $dW_j(t)$ are Wiener increments and \eqref{langevinform} is to be interpreted according to the Ito calculus (see below). 

A Wiener increment is defined in terms of the Wiener process $W(t)$, which is the special case of \eqref{langevinform} with $j=1$, $A_1=0$ and $B_{1k}=\delta_{k1}$. The probability distribution of $W$ is governed by the Fokker-Planck equation (FPE)
\EQN{
\dada{P_W(W,t)}{t} = \frac{\partial^2P_W(W,t)}{\partial W^2}
,}
a special case of Brownian motion. The individual realizations of $W$ are continuous but not differentiable. The (infinitesimal) Wiener increment can, however, be defined, and is
\EQN{
  dW(t) = W(t+dt)-W(t)
,}
with $dt$ infinitesimal. This random quantity has the expectation values
\SEQN{\label{wiener}}{
\average{dW(t)} &=& 0\\
\average{dW(t)^2} &=& dt\\
\average{dW(t)dW(t')} &=& 0 \text{\qquad if $t\neq t'$}\\
\average{\prod_{j=1}^{\text{max}[j]>2} dW(t_j)} &=& 0
}
The Wiener increment is related to processes with white noise correlations such that if $\average{\xi(t)\xi(t')}=\delta(t-t')$ then one can write $\xi(t)=dW(t)/dt$. In \eqref{langevinform}, the Wiener increments are independent:
$\average{dW_j dW_{k\neq j}}=\average{dW_j}\average{dW_k}$ etc.
In a numerical calculation, the Wiener increment is usually implemented as independent Gaussian random variables $\Delta W_j$ at each time step of length $\Delta t$ with mean zero and variance \
\EQN{\label{deltaWjk}
\average{\Delta W_j \Delta W_k} = \Delta t\delta_{jk}
,}
 although other choices of the distribution of $\Delta W_j$ are possible provided only that the discrete step analogues of \eqref{wiener} are satisfied, as in \eqref{deltaWjk}.

An equation \eqref{langevinform} in the Ito calculus is equivalent to the Stratonovich calculus equation
\EQN{
dx_j(t)  &=&  A_j(\bo{x},t)\,dt + \sum_k B_{jk}(\bo{x},t)\,dW_k(t) + S_j(\bo{x},t)\nonumber\\
&=& dx_j^{\rm Ito}(t) + S_j(\bo{x},t)
,}
where the \textit{Stratonovich correction} is 
\EQN{\label{stratcorrection}
S_j(\bo{x},t) = -\half\sum_{kl} B_{lk}(\bo{x},t)\dada{B_{jk}(\bo{x},t)}{x_l} 
.} 
These two forms arise from different ways of defining the integral of the differential equations, both useful.
For practical purposes, the main differences are that:
\begin{itemize}
\item In the Ito calculus, the time-dependent variables $x_j(t)$ are independent of the same-time Wiener increments $W_k(t)$, while this is not the case in the Stratonovich calculus.  
\item In the Stratonovich calculus, the $x_j$ obey the usual rules of deterministic calculus, in particular the chain rule. In the Ito calculus one instead has
\EQN{
  dy(\bo{x}) = \sum_j  dx_j(\bo{x},t) \dada{y(\bo{x})}{x_j} + \frac{dt}{2}\sum_{jkl}B_{jk}(\bo{x},t)B_{lk}(\bo{x},t)\frac{\partial^2y(\bo{x})}{\partial x_j\partial x_l} 
.}
\end{itemize}
Because of the first point above, the Ito calculus corresponds to a discrete step integration algorithm with the derivative approximated using values at the beginning of the time interval:
\EQN{\label{zerokappa}
x_j(t+\Delta t) = x_j(t) + A_j(\bo{x}(t),t)\,\Delta t + \sum_kB_{jk}(\bo{x}(t),t)\,\Delta W_k(t)
.}
Implicit integration algorithms estimate the derivative using quantities evaluated at later times\footnote{A common algorithm for implicitly estimating the $\bo{x}(t_{\kappa})$ is to iterate \eqref{kappaeqn} several times using the appropriate smaller timestep $\Delta t\to\kappa\Delta t$.}
\begin{subequations}\label{tkappa}\EQN{
t_{\kappa}=t+\kappa\Delta t
}
 (where $\kappa\in(0,1]$) during the timestep.  That is, 
\EQN{\label{kappaeqn}
x_j(t+\Delta t) = x_j(t) + A_j(\bo{x}(t_{\kappa}),t_{\kappa})\,\Delta t + \sum_kB_{jk}(\bo{x}(t_{\kappa}),t_{\kappa})\,\Delta W_k(t)
.}\end{subequations}
In this case, one must use a Stratonovich (or Stratonovich-like) form of the drift:
\EQN{
dx_j(t)  &=& dx_j^{\rm Ito}(t) + 2\kappa S_j(\bo{x},t)
}
to ensure that \eqref{kappaeqn} is the same as \eqref{zerokappa} up to lowest deterministic order $\Delta t$. This follows by use of \eqref{deltaWjk} on \eqref{kappaeqn}.
For multiplicative noise it has been found that using a semi-implicit method ($\kappa=\half$) integration method gives superior numerical stability\cite{DrummondMortimer91}.

\chapter{Quotients of means or random variables}
\label{APPC}
   In the gauge P representation, estimates of the expectation value of an observable $\op{O}$  are made by evaluating the expression \eqref{observables} for a finite ensemble of $\mc{S}$ samples. This has the form of a quotient
\EQN{\label{quotient}
\langle\op{O}\rangle \approx \bar{O}=\frac{\average{f}}{\average{\re{\Omega}}}=\frac{\average{f_N}}{\average{f_D}}
.}
Error estimates are made by subensemble averaging as explained in Section~\ref{CH3StochasticAccuracy}. While for dynamics calculations, one is assured of $\average{\re{\Omega}}=1$ in the limit $\mc{S}\to\infty$ at all times, so that the denominator can be ignored, thermodynamics calculations require the evaluation of both denominator $\average{f_D}$ and numerator in \eqref{quotient}. In particular, the  error estimate \eqref{subensembleuncertainty} using subensembles requires some care. 

\begin{figure}[tb]
\center{\includegraphics[width=350pt]{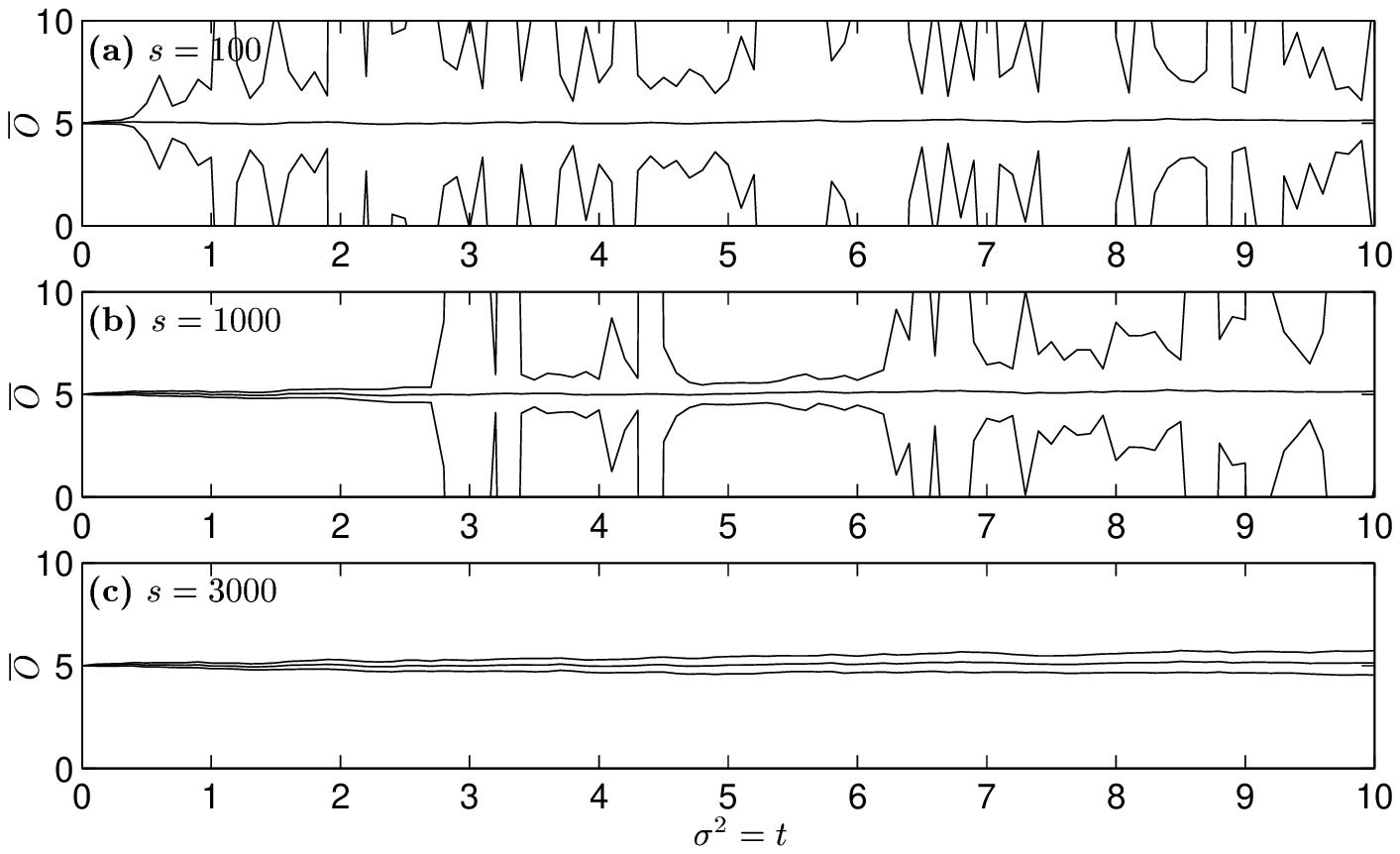}}\vspace{-8pt}\par
\caption[Error estimates with several subensemble sizes]{\label{FIGUREsens}\footnotesize
\textbf{Error estimates with different subensemble sizes $s$}. {\scshape solid} lines show the means and error estimates obtained using \eqref{subensembleuncertainty} for the quantity $\bar{O}=1/\left(\average{v}+0.2\right)$, where $v(t)$ obeys the Brownian motion stochastic equation $dv=dW(t)$ ($dW(t)$ is a Wiener increment). The standard deviation of $v$ is $\sigma=\sqrt{t}$. 
In each simulation there were $\mc{S}=10^5$ trajectories in all. Expression \eqref{slimit} suggests $s\gtrsim 2250$ is required at $t=10$.
\normalsize}
\end{figure}

The problem is that if there are not enough samples $s$ in (say) the $j$th  subensemble, then the $j$th denominator average may have the wrong sign or be close to zero. The result is that this $j$th subensemble average $\bar{O}^{(j)}$ takes on very large or non-realistic values, which are nowhere near $\langle\op{O}\rangle$. Since their absolute values can be very large if the denominator was close to zero, these outlier $\bar{O}^{(j)}$ strongly influence the final error estimate. 
An example of such excessive error estimates is shown in Figure~\ref{FIGUREsens}, and typically goes in hand with spiking in the uncertainty $\Delta\bar{O}$.

Roughly, one expects that if there are $\mc{S}_E\approx 1000$ subensembles, then the farthermost outlier will be at about $3\sigma$ from the mean. Thus, excessive uncertainty is likely to occur once  $3\sqrt{\vari{\bar{f}_D^{(j)}}}\gtrsim\average{f_D}$, where $\bar{f}_D^{(j)}$ is the denominator average for the $j$th subensemble.  For subensembles with $s$ elements, this gives the rough limit
\EQN{\label{slimit}
s \gtrsim 9\frac{\vari{f_D}}{\average{f_D}^2}
.}
Often the variation in the weight  $\Omega=e^{z_0}$ is mainly due to a close-to-gaussian distributed $\re{z_0}$. In this case, using \eqref{mexp} and \eqref{varexp},  the requirement \eqref{slimit} becomes 
\EQN{
s\gtrsim 9\left(e^{\vari{\re{z_0}}}-1\right)
.}


\end{appendix}
\end{document}